%% file: main.tex
\documentclass[12pt]{book}
\usepackage{psfrag}
\usepackage{epsfig}
\usepackage{amsmath}
\usepackage{mathrsfs}
\usepackage{amssymb}
\usepackage{epsf}
\usepackage{graphicx}
\usepackage{epstopdf}
\usepackage{indentfirst}
\usepackage{subfigure}
\usepackage{multirow}
\usepackage{color}
\usepackage{rotating}
\usepackage{tabulary}                                                           
\usepackage{hyperref}                                                         
\usepackage[numbers,sort,compress]{natbib} 
\hypersetup{
    colorlinks=true,
    breaklinks,
    citecolor=blue,
    frenchlinks=true,
    linkcolor=blue
}

\RequirePackage{lineno}
\include{mypagesty}

\input{define/newcmd.tex}

\input{define/definition.tex}

\begin{document}

%\setpagewiselinenumbers
\modulolinenumbers[2]
%\linenumbers

\input{define/prowds.tex}
\input{define/prolist.tex}

\setcounter{page}{0}
\pagenumbering{arabic}
\setcounter{chapter}{0}
\setcounter{part}{0}

%% Haibo LI and Xiao-Rui LYU for the BESIII white paper 
%
%%%
\input{Introduction/introduction.tex}
\input{Lighthadron/lighthadron.tex}
\input{Charmonium/charmonium.tex}
\input{QCD/qcd.tex}
\input{Charm/charm.tex}
\input{New_physics/new_physics.tex}
\input{Sum/sum.tex}
\clearpage
\input{define/Acknowledgements.tex}
\end{document}

%% file: define/mypagesty.tex
\usepackage{fancyheadings}

%% file: define/newcmd.tex
\newcommand{\BR}{{\cal B}}

\newcommand{\pip}{\pi^+}
\newcommand{\pim}{\pi^-}

\newcommand{\kl}{K_L^0}

\newcommand{\etap}{\eta^{\prime}}

\newcommand{\psip}{\psi(3686)}

\newcommand{\jpsi}{J/\psi}

\newcommand{\EE}{e^+e^-}

\newcommand{\kk}{K^+K^-}

\newcommand{\ppjpsi}{\pi^+\pi^- J/\psi}

\newcommand{\ra}{\rightarrow}

\newcommand{\beq}{\begin{equation}}
\newcommand{\eeq}{\end{equation}}
\newcommand{\beqn}{\begin{eqnarray}}
\newcommand{\eeqn}{\end{eqnarray}}
\newcommand{\beqns}{\begin{eqnarray*}}
\newcommand{\eeqns}{\end{eqnarray*}}
\newcommand{\bfg}{\begin{figure}}
\newcommand{\efg}{\end{figure}}
\newcommand{\bitm}{\begin{itemize}}
\newcommand{\eitm}{\end{itemize}}
\newcommand{\bnum}{\begin{enumerate}}
\newcommand{\enum}{\end{enumerate}}
\newcommand{\btbl}{\begin{table}}
\newcommand{\etbl}{\end{table}}
\newcommand{\btbu}{\begin{tabular}}
\newcommand{\etbu}{\end{tabular}}
\newcommand{\egeg} {{\it e.g.}}
\newcommand{\ieie} {{\it i.e.}}

\newcommand{\eelclc}{e^{+}e^{-} \rightarrow \Lambda_{c}^{+} \bar{\Lambda}_{c}^{-}}

%% file: define/definition.tex
\long\def\singleMnfA#1#2#3#4{
\begin{figure}[p]
   \epsfxsize10cm
   \centerline{\epsffile{figures/#1}}
\vspace{-0.5cm}
   \caption[.]{\label{fig:#2}#3}
\vspace{-0.1cm}
}
\long\def\singleMnfB#1#2#3#4{
   \epsfxsize10cm
   \centerline{\epsffile{figures/#1}}
\vspace{-0.5cm}
   \caption[.]{\label{fig:#2}#3}
\end{figure}
}
\long\def\inst#1{\par\nobreak\kern 4pt\nobreak
    {\itshape #1}\par\vskip 10pt plus 3pt minus 3pt}
\RequirePackage{xspace}
\def\bes3{BESIII}

\def\Kbar    {\kern 0.18em\bar{\kern -0.18em K}{}\xspace}

\def\Kz      {\ensuremath{K^0}\xspace}
\def\Kzb     {\ensuremath{\Kbar^0}\xspace}
\def\KzKzb   {\ensuremath{\Kz {\kern -0.16em \Kzb}}\xspace}

\def\Ks     {\ensuremath{K_S}\xspace}
\def\Kl     {\ensuremath{K_L}\xspace}
\def\KsKs   {\ensuremath{\Ks {\kern -0.16em \Ks}}\xspace}
\def\KlKl   {\ensuremath{\Kl {\kern -0.16em \Kl}}\xspace}
\def\KsKl   {\ensuremath{\Ks {\kern -0.16em \Kl}}\xspace}
\def\KlKs   {\ensuremath{\Kl {\kern -0.16em \Ks}}\xspace}
\def\Dbar    {\kern 0.18em\bar{\kern -0.18em D}{}\xspace}

\def\Dz      {\ensuremath{D^0}\xspace}
\def\Dzb     {\ensuremath{\Dbar^0}\xspace}
\def\DzDzb   {\ensuremath{\Dz {\kern -0.16em \Dzb}}\xspace}
\def\DsP      {\ensuremath{D_S^+}\xspace}
\def\DsM     {\ensuremath{D_S^-}\xspace}
\def\DspDsm    {\ensuremath{\DsP {\kern -0.16em \DsM}}\xspace}
\def\Dp      {\ensuremath{D^+}\xspace}
\def\Dm     {\ensuremath{D^-}\xspace}
\def\DpBm    {\ensuremath{\Dp {\kern -0.16em \Dm}}\xspace}
\def\Bbar    {\kern 0.18em\bar{\kern -0.18em B}{}\xspace}

\def\Bz      {\ensuremath{B^0}\xspace}
\def\Bzb     {\ensuremath{\Bbar^0}\xspace}
\def\BzBzb   {\ensuremath{\Bz {\kern -0.16em \Bzb}}\xspace}
\def\Bu      {\ensuremath{B^+}\xspace}
\def\Bub     {\ensuremath{B^-}\xspace}

\def\BpBm    {\ensuremath{\Bu {\kern -0.16em \Bub}}\xspace}

\newcommand{\optbar}[1]{\shortstack{{\tiny (\rule[.4ex]{1em}{.1mm})}
  \\ [-.7ex] $#1$}}
\def\BorBbar    {\kern 0.18em\optbar{\kern -0.18em B}{}\xspace}
\def\DorDbar    {\kern 0.18em\optbar{\kern -0.18em D}{}\xspace}
\def\KorKbar    {\kern 0.18em\optbar{\kern -0.18em K}{}\xspace}
\def\cp                {\ensuremath{C\!P}\xspace}

\def\bepc2{BEPCII}
\mathchardef\Upsilon="7107
\def\Y#1S{\ensuremath{\Upsilon{(#1S)}}\xspace}% no space before {...}!

\def\st{\scriptstyle}
\def\sst{\scriptscriptstyle}
\def\mco{\multicolumn}
\def\epp{\epsilon^{\prime}}
\def\vep{\varepsilon}
\def\ra{\rightarrow}
\def\ppg{\pi^+\pi^-\gamma}
\def\vp{{\bf p}}
\def\ko{K^0}
\def\kb{\bar{K^0}}
\def\al{\alpha}
\def\ab{\bar{\alpha}}
\def\be{\begin{equation}}
\def\ee{\end{equation}}
\def\bea{\begin{eqnarray}}
\def\eea{\end{eqnarray}}

%% file: define/prowds.tex
\renewcommand{\thefootnote}{\fnsymbol{footnote}}
\setcounter{footnote}{4}
\title{\bf \Huge Future Physics Programme of \bes3}
%\vspace{2cm}
%\author{{\bf \large The \bes3 collaboration}\footnote{Author list shown on the following pages}\vspace{0.2cm} \\
\author{
{\large IHEP-Physics-Report-\bes3-\number\year-\number\month-\number\day} \\
%IHEP-Physics-Report-\bes3-2019-12-13 \\
\\
Published in Chinese Physics C {\bf 44},  040001 (2020)
}
\date{}
\maketitle

\input{define/authors.tex}

\newpage 

\begin{center} \Large{\bf Abstract } \end{center}
\addcontentsline{toc}{chapter}{Abstract}
There has recently been a dramatic renewal of interest in the subjects of hadron spectroscopy and charm physics.
This renaissance has been driven in part by the discovery of 
a plethora of
charmonium-like $XYZ$ states at \bes3 and $B$ factories,
and the observation of an intriguing proton-antiproton threshold
enhancement and the possibly related $X(1835)$ meson state at
\bes3, as well as the threshold measurements of charm mesons and charm baryons.    

We present a detailed survey of the important topics in tau-charm physics and hadron physics that can be further explored at BESIII over the remaining lifetime of BEPCII operation. 
This survey will help in the optimization of the data-taking plan over the coming years,  and provides physics motivation for the possible upgrade of BEPCII to higher luminosity.

\cleardoublepage
%\setcounter{equation}{0}
%\addchap[Abstract]{ }
%%%%%%%%%%%new page%%%%%%%%%%%%%%%
%\newpage
%\centerline{\bf \Large The \bes3 collaboration}
%\vspace{1cm}
%\input{define/authors}
%
%%%%%%%%%%%new page%%%%%%%%%%%%%%%
\newpage

\begin{center}  
~~
\end{center}  

\vspace{3.0cm}

%\begin{flushleft}
%{{\Large \bf Editor:} \large  ~~~Hai-Bo Li,~~~Xiao-Rui Lyu} \\
%\end{flushleft}
%
\vspace{1.0cm}

\begin{center}  
{\Large \bf Working Group and Conveners}
\end{center}

\vspace{1.0cm}

\noindent {\large \bf Part One: Introduction}\\
 \hspace*{3cm} Conveners:  Mingyi Dong, Hai-Bo Li,\\  
\hspace*{5.2cm} Shengsen Sun, Ulrich Wiedner \\
\noindent {\large \bf Part Two: Light Hadron Physics }\\ 
 \hspace*{3cm} Conveners: Shuangshi Fang, Beijiang Liu,  Marc Pelizaeus, \\
\noindent {\large \bf Part Three:  Charmonium Physics}\\  
 \hspace*{3cm} Conveners: Wolfgang Kuehn, Ryan Mitchell, \\
 \hspace*{5.2cm}  Changzheng Yuan, Kai Zhu\\
\noindent {\large \bf Part Four: R values, QCD and $\tau$ Physics} \\ 
 \hspace*{3cm} Conveners: Achim Denig, Rinaldo Baldini Ferroli, Xiaohu Mo,\\
\hspace*{5.2cm} Christoph Florian Redmer, Karin Schoenning,\\
\hspace*{5.2cm} Wenbiao Yan, Jianyong Zhang\\
\noindent {\large \bf Part Five: Charm Physics}  \\ 
 \hspace*{3cm} Conveners: Hai-Bo Li, Jim Libby, Xiao-Rui Lyu, Hailong Ma,\\
\hspace*{5.2cm}  Hajime Muramatsu,  Karin Schoenning\\
\noindent {\large \bf Part Six: Exotic Decays and New Physics} \\ 
 \hspace*{3cm} Conveners: Shenjian Chen, Alexey Petrov, Dayong Wang,\\
\noindent {\large \bf Part Seven: Summary} \\ 
\hspace*{3cm} Conveners:  Hai-Bo Li, Xiao-Rui Lyu, Xinchou Lou\\
\cleardoublepage

%% file: define/authors.tex
{\centering
M.~Ablikim$^{1,\S}$, M.~N.~Achasov$^{10,d,\S}$, P.~Adlarson$^{58,\S}$, S. ~Ahmed$^{15,\S}$, M.~Albrecht$^{4,\S}$, M.~Alekseev$^{57A,57C,\S}$, A.~Amoroso$^{57A,57C,\S}$, F.~F.~An$^{1,\S}$, Q.~An$^{54,42,\S}$, Y.~Bai$^{41,\S}$
% ...
\vspace{0.5cm}\\
{\it
$^{1}$ Institute of High Energy Physics, Beijing 100049, People's Republic of China\\
$^{2}$ Beihang University, Beijing 100191, People's Republic of China\\
$^{3}$ Beijing Institute of Petrochemical Technology, Beijing 102617, People's Republic of China\\
$^{4}$ Bochum Ruhr-University, D-44780 Bochum, Germany\\
$^{5}$ Carnegie Mellon University, Pittsburgh, Pennsylvania 15213, USA\\
$^{6}$ Central China Normal University, Wuhan 430079, People's Republic of China\\
$^{7}$ China Center of Advanced Science and Technology, Beijing 100190, People's Republic of China\\
$^{8}$ COMSATS University Islamabad, Lahore Campus, Defence Road, Off Raiwind Road, 54000 Lahore, Pakistan\\
$^{9}$ Fudan University, Shanghai 200443, People's Republic of China\\
$^{10}$ G.I. Budker Institute of Nuclear Physics SB RAS (BINP), Novosibirsk 630090, Russia\\
$^{11}$ GSI Helmholtzcentre for Heavy Ion Research GmbH, D-64291 Darmstadt, Germany\\
$^{12}$ Guangxi Normal University, Guilin 541004, People's Republic of China\\
$^{13}$ Guangxi University, Nanning 530004, People's Republic of China\\
$^{14}$ Hangzhou Normal University, Hangzhou 310036, People's Republic of China\\
$^{15}$ Helmholtz Institute Mainz, Johann-Joachim-Becher-Weg 45, D-55099 Mainz, Germany\\
$^{16}$ Henan Normal University, Xinxiang 453007, People's Republic of China\\
$^{17}$ Henan University of Science and Technology, Luoyang 471003, People's Republic of China\\
$^{18}$ Huangshan College, Huangshan 245000, People's Republic of China\\
$^{19}$ Hunan Normal University, Changsha 410081, People's Republic of China\\
$^{20}$ Hunan University, Changsha 410082, People's Republic of China\\
$^{21}$ Indian Institute of Technology Madras, Chennai 600036, India\\
$^{22}$ Indiana University, Bloomington, Indiana 47405, USA\\
$^{23}$ (A)INFN Laboratori Nazionali di Frascati, I-00044, Frascati, Italy; (B)INFN and University of Perugia, I-06100, Perugia, Italy\\
$^{24}$ (A)INFN Sezione di Ferrara, I-44122, Ferrara, Italy; (B)University of Ferrara, I-44122, Ferrara, Italy\\
$^{25}$ Institute of Physics and Technology, Peace Ave. 54B, Ulaanbaatar 13330, Mongolia\\
$^{26}$ Johannes Gutenberg University of Mainz, Johann-Joachim-Becher-Weg 45, D-55099 Mainz, Germany\\
$^{27}$ Joint Institute for Nuclear Research, 141980 Dubna, Moscow region, Russia\\
$^{28}$ Justus-Liebig-Universitaet Giessen, II. Physikalisches Institut, Heinrich-Buff-Ring 16, D-35392 Giessen, Germany\\
$^{29}$ KVI-CART, University of Groningen, NL-9747 AA Groningen, The Netherlands\\
$^{30}$ Lanzhou University, Lanzhou 730000, People's Republic of China\\
$^{31}$ Liaoning University, Shenyang 110036, People's Republic of China\\
$^{32}$ Nanjing Normal University, Nanjing 210023, People's Republic of China\\
$^{33}$ Nanjing University, Nanjing 210093, People's Republic of China\\
$^{34}$ Nankai University, Tianjin 300071, People's Republic of China\\
$^{35}$ Peking University, Beijing 100871, People's Republic of China\\
$^{36}$ Shandong University, Jinan 250100, People's Republic of China\\
$^{37}$ Shanghai Jiao Tong University, Shanghai 200240, People's Republic of China\\
$^{38}$ Shanxi University, Taiyuan 030006, People's Republic of China\\
$^{39}$ Sichuan University, Chengdu 610064, People's Republic of China\\
$^{40}$ Soochow University, Suzhou 215006, People's Republic of China\\
$^{41}$ Southeast University, Nanjing 211100, People's Republic of China\\
$^{42}$ State Key Laboratory of Particle Detection and Electronics, Beijing 100049, Hefei 230026, People's Republic of China\\
$^{43}$ Sun Yat-Sen University, Guangzhou 510275, People's Republic of China\\
$^{44}$ Tsinghua University, Beijing 100084, People's Republic of China\\
$^{45}$ (A)Ankara University, 06100 Tandogan, Ankara, Turkey; (B)Istanbul Bilgi University, 34060 Eyup, Istanbul, Turkey; (C)Uludag University, 16059 Bursa, Turkey; (D)Near East University, Nicosia, North Cyprus, Mersin 10, Turkey\\
$^{46}$ University of Chinese Academy of Sciences, Beijing 100049, People's Republic of China\\
$^{47}$ University of Hawaii, Honolulu, Hawaii 96822, USA\\
$^{48}$ University of Jinan, Jinan 250022, People's Republic of China\\
$^{49}$ University of Manchester, Oxford Road, Manchester, M13 9PL, United Kingdom\\
$^{50}$ University of Minnesota, Minneapolis, Minnesota 55455, USA\\
$^{51}$ University of Muenster, Wilhelm-Klemm-Str. 9, 48149 Muenster, Germany\\
$^{52}$ University of Oxford, Keble Rd, Oxford, UK OX13RH\\
$^{53}$ University of Science and Technology Liaoning, Anshan 114051, People's Republic of China\\
$^{54}$ University of Science and Technology of China, Hefei 230026, People's Republic of China\\
$^{55}$ University of South China, Hengyang 421001, People's Republic of China\\
$^{56}$ University of the Punjab, Lahore-54590, Pakistan\\
$^{57}$ (A)University of Turin, I-10125, Turin, Italy; (B)University of Eastern Piedmont, I-15121, Alessandria, Italy; (C)INFN, I-10125, Turin, Italy\\
$^{58}$ Uppsala University, Box 516, SE-75120 Uppsala, Sweden\\
$^{59}$ Wuhan University, Wuhan 430072, People's Republic of China\\
$^{60}$ Xinyang Normal University, Xinyang 464000, People's Republic of China\\
$^{61}$ Zhejiang University, Hangzhou 310027, People's Republic of China\\
$^{62}$ Zhengzhou University, Zhengzhou 450001, People's Republic of China\\
\vspace{0.2cm}
$^{I}$Aix-Marseille Univ, Universit\'e de Toulon, CNRS, CPT, Marseille, France\\
$^{II}$ Budker Institute of Nuclear Physics, SB RAS, Novosibirsk, 630090, Russia \\
$^{III}$ Laboratoire de l'Acc\'{e}l\'{e}rateur Lin\'{e}aire, IN2P3-CNRS et Universit\'{e} Paris-Sud 11, F-91898, Orsay Cedex, France\\
$^{IV}$ Laboratoire de Physique Th\'{e}orique, UMR 8627, CNRS, Univ. Paris-Sud, Universit\'e Paris-Saclay, 91405 Orsay Cedex, France\\
$^{V}$ Lebedev Physical Institute RAS, 119991 Moscow, Russia \\
$^{VI}$ Institute of Physics, Academia Sinica, Taiwan 115, Republic of China \\
$^{VII}$ Institute of Theoretical Physics, Beijing 100190, People's Republic of China\\
$^{VIII}$ Novosibirsk State  University, Novosibirsk, 630090, Russia\\
$^{IX}$ University of Chicago, 5620 S. Ellis Avenue, Chicago, IL 60637, USA\\
$^{X}$ Wayne State University, Detroit, MI 48201, USA\\
\vspace{0.2cm}
$^{a}$ Also at Bogazici University, 34342 Istanbul, Turkey\\
$^{b}$ Also at the Moscow Institute of Physics and Technology, Moscow 141700, Russia\\
$^{c}$ Also at the Functional Electronics Laboratory, Tomsk State University, Tomsk, 634050, Russia\\
$^{d}$ Also at the Novosibirsk State University, Novosibirsk, 630090, Russia\\
$^{e}$ Also at the NRC "Kurchatov Institute", PNPI, 188300, Gatchina, Russia\\
$^{f}$ Also at Istanbul Arel University, 34295 Istanbul, Turkey\\
$^{g}$ Also at Goethe University Frankfurt, 60323 Frankfurt am Main, Germany\\
$^{h}$ Also at Key Laboratory for Particle Physics, Astrophysics and Cosmology, Ministry of Education; Shanghai Key Laboratory for Particle Physics and Cosmology; Institute of Nuclear and Particle Physics, Shanghai 200240, People's Republic of China\\
$^{i}$ Also at Government College Women University, Sialkot - 51310. Punjab, Pakistan. \\
$^{j}$ Also at Key Laboratory of Nuclear Physics and Ion-beam Application (MOE) and Institute of Modern Physics, Fudan University, Shanghai 200443, People's Republic of China\\
$^{k}$ Also at Harvard University, Department of Physics, Cambridge, MA, 02138, USA \\
}}
\vspace{0.2cm}
\begin{flushleft}
$^{\bigstar}$ Editor\\
$^{\S}$ \bes3 collaborator\\
$^{\P}$ External contributing author\\
\vspace{0.2cm}
$^{\dagger}$ Correspondence: gradl@uni-mainz.de \\
$^{\ddagger}$ Correspondence: xiaorui@ucas.ac.cn \\
$^{\sharp}$ Correspondence: yuancz@ihep.ac.cn \\
\end{flushleft}

%% file: define/prolist.tex
%\large 
\normalsize
%\pagenumbering{roman}
%\setcounter{page}{1}
%\pagestyle{fancy}
%\fancyhead[RE,LO]{\slshape Content}
\tableofcontents
\clearpage{\pagestyle{empty}\cleardoublepage}
%%\cleardoublepage

%\fancyhead[RE]{\slshape \leftmark}
%\fancyhead[LO]{\slshape \rightmark}

%\listoffigures
\clearpage{\pagestyle{empty}\cleardoublepage}
%%\cleardoublepage

%\listoftables
\clearpage{\pagestyle{empty}\cleardoublepage}
%\cleardoublepage

%% file: Introduction/introduction.tex
\chapter[Introduction]{Introduction}
\label{chapter:Introduction}

\input{Introduction/introduction_main.tex}

\input{Introduction/bib.tex}

%% file: Introduction/introduction_main.tex
\section{Motivation}

The purpose of this White Paper is to examine the BESIII program~\cite{Asner:2008nq}, to consider further physics opportunities, and to plan for possible upgrades to the BEPCII accelerator and the BESIII detector~\cite{Ablikim:2009aa}, in order to fulfill the physics potentials with the \bes3 experiment. The \bes3 Yellow Book~\cite{Asner:2008nq}  documented the original plan for the BESIII physics program before its commission. The discovery of the $Z_c$(3900)~\cite{Ablikim:2013mio1}, followed by many experimental results on the $XYZ$ hadrons~\cite{Ablikim:2013wzq1,Ablikim:2013emm1,Ablikim:2013xfr1} by \bes3, were pleasant surprises, which were not foreseen in the Yellow Book. 
Another surprise came from the first systematical absolute measurements of the $\Lambda_c^+$ decay properties based on threshold $\Lambda_c^+\bar{\Lambda}{}_c^-$ pair production~\cite{Ablikim:2015prg1,Ablikim:2015flg1}.  
The physics related to the $XYZ$  hadrons and (heavier) charmed baryons have also become focal points for the Belle II and LHCb experiments, and present exciting area for the BESIII experiment going forward. In addition to these, a full spectrum of other important experimental opportunities, as will be discussed in this White Paper, will be continually pursued by BESIII, such as light hadron spectroscopy and charmed meson physics.

The integration of quantum theory and Maxwell's electrodynamics leads to a new, powerful theoretical scenario, \ieie, 
of quantum electrodynamics (QED), which was the first building block of what is called today the Standard Model (SM) of 
particle physics. 
Experimental progress led to the discoveries of new particles, and characterizations of their properties, which helped to develop the theoretical framework further to a common understanding of the weak and electromagnetic interactions, called electroweak theory. The modern theory of the strong interaction, called quantum chromodynamics (QCD), was modeled in a similar way on the basis of exact color  SU(3) symmetry for the quarks and gluons. 

Despite being so successful, several issues remain un-answered in the SM. The strong interaction only allows the
existence of composite objects; free quarks and gluons have never been observed. This is called confinement, but it is 
far from being theoretically understood due to its non-perturbative nature. A detailed study of composite objects and 
their properties will shed light on this part of QCD. Furthermore, one might suspect that additional features or 
underlying symmetries beyond the SM might have not been discovered yet, which is usually summarized by the phrase 
`new physics'.

The hadron physics experiments in the 1970's and 1980's concentrated on studying the spectroscopy of the newly 
discovered hadrons containing relatively heavy charm and bottom quarks, or tried to understand specific questions in the 
light-hadron sector with dedicated experiments. For the heavy-quark mesons, no clearly superfluous or ambiguous hadron 
states have been reported. Recent discoveries of `exotic' charmonium-like states have made the picture more complicated~\cite{ch1_reviews}.  
Furthermore, the situation has always been less straightforward for light mesons and baryons containing solely light 
quarks~\cite{klempt}. Here, the high density of states and their broad widths often make the identification and interpretation of 
observed signals rather ambiguous. 
So far the unambiguous identification and understanding of gluonic hadrons is clearly missing.
However, the self-interaction of gluons is central to QCD and leads to a flux tube of gluons binding the quarks inside a hadron together. Due to the self-interaction, bound states of pure gluons (named glueballs) or their mixing with conventional mesonic state, should exist as well as so-called hybrids, where quarks and gluonic excitations contribute explicitly to the quantum numbers.

The energy regime that \bes3 is operating in and the detector design allow for a detailed study of the charmonium and
light-quark region. Charmonium physics received a major renewal of interest when in the 2000's many new, unexpected 
resonances, called $X$, $Y$ and $Z$ states~\cite{ch1_reviews}, were discovered, that could not be accommodated by  the quark model. Many of 
those were found by the Belle, BaBar, CDF, D0 and later the LHC experiments, but only \bes3 is dedicated to the energy 
region where most of these states appear. It is therefore no surprise that detailed studies with much higher statistics 
can only be performed at \bes3. Nowadays, \bes3 is one of the main contributors to the understanding of these $XYZ$ 
states. At the same time, the high production cross section of charmonia at BEPCII together with a modern, almost 
hermetic detector for charged and neutral particles, allows also for high-precision studies of light-quark hadrons in 
the decay of charmonia. Since this decay into light quarks proceeds via gluons, it is likely that the desired studies of 
gluonic excitations may be performed at \bes3, as will be shown in this White Paper.  

Despite the discovery of charm quark more than 40 years ago, many questions about
charmed particles still remain unsolved~\cite{ch1_reviews}.  An upgraded BEPCII and \bes3 can make key contributions to lepton
flavor universality, the unitarity of the Cabibbo-Kobayashi-Maskawa (CKM) matrix,
validity of lattice QCD (LQCD), as well as theories of decay constants and form
factors by studying leptonic and semileptonic decays of charmed particles.  It can give insights into the applicability of QCD in low-energy nonperturbative
contexts and can greatly expand our knowledge of charmed baryon properties.
Open questions here include missing $\Lambda_c$ decay modes (\egeg, those with
as yet undetected neutral or excited final-state baryons) and baryon
electromagnetic structure.

\section{The \bes3 detector and its upgrades}

The \bes3 detector and BEPCII accelerator represent
major upgrades over the previous version of BES~\cite{Bai:1994zm,Bai:2001dw} and BEPC~\cite{Ye:1987nh}; the facility is used for studies of hadron physics and $\tau$-charm physics. The BEPCII collider,
installed in the same tunnel as BEPC, is a double-ring multi-bunch collider with a design luminosity of $1\times 10^{33}~\text{cm}^{-2} \text{s}^{-1}$ optimized at a center-of-mass (cms) energy of $2 \times 1.89$ GeV, an increase of a factor of 100 over its predecessor. The design luminosity was  reached in 2016, setting a new world record for the accelerator in this energy regime~\cite{BEPCII-news}.

\begin{figure}[htbp]
\centerline{\includegraphics[angle=90,width=14cm,height=10cm]{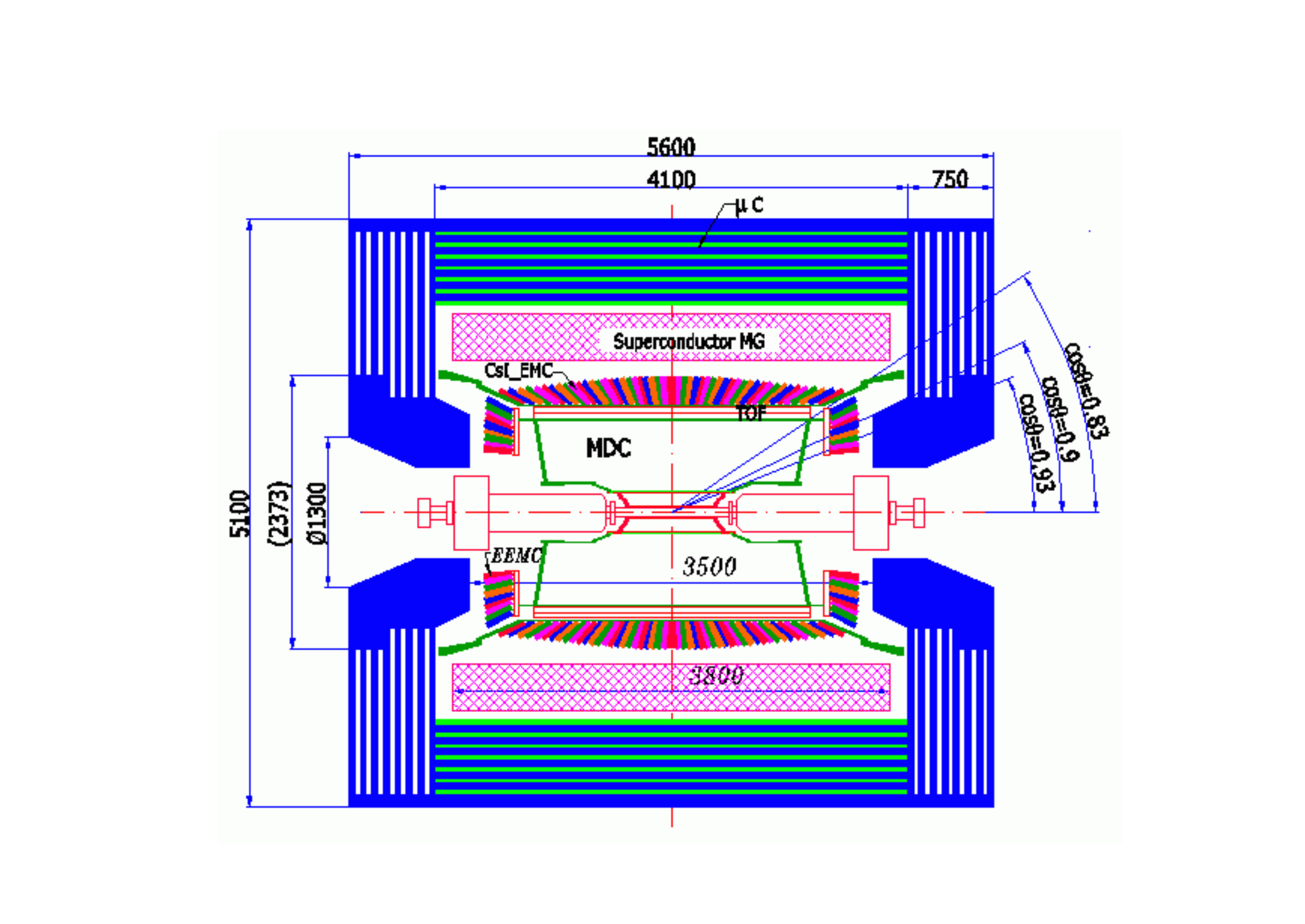}}
\caption{An overview of the \bes3 detector.}
\label{fig:r99bes}
\end{figure}

The \bes3 detector is designed to fulfill the physics requirements, and 
the technical requirements for a high 
luminosity, multi-bunch collider. Detailed descriptions of the \bes3 
detector can be found in Ref.~\cite{Ablikim:2009aa}. 
Figure~\ref{fig:r99bes} shows a schematic view of the
\bes3 detector, which covers 93\% of 4$\pi$ solid angle. It consists of the following components:
\begin{itemize}
\item A Helium-gas based drift chamber (MDC) with a single wire resolution 
that is better than 120 $\mu$m and 
a d$E$/d$x$ resolution that is better than 6\%. The 
momentum resolution in the 1.0 T magnetic field is better than 0.5\% 
for charged tracks with a momentum of 1~GeV/$c$. 
\item A CsI(Tl) crystal calorimeter with an energy resolution 
that is better than 2.5\% and position resolution better 
than 6~mm for 1~GeV electrons and gammas.
\item A Time-of-Flight (TOF) system with an intrinsic timing resolution 
of 68 ps in barral part and resolution of 110 ps in end-cap part.
\item A super-conducting solenoid magnet with a central field of 1.0 
Tesla.
\item A 9-layer RPC-based muon chamber system with a spatial resolution 
that is better than 2~cm.
\end{itemize}
Details of each sub-detector and their performance, together with the 
trigger system, are discussed in Ref.~\cite{Ablikim:2009aa}. 

The \bes3 detector has been operating since 2009, and BEPCII has delivered around 30 fb$^{-1}$ of integrated luminosity at different cms energies.  The experiment has received several upgrades, and new upgrades for both the detector and accelerator are being considered.

\subsection{Upgrade of ETOF}
 
\begin{figure}[htbp]
\centerline{\includegraphics[width=8cm]{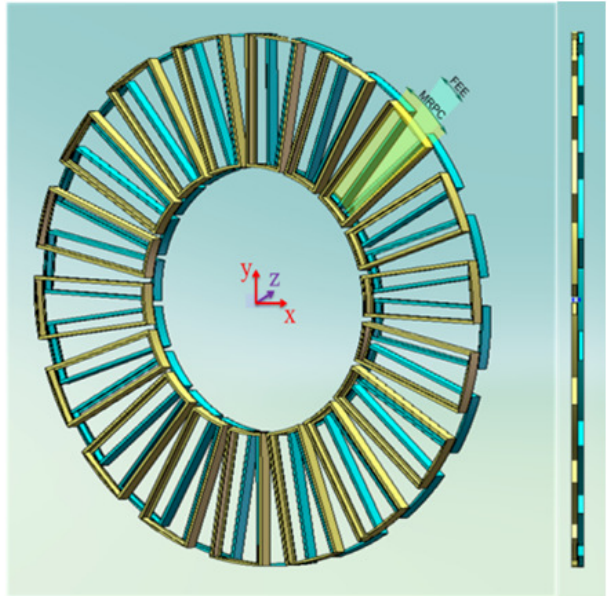}}
\caption{Schematic drawing of MRPC ETOF at \bes3.}
\label{fig:r99etof1}
\end{figure}

\begin{figure}[htbp]
\centerline{\includegraphics[width=8cm]{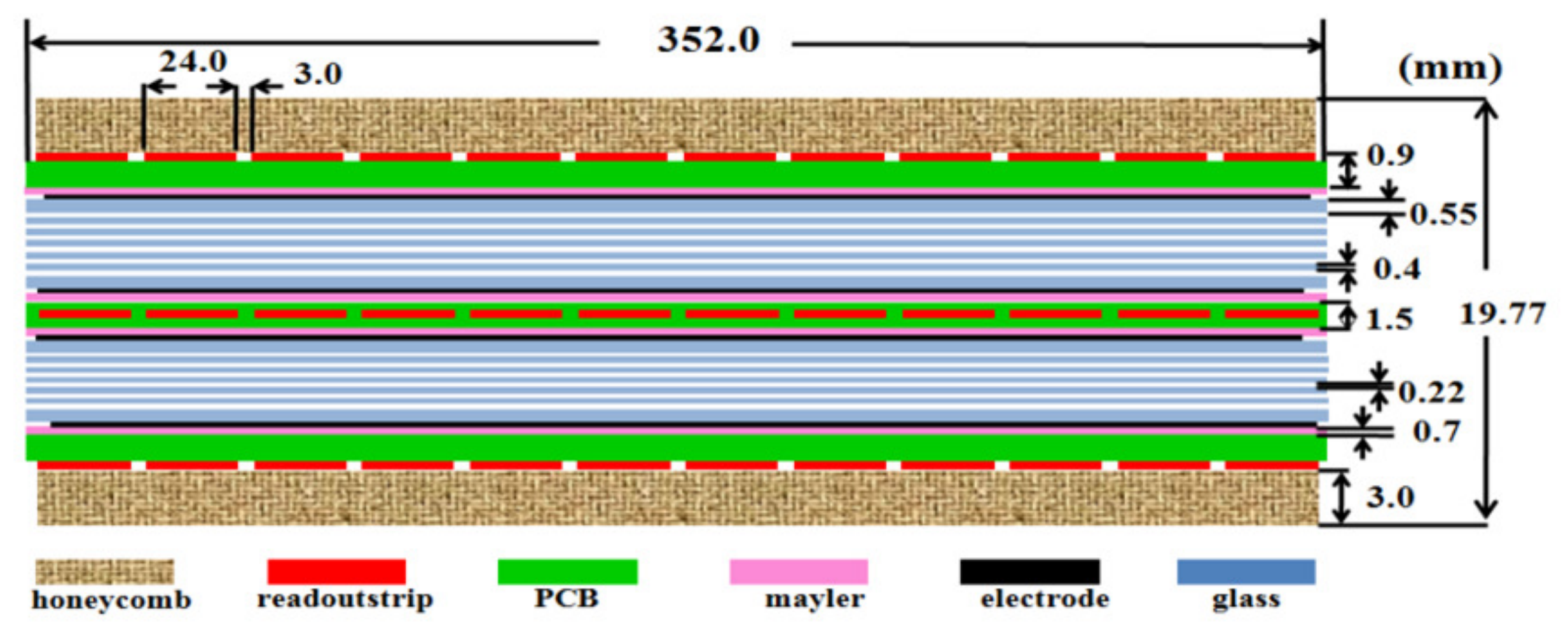}}
\caption{The cross-sectional view of a MRPC module along its length for the ETOF at \bes3.}
\label{fig:r99etof2}
\end{figure}

In order to improve the capability of particle identification of the \bes3 experiment, the end-cap time-of-flight (ETOF) detector was upgraded with multi-gap resistive plate chamber (MRPC) technology in 2015~\cite{ETOF}.
The MRPC is a new type of gaseous detector that has been successfully used as TOF detectors in several experiments.
The new ETOF system of  \bes3 consists of two end-caps; each end-cap station has 36 trapezoidal shaped MRPC modules arranged in circular double layers as shown in Figs.~\ref{fig:r99etof1} and~\ref{fig:r99etof2}.
Each MRPC is divided into 12 readout strips which are read out from both ends in order to improve the timing resolution.
The readout electronics system of MRPC detectors consists of FEE boards, time-to-digital conversion modules, calibration-threshold-test-power board, fast control module and a clock module in NIM crates that communicates with and is controlled by the data acquisition  system.
A multi-peak phenomenon in the time-over-threshold distribution is observed, and the reflection of the inductive signal at the ends of the strip is the main contribution.
An empirical calibration function based on the analysis of the correlation of raw measured time, time-over-threshold and extrapolated hit position of the charged particle, is implemented using the real data of Bhabha events.
Performance checks show the overall time resolution for pions with momentum around 0.8 GeV/$c$ is  about 65 ps, which is better than the original design goals.

\subsection{Upgrade of Inner MDC with a CGEM inner tracker}

\begin{figure}[htbp]
\centerline{\includegraphics[width=8cm]{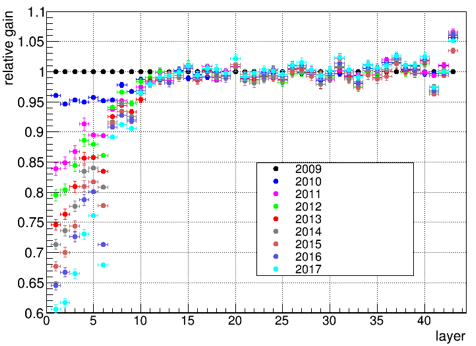}}
\caption{Relative gain decrease of the cells in each year as a function of the MDC layer.}
\label{fig:mingyi-1}
\end{figure}

The MDC is the main tracker of the \bes3 with the capability of accurate measurements for the position and the momentum of charged particles produced in $e^+ e^-$ collisions, as well as the charged particles identification by measuring d$E$/d$x$. The MDC is a low-mass cylindrical wire chamber with small-cell geometry, using helium-based gas and operating in a 1 T magnetic field. It consists of an inner chamber (8 layers) and an outer chamber (35 layers), which are jointed together at the endplates and share a common gas volume. After many years of running since 2009, the MDC is suffering from ageing problems due to beam-induced background with a hit rate up to 2 kHz/cm$^2$~\cite{part1:dongmy}, which has caused the cell gains of the inner chamber to drop dramatically (about 39\% drop for the first layer cells in 2017 as shown in Fig.~\ref{fig:mingyi-1}), and furthermore has led to a degradation of the spatial resolution and reconstruction efficiency. Because of the radiation damage in the inner chamber, a cylindrical gas electron multiplier (CGEM) has been selected as one of options of the upgrade, due to its attractive features of a high counting rate capability and low sensitivity to ageing. The CGEM inner tracker (CGEM-IT) project deploys a series of innovations and special attributes in order to cope with the requirements of \bes3, as listed in Table~\ref{tab:CGEM-1}. 

\begin{figure}[tp]
\centerline{\includegraphics[width=8cm]{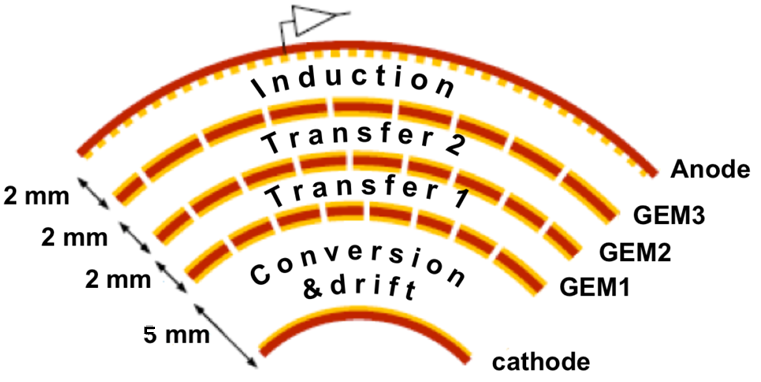}}
\caption{Cross-section of the triple GEM detector used for the \bes3 CGEM-IT.}
\label{fig:CGEM-1}
\end{figure}

The CGEM-IT consists of three layers of triple cylindrical GEM~\cite{part1:cgem1}, shown in Fig.~\ref{fig:CGEM-1}. Each layer is assembled with five cylindrical structures: one cathode, three GEMs and the anode readout (see Fig.~\ref{fig:CGEM-1})~\cite{part1:cgem11}. The GEMs and electrode foils are produced in planes and then shaped as cylinders. The assembly is performed inside a vertical inserting machine. To minimize the material budget, no support frames are used inside the active area and the GEM foils are mechanically stretched, being glued to Permaglass rings at their ends. The Permaglass rings are used only outside the active area to operate as gas sealing structure and gap spacers. A sandwich of a PMI foam, called Rohacell, and kapton is used in order to provide mechanical rigidity to the anode and the cathode electrodes. Rohacell is a very light material that limits the material budget to 0.3\% of radiation length ($X_0$) per layer.

\begin{table}[tp]
\centering
\caption{\label{tab:CGEM-1} List of the requirements for the new inner tracker.}
\begin{tabular}{|c|c|}
  \hline
  Value & Requirements  \\
    \hline
  $\sigma_{xy}$ & $\leq 130$ $\mu$m \\
  $\sigma_{z}$ & $\leq 1$ mm \\
  d$p$/$p$ for 1~GeV/$c$ &  0.5\% \\
  Material budget & $\leq 1.5$\% $X_0$ \\
  Angular Coverage & 93\% $\times 4\pi$ \\
  Hit Rate & $10^4$ Hz/cm$^2$ \\
 Minimum Radius &  65.5 mm\\
 Maximum Radius & 180.7 mm \\
  \hline
\end{tabular}
\end{table}

The readout anode circuit is manufactured with the 5 $\mu$m copper clad, 50 $\mu$m thick polyimide substrate. Two foils with copper segmented in strips are used to provide two-dimensional readout. The strip pitch is 650 $\mu$m, with 570 $\mu$m wide X-strips parallel to CGEM axis, providing the $r-\phi$ coordinates, while the V-strips, having a stereo angle with respect to the X-strips, are 130 $\mu$m wide and, together with the other view, gives the $z$ coordinate. The stereo angle depends on the layer geometry. A jagged-strip layout is used to reduce the inter-strip capacitance up to 30\%. An innovative readout based on analogue information and data-pushing architecture has been developed. A dedicated ASIC has been developed to provide time and charge information for each strip. 

\begin{figure}[tp]
\centerline{\includegraphics[width=0.8\textwidth]{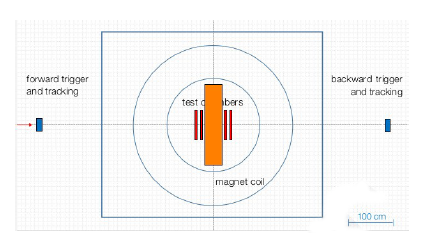}}
\caption{Sketch of the setup for the CGEM test beam.}
\label{fig:CGEM-2}
\end{figure}

In order to verify that the CGEM-IT can reach the required performance, an extensive series of  beam tests have been conducted in the last few years within the test beam activities of the RD51 Collaboration of CERN. The tests were performed both on 10 $\times$10 cm$^2$ planar GEM chambers and on a cylindrical prototype with the dimension of the second layer of the final CGEM-IT~\cite{part1:cgem2}. All the tests were performed in the H4 line of the SPS, in CERN North Area. Since the CGEM-IT will operate in a magnetic field, all the test chambers were placed inside Goliath, a dipole magnet that can reach up to 1.5 T in both polarities. Pion and muon beams with  momentum of 150 GeV/$c$ were used. Two scintillators were placed upstream and downstream the magnet to operate as a trigger. A typical setup using the cylindrical prototype is shown in Fig.~\ref{fig:CGEM-2}.

\begin{figure}[tp]
\centerline{\includegraphics[width=8cm]{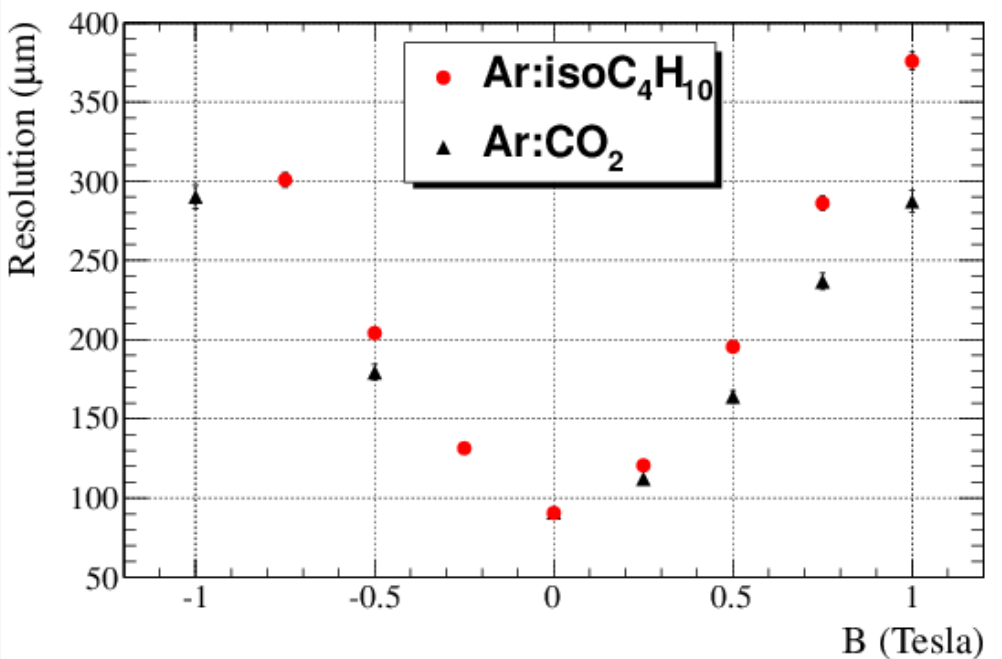}}
\caption{Resolution with respect to the magnetic field strength for Ar/iC$_{4}$H$_{10}$(90/10) and Ar/CO$_{2}$(70/30).}
\label{fig:CGEM-3}
\end{figure}

\begin{figure}[tp]
\centerline{\includegraphics[width=8cm]{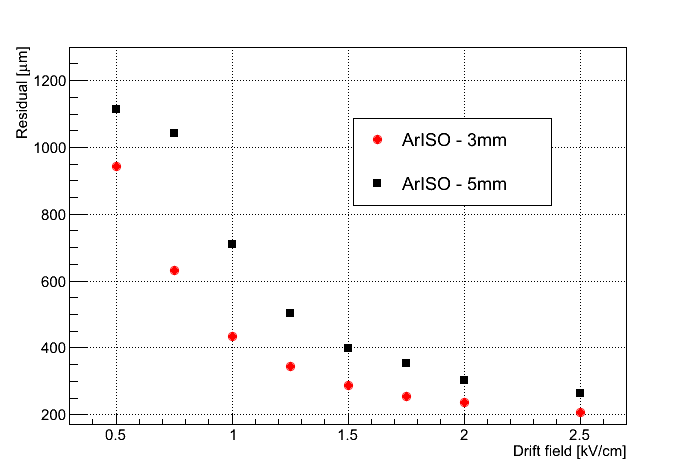}}
\caption{ Resolution with respect to drift field in 1 T magnetic field for two different drift gaps: 3 mm drift gap and 5 mm drift gap.}
\label{fig:CGEM-4}
\end{figure}

\begin{figure}[htbp]
\centerline{\includegraphics[width=8cm]{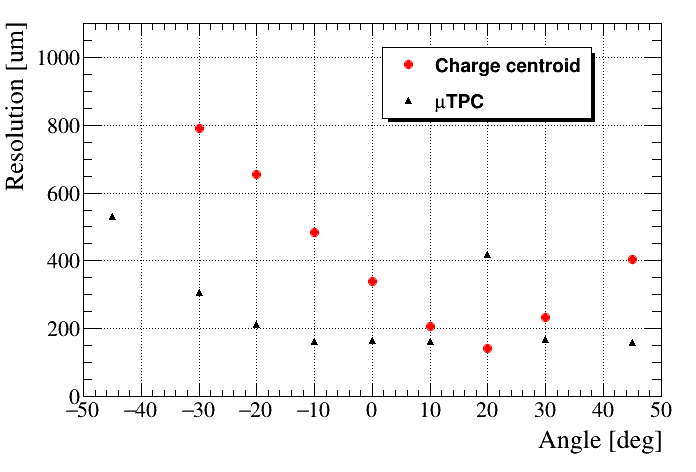}}
\caption{ Resolution comparison between charge centroid and $\mu$-TPC with respect to the incident angle of the track in 1 T magnetic field.}
\label{fig:CGEM-5}
\end{figure}

The performance of the planar GEM chambers in magnetic field was studied with the charge centroid method.  The presence of an external magnetic field induces a deformation of the avalanche shape at the anode due to the Lorentz force: the charge centroid method performance degrades almost linearly with the magnetic field strength as shown in Fig.~\ref{fig:CGEM-3}. It is still possible to improve the performance by a proper optimization of the drift field as shown in Fig.~\ref{fig:CGEM-4}. With the proper choice of gas mixture (Ar/iC$_4$H$_{10}$(90/10)) and drift field (2.5 kV/cm) it is possible to achieve a resolution of 190 $\mu$m in a 1 T magnetic field.

The $\mu$-TPC is another available method for track reconstruction. It is an innovative approach that exploits the few millimeters drift gap as a Time Projection Chamber. Indeed, the time of arrival of the induced charge on the strip can be used to reconstruct the first ionization position in the drift gap and thus improve the spatial resolution. The $\mu$-TPC method can improve and overcome the charge centroid limits granting a spatial resolution lower than 200 $\mu$m for a large angle interval, as shown in Fig.~\ref{fig:CGEM-5}. Further studies are ongoing. By merging these two methods it will be possible for the spatial resolution of the CGEM-IT  to satisfy the requirements of BESIII.

The construction of the CGEM-IT  will be completed, and then a long term cosmic-ray test will be performed to evaluate the performance of the whole CGEM-IT before the replacement of the inner chamber of the MDC.

\subsection{Upgrade of Inner Chamber with an improved inner MDC }

In addition to the construction of the CGEM-IT, an improved new inner MDC has been built that can replace the aged inner part of the MDC if needed~\cite{part1:cdc1}. 

\begin{figure}[htbp]
\centerline{\includegraphics[width=8cm]{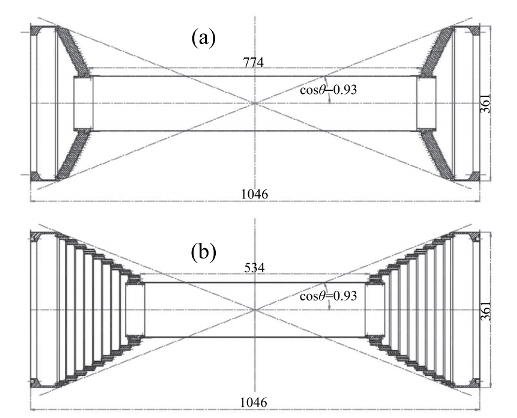}}
\caption{Overview of the mechanical structure of the inner MDC. (a) The old inner chamber. (b) The new inner chamber.}
\label{fig:inner-chamber-1}
\end{figure}

The new inner MDC is designed with multi-stepped end-plates. 
Each step contains one sense-wire layer and one field-wire layer, which can shorten the wire length exceeding the effective detection sold angle, and minimize the ineffective area in the very forward and backward region and hence reduce the background event rate of all cells, as shown in Fig.~\ref{fig:inner-chamber-1}. The maximum reduction of the rate of background events is more than 30\% for the first layer cells. With this design, the new inner MDC is expected to have a longer lifetime and improved performance due to the lower occupancy.

\begin{figure}[htbp]
\centerline{\includegraphics[width=8cm]{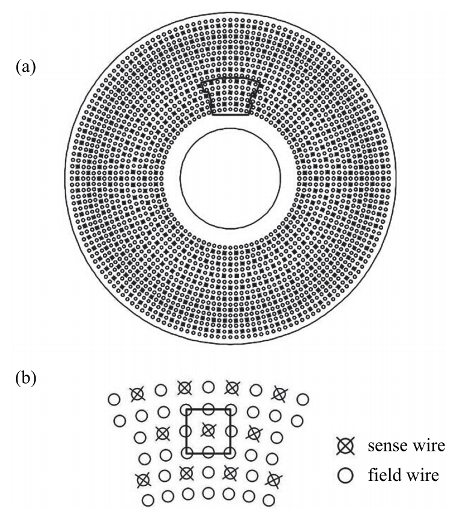}}
\caption{End view of the new inner chamber and the layout of the cells. (a) End view of the new inner chamber. (b) The drift cells of the chamber.
}
\label{fig:inner-chamber-2}
\end{figure}

The new inner MDC is mainly composed of two multi-stepped endplates and an inner carbon fiber cylinder. The length of the new inner chamber is 1092 mm, and the radial extent is from 59 mm to 183.5 mm, including 8 stereo sense wire layers, comprising 484 cells in total. Similar to the old chamber, the drift cells of the new chamber have a nearly square shape, as shown in Fig.~\ref{fig:inner-chamber-2}. The size of each cell is about 12 mm$\times$12 mm  with a sense wire located in the center, surrounded by eight field wires. The sense wires are 25 $\mu$m gold-plated tungsten wires, while the field wires are 110 $\mu$m gold-plated aluminum wires.

\begin{figure}[htbp]
\centerline{\includegraphics[width=8cm]{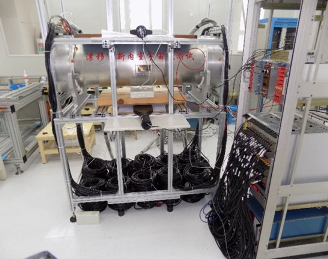}}
\caption{The cosmic-ray test of the new inner MDC.}
\label{fig:inner-chamber-3}
\end{figure}

\begin{figure}[htbp]
\centerline{\includegraphics[width=8cm]{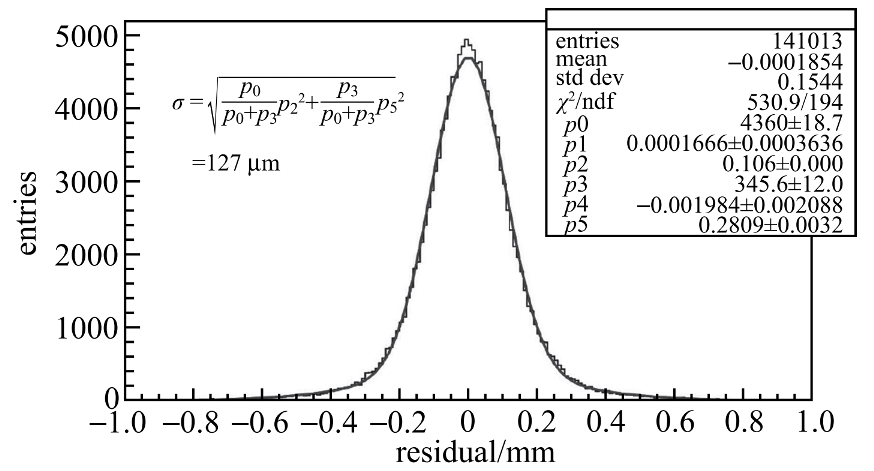}}
\caption{Residual distribution of the new inner MDC, showing the results of a fit with a double-Gaussian function.}
\label{fig:inner-chamber-4}
\end{figure}
\begin{figure}[htbp]
\centerline{\includegraphics[width=8cm]{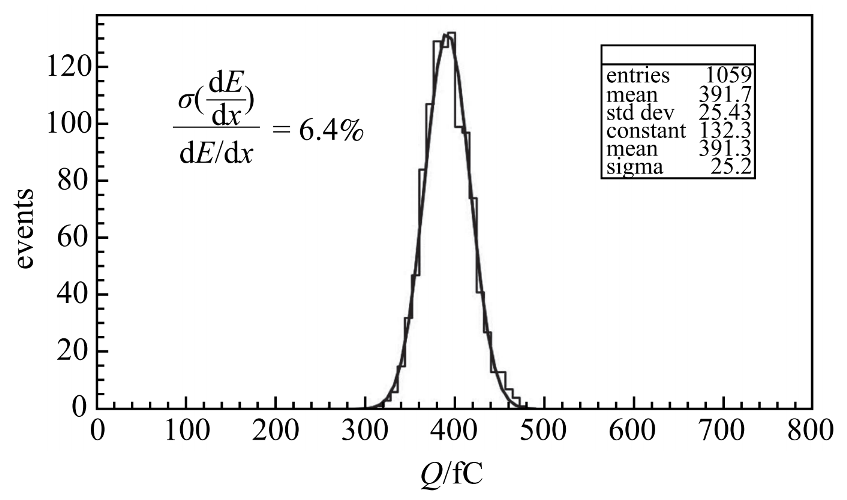}}
\caption{The d$E$/d$x$ resolution with the new inner MDC, showing the results of a fit with a single Gaussian function.}
\label{fig:inner-chamber-5}
\end{figure}

In the construction of the new inner chamber, two aluminum endplates were manufactured with an eight-step structure for each one. A total of 2096 wire holes with a diameter of 3.2 mm were drilled in each endplate with the mean value of the tolerances of 14 $\mu$m. A carbon fiber inner cylinder with a thickness of 1.0 mm, was covered with two layers of 100 $\mu$m thick aluminum foils on its inner surface and outer surface respectively for electromagnetic shielding. The endplates and the cylinder were assembled with a precision better than 30 $\mu$m. Wire stringing was performed after the mechanical structure was assembled. Good quality of wire stringing was achieved by monitoring the wire tension and leakage current during the stringing. The non-uniformity of wire tension was less than 10\%, and the leakage current was lower than 2 nA for each wire.

After the completion of the construction of the new chamber, a cosmic-ray test without magnetic field was carried out to evaluate its performance, shown in Fig.~\ref{fig:inner-chamber-3}. The results of the cosmic-ray test show that the new inner chamber achieves a spatial resolution of 127 $\mu$m and a d$E$/d$x$ resolution of 6.4\%, shown in Fig.~\ref{fig:inner-chamber-4} and Fig.~\ref{fig:inner-chamber-5}, which satisfy the design specifications. These performances verify the successful construction of the new chamber, and now the new inner chamber is ready to be used if needed.

A decision on whether to install the CGEM or new inner MDC will be made, according to the results of their beam and cosmic-ray tests.

\section{BEPCII upgrades}

BEPCII delivered its first physics data in 2009 on the $\psip$ resonance. 
Since then, \bes3 has collected about 30 fb$^{-1}$ integrated luminosity at different 
 energy points from 2.0 to 4.6 GeV. By using these data samples, the \bes3 collaboration has published more than 270 papers, which have been making significant contributions to hadron spectroscopy, tests of various aspects of QCD, charmed hadron decays, precision test of the SM,  probes of new physics beyond the SM, as well as $\tau$ mass measurement. Nowadays, the \bes3 experiment plays a leading role in the study of the $\tau$-charm energy region. 

During the past  10 years of successful running, better understanding of the machine is achieved. With the increasing physics interest, two upgrade plans of BEPCII were proposed and approved. The first one is to increase the maximum beam energy up to 2.45 GeV, to expand the energy territory. The second is the top-up injection to increase the data taking efficiency. 
%The top-up injection has been implemented in the dedicated synchrotron radiation (SR) mode, while it is not the case in collision mode. 
The activities of these two upgrades begin in 2017.

\begin{figure}
  \centering
  \includegraphics[width=0.6\textwidth]{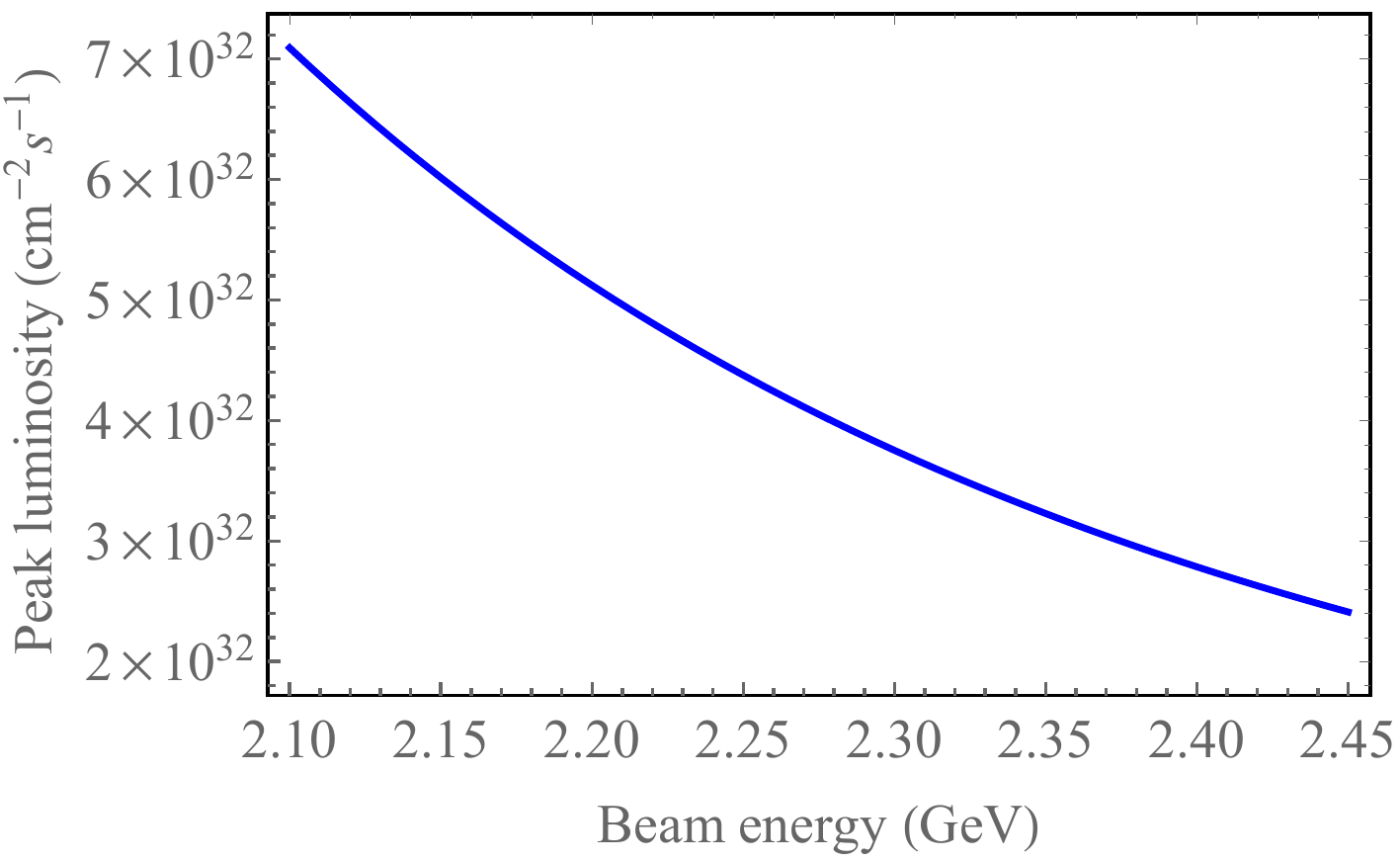}
  \caption{The estimation of peak luminosity of BEPCII in the energy region above 2.1 GeV.}\label{fig:lum-above-2.1GeV}
\end{figure}

Before 2019, the beam energy of BEPCII ranges from 1.0 to 2.3 GeV. In order to extend
the physics potential of BESIII, an upgrade project to increase the
beam energy up to 2.45 GeV was initiated. In order to achieve this goal, some hardware modification are necessary, including the power supplies of dipole magnet, power supplies of special magnets in the interaction region, septum magnet and its water cooling system.
These hardware modifications were completed during the summer shutdown in 2019, while the commissioning will be finished in the end of 2019. However, as expected, running at high energy region above 1.89 GeV, the beam current will decrease due to the limitation
of Radio Frequency (RF) power and the difficulty in controlling the bunch length and emittance. Hence, the
peak luminosity will go down with the increasing beam energy, which is shown in Fig.~\ref{fig:lum-above-2.1GeV}.
In the future, it would be interesting to investigate the possibilities of slight increasing
of the beam energy up to 2.5 GeV and slight decreasing down to 0.9 GeV, for the interests of studying $\Xi_c$ and nucleon productions, respectively.

The top-up injection is a highly efficient operation scheme for the accelerator~\cite{ref:top-up},  which provides a nearly constant
beam current. As there is no stop for beam refilling, the integrated luminosity can be increased by 20\% to 30\% for long data taking runs. The BEPCII upgrade of the top-up injection for the collision mode
has started in September 2017. In order to obtain stable online luminosity, the
beam current fluctuation is controlled within 1.5\% with one $e^+$ injection and
two $e^-$ injections every 90 seconds, so that the variation of the instantaneous luminosity is less than 3\% of its nominal value. The injection rates of the $e^+$ and $e^-$ bunches must be
higher than 60 mA/min and 180 mA/min, respectively. The commissioning of the top-up
injection began after the summer shutdown in 2019 and finished by the end of year.

There are also discussions on further machine upgrades on luminosity. The recently proposed crab-waist collision scheme~\cite{ref:crab-waist} is believed to be essential for the luminosity challenge in the next-generation high luminosity $e^+e^-$ collider. 
The possibility of crab-waist scheme at BEPCII has been considered since 2007.  However, it was found impossible if only minor changes on the current design are allowed. A recent upgrade proposal of BEPCII based on crab-waist scheme was discussed in detail in Ref.~\cite{ref:anton-tau},
which presents an upgrade project with the peak luminosity of $6.0\times 10^{33}$ cm$^{-2}s^{-1}$. This is 10 times
higher than the achieved luminosity of BEPCII at the beam energy of 2.2 GeV. The crab-waist collision with large Piwinski angle is suggested to be adopted with
the parameter modifications of the BEPCII. The $\beta$ functions at the interaction point
would be modified from 1.0 m/1.5 cm to 0.14 m/0.8 cm in the horizontal and vertical
plane, respectively. The emittance is reduced from 140 nm to 50 nm with the damping wigglers.
Regarding to this proposal, a detailed design with crab-waist scheme has been studied. From the study, many physical and technical issues are investigated, such as injection, dynamic aperture, emittance coupling, high power RF, super-conducting quadrupoles/wigglers, and strong crab sextupoles etc. Finally, it is found that the crab-waist scheme is a complicated and time-consuming project and not practical with the present BESIII detector. 

\begin{figure}
  \centering
  \includegraphics[trim={0 -1cm 3.5cm 2.1cm},clip,width=0.68\textwidth]{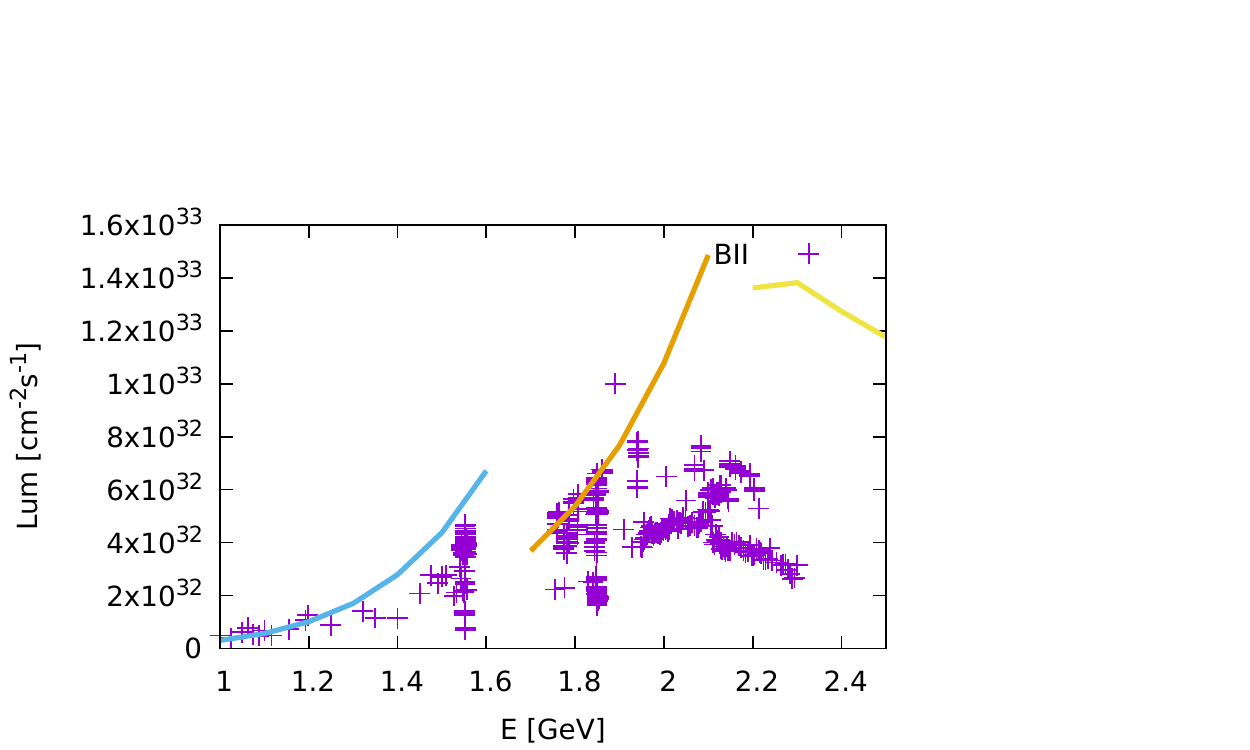}
  \includegraphics[trim={0 0 3.5cm 2.1cm},clip,width=0.68\textwidth]{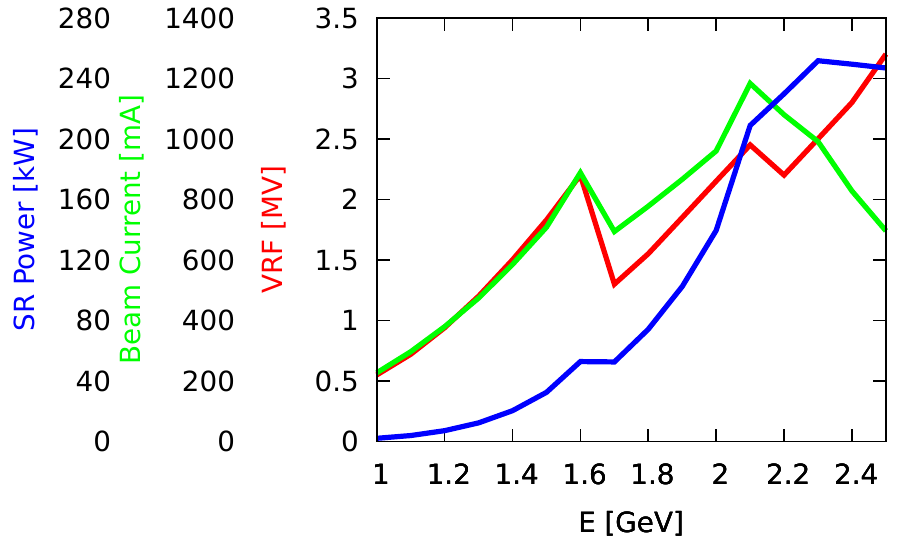}
  \caption{Scenario of BEPCII upgrade by increasing beam current. The lines show the expected performances after upgrade. The points in the upper plot show the achieved values of the current BEPCII.}\label{fig:bepc2-upgrade}
\end{figure}

Another economic way to increase luminosity is by raising up beam current, which would  potentially gain a factor of 2 improvement of  peak luminosity. For this purpose, one needs to suppress bunch lengthening, which require higher RF voltage. The scenario of expected luminosity, beam current and SR power  is shown in Fig.~\ref{fig:bepc2-upgrade}. The RF, cryogenic and feedback systems need to  be upgraded accordingly according to the higher beam current. Nearly all the photon absorbers along the ring and some vacuum chambers also need to be replaced in order to protect the machine from heat of SR. The budget is estimated to be about 100-200 million CNY and it will take about 3 years to prepare the upgraded components and 1 year of installation and commissioning. This upgrade scheme with higher beam current is more realistic at present than the crab-waist scheme.

%% file: Lighthadron/lighthadron.tex
\chapter[Light Hadron Physics]{Light Hadron Physics}
\label{chapter:LH}

\input{Lighthadron/lighthadron_main.tex}

\input{Lighthadron/bib.tex}

%% file: Lighthadron/lighthadron_main.tex
\section{Introduction}

The generally accepted theory for the strong interaction, quantum chromodynamics (QCD), remains a challenging part of the standard model in the low- and medium-energy regime.
In the high energy regime, asymptotic freedom of the partons constituting hadrons allows systematic calculations in QCD
using perturbation theory. In the low-energy regime
where the energies are (much) smaller than a
typical strong interaction scale,  there are well-established
theoretical methods, chiral perturbation theory (ChPT). In the intermediate-energy regime, the non-Abelian character of QCD requires a non-perturbative approach
which must rely either on lattice QCD (LQCD) or on QCD-inspired models.
Therefore, the study of light hadrons is central to the understanding of
confinement physics.

Significant progress in the light-quark sector appeared in the last few years due to unprecedented high-statistics data sets from
experiments at both electron and hadron machines. Due to the non-Abelian structure of QCD bound states
beyond the constituent
quark degrees of freedom,~{\it e.g.}, multiquark states or states with
gluonic degrees of freedom (hybrids, glueballs) are expected. Their unambiguous identification and systematic study would provide a validation of and valuable input to the quantitative understanding of QCD.  Over the last decade, there have been several relevant reviews ~\cite{Brambilla:2014jmp,Meyer:2010ku,Crede:2008vw,Klempt:2007cp,Amsler:2004ps,Godfrey:1998pd} on this subject, which cover in great detail both meson spectroscopy and baryon spectroscopy.

Data with unprecedented statistical accuracy and clearly defined initial and final state properties offer \bes3 great opportunities to investigate hadron spectroscopy and led to significant advances in recent years. The road map for the light hadron physics program at \bes3 has already been defined, following the trajectory of endeavors over last decades~\cite{Kopke:1988cs,bes3}. In this document we reiterate the physics case on the basis of achieved results.

\section{Meson spectroscopy and the search for QCD exotics}
Confinement is a unique property of QCD. The quark model describs mesons as bound states of
quarks and antiquarks. LQCD and QCD-motivated models for
hadrons, however, predict a richer spectrum of mesons that takes into
account not only the quark degrees of freedom but also
the gluonic degrees of freedom.
A primary goal of the \bes3 experiment is to search for and study those QCD exotics or states with a composition that is different from normal mesons and baryons. Understanding these will provide critical information on the quantitative understanding of confinement.

\subsection{Glueballs}

The spectrum of glueballs is predicted by quenched LQCD~\cite{Bali:1993fb,Morningstar:1999rf,Chen:2005mg} with the lightest one having scalar quantum
numbers $0^{++}$ and a mass between 1.5~GeV/$c^2$ and 1.7~GeV/$c^2$.
Also the next-higher glueball
states have nonexotic quantum numbers,
$2^{++}$ (mass 2.3--2.4~GeV/$c^2$)
and $0^{-+}$ (mass 2.3--2.6~GeV/$c^2$), and hence will be mixed into the conventional meson spectrum and difficult to be identified experimentally.
It requires systematic studies  to identify a glueball by
searching for outnumbering of conventional quark model states and comparing a candidate¡¯s properties to the expected
properties of glueballs and conventional mesons.

In a simple constituent-gluon
picture, these three states correspond to two-gluon systems in a
relative $S$~wave, with different combinations of
helicities. Table~\ref{tab:lqcd.glueballs} summarizes the quenched and
unquenched lattice results for the masses of the lightest glueballs.
The masses of scalar and tensor glueballs from quenched LQCD are consistent with those obtained by $N_f=2$~\cite{Sun:2017ipk} and
$N_f=2 + 1$~\cite{Gregory:2012hu} unquenched LQCD.

\begin{table*}[tbp]
\caption{Glueball masses (in units of MeV/$c^2$) from the quenched~\cite{Morningstar:1999rf, Chen:2005mg}
and unquenched~\cite{Gregory:2012hu, Sun:2017ipk} lattice QCD studies. }
\label{tab:lqcd.glueballs}
\begin{center}
\begin{tabular}{lclll}
\hline\hline
                              &      $m_\pi$    & $m_{0^{++}}$   &   $m_{2^{++}}$     &   $m_{0^{-+}}$     \\
\hline
quenched Ref.~\cite{Morningstar:1999rf} &        ---          &    1710(50)(80)&    2390(30)(120)        &    2560(35)(120)         \\
quenched Ref.~\cite{Chen:2005mg}        &        ---          &    1730(50)(80)&    2400(25)(120)        &    2590(40)(130)         \\
&&&&\\
unquenched $N_f=2$~\cite{Sun:2017ipk}                       &  $938$  &    1417(30)    &    2363(39)             &   2573(55)               \\
                              &  $650$  &    1498(58)    &    2384(67)             &   2585(65)               \\
unquenched $N_f=2+1$~\cite{Gregory:2012hu}&  $360$ &    1795(60)    &    2620(50)             &    ---                    \\
\hline\hline
\end{tabular}
\end{center}
\end{table*}

Glueballs are expected to appear in so-called gluon-rich environments. The radiative decays of the $J/\psi$ meson provide such a gluon-rich environment and are therefore regarded as one of the most
promising hunting grounds for glueballs. Recent LQCD calculations predict that the partial width of $J/\psi$ radiatively decaying into the pure gauge scalar glueball is 0.35(8) keV, which corresponds to a branching ratio of $3.8(9)\times 10^{-3}$~\cite{Gui:2012gx}; the partial decay width for a tensor glueball is estimated to be 1.01(22)(10) keV which corresponds to a large branching ratio $1.1(2)(1)\times10^{-2}$~\cite{Yang:2013xba}; the partial decay width for a pseudoscalar glueball is estimated to be 0.0215(74) keV which corresponds to a branching ratio $2.31(80)\times10^{-4}$~\cite{Gui:2019dtm}. With the unique advantage of high-statistics $J/\psi$ sample, a systematics research program of glueballs has been performed at \bes3. The pseudoscalar sector draws special attention due to the small number of expected resonances of the quark model. However, experimental input is very limited for now and the hope is in \bes3 for significant improvements. In the scalar sector and tensor sector already a large number of resonances is observed. However, the nature of all of these states is still controversial, but the program at \bes3 can provide crucial information to map out the scalar and tensor excitations.

\subsubsection{The scalar mesons}

A related review on the topic can be found in the section ``Note on Scalar Mesons below 2~GeV'' in PDG~\cite{PDG}. The most striking observation is that the $f_0(1370)$,  $f_0(1500)$ and  $f_0(1710)$ appear to be supernumerary. Many papers interpret the existence of these three scalars as a manifestation of the underlying light quarkonium nonet and the lowest-mass scalar glueball.

Challenges in the interpretation of the scalar sector involve both experimental and theoretical efforts. The following key
questions account for the major differences in the models on scalar mesons and need to be addressed in the future:
\begin{itemize}

  \item Is the $f_0(1370)$ a true $q\bar{q}$ resonance or generated by $\rho\rho$ molecular dynamics?
  \item Even though the fact of a supernumerary  state
is suggestive, the decay rates and production mechanisms are also needed to unravel the quark content of $f_0(1500)$ and $f_0(1710)$. In the partial wave analysis (PWA) of $J/\psi\to\gamma\eta\eta$~\cite{Ablikim:2013hq} and $J/\psi\to\gamma K_S K_S$~\cite{Ablikim:2018izx} at BESIII, the branching fractions of the $f_0(1710)$ are one order of magnitude larger than those of the $f_0(1500)$. With the new measurements from BESIII, the known branching fraction of $J/\psi\to\gamma f_0(1710)$~\cite{PDG} is up to $1.7\times 10^{-3}$, which is already comparable to the LQCD calculation of scalar glueball ($3.8(9)\times 10^{-3}$~\cite{Gui:2012gx}). The production property suggests $f_0(1710)$ has large gluonic component than $f_0(1500)$. More precise measurements of the partial decay widths in the future will improve the understanding of the internal structure of these states. A significant
property of glueball decays is their expected flavor symmetric coupling to final-state hadrons, even though some modifications from phase space, the glueball wave function, and the decay mechanism  are expected.

  \item How many distinct resonances exist around 1.7 GeV/$c^2$? The $f_0(1710)$ and $f_0(1790)$ were observed at BES and the $X(1810)$ was observed in $J/\psi\to\gamma\omega\phi$~\cite{Ablikim:2012ft}.
  \item What is the nature of the $f_0(2100)$? Besides $f_0(1710)$, large production of scalars around 2.1 GeV/$c^2$ is also observed in  $J/\psi\to\gamma\eta\eta$~\cite{Ablikim:2013hq}, $J/\psi\to\gamma\pi^0\pi^0$~\cite{Ablikim:2015umt} and  $J/\psi\to\gamma K_S K_S$~\cite{Ablikim:2018izx}. The pattern of production in the gluon-rich radiative $J/\psi$ decays agrees well with that of the ground state glueball and its first excitation as predicted by LQCD. It is notable that the $f_0(2100)$ is also largely produced in $p\bar{p}$ annihilation. An additional
way to unveil its nature is to measure the ratio of its decay modes into $\eta\eta'$ and $\eta'\eta'$. Furthermore, the number of existing scalars in the $f_0(2100)$ region needs clarification.
\end{itemize}

       \subsubsection{The tensor mesons}
\begin{figure}[tbp]
  \begin{center}
    \includegraphics[width=0.48\textwidth]{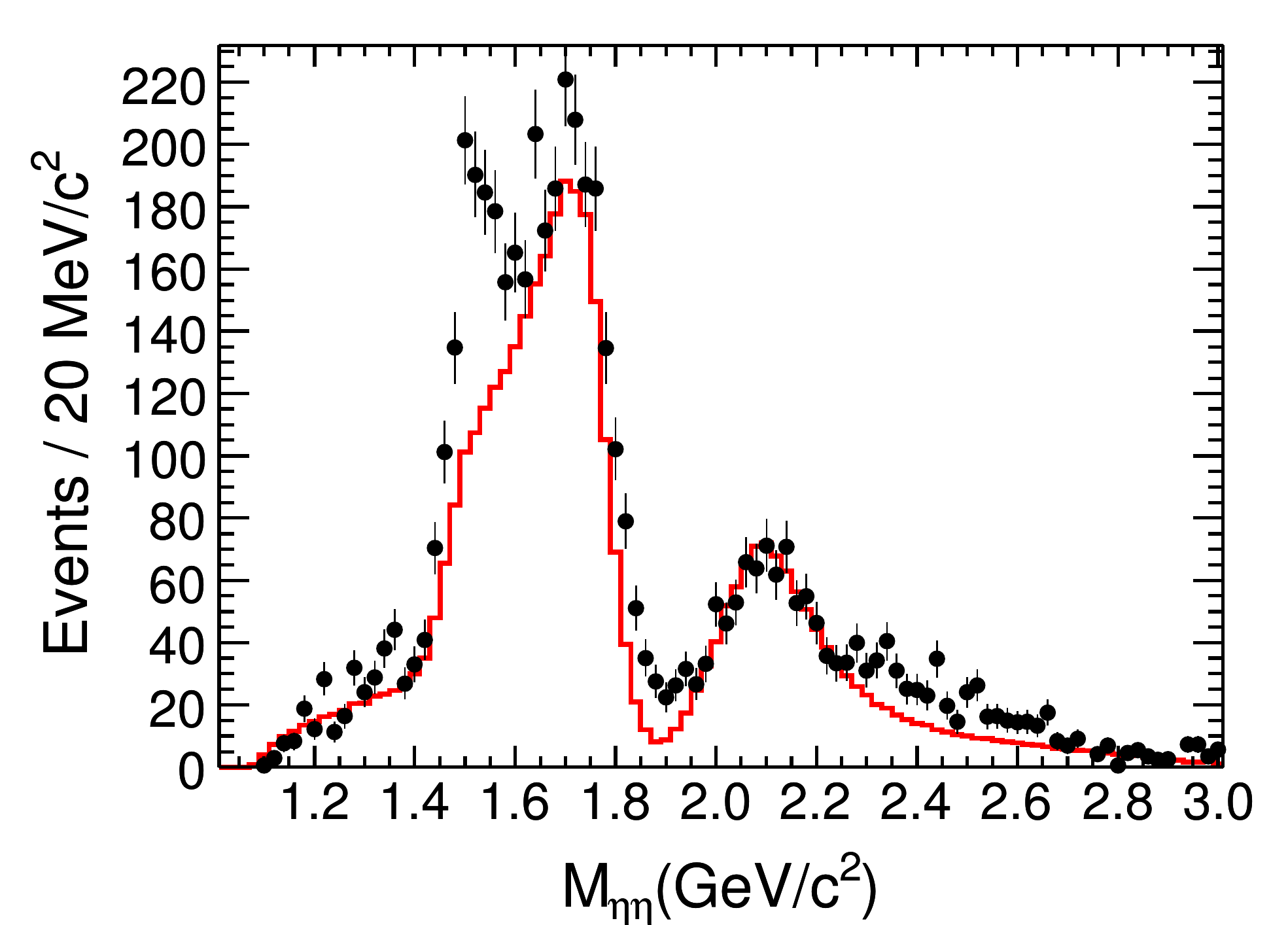}
    \includegraphics[width=0.48\textwidth]{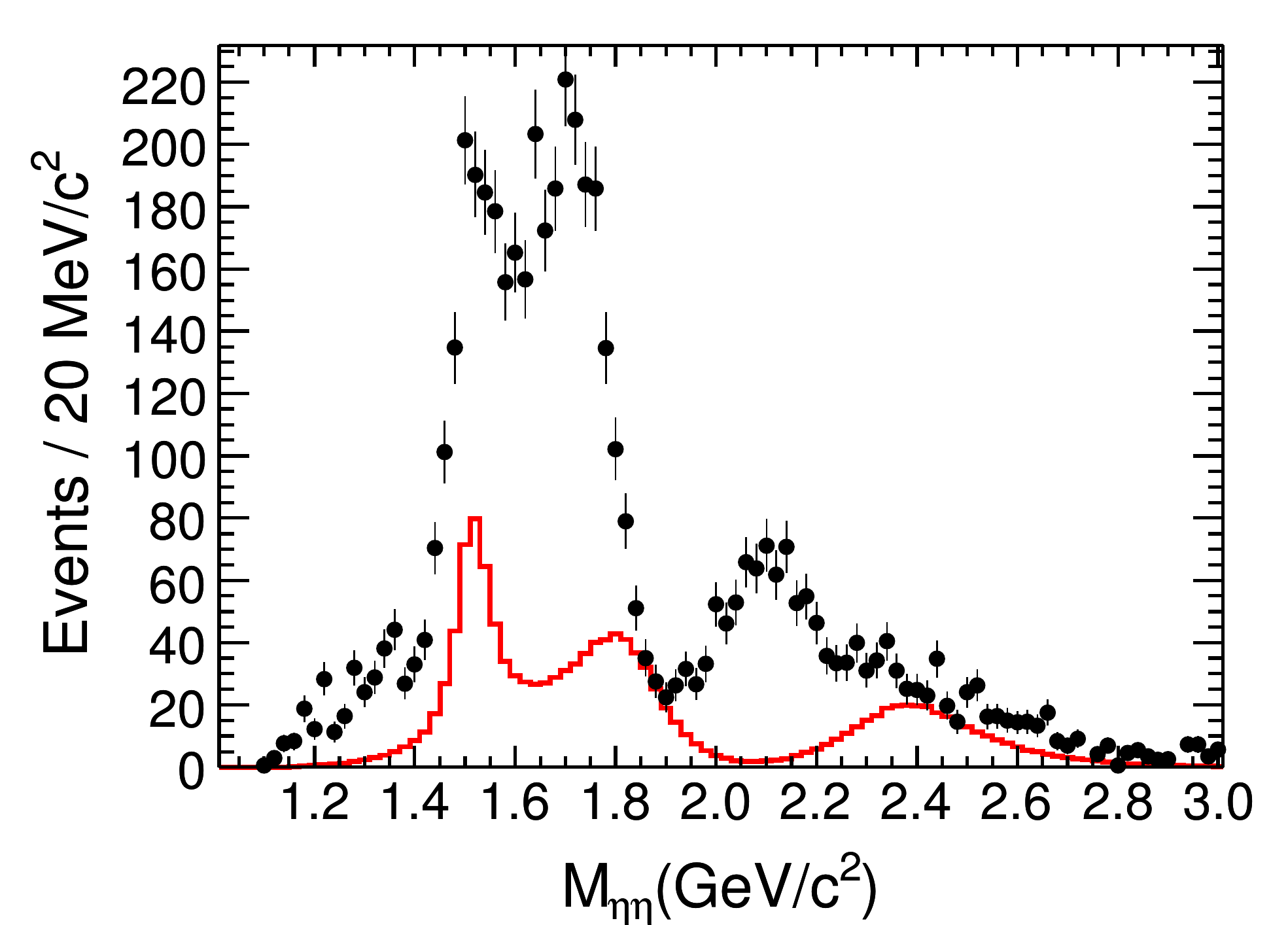}
    \caption{\label{fig:etaeta}Invariant mass distribution of $\eta\eta$ from
      $J/\psi\rightarrow \gamma\eta\eta$, and the projection of the
      PWA fit from \bes3: (a) the $0^+$ component, (b) the $2^+$ component \cite{Ablikim:2013hq}. Dots with error bars are data. Solid histograms are the projections of the PWA fit.}
  \end{center}
\end{figure}

\begin{figure}[tbp]
  \begin{center}
    \includegraphics[width=0.5\columnwidth]{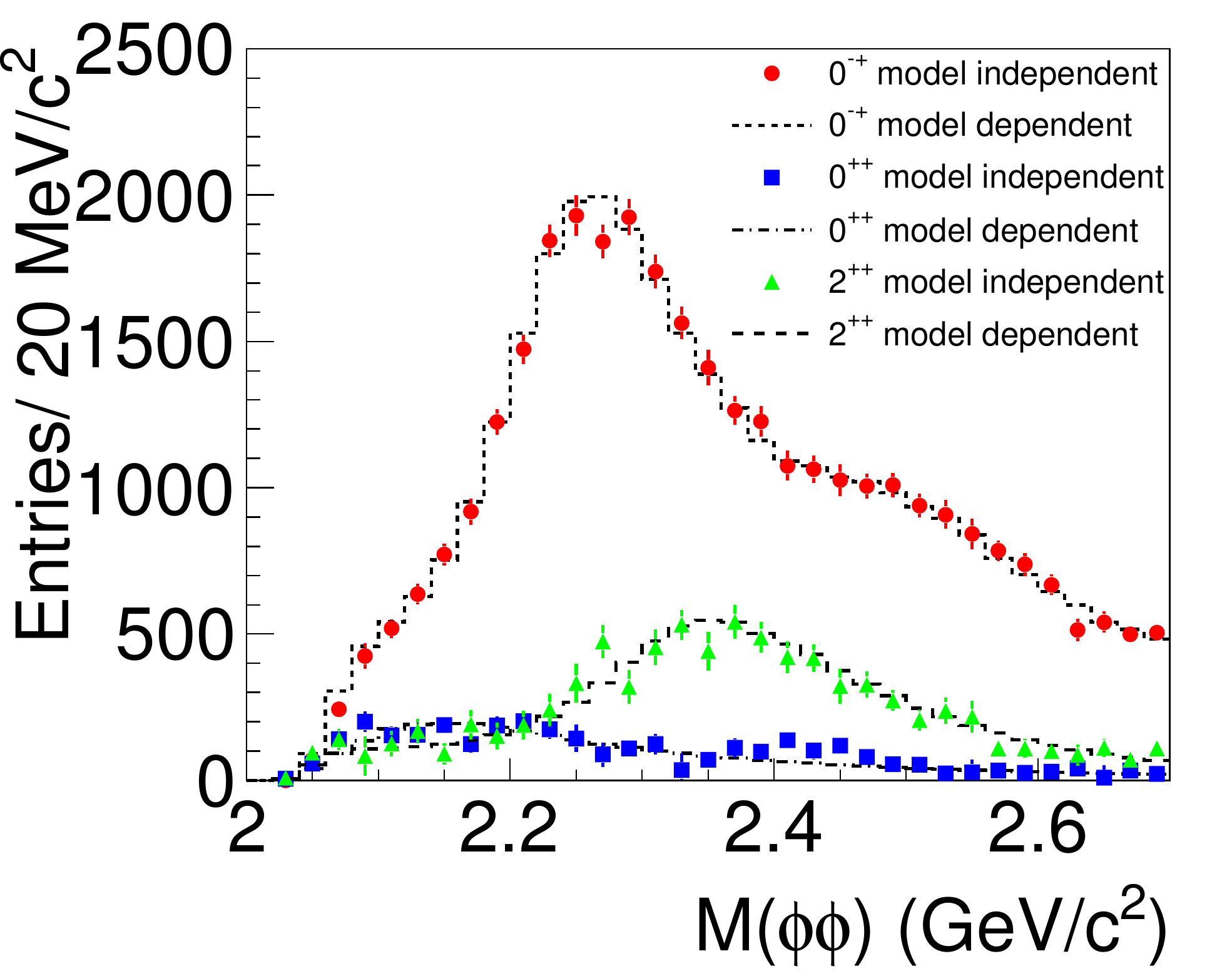}
    \caption{\label{fig:phiphi}Intensities of individual $J^{PC}$ components from the model-independent fit in the mass distribution of $\phi\phi$ from
      $J/\psi\to \gamma\phi\phi$, comparing with the projection of the
      model-dependent PWA fit from \bes3 \cite{Ablikim:2016hlu}. }
  \end{center}
\end{figure}

Within the quark model, there are two quark configurations, the $^3P_2(L=1,S=1,J=2)$  and the $^3F_2(L=3,S=1,J=2)$ nonets. Hence, the tensor sector is extremely busy and there are a large number of tensor states appear in the PDG~\cite{PDG}.
The three tensors $f_2(2010)$, $f_2(2300)$ and $f_2(2340)$ observed
in $\pi^- p\rightarrow \phi\phi n$~\cite{bibpiN} are also observed
in $J/\psi\rightarrow\gamma\phi\phi$~\cite{Ablikim:2016hlu}.
Figure~\ref{fig:etaeta} and~\ref{fig:phiphi} show the resulting PWA fit result of the
$\eta\eta$ and $\phi\phi$
invariant mass spectra.  The large production rate of the $f_2(2340)$
in $J/\psi\rightarrow\gamma\phi\phi$ and $J/\psi\rightarrow\gamma\eta\eta$~\cite{Ablikim:2013hq} indicates $f_2(2340)$ is a good candidate of tensor glueball.  Significant tensor contribution around 2.4~GeV/$c^2$ also presents in $J/\psi\to\gamma\pi^0\pi^0$~\cite{Ablikim:2015umt} and  $J/\psi\to\gamma K_S K_S$~\cite{Ablikim:2018izx}. However, the measured production rate of $f_2(2370)$ appears
to be substantially lower than the LQCD calculated
value~\cite{Yang:2013xba}. It is desirable to search for more decay modes  to establish
and characterize the lowest tensor glueball.

    \subsubsection{The pseudoscalar mesons}

  The $I=0$ $J^{PC}=0^{-+}$ ground states are the $\eta$ and the $\eta^\prime$. Only radial excitations are expected in the quark model for $0^{-+}$ states. The small number of expected pseudoscalars in the quark model provide a clean and promising environment for the search of glueballs.

\begin{figure}[tbp]
  \centering
    \includegraphics[width=0.48\textwidth]{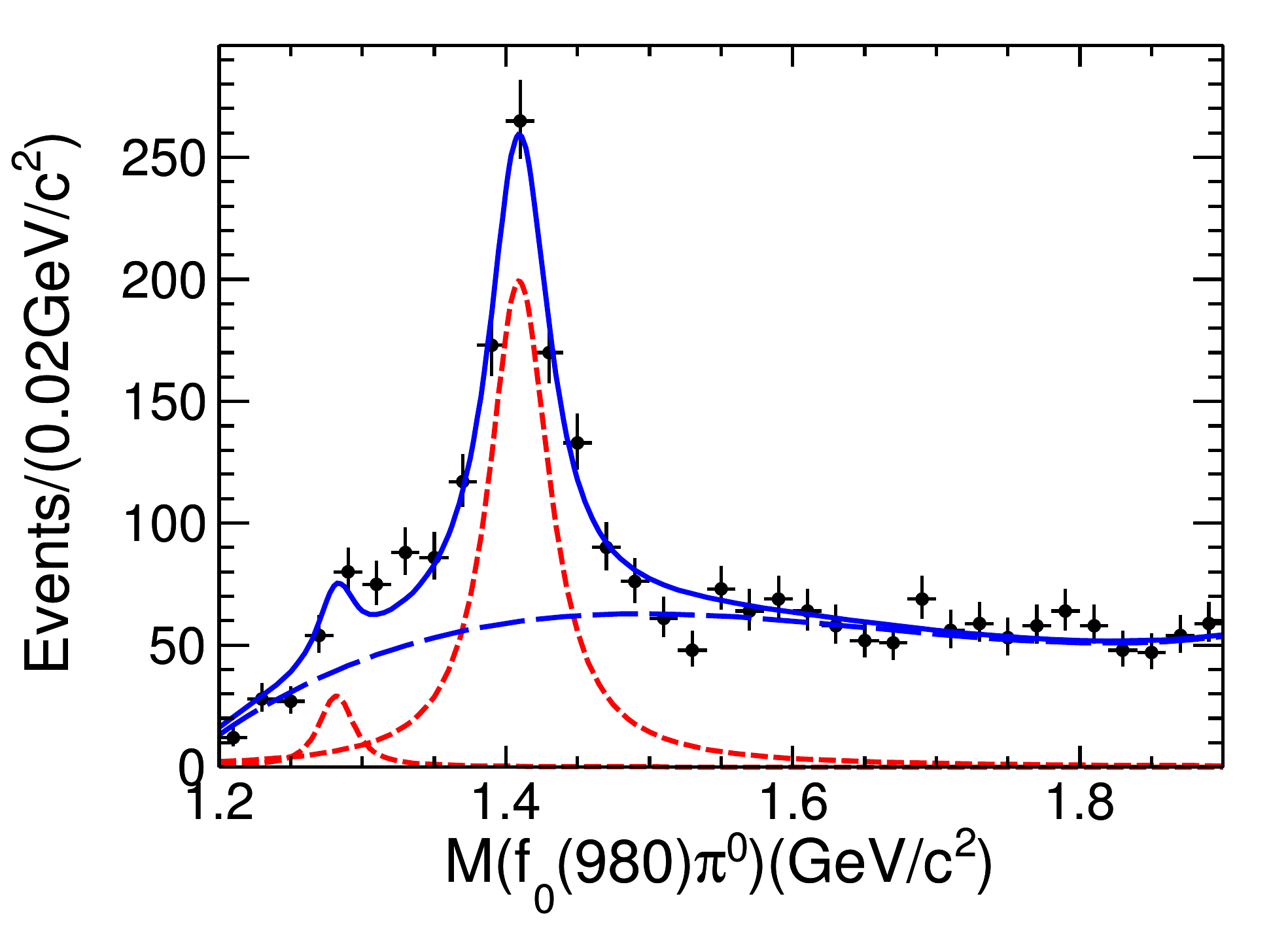}
    \includegraphics[width=0.48\textwidth]{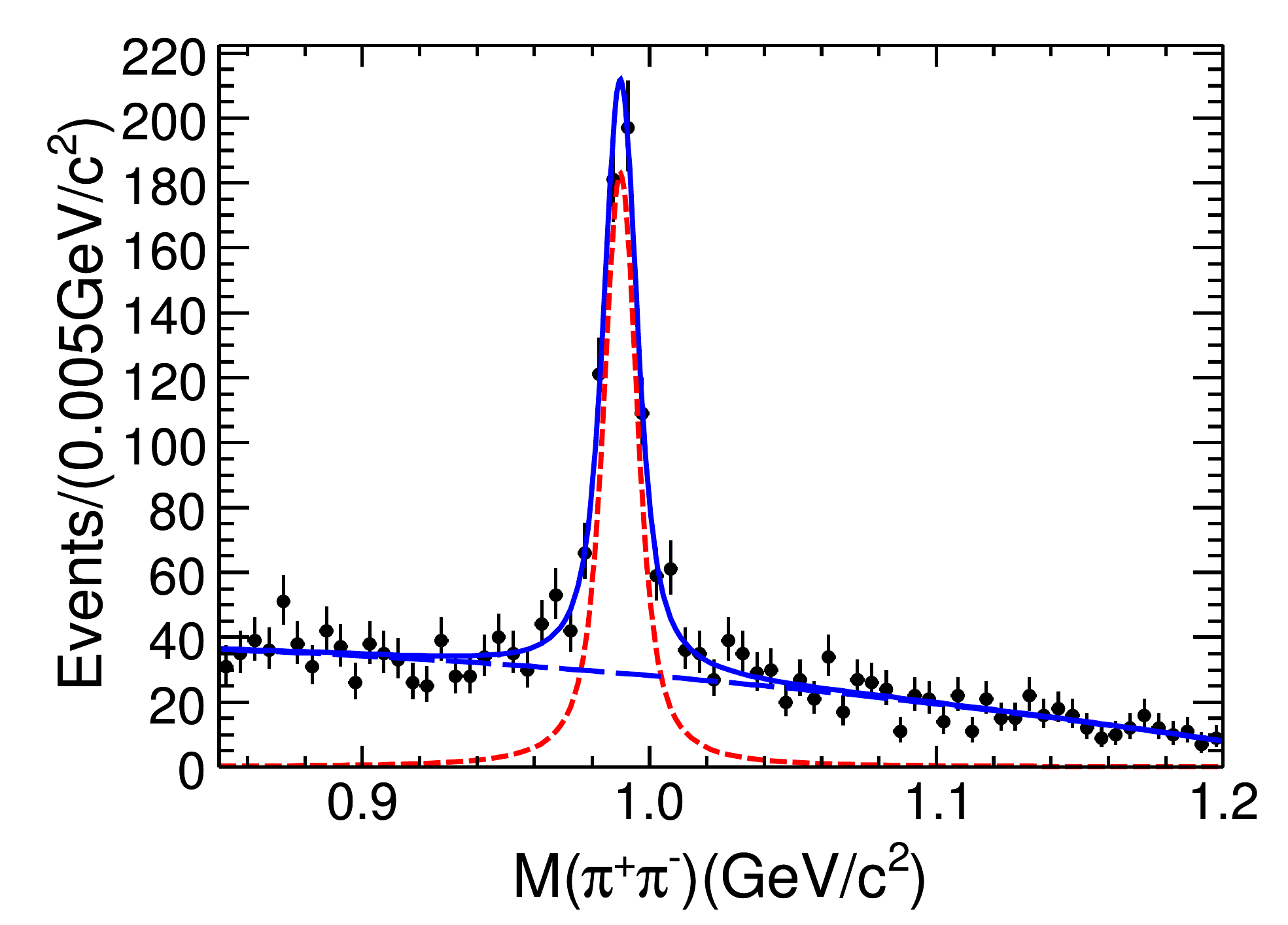}
  \caption{(a) Invariant mass of $f_0(980)\pi^0$; (b) Invariant mass of $\pi^+\pi^-$ with the
      $\pi^+\pi^-\pi^0$ ($3\pi^0$) mass in the $\eta(1405)$ mass
      region, measured at \bes3 \cite{BESIII:2012aa}. }
  \label{eta1405}
\end{figure}

A striking observation is that there are two pseudoscalar states near 1.4 GeV/$c^2$, the $\eta(1405)$ and the $\eta(1475)$, listed in the PDG~\cite{PDG}. The
$\eta(1405)$ as a supernumerary state was  proposed to be the ground state pseudoscalar glueball candidate. However, the extra state appears to be
relatively far from the expected mass of the pseudoscalar glueball of  2 GeV/$c^2$.
This is known as the long standing ``E-$\iota$ puzzle''~\cite{Masoni:2006rz}.  \bes3
reported the first observation of the isospin-violating decay $\eta(1405)
\rightarrow\pi^0
f_0(980)$ in $J/\psi\rightarrow\gamma 3\pi$
\cite{BESIII:2012aa}, together with an anomalous
lineshape of the $f_0(980)$, as shown in Fig.~\ref{eta1405}.
The $f_0$ mass, deduced from a Breit-Wigner
fit to the mass spectra, is slightly shifted compared to its nominal
value, with a width of $<11.8$~MeV ($90\%$ C.L.), much smaller than its
nominal value. The observed isospin violation is $(17.9\pm 4.2)\%$,
too large to be explained by $f_0(980)$-$a_0(980)$ mixing, also observed by \bes3~\cite{Ablikim:2010aa, Ablikim:2018pik}.
Based on this observation, Wu et al.\ \cite{Wu:2011yx} suggest that
a triangular singularity mixing $\eta\pi\pi$ and $K^* \bar{K}$ could be
large enough to account for the data. The splitting of $\eta(1405)$ and $\eta(1475)$ could also be due to this triangle anomaly.

It is also crucial to examine carefully  the existence of the $\eta(1295)$, in order to nail down the first excitation of $\eta$.
Alternatively, the $\eta(1295)$ could be explained as a misidentified $f_1(1285)$.

The $X(1835)$ observed in $J/\psi\to\gamma \eta^\prime \pi^+ \pi^-$, which will be discussed in Sec.~\ref{sec:multiquarks}, is determined to have $0^{-+}$ quantum numbers. In the same reaction, two additional structures, the $X(2120)$ and the $X(2370)$, are observed. However, the spin-parity of these two new structures are not yet determined. X(2370) is also observed in $J/\psi\to\gamma \eta^\prime K \bar K$. It is crucial to explore other decay modes of X(2370) and establish its spin-parity.

Aside from the $\eta(2225)$, very little is known in the pseudoscalar sector above 2 GeV/$c^2$ where the lightest pseudoscalar glueball is expected to be by LQCD calculations. A PWA of the decay $J/\psi\rightarrow \gamma \phi \phi$~\cite{Ablikim:2016hlu} is performed in order to study the intermediate states. The most remarkable feature of the PWA results is
that $0^{-+}$ states are dominant. The existence of the $\eta(2225)$
is confirmed and two additional pseudoscalar states, $\eta(2100)$ and X(2500), are found.

Besides the radiative $J/\psi$  decays, flavor filter reactions could also play an important role in unraveling the quark content of the pseudoscalars  ({\it e.g.}, $J/\psi\to\gamma X, X\to\gamma V$ and $J/\psi\to V X$, where $V$ stands for $\rho$, $\omega$, $\phi$ and $X$ stands for the pseudoscalars). $\eta(1475)\to\gamma\phi$ and $X(1835)\to\gamma\phi$ are observed in the decay of $J/\psi\to\gamma\gamma\phi$ at \bes3 ~\cite{Ablikim:2018hxj}, which indicates that both $\eta(1475)$ and $X(1835)$ contain a sizeable $s\bar{s}$ component.

  \subsection{Hybrids}
\begin{figure*}[tbp]
  \includegraphics[width=\textwidth]{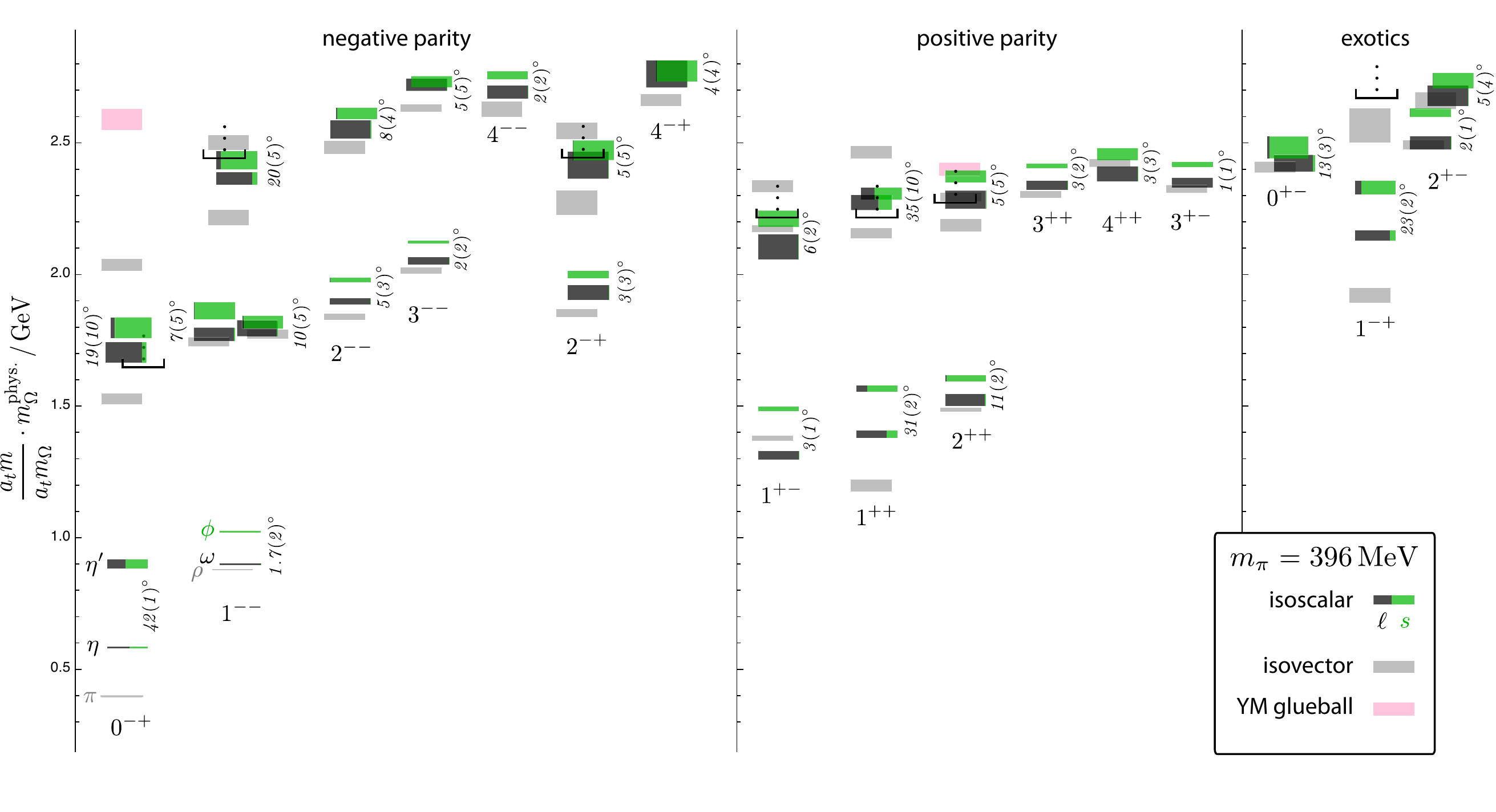}
  \caption{Light-quark nonstrange meson spectrum resulting from
    LQCD \cite{Dudek:2011tt}, sorted by the quantum
    numbers $J^{PC}$. Note that these results have been obtained with
    an unphysical pion mass, $m_\pi=396$ MeV/$c^2$.}
  \label{fig:hybrid}
\end{figure*}

Since the expected quantum numbers of low-lying glueballs are not exotic, they should
manifest themselves as additional states that cannot be
accommodated within the $q\bar{q}$ nonets. Their unambiguous identification is complicated by the fact that they can mix when overlapping with $q\bar{q}$ states of the same quantum numbers. An easy way to avoid mixing with regular mesons are additional degrees of freedom leading to exotic quantum numbers. Such degrees of freedom could arise from explicit gluonic contributions. Those particles carry a special name and are called hybrids.
In many models, some of the hybrid mesons can have a
unique signature, the already mentioned exotic (not allowed in a simple $q\bar{q}$ system) $J^{PC}$ quantum numbers. This signature simplifies
the spectroscopy of such exotic hybrid mesons because
they do not mix with conventional $q\bar{q}$ states. LQCD calculations support the existence
of exotic-quantum-number states within the meson spectrum, independent of specific models. As shown in Fig.~\ref{fig:hybrid}~\cite{Dudek:2011tt}, LQCD calculations consistently show that the $J^{PC} = 1^{-+}$ nonet is the lightest hybrid.

Currently, there are
three
experimental candidates for a light $1^{-+}$ hybrid: the
$\pi_1(1400)$ and the $\pi_1(1600)$, observed in diffractive reactions
and $\overline{p}N$ annihilation, and the $\pi_1(2015)$, seen only in
diffraction.
The $\pi_1(1400)$ has only been observed in its decay into the $\pi\eta$ final state, and
is generally considered too light to be a hybrid meson. Reviews  like ~\cite{Meyer:2010ku,Klempt:2007cp} provide a summary of the experimental studies.

An amplitude analysis of $\chi_{c1}\rightarrow\eta\pi^+\pi^-$ or
$\chi_{c1}\rightarrow\eta'\pi^+\pi^-$ was performed at \mbox{CLEO-c}~\cite{Adams:2011sq} and \bes3~\cite{Kornicer:2016axs}. For these final states, the only allowed $S$-wave
decay of the $\chi_{c1}$ goes through the spin-exotic $1^{-+}$ wave,
which then decays to $\eta(')\pi$. It turns out that a significant
contribution of an exotic $1^{-+}$ wave is needed to describe the data in
the $\eta'\pi^+\pi^-$ channel, but not in the $\eta\pi^+\pi^-$ channel.

While there is evidence for the isovector member of
the $J^{PC} = 1^{-+}$ nonet, we also expect two isoscalar states
($\eta_1$ and $\eta_1^\prime$
), which are crucial to establish the nonet nature of the $\pi_1$ states. However, there is still no experimental
evidence for these two isoscalar states. The gluon-rich radiative $J/\psi$ decays may provide an ideal laboratory for the search for such isoscalar $1^{-+}$ states. Model predictions for their decay modes are $f_1\eta$, $a_1\pi$ and $\eta\eta^\prime$, etc~\cite{Isgur:1985vy, Page:1998gz, Huang:2010dc, Chen:2010ic}.

  \subsection{Multiquarks}\label{sec:multiquarks}

An early quark model prediction was the existence of multiquark states, specifically
bound meson-antimeson molecular states. In the light quark sector the
$f_0(980)$ and $a_0(980)$ are considered to be strong candidates for $K\bar{K}$ molecules.
However, in general, it is challenging to definitively identify a light multiquark
state in an environment of many broad and often overlapping conventional states.

Two generic types of multiquark states have been often discussed. Molecular states are a loosely
bound state of a pair of mesons near threshold. Tetraquarks are tightly
bound diquark-antidiquark states. A prediction of tetraquark models is that they come in flavour multiplets. In addition,  an enhancement near threshold may also arise from threshold effects due to rescattering of the two outgoing mesons close to the
threshold. This could result in mass shifts due
to the thresholds. Couple-channel effects result in the mixing of two-meson
states with $q\bar{q}$ resonances.

In general, a multiquark state is expected to be broad since it can easily
decay into mesons and/or baryons when its mass is above the mass threshold
for producing these hadrons. Multiquark states may only be experimentally observable
when their masses are near these mass thresholds either below or just above them;
otherwise the multiquark states might be too wide to be experimentally
distinguishable from non-resonant background.

\begin{figure}[tp]
  \begin{center}
    \includegraphics[width=0.8\textwidth]{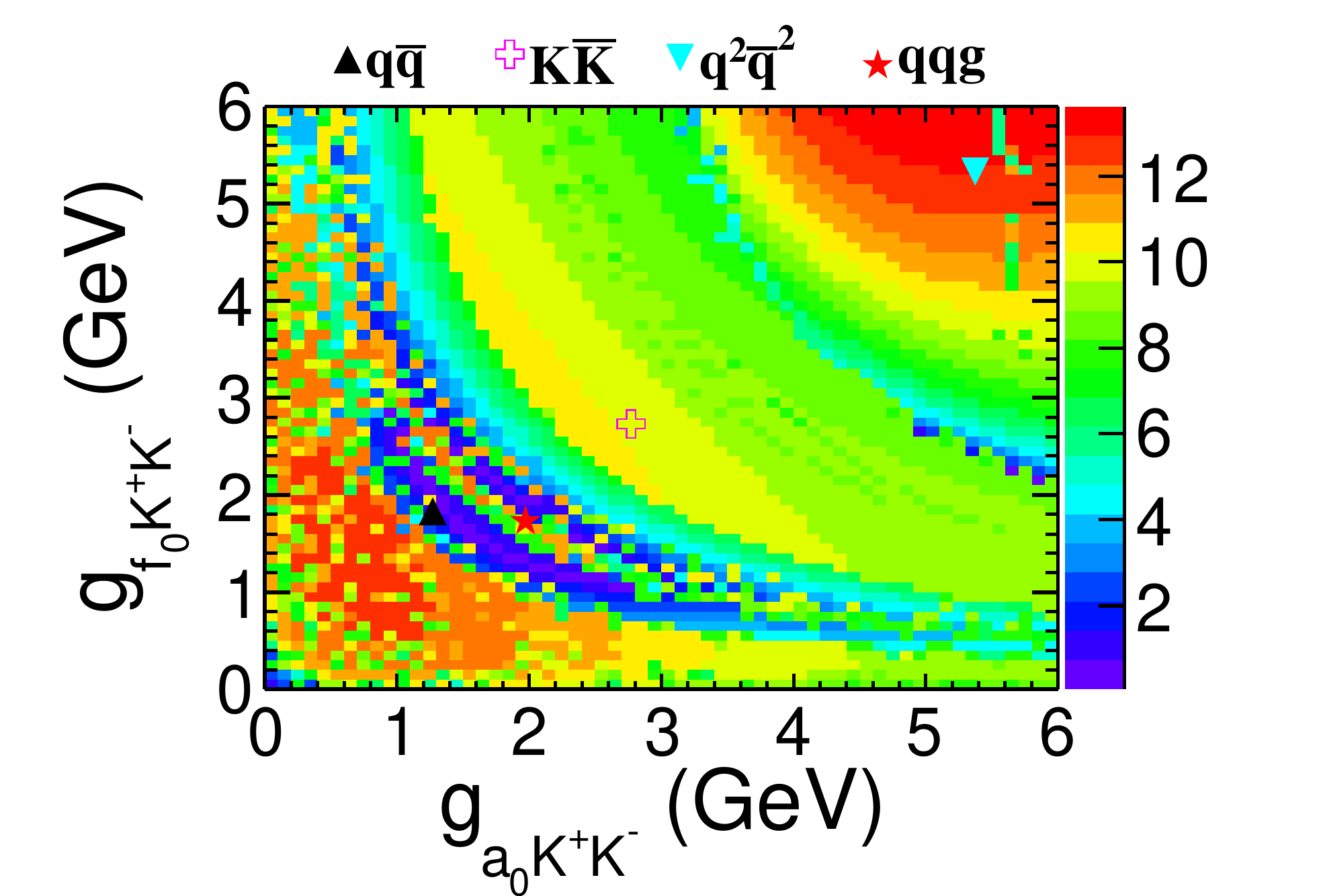}
    \caption{\label{signif}The statistical significance of the signal scanned in the two-dimensional space of $g_{a_{0}K^{+}K^{-}}$ and $g_{f_{0}K^{+}K^{-}}$. The regions with higher statistical significance indicate larger probability for the emergence of the two coupling constants. The markers indicate predictions from various illustrative theoretical models.}
  \end{center}
\end{figure}

After the discoveries of $a_0(980)$ and $f_0(980)$ several decades ago, explanations about the nature of these two light scalar mesons have still been controversial. These two states, with similar masses but different decay modes and isospins, are difficult to accommodate in the traditional quark-antiquark model, and many alternative formulations have been proposed to explain their internal structure, including tetra-quarks~\cite{Jaffe:1976ig,Alford:2000mm,Maiani:2004uc,Maiani:2007iw,Hooft:2008we}, $K\bar{K}$ molecule~\cite{Weinstein:1990gu}, or quark-antiquark gluon hybrid~\cite{gluon}. Further insights
into $a_0(980)$ and $f_0(980)$ are expected from their mixing~\cite{Achasov:1979xc}.  The mixing mechanism in the system of $a_0(980)-f_0(980)$ is considered to be a sensitive probe to clarify the nature of these two mesons. In
particular, the leading contribution to the isospin-violating mixing
transition amplitudes for $f_{0}(980)\to a^{0}_{0}(980)$ and
$a^{0}_{0}(980)\to f_{0}(980)$, is shown to be dominated by the
difference of the unitarity cut which arises from the mass difference
between the charged and neutral $K\bar{K}$ pairs. As a consequence, a
narrow peak of about 8 MeV$/c^2$ is predicted between the charged and
neutral $K\bar{K}$
thresholds.
The corresponding signal is predicted
in the isospin-violating processes of $J/\psi\to\phi a^{0}_{0}(980)$~\cite{Wu:2007jh,Hanhart:2007bd} and $\chi_{c1}\to\pi^{0} f_{0}(980)$~\cite{Wu:2008hx}.
The signals of $f_0(980)\to a^0_0(980)$ and $a^0_0(980)\to f_0(980)$ mixing are first observed in $J/\psi\to\phi f_{0}(980)\to\phi a^{0}_{0}(980)\to\phi\eta\pi^{0}$ and $\chi_{c1}\to\pi^{0} a^{0}_{0}(980)\to\pi^{0} f_{0}(980)\to\pi^{0}\pi^{+}\pi^{-}$ at \bes3~\cite{Ablikim:2010aa, Ablikim:2018pik}.
The statistical significance of the signal versus the values of $g_{a_{0}K^{+}K^{-}}$ and $g_{f_{0}K^{+}K^{-}}$ is shown in Fig.~\ref{signif}. The regions with higher statistical significance indicate larger probability for the emergence of the two coupling constants. This direct measurement of $a_0(980)-f_0(980)$ mixing is a sensitive probe to the internal structure of those ground state scalars and sheds important light on their nature. The new results from \bes3 provide critical constraints to the development of theoretical models for $a_0(980)$ and $f_0(980)$. 

\begin{figure}[htbp]
  \centering
    \includegraphics[width=0.48\textwidth]{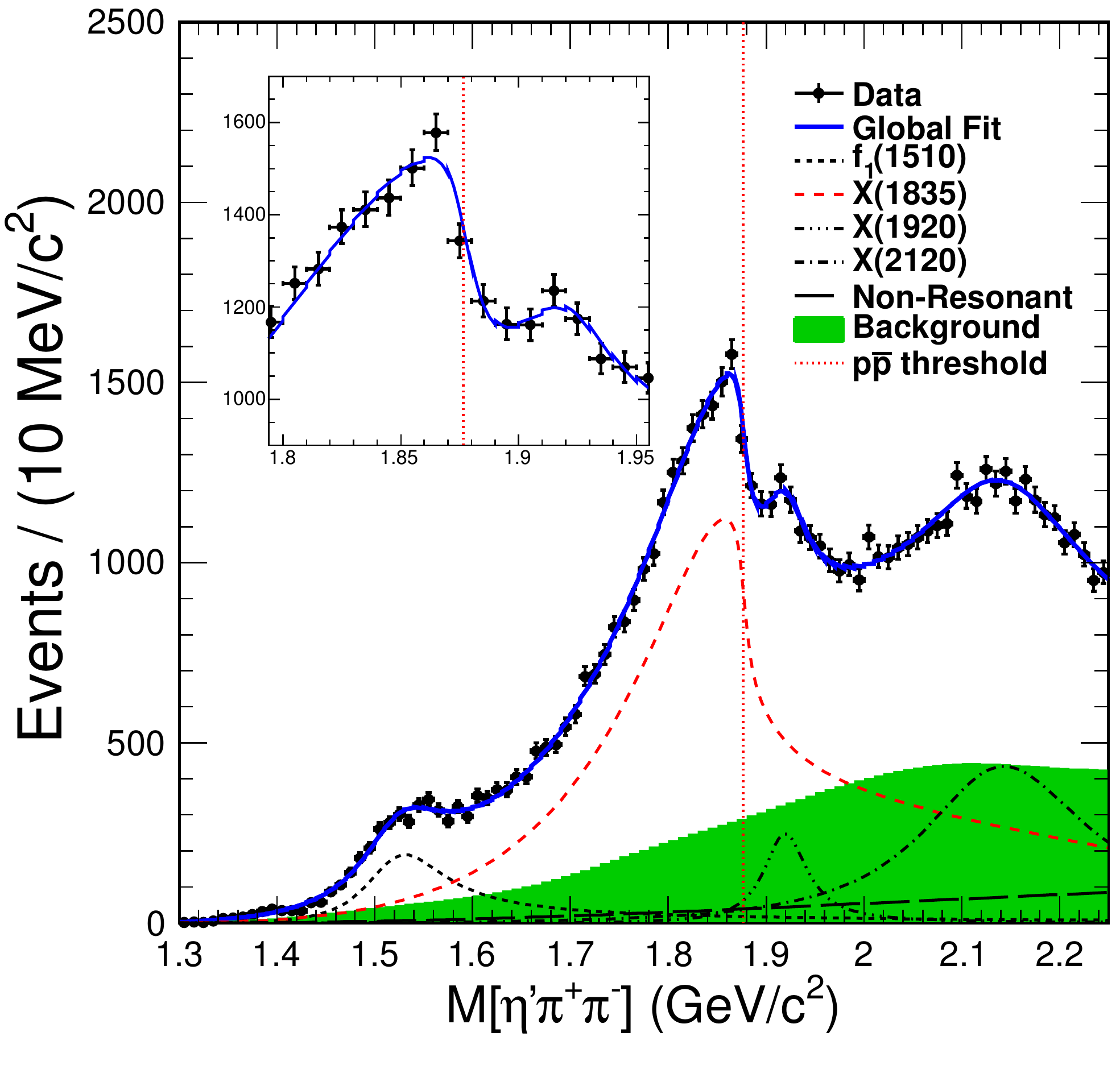}
  \includegraphics[width=0.48\textwidth]{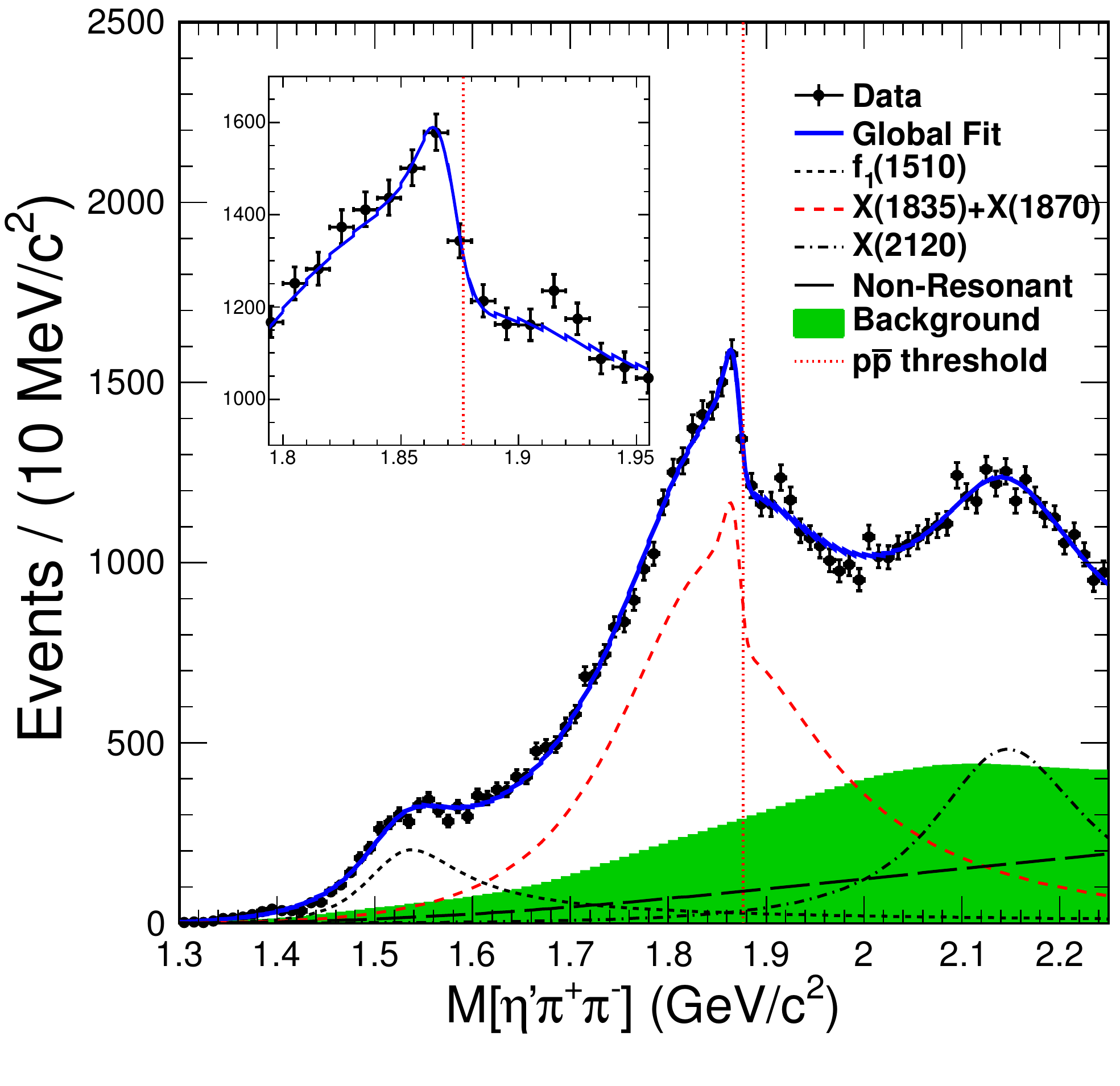}
  \caption{An anomalous line shape of the $\eta^{\prime}\pi^{+}\pi^{-}$ mass spectrum near the $p\bar{p}$ mass threshold in $J/\psi\rightarrow\gamma\eta^{\prime}\pi^{+}\pi^{-}$. (a) shows
  the fit results with a Flatt\'{e} formula and (b) shows the fit results with the coherent sum of two Breit-Wigner amplitudes }
  \label{x1835}
\end{figure}

The state $X(1835)$ was first observed by the BES experiment as a peak
in $J/\psi\rightarrow\gamma\eta^{\prime}\pi^{+}\pi^{-}$ decays~\cite{x1835_bes2}. This observation was later
confirmed by \bes3~\cite{x1835_bes3}. The $X(1835)$ was also observed in the $\eta K^{0}_{S} K^{0}_{S}$ channel, where its spin-parity was determined to be $J^{P}=0^{-}$ by a PWA~\cite{x1835_qiny}.
An anomalously strong enhancement at the proton-antiproton ($p\bar{p}$)
mass threshold, dubbed $X(p\bar{p})$,
was first observed by BES in $J/\psi\rightarrow\gamma p\bar{p}$ decays~\cite{xpp_bes2};
this observation was confirmed by \bes3~\cite{xpp_bes3} and CLEO~\cite{xpp_cleo}.
This enhancement was subsequently determined to have spin-parity
$J^{P}=0^{-}$ by \bes3~\cite{xpp_bes3pwa}.
Using a high-statistics sample of $J/\psi$ events, \bes3 studied the $J/\psi\rightarrow\gamma\eta^{\prime}\pi^{+}\pi^{-}$ process and
  observed a significant abrupt change in the slope of the $\eta^{\prime}\pi^{+}\pi^{-}$ invariant mass distribution at the
  proton-antiproton ($p\bar{p}$) mass threshold~\cite{Ablikim:2016itz}.
  Two models are used to characterize the $\eta^{\prime}\pi^{+}\pi^{-}$ line shape
  around 1.85~GeV$/c^{2}$: one which explicitly incorporates the opening of a
  decay threshold in the mass spectrum (Flatt\'{e} formula) (Fig.~\ref{x1835}(a)), and another
  which is the coherent sum of two resonant amplitudes(Fig.~\ref{x1835} (b)).
  Both fits show almost equally good agreement with data,
  and suggest the existence of either a broad state with strong couplings to the final state $p\bar{p}$
  or a narrow state just below the $p\bar{p}$ mass threshold.
  The goodness-of-fit are equivalent for both fits. One supports the existence of a $p\bar{p}$ molecule-like
  state and the other a bound state with greater than $7\sigma$ significance. Further study of the fine lineshape of $X(1835)$ in other decay modes will provide conclusive information on the nature of the state.

An analysis of  $J/\psi\to pK^-\bar\Lambda$ was performed at BES~\cite{pkl}.
Enhancements both at  the $p\bar \Lambda$  and the $K^-\bar\Lambda$ mass thresholds
are observed. Further investigations in other decay modes and production mechanisms are needed to clarify the nature of those structures.

\section{Baryon spectroscopy}

Baryons are the basic building blocks of our world. Since baryons represent the
simplest system in which all the three colors of QCD neutralize into
colorless objects, understanding the baryon structure is absolutely necessary
before we claim that we really understand QCD.
Given many recent experimental results, our present understanding of baryon spectroscopy is clearly incomplete.
Many fundamental issues in baryon spectroscopy are still not well
understood~\cite{Capstick:2000dk, Klempt:2009pi}.
Most important among them is the
problem of missing resonances:
in quark models based on approximate flavor SU(3) symmetry it is expected
that resonances form multiplets; many excited
states are predicted which have not been observed (for a review see Ref.~\cite{Capstick:2000qj}).
More recently, LQCD calculations~\cite{Edwards:2011jj} have also predicted
a similar pattern as quark models.
The possibility of new, as yet unappreciated,
effective symmetries could be addressed with the accumulation of more data. The new
symmetries may not have obvious relation with QCD, just like nuclear
shell model and collective motion model.

In addition to baryons made of $u$ and $d$ quarks, the search for hyperon resonances remains an important
challenge. Some of the lowest excitation resonances have not yet been
 experimentally established, which are necessary to establish the spectral pattern
of hyperon resonances.

\begin{figure}[htbp]
\vspace{-0.8cm}
\hspace{1.5cm}\includegraphics[width=0.9\textwidth]{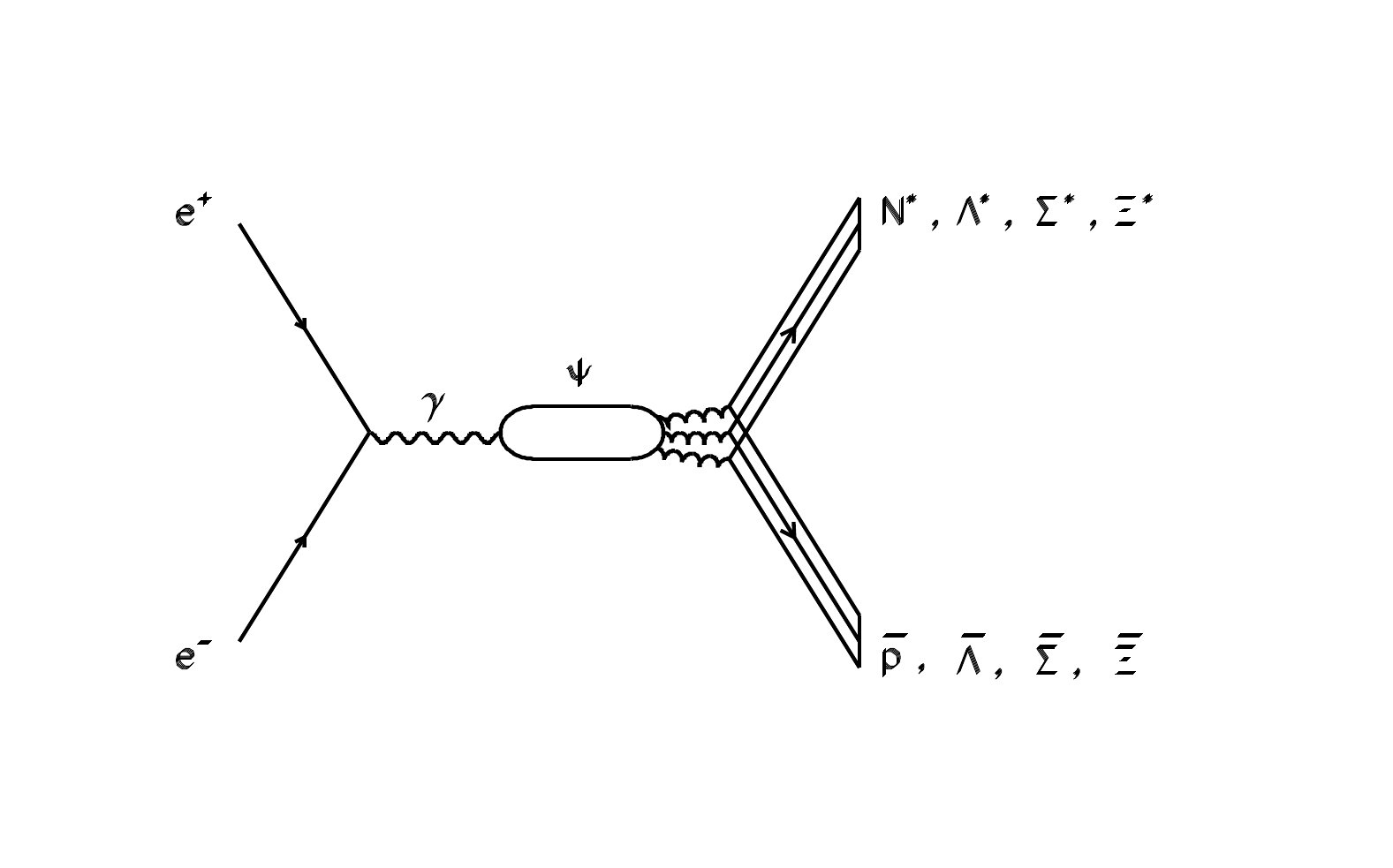}
\vspace{-1.1cm}
\caption{$\bar pN^*$, $\bar\Lambda\Lambda^*$,
$\bar\Sigma\Sigma^*$ and $\bar\Xi\Xi^*$ production
from $e^+e^-$ collision through $\psi$ meson.}
\label{fig:baryon}
\end{figure}

Charmonium decays provide an
excellent place to study excited nucleons and hyperons -- $N^*$,
$\Lambda^*$, $\Sigma^*$ and $\Xi^*$ resonances~\cite{Zou:2000wg}.
The corresponding Feynman graph for the production of these excited
nucleons and hyperons is shown in Fig.~\ref{fig:baryon} where $\psi$
represents charmonium.

Complementary to other facilities, the baryon program at \bes3 has several advantages~\cite{Zou:2001uc}. For instance, $\pi N$ and $\pi\pi N$ systems from
$\psi\to\bar NN\pi$ and $\bar NN\pi\pi$ processes have an isospin of 1/2 due to isospin conservation; $\psi$ mesons decay to baryon-antibaryon pairs through three or
more gluons providing a favorable place for producing hybrid (qqqg) baryons,
and for searching for some ``missing'' $N^*$ resonances which have weak
coupling to both $\pi N$ and $\gamma N$, but stronger coupling to $g^3N$. The phase space of $\psi(3686)$ decays is larger than that of $J/\psi$ decays, thus having more potential to investigate higher excitations of baryons and hyperons.
In a PWA of $\psi(3686)\to p \bar{p} \pi^0$~\cite{Ablikim:2012zk}, two new $N^*$ resonances $N(2300)$ and $N(2570)$ have been observed with $J^P$ assignments of $1/2^+$ and $5/2^-$, respectively.

Besides the high-statistics data samples of charmonium events, the large number of $\Lambda_c$ hadronic decays collected at \bes3 provide a novel laboratory for studying light baryon excitations.

\section{\bes3 amplitude analysis}

The basic task of experimental study of hadron spectroscopy is to systematically map out all the resonances with the determination of their properties like mass, width, spin-parity as well as partial decay widths with high sensitivity and accuracy. Extracting resonance properties from experimental data is however far from straightforward; resonances tend to be broad and plentiful, leading to intricate interference patterns, or they are buried under a background in the same and in other waves. The key to success lies in high statistical precision complemented with sophisticated analysis methods. PWA or amplitude analysis techniques are the state-of-the-art way to disentangle contributions from
individual, and even small, resonances and to determine
their quantum numbers. Multiparticle decays are usually
modeled using the phenomenological approach of the isobar
model, which describes multiparticle final states by
sequential two-body decays into intermediate resonances
(isobars), that eventually decay into the final state observed
in the experiment. Event-based fits allow one to
take into account the full correlation between final-state
particles.

Facing the extremely high statistics at \bes3, the PWA fits will be computationally very expensive. The pioneer approach of harnessing GPU parallel acceleration in PWA was performed in the
framework of \bes3~\cite{Battaglieri:2014gca}. The GPUPWA framework~\cite{Berger:2010zza} provides facilities for amplitude calculation, minimization and plotting and is widely used for analyses at \bes3. A high performance computing cluster has been established in the computing center at IHEP, which is equipped with hundreds of GPUs. In addition, for baryon spectroscopy analyses, the amplitudes can be extremely complicated. FDC-PWA, a package for automatic Feynman diagram calculation~\cite{Wang:2004du}, has been extensively used to generate a complete set of Fortran source code for PWA amplitudes.

One notoriously difficult problem is the parameterization of the dynamical properties of resonances, especially for those resonances related to thresholds. Coupled-channel analyses are mandatory to extract the partial width and other pole properties. The correct analytical properties of the amplitude are essential for
an extrapolation from the experimental data into the complex plane in order to determine the pole
positions. The proper implementation of dynamical function in PWA requires cooperation between theorists and experimentalists. In the PWA of $J/\psi\to\gamma\pi^0\pi^0$~\cite{Ablikim:2015umt},  $J/\psi\to\gamma K_S K_S$~\cite{Ablikim:2018izx} and $J/\psi\rightarrow\gamma\phi\phi$~\cite{Ablikim:2016hlu}, the results of  mass-independent amplitude analysis are provided, which extract a piecewise function that describes the dynamics of the two-body meson system while making minimal
assumptions about the properties and number of poles in the amplitude. Such a model-independent description allows the experiment-theory cooperation to develop phenomenological models, which can then be used to fit experimental data in the future. Global studies in different reactions and kinematics across experiments are also needed to clarify the underlying production mechanisms.

\section{Other physics opportunities}

\subsection{Light meson decays}

Light meson decays are an important tool for studies
of the strong interaction in the non-perturbative region and for
determination of some SM parameters.
As the neutral members of the ground state pseudoscalar nonet, both $\eta$ and $\eta^\prime$
play an important role in understanding low energy quantum
chromodynamics (QCD). Decays of the $\eta/\eta^\prime$ probe a wide variety of
physics issues, {\it e.g.}, $\pi^0-\eta$ mixing, light quark masses
and pion-pion  scattering.
 In particular the $\eta^\prime$ meson, much heavier than the
Goldstone bosons of broken chiral symmetry, plays a special role
as  the predominant singlet state arising from
the strong axial $U(1)$ anomaly. In addition, the decays of
both $\eta$ and $\etap$ mesons are
used to search for
processes beyond the Standard Model (SM) and to test fundamental
discrete symmetries.

The main decays of the $\eta/{\eta}^{\prime}$ mesons
are hadronic and radiative.
Alternatively, one can divide the decays
into the two following
classes.  The first class consists of hadronic decays into three
pseudoscalar mesons, such as ${\eta}^{\prime}$ ${\to}$
${\eta}{\pi}{\pi}$. Those processes are already included
in the lowest order, ${\cal O}(p^2)$, of chiral perturbation
theory  (ChPT)~\cite{Gasser:1983yg}.
The second class includes anomalous processes involving an
odd number of pseudoscalar mesons,
such as ${\eta}^{\prime} \to \rho^0 \gamma$ and ${\eta}^{\prime} \to \pi^+\pim\pi^+\pim$.
They are
driven by the ``Wess-Zumino-Witten'' (WZW) term \cite{Wess:1971yu,Witten:1983tw}
which enters at  order ${\cal O}(p^4)$ order~\cite{Bijnens:1989jb}.
The dynamics of $\eta^\prime$ decays remains a subject of extensive studies aiming at precision tests of ChPT in the $SU_L(3)\times SU_R(3)$ sector
({\it i.e.}, involving an $s$ quark).
Model-dependent approaches for describing low-energy meson
interactions, such as vector meson dominance
(VMD)~\cite{Sakurai:1960ju,Landsberg:1986fd},
and the large number of colors, $N_C$, extensions of ChPT~\cite{Kaiser:2000gs},
together with dispersive methods, could be extensively
tested in ${\eta}^{\prime}$ decays.

Due to the high
production rate of light mesons in charmonium ({\it e.g.}, $J/\psi$) decays, the \bes3 experiment also offers a unique
possibility to investigate the light meson decays. The decays of
$J/\psi\to\gamma\eta(\eta^\prime)$ and
$J/\psi\to\phi\eta(\etap)$
provide clean and efficient sources of $\eta/\etap$ mesons
for the decay studies.
During several run periods from 2009 to 2019, a total data sample
 of $10^{10}$ J/psi events was collected with the \bes3 detector~\cite{Ablikim:2012cn,Ablikim:2016fal}.
In recent years much important progress on $\eta/\eta^\prime$ decays was achieved at the \bes3
experiment.  In addition to the improved accuracy of the
branching fractions of $\eta^\prime$,  observations of
new $\eta^\prime$  decay modes, including $\eta^\prime\rightarrow\rho^\mp\pi^\pm$~\cite{Ablikim:2016frj} and $\eta^\prime\rightarrow \gamma
e^+e^-$~\cite{Ablikim:2015wnx}, $\eta^\prime\rightarrow\pi^+\pi^-\pi^+\pi^-$,
and $\eta^\prime\rightarrow\pi^+\pi^-\pi^0\pi^0$~\cite{Ablikim:2014eoc}, have been reported for the first time.
The precision of the $\eta'\to \pi^+\pi^-\gamma$  $M(\pi^+\pim)$ distribution
from \bes3~\cite{Ablikim:2018rho} with clear $\rho^0-\omega$ interference is comparable
to the $\EE\to\pi^+\pim$ data and allows comparison of these two reactions
in both model-dependent and model-independent ways~\cite{Stollenwerk:2011zz, Kubis:2015sga}.  A further investigation on
the extra contribution,   i.e., the box anomaly or $\rho(1405)$, is necessary besides the contributions from $\rho^0(770)$ and $\omega$.
In particular
a competitive extraction of the $\omega\to\pi^+\pim$ branching fraction is possible~\cite{Hanhart:2016pcd}.
%It is found
%that an extra contribution is necessary to describe the data besides the
%contributions from $\rho^0(770)$ and $\omega$.

\begin{table}
\begin{center}
%\centering
 \caption{\label{tab:eta} The available $\eta/\eta^\prime$ decays calculated with the expected $1\times 10^{10}$ $J/\psi$ events at \bes3.}
 \begin{tabular}{l c c c }\hline\hline
        Decay Mode    &       $\mathcal{B}$ ($\times 10^{-4}$) ~\cite{PDG}   & $\eta/\eta^\prime$ events \\ \hline
      $J/\psi\rightarrow\gamma\eta^\prime$ &$51.5\pm1.6$ &    $5.2\times 10^7$ \\ \hline
          $J/\psi\rightarrow\gamma\eta$ &$11.04\pm0.34$ &$1.1\times 10^7$\\  \hline
       $J/\psi\rightarrow\phi\eta^\prime$ & $7.5\pm0.8$ &    $7.5\times 10^6$ \\ \hline
          $J/\psi\rightarrow\phi\eta$ & $4.5\pm0.5$&   $4.5\times 10^6$ \\  \hline
 $J/\psi\rightarrow\omega\eta$& $17.4\pm2.0$ & $1.7\times 10^7$\\ \hline
  $J/\psi\rightarrow\omega\eta^\prime$& $1.82\pm0.21$ & $1.8\times 10^6$\\ \hline

                  \end{tabular}
                  \end{center}
\end{table}

Despite the impressive progress, many $\eta/\eta^\prime$ decays are
still to be observed and explored.  With a sample of $10^{10}$ $J/\psi$ events collected, the available $\eta$ and $\eta^\prime$ events
from radiative decays of $J/\psi\to \gamma\eta$, $\gamma\eta^\prime$, and hadronic decays of $J/\psi\rightarrow\phi(\omega)\eta$, $\phi(\omega)\eta^\prime$, are summarized in Table~\ref{tab:eta}, making further more detailed
$\eta/\eta^\prime$ studies possible. A list of the specific decay channels
where the new data are expected to have an important impact is shown in Table \ref{tab:phseta}.

\begin{table}[htb]
\caption{\label{tab:phseta}A few topics for the $\eta$ and $\eta^\prime$ programs. }

\begin{tabular}{|c|c|c|c|}
  \hline
   %after \\: \hline or \cline{col1-col2} \cline{col3-col4} ...
  $\eta$ decay mode & physics highlight & $\eta^\prime$ mode & physics highlight \\
   \hline
  $\eta\to\pi^0 2\gamma$ & ChPT & $\eta^\prime\to\pi\pi$ & CPV \\
  $\eta\to\gamma B$ & leptophobic dark boson & $\eta^\prime\to 2\gamma$ & chiral anomaly \\
  $\eta\to 3\pi^0$ & $m_u - m_d$ & $\eta^\prime\to\gamma\pi\pi$ & box anomaly, form factor \\
  $\eta\to \pi^+\pi^-\pi^0$ & $m_u - m_d$, CV & $\eta^\prime\to\pi^+\pi^-\pi^0$ & $m_u - m_d$, CV \\
  $\eta\to 3\gamma$ & CPV & $\eta^\prime\to \pi^0\pi^0\eta$ & cusp effect~\cite{Kubis:2009sb} \\

  \hline
\end{tabular}

\end{table}

In addition, the high production rate of $\omega$ in $J/\psi$ hadronic decays, {\it e.g.}, $\mathcal{B}(J/\psi\rightarrow\omega\eta)=(1.74\pm0.20)\times 10^{-3}$~\cite{PDG},  allows one to select a clean sample of $1.7\times 10^7$ events, which offers a unique opportunity to test the theoretical calculations~\cite{theory1,theory2,theory3} by investigating the Dalitz plot of $\omega\rightarrow \pi^+\pi^-\pi^0$.

\subsection{Two photon physics}

Production of resonances by two-photon fusion, as well as decays of resonances into two photons, provide important information on hadron structure. The anomalous magnetic moment of the muon $\alpha_\mu\equiv (g-2)_\mu/2$ is a precision observable of the Standard Model. The accuracy of the SM prediction of $(g-2)_\mu$ is currently limited by the knowledge of the hadronic light-by-light contribution.  Not only contributing to the $(g-2)_\mu$ studies, the two-photon width can be used to identify non-$q\bar{q}$  states, because $\gamma\gamma$ decay of non-$q\bar{q}$ mesons like glueballs, hybrids, multiquark objects, or mesonic molecules is expected to be suppressed in various models. \bes3 offers some good opportunities for precision measurements of the production of low-mass hadronic systems in two-photon collisions with the two-photon invariant mass region accessible up to 3 GeV/$c^2$.

\section{Prospects}
Although years of continuous experimental efforts have been made to search for QCD exotic hadrons beyond quark model, no compelling evidence has been unambiguously established yet.
The experimental search for QCD exotics (glueballs, hybrids, multiquarks) continues to be an exciting problem, which is limited by the current data. Recent progress in LQCD reaffirms the existence of glueballs and hybrids.
Other LQCD calculations indicate that radiative $J/\psi$ decays are a promising hunting ground for glueballs. \bes3 will continue playing a leading role in this search.

High precision data and systematic studies with various production mechanisms and decay modes are needed to determine resonance properties.
The primary requirement for the data taking of light hadron program at \bes3 is to have sufficient high-statistics $J/\psi$ events for the systematic study of glueballs. The pseudoscalar glueball is clearly the main focus of research with its production rate in radiative $J/\psi$ decays predicted to be ${\cal O}(10^{-4})$ by LQCD. A major difficulty for the identification of glueballs is the lack of first-principle theoretical predictions of glueball couplings and decay rates. Due to the ''flavour-blindness'' of gluons, there will be no dominant decay mode of a glueball. The decay rate of a glueball into a certain final state may be estimated to be at the level of ${\cal O}(10^{-3})\sim {\cal O}(10^{-2})$ in analogy to $\eta_c$ or $\chi_{c0,2}$ decays. But mixing with ordinary mesons complicates the situation in the light-quark mass range. The detection efficiency of a typical decay mode of a pseudoscalar glueball is estimated to be a few percent from simulation studies. Recently, a new decay mode of $X(2370)$ has been observed in the spectrum of $\eta^\prime K\bar{K}$ of $J/\psi\to\gamma\eta^\prime K\bar{K}$. We performed a feasibility study for determination of the spin-parity of $X(2370)$. The neutral channel $J/\psi\to\gamma\eta^\prime K^0_S K^0_S$ provides a clean environment to perform the amplitude analysis as it does not suffer from significant backgrounds such as $J/\psi\to\pi^0\eta^\prime K^0_S K^0_S$
, which are present in the charged channel $J/\psi\to\pi^0\eta^\prime K^+K^-$. MC samples with statistics equivalent to current data and 10 billions of $J/\psi$ are generated with a certain set of amplitude parameters. Amplitude analyses have been performed to the MC samples with various hypothesis. Table~\ref{tab:pwa} shows that the spin-parity of $X(2370)$ can be unambiguously determined with higher statistics of data.
\bes3 accumulated 10 billion $J/\psi$ events, which are mandatory for mapping out the spectrum of light hadrons in $J/\psi$ decays. In addition, $\psi(3686)$ decays have larger phase space for studying mesons and baryons with higher mass, even though the production rates of light hadrons are typically suppressed with respect to $J/\psi$. $\eta_c$ and $\chi_c$ events from $\psi(3686)$ decays can also provide an opportunity for investigating QCD exotics. Currently, BESIII collected 450 million of $\psi(3686)$ events. The light hadron physics program can be benefit from the high statistics $\psi(3686)$ data set for the charmonium program in the future.

\begin{table}[htb]
\caption{\label{tab:pwa}A feasibility study for determination of the spin-parity of $X(2370)$ in amplitude analysis of $J/\psi\to\gamma\eta^\prime K^0_S K^0_S$. The significance is obtained by comparing the likelihoods of amplitude analyses with different spin-parity hypothesis of $X(2370)$.}
\begin{center}
\begin{tabular}{|c|c|c|}
  \hline
   %after \\: \hline or \cline{col1-col2} \cline{col3-col4} ...
 & current statistics& expected statistics \\
 & (1.3 billions of $J/
\psi$)& (10 billions of $J/\psi$)\\
   \hline
 significance for $0^{-+}$  &  & \\
  assignment of $X(2370)$ & 3.2$\sigma$ & 13.2$\sigma$ \\
  \hline
\end{tabular}
\end{center}
\end{table}

In exploratory physics program for a future high luminosity $\tau$-charm experiment, electromagnetic couplings to glueball candidates and their form factors can be further extracted with higher accuracy, which are critical in understanding the nature of glueballs. High-statistics charmonium decays also provide an opportunity for investigating low-lying exotic hybrid nonets. In future, the available high-statistics light-meson events from decays can be used not only for some precision measurements of QCD at low energy, but also for probing physics beyond the SM.

In the next few years, many experiments (COMPASS, \bes3, etc.) will  continue to be active, while a number of new experiments (GlueX, Belle II, PANDA,~{\it etc.}) appear on the horizon. Definitive conclusions on the nature of confinement will need complementary studies with these experiments. Key features of these experiments are high statistics and high sensitivity to explore hadron spectroscopy.
COMPASS~\cite{COMPASS} has comprehensively studied $a_J$ and $\pi_J$-like mesons
 up to masses of 2 GeV/$c^2$ in diffractive scattering of hadron beams.
GlueX~\cite{GlueX} is designed to search for and measure the spectrum
of light-mass hybrid mesons. It begun its physics run in 2017 and will start a high luminosity run with an updated detector in 2019. An important advantage
of this experiment is the use of polarized
photons, which simplifies the initial states and production process.
PANDA~\cite{PANDA} is
designed for high-precision studies of the
hadron spectrum at cms energies between 2.3 and
5.5~GeV. It is scheduled to start data taking with full setup in 2026.
In $\overline{p}p$
annihilations, spin-exotic states ({\it e.g.}, oddballs) can be produced.
Belle II~\cite{belle2} will start collecting data in 2019, and
will accumulate 50~ab$^{-1}$ data at the $\Upsilon(4S)$ peak by
2027. Although not its primary goal as a next generation flavour factory, Belle II can also explore the light quark sector using the two-photon process,  because a glueball should have suppressed couplings to $\gamma\gamma$. Hadronic decays
of heavy hadrons may also serve as a well-defined source for light mesons.
\bes3 remains unique for studying and searching for QCD exotics and new excited baryons, as its high-statistics data sets of charmonia provide a gluon rich environment with clearly defined initial and final state properties.

%% file: Charmonium/charmonium.tex
\chapter[Charmonium Physics]{Charmonium Physics}
\label{chapter: Charmonium}

\input{Charmonium/charmonium_main.tex}

\input{Charmonium/bib.tex}

%% file: Charmonium/charmonium_main.tex
\section{Introduction}

\begin{figure}[h!tb]
\begin{centering}
\includegraphics*[width= 0.87\columnwidth]{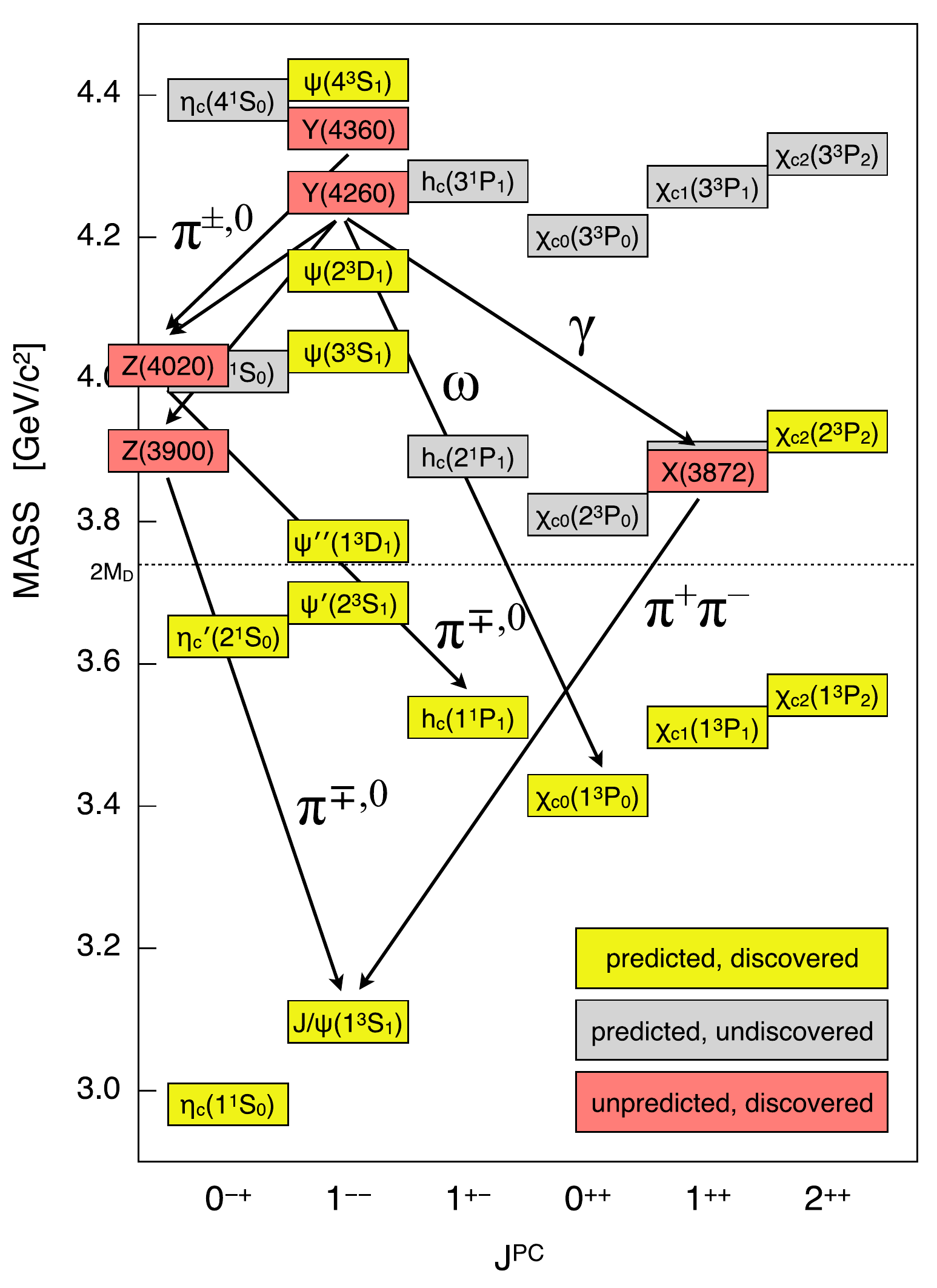}
\caption{\label{fig:ccbarspectrum} The charmonium spectrum.  Yellow boxes denote states predicted by the quark model~\cite{Barnes:2005pb} that have already been discovered; gray boxes are for predicted states that have not yet been discovered; and the red boxes are for states that were unexpectedly discovered -- likely pointing towards the existence of exotic hadrons.  All of these states have been studied at \bes3.  A few of the key transitions studied at \bes3 are indicated by black arrows.} 
\end{centering}
\end{figure}

\begin{figure}[h!tb]
\begin{centering}
\includegraphics*[width= 0.87\columnwidth]{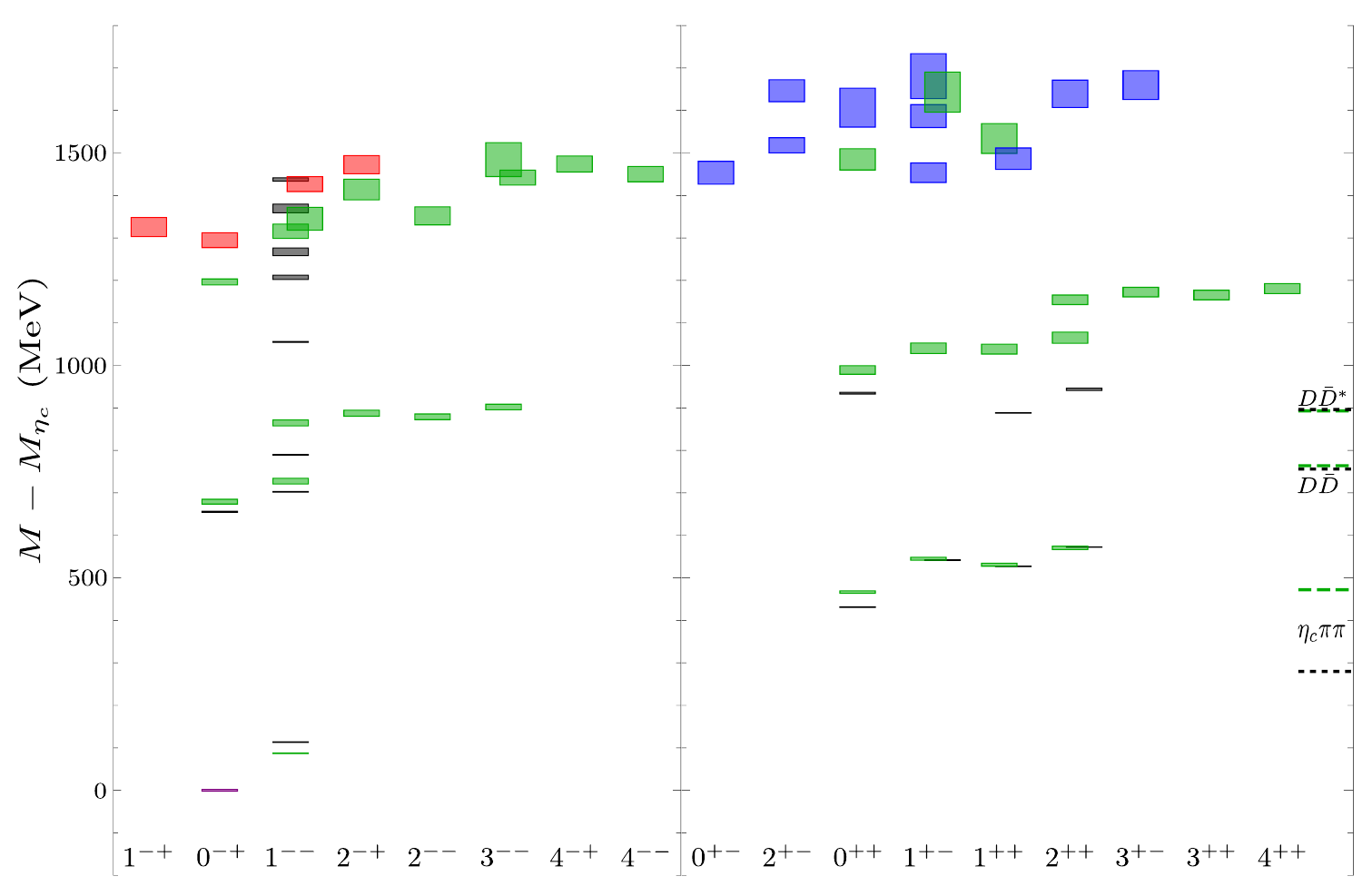}
\caption{\label{fig:latticeccbarspectrum} The charmonium and hybrid spectrum. Green boxes
  represent masses of charmonium states calculated by lattice QCD with $M_\pi\sim 240$
  MeV~\cite{Cheung:2016bym}; red and blue boxes represent the lightest and first-excited
  hybrid states; black boxes are world averaged experimental
  results in 2015~\cite{Agashe:2014kda}; dashed lines show the location of some lower thresholds
  for strong decay from computing (coarse green) or experiments (fine grey). The states
  are labeled with their quantum numbers $J^{PC}$ and their masses are shown with $\eta_c$
  mass subtracted. The vertical size of the boxes denote the standard uncertainty.}
\end{centering}
\end{figure}

Heavy quarkonia are frequently referred to as the ``positronium of QCD"~\cite{AppelPol} due to 
consistent, one-to-one correspondence of the level schemes that reflect Coulomb-like interactions at small distances. Within the \bes3 energy range, charmonium states both below and above the open charm threshold are accessible~(Figs.~\ref{fig:ccbarspectrum} and~\ref{fig:latticeccbarspectrum}). The spectrum of charmonium states with masses below the open charm threshold has been well-established for several decades.  These states can therefore be used to precisely test predictions based on various theoretical techniques, ranging from models (like the quark model) to approximations of QCD~(like non-relativistic QCD, described below) 
to numerical calculations of the full QCD Lagrangian (\ieie, LQCD). 
The fact that the energy scales range from the charmonium mass to 
$\Lambda_{\rm QCD}$ in charmonium processes make them a rich laboratory 
to probe both perturbative and non-perturbative QCD~\cite{Asner:2008nqch3}.

The relative simplicity of the lowest-lying~(least massive) charmonium states allows for
precision tests of QCD and QCD-inspired models in a region where both non-perturbative and
perturbative aspects of QCD play a role.  The higher-mass states, on the other hand, pose
serious challenges even to our qualitative understanding of mesons.  Several of these
states provide potential evidence for a wealth of exotic configurations of quarks and gluons, including: tetraquark states~(two quarks and two antiquarks), hadronic molecules~(two hadrons), hybrid mesons~(a quark and antiquark with an excited gluonic field), and so on.  

The charmonium group studies both of these regions of the charmonium spectrum, providing a
unique and important look at the dynamics of strong force physics. These studies include:
searching for new charmonium states, determining the internal structure of previously
established charmonium states, measuring masses and widths, precisely measuring
transitions (both radiative and hadronic) between charmonium states, and finding new decay
channels.  The capabilities of the \bes3 experiment are uniquely suited to the study of
both light and heavy charmonium states.  The lighter charmonium states are primarily
studied using large and clean samples of $\psip$ (\ieie, $\psi(2S)$ or $\psi'$) decays; the heavy charmonium states are produced using higher-energy collisions, where exotic charmonium states are either produced directly or through the decays of other states.

\setcounter{footnote}{3}
As noted above, the spectrum of states above open charm
threshold~(Figs.~\ref{fig:ccbarspectrum} and~\ref{fig:latticeccbarspectrum}) are clearly quite convoluted.  The past decade has seen the discovery of a large number of new states that are yet to be satisfactorily understood. These states have been named ``$X$'', ``$Y$'', and ``$Z$'' since their internal structure is still unclear\footnote{
The ``$Y$'' states usually have $J^{PC}=1^{--}$; the ``$Z$'' states are electrically charged; and the ``$X$'' states are the remainder.  Note that the naming scheme is revised in the latest edition of the Review of Particle Physics by PDG~\cite{pdg}, but we use the older scheme in this White Paper for consistency.
}.
One thing is clear: these new states cannot all be conventional bound states of a charm quark and antiquark. They are likely evidence for more exotic configurations of quarks and gluons (such as tetraquarks, meson molecules, or hybrid mesons, etc.).  Their existence therefore provides a crucial opportunity to study the dynamics of quarks and gluons in a new environment.  At \bes3, the accessible $e^+e^-$ cms energies allow us to directly produce a large number of these states, which in turn allows us to measure their masses, widths, and decay modes.  The effort to understand these new states (which we refer to as ``$XYZ$ physics'') is described further in Sec.~\ref{sec:ccabove}.

Future studies of the charmonium system, both above and below open-charm threshold, will
require additional data sets.  These requirements will be detailed throughout this chapter
and summarized in Sec.~\ref{sec:ccfuture}.  For further studies of charmonium below
open-charm threshold, around 3~billion $\psip$ decays are required, representing about an
order-of-magnitude increase in statistics over the current sample of about 450~million
$\psip$ decays.  Above open-charm threshold, we require three types of additional data
samples: (1)~a large number of additional data samples, each with an integrated luminosity
of approximately 500~pb$^{-1}$ and spread over a variety of cms energies, in order to
study the spectrum of the $Y$ states and to study the evolution of the $Z$ states; (2)~a
small number of larger data samples, each composed of approximately 5~fb$^{-1}$, to do
detailed studies of the $Z$ states; and (3)~samples of higher-energy data to explore the
poorly established mass region above 4.6~GeV/$c^2$, where mysterious peaks in the
$\Lambda_c^+\bar{\Lambda}_c^-$~\cite{Pakhlova:2008vn} and $\pi^+\pi^-\psi(2S)$ cross
sections~\cite{Wang:2007ea} have been observed.

\clearpage
\section{Charmonium States Below Open Charm Threshold}
\label{sec:ccbelow}

\begin{figure}[htp]
  \centering
  \includegraphics[width=1.0\textwidth]{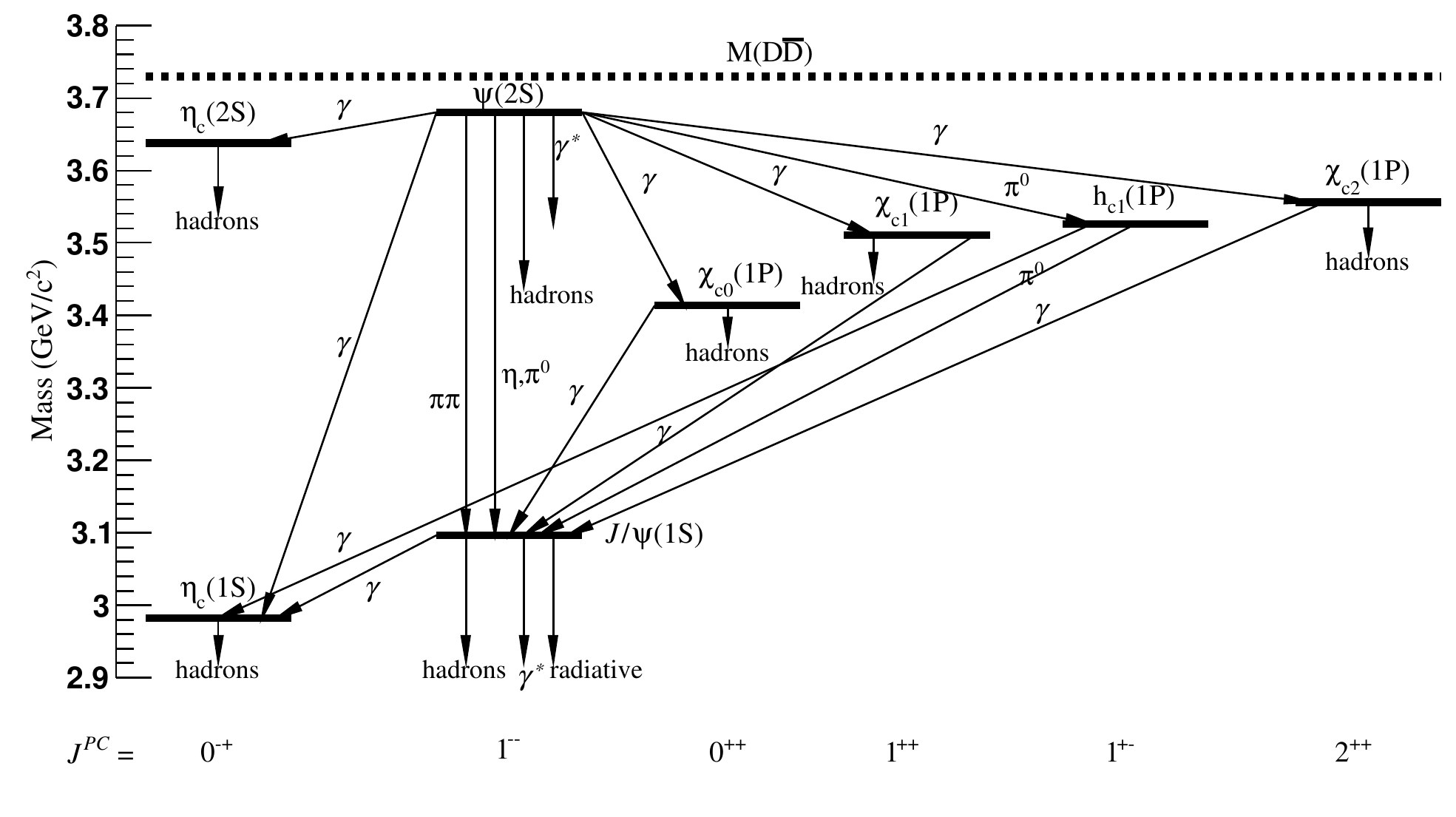}
  \caption{The low-lying charmonium ($c\bar{c}$) spectrum and some observed transitions.}
  \label{fig:chart}
  
\end{figure}

The goal of  \bes3 studies of charmonium states below the open charm threshold is to
investigate the spectroscopy, transitions, and decays of charmonium states by mainly
analyzing (but not  limited to) the $\psip$ data.  The $\psip$ data is especially well suited for the study of charmonium states due to many transitions between the $\psip$ and lower-lying states.  Thus, starting with a sample of $\psip$ data, one gains access to most of the  charmonium states below open-charm threshold. Figure~\ref{fig:chart} shows the low-lying charmonium~($c\bar{c}$) spectrum and some of the commonly observed transitions.
\hspace{1cm}
\subsection{The Theoretical Framework }

The charmonium meson system is an ideal, and to some extent unique, laboratory to investigate the boundary between perturbative and non-perturbative QCD in a controlled environment. Since reliable calculations of non-perturbative QCD are still difficult, several phenomenological models and effective theories have been proposed.  These models and calculational techniques can be tested and further developed using the phenomenology of charmonium states. For example, non-relativistic QCD (NRQCD)~\cite{PhysLett.B167.437, PhysRev.D46.4052, hep-ph/9407339}, 
expressed as a Pauli two-component field theory,
can be constructed in correspondence with the hierarchy of energy scales in charmonium.
NRQCD can be regarded as an effective theory that expands full QCD in powers of $v$, where $v$ is the relative velocity between the $c$ and $\bar{c}$ quarks. 

According to potential model calculations and lattice simulations, $v^2 \sim 0.3$ in the
charmonium system~\cite{EFT}. For small $v$, multitude energy scales are observed as the
hard $m_c$, soft $m_c v$, and ultra-soft $m_c v^2$, respectively. The effects at energy
scale $m_c$ can be integrated out explicitly. The resultant theory, NRQCD, reduced in the
number of dynamical degrees of freedom, is simpler than full QCD, and it turns out to be
very useful in calculating charmonium-relevant processes such as inclusive production, and
annihilation decays and spectroscopy.  Starting from NRQCD one can obtain the effective
theory potential NRQCD (pNRQCD)~\cite{hep-ph/9707481, hep-ph/9907240, hep-ph/0410047} by
integrating out the scale $m_c v$. Here the role of potentials and the quantum mechanical
nature of the problem can be reduced to the zeroth order Schr\"{o}dinger equation for the
two heavy quarks. These effective field theories, as well as potential models and LQCD,
make it possible to calculate a wide range of charmonium observables in a controlled and
systematic way, therefore allowing an investigation of one of the most elusive sectors of
the SM: low-energy QCD.

These charmonium observables can be taken from spectroscopy (\egeg, masses and widths),
transitions (\egeg, transition rates), leptonic and electromagnetic decays, radiative
decays, hadronic decays, rare and forbidden decays, and some miscellaneous topics such as
Bell inequalities in high energy physics and special topics in $B\bar{B}$ final states,
where $B$ refers to baryon. \bes3 is well suited to address the remaining experimental
questions that are related to the low-mass, i.e. below open-charm threshold, charmonium
spectrum, such as precise determinations of the mass and width of the $\eta_c$, $h_c$,
and $\eta'_c$. The QCD multipole expansion (QCDME)~\cite{hep-ph/0601044, hep-ph/0412158}
is a feasible approach to charmonium hadronic transitions. Its results can be examined via
observations at \bes3 such as $\pi \pi$ transitions of $S$-Wave ($P$-Wave or $D$-Wave)
charmonium states, the $\eta$ transition $\psip \to \eta J/\psi$, and iso-spin violating
$\pi^0$ transition $\psip \to \pi^0 h_c$.  Many radiative transition channels, including
$E$1 (electric dipole) and $M$1 (magnetic dipole) transitions, can be investigated with
the \bes3 $\psip$ data sample using the cascade transition chain. Other EM related
processes, such as charmonium leptonic and EM decays, can also be studied at \bes3. In
addition, studies of the hadronic decays will shed light on the $\rho-\pi$
puzzle~\cite{PhysRevLett.51.963} and reveal the inertial structure and decay dynamics of
charmonium states. The baryonic decays are a special topic since the structure of baryons
is comparatively more complicated than that of mesons. \bes3 can investigate the baryonic
decays via two-body, three-body, multi-body, and semi-inclusive modes. With a large
$\psip$ data sample, \bes3 can also search for rare and forbidden charmonium decays to
explore some interesting topics such as $CP$ violation or lepton flavor violation.

With the theory tools and the impressive number of collected charmonium states, \bes3 will
make a difference in this field allowing to carry on important investigation within the
SM and beyond it.

\subsection{Results with the Current $\psip$ Data Set}

\begin{figure}[hbtp]
  \centering
  \includegraphics[width=0.5\textwidth]{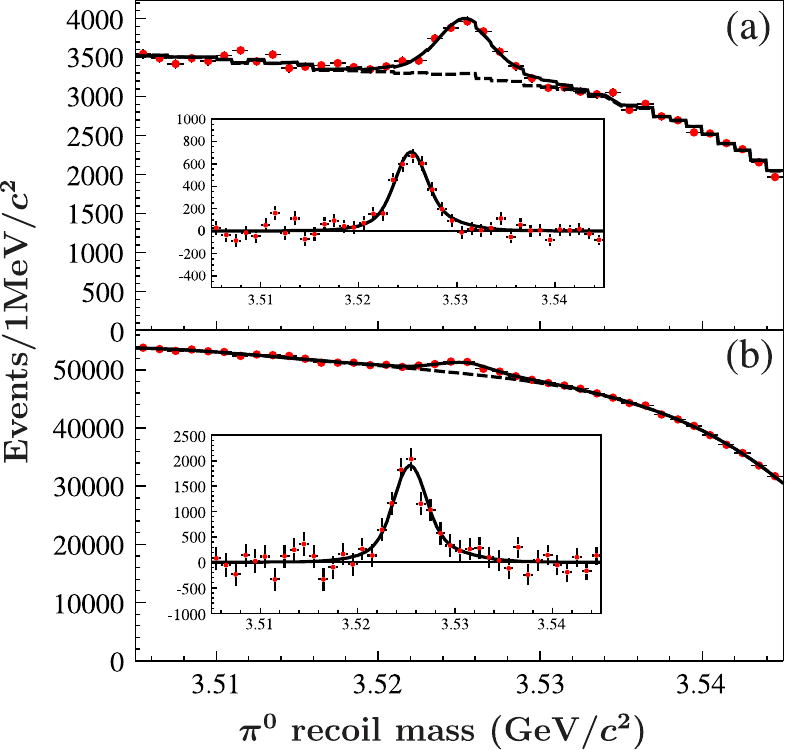}
  \caption{Measurements of the $h_c$ in $\psip$ decays. (a) The $\pi^0$ recoil-mass spectrum and fit for the $E$1-tagged analysis of $\psip \to \pi^0 h_c$, $h_c \to \gamma \eta_c$. (b) The $\pi^0$ recoil-mass spectrum and fit for the inclusive analysis of $\psip \to \pi^0 h_c$. Fits are shown as solid lines, background as dashed lines. The insets show the background-subtracted spectra.}
  \label{fig:hc}
\end{figure}

\begin{figure}[hbtp]
  \centering
  \includegraphics[width=0.5\textwidth]{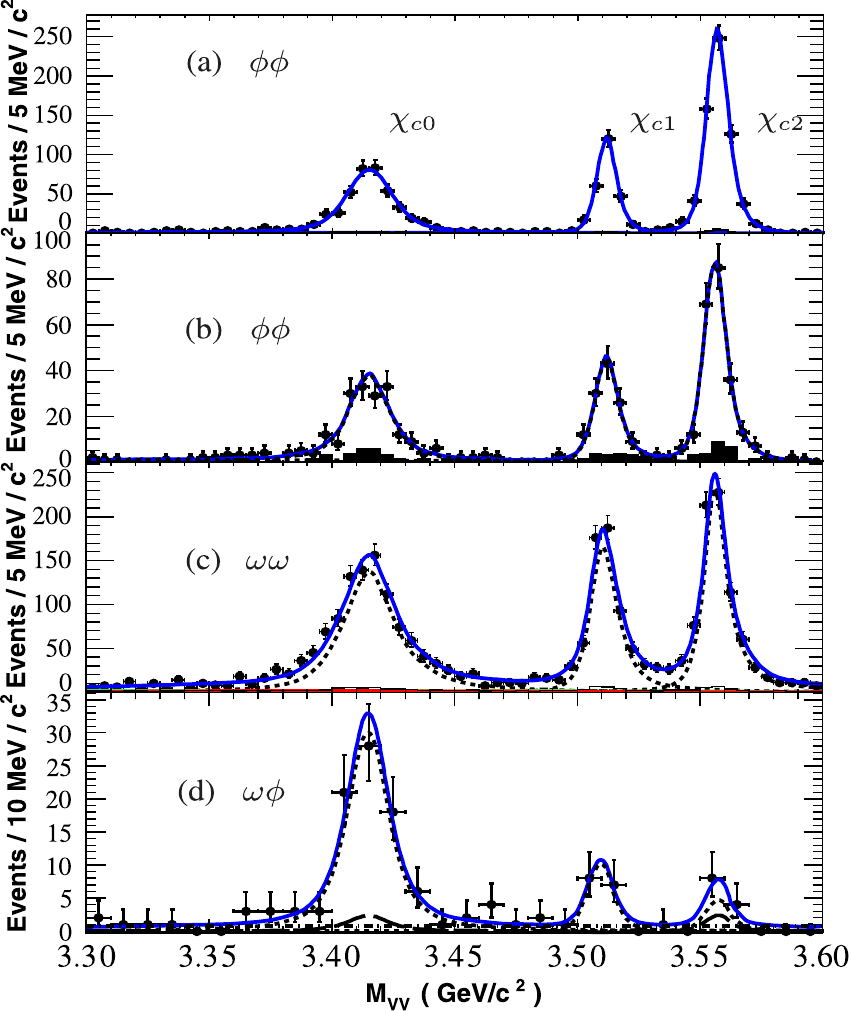}
  \caption{Observation of $\chi_{c1}$ decays into vector meson pairs. Invariant mass spectra of $VV$ for (a) $\phi\phi$ mode in the $\gamma 2(K^+K^-)$ final state, (b) $\phi \phi$ mode in the $\gamma \pi^+\pi^-\pi^0K^+K^-$ final state, (c) $\omega\omega$ mode in the $\gamma 2(\pi^+\pi^-\pi^0)$ final state, and (d) $\omega\phi$ mode in the $\gamma\pi^+\pi^-\pi^0K^+K^-$ final state.}
  \label{fig:chicj:v}
\end{figure}

\begin{figure}[hbtp]
  \centering
  \includegraphics[width=1.0\textwidth]{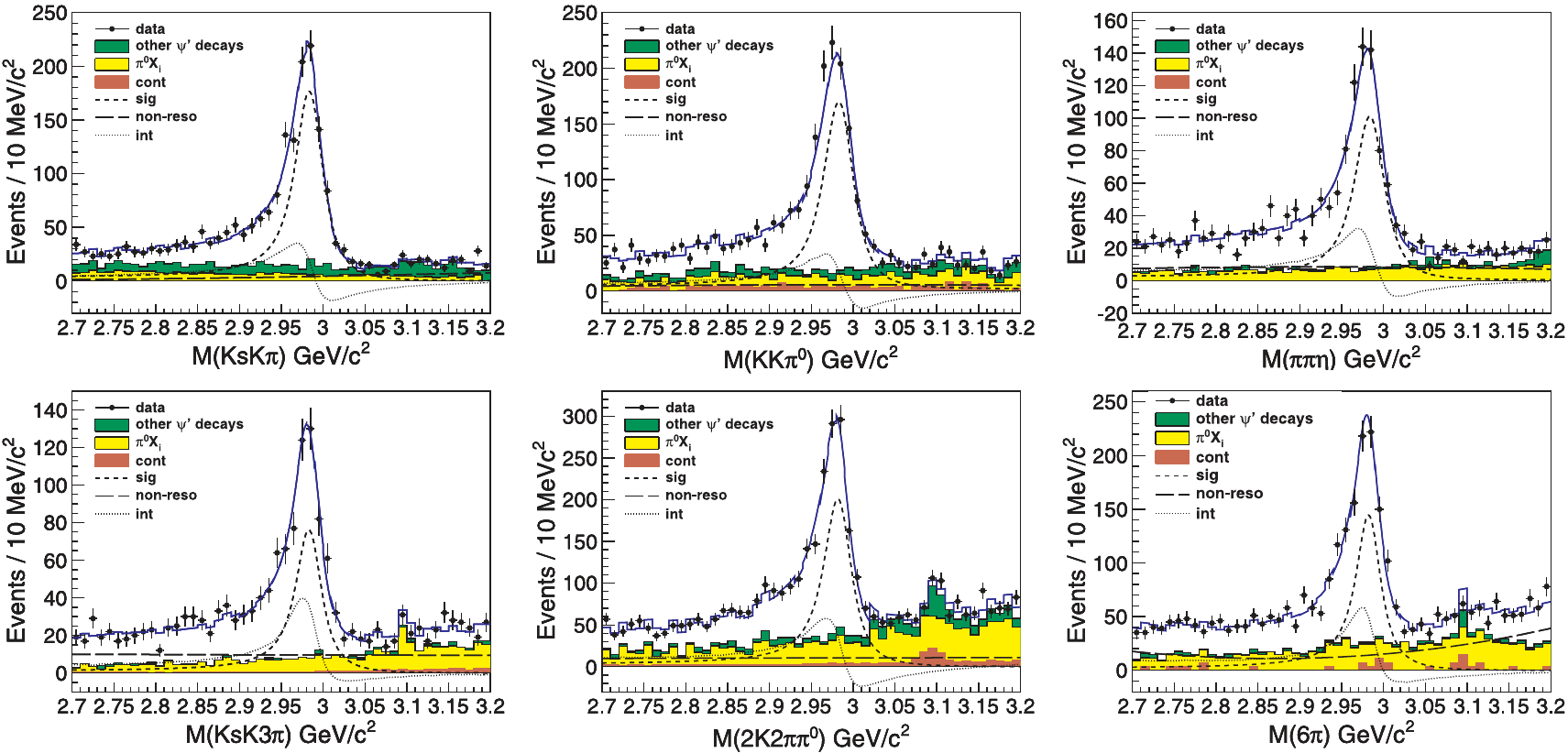}
  \caption{Measurement of mass and width of the $\eta_c$ using $\psip \to \gamma \eta_c$. The invariant mass distributions for the decays $K_S K^+ \pi^-$, $K^+K^-\pi^0$, $\eta\pi^+\pi^-$, $K_SK^+\pi^+\pi^-\pi^-$, $K^+K^-\pi^+\pi^-\pi^0$, and $3(\pi^+\pi^-)$, respectively, with the fit results (for the constructive solution) superimposed.}
  \label{fig:etac}
\end{figure}

\begin{figure}[hbtp]
  \centering
  \includegraphics[width=0.5\textwidth]{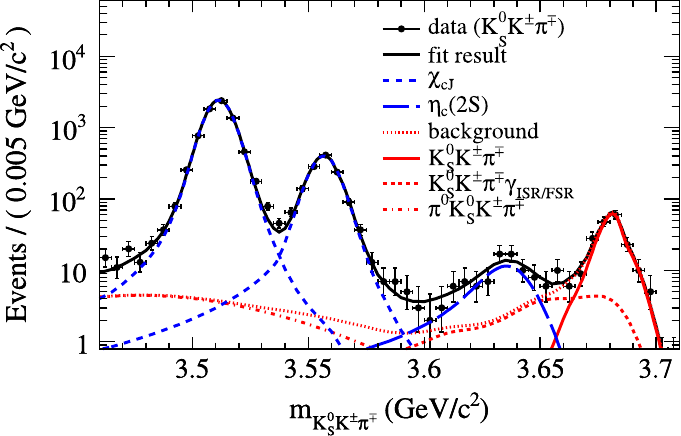}
  \caption{First observation of the $M$1 transition $\psip\to \gamma\eta_c(2S)$. The invariant-mass spectrum for $K^0_S K^\pm \pi^\mp$, and the simultaneous likelihood fit to the three resonances and combined background sources.}
  \label{fig:etac2s}
\end{figure}

\begin{figure}[hbtp]
  \centering
  \includegraphics[width=0.5\textwidth]{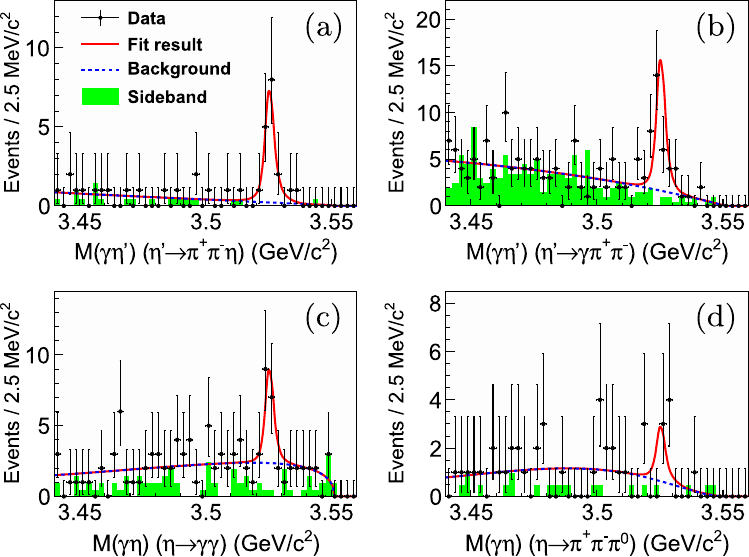}
  \caption{Observation of $h_{c}$ radiative decay $h_{c} \rightarrow \gamma \eta'$. Results of the simultaneous fits to the two invariant mass distributions of (top) $M(\gamma \eta')$ and (below) $M(\gamma \eta)$ for data. (a) $M(\gamma \eta')$ distribution for $h_c \to \gamma \eta'(\eta' \to \pi^+\pi^-\eta)$. (b) $M(\gamma \eta')$ distribution for $h_c \to \gamma \eta'(\eta' \to \gamma\pi^+\pi^-)$. (c) $M(\gamma \eta)$ distribution for $h_c \to \gamma \eta(\eta \to \gamma \gamma)$. (d) $M(\gamma \eta)$ distribution for $h_c \to \gamma \eta(\eta \to \pi^+\pi^-\pi^0)$.}
  \label{fig:hc:eta}
\end{figure}

 \bes3 has so far collected a total of 448~million $\psip$ decays~(including 107 million
 in 2009 and 341 million in 2012).  These data samples have led to the publication of more
 than 30~papers.  Some of the most important of these are as follows:
\begin{enumerate}
\item Measurements of the $h_c$ in $\psip$ decays~\cite{PhysRevLett.104.132002}. Clear signals are observed for $\psip\to \pi^0 h_c$ with and without the subsequent radiative decay $h_c \to \gamma \eta_c$. 
In addition, first measurements of the absolute branching ratios ${\cal B}(\psip \to \pi^0 h_c) = (8.4 \pm 1.3 \pm 1.0)\times 10^{-4}$ and ${\cal B}(h_c \to \gamma \eta_c) = (54.3 \pm 6.7 \pm 5.2)\%$ are presented. Figure~\ref{fig:hc} shows the recoil mass of the $\pi^0$ with or without the subsequent radiative decay $h_c \to \gamma \eta_c$.
\item Observation of  $\chi_{c1}$ decays into vector meson pairs~\cite{PhysRevLett.107.092001}. The first measurements of decays of $\chi_{c1}$ to vector meson pairs ($VV$) $\phi\phi$, $\omega\omega$, and $\omega\phi$ are presented. 
The branching fractions are measured to be $(4.4 \pm 0.3 \pm 0.5)\times 10^{-4}$, $(6.0 \pm 0.3 \pm 0.7)\times 10^{-4}$, and $(2.2 \pm 0.6 \pm 0.2)\times 10^{-5}$, for $\chi_{c1} \to \phi \phi$, $\omega\omega$, and $\omega \phi$, 
respectively, which indicates that the hadron helicity selection rule is significantly violated in $\chi_{cJ}$ decays. 
Figure~\ref{fig:chicj:v} shows the invariant mass spectra of $VV$ in different final states.
\item Measurement of the mass and width of the $\eta_c$ using $\psip \to \gamma \eta_c$~\cite{PhysRevLett.108.222002}. A novel model that incorporates full interference between the signal reaction, $\psip \to \gamma \eta_c$, and a nonresonant radiative background is used to successfully describe the line shape of the $\eta_c$. The $\eta_c$ mass is measured to be $2984.3 \pm 0.6 \pm 0.6$ MeV$/c^2$ and the total width to be $32.0 \pm 1.2 \pm 1.0$~MeV. 
Figure~\ref{fig:etac} shows the invariant mass distributions for the decays $K_S K^+ \pi^-$, $K^+K^-\pi^0$, $\eta\pi^+\pi^-$, $K_SK^+\pi^+\pi^-\pi^-$, $K^+K^-\pi^+\pi^-\pi^0$, and $3(\pi^+\pi^-)$, respectively, with the fit results (for the constructive solution) superimposed.
\item First observation of the $M$1 transition between the radially excited charmonium $S$-wave spin-triplet and the radially excited $S$-wave spin-singlet states: $\psip \to \gamma\eta_c(2S)$~\cite{PhysRevLett.109.042003}. Analyses of the processes $\psip \to \gamma \eta_c(2S)$ with $\eta_c(2S) \to K^0_S K^\pm \pi^\mp$ and $K^+K^-\pi^0$ give an $\eta_c(2S)$ signal with a statistical significance of greater than 10 standard deviations under a wide range of assumptions about the signal and background properties. The product branching fraction ${\cal B}\left(\psip \to \gamma \eta_c(2S)\right)\times {\cal B}\left(\eta_c(2S) \to K\bar{K}\pi\right)$ is measured to be $(1.30 \pm 0.20 \pm 0.30)\times 10^{-5}$. 
Figure~\ref{fig:etac2s} shows the invariant mass spectrum for $K^0_S K^\pm \pi^\mp$, and the simultaneous likelihood fit to the three resonances and combined background sources.
\item Observation of the $h_{c}$ radiative decay $h_{c} \rightarrow \gamma \eta'$~\cite{PhysRevLett.116.251802}. Events from the reaction channels $h_c \to \gamma \eta'$ and $\gamma \eta$ are observed with a statistical significance of $8.4 \sigma$ and $4.0\sigma$, respectively, for the first time. The branching fractions for $h_c \to \gamma \eta'$ and $h_c \to \gamma \eta$ are measured to be ${\cal B}(h_c \to \gamma \eta') = (1.52 \pm 0.27 \pm 0.29)\times 10^{-3}$ and ${\cal B}(h_c \to \gamma \eta) = (4.7 \pm 1.5 \pm 1.4)\times 10^{-4}$, respectively. 
Figure~\ref{fig:hc:eta} shows the results of the simultaneous fits to the invariant mass distributions of $M(\gamma \eta')$ and $M(\gamma \eta)$ for data. 
\end{enumerate}
The above list represents only part of the important results from the \bes3 $\psip$ data set; more results can be found in the \bes3 publication page~\cite{bes3C:pub-page}.

\subsection{Prospects for the Charmonium Program}

We continue to explore important physics topics via charmonium decays.  A few of these include: searches for new decay modes of the $h_c$, measurements of $M$1 transition processes and properties of the associated charmonium states (an $M$1 transition working group has been formed), and systematic analyses of baryon final states.  For many channels with larger statistics, amplitude analyses are being applied to extract more physics information. With more $\psip$ data, we can search for more decay modes of the $\eta_c$, $\eta_c(2S)$ and $h_c$, give more precise measurements of the masses and widths of the $\eta_c$, $\eta_c(2S)$ and $h_c$ (and better understand the line-shapes associated with their productions), perform partial-wave analysis (PWA) on more channels, and so on. Among these topics, some important measurements may require more $\psip$ data in order to achieve the desired precision.

For $h_c \to hadrons$, at present only three channels are observed and the sum of their branching fractions is only about $1.5\%$~\cite{Ablikim:2018ewr}. 
Many more hadronic decay modes are expected and will be searched for at \bes3. If we expect a $5 \sigma$ significant observation of a  $h_c$ decay channel with the branching fraction of $5\times 10^{-4}$, a $2\times 10^9$ $\psip$ sample is needed with the assumption that this channel is produced via the $\psip \to \pi^0 h_c$ transition, with $10\%$ efficiency for the detection of the final states, and a similar background level to the $\pi^+ \pi^- \pi^0$ channel. With this data sample, $h_c$ hadronic transitions such as the spin-flip transition $h_c \to \pi^+ \pi^- J/\psi$ and its radiative decays can also be searched for.

Compared with $E$1 transitions, the rates for $M$1 transitions between charmonium states are much lower. With the previously collected $\psip$ data sample, the first observation of the $M$1 transition $\psip\to \gamma\eta_c(2S)$ has been reported by \bes3~\cite{PhysRevLett.109.042003}. However, in order to understand more about the $\eta_c(2S)$ and measure its mass, width, and decays more precisely, more $\psip$ data is necessary. 
For example, for the observation of the $\eta_c(2S)$ decaying into some channel with the production branching fraction ${\cal B}\left(\psip \to \gamma \eta_c(2S)\right)\times {\cal B}\left(\eta_c(2S) \to K_S K 3\pi \right)$ of about $5\times 10^{-6}$, a $10^9$ $\psip$ sample is needed with the assumption that the detection efficiency is about $5\%$, where the low efficiency and high background levels are due to the softness of the transition photon. With this large sample, we could also study the $\eta_c$  charmonium state via $M$1 transition to improve the precision of ${\cal B}(\psip \to \gamma \eta_c)$ and its comparison with theoretical predictions. A measurement of ${\cal B}(\psip \to \gamma \eta_c(2S))$ could  be used to extract the absolute branching fractions for some specific $\eta_c(2S)$ decays. 
There are other radiative transitions such as $\eta_c(2S) \to \gamma J/\psi$, $\eta_c(2S) \to \gamma h_c$, $\chi_{c2} \to \gamma h_c$, and $h_c \to \gamma \chi_{c0,1}$ that are challenges for \bes3 even with  $10^9$  $\psip$ data sample because of low decay rates or difficulty in detecting the soft photon, but these rates can be calculated in the potential model~\cite{Asner:2008nq} and experimental searches are therefore important.

In addition to the radiative transitions, hadronic transitions are also very important and can be calculated better than the hadronic decays of charmonia. It is interesting to observe or find evidence for some hadronic transitions such as the spin-flip $\chi_{c1} \to \pi \pi \eta_c$ and $h_c \to \pi \pi J/\psi$, for which only upper limits have been set at present. We have estimated the needed statistics to see evidence for the transition $\chi_{c1} \to \pi^+ \pi^- \eta_c$ and found that at least $10^9$ $\psip$ events are needed according to the previous \bes3 results~\cite{PhysRev.D87.012002}. Note that in Ref.~\cite{PhysRev.D87.012002} only two $\eta_c$ hadronic decay channels were used. If more decay modes were included,  a smaller data set could satisfy the requirement. To uncover evidence for $h_c \to \pi \pi J/\psi$, at least a sample of $2\times 10^9$ $\psip$ decays will be needed with the assumption that there is no background. There are other hadronic transitions such as $\psip \to \eta J/\psi$ and $\psip \to \pi^0 J/\psi$, that have been observed with the present \bes3 $\psip$ data sample but will be improved with a larger one.

Besides transitions between charmonium states, there are also decays of charmonia that
should be measured. Both radiative decays and hadronic decays of the charmonium states
should be studied for a better understanding of charmonium decay dynamics. From
Ref.~\cite{hep-ph/0607278}, the typical branching fractions of $\chi_{cJ} \to \gamma V$,
where $V$ represents a vector resonance, is about $10^{-6}$. This means a $10^9$ $\psip$
data sample would be needed to observe the signal if we assume the intermediate product
branching fractions are $80\%$ and the selection efficiency is $30\%$. Since the branching
fractions for $\eta_c \to \gamma V$ are expected to be similar to $\eta_c \to \gamma
\gamma$, they should be at the $10^{-4}$ level. And the improved measurement of $\eta_c
\to \gamma \gamma$ will shed light on the effects of higher order QCD corrections as well
as provide validation of the decoupling of the hard and soft contributions in the NRQCD
framework due to its simplicity~\cite{1505.02665, 1707.05758}.  With the present $\psip$
data sample at \bes3, the processes $\psip \to \gamma \pi^0$ and $\gamma \eta$ have been
observed for the first time~\cite{PhysRevLett.105.261801}. There are also many studies of
$J/\psi$ and $\psip$ decays into $\gamma p \bar{p}$, $\gamma K^+ K^-$, $\gamma \pi^+
\pi^-$, etc. With a larger $\psip$ data sample, all of these measurements will be
improved.

More and better measurements of $\psip$ hadronic decays are crucial to help solve the long
standing $\rho$-$\pi$ puzzle~\cite{PhysRevLett.51.963}. The ratio of the branching
fractions of $J/\psi$ and $\psip$ is expected to hold in a reasonably good degree to
$12\%$, based on pQCD, for both inclusive and exclusive decays. But this relation is
observed severely violated for the $\rho\pi$ and several other decay channels. From
already obtained \bes3 results, it has become more clear that kinematic effects, which
have been previously ignored, might contribute significantly and the amplitude analysis
method might also be necessary to clarify the different dynamic processes between the
$J/\psi$ and $\psip$. For charmonium states beyond the open-charm threshold, the studies
of their decays to open-charm states will not only provide information of the decay
mechanism, but also serve for the spectroscopy study. Furthermore, it is interesting to
study the non-$D\bar{D}$ decays of $\psi(3770)$, such as $\psi(3770) \to p\bar{p}$,
$p\bar{p}\pi^0$, $\gamma \eta_c$, $ \gamma \chi_{cJ}$, $\gamma \eta_c(2S)$, and $\pi^+
\pi^- J/\psi$. These can be studied using data collected on and around the $\psi(3770)$
peak.

The two-body baryonic decays of the $\chi_{cJ}$ can provide information on color-singlet
and color-octet contributions. In addition, systematical studies of two-body baryonic decays
of the $J/\psi$ and $\psip$ will shed light on the relative angle and magnitude of the
electro-magnetic and strong interaction amplitudes~\cite{1505.03930}. The $\eta_c(2S)$ and
$h_c$ decaying into $p\bar{p}$ have been searched for based on $1\times 10^8$ $\psip$ decay sample at BESII and no obvious signal has
been observed~\cite{PhysRev.D88.112001}. To find evidences for both of  them, a data sample of $2\times 10^9$ $\psip$ decays will be required under the assumption that the efficiency is about $40\%$ and there is no background. Also, the similar data size is needed to study the
Bell inequality~\cite{PhysLett.A117.1}, SU(3) flavor symmetry~\cite{LKopke}, CP
violation~\cite{PhysRev.D47.1744, PhysRev.D49.4548}. Measurement of three-body baryonic
decays of higher excited charmonia will provide more information for excited baryonic
states, where amplitude analysis will be required.

Some rare decays can be searched for with larger $\psip$ data samples such as the $C$-violating process $J/\psi \to \gamma \gamma$ via $\psip \to \pi \pi J/\psi$. This has been discussed in the new physics part.

For the excited conventional charmonium states beyond the open-charm threshold there are
still many unsolved puzzles. There are only four $1^{--}$ states between $4.0$ and $4.6$
GeV predicted by potential models, but the number of observed vector resonances is
larger than that. The masses and partial widths of these states decaying into $e^+e^-$
($\Gamma_{ee}$) can be calculated via potential models assuming they are conventional
charmonia. However, strong coupling to open-charm meson pairs and relativistic corrections
significantly change both the spectrum and $\Gamma_{ee}$. Better measurements would be
very helpful to clarify the situation. \bes3 has observed the $\psi(1^3D_2)$ state
$X(3823)$ via its decay into $\gamma \chi_{c1}$~\cite{PhysRevLett.115.011803}, but
many predicted states, such as 1$F$, 3$S$, 2$D$, 1$G$, 3$P$, 2$F$, 4$S$, 3$D$ and
2$G$ have not been observed yet. The characteristics of the excited conventional
charmonia will be studied with the data sets collected for $XYZ$ states.

\bes3 has already collected scan samples around the $J/\psi$, $\psip$, and $\psi(3770)$
peaks, as well as those above $4.0$ GeV. These samples are dedicated to some specific
problems that we can do the best in the world, such as determinations of the resonance parameters,
line shapes, interference between strong and EM amplitudes in resonance decay, etc.

\begin{table}[tp]
\centering
  \caption{Some tentative measurements and correspondingly required statistic of $\psip$
    sample to achieve the desired precision on branching fraction. The observation and evidence are
    corresponding to $5\sigma$ and $3\sigma$ significance of the signal.}
  \label{chamonium:table}
  \begin{tabular}{lll}
    \hline
    \hline
   Measurement & Expected sensitivity & Needed $\psip$ sample ($\times10^9$) \\
    \hline
$h_c \to hadrons $ & Observation of $5\times 10^{-4}$ & $2$ \\
$\eta_c(2S) \to X$ & Observation of $1 \times 10^{-6}$ & $5$ \\
$\chi_{c1} \to \pi^+ \pi^- \eta_c$ & Evidence of $3 \times 10^{-3}$ & $>1$ \\
$h_c \to \pi^+ \pi^- J/\psi$ & Evidence of $2 \times 10^{-3}$ & $>2$ \\
$\chi_{cJ} \to \gamma V $ & Observation of $1 \times 10^{-6}$ & $1$ \\
$h_c \to p \bar{p}$ & Evidence of $2\times 10^{-4}$  & $>2$ \\

%$J/\psi \to \gamma \gamma$ & Upper limit at $1\times 10^{-8}$ & 3 \\
%$\psip \to \Lambda_c \bar{p} e^+ e^-$ & Upper limit at $1\times 10^{-7}$ & 8 \\

    \hline
    \hline
  \end{tabular}
\end{table}

Some tentative measurements and their correspondingly needed $\psip$ sample are listed in
Table~\ref{chamonium:table}. From this table a $3\times 10^9$ $\psip$ sample may be proper
to guarantee the expected precision and lead us to some exciting discoveries. But at present we
consider the $XYZ$ physics with higher priority, so we prefer to take the $\psip$ data
only after finishing the $XYZ$ proposal.

\section{$XYZ$ Physics}
\label{sec:ccabove}

The discovery of the $XYZ$ states has opened a new era in the study of the charmonium
spectrum~\cite{reviews,Chen:2016qju,Esposito:2016noz,Guo:2017jvc,Ali:2017jda,Olsen:2017bmm}.  Before the discovery of the $X(3872)$ in 2003~\cite{xdiscovery},
every meson in the mass region between 2.9 and 4.5~GeV/$c^2$ could be successfully described as
a $c \bar c $ bound state.  Simple potential models, using QCD-inspired potentials binding quarks and antiquarks, could reproduce the spectrum of charmonium states all the way from the $\eta_c(1S)$ (the ground state) up to the $\psi(4415)$~(usually considered to be the third radial excitation of the $J/\psi$)~\cite{Barnes:2005pb}.
This simple model of the charmonium spectrum has since broken down in a dramatic fashion.  Following the discovery of the $XYZ$ states~(Fig.~\ref{fig:ccbarspectrum}), we  now observe  exotic hadronic configurations containing charm and anticharm quarks.  These new configurations, such as tetraquarks, hadronic molecules, and hybrid mesons, allow us to probe the mysterious nonperturbative QCD which underlies the way quarks and gluons combining into larger hadronic composites.

The $X(3872)$, the first of the $XYZ$ states to be discovered, was discovered in 2003 by the Belle Collaboration in the process $B\to K X(3872)$ with $X(3872)\to\pi^+\pi^-J/\psi$~\cite{xdiscovery}.  Its unexpected appearance, combined with its narrow width and the fact that its mass is very close to the $D^0\bar{D}^{*0}$ mass threshold, immediately signaled that the $X(3872)$ is an unusual meson.  Since its discovery, many more of its properties have been determined~\cite{pdg}.  For example, its total width is less than 1.2~MeV/$c^2$, its mass is within 0.18~MeV/$c^2$ of the $D^0\bar{D}^{*0}$ threshold, 
it has $J^{PC}=1^{++}$, and in addition to
$\pi^+\pi^-J/\psi$, it decays to $D^0\bar{D}^{*0}$ with a large branching fraction, $\gamma J/\psi$, $\gamma\psip$ and $\omega J/\psi$.  It is currently believed to be a mixture of  $\chi_{c1}(2P)$ 
 and a tetraquark or meson molecule~\cite{reviews}.  However, many important open questions remain and further investigation is needed.

In 2013, \bes3 discovered that it has experimental access to the $X(3872)$ through the process $e^+e^-\to\gamma X(3872)$ using cms energies in the vicinity of 4.26~GeV~\cite{Ablikim:2013dyn}, which, in itself, provides an important hint at the relation between the $Y(4260)$ and $X(3872)$~\cite{Guo:2013nza}.  This discovery has initiated a new and vigorous program at \bes3 to search for new decay modes of the $X(3872)$, which will in turn offer new insight into its nature.

The discovery of the $X(3872)$ quickly led to the discovery of many more states with exotic configurations of quarks and gluons.  For example, the search for the decay $X(3872)\to\omega J/\psi$ led to the discovery of the $X(3915)$ in the process $B \to K X(3915)$ with $X(3915)\to\omega J/\psi$~\cite{xtoomegajpsi}.  This state was originally a candidate to be the quark model $\chi_{c0}(2P)$ state, but its mass and decay patterns were inconsistent with that interpretation.  Moreover, since that time a better candidate for  the $\chi_{c0}(2P)$ state, the $X(3860)$,  has been discovered~\cite{x3860} (see also in Ref.~\cite{Guo:2012tv}).  
This leaves the nature of the $X(3915)$ an unsettled question.  Searching for the $X(3915)$ in the process $e^+e^- \to \gamma X(3915)$ is currently an important topic at \bes3. In addition, it is also important to search for resonant signals for the broad  $X(3860)$ and the conjectured narrow $2^{++}$ $X_2$ with a mass around 4~GeV/$c^2$~\cite{Guo:2014ura}, the predicted spin partner of the $X(3872)$, in the process $e^+e^-\to \gamma D\bar D$. 

Also following the discovery of the $X(3872)$ was the discovery of the $Y(4260)$ by BaBar using the Initial State Radiation~(ISR) process $e^+e^-(\gamma_{\mathrm{ISR}})\to Y(4260)$ with $Y(4260)\to\pi^+\pi^-J/\psi$~\cite{babary4260}.  Because of its production mechanism, we can immediately infer that it has $J^{PC}=1^{--}$.  
However, its mass is inconsistent with any of the known or expected charmonium vector excitations. For example, its mass lies between the masses of the $\psi(4160)$ and $\psi(4415)$.  
It is therefore supernumerary.  LQCD calculations~\cite{latticey4260}, as well as other models~\cite{othery4260}, suggest the $Y(4260)$ could be a hybrid meson. Another intriguing model that considers the $Y(4260)$ to be strongly coupled  to $D_1(2420)\bar D$ predicts a nontrivial behavior for its line shape~\cite{Wang:2013cya}.
At \bes3, the $Y(4260)$ can be produced directly by simply tuning the cms energy of the $e^+e^-$ collisions to the mass of the $Y(4260)$.  In the same way, \bes3 can also directly produce the $Y(4360)$ (seen in $e^+e^- \to \pi^+\pi^-\psip$), and can search for new $Y$ states.  The \bes3 discovery of a multitude of new $Y$ states (or the discovery of a complicated coupled-channel system) will be further described  in the next section (Sec.~\ref{sec:ccpast}).

The final, and perhaps most interesting, class of exotic structures are the isovector $Z_c$ states.  Since they are known to contain an isosinglet  $c\bar{c}$ pair, they must also contain light quarks to account for the non-zero isospin.  One of the first of these states to be observed, the $Z_c(3900)$, was discovered by \bes3 in the process $e^+e^- \to \pi^{\mp} Z_c(3900)^\pm$ with $Z_c(3900)^\pm \to \pi^\pm J/\psi$~\cite{Ablikim:2013mio:3}.  Its two distinctive features are that it carries an electric charge and it has a mass near the $D\bar D^*$ threshold.  This suggests that it is a candidate for a meson molecule or tetraquark with rescattering effects due to the presence of the $D\bar D^*$ threshold and more complicated kinematical singularities that are also expected to be important.  The lineshape of the $Z_c(3900)$ and how that shape evolves with  cms energy will be one of the keys to its interpretation.  As described in more detail later, disentangling the nature of the $Z_c(3900)$ is one of the primary goals of the \bes3 $XYZ$ program.

Since the discovery of the $Z_c(3900)$, other $Z_c$ states have also been discovered.  For example, \bes3 discovered the $Z_c(4020)$ in the process 
$e^+e^- \to \pi^{\mp} Z_c(4020)^\pm$ with $Z_c(4020)^\pm \to \pi^\pm h_c(1P)$~\cite{Ablikim:2013wzq:3}.  Its mass is close to the $D^*\bar D^*$ threshold, suggesting it is closely related to the $Z_c(3900)$.

So far, the $Z_c(3900)$ and $Z_c(4020)$ have only been produced in $e^+e^-$ collisions with cms energies around 4.2--4.4~GeV with possible connection to the reported $Y$ states.  Other potential production mechanisms, such as $B\to K Z_c$ decays, appear to be insensitive to the $Z_c(3900)$ and $Z_c(4020)$ states.  A different class of $Z_c$ states has been discovered in $B$ decays, among which the most prominent one is the $Z_c(4430)$ seen in $B \to K Z_c(4430)$ with $Z_c(4430)^\pm \to \pi^\pm\psip$~\cite{bellezc4430}.  The reason that one class of $Z_c$ states is produced in $e^+e^-$ collisions and a different class is produced in $B$ decays remains a fascinating open question.

\subsection{Overview of \bes3 Accomplishments}
\label{sec:ccpast}

The goal of the $XYZ$ physics program at \bes3 is to understand the novel phenomena apparent in the spectrum of charmonium states with masses above open charm threshold, as outlined in the previous section.
The presence of these $XYZ$ states provides an ideal opportunity for \bes3 to study exotic and unexplored features of the strong force.  \bes3 is currently in a unique position to both directly access a large number of these states and to search for new states in their decays.  Note that a given $Y$ state, because it has $J^{PC}=1^{--}$, can be produced directly at \bes3 by the appropriate choice of the cms energy of the $e^+e^-$ collisions. 

\begin{figure}[tp]
\begin{centering}
\includegraphics*[width= 0.9\columnwidth]{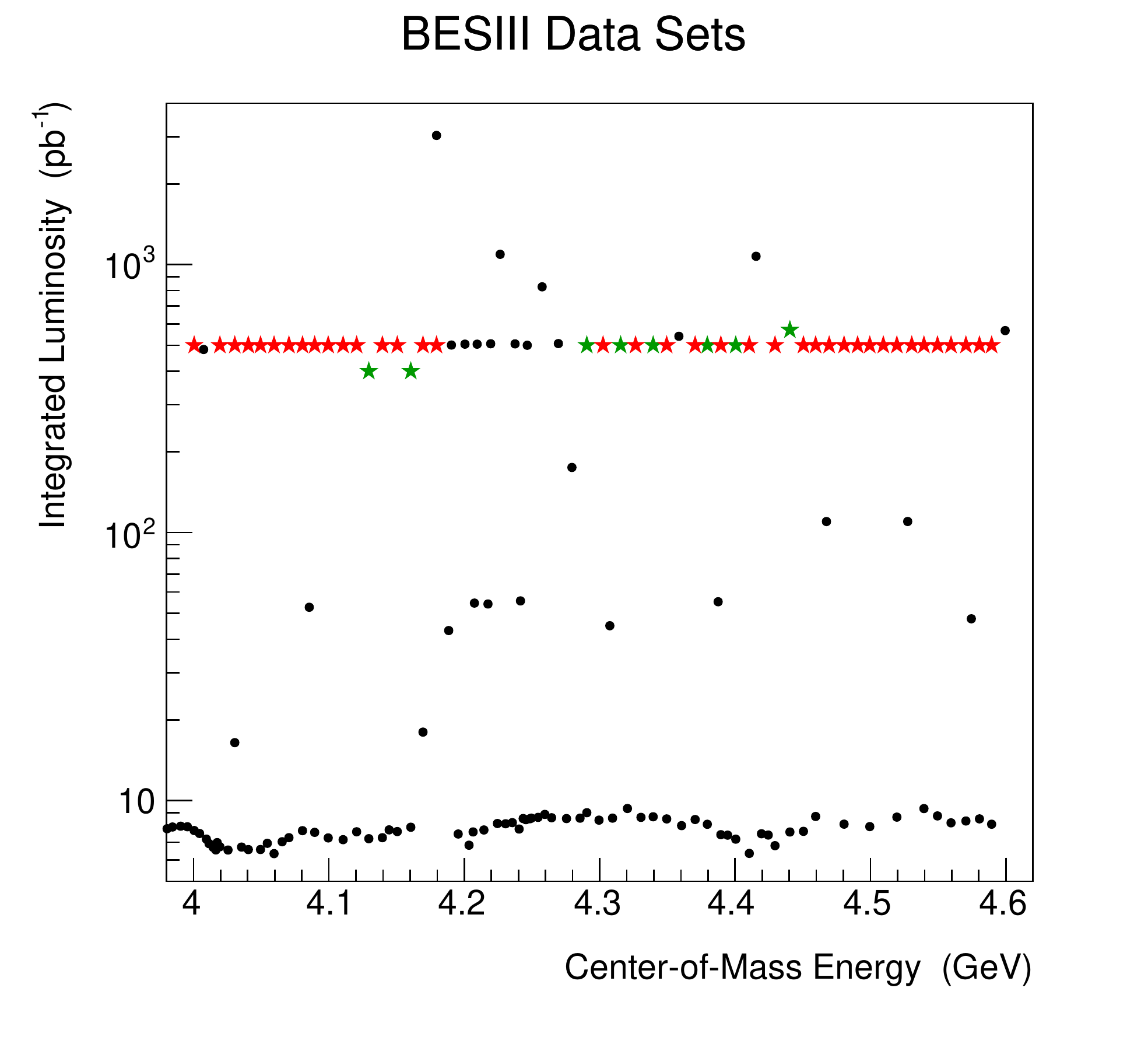}
\caption{\label{fig:XYZData} \bes3 data sets that are relevant for $XYZ$ physics.  The data sets collected prior to 2019 are shown in black; those collected in 2019 are in green; and those considered for potential future measurements are shown in red.  
}
\end{centering}
\end{figure}

\bes3 has already made significant progress in studies of the $XYZ$ states.  Existing data sets are shown in Fig.~\ref{fig:XYZData}.  We originally had ``large'' data sets (with integrated luminosity at or above 500~pb$^{-1}$) at only a few $e^+e^-$ cms energies: 4.01, 4.18~(primarily used for $D_s$ physics), 4.23, 4.26, 4.36, 4.42 and 4.6~GeV. In 2017, we collected large samples at seven additional points between 4.19 and 4.27~GeV,  and in 2019 eight additional energies between 4.28 and 4.44 GeV. 

\begin{figure}[tbp]
\begin{centering}
\includegraphics*[width= 0.85\columnwidth]{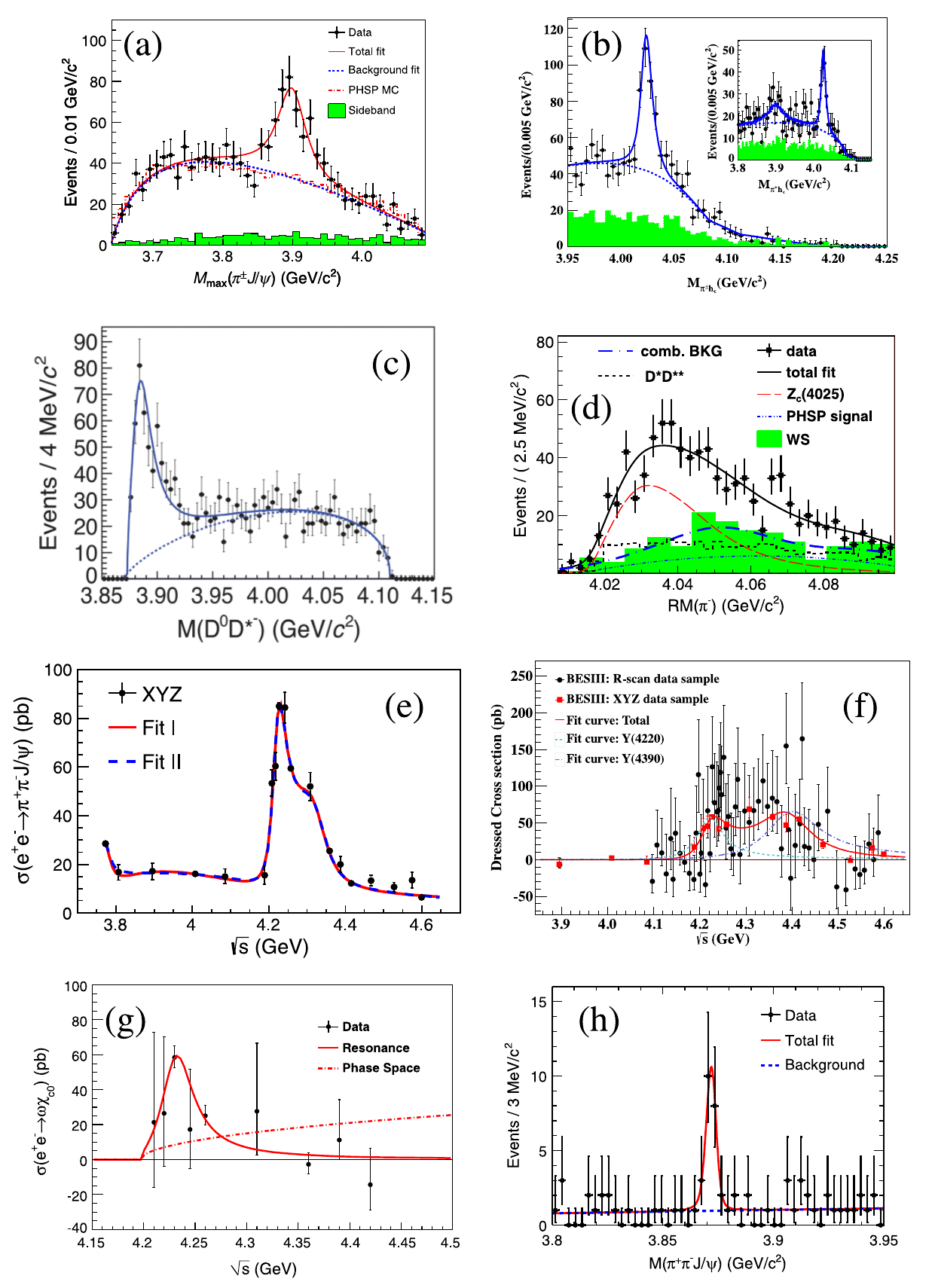}
\caption{\label{fig:XYZHighlights} A selection of $XYZ$ results from \bes3:
(a)~discovery of the $Z_c(3900)$~\cite{Ablikim:2013mio:3};
(b)~discovery of the $Z_c(4020)$~\cite{Ablikim:2013wzq:3};
(c)~discovery of open-charm decays of the $Z_c(3900)$~\cite{Ablikim:2013xfr};
(d)~discovery of open-charm decays of the $Z_c(4020)$~\cite{Ablikim:2013emm:3};
(e)~observation of $Y(4260)\to \pi^+\pi^- J/\psi$~\cite{pipijpsi};
(f)~discovery of two peaks in $e^+e^- \to \pi^+\pi^- h_c$~\cite{BESIII:2016adj};
(g)~discovery of a peak in $e^+e^- \to \omega \chi_{c0}$~\cite{Ablikim:2014qwy};
(h)~discovery of $e^+e^- \to \gamma X(3872)$~\cite{Ablikim:2013dyn}.
}
\end{centering}
\end{figure}

Using the limited amount of data collected prior to 2017, we have already made a number of important discoveries and measurements.  Here we highlight eight of the most important measurements:
\begin{enumerate}
\item We discovered the $Z_c(3900)$ in the process $e^+e^- \to \pi^\mp Z_c^\pm$ with $Z_c^\pm \to \pi^\pm J/\psi$~\cite{Ablikim:2013mio:3} in Fig.~\ref{fig:XYZHighlights}(a).  
Since the $Z_c(3900)$ has a mass in the charmonium mass region and it decays to $J/\psi$, we know it includes $c\bar{c}$ quarks.  But since it has an electric charge, we know it must include additional quarks.  The $Z_c(3900)$ is therefore a good candidate to be a tetraquark or meson molecule. 

With 680~citations (as of Oct.~9,~2019), this publication is the most-cited of all of BES papers (including BES and \bes3).  Additional work is ongoing to better determine the internal structure of the $Z_c(3900)$.  
\item We discovered the $Z_c(4020)$ in the process $e^+e^-\to \pi^\mp Z_c^\pm$ with $Z_c^\pm \to \pi^\pm h_c$~\cite{Ablikim:2013wzq:3} in Fig.~\ref{fig:XYZHighlights}(b).
Like the $Z_c(3900)$, the $Z_c(4020)$ contains $c\bar{c}$ quarks and is electrically charged.  
While the $Z_c(3900)$ is near the $D\bar{D}^{*}$ threshold, the $Z_c(4020)$ is near the $D^{*}\bar{D}^{*}$ threshold.
With  337~citations (as of Oct.~9,~2019), this publication is the second-most-cited  \bes3 papers.

\item We observed the $Z_c(3900)$ decaying to the open-charm channel $(D\bar{D}^* + c.c.)^\pm$~\cite{Ablikim:2013xfr} in Fig.~\ref{fig:XYZHighlights}(c).  With  258~citations (as of Oct.~9,~2019), this publication is the fourth-most-cited \bes3 papers.

\item Similarly, we observed the $Z_c(4020)$ decaying to the open-charm channel $(D^*\bar{D}^*)^\pm$~\cite{Ablikim:2013emm:3} in Fig.~\ref{fig:XYZHighlights}(d).  Observation of the $Z_c(3900)$ and $Z_c(4020)$ decays to these open charm channels (and the lack of evidence for $Z_c(4020)$ decays to $(D\bar{D}^* + c.c.)^\pm$), combined with their masses being close to threshold, suggest that the nature of the $Z_c(3900)$ and $Z_c(4020)$ is somehow intimately tied to these channels.
With  291~citations (as of Oct.~9,~2019), this publication is the third-most-cited  \bes3 papers.

\item We precisely measured the cross section for $e^+e^- \to \pi^+\pi^- J/\psi$~\cite{pipijpsi}.  Since this is the reaction that led to the discovery of the $Y(4260)$~(although here Initial State Radiation is not used), the aim of our measurement was to precisely determine the parameters of the $Y(4260)$.  Instead, what we observed was a cross section that was inconsistent with that arising from a single ordinary resonance, as shown in Fig.~\ref{fig:XYZHighlights}(e).  The cross section could be successfully fit using two resonances: one with a narrow total width and mass around 4.22~GeV/$c^2$and one with a wider total width and mass around 4.32~GeV/$c^2$.  Thus, the $Y(4260)$, one of the first of the $XYZ$ to be discovered, is either composed of two resonances or has a highly nontrivial line-shape due to other dynamics such as a strong coupling to the $D_1(2420)\bar D$.  
In addition, we observed the decay $Y(4260)\to \pi^0\pi^0 J/\psi$ and the neutral $Z_c(3900)$ in $e^+e^- \to  \pi^0 Z_c^0 \to \pi^0 \pi ^0 J/\psi$~\cite{Ablikim:2015tbp}.  This further cements the case that the $Z_c(3900)$ is an isospin-1 multiquark state of charmonium.

\item We discovered two peaks in the cross section for $e^+e^- \to \pi^+\pi^- h_c$~\cite{BESIII:2016adj} in Fig.~\ref{fig:XYZHighlights}(f).  This, along with the above-described measurement of  $e^+e^- \to \pi^+\pi^- J/\psi$, indicates that the family of $Y$ states is more complicated than was originally thought. In particular, these two final states have different total spins for the charm and anticharm quark pair. The masses and widths of the peaks in $\pi^+\pi^- h_c$ and $\pi^+\pi^- J/\psi$ are inconsistent.

\item We discovered a peak in $e^+e^- \to \omega \chi_{c0}$~\cite{Ablikim:2014qwy} in Fig.~\ref{fig:XYZHighlights}(g) that is inconsistent with the reported parameters for the $Y(4260)$.  This suggests that either there is yet another $Y$ state in this mass region or that the mass of the $Y(4260)$ should be shifted downwards.
\item We discovered the process $e^+e^- \to \gamma X(3872)$~\cite{Ablikim:2013dyn} in Fig.~\ref{fig:XYZHighlights}(h), which hints at the existence of the radiative decay $Y(4260)\to \gamma X(3872)$.  This is a unique transition in the sense that it is a transition between two states both of whose natures are unclear.  This suggests there is an intimate connection between the $Y(4260)$ and the $X(3872)$.
\end{enumerate}

There is currently a great amount of community interest in $XYZ$ physics, as is  clear from the high profile of the \bes3 papers mentioned above.  Since the natures of most of these phenomena are yet to be understood, the theory community has been especially active developing new techniques to aid in the interpretation of the $XYZ$~\cite{reviews}.  This has led to many innovations.  What is currently needed, however, is more high precision data.  Therefore,  new data that \bes3 will provide is much anticipated and will be put to immediate use.

\subsection{Broad Problems in $XYZ$ Physics}

The $XYZ$ results from \bes3 have helped uncover several broad problems in the field, and these are the subjects of intense studies at \bes3.  Below, these are labeled the ``$Y$ problem,'' the ``$Z$ problem,'' and the  ``$X$ problem.''  With more data, \bes3 is in the unique position to definitively address all three.  This section includes descriptions of these problems and indicates a variety of the ways they can be addressed at \bes3.

\subsubsection{The $Y$ Problem}

\begin{figure}[tbp]
\begin{center}
\includegraphics[width=1.\textwidth]{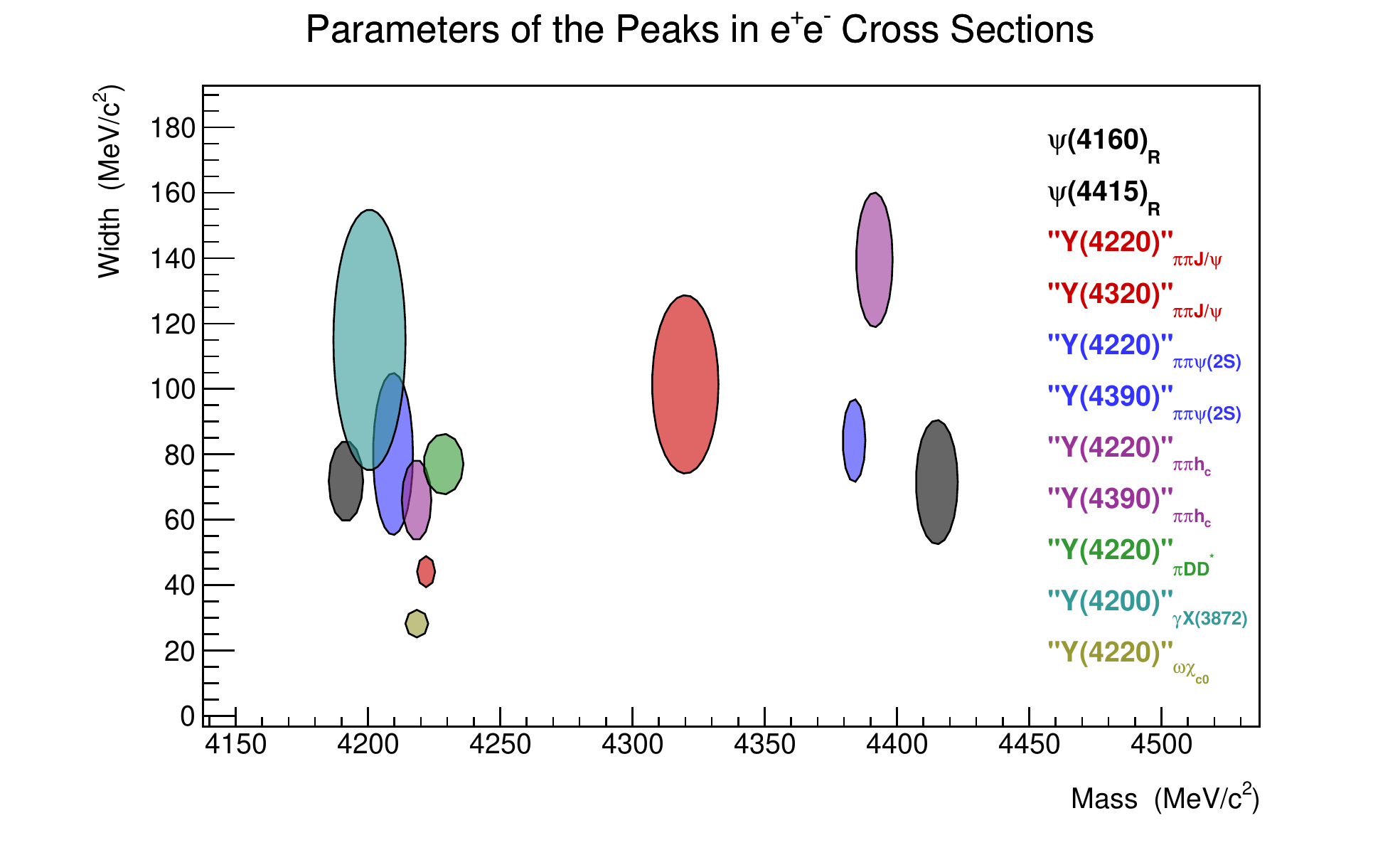}
\caption{A collection of resonance parameters as determined from fits to $e^+e^-$ cross sections.  Each ellipse encloses 1$\sigma$ around the measurements of resonance parameters, where the size of the ellipses are determined by adding together the statistical and systematic uncertainties in quadrature. The parameters are from inclusive hadronic cross sections~\cite{Ablikim:2007gd}; $\pi^+\pi^-J/\psi$~\cite{pipijpsi};  $\pi^+\pi^-\psi(3686)$~\cite{Ablikim:2017oaf}; $\pi^+\pi^-h_c$~\cite{BESIII:2016adj}; $\pi D\bar{D}^*+c.c.$~\cite{Ablikim:2018vxx}; $\gamma X(3872)$~\cite{Ablikim:2019zio}; and $\omega \chi_{c0}$~\cite{Ablikim:2019apl}.} \label{fig:yparameters}
\end{center}
\end{figure}

Exclusive $e^+e^-$ cross sections have shown surprisingly complex behavior as a function of cms energy.
The $Y(4260)$ is more complex than a single ordinary resonance, as shown by the complicated lineshape in the $e^+e^- \to \pi^+\pi^- J/\psi$ cross section in Fig.~\ref{fig:XYZHighlights}(e);
the $Y(4360)$ and $Y(4660)$ are seen in $e^+e^- \to \pi^+\pi^-\psip$;
two other peaks are seen in $e^+e^- \to \pi^+\pi^- h_c$ in Fig.~\ref{fig:XYZHighlights}(f);
the $Y(4220)$ is seen in $e^+e^- \to \omega \chi_{c0}$ in Fig.~\ref{fig:XYZHighlights}(g) and so on.  
A summary of the masses and widths of resonances extracted from recent \bes3 results is shown in Fig.~\ref{fig:yparameters}.
There is currently very little consistency between different reactions. Furthermore, none of these complicated features are apparently present in the inclusive $e^+e^-$ cross section, which only shows evidence for the $\psi(3770)$, $\psi(4040)$, $\psi(4160)$, and $\psi(4415)$~\cite{Ablikim:2007gd}.  This is the ``$Y$'' problem.  Are the many peaks seen in $e^+e^-$ cross sections really new states?  Or are they the results of more subtle effects?  With new data, will new patterns emerge?
What are their exact line shapes? Will they match theoretical predictions, such as a very asymmetric line shape for the Y(4260) obtained within a molecular frame~\cite{Cleven:2013mka}?
With our limited number of data points (cms energies), there is little hope in resolving the issue.  We require (1)~more data spread over a variety of cms energies, and (2)~a global and simultaneous analysis of many final states.  This latter effort will likely require close collaboration with the theory community, in particular with the view on amplitude analysis.

\subsubsection{The $Z$ Problem}

\bes3 discovered the $Z_c(3900)$ in $e^+e^- \to \pi^+\pi^- J/\psi$ events at a single cms energy of 4.26~GeV in Fig.~\ref{fig:XYZHighlights}(a).  Higher-energy data, however, has revealed more complex structures. There is a similar situation in $e^+e^- \to \pi^+\pi^-\psip$, where the lower-energy data appears relatively simple, but the Dalitz plots at higher energies are more complex~\cite{Ablikim:2017oaf}.  This is the ``$Z$'' problem.  How do the $Z_c$ structures produced in $e^+e^-$ annihilation change as a function of $e^+e^-$ cms energy?  Will their properties remain constant, as would be expected for a true resonance, or will their properties change, which would indicate the discovery of other effects (which are also important to study)?  Again, \bes3 is in a unique position to address this problem.  Again, we require large data sets at a variety of energies, and we require closer cooperation with the theory community.  

Using existing data samples, \bes3 has successfully determined the quantum numbers of the $Z_c(3900)$ to be $J^P = 1^+$~\cite{Collaboration:2017njt}.  However, the data samples have so far been insufficient to measure the phase of the $Z_c(3900)$ with respect to the other amplitudes present in $e^+e^- \to \pi^+\pi^- J/\psi$.  Such an effort to measure the Argand diagram for the $Z_c(3900)$ is currently being explored.  Determining the lineshape of the $Z_c(3900)$ and studying its phase motion would allow us to distinguish between different models for the nature of the $Z_c(3900)$~\cite{Swanson:2015bsa}.  There are at least two challenges.  First, a larger data sample at a single cms energy is required.  Second, and perhaps more importantly, we must understand all of the other resonances that are present in the $\pi^+\pi^- J/\psi$ Dalitz plots.  The measurement of the phase of the $Z_c(3900)$ (and even the magnitude of its amplitude) is only as good as our understanding of the parameterization of the rest of the Dalitz plot.  This is a challenging issue.

Another route for understanding the nature of the $Z_c(3900)$ would be to look for patterns of other $Z_c$ states.  \bes3 has already discovered the $Z_c(4020)$, which seems to indicate that the $D\bar D^*$ and $D^*\bar D^*$ thresholds play important roles in the characteristics of the $Z_c$. Another important subject to study might be  the putative  strange hidden-charm states, like the $Z_{cs}$ which replaces up or down quark in $Z_c$ with strange quark .  If a $Z_{cs}$ state exists, perhaps decaying to $K J/\psi$, this would provide another crucial handle on the nature of the $Z_c$ states.  To search for the heavier $Z_{cs}$ states, larger samples of higher-energy data samples will be required.

\subsubsection{The $X$ Problem}

The interpretation of the $X(3872)$ is intimately related to the problem of determining the parameters of the conventional $2P$ $c\bar{c}$ states -- the spin-triplet $\chi_{cJ}(2P)$ and the spin-singlet $h_c(2P)$.  This is because the $X(3872)$, with $J^{PC}=1^{++}$, certainly contains some admixture from the $\chi_{c1}(2P)$ state.  Sorting out the spectrum of states in this region, and determining which are exotic and which are conventional and how they mix, is the ``$X$" problem.

One important piece of information in this program to sort out the $2P$ states is the mass of the $h_c(2P)$.  Once this state is found, it will indicate where the $\chi_{cJ}(2P)$ states are located.  Then states deviating from this pattern can be identified as exotic.  Since there is already clear evidence for   $e^+e^- \to \pi^+\pi^-\psi(1S,2S)$ and $\pi^+\pi^-h_c(1P)$ at \bes3, one  also expects to produce the process $\pi^+\pi^-h_c(2P)$.  Search is underway, but larger data sets at higher cms energies are needed.

At least in principle, \bes3 also has access to a variety of other $X$ states through radiative transitions, $e^+e^-\to \gamma X$, in the similar way the $X(3872)$ is produced.  For example, this is being used to search for the $X(3915)$ in the process $e^+e^- \to \gamma \omega J/\psi$ and to search for the $X(4140)$ in the process $e^+e^- \to \gamma \phi J/\psi$.  

\subsection{Possibilities for $XYZ$ Data Taking}

The $X$, $Y$ and $Z$ problems are all Tier~A physics priorities within the charmonium group.  All could benefit from additional data sets.  Three qualitatively different types of data-taking plans are discussed below, each targeting different topics.

\subsubsection{(1) High-Statistics Scan from 4.0 to 4.6~GeV}

\begin{figure}[tbp]
\begin{centering}
\includegraphics*[width= 0.32\columnwidth]{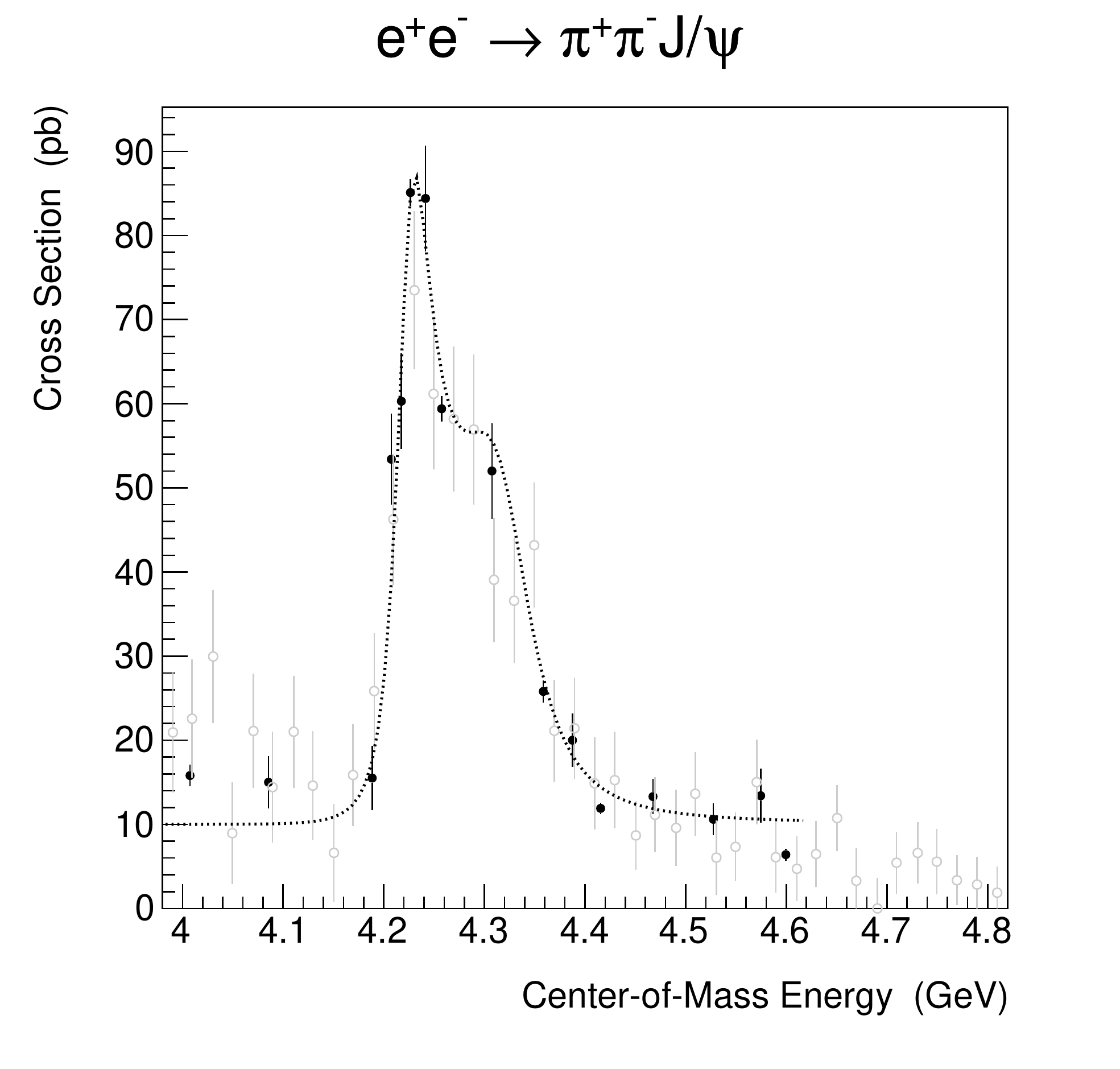}
\includegraphics*[width= 0.32\columnwidth]{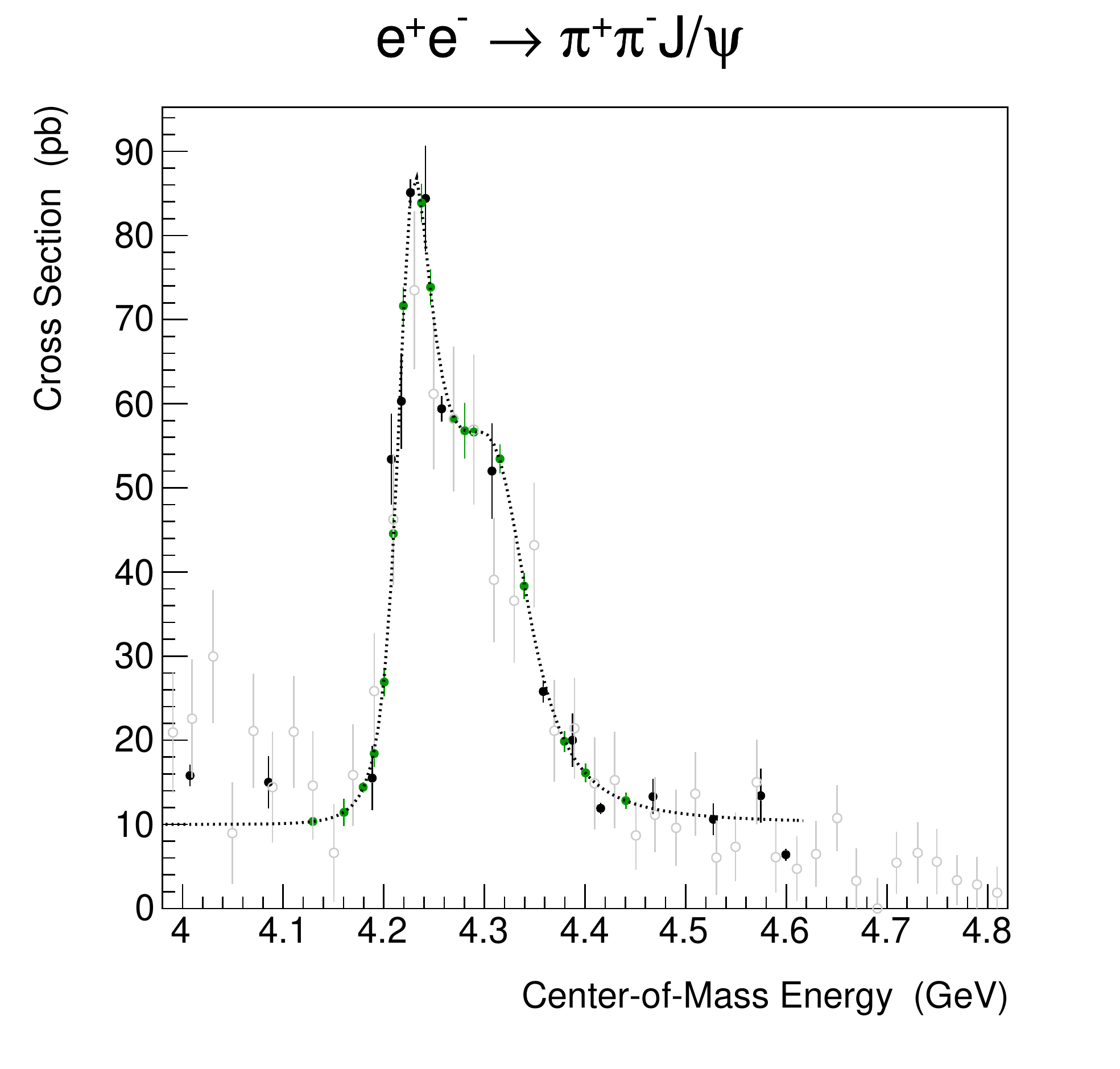}
\includegraphics*[width= 0.32\columnwidth]{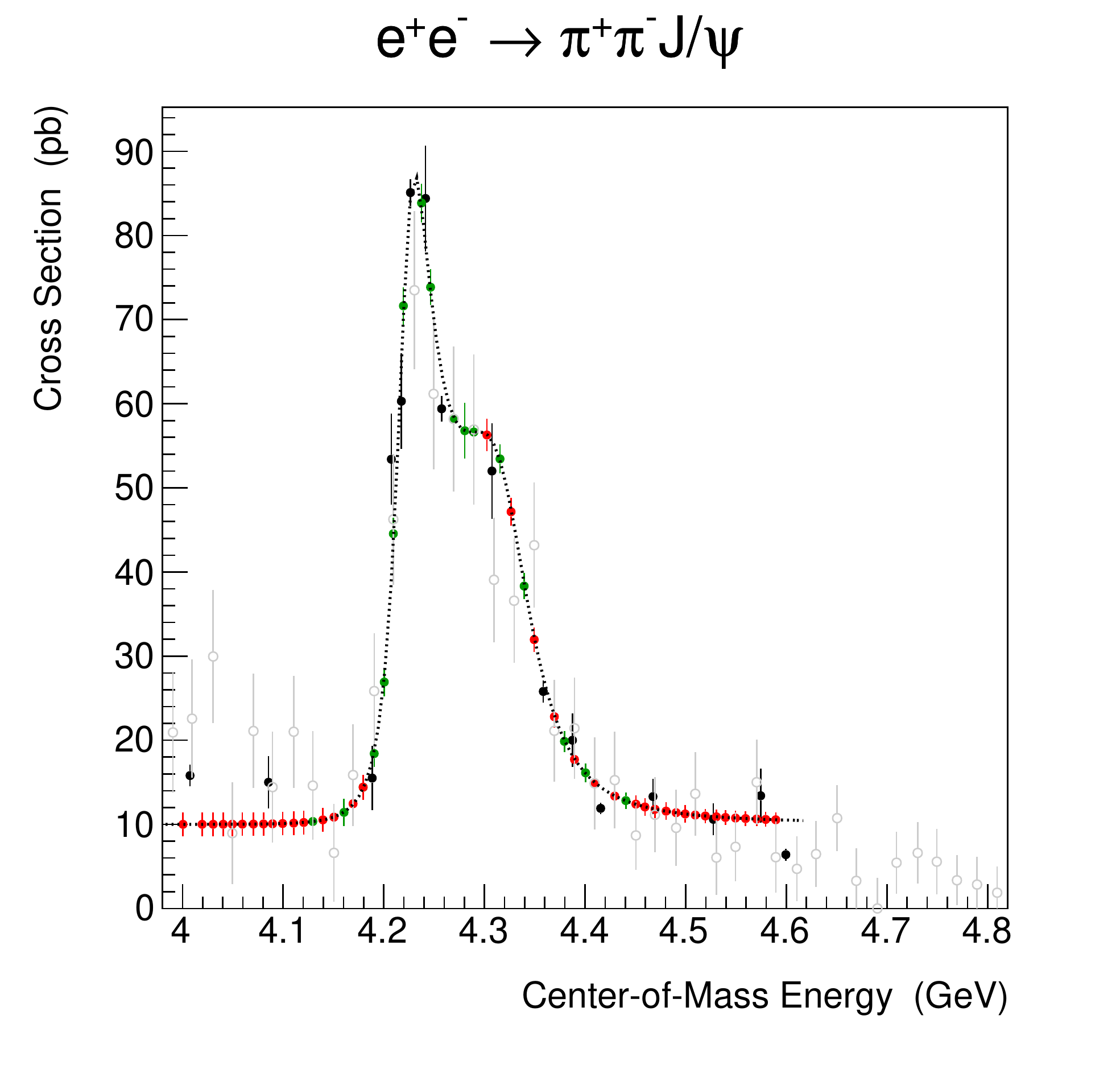} \\
\includegraphics*[width= 0.32\columnwidth]{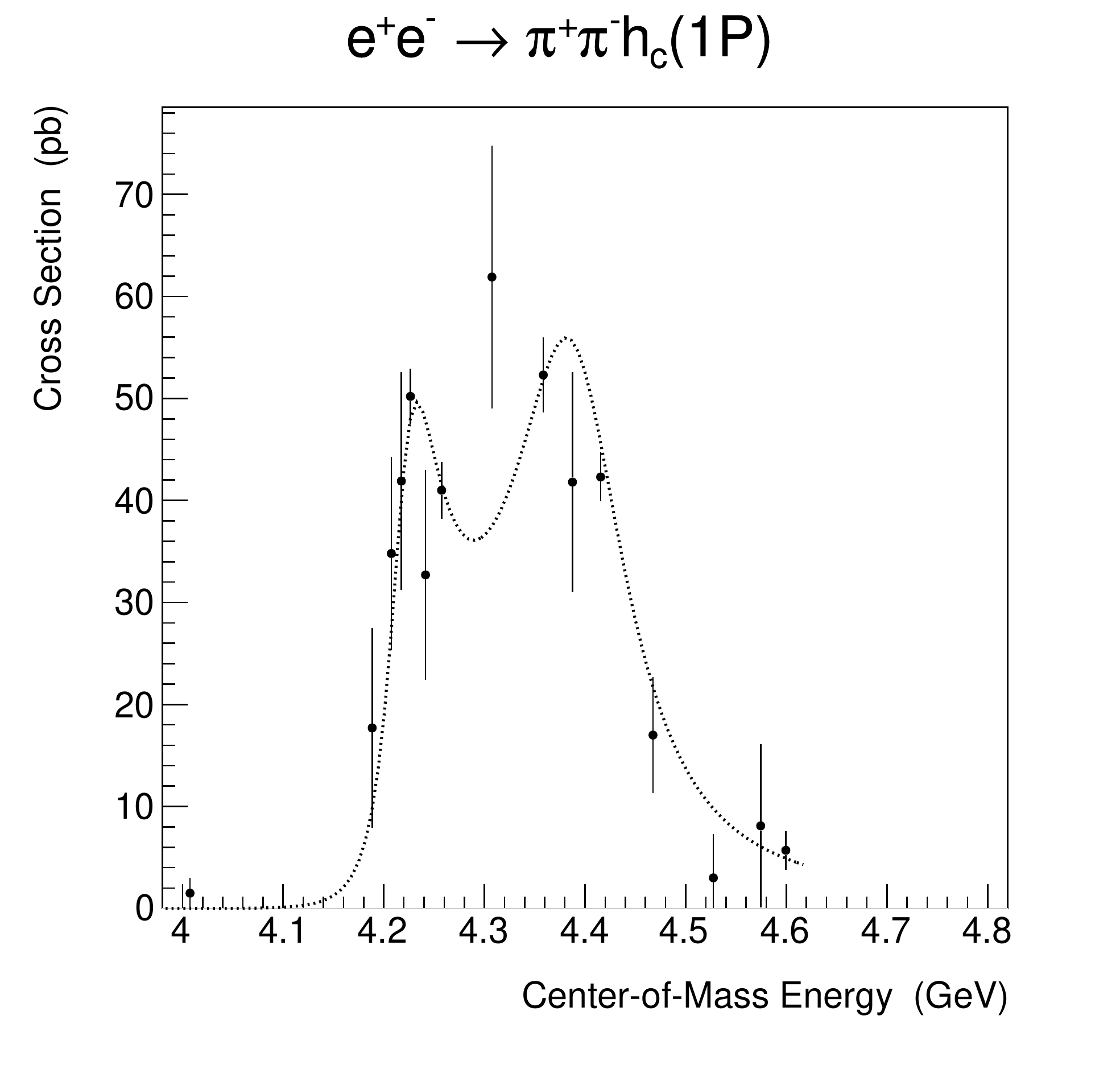}
\includegraphics*[width= 0.32\columnwidth]{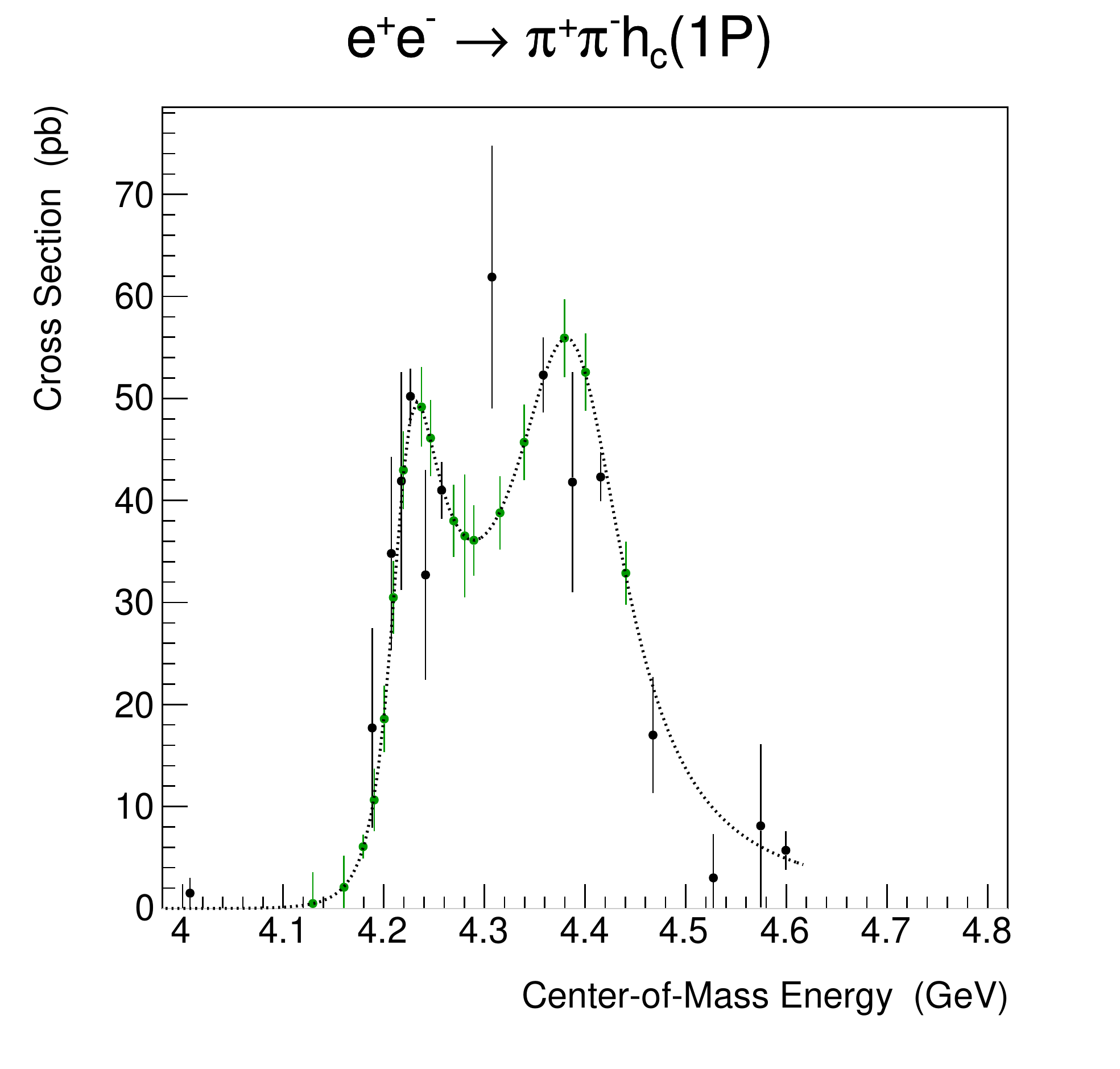}
\includegraphics*[width= 0.32\columnwidth]{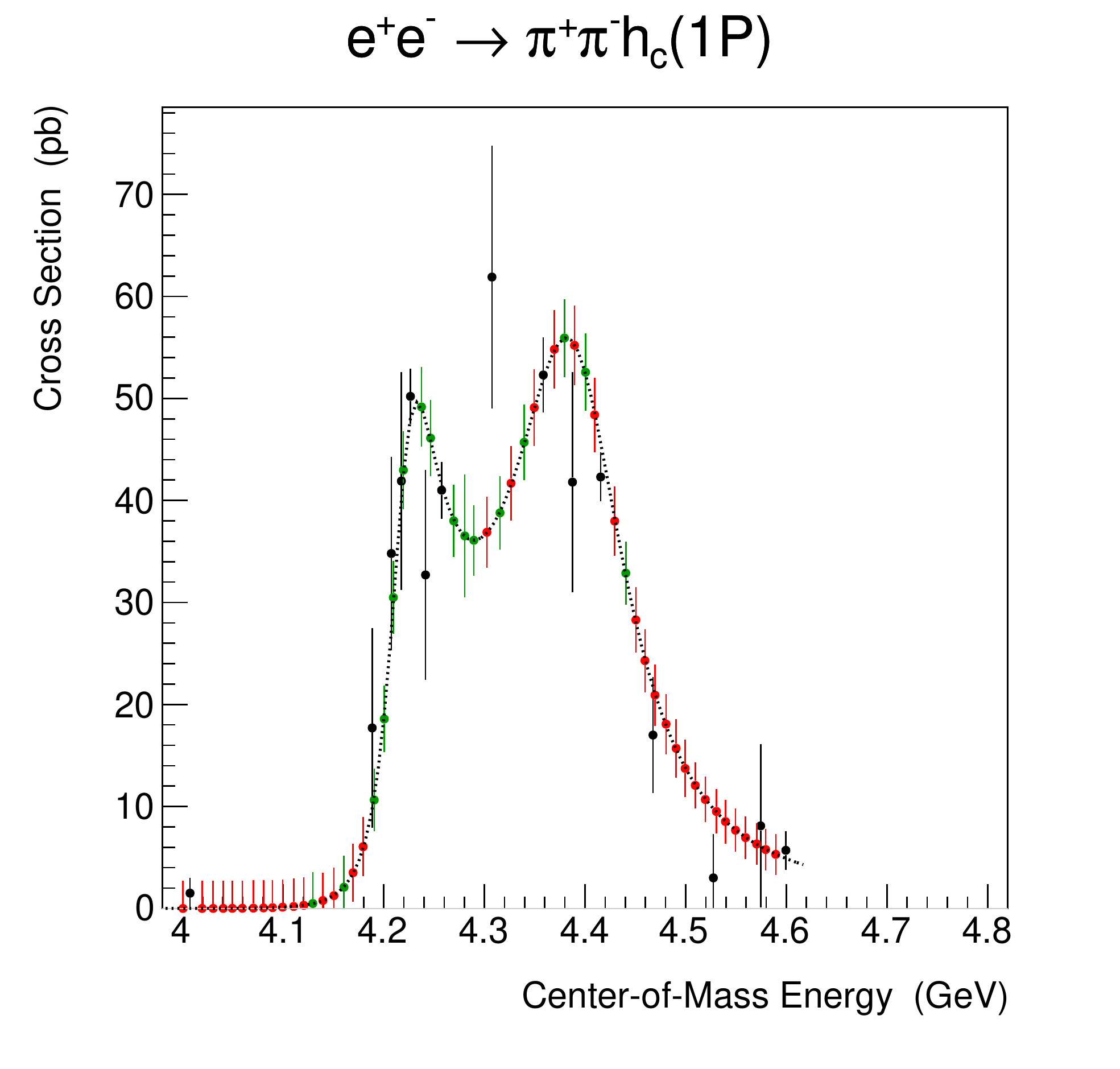} \\
\includegraphics*[width= 0.32\columnwidth]{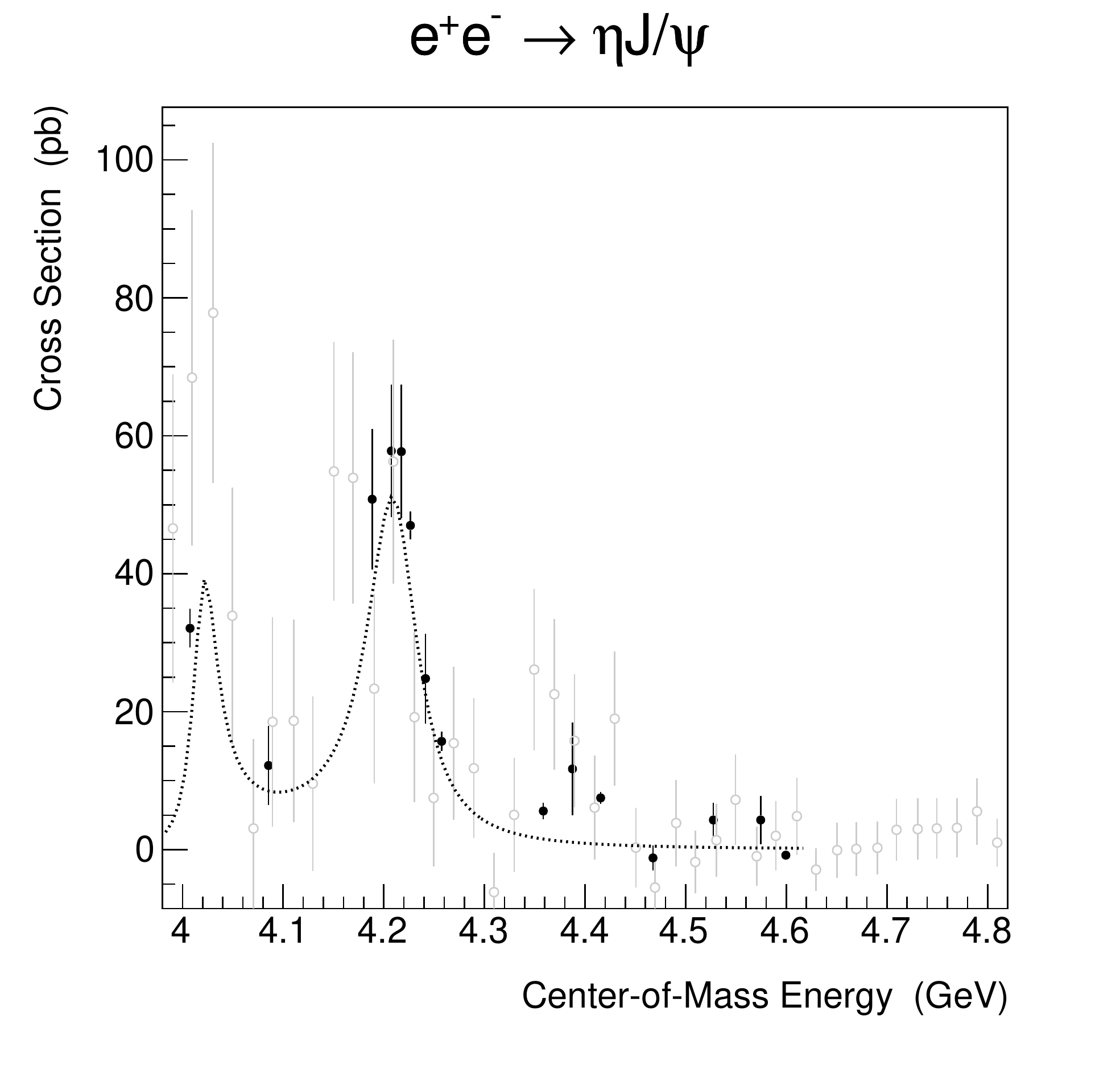}
\includegraphics*[width= 0.32\columnwidth]{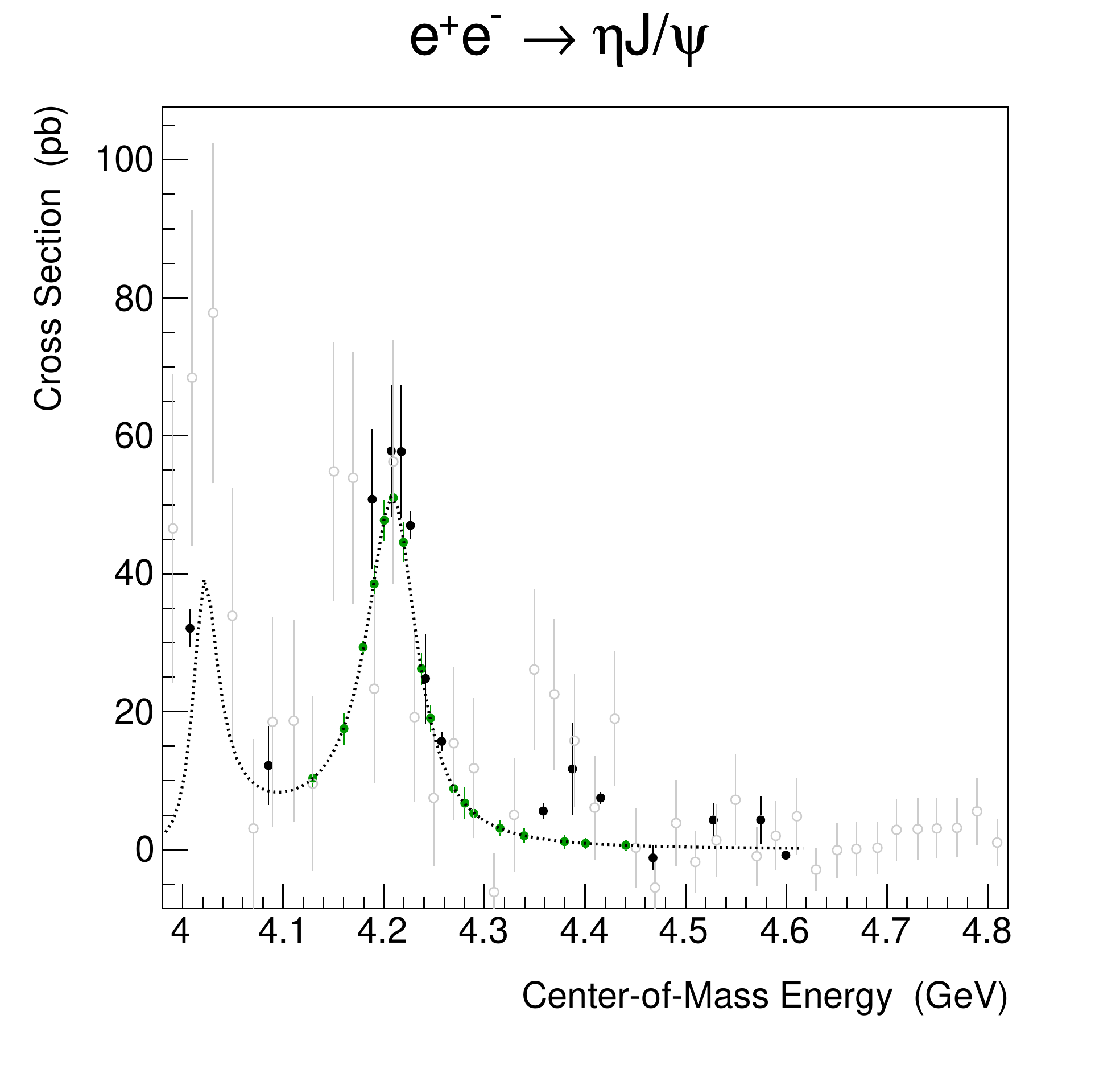}
\includegraphics*[width= 0.32\columnwidth]{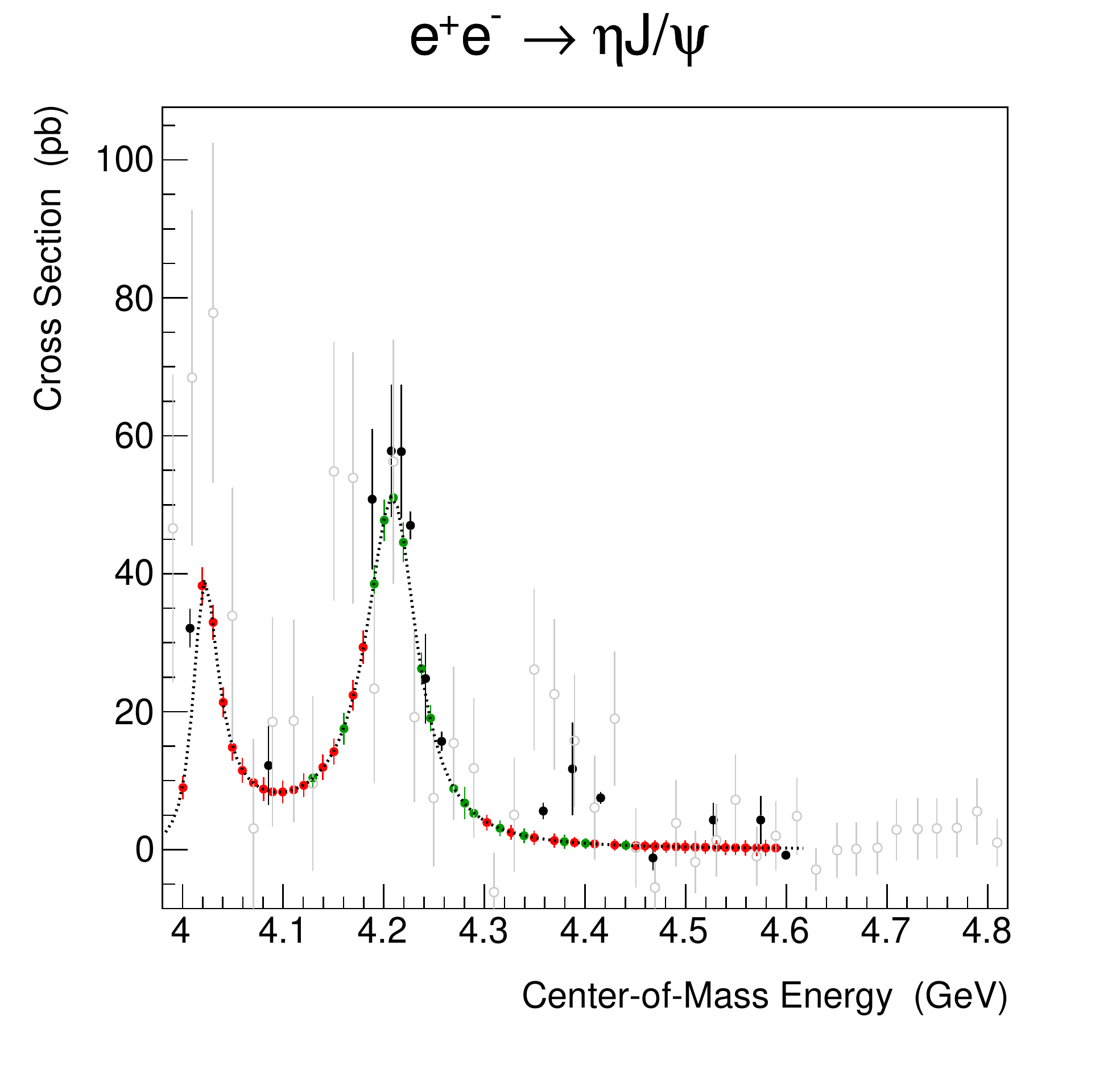} \\
\includegraphics*[width= 0.32\columnwidth]{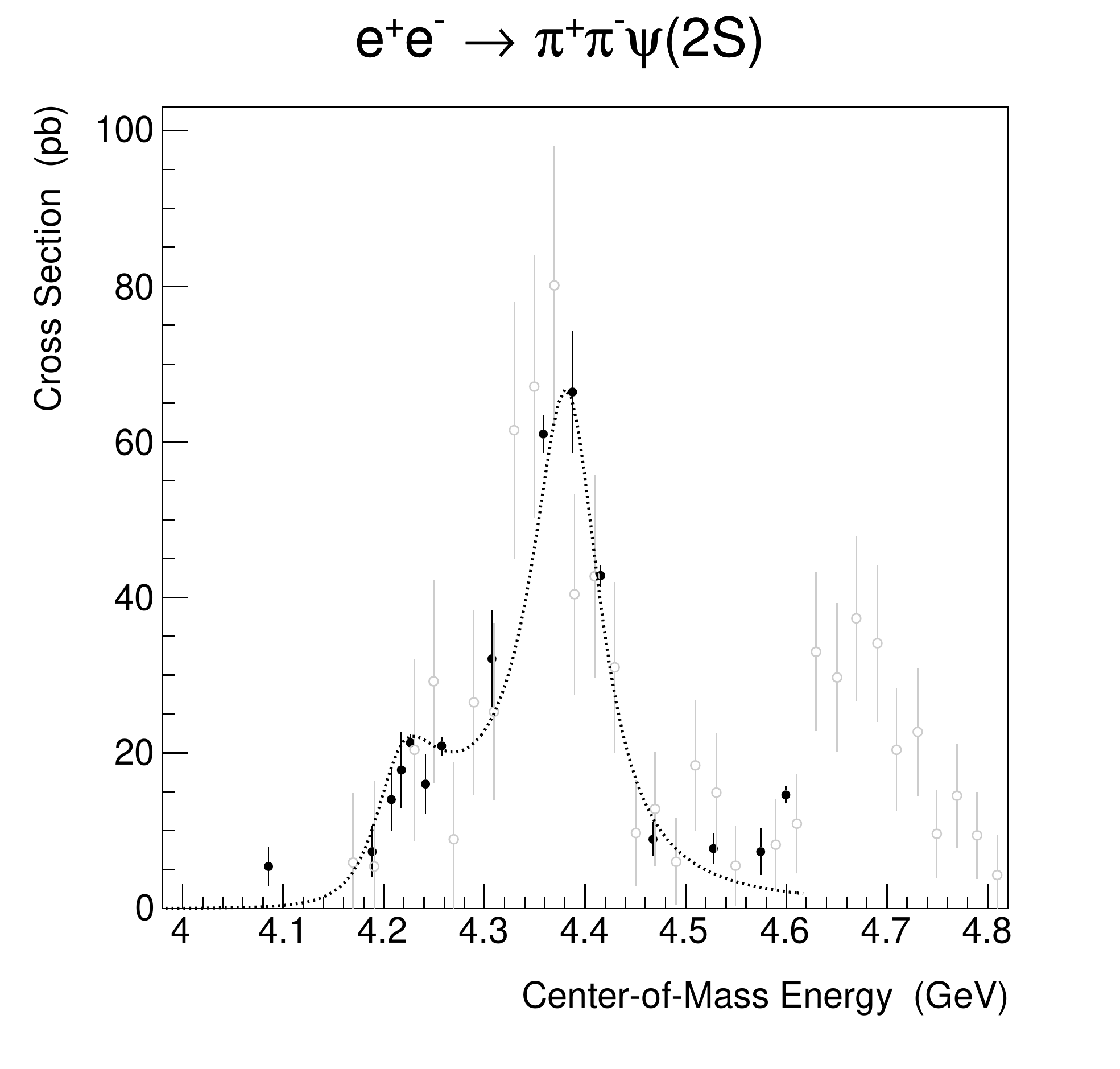}
\includegraphics*[width= 0.32\columnwidth]{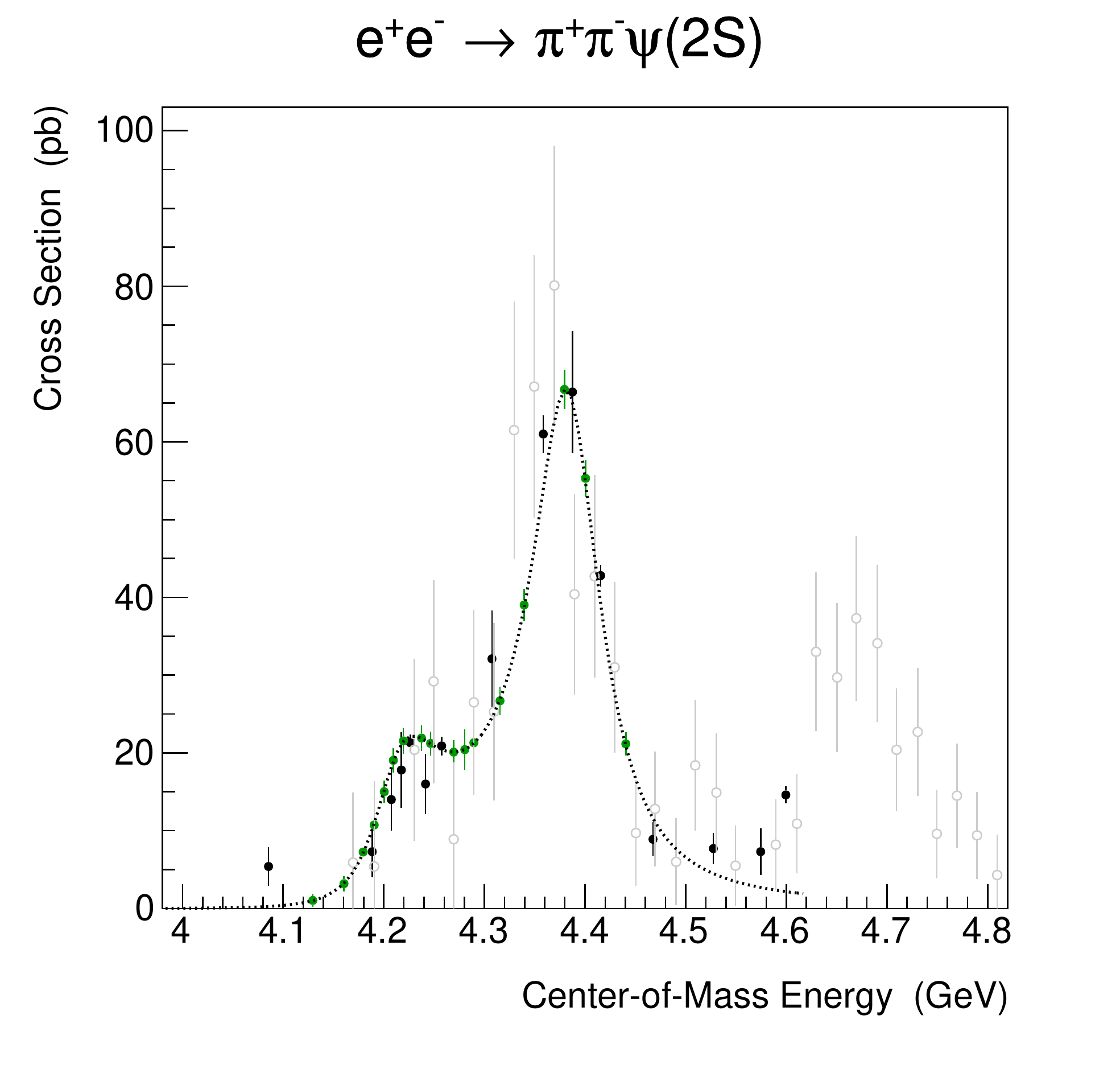}
\includegraphics*[width= 0.32\columnwidth]{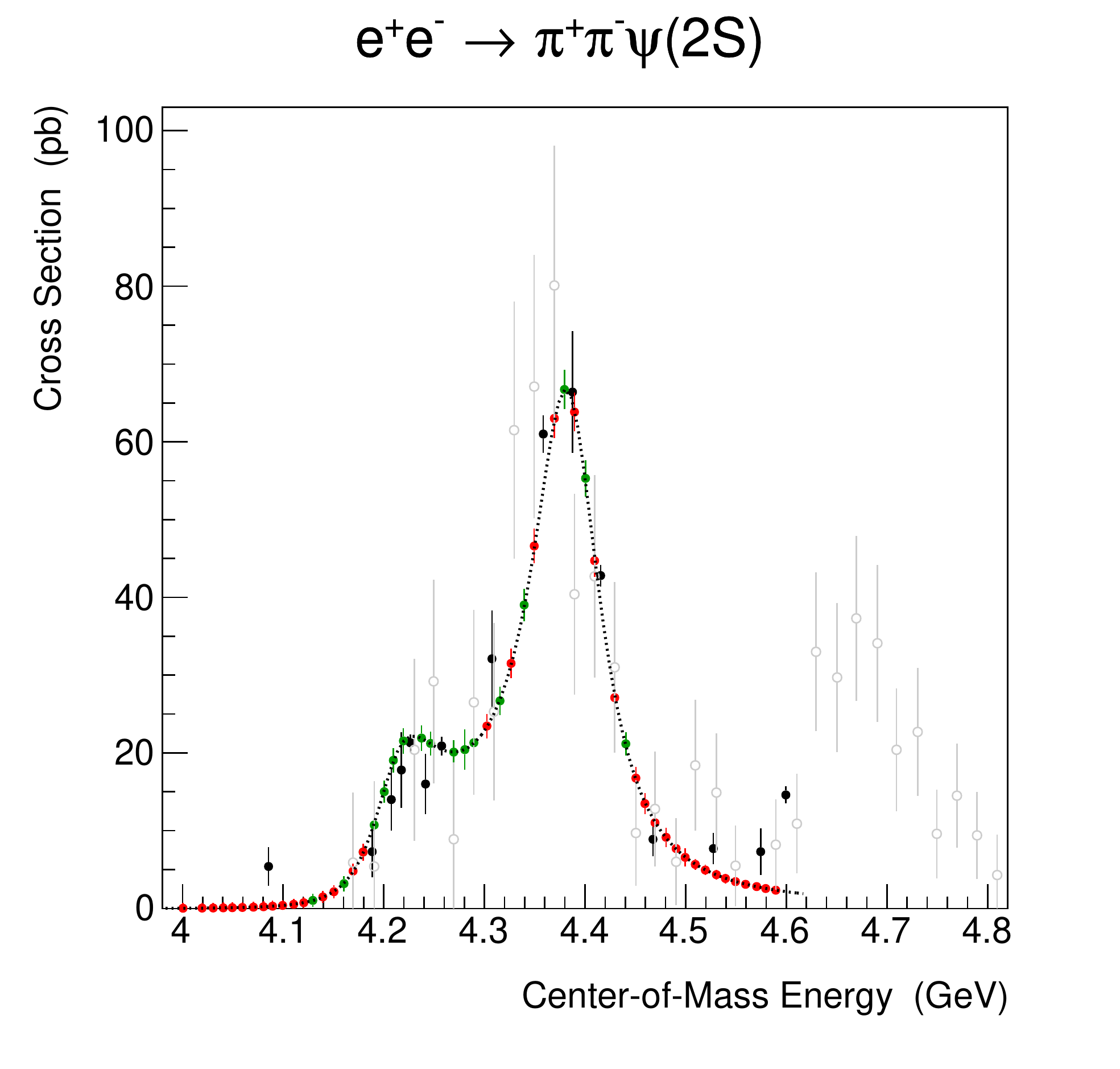} \\
\caption{\label{fig:XYZSimulations}Simulations for different exclusive $e^+e^-$ cross
  sections.  The black points are already measured (before 2019); the green points are
  projections for 2019; the red points are projections for the proposal from 4.0 to
  4.6~GeV, and the gray points are from Belle. The top row shows $e^+e^-\to \pi^+\pi^-
  J/\psi$; the second row shows $e^+e^-\to \pi^+\pi^- h_c$; the third row shows $e^+e^-\to
  \eta J/\psi$; the bottom row shows $e^+e^-\to \pi^+\pi^- \psip$.}
\end{centering}
\end{figure}

To study the $Y$ problem, we need to do a detailed scan of cross sections between 4.0 and 4.6~GeV.  The range is chosen in order to study a wide range of channels, important to achieve a more global picture in this region.  We propose 500~pb$^{-1}$ per point, for points spaced at 10~MeV intervals.  The intervals were chosen to cover the possibility for narrow features -- the narrowest of the $XYZ$ states discovered by \bes3 is currently the $Z_c(4020)$ with a width of 7~MeV, and the $\pi\pi J/\psi$ cross section shows rapid changes between 4.20 and 4.23~GeV.  Other cross sections are likely to include other rapidly changing features.
A series of simulations are shown in Fig.~\ref{fig:XYZSimulations}.  Background rates and efficiencies are based on existing data.  The cross section shapes are only guesses; they need to be measured.  Statistical fluctuations are not included.  The error bars, however, are reliable estimates.
The top part of Table~\ref{tab:xyzreq} shows the precision with which cross sections could be measured for a few select channels at 4.30~GeV.

This high-statistics scan would also allow us to study the way the $Z_c$ states evolve with cms energy.  Changes in the shapes of the peaks and/or their production cross sections would provide clues about the nature of the $Z_c$.  By combining data sets at different cms energies, we could also better explore the $X$ problem.

\begin{table}[tbp]
  \caption{Requirements for $XYZ$ data-taking for a few select channels.}
  \label{tab:xyzreq}
  \begin{tabular}{lcccc}
    \hline
    \hline
   channel & data plan  & luminosity & cross section precision & \# of events \\
    \hline
$\pi^+\pi^-J/\psi$ & (1) & 500 pb$^{-1}$ at 4.30~GeV & 3\%  & 1270  \\
$\pi^+\pi^-h_c(1P)$ & (1) & 500 pb$^{-1}$ at 4.30~GeV & 9\%  & 220  \\
$\eta J/\psi$ & (1) & 500 pb$^{-1}$ at 4.30~GeV & 30\%  & 28  \\
$\pi^+\pi^-\psip$ & (1) & 500 pb$^{-1}$ at 4.30~GeV & 3\%  & 230  \\
    \hline
$\pi^+\pi^-J/\psi$ & (2) & 5 fb$^{-1}$ at 4.23~GeV & $<$1\%  & 18k  \\
$\pi^+\pi^-J/\psi$ & (2) & 5 fb$^{-1}$ at 4.42~GeV & 3\%  & 3k  \\
$\pi^+\pi^-\psip$ & (2) & 5 fb$^{-1}$ at 4.42~GeV & 2\%  & 4k  \\
    \hline
    \hline
  \end{tabular}
\end{table}

\subsubsection{(2) A Series of Higher-Statistics Points}

For the $Z$ problem, especially detailed studies of Argand diagrams, we require very high statistics samples at a few points.  We currently have 1~fb$^{-1}$ of data for $e^+e^-$ cms energy at 4.42~GeV.  For the $\pi\pi J/\psi$ channel, this is not adequate to definitely resolve the substructure.  We require on the order of 5~fb$^{-1}$ or more per point to have adequate statistics for unambiguous Dalitz plot analyses.  Three or four of these high-statistics points would likely reveal the nature of the energy dependence of the Dalitz plot.  A working group is studying the substructure within various channels using the 4.42~GeV data.  As the working group progresses, the requirements on future data-taking will become more clear.  The lower part of Table~\ref{tab:xyzreq} lists the expected numbers of events for a few channels given future 5~fb$^{-1}$ data samples.

\subsubsection{(3) Possibilities for Data Above 4.6~GeV}

If \bes3 will take data above 4.6~GeV, a number of new exciting possibilities would become accessible.  Here we list five cases. (1)~We know there is an unexplained peak at 4.66~GeV in the $e^+e^- \to \pi^+\pi^-\psip$ cross section.  \bes3 would be able to produce and study it directly.  (2)~We could also study the peak at 4.63~GeV in the $\Lambda_c^+ \bar{\Lambda}_c^-$ cross section.  (3)~We could search for new peaks; it seems likely that more than those two exist.  These three goals would extend our study of the $Y$ problem.  
(4)~Production of a $Z_{cs}$ tetraquark candidate in the process $e^+e^-\to K Z_{cs}$ with
$Z_{cs} \to K J/\psi$ would require cms energies above 4.6~GeV if the $Z_{cs}$ is near the
$D_s \bar{D}^*$ or $D_s^*\bar{D}$ threshold.  
(5)~Data significantly above 4.6~GeV provide access to additional charmed baryon thresholds.  \bes3 would thus be able to study charmed baryons in a uniquely clean environment. (6) Data above 4.6~GeV would provide a unique opportunity to search for the excited $1^{+-}$ $h_c$ state, expected to be around 3.9~GeV and show up in $e^+e^-\to \eta D\bar D^*$. The identification of this state would likely help clarify many of the $J^{++}$ states between 3.8 and 4.0~GeV.

\subsection{Comparisons with Other Experiments}

\begin{figure}[tbp]
\begin{center}
\includegraphics[width=0.7\textwidth]{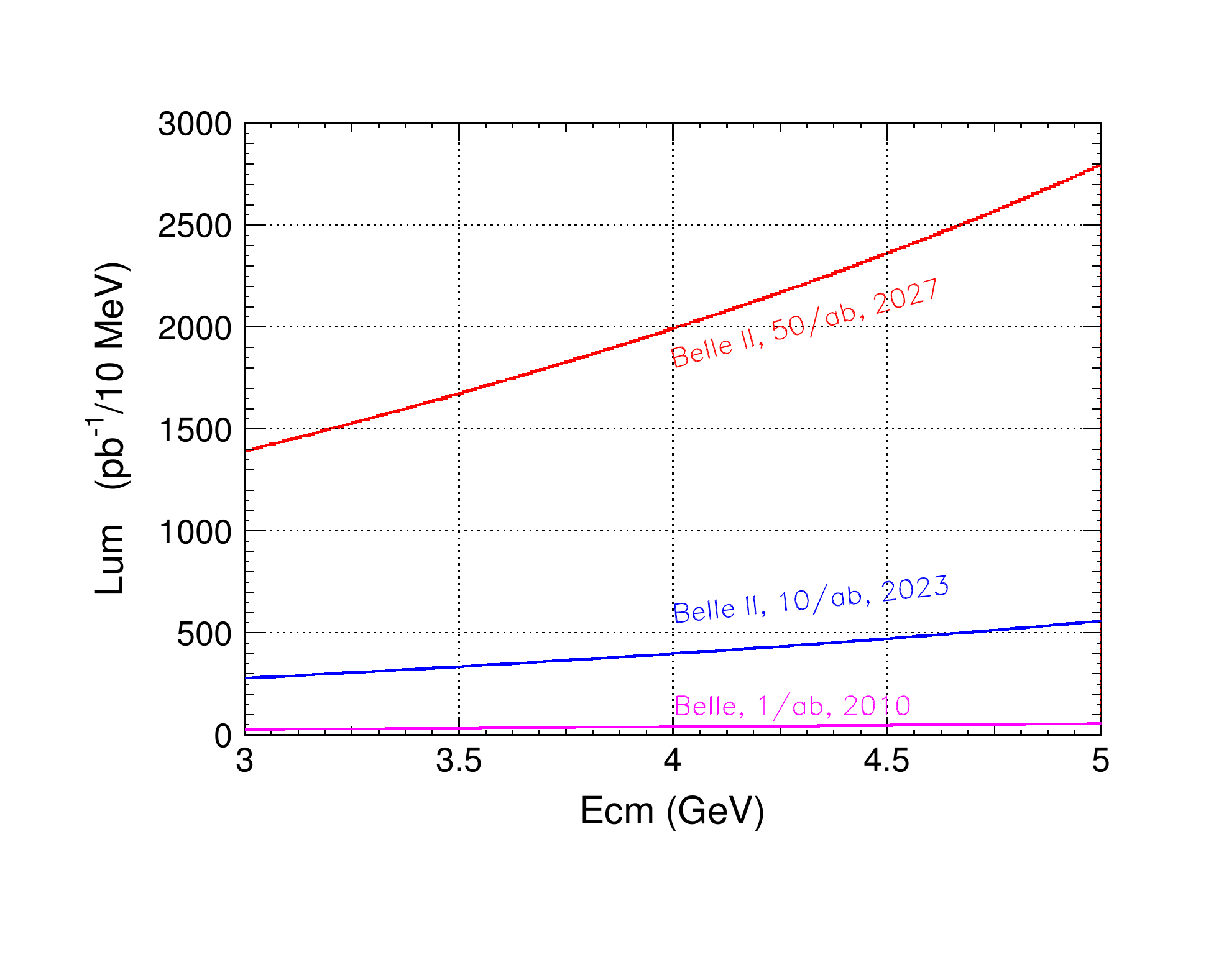} \\
\caption{Effective luminosity at low energy in the Belle and Belle
II $\Upsilon(4S)$ data samples.
} \label{lum_belle2}
\end{center}
\end{figure}

Belle~II~\cite{belle2} started collecting data in 2019, and
will accumulate 50~ab$^{-1}$ data at the $\Upsilon(4S)$ peak by
2027. These data samples can be used to study the $XYZ$ and
charmonium states in many different ways~\cite{PBFB}, among which
ISR can produce events in the same energy range covered by \bes3.
Figure~\ref{lum_belle2} shows the effective luminosity at BEPCII
energy in the Belle~II data samples. We can see that 50~ab$^{-1}$ of
Belle~II data corresponds to 2,000--2,800~pb$^{-1}$ data for each
10~MeV interval between 4--5~GeV, similar statistics will be accumulated for modes
like $\EE\to \ppjpsi$ at Belle~II and \bes3, taking into account
the fact that Belle~II has lower efficiency. 

Table~\ref{tab:belle} also lists the expected ratios of numbers of events for \bes3 and Belle~II given \bes3 data sets with 500~pb$^{-1}$ per point and a Belle~II sample of 50~ab$^{-1}$.
Belle~II has the advantage that data at different energies will be accumulated at
the same time, making the analysis simpler than  \bes3 scans over
many data points.
On the other hand, Belle~II needs to integrate over large energy bins, while the \bes3 data is collected at individual energy points with an energy resolution spread of around 2~MeV.

The LHCb experiment has also made large contributions to our understanding of the $XYZ$
mesons.  The strength of LHCb for studies of $B$ decays to final states with all charged
particles cannot be matched by $e^+e^-$ facilities.  For example, LHCb has made precision
measurements of the $X(3872)$ in the process $B\to K X$ with $X \to \pi^+ \pi^- J/\psi$
and $J/\psi \to \mu^+\mu^-$~\cite{lhcbx}.  The discovery of the process $B \to K Y(4260)$
with $Y(4260) \to \pi Z_c(3900)$ at D0~\cite{Abazov:2018cyu,D0:2019zpb} may give LHCb an
opportunity to study the $Y(4260)$ and $Z_c(3900)$ in detail.  However, since most of the
$XYZ$ particles decay to final states with neutral particles, the LHCb experiment is
mostly complementary to the \bes3 experiment. In contrast to BESIII, LHCb does not feature
a high quality electromagnetic calorimeter. Thus, the identification of soft neutral pions
and photons is difficult.

Finally, it should be noted that all three experiments have rather different systematics, which emphases their complementarity.

\begin{table}[tbp]
\centering
  \caption{A comparison of \bes3 and Belle~II for various channels at a few different $e^+e^-$ cms energies assuming 50~ab$^{-1}$ of Belle~II data in 10~MeV energy bins. $L$ and $N$ denote the expected luminosity and number of observed signals, respectively. }
  \label{tab:belle}
  \begin{tabular}{lccc}
    \hline
    \hline
   ISR mode & $L_{\mathrm{BESIII}}$/$L_{\mathrm{Belle~II}}$ & $\varepsilon_{\mathrm{BESIII}}/\varepsilon_{\mathrm{Belle~II}}$ & $N_{\mathrm{BESIII}}/N_{\mathrm{Belle~II}}$  \\
    \hline
$\pi^+\pi^-J/\psi$ at 4.26 GeV & 0.5 fb$^{-1}$ / 2.2 fb$^{-1}$ & 46\% / 10\% & 1.07   \\
\hline
$\pi^+\pi^-\psip$ at 4.36 GeV & 0.5 fb$^{-1}$ / 2.3 fb$^{-1}$ & 41\% / 5\% & 1.82   \\
$\pi^+\pi^-\psip$ at 4.66 GeV & 0.5 fb$^{-1}$ / 2.5 fb$^{-1}$ & 35\% / 6\% & 1.19   \\
\hline
$\pi^+\pi^- h_c$ at 4.26 GeV & 0.5 fb$^{-1}$ / 2.2 fb$^{-1}$ & 2.7\% / --\% & $>5$   \\
%$\pi^+\pi^- h_c$ at 4.36 GeV & 0.5 fb$^{-1}$ / 2.3 fb$^{-1}$ &   & 1.19   \\
\hline
$K^+K^- J/\psi$ at 4.6 GeV & 0.5 fb$^{-1}$ / 2.4 fb$^{-1}$ & 29\% / 7.5\% & 0.81   \\
$K^+K^- J/\psi$ at 4.9 GeV & 0.5 fb$^{-1}$ / 2.7 fb$^{-1}$ & 
$\approx$29\% / 10\% & 0.54   \\
\hline
$\Lambda_c^+\bar{\Lambda}_c^-$ at 4.6 GeV & 0.5 fb$^{-1}$ / 2.4 fb$^{-1}$ & 51\% / 7.5\% & 1.42   \\
$\Lambda_c^+\bar{\Lambda}_c^-$ at 4.9 GeV & 0.5 fb$^{-1}$ / 2.7 fb$^{-1}$ & $\approx$37\% / 7.5\% & 0.91   \\
    \hline
    \hline
  \end{tabular}
\end{table}

\section{Summary of Data Requirements}
\label{sec:ccfuture}

Our data-taking requirements are summarized in Table~\ref{tab:plan}.  Sensitivities to various channels in $XYZ$ physics are given in Table~\ref{tab:xyzreq}; sensitivities to charmonium studies are listed in Table~\ref{chamonium:table}.

\begin{table}[tbp]
  \caption{Data-taking requirements for $XYZ$ physics~(top) and charmonium physics~(bottom).}
  \label{tab:plan}
  \begin{tabular}{ll}
    \hline
    \hline
   Plan & Data Sets  \\
    \hline
$XYZ$ plan (1) & 500 pb$^{-1}$ at a large number of points between 4.0 and 4.6~GeV  \\
$XYZ$ plan (2) & 5 fb$^{-1}$ at 4.23, 4.42~GeV for large $Z_c$ samples \\
$XYZ$ plan (3) & 5 fb$^{-1}$ above 4.6~GeV \\
\hline
charmonium plan & $3\times 10^9$ $\psip$ decays \\
    \hline
    \hline
  \end{tabular}
\end{table}

%% file: QCD/qcd.tex
\chapter[$R$ values, QCD and $\tau$ Physics]{$R$ values, QCD and $\tau$ Physics}
\label{chapter:qcd}

\input{QCD/qcd_main.tex}

\input{QCD/bib.tex}

%% file: QCD/qcd_main.tex
%\section{The physics programme}

%\section{Precision tests of the Standard Model}

\input{QCD/introduction}

\input{QCD/gminus2}

\input{QCD/ISR}

% R value
\input{QCD/rvalue}
%\input{QCD/alpha}

% hadron form factor
\input{QCD/form}

%baryon form factor
\input{QCD/baryon}

% hadron production
\input{QCD/hadron}

%tau mass
\input{QCD/tau}

%phase difference
\input{QCD/phase}

% meson specctum
\input{QCD/meson}

% requirements for data
\input{QCD/data}

%% file: QCD/introduction.tex
% introduction
\section{Introduction}

The $R$-QCD-$\tau$ working group deals with various aspects of QCD, the accepted theory of strong interactions within the Standard Model of particle physics. The cms energy of BEPCII is ideally suited to study the transition region between non-perturbative and perturbative aspects of QCD~\cite{Asner:2008nq}. It offers a unique laboratory to test not only the validity of QCD in the few-GeV energy range, but also the validity of effective theories and hadronic models towards high energies.

\begin{figure}[htb]
 \centerline{\includegraphics[width=0.8\textwidth]{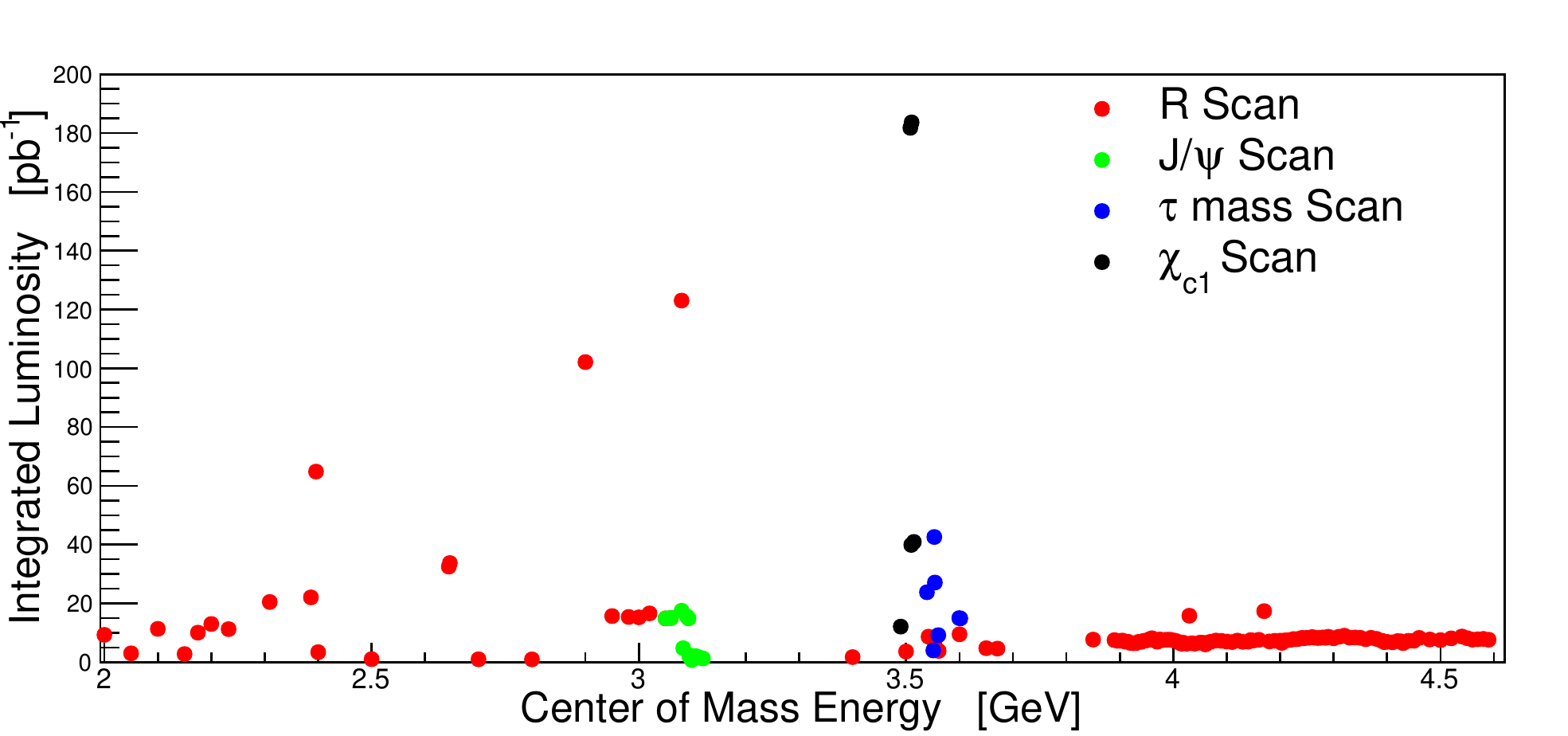}}
 \caption{\label{fig:TauQCDdata}Overview of data sets and their integrated luminosities acquired for the $R$-QCD-$\tau$ working group. $R$ scan data (red) were collected from 2012-2015, the $J/\psi$ line shape scan (green) was performed in 2012, a search for the production of $\chi_{c1}$ (black) was carried out in 2017, and the latest $\tau$ mass scan data (blue) were acquired in 2018.}
\end{figure}

\bes3 has already demonstrated that the high statistics and accuracy of the data, illustrated in Fig.~\ref{fig:TauQCDdata}, allow to measure form factors of mesons and baryons with unprecedented accuracy and with paramount impact on hadron structure investigations. These form factors are not only accessible in the timelike domain via electron-positron annihilations, but can also be determined in the spacelike domain in two-photon scattering.

As is well known, the measurement of electron-positron annihilation into hadrons is of utmost importance for precision tests of the SM. While the measurement of exclusive hadronic channels is needed for the hadronic vacuum polarization contribution of the anomalous magnetic moment of the muon, $(g-2)_\mu$, the inclusive measurement of the hadronic cross section above 2\,GeV can be used to improve the knowledge of the fine structure constant at the $Z$-pole. The latter is currently limiting precision tests in the electroweak sector of the SM. Thus, its improvement is of central importance for the physics programs at future high-energy electron-positron colliders, such as Higgs factories~\cite{CEPCStudyGroup:2018ghi,Abada:2019zxq}. The exclusive channels can be measured at \bes3 via the initial state radiation (ISR) technique, while the accuracy of the inclusive $R$-measurement will be further improved with respect to measurements at BES and KEDR. Regarding $(g-2)_\mu$, \bes3 can also measure the transition form factors of pseudoscalar mesons, as motivated by the hadronic light-by-light scattering contributions to $(g-2)_\mu$.

The analysis of the high-statistics energy scan between 2.0 and 4.6\,GeV allows for a first precision measurement of neutron and hyperon form factors. In the case of hyperons, their self-analyzing decay does not only provide access to the absolute values of the electric and magnetic form factors, but it allows even to extract the relative phase between them. A complementary view on hadron structure is possible through the measurement of Collins asymmetries. The recent result obtained at \bes3 extends previously available knowledge, which was restricted to measurements at $B$ factories, towards lower energies. The picture of the non-perturbative features of QCD can be further extended by the investigation of fragmentation functions. 

Naturally, the research program of the $\tau$-QCD working group comprises measurements of such fundamental parameters of the SM as the mass of the $\tau$ lepton. It can be determined via an energy scan in its threshold region. The first result of \bes3 is already the world's most precise measurement. Further improvement is expected with new data. Moreover, the mass of the charm quark is accessible by determining the cross section of $e^+e^- \to c\bar{c}$, which relates the physics of hadronic cross section measurements to yet another precision observable of the SM. 

Finally, the precise scan measurements for the $R$ value and $\tau$ mass measurements allow for detailed investigations of the production of charmonium resonances. The line shape of vector charmonium resonances might reveal through a relative phase of strong and electromagnetic decay amplitudes a more detailed picture of their internal structure. Even the production of non-vector resonances like the $\chi_{c1}$ in $e^+e^-$ annihilations might be observed through interference patterns with the continuous background. Another exciting spin-off of the \bes3 $R$-program are investigations of the light hadron spectrum, enabling studies of potentially exotic resonances, such as the $\phi(2170)$.

%% file: QCD/gminus2.tex
\section[\bes3 measurements related to precision variables $(g-2)_\mu$ and $\alpha_{\rm em}(s)$]{\bes3 measurements related to precision variables $\boldsymbol{(g-2)_\mu}$ and $\boldsymbol{\alpha_{\rm em}(s)}$ }
\label{bes_g-2}

\subsection{The anomalous magnetic moment of the muon, $(g-2)_\mu$}
At present, both the SM prediction, as well as the experimental value of the anomalous magnetic moment of the muon, $a_\mu=(g-2)_\mu/2$~\cite{Jegerlehner:2017gek} are determined with a relative uncertainty of about one half part per million: 
\begin{align}
  a_\mu^{\rm SM}& = ( 11\,659\,182.04\pm 3.56 ) \times 10^{-10}& &\text{[Keshavarzi {\it et al.}]~\cite{ref_teubner18},} 
\\
  a_\mu^{\rm exp}& = ( 11\,659\,208.9\;\,\pm 6.3 ) \times 10^{-10}& &\text{[BNL-E821]~\cite{ref_e821}.}
\label{eq:amu_status}
\end{align}

\begin{figure}[tp]
 \centerline{ \includegraphics*[width=13cm]{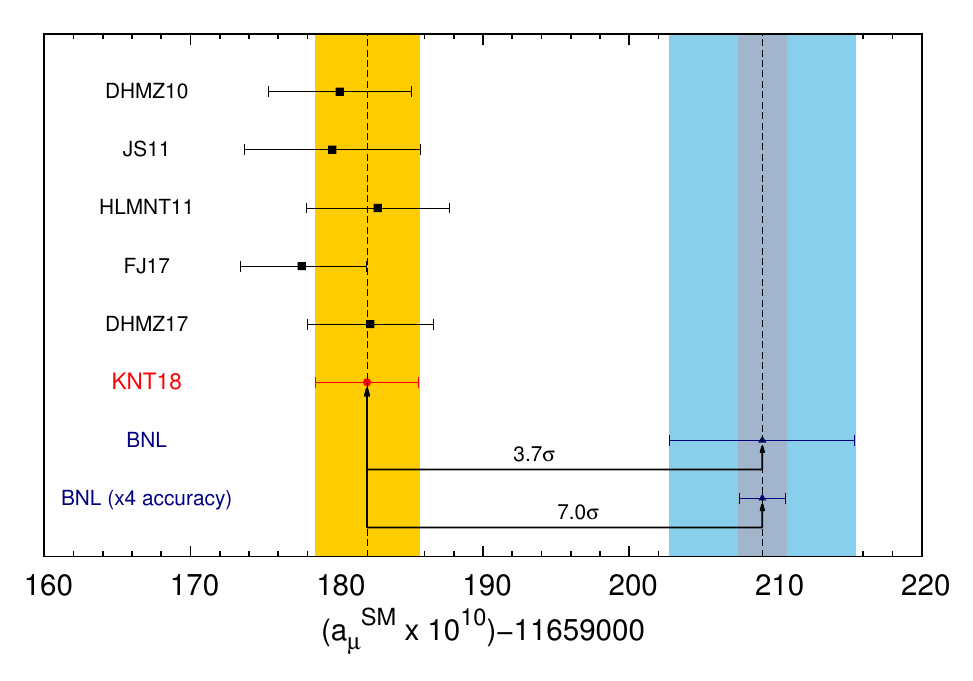} }
 \caption{\label{fig:amucompilation}Compilation of recent results for $a_\mu^{\rm SM}$ (figure taken from 
Ref.~\cite{ref_teubner18}). The ``DHMZ17'' and ``KNT18'' points correspond to Refs.~\cite{Davier:2017zfy}  
and~\cite{ref_teubner18}, respectively. Also shown is the accuracy of the upcoming new FNAL measurement of $a_\mu^{\rm 
exp}$ under the hypothetical assumption that the central mean value of $a_\mu^{\rm exp}$ will remain unchanged.}
\end{figure}

A second evaluation of $a_\mu^{\rm SM}$ by Davier {\it et al.} finds $a_\mu^{\rm SM} = ( 11\,659\,182.3 \pm 4.3 ) \times 10^{-10}$~\cite{Davier:2017zfy}. Depending on the value used for $a_\mu^{\rm SM}$, a tension of 3.5 -- 3.7~standard deviations between the SM prediction and the direct measurement of $a_\mu$ is observed, as illustrated in Fig.~\ref{fig:amucompilation}. 

Currently, the SM prediction of $a_\mu$ is slightly more precise than the experimental value, but the situation will change soon~\cite{ref_e989a}. New direct measurements have been started at FNAL~\cite{fnal} and a complementary project is under construction at J-PARC~\cite{jparc}. Both aim to reduce the experimental uncertainty by a factor of four. The light grey band in Fig.~\ref{fig:amucompilation} represents the hypothetical situation of the new FNAL measurement yielding the same mean value for $a_\mu^{\rm exp}$ with fourfold improved accuracy.

\begin{figure}[tb]
 \centerline { \includegraphics[width=3.8cm]{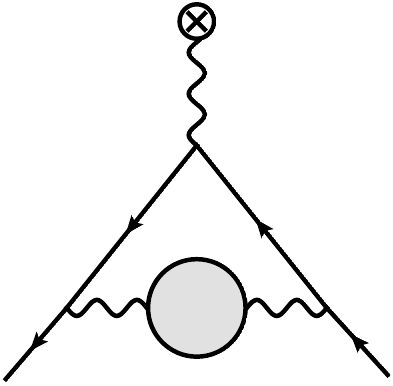}\hspace{1cm}%
                     \includegraphics[width=4.3cm]{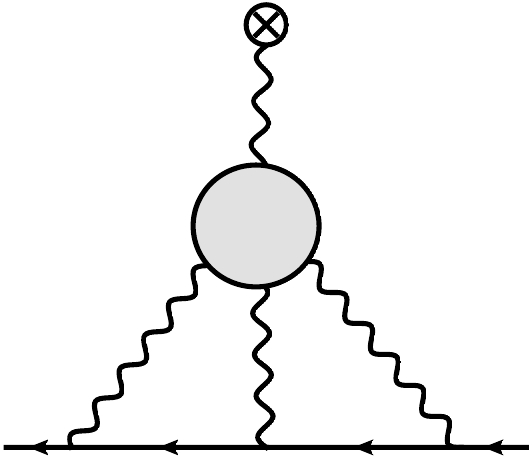} }
 \caption{\label{fig:hvphlblfeyn}The HVP (left panel) and the HLbL (right panel) contributions to the anomalous magnetic moment of the muon.}
\end{figure}

The main uncertainties in $a_\mu^{\rm SM}$ originate from hadronic effects, in particular the contributions from hadronic vacuum polarization (HVP), $a_\mu^{\rm HVP}$, and hadronic light-by-light scattering (HLbL), $a_\mu^{\rm HLbL}$, as shown in Fig.~\ref{fig:hvphlblfeyn}. It is of vital importance to investigate whether physics models beyond the SM~\cite{ref_stockinger} or poorly understood hadronic effects are responsible for the observed tension. This is one of the main goals of the $R$-QCD research program.

The current estimate of $a_\mu^{\rm HVP}$, which enters the SM prediction, is based on dispersion theory, in which experimental measurements of the annihilation cross section $e^+e^- \to {\rm  hadrons}$ are used as input for the evaluation of a dispersion integral:

\begin{equation}
a_{\mu}^{\rm HVP} = (\frac{\alpha m_{\mu}}{3\pi})^2 \int_{4m_{\pi}^2}^{\infty}ds\frac{R_{\rm had}(s) K(s)}{s^{2}},
\label{eq:amuhad}
\end{equation}
where $K(s)$ is a kernel varying from 0.63 at $s=4m_\pi^2$ to 1.0 at $s=\infty$,  and $m_\mu$($m_\pi$) is the nominal mass of the pion(muon).

\begin{figure}[tbp]
 \centerline{ \includegraphics*[width=0.48\textwidth]{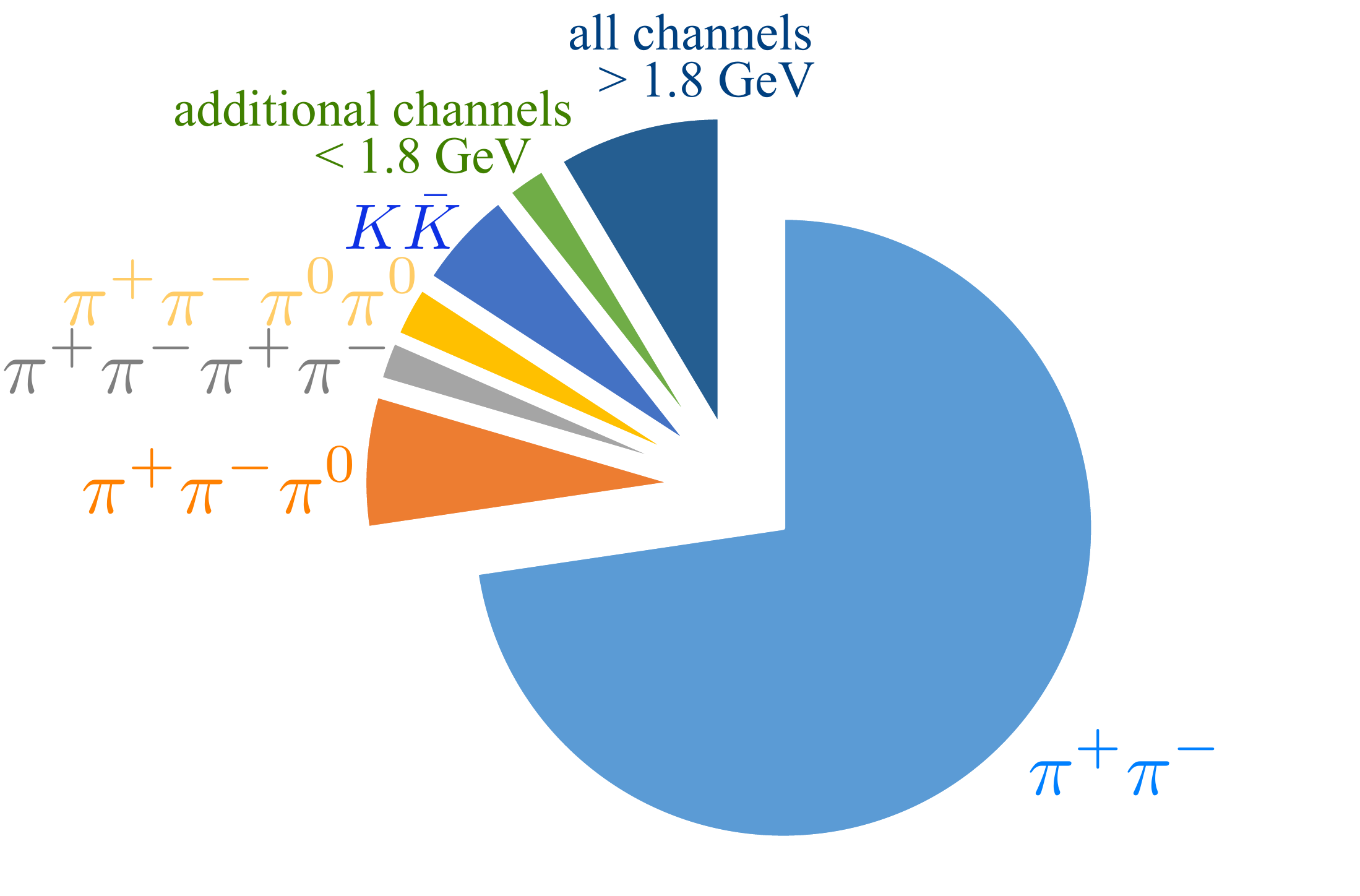}\hfill%
              \includegraphics*[width=0.48\textwidth]{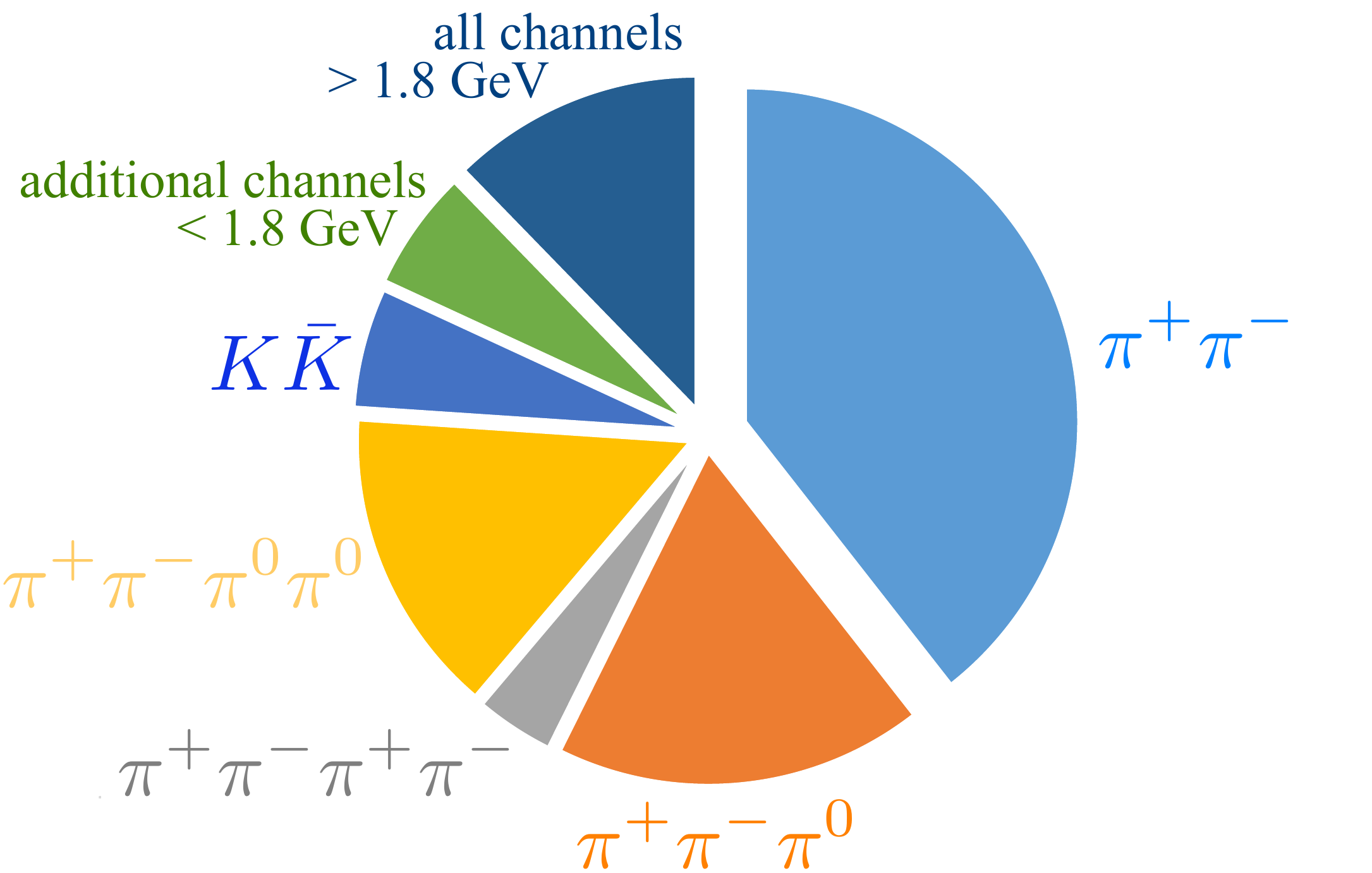}
  } 
 \caption{\label{fig:amupie}Contributions of different channels and energy ranges to the absolute value (left) and the uncertainty (right) of $a_\mu^\text{HVP}$, demonstrating the importance of the processes $e^+e^- \to \pi^+\pi^-$, $e^+e^- \to \pi^+\pi^-\pi^0$, $e^+e^- \to \pi^+\pi^-2\pi^0$, and $e^+e^-\to K\bar{K}$ (numbers taken from Ref.~\cite{ref_teubner18}).}
\end{figure}

Within Ref.~\cite{ref_teubner18}, the leading-order hadronic contribution is calculated to be $a_\mu^{\rm HVP} = (684.68 \pm 2.42) \times 10^{-10}$. At \bes3, we are aiming at a further reduction of the uncertainty of $a_{\mu}^\text{HVP}$ by measuring exclusive channels of $R_{\rm had}$ in the most relevant energy range. Figure~\ref{fig:amupie} shows the contributions of various exclusive hadronic channels to the absolute value (left) and the uncertainty (right) of $a_\mu^\text{HVP}$. As can be seen, the channels with $\pi^+\pi^-$, $\pi^+\pi^-\pi^0$, $\pi^+\pi^-\pi^0\pi^0$, and $K\bar{K}$ are most relevant, which the current efforts at \bes3 are focusing on exactly. A major program of hadronic cross section measurements, which is presented in detail in Sec.~\ref{sec:ISR}, has been launched. 

Beyond HVP, the next important contribution to the uncertainty of $(g-2)_\mu$ is given by the HLbL contribution shown in the right panel of Fig.~\ref{fig:hvphlblfeyn}. The leading contribution to the HLbL diagram is given by the coupling of photons to the pseudoscalar mesons $\pi^0, \eta$, $\eta^\prime$ as well as channels like $\pi\pi$ and $\pi\eta$, as shown in Fig.~\ref{fig:LbL}. So far, hadronic models have been used for the calculation of the HLbL diagram. Although most groups report similar values for the absolute size of the HLbL contribution, the assumed uncertainties differ largely. The calculation with the lowest uncertainties stems from Prades, de Rafael, and Vainshtein~\cite{M1_Prades:2009tw}. They find the following value: $a_\mu^{\rm HLbL}=(10.5 \pm 2.6) \times 10^{-10}$. 

\begin{figure}[tbp]
 \centerline{ \includegraphics*[width=9cm]{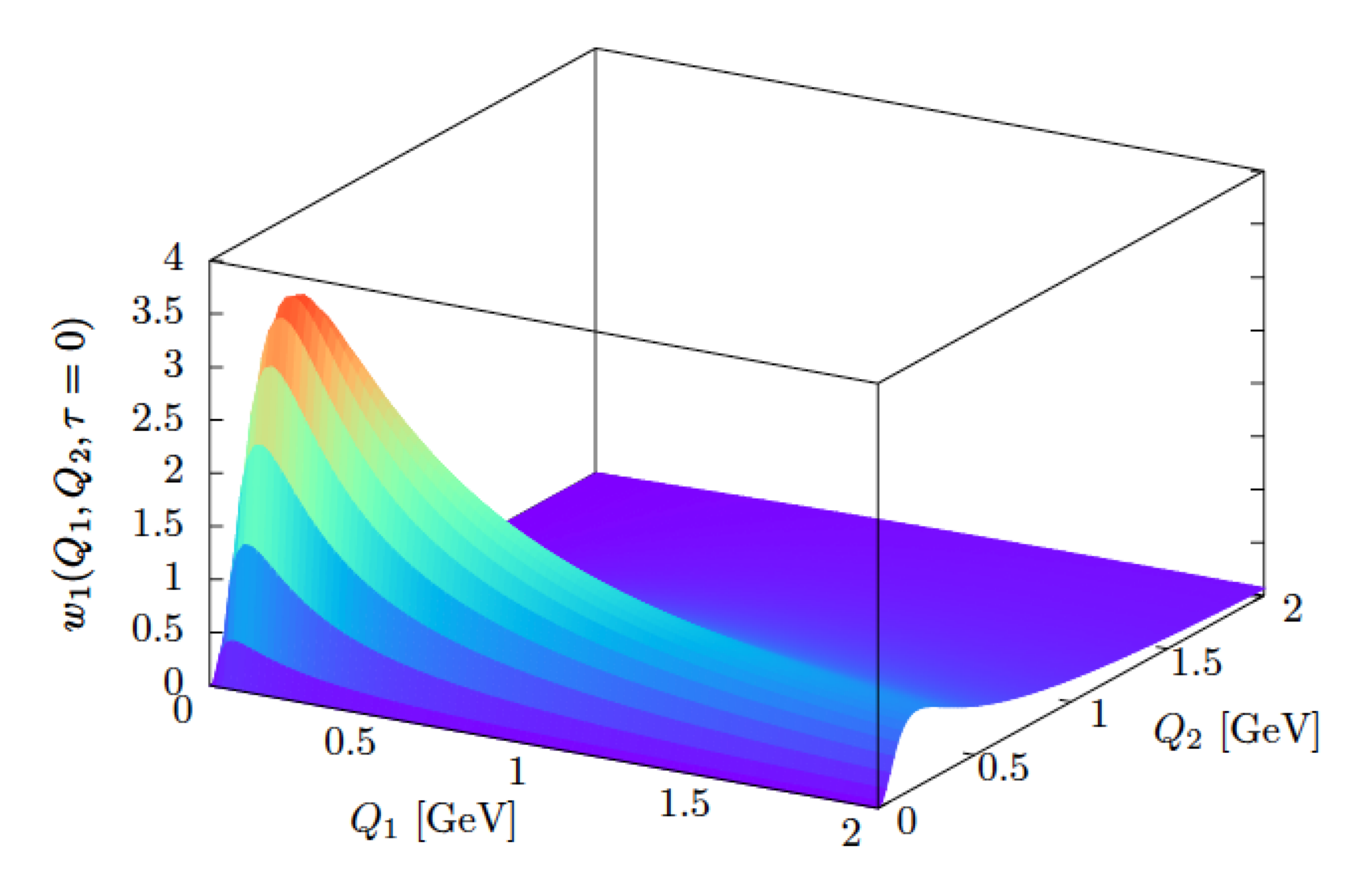} }
 \caption{\label{fig:hlblweight}Weight of the transition form factors of the lightest pseudoscalar mesons in the calculation of $a_\mu^\text{HLbL}$ as a function of the virtualities of the two photons. Figure taken from Ref.~\cite{Nyffeler:2016gnb}.}
\end{figure}

Very recently, new theoretical approaches have been proposed by groups from Bern and Mainz~\cite{M1_Colangelo:2014dfa,M1_Colangelo:2014pva,Colangelo:2015ama,Colangelo:2017fiz,Colangelo:2017qdm,Pauk:2014rfa,Pauk:2014jza} in order to attack the HLbL calculation, namely by exploiting dispersion relations. Form factor measurements of the two-photon coupling $\gamma\gamma \to P$, where $P$ is a one-hadron or two-hadron system, are of special interest. The Belle and BaBar collaborations have determined these couplings referred to as meson TFF for the lightest pseudoscalar mesons. However, the results of the $B$-factories have only been obtained at very large momentum transfers above 2\,GeV, while for the HLbL contribution measurements at low momentum transfers are required. This is illustrated by Fig.~\ref{fig:hlblweight}, which shows the weight of the light pseudoscalar TFFs in the HLbL calculation~\cite{Nyffeler:2016gnb} as a function of the two virtualities $Q_1$ and $Q_2$ of the photons. It can clearly be seen that the region of small momentum transfers is most relevant. This is exactly where \bes3 can provide precision result.

To summarize, an improvement of the SM prediction of $(g-2)_\mu$ is urgently needed in view of two upcoming direct measurements of $a_\mu$ at Fermilab and J-PARC with a fourfold improved precision. In the white paper of Ref.~\cite{ref_whitepaper} in 2013, it was argued that new experiments, like the ones carried out at \bes3, should lead to a final reduction of the SM uncertainty of $a_\mu$ down to $3.5 \times 10^{-10}$. The recent evaluations in Refs.~\cite{ref_teubner18,Davier:2017zfy,Colangelo:2018mtw} show that such an accuracy has almost been achieved already. \bes3 have contributed to this achievement. The future program of \bes3 together with new analyses at BaBar, Belle II, and elsewhere will lead to an additional significant reduction. The midterm goal is to reduce both the HVP and HLbL uncertainties to a level similar to the future experimental uncertainty, \ieie, $1.6\times10^{-10}$. It should also be mentioned that since 2017 the theoretical and experimental work in view of $a_\mu^{\rm SM}$ is coordinated by the {\it $g-2$ Theory Initiative}~\cite{g-2theoryinitiative}, which is a consortium of theoretical and experimental physicists working towards an improved SM prediction. Members of the \bes3 collaboration are part of this consortium.

%-----------------------------------------------------------------------------------------

\subsection[The running of the electromagnetic fine structure constant,
$\alpha_{\rm em}(s)$]{The running of the electromagnetic fine structure constant,
$\boldsymbol{\alpha_{\rm em}(s)}$}

Due to vacuum polarization effects, the electromagnetic fine structure constant $\alpha_{\rm em}$ is a ``running'' quantity. Its  value increases with increasing momentum transfer $s$ of the scattering process. The effective running of the fine structure constant as a function of $s$ is usually parametrized in SM in the following way:%
\begin{eqnarray}
  \alpha_{\rm em}(s) & = & \frac{\alpha(0)}{(1 - \Delta \alpha_{\rm em}(s))},
\\
  \Delta \alpha_{\rm em}(s) & = & \Delta \alpha_{\rm em}^{\rm lept}(s) + \Delta
\alpha_{\rm em}^{\rm had(5)}(s) + 
  \Delta \alpha_{\rm em}^{\rm top}(s).
\label{eq:alpha_qed}
\end{eqnarray}
In the above formula, $\Delta\alpha_{\rm em}^{\rm lept}$ denotes the vacuum polarization effects due to lepton loops, $\Delta \alpha_{\rm em}^{\rm had(5)}$ accounts for the effects due to loop contributions of the five lightest quarks, and the loop contributions due to the top quark are given by $\Delta \alpha_{\rm em}^{\rm top}$. The leptonic contribution can be computed in QED with very high precision, while the top-quark loop contribution is very small. Therefore, the total uncertainty of $\Delta \alpha_{\rm em}$ is entirely limited by $\Delta \alpha_{\rm em}^{\rm had(5)}$. As in the case of $(g-2)_\mu$, a dispersion relation can be used to relate experimental hadronic cross section data with $\Delta \alpha_{\rm em}^{\rm had(5)}$.

Of special interest is the knowledge of $\Delta \alpha_{\rm em}$ for $s=M_Z^2$ since most of the electroweak precision 
tests have been performed at the $Z^0$ peak at LEP.  The total correction to the fine structure constant amounts 
to~\cite{ref_teubner18}
\begin{equation}
  \Delta \alpha_{\rm em}(M_Z^2) = (276.11 \pm 1.11)\times 10^{-4},
\label{eq:delta_alpha_sm1}
\end{equation}
and the value of $\alpha_{\rm em}$ at the $Z$ pole mass is therefore known to be
\begin{equation}
  \alpha^{-1}_{\rm em}(M_Z^2) = 128.947 \pm 0.012.
%\label{eq:delta_alpha}
\end{equation}
A similar result of $ \alpha^{-1}_{\rm em}(M_Z^2) = 128.946 \pm 0.015$ is found in Ref.~\cite{Davier:2017zfy}. The current uncertainty of $\Delta \alpha_{\rm em}$ represents a severe limitation for electroweak precision fits to the SM (see a review on {\it Electroweak model and constraints on new physics} in Ref.~\cite{pdg2018}). Typically these fits are performed using three independent input variables, such as $\alpha_{\rm em}(M_Z^2)$, the Fermi constant $G_\mu$, and $M_Z$. Among these three quantities, $\alpha_{\rm em}(M_Z^2)$ is known with the least precision. Its relative uncertainty is $1\times 10^{-4}$, while $G_\mu$ and $M_Z$ are known with uncertainties of $5 \times 10^{-7}$ and $2 \times 10^{-5}$, respectively. In the past, the insufficient knowledge of $\Delta \alpha_{\rm em}$ led to imprecise predictions of the Higgs mass. Now that the Higgs mass is known with high accuracy, any new precision measurement of an electroweak observable (as for instance the electroweak mixing angle sin$^2(\Theta_W)$) establishes a significant test of the electroweak SM. The smaller the uncertainty of $\Delta \alpha_{\rm em}(M_Z^2)$ is, the more powerful this test will be. As will be discussed in Sec.~\ref{sec:Rincl}, the goal of the new \bes3 measurement is to reduce the uncertainty of the $R$ measurement to 3\%, which allows to improve the accuracy of the predition of $\Delta \alpha_{\rm em}^{\rm had(5)}$.

%% file: QCD/ISR.tex
\subsection{Measurement of exclusive hadronic channels via ISR}\label{sec:ISR}
A major campaign of ISR~\cite{ref_binner,Benayoun:1999hm,Druzhinin:2011qd} measurements was launched at \bes3, mainly in order to improve the HVP contribution to $(g-2)_\mu$ and of $\Delta\alpha_{\rm em}^{\rm had(5)}$. The ISR technique allows for precision measurements of the hadronic cross section at high-luminosity electron-positron colliders by using events in which either the incoming electron or positron has emitted a high-energetic photon. In such a way, the available cms energy for the hadronic system is varied depending on the energy of the ISR photon, which is the reason that the technique is also called Radiative Return.

From the measurement of the hadronic cross section $\sigma(e^+e^- \to {\rm hadrons} + \gamma_{\rm ISR})$, the non-radiative cross section $\sigma(e^+e^- \to {\rm hadrons})$ can be extracted using a radiator function from QED theory. This radiator function is known with a precision of 0.5\% within the Monte Carlo (MC) generator \textsc{Phokhara}~\cite{Rodrigo:2001kf,Czyz:2008kw,Czyz:2009vj}, which also simulates the most relevant exclusive final states in view of the HVP contribution to $(g-2)_\mu$.

At \bes3, the data set taken at a cms energy of 3.773\,GeV is currently used for most of the ISR analyses. The total integrated luminosity available at this cms energy amounts to 2.9\,fb$^{-1}$. Using this data set, the statistics of ISR hadronic events is superior to the ISR statistics of BaBar above hadronic masses of approximately 1.5\,GeV. With upcoming new data and by including the already available data sets, the available ISR statistics will be similar to the BaBar statistics also at masses below 1.5\,GeV.

Both the tagged and  untagged ISR approaches are currently carried out at \bes3. The tagged approach requires the explicit detection of the ISR photon in the calorimeter, and allows for studies of the full hadronic mass range starting from the dipion mass threshold. An untagged measurement corresponds to the usage of ISR events, in which the ISR photon is emitted at very small polar angles, essentially collinear with the initial electron beams. Although tagging of those photons is not feasible, the momentum information can be extracted from the missing momentum of the fully reconstructed hadronic system. For kinematic reasons, the untagged approach is limited to the energy with hadron productions above approximately 1\,GeV. Above 1.5\,GeV, it provides significantly improved statistics compared to the tagged measurement, and furthermore, guarantees low background conditions. 

The main results of the ISR program at \bes3 can be summarized as follows:

\begin{itemize}
\item {\bf {Timelike pion form factor: \boldmath $e^+e^- \to \pi^+\pi^-$}}

A new measurement of the hadronic cross section of the channel $e^+e^- \to \pi^+\pi^-$ was performed in the energy range between 600\,MeV and 900\,MeV, which corresponds to the peak region of the $\rho(770)$ resonance~\cite{Ablikim:2015orh}. In this energy range the two-pion channel indeed contributes more than 50\% to $a_\mu^{\rm HVP}$, and the uncertainty of this exclusive channel is therefore decisive for the SM error of $(g-2)_\mu$. Thus, this is \emph{the} flagship analysis of the BESIII ISR program. A total systematic uncertainty of 0.9\% for the cross-section measurement has already been achieved. The two limiting contributions to the total systematic uncertainty are from the luminosity measurement and the theoretical radiator function, with 0.5\% for each. As will be discussed below, with improved calculations of the radiator function, by including larger data sets, and by an overall improved understanding of the detector performance, the errors of both these contributions can be reduced from 0.5\% to 0.2\% for each, yielding a reduced total systematic uncertainty. The mass range studied in this analysis is accessible via the tagged ISR approach only. The same holds for the mass range from the dipion mass threshold to 600\,MeV, which will be studied in future, along with the mass range above 900\,MeV, where the investigations can be performed in the untagged ISR approach as well.

\begin{figure}[htb]
 \centerline{ \includegraphics*[width=14cm]{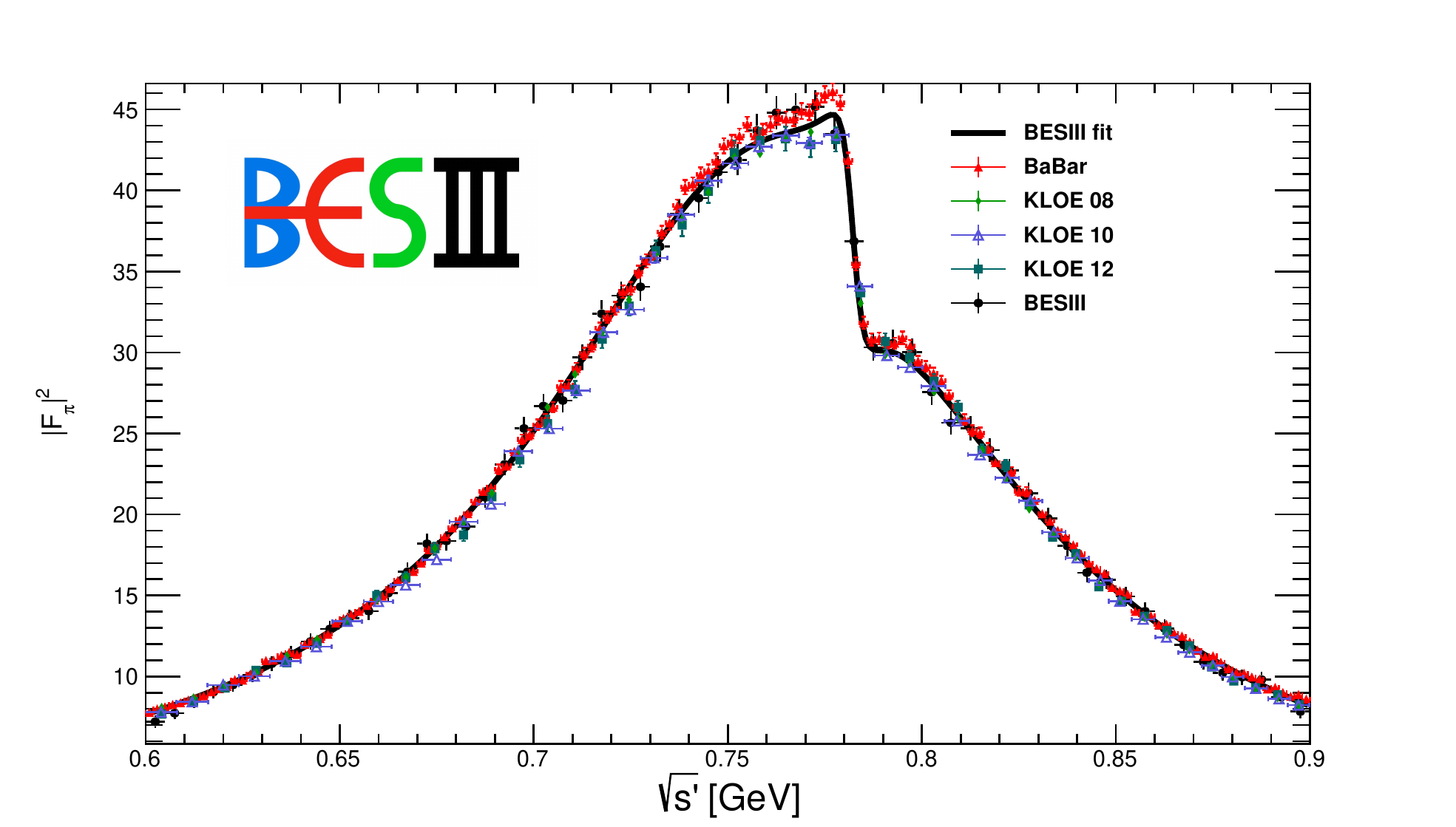} }
 \caption{Pion form factor measurements at \bes3~\cite{Ablikim:2015orh}, 
KLOE~\cite{ref_kloe8a,ref_kloe11a,ref_kloe12a}, and BaBar~\cite{ref_babar2pia}. The black line is a fit to the \bes3 
spectrum according to the Gounaris-Sakurai parametrization. }
\label{pionff_bes}
\end{figure}

In this analysis all the experimental detection efficiencies, the luminosity, as well as the knowledge of background channels need to be determined at the per mil level. A separation of pion tracks from muon tracks turns out to be the major challenge of the analysis. As a consequence, a multivariate analysis technique is applied~\cite{TMVA2007} by training an artificial neural network (ANN) for particle identification. When selecting muon events rather than pion events, the efficiency of the technique can be tested by comparing the absolute yield of $e^+e^-\to \mu^+\mu^-\gamma$ events with the \textsc{Phokhara}~\cite{ref_czyz_muons} MC prediction. An agreement between data and MC simulations at the level of $(0.5 \pm 0.3)\%$ is observed, demonstrating the excellent performance of the ANN.

From the cross section measurement, the timelike pion form factor $|F_\pi|$ is extracted, as shown in Fig.~\ref{pionff_bes}. An agreement between the \bes3 data and the three KLOE~\cite{ref_kloe8a,ref_kloe11a,ref_kloe12a} analyses is found up to the peak of the $\rho$ resonance. At the same time, the \bes3 data are systematically lower with respect to BaBar's results~\cite{ref_babar2pia} in this mass range. Above the $\rho$ peak the situation is reversed. In this case, the \bes3 spectrum comes out to be in good agreement with that of BaBar, while the KLOE results are systematically lower. In Fig.~\ref{fig:amu2pi}, we show the impact of the new \bes3 measurement on $a_\mu^{\rm HVP}$. It agrees with all three KLOE analyses in the mass range $600 - 900$\,MeV, while deviations from BaBar's results are observed. 

\begin{figure}[tp]
 \centerline{ \includegraphics*[width=13cm]{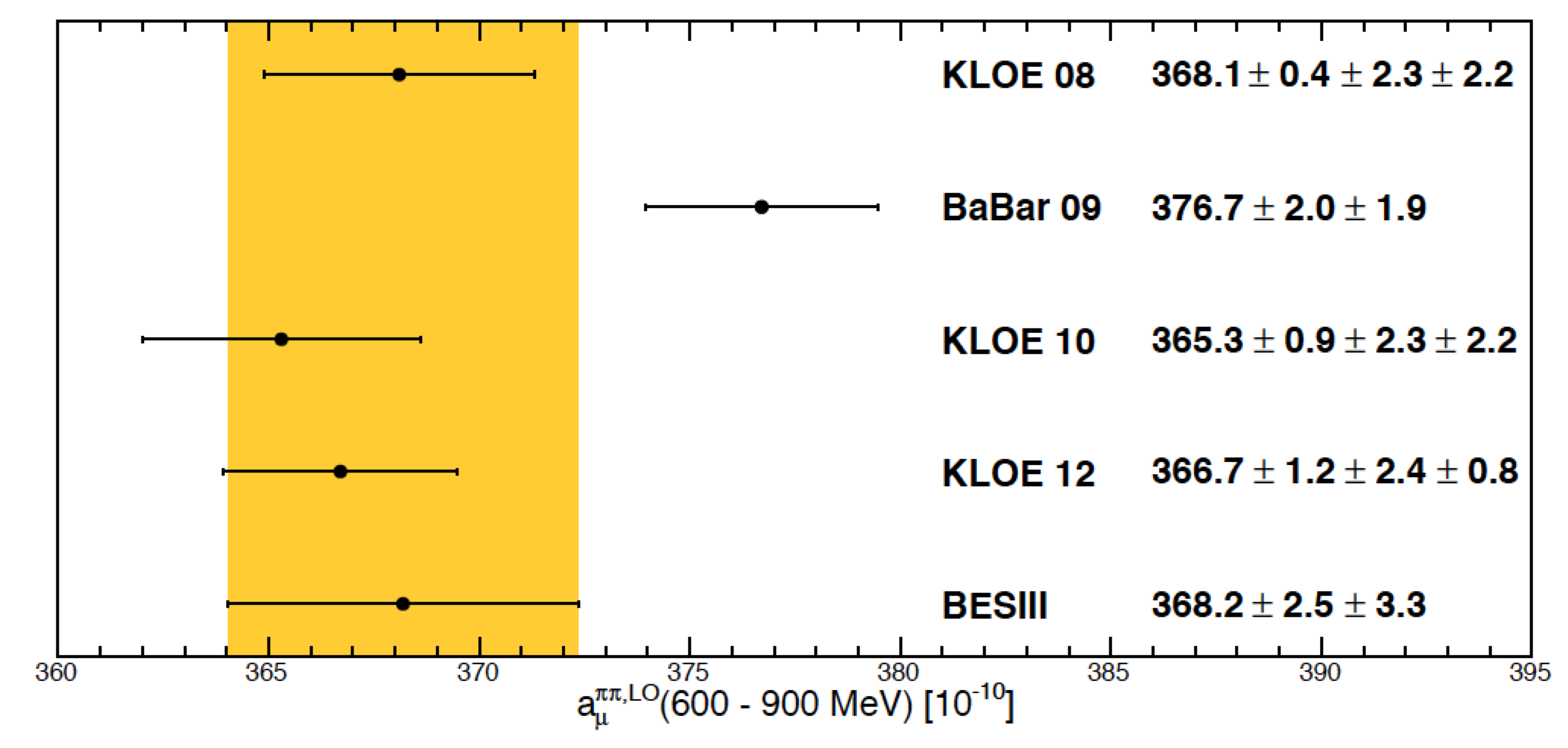} }
 \caption{\label{fig:amu2pi}Two-pion contributions to the hadronic vacuum polarization contribution to $a_\mu$ in the 
energy range between 600\,MeV and 900\,MeV.}
\end{figure}

The good understanding of the radiative muon sample, which was achieved in this analysis, also led to two additional 
publications, which made use of the high statistics and quality of the muon data. In a first paper, the electronic width
of the $J/\psi$ resonance was determined with world-leading accuracy~\cite{Ablikim:2016xbg}. In a second paper a 
competitive dark photon limit was achieved by looking for an enhancement of events in the dimuon invariant 
mass~\cite{ref_bes3dark} between 1.5 and 3.4\,GeV/$c^2$.

\item {\bf Cross section of \boldmath $e^+e^- \to \pi^+\pi^-\pi^0$}

The hadronic channel $e^+e^- \to \pi^+\pi^-\pi^0$, which is dominated at low energies by the $\omega$(782) and 
$\phi$(1020) resonances, has been measured by the Novosibirsk experiments CMD-2~\cite{ref_cmd23pi} and 
SND~\cite{ref_snd3pi} below 1.4\,GeV. Their results show obvious scatter, although within claimed accuracy, and would 
benefit from improved measurements. Above the $\phi(1020)$ resonance, BaBar has also performed a measurement of this 
channel~\cite{ref_babar3pi} and has observed structures which are interpreted as two excited $\omega$ states. This 
BaBar result is in conflict with an old DM2 measurement~\cite{ref_dm23pi}. 

In contrast to the two-pion analysis discussed above, both the tagged (in the full mass range) and the untagged ISR 
methods (above 1\,GeV) were analyzed by \bes3. Preliminary results for this ISR measurements are already available. For 
the final spectrum, the tagged and  untagged spectra were averaged and a systematic uncertainty of better than 3\% 
was achieved in a wide mass range from threshold up to the $J/\psi$ resonance. In a fit to the mass spectrum assuming 
vector meson dominance the mass and width of the resonances, $\omega(1420)$ and $\omega(1650)$, could be obtained 
with unprecedented accuracy. Furthermore, the branching fraction of the $J/\psi$ decays to three pions was measured 
precisely. 

\item {\bf Cross section of \boldmath $e^+e^- \to \pi^+\pi^-\pi^0\pi^0$}

In the channel $e^+e^- \to \pi^+\pi^-\pi^0\pi^0$ some deviations between the two Novosibirsk experiments CMD-2~\cite{ref_cmd24pi} and SND~\cite{ref_snd4pi} were observed below 1.4\,GeV. Even larger deviations are seen in 
comparison with $\tau$ spectral functions which can be related to the cross section via an isospin relation. It has 
been speculated whether large isospin violating effects might be the reason for this observation. Above approximately
1\,GeV, the BaBar collaboration has published an analysis, in which the world data set in terms of statistical and 
systematic precisions are exceeded by a large amount~\cite{babar_4pi}. It is therefore the goal of the \bes3 analysis to 
provide an independent high-accuracy data set besides BaBar. Furthermore, the channel is extremely interesting from the 
spectroscopy point of view. It has a rich internal structure, where the $\omega\pi^0$, $a_1\pi$,  $\rho^+\rho^-$, 
and many other intermediate states play significant roles (including the $f_0(500)$ and $f_0(980)$). 

Also for this channel preliminary results exist at \bes3 using the tagged and untagged ISR approach. The mass range from threshold up to 3.4\,GeV is covered and the cross section is determined with a systematic uncertainty of approximately 3\%. Besides the  cross section of $e^+e^- \to \pi^+\pi^-\pi^0\pi^0$,  the cross section of the intermediate state $e^+e^-\to\omega$(782)$\pi^0$ is measured, and the branching fraction of the decay $J/\psi \to \pi^+\pi^-\pi^0\pi^0$ is extracted. 

\end{itemize}

The existing results on $e^+e^- \to \pi^+\pi^-\pi^0$ and $e^+e^- \to \pi^+\pi^-\pi^0\pi^0$ are obtained using the
$2.9\,\textrm{fb}^{-1}$ data at $\sqrt{s}=3.773\,\textrm{GeV}$. As discussed in the case of the $e^+e^- \to \pi^+\pi^-$ analysis, including the already existing and upcoming new data sets, the systematic uncertainties, which are already at this point at world-class level, can be further reduced.

%% file: QCD/rvalue.tex
\subsection{Inclusive $R$ scan data}\label{sec:Rincl}
\begin{figure}[tb]
 \centering
 \includegraphics[width=11cm,clip]{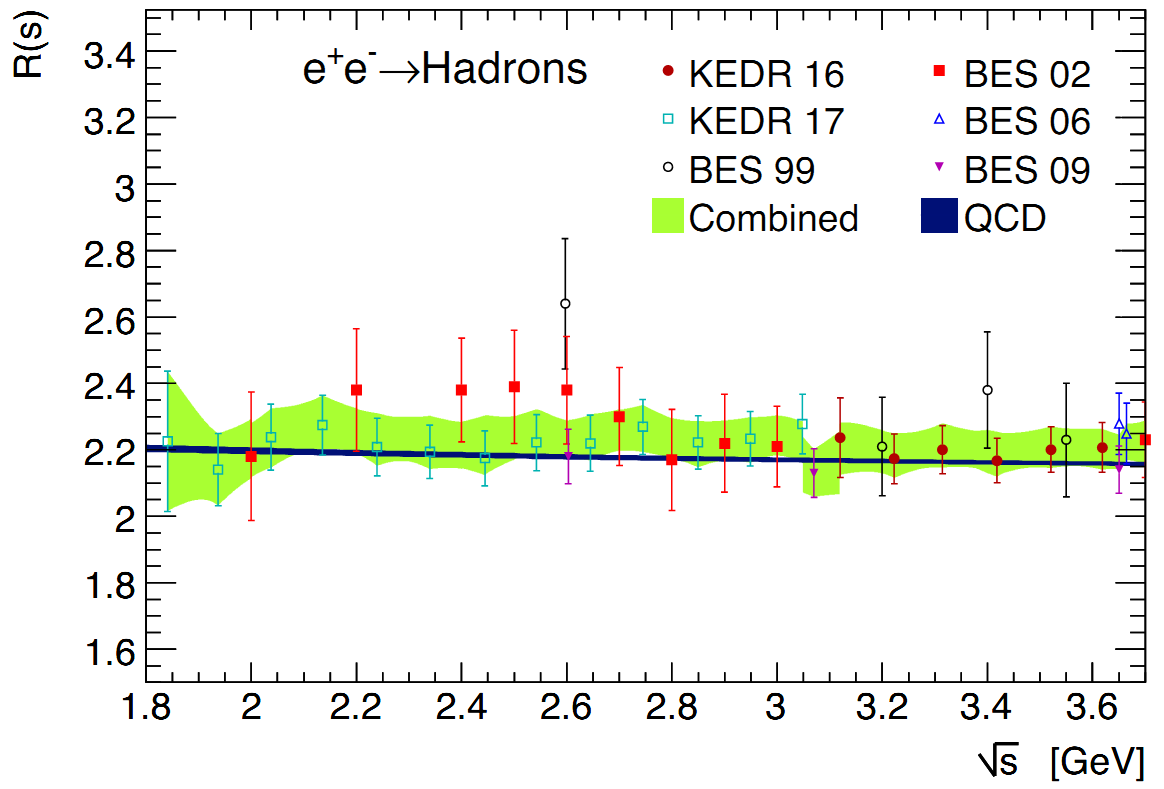}
 \caption{\label{rpqcd}The hadronic $R(s)$ ratio in the continuum region below the $D\bar{D}$ threshold. Shown are the 
results from BES and KEDR as well as their average (shaded band). The solid line shows the pQCD 
prediction. Plot taken from Ref.~\cite{Davier:2017zfy}.}
\end{figure}

Up to an energy of 2.0\,GeV, the $R$ value is determined by the sum of measured exclusive hadronic cross sections, either via the energy scan or the ISR technique. In case of unmeasured exclusive channels, isospin invariance is assumed in that energy range. At larger values of $\sqrt{s}$, more exclusive channels open up, so inclusive $R$ measurements are necessary. The energy region between 2.0 and 4.6\,GeV is rich of resonances and has transitions between the smooth continuum regions and the resonances. Figure~\ref{rpqcd} shows a comparison of BES~\cite{rbes2} and KEDR data~\cite{rkedr}, as well as the perturbative QCD (pQCD) prediction between 2.0 and 3.7\,GeV. Agreement within uncertainties is found. It should be noted that the most recent KEDR analysis~\cite{Anashin:2018vdo} in the energy interval between 3.08 and 3.72\,GeV is not yet included.  

Regarding the hadronic vacuum polarization contribution to $(g-2)_\mu$, theorists deal with the $R$ values in the energy region above 1.8\,GeV in different ways. Ref.~\cite{ref_teubner18} uses in the dispersive evaluation inclusive $R$ values measured above 1.937\,GeV. In the energy interval from 1.8 to 3.7\,GeV,  the contribution to the uncertainty of the muon anomaly $a_\mu$ is found to be $0.56\times 10^{-10}$, which is roughly a factor of 3 smaller than the expected accuracy $\delta a_\mu(\rm exp)=16\times 10^{-11}$ of the new direct measurements of $a_\mu$. In Ref.~\cite{Davier:2017zfy} four-loop pQCD is used in the energy region between 1.8 and 3.7\,GeV, resulting in a theory uncertainty of $0.65 \times 10^{-10}$. The open charm region between 3.7 and 5\,GeV is governed by broad resonances. Its contribution to $a_\mu$ is computed with experimental data. However, the contribution is found to be very small ($0.11 \times 10^{-10}$ in Ref.~\cite{ref_teubner18} and $0.03 \times 10^{-10}$ in Ref.~\cite{Davier:2017zfy}). At even higher energies, either experimental data (for instance in the bottomonium region) or pQCD are used for the evaluation of the hadronic vacuum polarization. The uncertainties of these contributions are below the  level of $0.10 \times 10^{-10}$.

The transition region between sum of exclusive channels
and inclusive $R$ data, is of interest and deserves re-examination. 
Table \ref{transition} summaries the $a_{\mu}^{\rm HVP, LO}$ at [1.841,
2.0]\,GeV by different inputs in this region ~\cite{ref_teubner18}. 
The results from inclusive data and pQCD calculations agree within uncertainty, and 
disagree with that from exclusive sum. Hence, precise measurements 
of $R$ data with exclusive sum and inclusive method are important to 
choose transition point between sum of exclusive channels and 
inclusive $R$ data, and test pQCD prediction on $R$ value in transition region.  
A new scan measurement between 1.8 and 2.0\,GeV is useful to answer
these questions.

\begin{table}[tb]
\begin{center}
\caption{\label{transition} The $a_{\mu}^{\rm HVP, LO}$ for $\sqrt{s}$ in [1.841, 2.0]\,GeV by
different inputs in this region. Numbers are taken from Ref.~\cite{ref_teubner18}.}
\begin{tabular}{l|c} \hline \hline
input & $a_{\mu}^{\rm HVP, LO}$ ($\times 10^{10}$) \\ \hline
Exclusive sum   & 6.06 $\pm$ 0.17 \\ \hline
Inclusive data  & 6.67 $\pm$ 0.26 \\ \hline
pQCD            & 6.38 $\pm$ 0.11 \\ \hline
Exclusive($<$ 1.937 GeV) + inclusive($>$ 1.937 GeV) & 6.23 $\pm$ 0.13 \\ \hline
\end{tabular}
\end{center}
\end{table}

Compared to $(g-2)_\mu$, the impact of inclusive $R$ data on the running of the electromagnetic fine structure constant is much more pronounced as higher energy scales are very relevant in the dispersion integral for $\Delta \alpha_{\rm em}$. In fact, in the case of $\Delta \alpha_{\rm em}$, the total uncertainty of $1.11 \times 10^{-4}$ cited in Ref.~\cite{ref_teubner18} stems almost entirely from the energy range between 1.19 and 11.20\,GeV, which amounts to $(82.82 \pm 1.05) \times 10^{-4}$. While Ref.~\cite{ref_teubner18} follows a more data-driven approach to calculate $\Delta \alpha_{\rm em}$, the evaluation by Davier \textit{et al.} in Ref.~\cite{Davier:2017zfy} relies on pQCD calculations for the $R$ value in the energy range between 1.8 and 3.7\,GeV. Above 3.7\,GeV, up to 5.0\,GeV, experimental information on $R$ is used. In this energy range an experimental uncertainty of $0.67 \times 10^{-4}$ out of a total uncertainty for $\Delta \alpha_{\rm em}$ of $0.9\times 10^{-4}$ is found. This strongly motivates new data on the inclusive $R$ ratio in the energy range covered by \bes3. The role of \bes3 is twofold: on the one hand the data can prove the validity of pQCD in the description of $R$ as required by a theory-based evaluation of $\Delta\alpha_{\rm em}$; On the other hand the data can be directly used as input in the dispersion integral in a data-driven approach.

In order to improve the knowledge of $R$, the \bes3 collaboration has 
recently carried out a series of energy scans in the range between 2.0 
and 4.6\,GeV with in total 130 energy points with total integrated 
luminosity about $1300 ~{\rm pb^{-1}}$. The total hadronic event 
yield exceeds $10^5$ events at each energy, such that the accuracy 
of the data will be entirely dominated by systematic uncertainties. 
The goal of the \bes3 experiment is to arrive at a total accuracy 
of the hadronic $R$ ratio of at least 3\%. A similar systematic 
accuracy has already been achieved in Ref.~\cite{rkedr} at KEDR.
The analysis of the data is currently ongoing, and preliminary 
result showes dominant uncertainy is by hadronic event generator, 
which is also a major challenge to describe $10^5$ hadronic events 
at each energy. MC simulation programs based on theoretical
descriptions of string fragmentation functions exist, 
like the \textsc{LuArLw} generator~\cite{Andersson:1999ui}.
The KEDR collaboration used \textsc{LuArLw} generator, which was
employed by the BES collaboration. With $10^5$ hadronic events at 
each energy, the \textsc{LuArLw} generator is optimized and 
tunned. At the same time, a precise description of the total hadronic
events in one event generator has been proposed. Making use of the existing measurements of many exclusive 
hadronic final states, event generator \textsc{ConExc}~\cite{conexc}, which can deal with imprecise exclusive
part of the \textsc{LuArLw} generator, have been considered
as an alternative. Hence, a more data-driven approach to the
description of the total hadronic events is identified.

The energy range between 3.85 and 4.6\,GeV have rich charmonium and
charmonium-like states. Because the collected scan data have small 
energy step, we could extract resonaces parameters by precisely 
measured $R$ values. BES collaboration did similar work ~\cite{rbes2}, 
but results have large uncertainty and model dependence. With more
studies on these charmonium and charmonium-like states, $R$ results could be further
improved.
 

%% file: QCD/form.tex
\subsection{Measurements of meson transition form factors}

The main motivation for the proposed program of precision measurements of meson TFFs at BESIII is to constrain the HLbL contribution to the level set by the forthcoming $(g-2)_\mu$ experiments at FNAL and J-PARC of $\delta a_\mu = 1.6\times10^{-10}$, in order to allow for a meaningful interpretation of these new measurements. Depending on the analysis of the hadronic contributions~\cite{Davier:2017zfy,M1_Jegerlehner:2013sja} the present SM uncertainty amounts to $\delta a_\mu({\rm SM})=\pm(49 - 58) \times 10^{-11}$, which significantly exceeds the future experimental accuracy. This motivates an intense activity to reliably estimate contributions of hadrons to $a_\mu$.

\begin{figure}[tbp]
 \centerline{ \includegraphics[width=4.3cm]{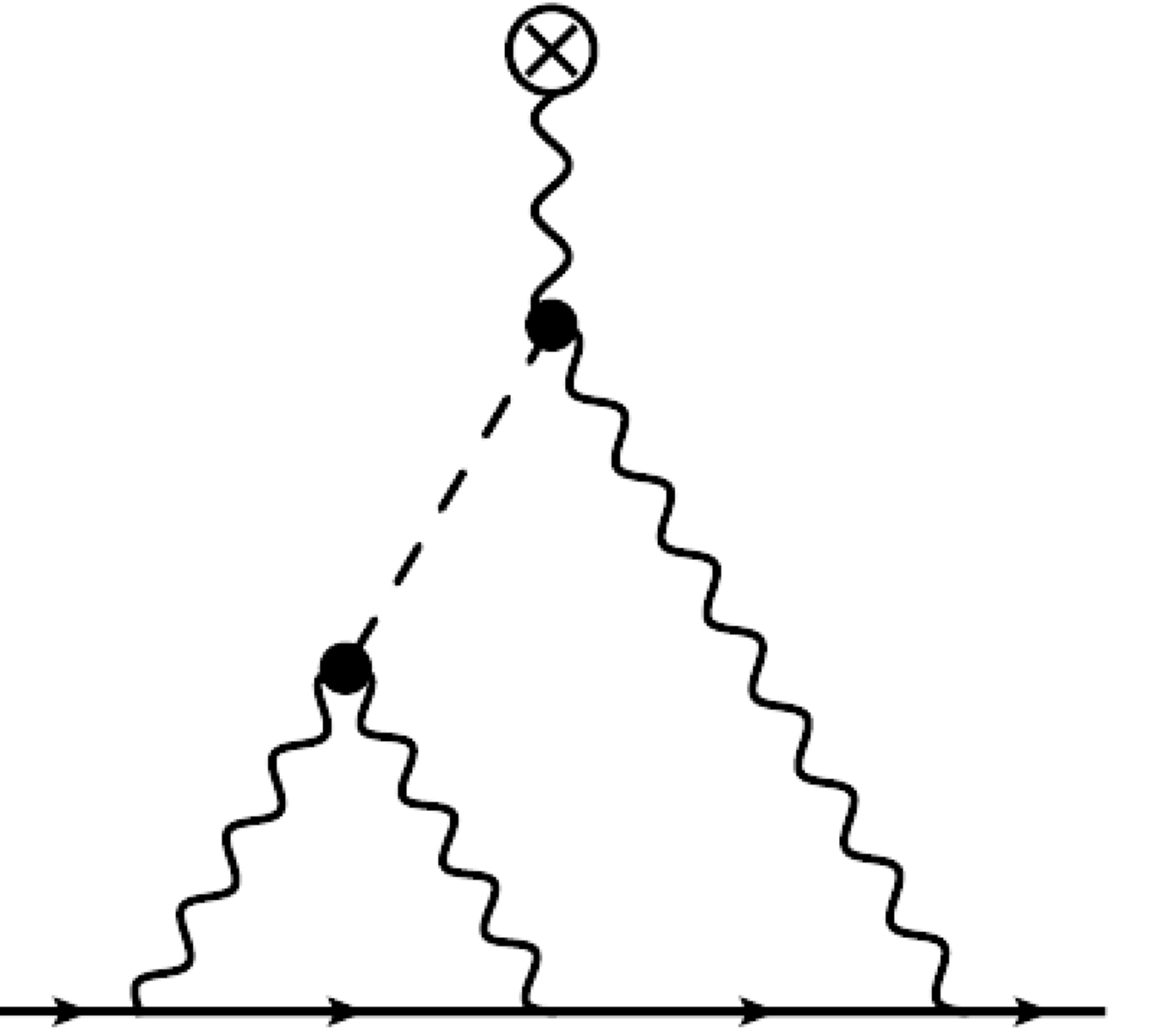} }
  \caption{\label{fig:LbL}The leading terms in the hadronic light-by-light scattering (HLbL) contribution to the
anomalous magnetic moment of the muon are given by the exchange of light pseudoscalar mesons as depicted by the dashed
line.}
\end{figure}

The leading diagram of the HLbL contribution is given by pseudoscalar meson exchange as shown in Fig.~\ref{fig:LbL}. Unlike the HVP contribution, in most of the existing estimates of the HLbL contribution, the description of the non-perturbative light-by-light matrix element is based on hadronic models rather than determined from data. These approximations are based on a requirement of consistency with the asymptotic constraints of QCD, and predict that the hadronic corrections are dominated by long-distance physics, namely due to exchange of the lightest pseudoscalar states. Unfortunately, a reliable estimate based on such models is possible only within certain kinematic regimes. This results in a large, mostly uncontrolled uncertainty of $a_\mu$. 

In order to reduce the model dependence, data-driven approaches for the HLbL contribution to $a_\mu$ have been proposed. Sum rules and a dispersive formalism can furthermore provide powerful constraints on the hadronic light-by-light scattering and its contribution to $a_\mu$. Measurements of meson TFFs are used as input in such data-driven approaches. As will be discussed below, essentially all the relevant channels in the spacelike and timelike regions can be studied at \bes3.

TFFs describe the effect of the strong interaction on the $\gamma^*\gamma^*M$ vertex, where $M=\pi^0, \eta, \eta^\prime, \eta_c\ldots$. They are represented by functions $F_{M\gamma^*\gamma^*}(q_1^2,q_2^2)$ of the photon virtualities $q_1^2$ and $q_2^2$. For the case of pseudoscalar mesons, there is one such function~\cite{Aubert:2006cy,BaBar:2018zpn}. For scalar, axial-vector, or tensor mesons, the $\gamma^*\gamma^*M$ vertex contains in general several such TFFs.

\begin{figure}[tbp]
 \centerline{ \includegraphics[height=3.0cm]{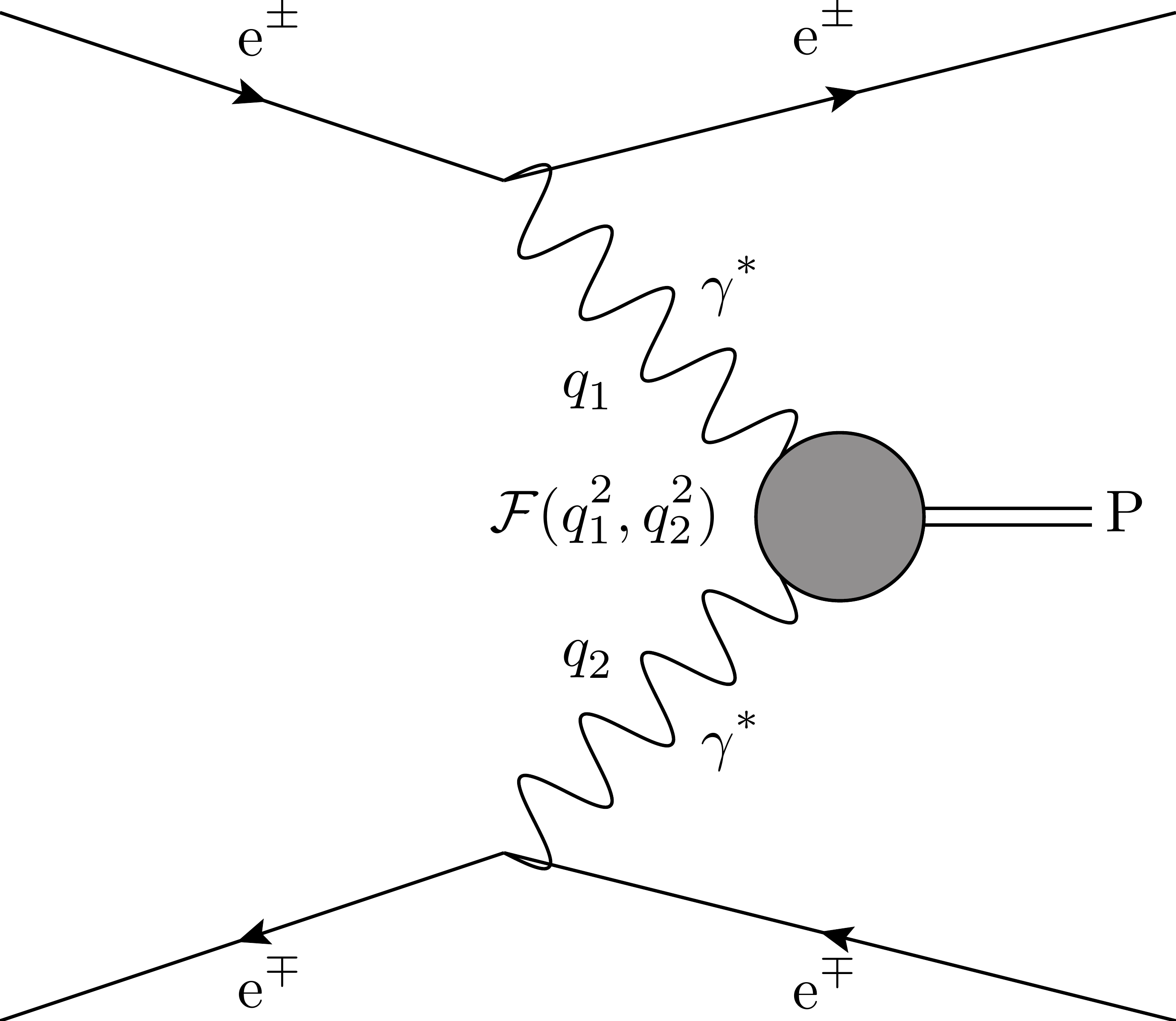}\hspace*{2mm}%
              \includegraphics[height=3.0cm]{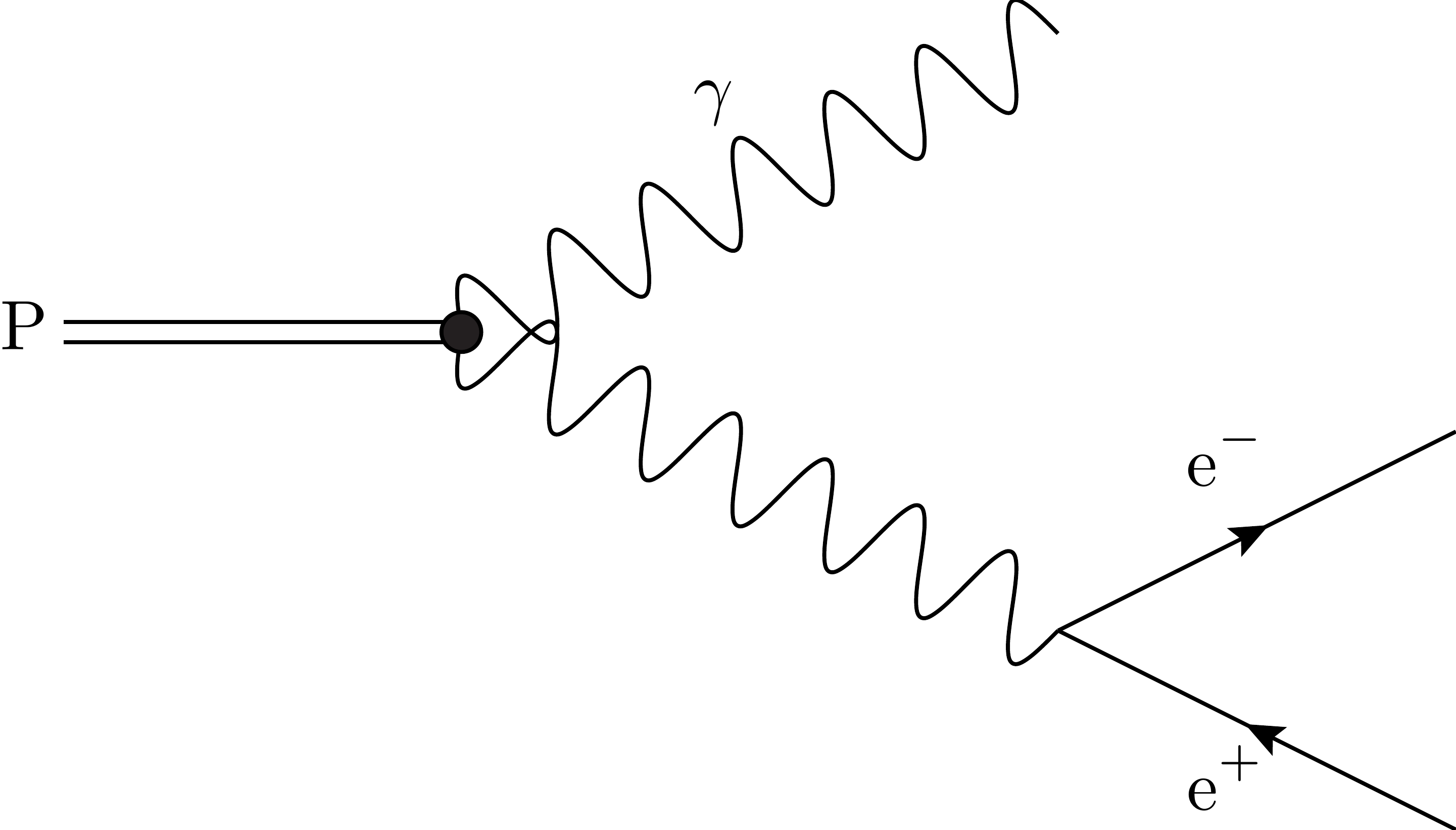}\hspace*{2mm}%
              \includegraphics[height=3.0cm]{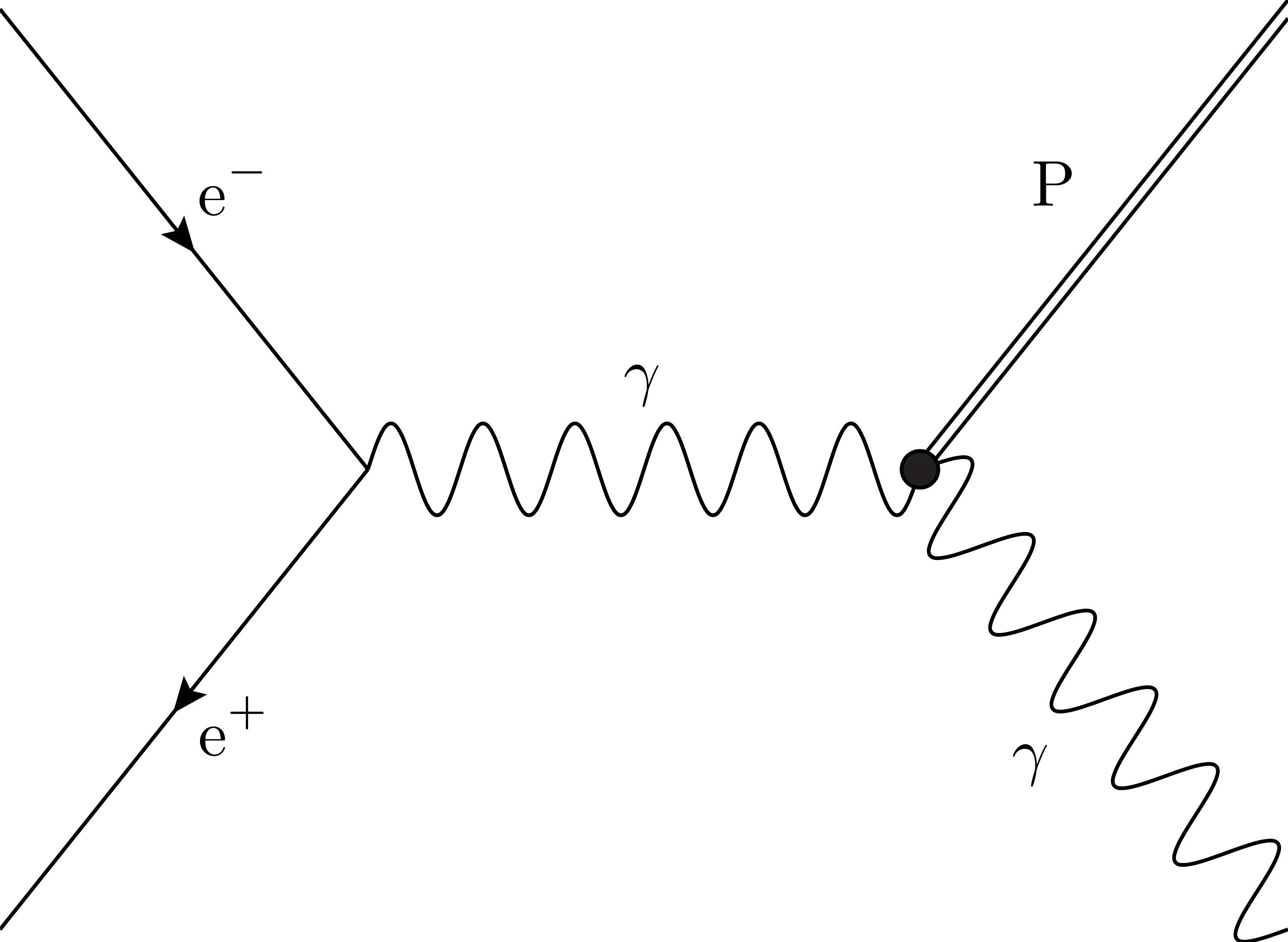} }
 \caption{\label{fig:TFF}The meson TFFs in the spacelike region (left panel) as accessed through the $\gamma^\ast \gamma^\ast$ annihilation process, and in the timelike region as accessed through the $\gamma e^+e^-$ Dalitz decay (middle panel) and through $e^+e^-$ annihilation into $P \gamma$ (right panel), where $P$ stands for a pseudoscalar meson.}
\end{figure}

The spacelike region of the TFFs is accessed at $e^+e^-$ colliders by means of the two-photon-fusion reaction $e^+e^-\to e^+e^-M$  (left panel in Fig.~\ref{fig:TFF}), where at present the measurement of both virtualities is still an experimental challenge. The common practice is to extract the TFFs when one of the outgoing leptons is tagged and the other is assumed to escape detection along the beam axis (single-tag method). The tagged lepton emits a highly off-shell photon with a transferred momentum $q_1^2\equiv -Q^2$ and is detected, while the other, untagged, is scattered at a small angle with $q_2^2\simeq 0$. The TFF extracted from the single-tag experiment is then $F_{M\gamma^\ast\gamma^\ast}(Q^2,0)\equiv F_{M\gamma^\ast\gamma}(Q^2)$. The timelike region of the TFFs can be accessed at meson facilities through the single Dalitz decay processes $M \to l^+l^- \gamma$, which contain a single virtual photon with a transferred momentum in the range $4m_l^2<q_1^2<m_M^2$ (with $m_l$ the lepton mass and $m_M$ the meson mass) whereas $q_2^2 = 0$ (middle panel in Fig.~\ref{fig:TFF}). To complete the timelike region, $e^+e^-$ colliders provide access to the values $q^2>m_M^2$ through the $e^+e^- \to M\gamma$ annihilation processes (right panel in Fig.~\ref{fig:TFF}).

Besides their relation to the anomalous magnetic moment of the muon, the pseudoscalar meson TFFs provide a unique window
on several symmetry-breaking mechanisms in QCD. In the limit of massless light quarks ($u, d, s$), \ieie, the chiral 
limit, the QCD Lagrangian exhibits an ${\rm SU}(3)_{\rm L} \times {\rm SU}(3)_{\rm R}$ chiral symmetry which is spontaneously broken to 
${\rm SU}(3)_V$, giving rise to 8 pseudoscalar Goldstone bosons ($\pi, K, \eta$). The QCD Lagrangian has in addition two 
other global symmetries: the ${\rm U}(1)_V$ symmetry leading to the conservation of baryon number, and the ${\rm U}(1)_A$ symmetry 
which is anomalous. Since the flavor-singlet axial-vector current is not conserved in the presence of this ${\rm U}(1)_A$ 
anomaly, the $\eta^\prime$ mass does not vanish in the chiral limit. In the massless $u, d, s$ quark world (with the 
other three quarks infinitely heavy), the massive $\eta^\prime$ would be a pure flavor-singlet state $\eta^0 \equiv (u 
\bar u + d \bar d + s \bar s)/\sqrt{3}$. In the real world, however, the ${\rm SU}(3)_V$ flavor symmetry is explicitly broken 
by the quark masses, which causes a mixing among $\pi^0, \eta$, and $\eta^\prime$~\cite{M1_Feldmann}. In the isospin 
limit ($m_u = m_d$), the $\pi^0$ can be identified as a pure isotriplet state $(u \bar u - d \bar d)/\sqrt{2}$. In the 
absence of the ${\rm U}(1)_A$ anomaly (large-$N_c$ limit of QCD), the two isosinglet pseudoscalar mass eigenstates would 
consist of $(u \bar u + d \bar d)/\sqrt{2}$ and $s \bar s$ (so-called {\it ideally mixed} states). The ${\rm U}(1)_A$ anomaly 
mixes these quark flavor states towards the physical $\eta$ and $\eta^\prime$ mesons, which are closer to the flavor 
octet   $\eta^8 \equiv (u \bar u + d \bar d -2 s \bar s)/\sqrt{6}$ and flavor singlet $\eta^0$ states, respectively. 
This mixing in the $\eta-\eta^\prime$ system is probing the strange quark content of the light pseudoscalars as well as 
the non-perturbative gluon dynamics of QCD, responsible for the ${\rm U}(1)_A$ anomaly. The mixing can also be related to 
physical observables~\cite{M1_Feldmann, M1_Donoghue:1986wv, M1_Leutwyler}, in particular through the $M \to \gamma 
\gamma$ decay widths and the $\gamma^\ast \gamma^\ast M$ TFFs.

\vspace{9.5mm}
\noindent Based on the experience obtained so far at \bes3, world-leading results in the field of meson TFFs have already been obtained with the existing data, as will be elaborated in the following. The overall goal of the \bes3 program is to provide the first precision measurements of the TFFs of pseudoscalar mesons, of the $\pi\pi, \pi\eta$ and $\eta\eta$ systems, as well as axial and tensor mesons at small momentum transfers. A future data set of additional $20\,\textrm{fb}^{-1}$ taken at $\sqrt{s}=3.773$\,GeV will make a first measurement of the double-virtual TFF of the lightest pseudoscalar mesons with high accuracy possible.

\begin{itemize}

 \item{\bf Single-tag pseudoscalar TFFs, spacelike}\\
The first \bes3 publication of the spacelike TFF of the $\pi^0$ meson will be based on the 2.9\,fb$^{-1}$ data sample obtained at a cms energy of 3.773\,GeV. Preliminary results for the analysis are presented in Fig.~\ref{fig:spacelike_pi0TFF_BES}. As described above, for single-tag events, one of the beam particles (electron or positron) is tagged in the detector. By detecting two photons from the $\pi^0$ decay, the missing momentum can be derived for the missing positron (electron), which is further required to be scattered at very small angles. By fitting the $\gamma \gamma$ invariant mass distribution in bins of $Q^2$, one obtains the differential cross section d$\sigma$/d$Q^2$ for the signal process,which is proportional to $|F(Q^2)|^2$. 

Figure~\ref{fig:spacelike_pi0TFF_BES} shows the product $Q^2 |F(Q^2)|$ as measured by \bes3 together with existing data from CELLO \cite{M1_Behrend:1990sr} and CLEO \cite{M1_Gronberg:1997fj}. Note that recent BaBar~\cite{Aubert:2009mc} and Belle~\cite{Uehara:2012ag} data could only access the $Q^2$ range above 4\,GeV$^2$, and hence, are not displayed. The \bes3 analysis covers the entire $Q^2$ range between 0.3\,GeV$^2$ and 3.1\,GeV$^2$, which significantly improves the accuracy upon existing data sets. The accessible range of momentum transfer is limited by the detector acceptance for the decay photons of $\pi^0$ at lowest values of $Q^2$ and by statistics at largest values of $Q^2$. Existing data taken in the scope of the $R$ measurement, discussed in Sec.~\ref{sec:Rincl}, will be used to extend the covered regions of momentum transfers down to approximately $0.1\,\textrm{GeV}^2$. A future data set of additional $20\,\textrm{fb}^{-1}$ at 3.773 GeV will help to extend the covered range of momentum transfer up to approximately $10\,\textrm{GeV}^2$, which will allow to test the discrepancy in the results of the $B$-factories, \ieie, the so-called BaBar-Belle puzzle.

The current preliminary results are in very good agreement with recent theoretical calculations using dispersion relations~\cite{Hoferichter:2018dmo} or LQCD~\cite{Gerardin:2019vio}. The data have also been considered in a recent review of the HLbL contribution to $a_\mu$~\cite{Danilkin:2019mhd}, where it was demonstrated that they contribute to a significant reduction of the uncertainties.

First feasibility studies of the $\eta$ and $\eta^\prime$ TFF measurements have been performed at \bes3. So far only individual decay modes have been considered to reconstruct the mesons from the same data that is used for the pion TFF measurement. The covered range of momentum transfer at \bes3 is found similar to the $\pi^0$ TFF result with an accuracy similar to what has been reported from previous measurements~\cite{M1_Behrend:1990sr,M1_Gronberg:1997fj}. A final result will make use of all major decay modes of both mesons and combine the existing and future data sets to obtain results of highest accuracy and impact.

\begin{figure}[tbp]
 \centerline{ \includegraphics[width=0.5\textwidth]{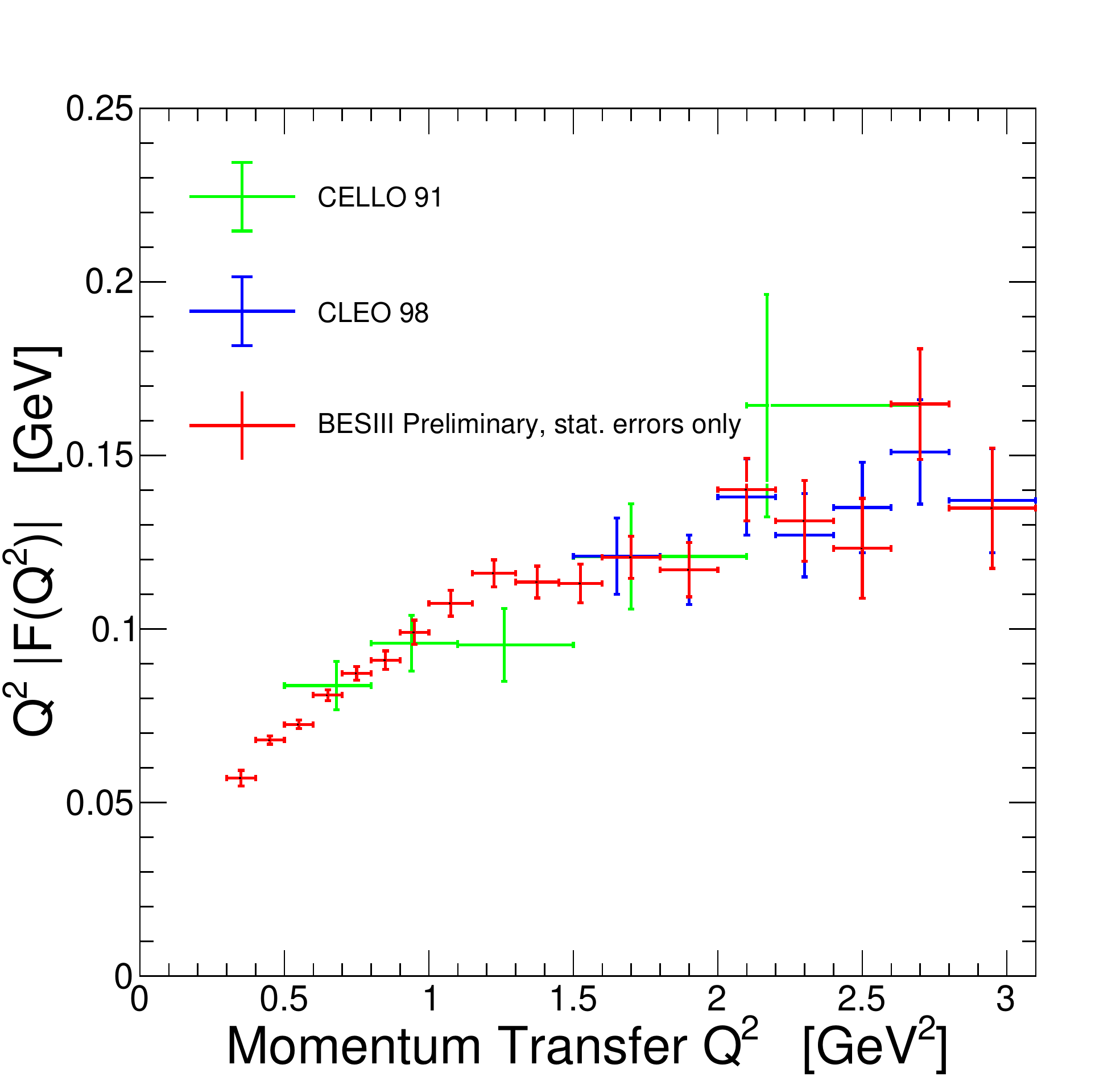} }
 \caption{\label{fig:spacelike_pi0TFF_BES}
 Preliminary \bes3 measurement of the spacelike $\pi^0$ TFF in comparison with data from CELLO \cite{M1_Behrend:1990sr} 
and CLEO \cite{M1_Gronberg:1997fj}.
 }
\end{figure}

 \item{\bf Single-tag \boldmath{$\pi\pi$, $\pi\eta$, $\eta\eta$} TFFs, spacelike}\\
Besides the lowest-lying pseudoscalar mesons, $\gamma \gamma^\ast$ processes also allow to access the structure of scalar, axial-vector and tensor mesons through the production of multi-meson final states. A first measurement at \bes3 focuses on the investigation of the $\gamma^\ast \gamma \to \pi^+ \pi^-$ channel, with one virtual photon. Data at cms energies between 3.773 and 4.6\,GeV are combined to perform the studies on an integrated luminosity of $7.5\,\textrm{fb}^{-1}$. Figure~\ref{fig:ggtwopi} shows the full simulation of $1\,\textrm{fb}^{-1}$ at $\sqrt{s}=4.23$\,GeV of the single-tagged analysis of the $\gamma\gamma^\ast\to\pi^+\pi^-$ after event selection in the three relevant kinematic variables to study the two-pion system. The dominating background of muon production in the reaction $e^+e^-\to e^+ e^-\mu^+\mu^-$ is rejected by adapting the machine learning techniques, successfully used in the pion form factor measurement discussed in Sec.~\ref{sec:ISR}. The irreducible background contribution due to the timelike amplitude of $e^+e^-\to e^+ e^-\pi^+\pi^-$ is subtracted using MC-derived distributions. The investigation at \bes3 has triggered an improvement of the \textsc{Ekhara 3.0} event generator~\cite{Czyz:2018jpp}, which allows to properly take into account interference effects between signal and background processes. Thus, a first high-statistics result, at masses starting from the two-pion mass threshold, at small momentum transfers, and with a full coverage of the helicity angle of the pion system is obtained at \bes3. While the results are awaited by the $(g-2)_\mu$ theory community~\cite{Hoferichter:2019nlq,Danilkin:2019opj}, in the threshold region, such data will also provide new empirical information on the pion polarizabilities. 

\begin{figure}[tb]
  \centerline{ \includegraphics[width=0.49\textwidth]{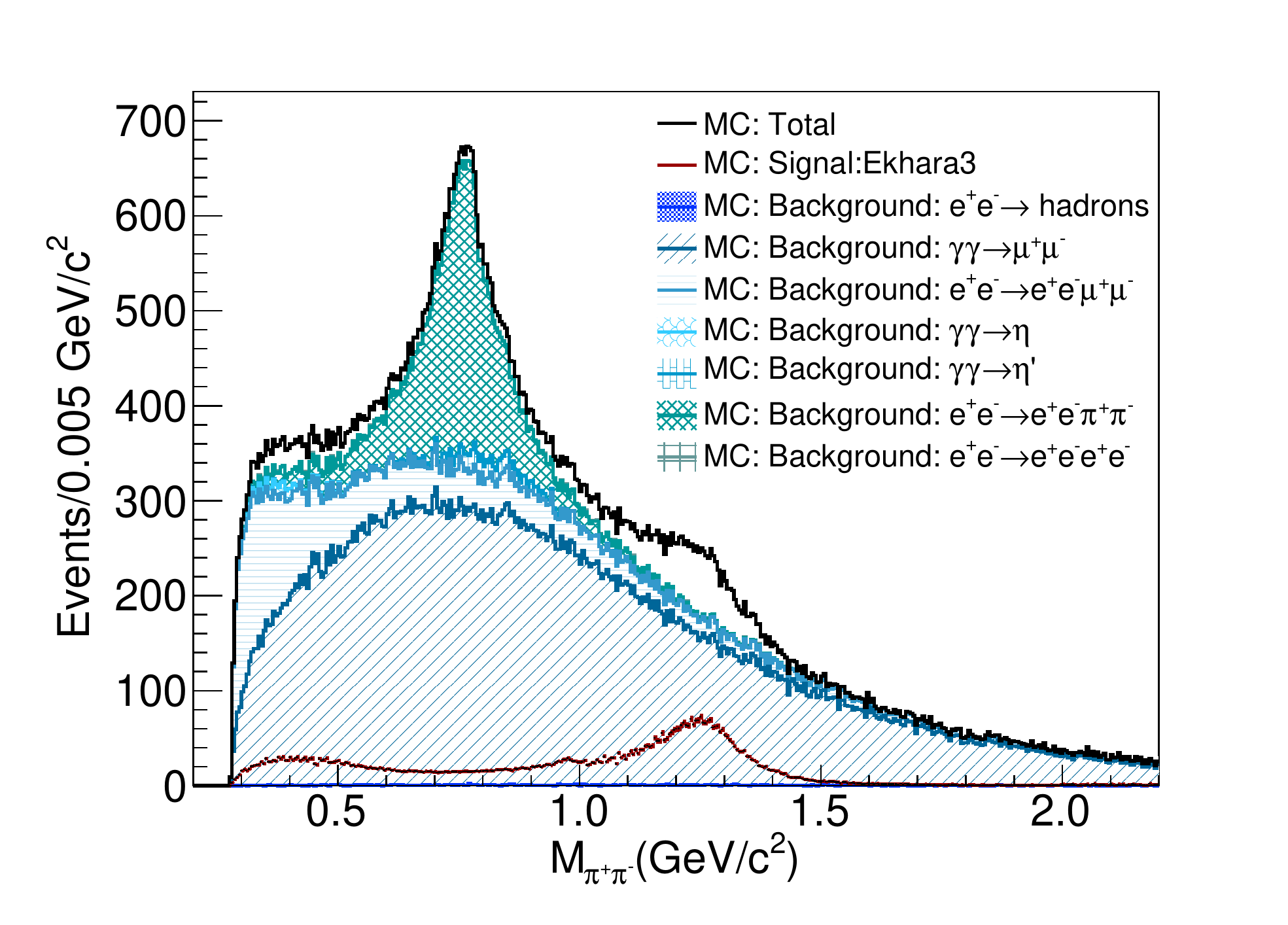}\hspace*{0.02\textwidth}%
               \includegraphics[width=0.49\textwidth]{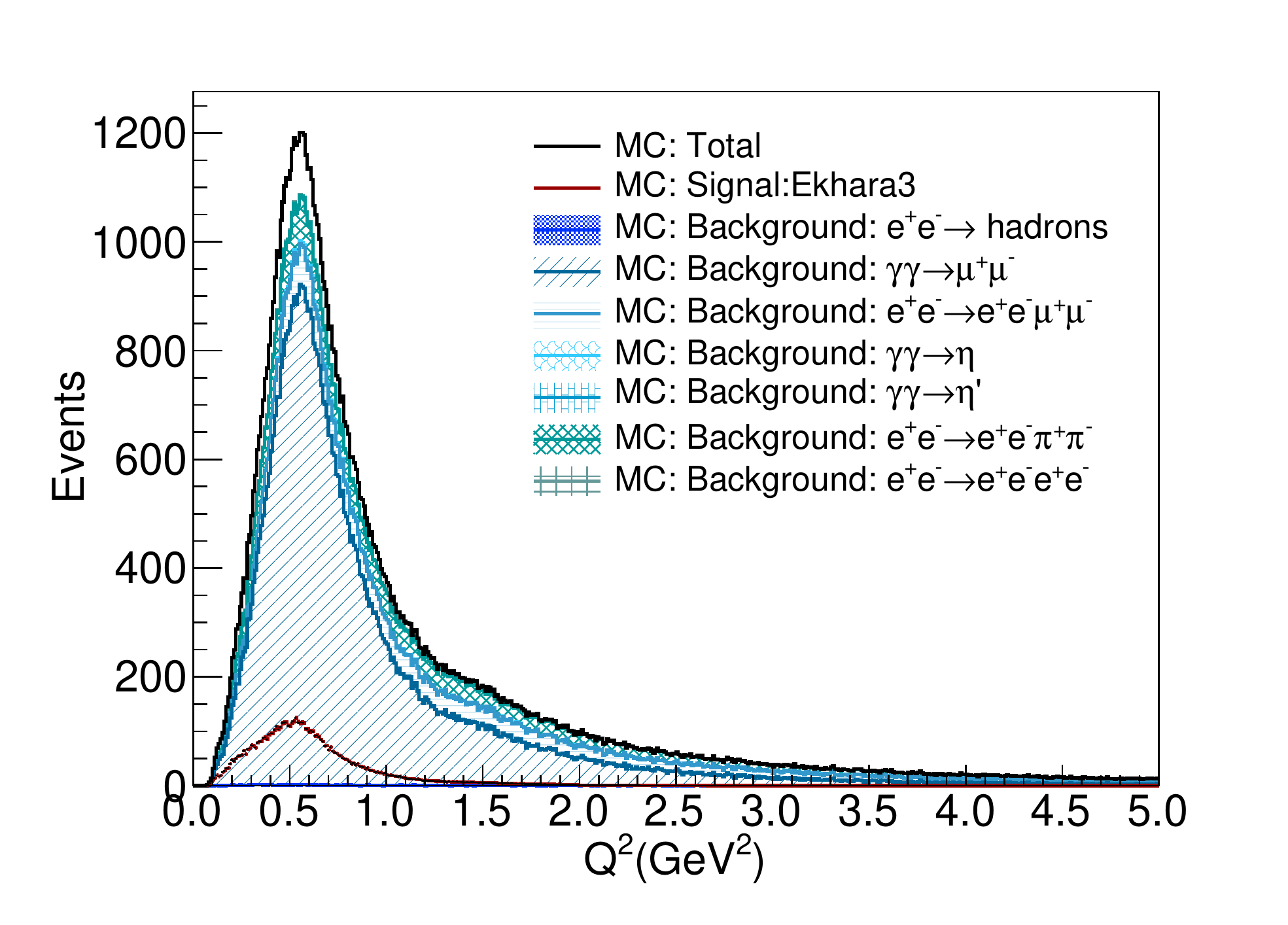} }
  \centerline{ \includegraphics[width=0.49\textwidth]{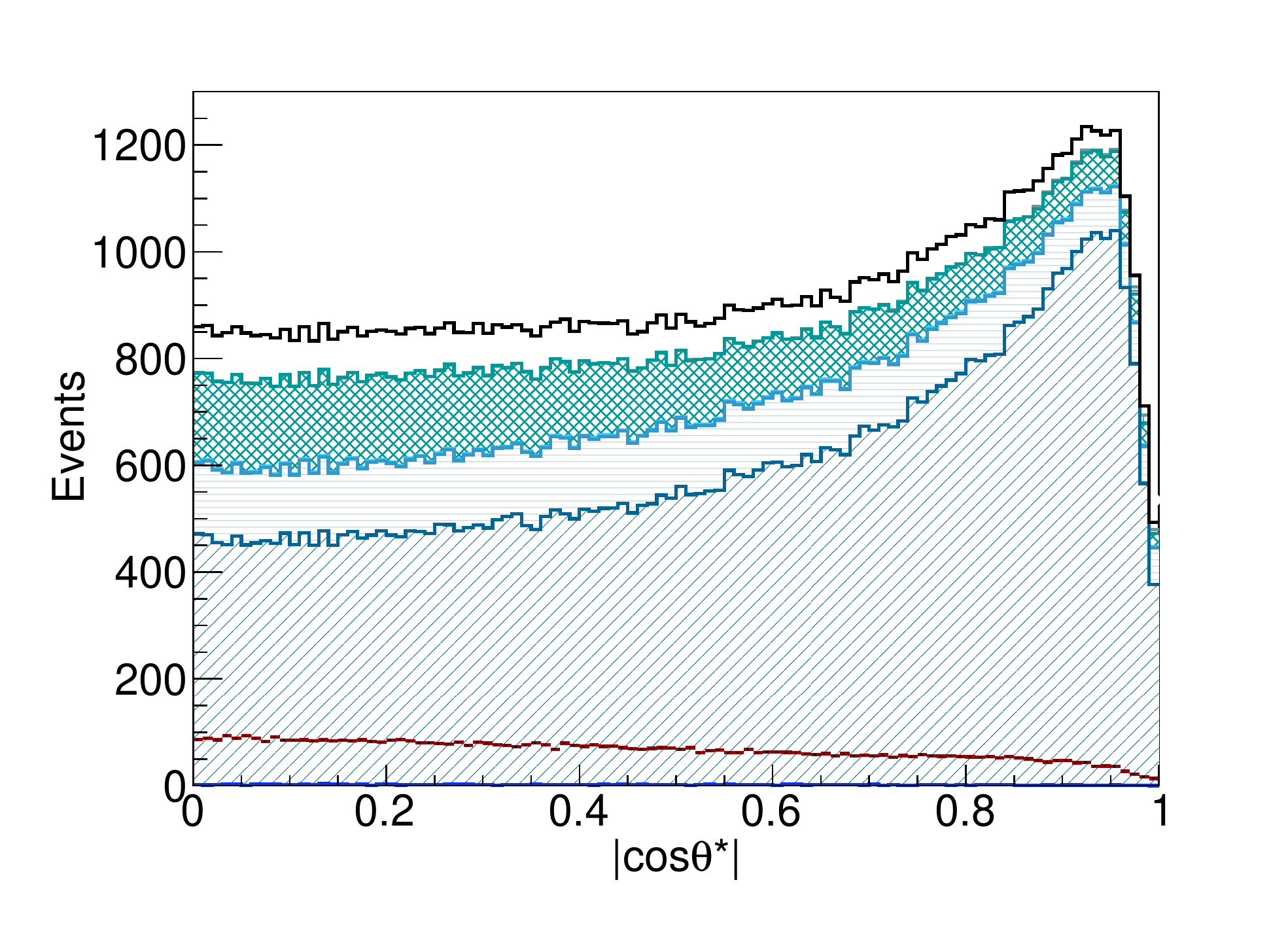} }
 \caption{\label{fig:ggtwopi}Full simulation of the single-tagged analysis of $\gamma\gamma^\ast\to\pi^+\pi^-$ at $\sqrt{s}=4.23\,\textrm{GeV}$. The three relevant distributions of the two-pion system are shown: invariant mass of the pions (top left), momentum transfer $Q^2$ of the tagged lepton (top right), and helicity angle of the pions $\cos\theta^\ast$ (bottom).
 }
\end{figure}

The analysis of the two-pion system is currently extended to the neutral pion system, where a complementary result to the recent Belle measurement at large momentum transfers~\cite{Masuda:2015yoh} is expected. Additional, future data will allow to extend the investigations also to $\pi\eta$ and $\eta\eta$ systems with high statistics. Furthermore, even higher multiplicity final states $\gamma\gamma^*\to 3\pi, \gamma\gamma^*\to 4\pi$ or $\gamma\gamma^\ast\to \eta\pi\pi$ can be performed to provide relevant information on axial and tensor mesons, like the $f_1(1285)$ and $a_2(1320)$ with high accuracy, needed as input for the calculations of the HLbL contribution to $(g-2)_\mu$.

 \item{\bf Double-tag \boldmath{$\pi^0$} TFF, spacelike}\\
In the case that both scattered leptons are identified in the detector, the doubly-virtual TFF can be accessed. First event sample of this kind has been identified in the 2.9\,fb$^{-1}$ data set at a cms energy of 3.773\,GeV. With improved statistics, the measurement of the doubly-virtual TFF can be used to compute the pion-pole contribution to HLbL in a completely model-independent way. Based on a MC simulation restricted by the geometric acceptance of the \bes3 detector only, the result is expected to cover momentum transfers in the range of $0.3\leq(Q_1^2,Q_2^2)\,\leq2.2
~\textrm{GeV}^2$. The provided information is crucial to constrain the precision of data driven calculations of the HLbL contribution to $(g-2)_\mu$~\cite{Nyffeler:2016gnb}. In addition to the doubly-virtual TFF of $\pi^0$, the same information of $\eta$ and $\eta^\prime$ mesons will be measured. The \bes3 result will be complementary to the recent measurement of the doubly-virtual $\eta^\prime$ TFF by the BaBar collaboration~\cite{BaBar:2018zpn}, which has been obtained at large momentum transfers only.

 \item{\bf Pseudoscalar TFFs at high \boldmath{$Q^2$}, timelike}\\
Up to now, at high momentum transfer the timelike $\pi^0$ TFF has not yet been extracted from the annihilation reaction $e^+e^- \to \pi^0 \gamma$. At \bes3 it is possible to access the timelike $\pi^0$ TFF in the $Q^2$ range of approximately 20\,GeV$^2$, which allows for a comparison with the spacelike data by BaBar~\cite{Aubert:2009mc} and Belle~\cite{Uehara:2012ag}. In the case of $\eta$ and $\eta^\prime$, in addition to the comparison to spacelike measurements, the data can be compared to a timelike BaBar measurement at $Q^2=$ 112 GeV$^2$~\cite{Aubert:2006cy}. Comparisons of this kind allow to test predictions of pQCD, in which it is assumed that the spacelike and timelike TFFs become identical and in both cases feature a $Q^{-2}$-dependence behavior~\cite{Aubert:2006cy}. 

\end{itemize}

We stress once more that the \bes3 program on meson TFF measurements is of highest relevance to constrain the HLbL contribution to $(g-2)_\mu$. In a recent review paper~\cite{Danilkin:2019mhd}, the $\pi^0$ pole exchange contribution is evaluated in a simplistic model, which fulfills the basic theory constraints, considering already the new preliminary data from \bes3. The data allow to reduce the uncertainty of this most relevant contribution to $\pm0.2\times10^{-10}$, while previous model dependent estimates have uncertainties in the range of $\pm(0.9-1.2)\times10^{-10}$ (see Tab.~2 in Ref.~\cite{Danilkin:2019mhd}). More elaborate and model independent evaluations of the $\pi^0$ contribution, as the dispersive calculations of~\cite{Hoferichter:2018dmo} or the lattice calculations of~\cite{Gerardin:2019vio}, which do not consider the \bes3 data, already achieve uncertainties on the level of $\pm(0.3-0.4)\times10^{-10}$. These results can be cross checked, when being confronted with present and future \bes3 data. As mentioned before, the goal is to achieve a total uncertainty of all hadronic contributions to $a_\mu$ on the level of $1.6\times10^{-10}$.

%% file: QCD/baryon.tex
\section{Baryon form factors}

Baryons provide a unique window to the strong interaction, since they constitute the simplest system for which the non-Abelian nature of QCD is manifest~\cite{isgur}. The most well-known baryon species is the nucleon. However, despite being known for more than a century, and despite its importance as the main contributor to the mass of our visible Universe, fundamental properties like its mass, spin~\cite{ashman, aidala} and structure~\cite{Miller} are difficult to describe from first principles. This difficulty is a consequence of the non-perturbative interactions between the quarks inside the nucleon. At this scale, the break-down of pQCD calls for quantitative predictions from \egeg, Chiral Perturbation Theory~\cite{ChPT1,ChPT2}, Lattice QCD~\cite{LQCD} or phenomenological models, \egeg, Skyrme models~\cite{Skyrme,Witten}.

The inner structure of baryons can be described and studied experimentally on a common footing through electromagnetic form factors (EMFFs), probed by processes involving hadrons interacting with virtual photons. The EMFFs are fundamental observables of non-perturbative QCD and quantify the deviation from the pointlike case. If one-photon exchange is assumed,  the momentum transfer squared $q^2$ of the virtual photon is given by $q^2=(p_i-p_f)^2<0$. Elastic, or spacelike, form factors ($q^2 < 0$) have been studied since the 1960's in electron-nucleon scattering~\cite{Punjabi}. Spin $\frac{1}{2}$ baryons have two form factors, often referred to as the electric $G_E$ and the magnetic $G_M$ form factor. In the so-called Breit frame, these are Fourier transforms of the charge- and magnetization density, respectively. The measured charge density of the neutron is particularly intriguing: though being negative near the center whilst positive further out can be explained in the simple quark model by $d$-quarks clustering in the center surrounded by the $u$-quark, the drop to negative values at even larger distances from the center requires more elaborate models involving \egeg, pion clouds~\cite{Miller}. One way to gain further insights into this puzzle is to replace one or several of the light quarks in the nucleon by heavier ones, \ieie, forming hyperons, and study how the structure changes~\cite{granados}. However, since hyperons are unstable, they are unfeasible as beams or targets, and  therefore, do not easily lend themselves to electron scattering experiments. Instead, their structure can be studied in the timelike region ($q^2 > 0$) through the electron-positron annihilations, with the subsequent production of a baryon-antibaryon pair. 

The somewhat abstract timelike form factors can be related to the more intuitive spacelike form factors by dispersion relations~\cite{dispersion}. In particular, spacelike and timelike form factors should converge to the same value as $|q^2|$ reaches a certain scale~\cite{analyticity}. For nucleons, the onset of this scale can be tested by measuring the spacelike and timelike form factors with great precision and compare them. For hyperons, this is unfeasible due to the poor experimental access to the spacelike region. However, one can utilize the fact that timelike form factors can be complex, with a relative phase that polarizes the final
state~\cite{analyticity}. For a ground-state hyperon $Y$, this phase is accessible, thanks to the weak, self-analyzing decays. The daughter baryon $B$ will be emitted according to the spin of the mother hyperon $Y$, giving a decay angular distribution that depends on the polarization of $Y$:
\begin{equation}\label{eq:hypdecay}
  W(\cos\theta_B)=\frac{1}{4\pi}(1+\alpha_Y P_Y \cos\theta_B).
\end{equation}
Since spacelike form factors are real, the same must hold for timelike form factors at large $|q^2|$. Hence, the onset of the scale, at which spacelike and timelike form factors converge to the same value, can be obtained by finding the scale at which the phase goes to zero.  

Also at threshold, the relative phase must be zero. This is because the phase is a result of interfering amplitudes, \textit{e.g.}, $s$-waves and $d$-waves. At threshold, only $s$-waves can contribute which means that the phase is zero. Furthermore, the absence of other waves also imply that the ratio between the electric and the magnetic form factor, \ieie, $|G_E/G_M|$ is equal to 1 at threshold.

\subsection{Formalism}

For spin $\frac{1}{2}$ baryons produced \textit{via} one-photon exchange  in $e^+e^-\rightarrow\gamma^*\rightarrow B\bar{B}$, the Born cross section can be parameterized in terms of $G_E$ and $G_M$:
\begin{equation}\label{equ-borncs}
  \sigma_{B\bar{B}}(s) = \frac{4C\pi\alpha^2\beta}{3s}\left[|G_M(s)|^2+\frac{1}{2\tau}|G_E(s)|^{2}\right].
\end{equation}

\noindent Here, $\alpha$=1/137.036 is the fine-structure constant, $\beta=\sqrt{1-4m^2_Bc^4/s}$ the velocity of the produced baryon, \textit{c} the speed of light, \textit{s} the square of the cms energy, $m_B$ the mass of the baryon and $\tau = s/(4m^2_B)$. The Coulomb factor $C$ is a correction to the one-photon exchange and describes the electromagnetic interaction between the outgoing $B$ and $\bar{B}$. For neutral baryons, $C$ is 1 which in combination with the vanishing phase space $\beta$ factor, the cross section should be zero at threshold. For charged baryons, $C$ is typically assumed to have the value for pointlike charged fermions  $C=\varepsilon R_S$~\cite{Sakarov}, where $\varepsilon = {\pi \alpha_{\rm em}}/{\beta}$ is an enhancement factor. In this case, the two $\beta$ factors cancel and as a result, the cross section becomes non-zero at threshold. The so-called Sommerfeld resummation factor
\begin{equation}
R_S={\sqrt{1-\beta^{2}}}/(1-e^{-\pi \alpha_{\rm em} \sqrt{1-\beta^{2}}/\beta})
\end{equation}
causes the cross section to rise rapidly with the fermion velocity~\cite{Sommerfeld}. 

In many experiments, the data samples are too small to separate between $G_E$ and $G_M$ since that requires analysis of angular distributions. In order to compare production cross sections of different baryon-antibaryon pairs for equivalent kinematic conditions, the \textit{effective form factor} is defined:
\begin{equation}\label{equ-effectiveff}
\begin{split}
|G(s)| \equiv &\sqrt{\frac{\sigma_{B\bar{B}}(s)}{(1+\frac{1}{2\tau})(\frac{4\pi\alpha^2\beta }{3s})}} \\
\equiv &\sqrt{\frac{2\tau|G_M(s)|^2+|G_E(s)|^2}{2\tau+1}}.
\end{split}
\end{equation}
In order to extract the relative phase, a full spin decomposition of the reaction is needed. The formalism has been outlined in Refs.~\cite{faldt,faldtkupsc}. In particular, the $\Lambda$ transverse polarization $P_Y$ is given by:
\begin{equation}\label{eq_hyppol}
P_Y=\frac{\sqrt{1-\eta^2}\sin\theta\cos\theta}{1+\eta\cos^2\theta}\sin(\Delta\Phi),
\end{equation}
where $\eta$ is an angular distribution parameter related to the form factor ratio $\mathcal{R}$ by $\eta=(\tau-\mathcal{R}^2)/(\tau+\mathcal{R}^2)$. The polarization $P_Y$ can be extracted from Eq.~\eqref{eq:hypdecay}.

The formalism presented outlined so far is based on the assumption that one-photon exchange dominates the production mechanism. It has been discussed whether two-photon exchange contributes to the production mechanism, leading to interference effects that results in an additional term $\kappa\cos\theta\sin^2\theta$ in the angular distribution ~\cite{Gakh}. As a consequence, the scattering angle distribution of the produced baryon will be slightly asymmetric. This is quantified by the asymmetry
\begin{equation}
\mathcal{A}=\frac{N(\cos\theta>0)-N(\cos\theta<0)}{N(\cos\theta>0)+N(\cos\theta<0)}.
\end{equation}
It is related to the $\kappa$ parameter in the following way:
\begin{equation}\label{eq:Abeta}
\mathcal{A}=\frac{3}{4}\frac{\kappa}{3+\eta},
\end{equation}
where $\eta$ is the same as in Eq.~\eqref{eq_hyppol}~\cite{egle_asy}. The asymmetry $\mathcal{A}$ is straight-forward to measure and offers a convenient way to study the importance of two-photon exchange in the timelike region.

\subsection{State of the art}
Proton EMFFs have been studied extensively in the spacelike region~\cite{Punjabi}. In particular, the development of the polarization transfer technique~\cite{polarimeter} in the late 1990's led to a veritable leap forward of the field, enabling a model independent separation of electric and magnetic form factors. The best precision of the ratio $|G_{E}/G_{M}|$ obtained with the new technique is  $\sigma_{p}|G_{E}/G_{M}|=1.7\%$, achieved at JLab~\cite{jlab}. The previously employed Rosenbluth separation technique~\cite{Rosenbluth} relies on one-photon exchange and comparing data obtained with the two methods shows a large and energy-dependent disagreement. The leading explanation is the effect from two-photon exchange~\cite{guichon,melnitchouk}. 

\begin{figure}[tp]
 \centering
 \includegraphics[height=5.5cm]{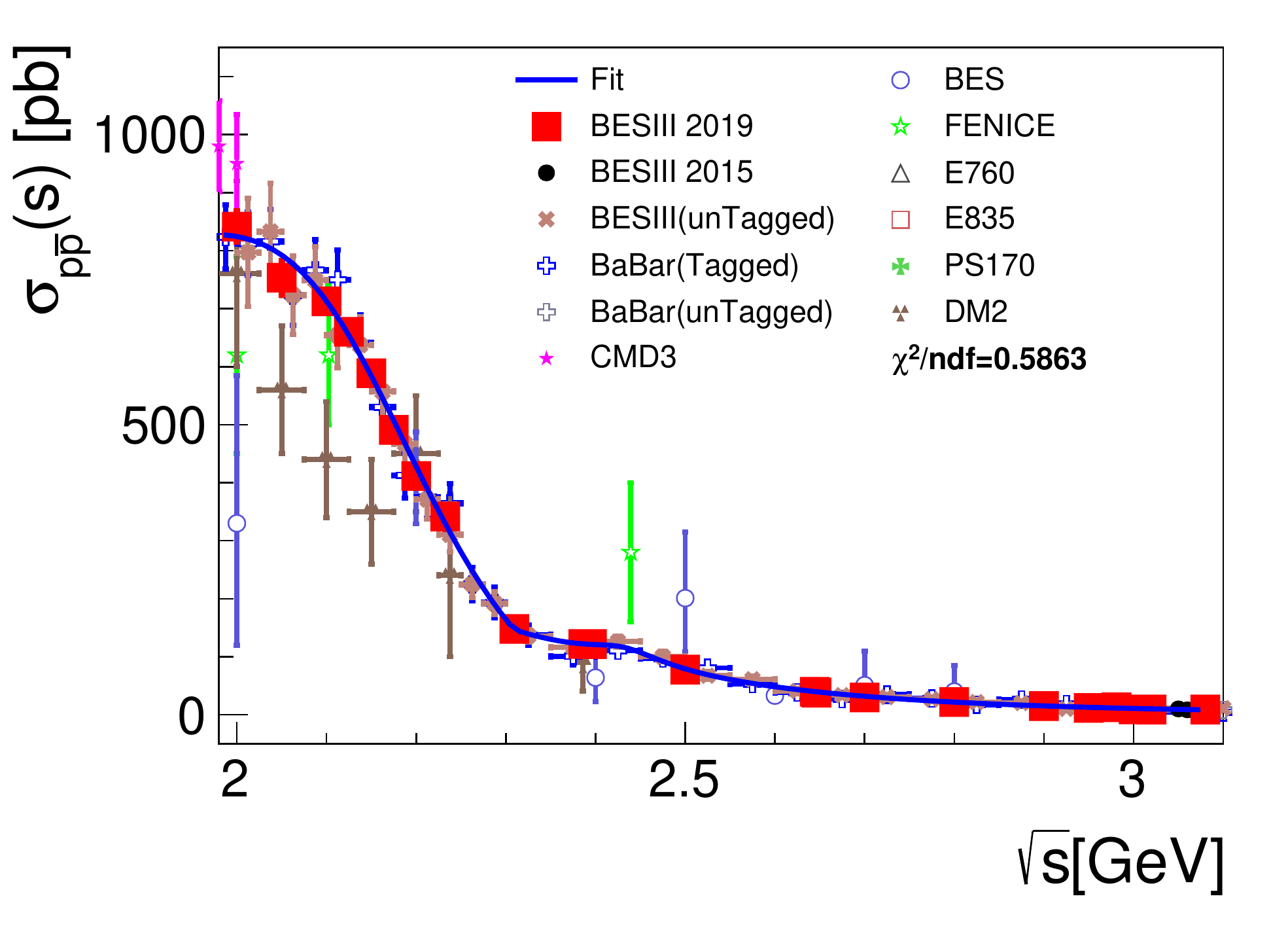}\hspace{0.5cm}%
 \includegraphics[height=5.5cm]{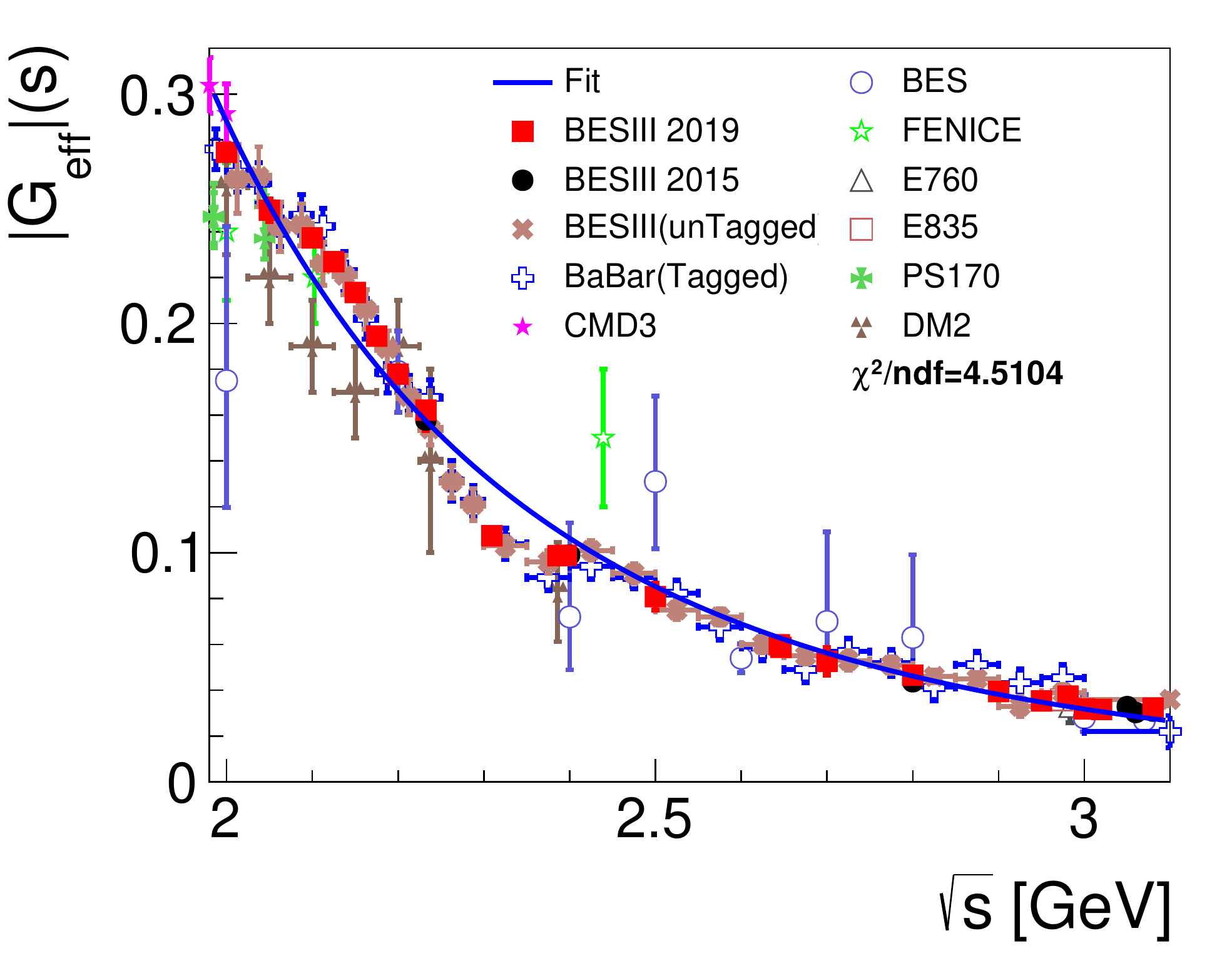}\hspace{0.5cm}%
 \includegraphics[height=5.5cm]{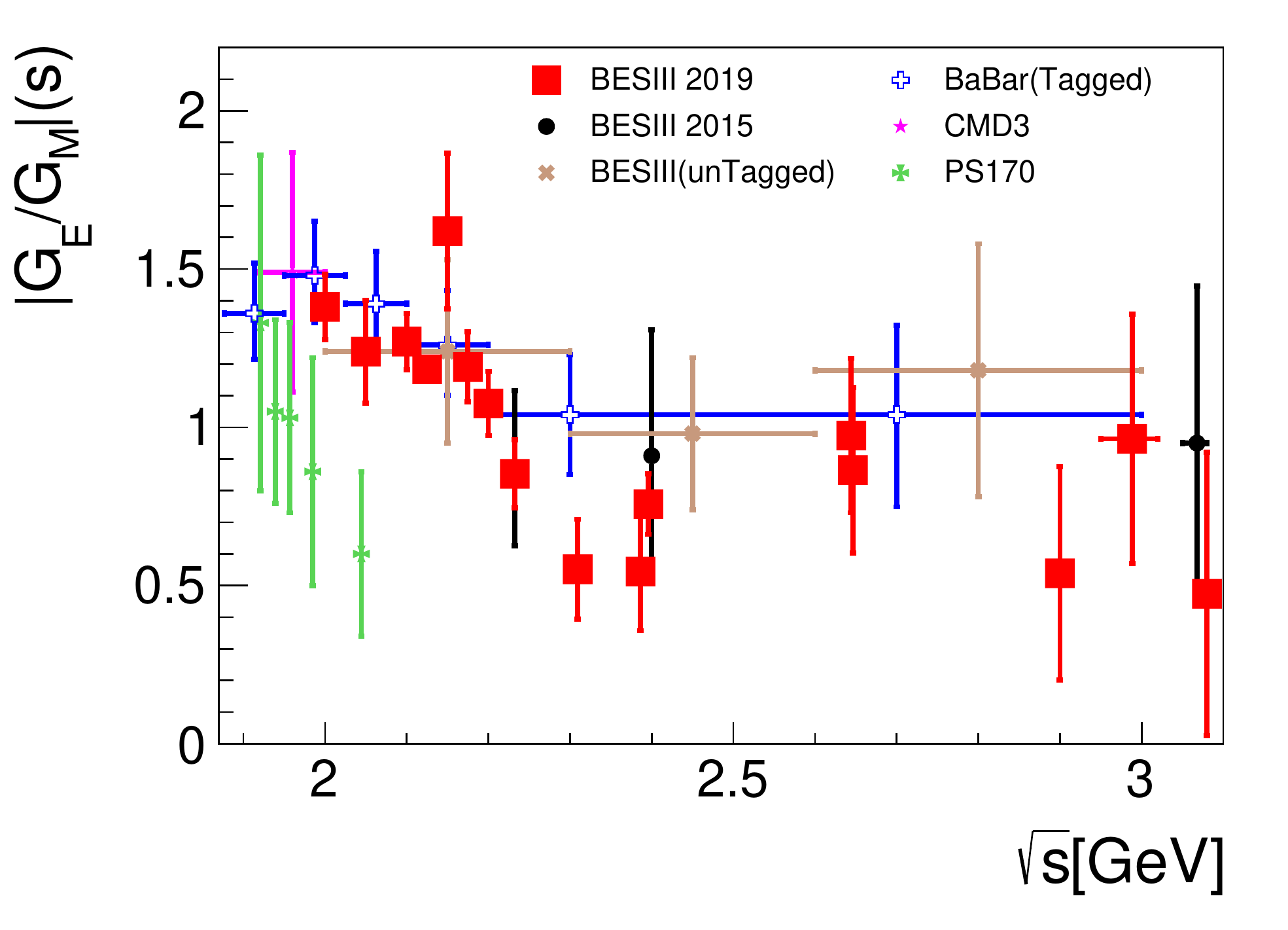}\hspace{0.5cm}%
 \includegraphics[height=5.5cm]{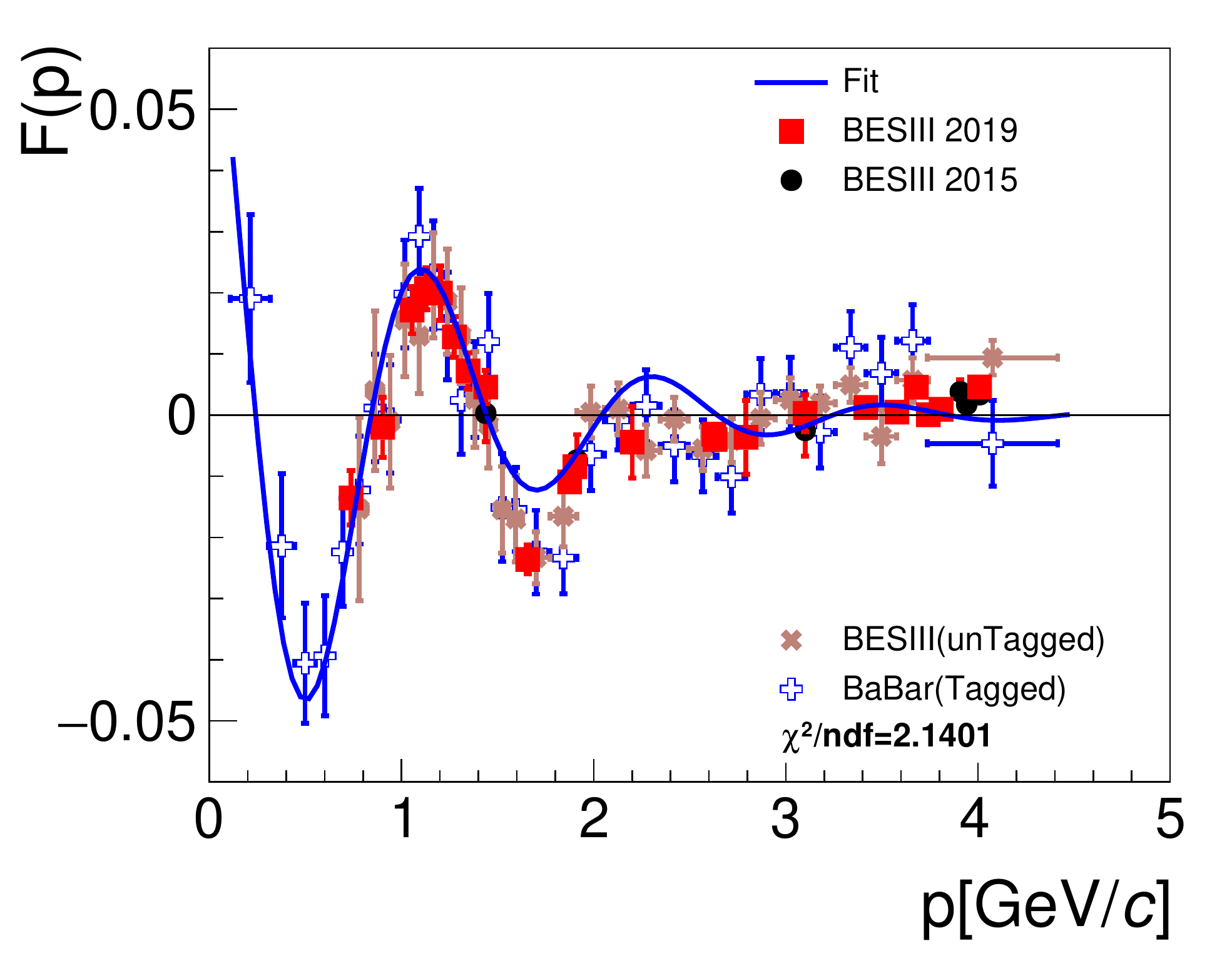}
 \caption{\label{fig:proton}Top left: world data on $e^+e^- \to p\bar{p}$ cross section. Top right: the effective proton timelike form factor. Bottom left: world data on $R=|G_E/G_M|$. Bottom right: effective form factor after subtracting the fitted line in the top right panel. The data are from BESIII~\cite{bes3ppbar2012,bes3ppbarisr}, BaBar~\cite{babarppbar}, CMD3~\cite{CMD3ppbar}, BES~\cite{BESppbar}, FENICE~\cite{feniceppbar}, E760~\cite{E760ppbar}, E835~\cite{E835ppbar}, PS170~\cite{ps170ppbar} and DM2~\cite{DM2ppbar}.}
\end{figure}

The timelike region became accessible with the advent of high-precision, high-intensity electron-positron colliders at intermediate energies. The world data on $e^+e^- \to p\bar{p}$ are shown in Fig.~\ref{fig:proton}. An advantage of electron-positron annihilations is the charge symmetry, which makes the measurements less sensitive to higher order processes such as two-photon exchange. However, the precision acquired so far has until recently not been compatible with that of the spacelike region~\cite{nucleondis}. The most precise measurements from BaBar (in $e^+e^- \to p\bar{p}$~\cite{babarppbar}) and PS170 (in $\bar{p}p \to e^+e^-$~\cite{ps170ppbar}) achieved $\sigma_{p}|G_{E}/G_{M}|\approx 10\%$ and differ by more than $3\sigma$ (bottom left panel of Fig.~\ref{fig:proton}). Recent data from BESIII obtained with a beam energy scan~\cite{bes3ppbar2012} and the radiative return or ISR method~\cite{bes3ppbarisr} agree with the BaBar measurements. New data from BESIII, collected in a high-precision energy scan in 2015, will offer improved precision over a large $q^2$ range. 

The cross section and effective form factor show interesting features, as can be seen in the top and bottom left panels of Fig.~\ref{fig:proton}. The BaBar collaboration reported an oscillating behavior~\cite{babarppbar,Egle} that was recently confirmed by BESIII~\cite{bes3ppbarisr}. This becomes particularly striking when being studied as a function of the relative momentum between outgoing proton and antiproton. More high-precision data are needed to establish this elusive feature at the level of many standard 
deviations.

The large amount of high-quality data on proton EMFFs has inspired the theory community to develop various approaches to nucleon structure, based on Chiral Perturbation Theory~\cite{chiral}, Lattice QCD~\cite{lattice}, Vector Meson Dominance~\cite{VMD}, relativistic Constituent Quark Models~\cite{CQM} and pQCD~\cite{pQCD}.

While experiencing an era of great progress on proton timelike EMFFs, corresponding data for its isospin partner, the neutron, remain a challenge. This is primarily due to the difficulty in identifying and reconstructing the neutron and antineutron from the  $e^+ e^- \to n \bar{n}$ process. The cross section has been measured by the SND~\cite{snd-nn} experiment up to 2\,GeV and by the FENICE experiment~\cite{fenice-nn} between 2 and 3\,GeV. FENICE collected only a small amount of data and identified the $\bar{n}$ by the time-of-flight method and the annihilation pattern in the detector. The ratio $R_{np}=\sigma(e^+ e^- \to n \bar{n})$ / $\sigma(e^+ e^- \to p \bar{p})$ is expected to be close to 1 if the process is dominated by the isoscalar or isovector amplitude. In a picture where the production cross section is proportional to the square of the leading quark charge ($u$ quark in the case of the proton and $d$ quark in the case of the neutron), the ratio should instead be close to 0.25. More elaborate predictions are presented in Refs.~\cite{elliskarliner,karlinernuss}. More precise data could shed further light on this issue.

Also hyperon EMFFs has been a fairly unexplored territory until recently. The cross section of the $e^+e^- \to Y\bar{Y}$ process ($Y$ referring to various ground-state hyperons) has been studied by DM2~\cite{hyperondm2}, BaBar~\cite{babarllbar}, BESIII~\cite{bes3hyp2012} and CLEO~\cite{cleocllbar}. The latter measurement compared the production of several different ground-state hyperons, including the $\Omega^-$, and interpreted the results in terms of di-quark correlations. The idea that certain configurations of  flavor, spin and isospin of two quarks inside a hadron have important impacts on the structure of hadrons, which has  been discussed since long~\cite{anselmino}. In particular, the effects on $\Lambda$ and $\Sigma$ structures are outlined in Ref.~\cite{wilczekjaffe}.  However, it is difficult to draw definite conclusions from CLEO, since all data points coincided with charmonium resonances
($\psi(3686)$, $\psi(3770)$ and $\psi(4170)$). Hence, interference effects may be important, which makes an unambiguous interpretation difficult. 

\begin{figure*}[tb]
\begin{center}
 \includegraphics[width=0.485\textwidth]{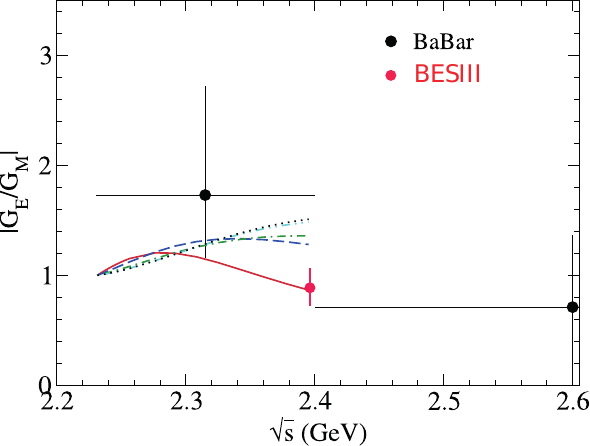}\hfill
 \includegraphics[width=0.51\textwidth]{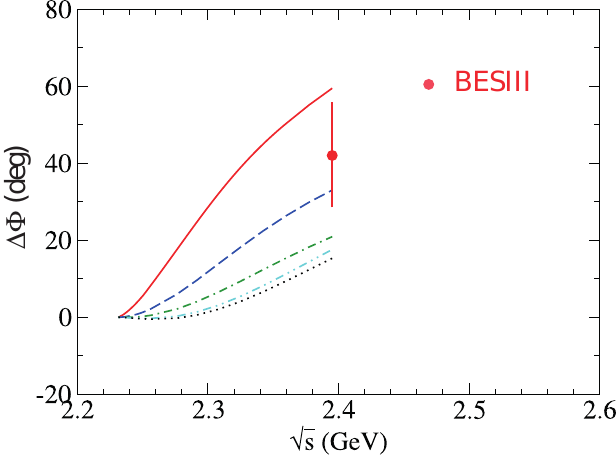}
 \caption{BESIII results of $|G_E/G_M|$ and $\Delta\Phi$ compared with theoretical predictions. Here, the values obtained using the PDG value for $\alpha_{\Lambda}$ has been used since that was used for the theory predictions. The five lines are different $\Lambda\bar{\Lambda}$ potentials, seen in~\cite{haidenbauerllbar}. Black dots are results of BaBar~\cite{babarllbar}, and red dots are measurements by BESIII.}
\label{fig:llbartheory}
\end{center}
\end{figure*}

Due to limited sample sizes, the electric and the magnetic form factors could not be separated with any conclusive precision in neither of the aforementioned experiments. However, in a more recent measurement from BESIII, dedicated data at $q=$ 2.396\,GeV/$c$ enabled a complete spin decomposition of the $e^+e^- \to \Lambda\bar{\Lambda}$ reaction, including measurement of the polarization and spin correlations. From this, it was possible to not only separate between the $\Lambda$ electric and magnetic form factors, but also to determine the relative phase between $|G_E|$ and the $|G_M|$ for the first time. It was found to be significantly different from zero~\cite{bes3hyp2015}. The prospect of measuring the relative phase $\Delta\Phi$ of the $\Lambda$ hyperon by \bes3 have triggered the first theory predictions based on various $\Lambda\bar{\Lambda}$ potential models~\cite{haidenbauerllbar}. It was found that the phase is more sensitive to the potential than the $|G_E/G_M|$ ratio, that in turn is more sensitive than the cross section. The theory predictions, as well as the measured values, are shown in Fig.~\ref{fig:llbartheory}. Note that the theory predictions were made assuming the old PDG value (from before 2019) of the $\Lambda$ decay asymmetry parameter, $\alpha_{\Lambda}$ = 0.642. The data are therefore rescaled to the old value. In Ref.~\cite{bes3hyp2015}, the phase is obtained with the 2019 update of the $\alpha_{\Lambda}$ from PDG, $\alpha_{\Lambda}$ = 0.750.

\begin{figure}[tp]
 \centering
 \begin{tabular}{p{0.5\textwidth} p{0.5\textwidth}}
  \vspace{0pt} \includegraphics[trim=0mm 0mm 0mm 10mm ,clip,width=0.49\textwidth]{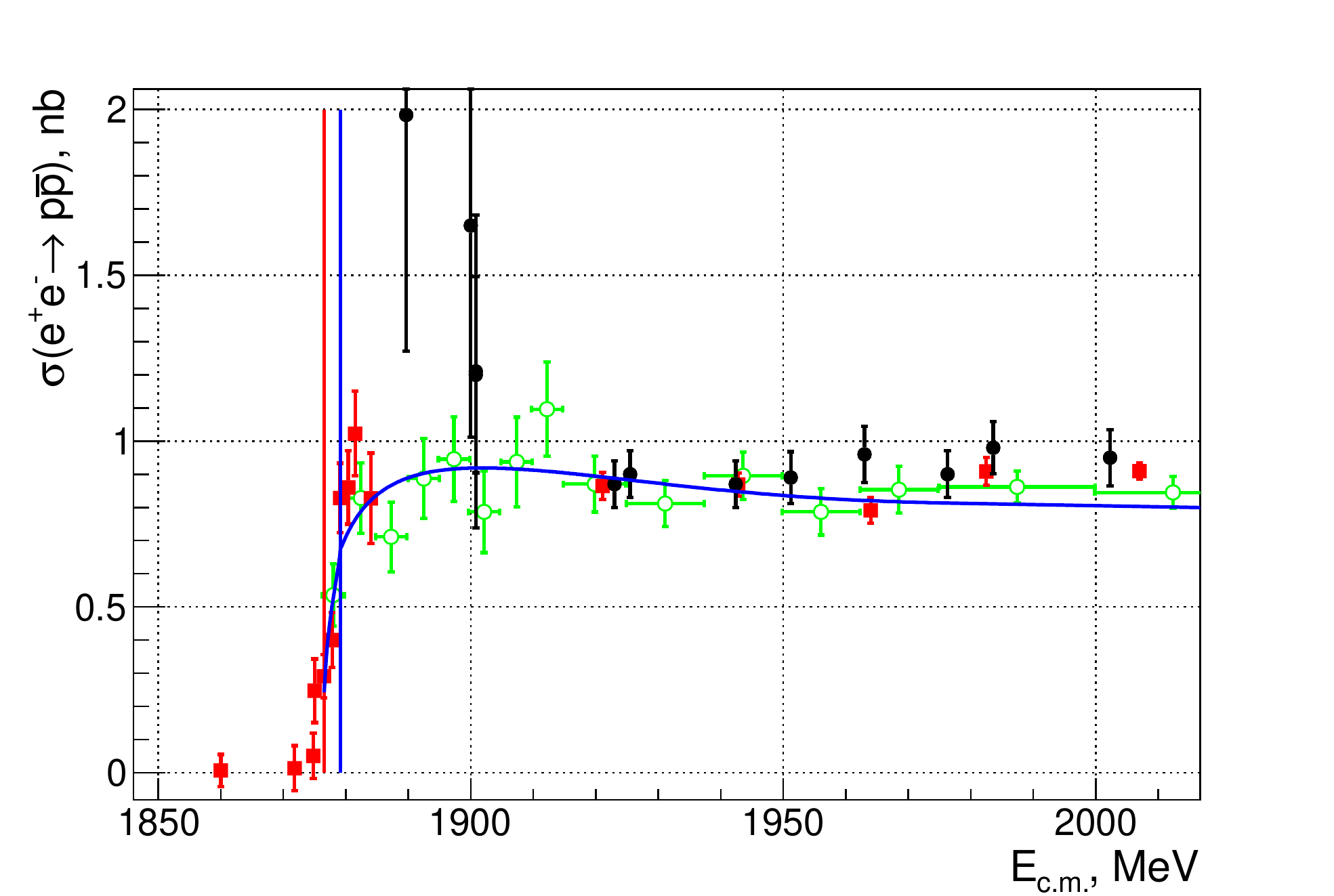} &
  \vspace{0pt} \includegraphics[width=0.49\textwidth]{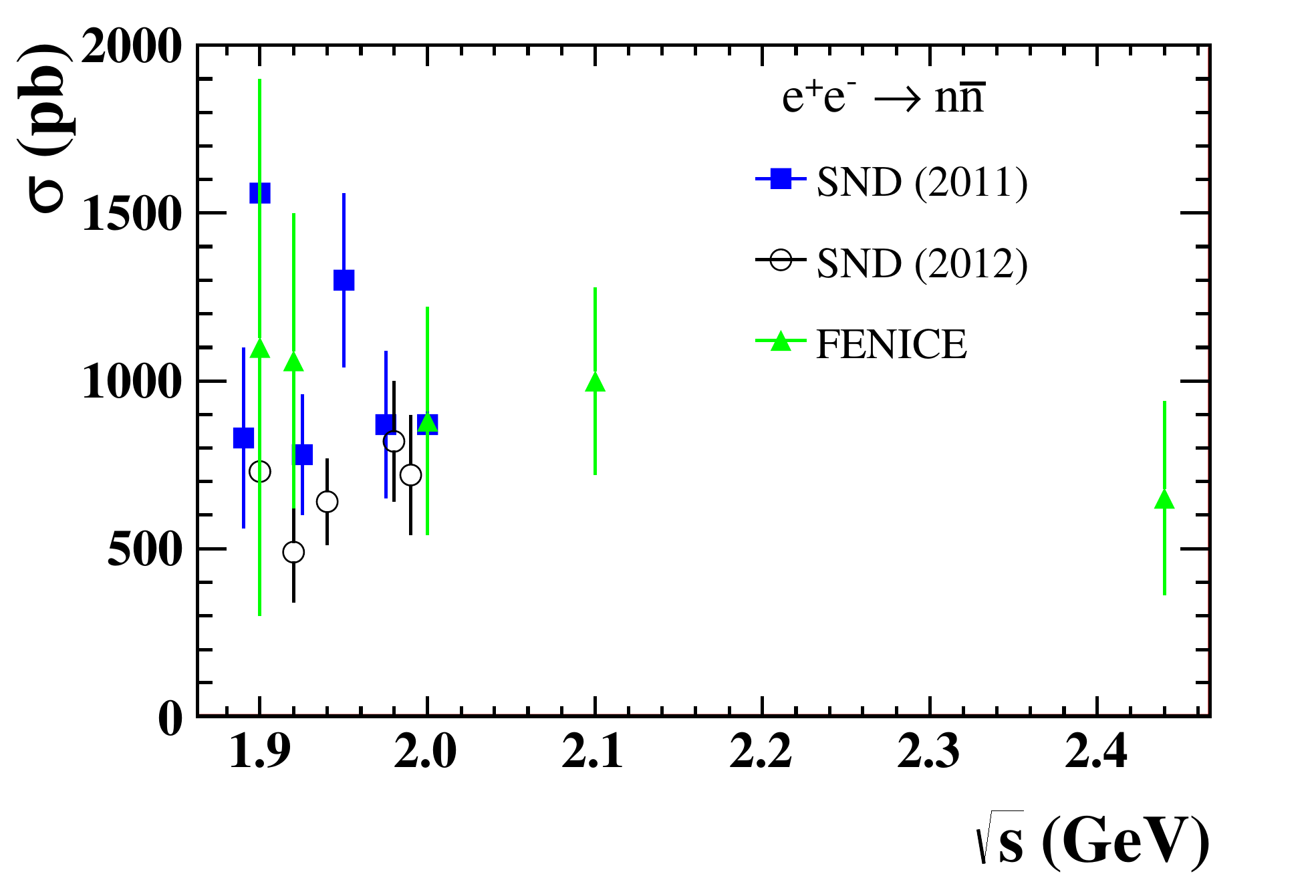}\\
  \vspace{0pt} \includegraphics[width=0.49\textwidth]{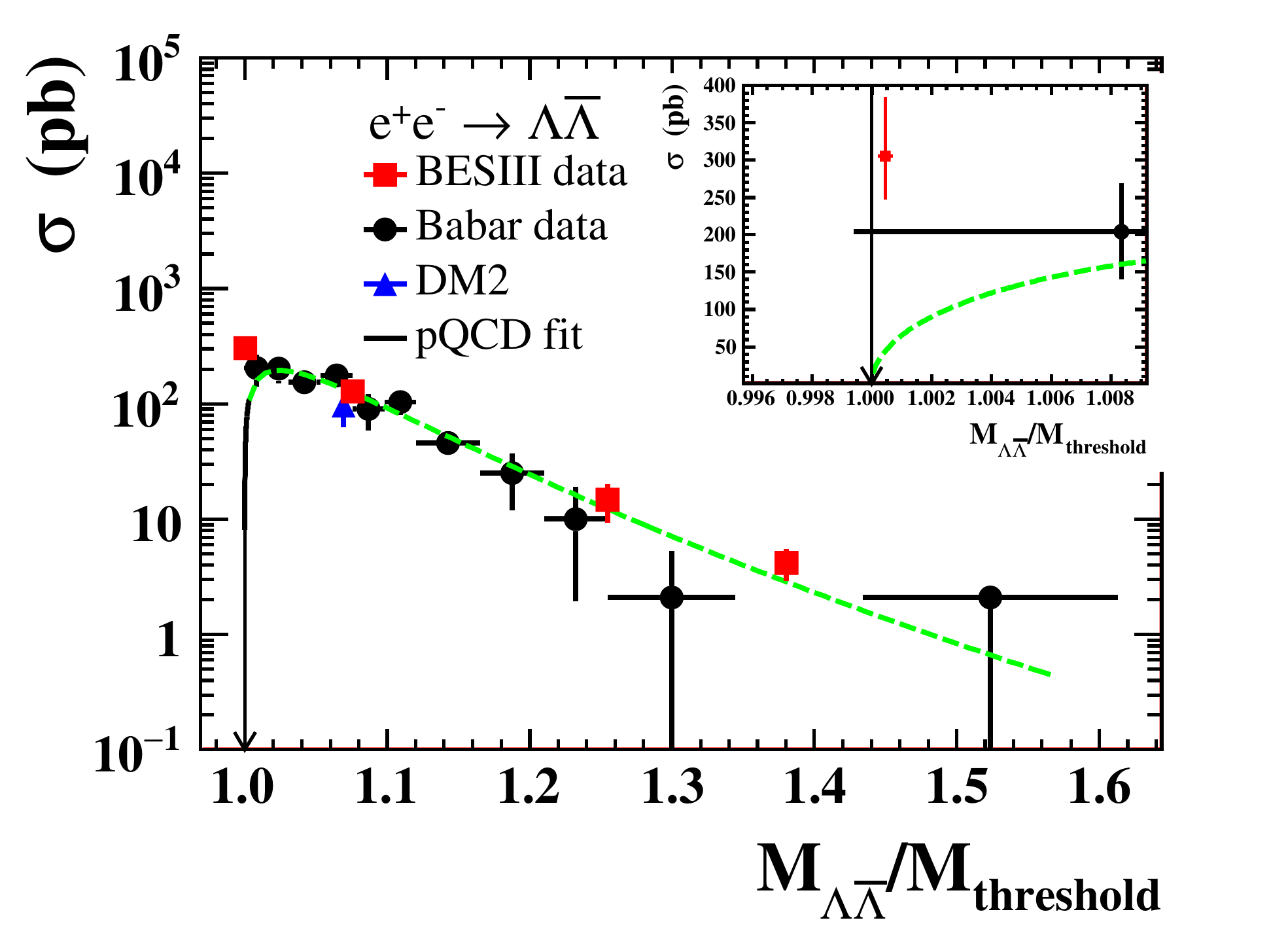} &
  \vspace{0pt} \includegraphics[width=0.49\textwidth]{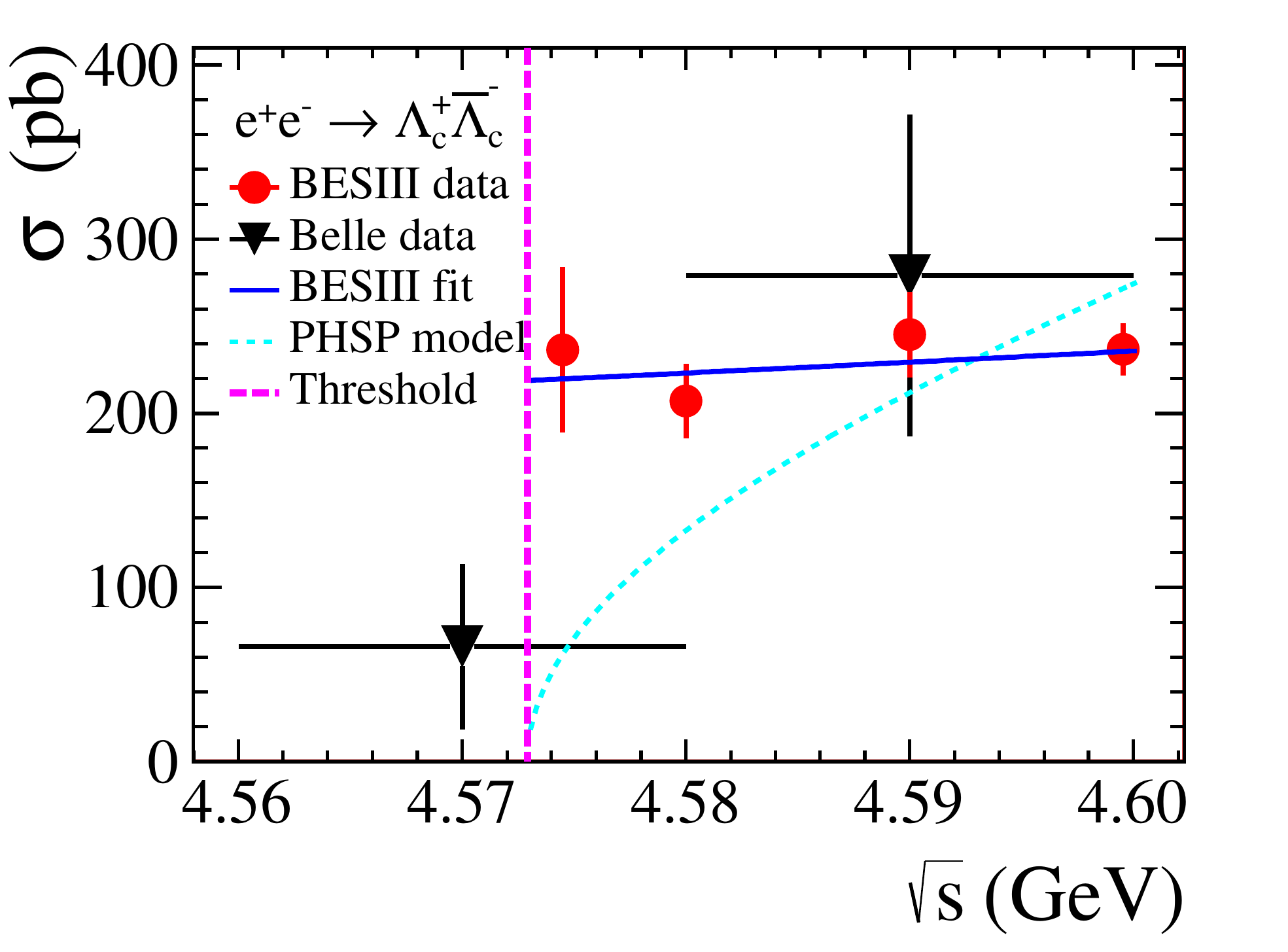}
 \end{tabular}
 \caption{\label{fig:baryonresults}Top left: the $e^+e^- \to p\bar{p}$ Born cross sections measured with CMD-3 (black dots from Ref.~\cite{CMD3ppbar}, red squares from Ref.~\cite{CMD2019}) and BaBar~\cite{babarppbar} (green open circles). The solid curve shows the result of the prediction from Ref.~\cite{ref:cmd-3-ref}. The vertical lines show the $p\bar{p}$ and $n\bar{n}$ thresholds. Top right: world data on $e^+e^- \to n\bar{n}$ (blue and white points from SND~\cite{snd-nn}, and green points from FENICE~\cite{feniceppbar}). Bottom left: the $e^+e^- \to \Lambda\bar{\Lambda}$ cross sections near threshold, from BaBar~\cite{babarllbar} and BESIII~\cite{bes3hyp2012}. Note that the scale on the $y$-axis is logarithmic. Bottom right: the $e^+e^- \to \Lambda^+_c\bar{\Lambda}_c^-$ cross sections from Belle~\cite{bellelc} and BESIII~\cite{bes3lc}.}
\end{figure}

The effective form factor of the $\Lambda_c^+$ EMFFs have been obtained from measurements at Belle~\cite{bellelc} and BESIII~\cite{bes3lc}. Furthermore, BESIII also measured the ratio $\mathcal{R}$ between the $\Lambda_c^+$ electric and magnetic form factors. In this measurement, as well as in measurements of the proton and $\Lambda$ hyperon EMFFs near threshold, interesting features can be discerned: the cross section undergoes a sharp rise close to threshold, followed by a plateau in the case of the positively charged proton~\cite{babarppbar,CMD3ppbar,CMD2019} and $\Lambda_c^+$~\cite{bes3lc}, as shown in Fig.~\ref{fig:baryonresults}. Various FSI models have been employed to explain this behavior~\cite{chiral}. In the case of the $\Lambda_c^+$, where there is some discrepancy between data from Belle and BESIII,  the cross section has been studied with a theory model taking into account the $Y(4630)$ resonance~\cite{dai}. The threshold of the neutral $\Lambda$ is particularly interesting. The cross section of $e^+e^- \to \Lambda\bar{\Lambda}$ decreases after its sharp rise~\cite{bes3hyp2012}. This sharp rise, that is similar to the behavior of the $e^+e^- \to n\bar{n}$~\cite{snd-nn}, is difficult to explain with Eq.~\eqref{equ-borncs}, since the $\Lambda$ is neutral which means that $C = 1$ and hence $\beta$ factor should cause the cross section to vanish at threshold. The nonzero cross section near threshold, as well as the wide plateau, have led to various interpretations, \egeg, final-state interactions~\cite{theoryfsi}, bound states or mesonlike resonances~\cite{theoryres}, and a gluon exchange contribution in the resummation factor~\cite{theorycoulomb}.

\subsection{Prospects with BESIII}

BESIII is uniquely suited for nucleon and strange hyperon form factor studies. It is currently the only running or planned $e^+e^-$ experiment that is optimized in the energy region where nucleon-antinucleon and strange hyperons are produced in abundance. The capability of detecting and identifying charged and neutral particles, including antineutron, is another advantage. Furthermore, the coming upgrade of the BEPCII collider up to cms energies of 4.9\,GeV will enable structure studies of single-charm hyperons.  The following topics can be addressed within BESIII in the near future:

\begin{itemize}

\item \textbf{EMFF phase measurements of octet hyperons}\\
Comparative studies of baryons with different isospin reveal the inner structure in terms of possible di-quark 
correlations~\cite{ref_diquark}. For example, while the quark content of the $\Lambda$ and the $\Sigma^0$ is the same, the isospin of the $ud$ pair is different. As a consequence, the spin structure should be different. In the case of the $\Sigma^+$, the isospin of the $uu$ pair should be the same as that of the $ud$ pair in the $\Lambda$. As a consequence, the cross sections of $e^+e^- \to \Lambda\bar{\Lambda}$ and $e^+e^- \to \Sigma^+\bar{\Sigma}^-$ should be similar, as observed by CLEO. However, measurements should be performed at energies that do not coincide with charmonium resonances in order to avoid interference effects. Furthermore, spin observables should be more sensitive to the underlying quark structure, and therefore, it would be illuminating to study the EMFF phase for different octet hyperons. For the $\Sigma$ triplet, simulation studies show that $q = 2.5$\,GeV/$c$ is optimal in terms of cross section and reconstruction efficiency to collect a sample of the required size. The simulation studies show that 100\,pb$^{-1}$ would yield a sufficient amount of events to extract the phase of $\Sigma^+$ and $\Sigma^0$. For the cascade doublet ($\Xi^-$ and $\Xi^0$), the corresponding optimal point is located at $q = 2.8$\,GeV/$c$.

\item \textbf{Energy dependence of the EMFF phase of the $\Lambda$ hyperon}\\
Preliminary calculations of the energy dependence of the $\Lambda$ EMFF phase have been performed~\cite{ref:Pacetti} in an attempt to predict the spacelike EMFFs. However, the calculations call for measurements in more data points in order to constrain the predictions. Presently, the only conclusive measurement is obtained at $q = 2.396$\,GeV/$c$. The analysis of these data, as well as simulations at other energy points, show that data samples of $\approx$100\,pb$^{-1}$ are required to achieve the necessary sample size. At low energies, collecting such data samples is time consuming and the points should therefore be chosen very carefully. The data samples at 2.5\,GeV and 2.8\,GeV, proposed in the previous bullet point, synergize well with the relevant criteria since the cross section at these points~\cite{babarllbar} should be sufficiently large for successful measurements.

\item \textbf{Neutron EMFFs}\\
The analysis of energy scan data from BESIII, collected in 2015, shows that the methods for identifying neutron-antineutron final states have relatively low efficiencies. In order to determine the EMFFs with good precision, yet larger data samples are required. This is in line with the proposal of the previous point. In combination with the data collected in 2015, that are currently being analyzed, this will give a comprehensive picture of the neutron structure in the timelike region and enable comparative studies of proton and neutron EMFFs.

\item \textbf{Energy dependence of $\Lambda_c^+$ EMFFs}
With an upgraded BEPCII ring, it will be possible to perform a precision scan from the $\Lambda_c^+\bar{\Lambda}_c^-$ threshold and up to $q=4.9$\,GeV/$c$. The observed discrepancy between Belle and BESIII can be investigated and the possible importance of the $Y(4630)$ resonance can be reviewed. Furthermore, it will be possible to study the EMFF ratio and phase, as outlined in the Charm chapter.

\item \textbf{Threshold behavior of octet baryons}\\
In addition to the samples collected during 2015 at each baryon-antibaryon threshold, currently being analyzed, additional samples are needed just below and slightly above each threshold. In this way, possible systematic effects can be detected and taken into account and the observed behavior can be cross-checked. By comparing baryons of different charge, deeper insights can be gained and the relation to possible resonances or interactions at quark level can be established. In particular, a scan around the $\Sigma^+\bar{\Sigma}^-$ threshold with about $20\,\textrm{pb}^{-1}$ per point would be very illuminating. 

\item \textbf{Contribution from two-photon exchange}\\
A contribution from two-photon exchange would show up as an asymmetry in the scattering angle distribution. This asymmetry has been measured in Ref.~\cite{bes3hyp2015} and was found to be consistent with zero, though with a large uncertainty. Other baryon channels, for example $p\bar{p}$, where the sample sizes are larger for a given integrated luminosity, can reach a precision where possible effects from two-photon exchange can be revealed. This is particularly true if more data can be collected at some off-resonance energies.

\item \textbf{Baryon form factor by ISR method}\\
The large amount of data that is planned to be collected at energies larger than 3.7\,GeV enables studies of baryon EMFFs using the ISR method. This possibility is particularly valuable close to threshold. Special event selection techniques are required in the analysis of energy scan data, due to the low momenta of the outgoing particles. Due to the boost of the ISR photon, the respective baryons have larger momenta in the laboratory frame and can be detected more easily. Additionally, it is currently technically challenging to acquire energy scan data in the threshold region of nucleon production, due to the design of the accelerator. However, away from threshold, the statistical precision is generally better when using the energy scan method, as can be seen for example in a comparison of the proton EMFF studies using the ISR method~\cite{bes3ppbarisr} and the energy scan method~\cite{bes3ppbarscan}. Measurements of the form factor ratio and its relative phase benefit from data collected at well-defined energies with the scan method. The uncertainty in energy that is inevitable with the ISR method, propagates to the uncertainty in the ratio and the phase.

\end{itemize}

%% file: QCD/hadron.tex
\section{Fragmentation function}

The fragmentation function $D_q^{h}(z)$ describes the probability of a hadron $h$ to be found in the debris of a quark (or antiquark), carrying the fraction $z$ of the quark energy. $D_q^{h}(z)$ is
an inherently non-perturbative object governing hadronization. It cannot be deduced from first principles, but can be extracted from experimental data~\cite{albino,radici}. Fragmentation functions are assumed to be universal, $i.e.$, they are not process dependent.

\begin{figure}[tbp]
 \centering
 \includegraphics[height=7.0cm]{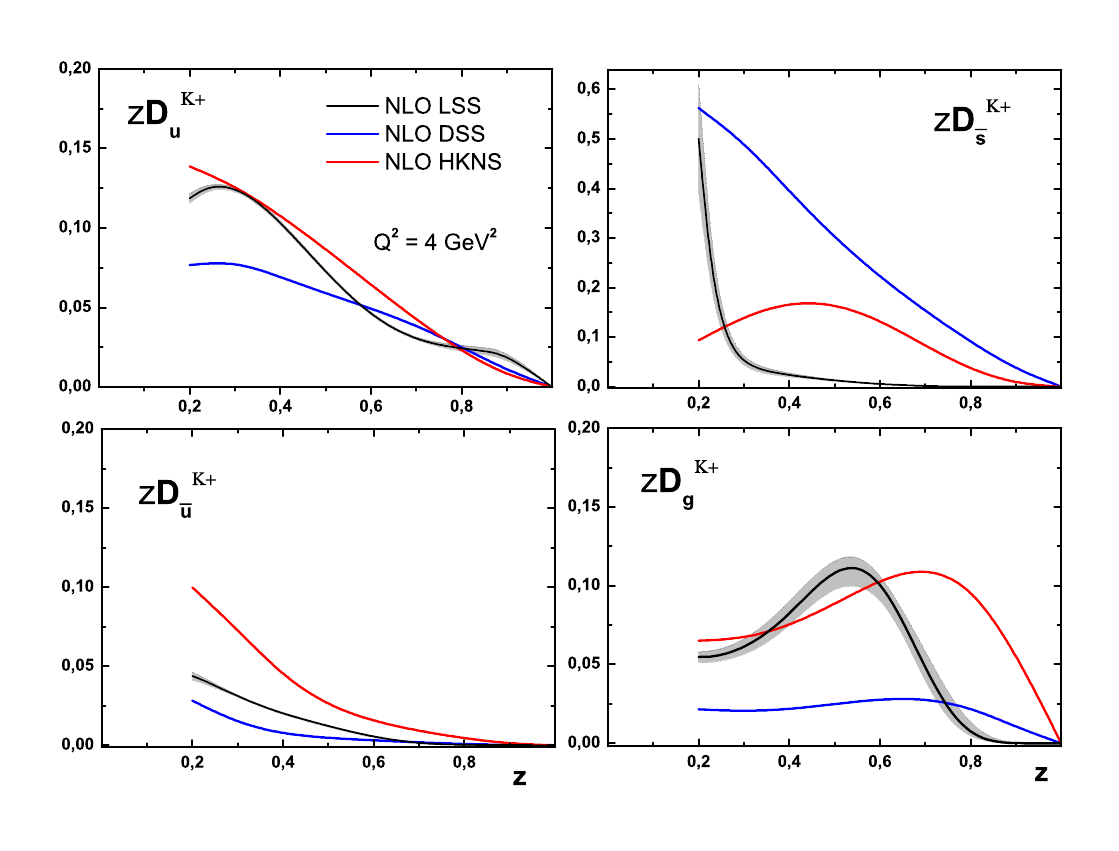}
 \caption{\label{kaonff}Comparison between NLO LSS, DSS and HKNS Kaon
 fragmentation functions at $Q^2=4$\,GeV$^2$, taken from
Ref.~\cite{kaonff}.}
\end{figure}

A large amount of data on inclusive hadron production from $e^+e^-$ collisions has been collected in a wide energy range $10 \leq \sqrt{s} \leq 200\,\textrm{GeV}$ and $0.005 \leq z \leq 0.8$~\cite{data}. These data sets, semi-inclusive deep-inelastic scattering (DIS) data and data sets from hadronic collisions are used to extract a fragmentation function by DSS~\cite{DSS}, HKNS~\cite{HKNS} and AKK08~\cite{AKK}. However, the number of experimental data points is small, and the uncertainty of experimental data points is large at the few GeV energy region. Figure~\ref{kaonff} compares the Kaon fragmentation function by different fragmentation function packages; the favored $D_{\bar s}^{K+}(z,Q^2)$ changes very rapidly between $z$=0.2 and $z$=0.3~\cite{kaonff}.

A precise determination of fragmentation functions could help to understand the internal structure of the nucleon. The 
strange quark polarization puzzle, $i.e.$, the polarization of the strange quark is positive in the measured region of 
Bjorken $x$ by semi-inclusive DIS analyses, where inclusive DIS yields significantly negative values of this quantity. 
It is pointed out in Ref.~\cite{strangepuzzle} that the polarization of the strange quark extracted from semi-inclusive 
DIS analyses is very sensitive to the input fragmentation functions. The semi-inclusive DIS process is used to study 
the proton spin at the upgraded JLab-12~\cite{jlab12} and the future Electron-Ion Collider~\cite{eic}. Thus, a better 
knowledge of fragmentation functions is needed.

DIS experiments have performed detailed studies of the transverse momentum dependence (TMD) by semi-inclusive processes.
They describe cross sections in terms of TMD parton density functions (PDF) and fragmentation functions. The TMD 
fragmentation functions $D_q^{h}(z, P_{h\perp}^2, Q^2)$ describes a fragmentation process of an unpolarized parton $q$ 
into an unpolarized hadron $h$, which carries the longitudinal-momentum fraction $z$ and transverse momentum 
$P_{h\perp}$ in the process. In order to understand TMD PDFs, knowledge of the TMD fragmentation functions is needed. 
Unfortunately, the determination of unpolarized TMD fragmentation functions is still missing~\cite{FFreview}. With data 
from the $B$ factories and BESIII, we could extract these required TMD fragmentation functions.

The Collins fragmentation function describes spin-dependent effects in fragmentation processes~\cite{collins}. It 
connects transverse quark spin with a measurable azimuthal asymmetry (so-called Collins effect) in the distribution of 
hadronic fragments along the initial quark's momentum. This azimuthal asymmetry has been reported in semi-inclusive DIS 
and $e^+e^-$ annihilation~\cite{FFreview}, where the Collins effect is studied in $e^+e^-$ annihilation by detecting 
simultaneously two hadrons, coming from fragmentation of quark and antiquark, in the process $e^+ e^- \to h_1 h_2 + X$. 
In this case, the observable is associated with two Collins fragmentation functions.

The $e^+e^-$ Collins asymmetries taken from Belle and BaBar correspond to higher $Q^2 (\approx 100\,{\rm GeV}^2)$ than 
the typical energy scale of existing DIS data (mostly $2-20\,{\rm GeV}^2$), which is similar to that of BESIII. 
Therefore, the very low cms energies reached at BESIII allow to investigate energy scaling. The results are crucial to 
explore the $Q^2$ evolution of the Collins fragmentation function and further uncertainty of extracted transversity, 
thus improving our understanding of both the Collins fragmentation function and transversity~\cite{peng}.

\begin{figure}[tbp]
 \centering
 \includegraphics[height=9.0cm]{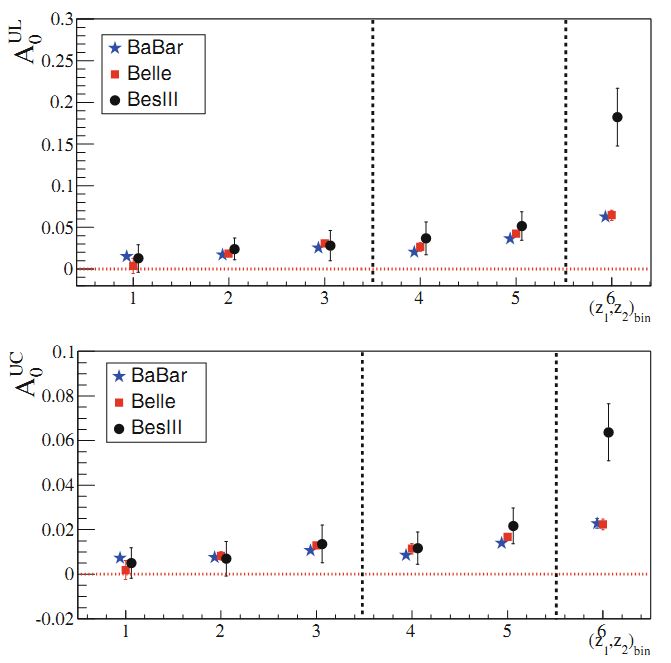}
 \caption{\label{collins}Comparison between the BESIII, BaBar and Belle measurements of the Collins effects. The 
detailed definition for the axis can be found in Ref.~\cite{collinbes3}.}
\end{figure}

The Collins asymmetries for pion pairs have been obtained at BESIII using data collected at $\sqrt{s} = 3.65$\,GeV which 
correspond to an integrated luminosity of 62\,pb$^{-1}$~\cite{collinbes3}. Figure~\ref{collins} compares the Collins 
effect measured by BESIII, BaBar and Belle. The measured asymmetries are almost consistent with zero for low $(z_1, 
z_2)$, and rise with increasing $z$. The BESIII asymmetry in the last interval, is about two/three times larger compared 
to measurements from the $B$ factories. Additional results are needed to confirm this observation
with about 250\,pb$^{-1}$ data at 3.65\,GeV. The Collins effect for strange quarks could be studied in $e^+ e^- \to \pi 
K + X$ and $e^+ e^- \to K K + X$. It is also interesting to study the Collins effect in $e^+ e^- \to \pi^0 \pi^0 + X$, 
$e^+ e^- \to \eta \eta + X$ and $e^+ e^- \to \pi^0 \eta + X$ for neutral hadrons. These results are useful to extract 
TMD PDFs. 

%% file: QCD/tau.tex
% Tau mass

\section{$\tau$ physics at BESIII}

Since the discovery of $\tau$ lepton in 1975 at SPEAR $\EE$ storage
ring~\cite{Perl:1975bf}, the study of $\tau$ lepton has been measured extensively. The properties of the $\tau$ lepton, including mass, lifetime and decays, have been tested sensibly~\cite{Asner:2008nq}. As a member of
the third fermion generation, it decays to the first and second generation fermions.
The pure leptonic or semileptonic character of $\tau$ decays provides a clean
laboratory to test the structure of the weak currents and the universality
of their couplings to the gauge bosons. Moreover, $\tau$ being the only lepton massive enough
to decay into hadrons, the semileptonic decays are an ideal tool for studying strong interactions.

The BEPCII is a $\tau$-charm factory with a cms energy ranging
from 2.0 to 4.9 GeV, and a design peak luminosity of $10^{33}$~cm$^{-2}$
s$^{-1}$ at the cms energy of 3.773 GeV. The great advantage of
BEPCII lies in running near threshold of $\tau$ pair, which provides us
an excellent opportunity for $\tau$ lepton physics. Comparing with other
machines, the threshold region makes possible a much better control of background and systematic uncertainty for measurements.
%the special kinematic feature of the threshold makes it possible 
%that the measurements with low systematic uncertainty could be obtained by
%the limited statistic.

There are three energy regions appreciated for researches. The first is
below the $\tau$ pair production threshold, say 3.50 GeV, where the light
quark background can be measured, and the result can be extrapolated to the
other energy regions. The second is at the production threshold, such as
3.55 GeV, where $\tau$ lepton pair are produced at rest. It is an unique energy to
study the $\tau$ decays. The third region is above the $\tau$ pair
production threshold, which can be further classified into two cases. The
one case is below the production of open charm, say 3.69 GeV, where the
background is same as described at the below threshold region; the other is
at 4.25 GeV, where the cross section of $\tau$ pair production is at the
maximum, while charm background should be considered in addition to
the light quarks background.

\begin{figure}[tbp]
\begin{center}
\includegraphics[height=9.0cm]{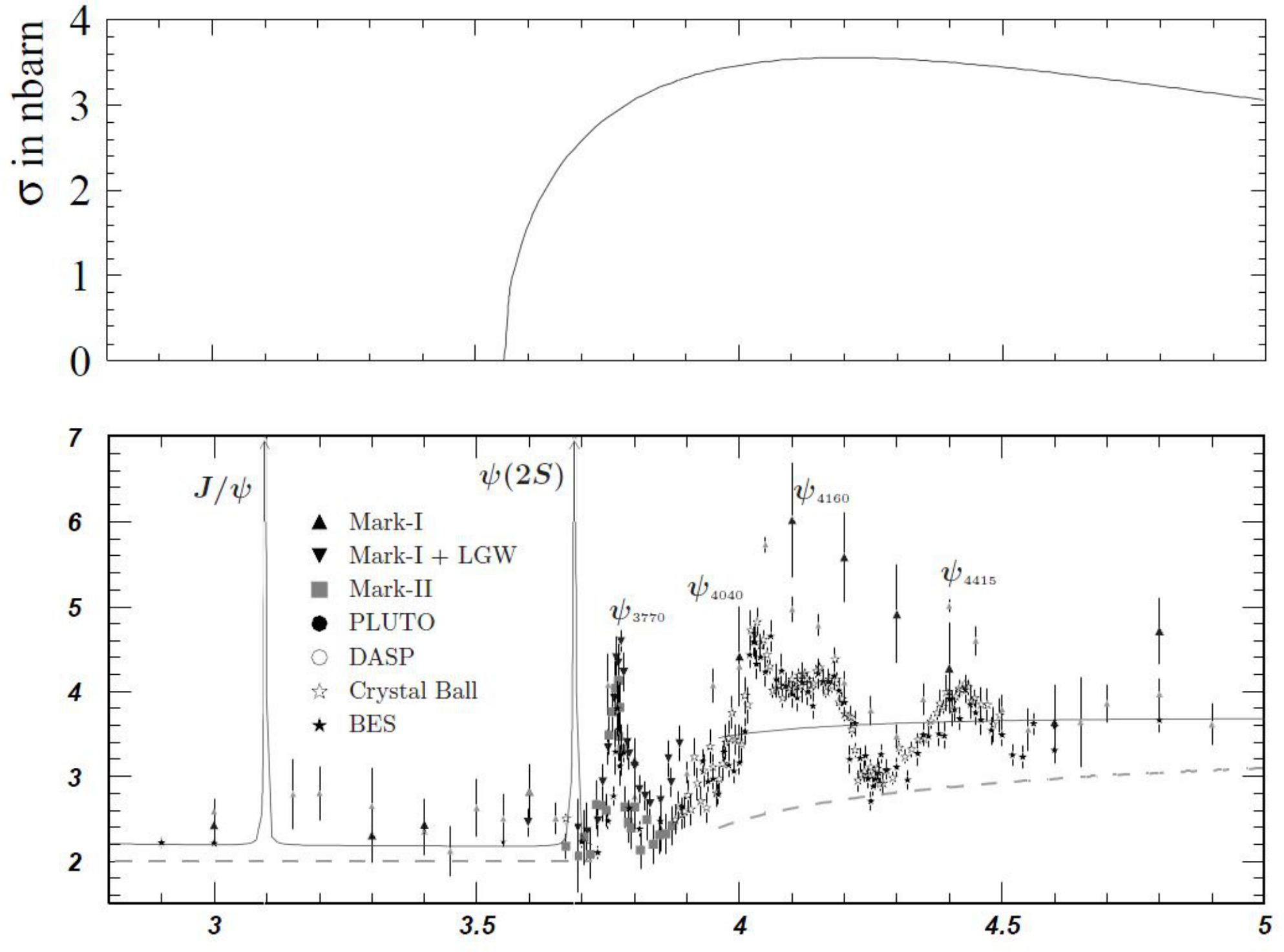}
\caption{\label{cs}
Top: Production cross section for $\tau$ pairs. Bottom: The R-ratio showing
the expected background from quark production (from PDG). Cross section and
$R$ values are shown versus the cms energy in GeV on identical
scales.}
\end{center}
\end{figure}

Figure~\ref{cs} shows the production cross section and $R$ ratio for $\tau$
pairs in the region of BEPCII. The luminosity at the $\tau$ lepton threshold
is about $0.3 \times 10^{33}$~cm$^{-2}$s$^{-1}$. Therefore several hundred
thousand $\tau$ lepton pairs will be produced in a running year (about six
months). In the past years, BESIII had taken about one month data near
$\tau$ lepton pair production threshold: about 150~pb$^{-1}$ has been
obtained. Comparing with the experiments at higher energy regions, such as $B$
factories and LHC, BESIII does not have advantage at statistics. An immense
amount of data have been obtained by high energy experiments, many
measurements are performed pretty precisely. The running energy of BEPCII is
low, the produced $\tau$ leptons are almost stationary, and the hardware
does not allow us to measure the life time of the $\tau$ lepton accurately.

The advantages of BEPCII are the experimental condition: the machine runs
near the threshold of $\tau$ pair production, the background is
pretty simple and the systematic uncertainties can be controlled easily. The
beam energy spread of BEPCII is small, about 1-2 MeV. Moreover, beam
energy measurement system (BEMS) was built to determine the energy and
energy spread accurately. Therefore, BESIII has unique advantage to perform
the $\tau$ mass measurement.

\subsection{Measurement of the $\tau$ mass}

The $\tau$ is one of three charged elementary leptons in nature, and its mass is an important parameter of the Standard Model.
The improvement of accuracy of $\tau$ mass ($m_{\tau}$) is needed in its own right. Listed as follows are the measured
mass values of three leptons according to PDG2012~\cite{PDG2012}:
\begin{equation}
\begin{array}{rrlll}
m_e      =&    0.510998910 &\pm 0.000000013\,\textrm{MeV} &~~~(\delta m_e/m_e &\approx 2.554 \times 10^{-8} )~, \\
m_{\mu}  =&  105.658367\phantom{000} &\pm 0.000004\,\textrm{MeV} &~~~( \delta m_{\mu}/m_{\mu} &\approx 3.786\times
10^{-8} )~,\\
m_{\tau} =& 1776.82\phantom{0000000} &\pm 0.16\,\textrm{MeV} &~~~( \delta m_{\tau}/m_{\tau} &\approx 9.568\times
10^{-5} )~.
\end{array}
\end{equation}
It can be seen that the accuracy of $m_{\tau}$ is almost four orders of magnitude lower than that of the other two
leptons. The accuracy of electron and muon masses is already at the level of $10^{-8}$.

One well-known motivation for obtaining an accurate value of $m_{\tau}$ is found in the test of lepton universality.
Lepton universality, a basic ingredient of the Standard Model, requires that the charged-current gauge coupling
strengths $g_{e}$, $g_{\mu}$, $g_{\tau}$ are identical: $g_e=g_{\mu}=g_{\tau}$. Comparing the electronic branching
fractions of $\tau$ and $\mu$, lepton universality can be tested as:
\begin{eqnarray}
    \left(\frac{g_{\tau}}{g_{\mu}}\right)^{2}=
\frac{\tau_{\mu}}{\tau_{\tau}}\left(\frac{m_{\mu}}{m_{\tau}}\right)^{5}\frac{B(\tau\rightarrow
            e\nu\bar{\nu})}{B(\mu\rightarrow
                e\nu\bar{\nu})}(1+F_{W})(1+F_{\gamma}),
\label{Eq.universaltest}
\end{eqnarray}
where $F_{W}$ and $F_{\gamma}$ are the weak and electromagnetic radiative corrections~\cite{LepUnivCor}. Note that
$(g_{\tau}/g_{\mu})^2$ depends on $m_{\tau}$ to the fifth power.

Furthermore, the precision of $m_{\tau}$ will also restrict the final sensitivity of $m_{\nu_{\tau}}$. The most
sensitive bounds on the mass of the $\nu_{\tau}$ is derived from the analysis of the  invariant-mass spectrum of
semi-hadronic $\tau$ decays. At present, the best limit of $m_{\nu_{\tau}}<18.2$\,MeV/$c^2$ ($95\%$
confidence level) is based on the kinematics of $2939$ ($52$) events of $\tau^{-}\to 2\pi^{-}\pi^{+}\nu_{\tau}$
($\tau^{-}\rightarrow 3\pi^{-}2\pi^{+}(\pi^{-})\nu_{\tau}$)~\cite{mntau}. This method depends on the determination of
the kinematic end point of the mass spectrum, thus, high precision on $m_{\tau}$ is needed.

Another test also depends only on the accuracy of $m_{\tau}$. 
An interesting formula related to the three leptons masses was discovered in 1981~\cite{Koide}:
\begin{equation}
m_{e}+m_{\mu}+m_{\tau}=\frac{2}{3}(\sqrt{m_{e}}+\sqrt{m_{\mu}}+\sqrt{m_{\tau}})^2~.
\label{Eq.koidefml}
\end{equation}
According to the error propagation formula ($f_m$ indicates the difference between the right and left sides of
Eq.~\eqref{Eq.koidefml})
\begin{equation}
\delta f_m =\sqrt{\sum_{k=e,\mu,\tau} \left[ m_k- \frac{2}{3}
(\sum_{i=e,\mu,\tau} \sqrt{m_k m_i}) \right]^2 \cdot \left(\frac{\delta
m_k}{m_k} \right)^2 }
\approx \frac{1}{3} \delta m_{\tau}~,
\end{equation}
which indicates that the test of Eq.~\eqref{Eq.koidefml} depends almost merely on the accuracy of $m_{\tau}$.

\begin{figure}[tbp]
\begin{center}
\begin{minipage}{8cm}
\includegraphics[height=8cm]{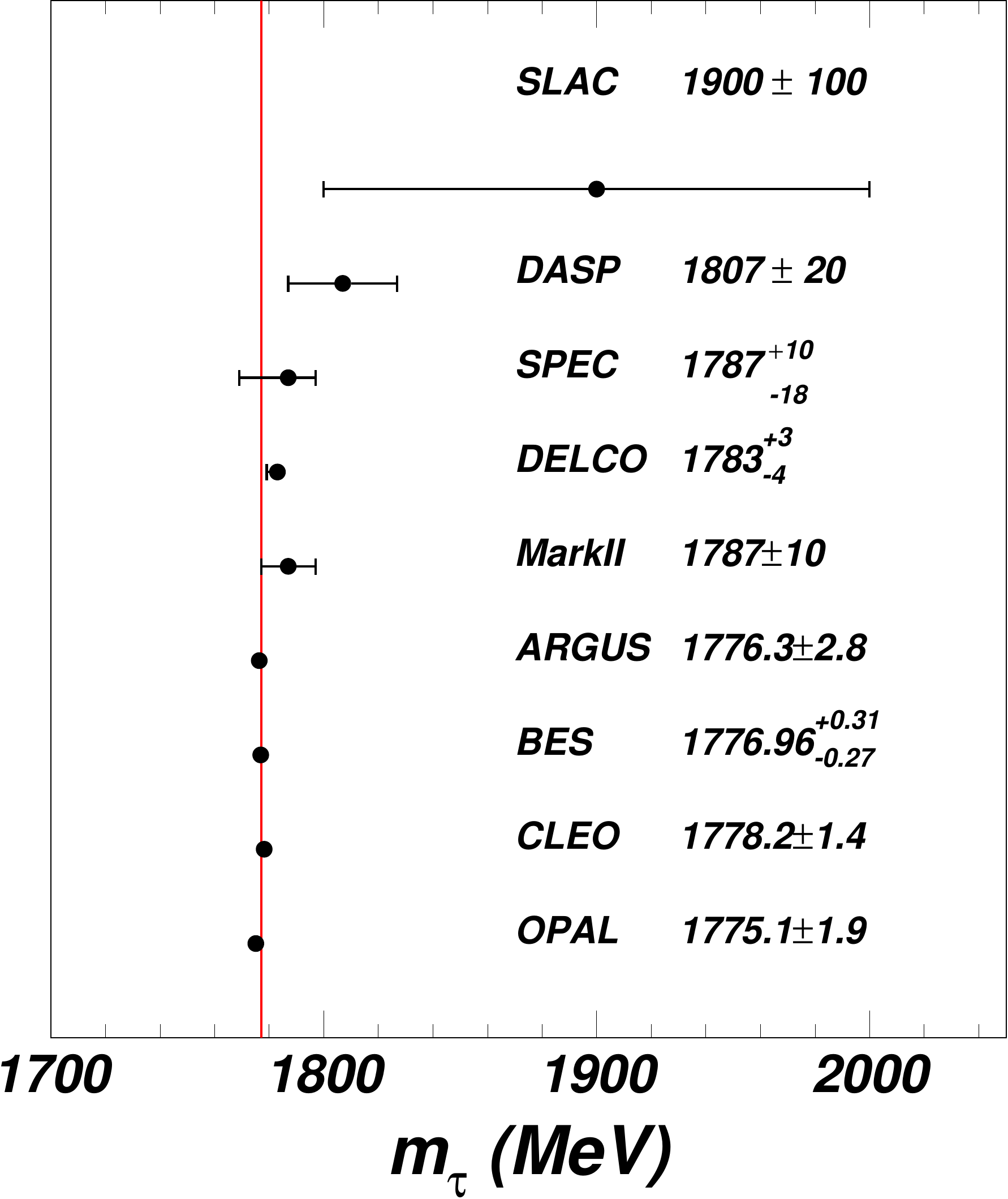}
\center (a) $m_{\tau}$ measured in the 20 century
\end{minipage}
\vskip 0.5cm
\begin{minipage}{8cm}
\includegraphics[height=8cm]{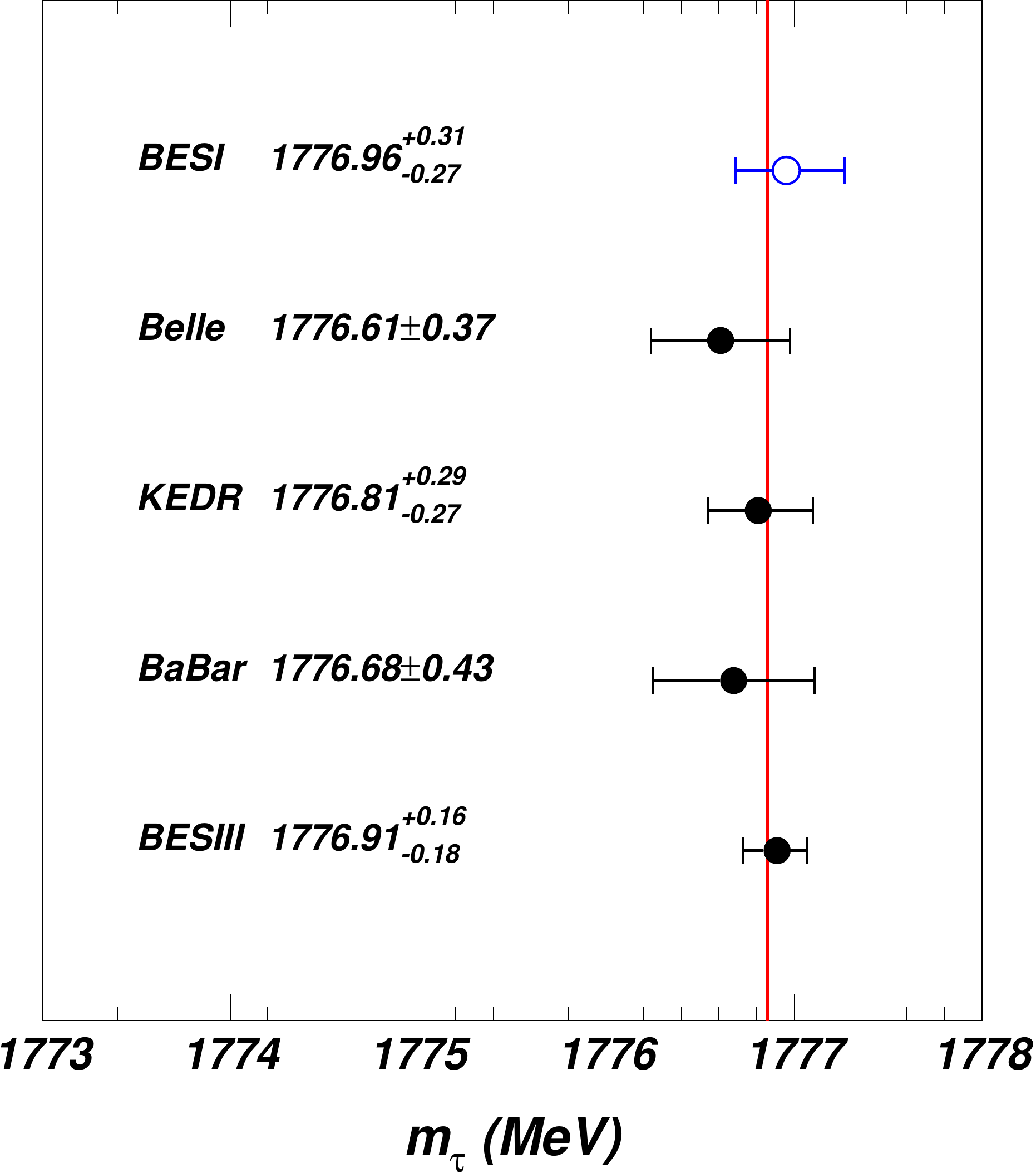}
\center (b) $m_{\tau}$ measured in the 21 century
\end{minipage}
\caption{\label{taumm}$m_{\tau}$ measured in the last and
this century. In (a) the red line indicates the average
value of $m_{\tau}$ in PDG2000~\cite{PDG2000}:
$m_{\tau} = 1777.03 ^{ + 0.30} _{-0.26}$\,MeV/$c^2$.
For comparison, the measured value from BES in 1996 is
also plotted in (b) but with blank blue circle for
distinction. The average value of $m_{\tau}$ in
PDG2015~\cite{PDG2014}: $m_{\tau} = 1776.86 \pm 0.12$\,MeV/$c^2$
is also indicated by the red line in (b). It should be
noted that since PDG1996~\cite{PDG1996,PDG1994}, the
results from experiments performed before 1990 were
removed except for the result of DELCO.}
\end{center}
\end{figure}

The measurement of the $\tau$ mass has a history of more than forty years. In the first
experimental paper on the $\tau$ lepton~\cite{Perl:1975bf}, $m_{\tau}$ is estimated to have a value in the range
from 1.6 to 2.0\,GeV/$c^2$. Since then, many experiments have been performed measurements of
$m_{\tau}$~\cite{Perl:1977se,Brandelik:1977xz,Bartel:1978ii,taudelco,Blocker1982,tauargus,taubes1,taubes2,taubes3,
taucleo1,taucleo2,tauopal,taubelle,taukedr,taubabar,Ablikim:2014uzh}, whose results are displayed in Fig.~\ref{taumm}.

The results of $m_{\tau}$ measurements in the 21$^{\rm st}$ century are summarized in Table~\ref{tab:taums21cn}, where
two results were acquired using the method of pseudo-mass, while other experiments used the method of threshold scan.
For the pseudo-mass method, the huge amount of data acquired at the $B$-factories is employed~\cite{taubelle}. Good
statistical accuracy is achieved, but large systematic uncertainty exists, which is mainly due to the absolute
calibration of the particle momentum. For the threshold-scan method, the value of $m_{\tau}$ was extracted from the
dependence of production cross section on the beam energy. In the KEDR experiment~\cite{taukedr}, both, the resonant
depolarization technique and the Compton backscattering technique~\cite{Bogomyagkov} are used to determine the beam
energy. All these techniques greatly decrease the uncertainty of beam energy.

\begin{table}[tb]
\caption{\label{tab:taums21cn}Measurement results of $m_{\tau}$
in the 21 century.}\centering
\begin{tabular}{l|l|l|l|l} \hline \hline
Measured $m_{\tau}$ (MeV/$c^2$)                 & Year & Exp. Group & Data sample
& Method  \\
       &      &            &
& \\  \hline
$1776.91\pm 0.12^{+0.10}_{-0.13}$  &2014  &BESIII~\cite{Ablikim:2014uzh}
& 23.26\,pb$^{-1}$ & Threshold-scan     \\
$1776.68\pm 0.12\pm 0.41           $   &2009  &BaBar~\cite{taubabar}
& 423\,fb$^{-1}$ & Pseudo-mass \\
$1776.81^{+0.25}_{-0.23}\pm 0.15$  &2007  &KEDR~\cite{taukedr}
& 6.7\,pb$^{-1}$ & Threshold-scan\\
$1776.61\pm 0.13\pm0.35        $   &2007  &Belle~\cite{taubelle}
& 414\,fb$^{-1}$ & Pseudo-mass \\
 \hline \hline
\end{tabular}
%\end{ruledtabular}
\end{table}

Although the accuracy of results from above two methods are at the comparable level, it is obvious that the pseudo-mass
method is already dominated by systematic errors. Additional data taking will not improve the result. For the
threshold-scan method, however, both the statistic and systematic errors still seem to have  room for  further
improvements. Therefore, the BESIII collaboration adopted the threshold-scan method to measure the $m_{\tau}$. In this
approach it is of utmost importance to determine the beam energy and the beam energy spread precisely. For this purpose,
starting from the year 2007, a high accuracy beam energy measurement system (BEMS) located at the north crossing point
(NCP) of BEPCII was designed, constructed, and finally commissioned at the end of
2010~\cite{principle,bems2009,bems2010,bems,bems2}. Two days were spent to perform a scan of the $\psi(3686)$
resonance. The mass difference between the PDG  value in 2010 and the one measured by BEMS is $1\pm36$\,keV, the deviation
of which indicates that the relative accuracy of BEMS is at the level of $2\times 10^{-5}$~\cite{bems}.

Prior to conducting the experiment, a study was carried out using Monte Carlo (MC) simulation and sampling to find out
the scheme that can provide the highest precision on $m_{\tau}$ for a specified period of data taking time or
equivalently for a given integrated luminosity. The main conclusions for the optimization study are summarized as
follow~\cite{Mo:2015gla,wangyk2007,wangyk2009,wangbq2012,wangbq2013}:
\begin{enumerate}
\item For $N$ free parameters, $N$ scan points are sufficient;
\item The optimal position can be obtained by single parameter scan;
\item Luminosity allocation can be determined analytically or by simulation method;
\item The uncertainty of $m_{\tau}$ is proportional to the inverse of square root of luminosity.
\end{enumerate}

By virtue of these conclusions, the optimal scan strategy is
designed as follows: firstly, a scan of the $J/\psi$ resonance (scan points 1-7), then a scan in the vicinity of the
$\tau$-pair threshold (scan points 8-12), finally a scan of the $\psi(3686)$ resonance (scan points 13-19). After that,
repeat (in $\tau$-region, the data are taken only at scan points 9 and 10). The two-round process is designed to
understand the stability of accelerator and detector. The key issue here is to aquire 100\,pb$^{-1}$ data in
$\tau$-region to guarantee an uncertainty of $m_{\tau}$ of less than 0.1\,MeV/$c^2$. The scan plan and parts of the offline
results are summarized in Table~\ref{scanplanresult}.

\begin{table}[tb]
\caption{Scan plan and offline results. Superscript ``plan''
indicates the
planed value for the scan;``exp'' indicates the experimental
measured
value of scan plan; while ``online'' denotes the value from
online record. $E_{\rm beam}$ and ${\cal L}$ represent the beam
energy and integrated luminosity respectively.
($\dagger$: the planed values are listed in last two columns;
$\ddagger$: the measured values are little different from the
expected ones due to the fluctuation of accelerator.)}
\centering
{\begin{tabular}{clccccccc} \hline \hline
Energy  & order &$E^{\rm plan}_{\rm beam}$&${\cal L}^{\rm plan}$ &$E^{\rm exp}_{\rm beam}$
&${\cal L}^{\rm online}$  &
 \multicolumn{3}{c}{$E^{\rm plan}_{\rm beam}$ (MeV)} \\
Region  &      & (MeV)     & (pb$^{-1}$ ) & (MeV)          &  (pb$^{-1}$ )
                                      &                    order
&  $J/\psi$   & $\psi(3686)$ \\ \hline
$ J/\psi$ & 1-7  &$\dagger$  &  $-$      & $\ddagger$  & 32.6
& 1/13 & 1544.0 & 1838.0 \\ \cline{1-6}
$\tau$  & 8    & 1771.0    & 14        & 1769.74     & 25.5
& 2/14 & 1547.8 & 1841.9 \\
        & 9    & 1776.6    & 14+25     & 1776.43     & 42.6
& 3/15 & 1548.2 & 1842.5 \\
        & 10   & 1777.0    & 14+12     & 1776.96     & 27.1
& 4/16 & 1548.6 & 1843.1 \\
        & 11   & 1780.4    & 7         & 1780.18     & 8.3
& 5/17 & 1549.0 & 1843.8 \\
        & 12   & 1792.0    & 14        & 1800.27     & 28.8
& 6/18 & 1549.4 & 1844.5 \\ \cline{1-6}
$\psi(3686)$  & 13-19& $\dagger$ & $-$       & $\ddagger$  & 67.2
& 7/19 & 1552.0 & 1847.0 \\ \hline \hline
\end{tabular}\label{scanplanresult}}
\end{table}

Based on all these full preparations, the BESIII collaboration performed the finer $\tau$ mass scan experiment from
April 14th to May 3rd, 2018. The actual data taking time is around 11.2 days (269 hours). The $J/\psi$ and $\psi(3686)$
resonances were each scanned at seven and nine energy points respectively, and data were collected at five scan
points near $\tau$ pair production threshold with cms energies of 3538.9, 3552.8, 3553.9, 3560.3, and
3599.5\,MeV. The first $\tau$ scan point is below the mass of $\tau$ pair~\cite{PDG2012}, while the other three are
above\footnote{The original two-round process was designed mainly to understand the stability of accelerator and
detector. During the period of $m_{\tau}$ scan, the status of accelerator and detector was fairly good, so two-round
process degenerated into no-circle process.}.

Using these data samples, the final goal of the finer scan is to obtain the $\tau$ lepton mass with high precision,
namely 0.1 MeV, which includes the statistical error and systematical uncertainties. It is an unprecedented accuracy, and as indicated in the Ref.~\cite{moxh2016mpla}, this is a challenging task that needs a considerable effort, energy, and great patience.

So far, the statistical uncertainty of $m_{\tau}$ measurement is roughly equivalent to the systematic error. Only after finding a better way to reduce the systematic error, BESIII would collect more data to improve the precision of $m_{\tau}$ further.

\subsection{Some $\tau$-physics topics at BESIII}

%\subsubsection{Second-class currents process}
As mentioned above, $\tau$ is the only lepton heavy enough to decay into
hadrons. Actually, its partial decay width involving hadrons in the final
state is about 65\%. The hadronic $\tau$ decays turn out to be a beautiful
laboratory for studying the non-perturbative regime of QCD, which will be
useful to understand the hadronization of QCD currents, to study form
factors and to extract resonance parameters.

Rare decays of the $\tau$ leptons are a very promising area because the
interaction is suppressed and the sensitivity to new physics might
eventually be enhanced. In SM, the suppression of rare
$\tau$ decays might be due to several reasons: i) Cabbibo suppression:
strange hadronic final states are suppressed with respect to non-strange
ones since the $|V_{us}|$ element of the CKM matrix enters the description $J^{PC}$
instead of $|V_{ud}|$; ii) phase space suppression: because of the larger
masses of Kaon and $\eta$ mesons in the final state, the phase space should always
be suppressed; iii) second-class currents: in hadronic $\tau$ decays, the
first-class currents have $J^{PC}$ = 0$^{++}$, 0$^{--}$, 1$^{+-}$ or
1$^{-+}$ and are expected to dominate. The second-class currents, which have
$J^{PC}$ = 0$^{+-}$, 0$^{-+}$, 1$^{++}$ or 1$^{--}$, are associated with a
matrix element proportional to the mass difference between up and down
quarks. They vanish in the limit of perfect isospin symmetry, but are not
prohibited by SM, which indicates the branching fractions of such $\tau$
decays at the order of $10^{-5}$.

The $\tau$ lepton provides a clean means to search for second-class
currents, through the decay mode $\tau^- \to \pi^- \eta \nu_{\tau} $. The
$\pi^- \eta$ final state must have either $J^{PC}$ = 0$^{+-}$ or $J^{PC}$ =
1$^{--}$, both of which can only be produced via second-class currents.

The CLEO collaboration analyzed 3.5 fb$^{-1}$ data taken at the
cms $\sqrt{s} = 10.6$ GeV, had produced the most stringent limit on
$\tau^- \to \pi^- \eta \nu_{\tau} $ decays, and set an upper limit of
$\mathcal{B}(\tau^- \to \pi^- \eta \nu_{\tau} ) < 1.4 \times 10^{-4}$ at the 95\%
confidence level~\cite{cleo2nd}. Also the BaBar collaboration analyzed 470
fb$^{-1}$ data taken at the cms $\sqrt{s} = 10.6$ GeV, studied the decay
$\tau^- \to \pi^- \eta \nu_{\tau} $, and set an upper limit of $\mathcal{B}(\tau^-
\to \pi^- \eta \nu_{\tau} ) < 0.99 \times 10^{-4}$ at the 95\% confidence
level~\cite{babar2nd}.

The research on the second-class currents is both promising and
challenging. On the one hand, the energy region of BEPCII is near the $\tau$
pair production threshold, and the background is relatively simple; on the other
hand, since our luminosity is pretty low, even if the $\tau$ data are taken at the
maximum cross section, say 4.25 GeV, to achieve considerable precision,
$10^{-4}$, BESIII has to run for ten years.

%\subsubsection{Three Kaons decay}
The decays of the $\tau$ lepton into three pseudoscalar particles can
provide information on hadronic form factors, the Wess-Zumino anomaly, and
also can be used for studies of CP violation in the leptonic sector. By
studying decays into final states containing three Kaons, it can provide a
direct determination of the strange quark mass and the
Cabibbo-Kobayashi-Maskawa (CKM) matrix element $|V_{us}|$.

The BaBar collaboration~\cite{babar3k} have studied the decay mode $\tau^-
\to K^- K^+ K^- \nu_{\tau} $, and obtained the branching fraction
$\mathcal{B}(\tau^- \to K^- K^+ K^- \nu_{\tau} $) = $(1.58 \pm 0.13 \pm 0.12)
\times 10^{-5}$ by meas of analyzing the integrated luminosity of 342
fb$^{-1}$ data at a cms energy near 10.58 GeV using the BaBar
detector at the SLAC PEP-II asymmetric-energy $\EE$ storage ring. The Belle
collaboration~\cite{bell3k} also measured the branching fraction of this
channel, their result is $(3.29 \pm 0.17 \pm 0.20) \times 10^{-5}$, based on
a data sample of 666 fb$^{-1}$ data collected with the Belle detector at the
KEKB, asymmetric-energy $\EE$ collider of 10.58 GeV. The difference between
the two measurements is larger than 3 standard deviations.

BESIII could perform this measurement if enough data are collected.

\subsection{Measurement of branching fraction of $\psi(3686)\to
\tau^+\tau^-$}

The $\psi(3686)$ provides a unique opportunity to compare the three lepton
generations by studying the leptonic decays
$\psi(3686)\to e^+e^-$, $\mu^+\mu^-$, and $\tau^+\tau^-$. The sequential
lepton hypothesis leads to a relationship between the branching fractions of
these decays, ${\cal B}_{e^+e^-}$, ${\cal B}_{\mu^+\mu^-}$, and ${\cal B}_{\tau^+\tau^-}$ given by

\begin{equation}
\frac{{\cal B}_{e^+e^-}}{v_e (\frac{3}{2}-\frac{1}{2} v_e^2)}=
\frac{{\cal B}_{e^+e^-}}{v_\mu (\frac{3}{2}-\frac{1}{2} v_\mu^2)}=
\frac{{\cal B}_{e^+e^-}}{v_\tau (\frac{3}{2}-\frac{1}{2} v_\tau^2)}
\label{Eq.fracrelation}
\end{equation}
with $v_l=[1-(4m_l^2/M^2_{\psi(3686)})]^{1/2}$, $l=e,\mu,\tau$. Substituting the
mass values for the leptons and the $\psip$ yields
\begin{equation}
{\cal B}_{e^+e^-}={\cal B}_{\mu^+\mu^-}=\frac{{\cal B}_{\tau^+\tau^-}}{0.3885}\equiv {\cal B}_{l^+l^-}.
\label{Eq.fracthree}
\end{equation}
BES once preformed such a study based on 3.96 million $\psi(3686)$ event
sample at the cms $\sqrt{s} = 3686.36$ MeV~\cite{Ablikim:2014uzh}. Now
with BESIII data sample, more detailed study can be expected.

In 2018, BESIII performed a fine $\psi(3686)$ scan at 10 points with
totally 67 pb$^{-1}$ data. Exploiting with this large sample, instead of using
Eq.~\eqref{Eq.fracrelation}, it is possible to measure each branching
fraction separately and test the the relation between them. Further, unlike
BES measurement at single energy point, with lots of scan points, the
interference between the continuum and resonance parts can be measured
directly, and even the phase angle between them. This will be a very technical
and interesting work related to $\tau$ lepton.

\subsection{Mass measurement for some hadrons}
The mass of $\tau$ lepton can be determined precisely by means of BEMS.
However not all masses of hadrons can be determined using BEMS. The details
of BEMS can be found in Refs.~\cite{bems, zhang}. Generally speaking, the
allowed region of beam energy for BEMS should not exceed 2 GeV. On the one
hand, to a high-purity germanium detector, the energy of upper limit for the
Compton backscattered photons is 10 MeV; on the other hand, there is no
suitable radiation sources to calibrate the detector for such high energy
backscattered photons. Moreover, the detection efficiency to high Compton
backscattered photons is pretty low. 
Hence, it is better to measure the masses for hadrons, such as $D_s^*$, $D_{s0}$,
$D_{s1}$, $\Lambda_{c}$, $\Sigma_{c}$ using the invariant mass method if we
have enough data.

\subsection{Discussion}
Experimentally speaking, measurements involving $\tau$ leptons can be divided
into three categories: the mass of $\tau$, the life time of $\tau$ and the
measurements related to the determination of branching fraction of $\tau$
decay. Except for $\tau$ mass measurement, due to the limitation of data
samples, BESIII lacks competitive power in other measurements. 

%% file: QCD/phase.tex
\section{Relative phase in vector charmonium decays}

For many years, there has been evidence of an unexpected phenomenon in the decay of narrow charmonium resonances. A relative phase of $|\Delta\Phi|=90^\circ$ between the strong and EM decay amplitudes of $J/\psi$ is observed~\cite{Ablikim:2018ege}. In case of the $\psip$, the available evidence is less compelling. However, $\psip$ scan data have been collected at \bes3 and are being analyzed, $e.g.$, $e^+e^- \to \mu^+\mu^-, K K, \eta \pi^+\pi^-$ and baryon pairs. Concerning the $\psi(3770)$, the phase difference has been measured in a few cases and $\Delta\Phi = -90^\circ$ has been established~\cite{Seth:2013eaa,Seth:2012nn}.

This phase difference, taking into account the aforementioned values, implies $\Gamma_P = \Gamma_{\rm EM} + \Gamma_{\rm strong}$, with $\Gamma_P$ being the total width of the vector charmonium $P$. This relation might suggest the unconventional hypothesis that the $J/\psi$ is a combination of two resonances, one decaying only through EM processes and another one decaying strongly (see \cite{Ablikim:2018ege}, and references therein).

Currently available data are barely sufficient to determine the absolute value of $\Delta\Phi$ for $J/\psi$ in a limited number of decay modes. The process $\eta \pi^{+} \pi^{-}$ is observed with low statistics in a $J/\psi$ line shape scan of 16 energy points with a total integrated luminosity of about 100\,pb$^{-1}$, where the fit procedure introduces 11\% systematic uncertainty~\cite{Ablikim:2018ege}. With ten times more of available scan data, we could obtain a more precise measurement  of $\Delta\Phi$. In a few of those cases it might already be possible to determine also the sign of the relative phase of EM and strong amplitudes. However, in $J/\psi$ decays into baryon pairs  the ``magnetic'' and ``electronic'' amplitudes might have different phases.

An intriguing situation has been found in $J/\psi\to\pi\pi$. Due to $G$-parity violation a purely strong amplitude should be suppressed. However, the EM amplitude is not sufficient to explain the partial with of $J/\psi\to\pi\pi$. As a remedy, an amplitude has been suggested with two gluons and one virtual photon as an intermediate state~\cite{Ferroli:2016jri}. In order to test the hypothesis, it is most interesting to determine the relative phase between the strong and EM decay amplitudes in the $J/\psi\to\pi\pi$ decay. The currently available data at \bes3, however, do not allow for such investigations yet.

In conclusion, much more data are needed to settle the question of the phase difference between the strong and EM decay amplitudes of vector charmonium states. If the aforementioned vector charmonium double nature is confirmed, something important has been missed in the understanding of the Zweig rule (see Ref.~\cite{2015zhu} and references therein).

%% file: QCD/meson.tex
% phi(2170)
\section{Study of $\bf{\phi(2170)}$ with the energy scan method}
\label{phi2170}

Quarkonia provide a unique platform to study QCD. Substantial progress has been made over the recent years from the investigation of charmonia ($c\bar{c}$) and bottomonia ($b\bar{b}$). A plethora of interesting new hadronic states were found. New types of hadronic matter, such as hybrids, multiquark states, and hadronic molecules with (hidden) charm and bottom quarks are considered in the interpretation~\cite{zhu2008,chen2016,olsen2018}. 

\begin{figure}[tbp]
 \centerline{\includegraphics[width=0.95\textwidth]{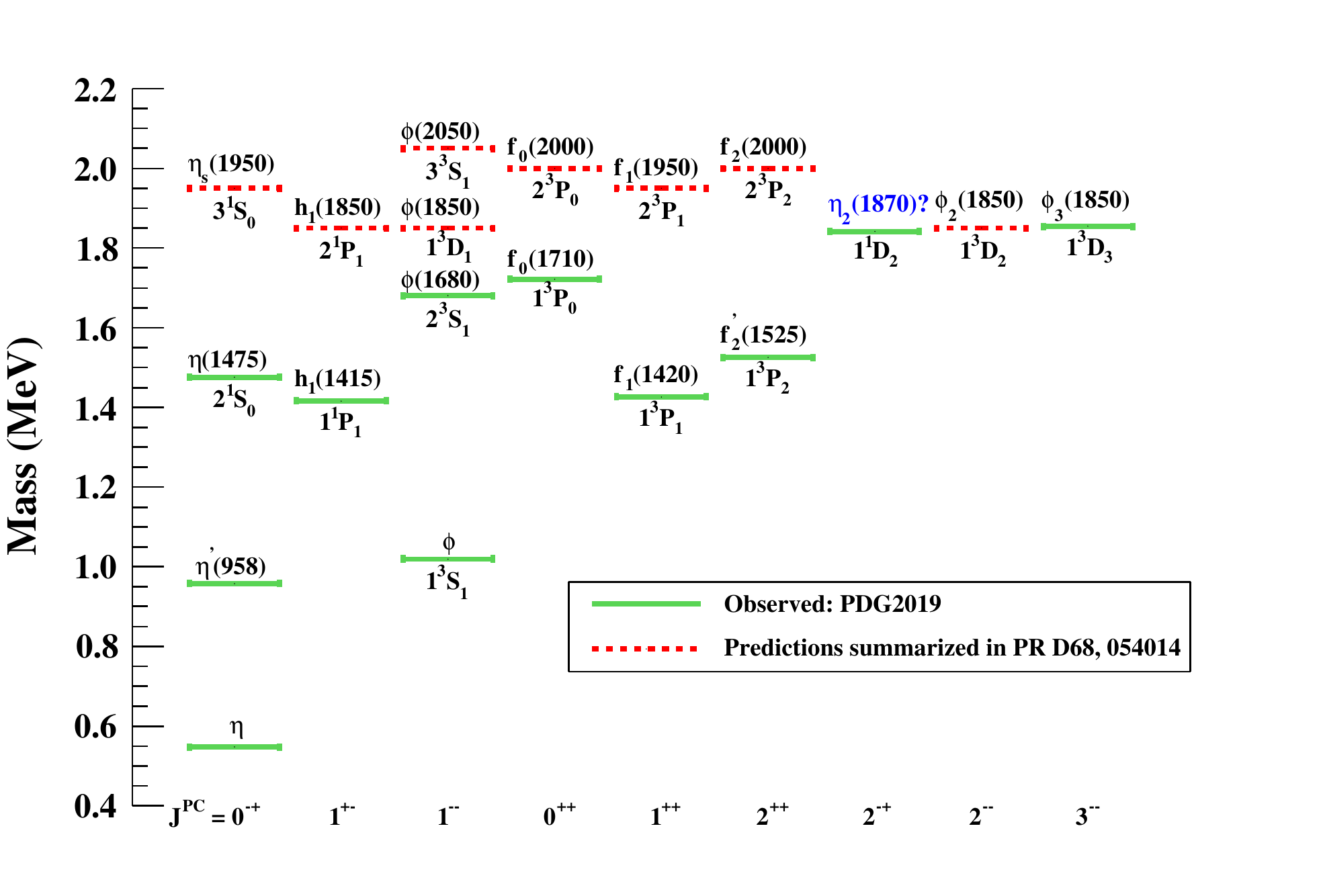}}
 \caption{\label{stangeonium} The strangeonium spectroscopy. The red dotted line corresponds to predicted strangeonium states summarized in 
Ref.~\cite{strange}, and the green line corresponds to observed $s\bar{s}$ resonances, taken from  PDG~\cite{pdg2018}.}
\end{figure}

The multitude of results from heavier quarkonia leads to the obvious question whether similar states should exists in the strange sector. However, experimental evidence for a rich spectrum of strangeonium ($s\bar{s}$) or new types of hadronic matter with strange quarks is scarce. Figure~\ref{stangeonium} shows predicted strangeonium states~\cite{strange} with identified $s\bar{s}$ resonances~\cite{pdg2018}. Only 10 probable $s\bar{s}$ resonances out of the 22 expected below 2.2\,GeV are established. A candidate for an exotic type of hadronic matter containing strange quarks is the $\phi(2170)$. At \bes3 the $\phi(2170)$ can be studied as an intermediate state in charmonium decays, exploiting the unique statistics of the $J/\psi$ data set, and by performing a dedicated energy scan around the mass of the $\phi(2170)$. Both methods provide unique opportunities in terms of precision and accuracy necessary to understand the nature of  the $\phi(2170)$.

\begin{figure}[tbp]
 \centerline{\includegraphics[height=6cm]{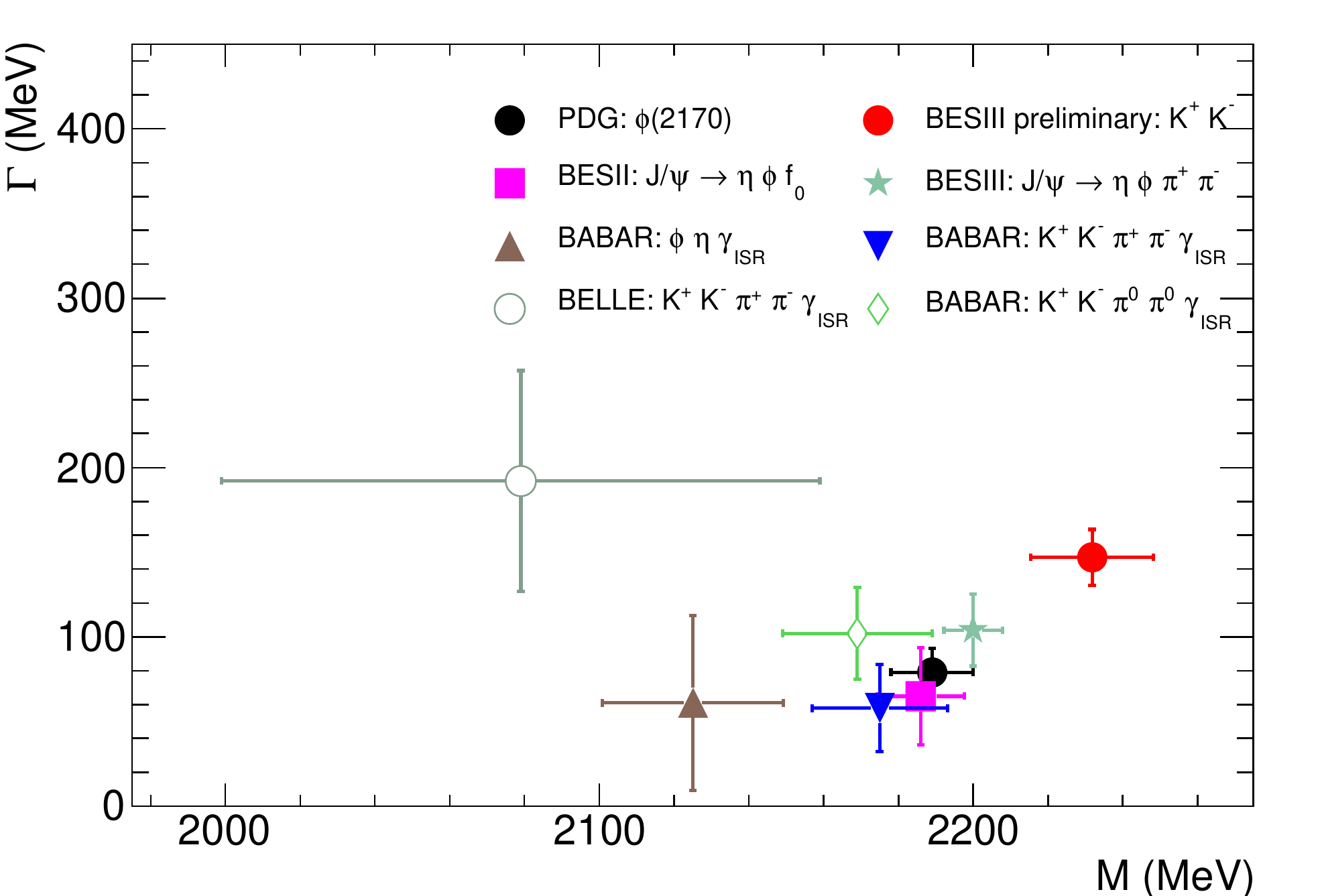}}
 \caption{\label{phi2170mass}Mass and width of the $\phi(2170)$.}
\end{figure}

The $\phi(2170)$, previously referred to as  the Y(2175), has been observed in $e^+ e^- \to \phi f_0(980)$ at $B$ factories using the ISR method~\cite{2170babar,2170belle}, in the charmonium decay $J/\psi \to \eta\phi f_0(980)$~\cite{2017bes}, and in $e^+ e^- \to \eta \phi f_0(980)$ at cms energies between 3.7 and 4.6\,GeV~\cite{2017bes3}. Theorists explain it as a traditional $3 ^3S_1 s \bar{s}$ or $2 ^3D_1 s \bar{s}$ state~\cite{2017ding,2017wang,2017afonin,2019pang}, as a $1^{--} s\bar{s}g$ hybrid ~\cite{2017ding2}, as a tetraquark state~\cite{2017wang2,2017chen,2017drenska,2019ke}, as a $\Lambda \bar{\Lambda}(^3S_1)$ bound state~\cite{2017zhao,2017deng,2017dong}, and as a $\phi K\bar{K}$ resonance state~\cite{2017oset}. According to PDG~\cite{pdg2018}, the $\phi(2170)$ has been observed in the final states $\phi f_0(980)$, $\phi\eta$, $\phi\pi\pi$, and $K^+K^- f_0(980)$. For the final state $K^{*0}\bar{K}^{*0}$ only an upper limit has been reported~\cite{2010bes}. Hence, the number of experimentally established decay modes of $\phi(2170)$ is limited. Additionally, as shown in Fig.~\ref{phi2170mass}, there is a large scatter of the measured values of mass and width of the $\phi(2170)$. So far, none of above theory models are either estimated or ruled out by experimental results. 

Describing the $\phi(2170)$ as a $3 ^3S_1 s \bar{s}$ or $2 ^3D_1 s \bar{s}$ state favors the decay modes $K^*{\bar K}^*$, $K{\bar K}(1460)+ c.c.$, and $\eta h_1(1380)$, while these modes are forbidden for a $1^{--} s\bar{s}g$ state. Instead, the $K{\bar K}_1(1400)+c.c.$ and $K{\bar K}_1(1270)+c.c.$ are favored for $1^{--} s\bar{s}g$ state~\cite{strange,2017ding,2017ding2}. However, all of the four previously mentioned decays produce the same final state: $K{\bar K}\pi\pi$. Originally, the $\phi(2170)$ was observed in the $\phi\pi\pi$ final state, an intermediate state of $K{\bar K}\pi\pi$. The fitted $\phi(2170)$ resonance parameters $m=(2079 \pm 13)$\,MeV/c$^2$ and $\Gamma=(192 \pm
23)$\,MeV in the $\phi\pi\pi$ mode are slightly lower than $m=(2163 \pm 32)$\,MeV/c$^2$ and $\Gamma=(125 \pm
40)$\,MeV in the $\phi f_0(980)$ mode~\cite{2170belle}, which is consistent with available measurements in the $\phi f_0(980)$ mode (see Fig.~\ref{phi2170mass}). A PWA analysis of $e^+e^- \to K{\bar K}\pi\pi$ at BESIII is useful to distinguish $s\bar{s}$ and $s\bar{s}g$, and understand results obtained from the $\phi\pi\pi$ and $\phi f_0(980)$ modes.

The $e^+ e^- \to \phi\eta$ and $e^+ e^- \to \phi\eta^\prime$ processes are well suited to study excited $\phi$ states, and useful to estimate mass and width of strange quarkonia due to the OZI rule. The main decay modes in tetraquark pictures~\cite{2017drenska} are $\phi\eta$ and $\phi\eta^\prime$ modes; the predicted decay width of a $1^{--} s\bar{s}g$ state to $\phi\eta$ is much larger than that of $\phi\eta^\prime$~\cite{2017ding2,1999page}, whereas $\phi\eta$ is forbidden for $2 ^3D_1 s \bar{s}$ states~\cite{2017ding}. However, the uncertainty of the cross section of $\phi\eta$ around $\sqrt{s} = 2.2$\,GeV measured by BaBar is at the 50$\%$ level~\cite{phieta2007}. A clear signal of $\phi \eta^\prime$ is observed by BaBar in Ref.~\cite{phietaprime}. A precise measurement of $\phi \eta$ and $\phi \eta^{\prime}$ modes at BESIII is useful to distinguish tetraquark, hybrid and $2 ^3D_1$ pictures of the $\phi(2170)$. Besides $\phi(2170)$, there is another possible structure for $\phi f_0(980)$ and $K^+K^-f_0(980)$ around 2.4\,GeV~\cite{2170babar,2170belle,2017bes}. BaBar determined its mass and width as ($2.37 \pm 0.07$)\,GeV/$c^2$ and ($77 \pm 65)$\,MeV, while Ref.~\cite{cpcshen} obtained ($2436 \pm 34$)\,MeV/$c^2$ and ($99 \pm 105)$\,MeV, respectively, whose statistical significance is smaller than $3\sigma$. A QCD sum rule also predicts two $J^{PC} = 1^{--}$ resonances $2.34 \pm 0.17$\,GeV and $2.41 \pm 0.25$\,GeV, both around 2.4\,GeV~\cite{2018chen}. In the $\phi K^+ K^-$ mode two broad structures were observed at 2.3\,GeV and 2.7\,GeV,
respectively, albeit in a different $m(K^+K^-)$ range~\cite{2170babar}. A scan in 40\,MeV steps in the range of [2.35, 2.83]\,GeV with integrated luminosities of $20\,\textrm{pb}^{-1}$ at each energy point, is called for to search for (possible) new resonances. 

Last but not least, Refs.~\cite{2017wang,milena2017} predict a broad resonance with mass around 1.9\,GeV regarded as the $1D$ state of the $\phi$ family. A new scan measurement between 1.8 and 2.0\,GeV is needed to search for $1 ^3D_1 s \bar{s}$ (shown in Fig.~\ref{stangeonium}), to investigate the threshold production of nucleon pairs and a structure around 1.9\,GeV, shown in Ref.~\cite{Lichard2018}. The scan should be performed with 10\,MeV steps between 1.8 and 2.0\,GeV and an integrated luminosity $4\,\textrm{pb}^{-1}$ at each energy point.  

Based on collected data for $R$ values in the energy range between 2.0 
and 3.08\,GeV,  we already studied the final states $K^+K^-$, $\phi K^+K^-$, and $K^+K^-\pi^0\pi^0$, and the investigation of further possible decay modes is ongoing. However, more data at additional energy points are needed, which are listed in Table \ref{tqcddata}.

%% file: QCD/data.tex
\section{Prospects}

The different physics aspects addressed in this chapter cannot be studied from a single data set. While studies based on 
ISR or two-photon collisions require large data samples, taken at masses well above the actual mass range of interest, 
precise studies of line shapes or threshold effects are best performed on several smaller data sets in an energy scan.

The existing ISR studies were mostly performed on $2.9~\textrm{fb}^{-1}$ taken at the $\psi(3770)$ peak. The results
are already competitive with previous measurements of hadronic cross sections. However, one of the key issues, the
hadronic contributions to $a_\mu$, is still unsettled. The published BESIII result on the dominating $\pi^+\pi^-$
contribution is currently dominated by systematics, where the largest contributions stem from the uncertainties of the
luminosity determination and the theoretical uncertainty of the radiator function. An alternative approach to the 
normalization of the cross section can reduce the systematic uncertainty. Using the ratio to the muon cross section
cancels the two uncertainties mentioned above. However, due to an insufficient muon yield in the current data set, the 
statistical uncertainty becomes dominant. Based on the performance of the published work, it is estimated that an 
additional data set of $20~\textrm{fb}^{-1}$, taken at the $\psi(3770)$ peak, 
and data set listed in Chapter 3 for XYZ physics,
will allow to collect sufficient 
statistics to perform the normalization of the hadronic cross sections with respect to the muon yield. The systematic 
uncertainty of the pion form factor measurement could be reduced to $\mathcal{O}(0.5\%)$, making the BESIII result not 
only comparable, but competitive to the KLOE, the BaBar, and the announced CMD-3 measurements as well as potential 
future analyses at Belle~II.

At the same time, the data can be used to study TFFs in two-photon collisions. While $\pi^0$, $\eta$, $\eta^\prime$, as 
well as pion pairs can be studied very well in a single-tag measurement using the existing data, the statistics is 
rather scarce for higher mass resonances, or a double-tagged measurement of the lightest pseudoscalar mesons. However, 
the latter two are of special interest in order to understand the hadronic light-by-light contribution to $a_\mu$ at the
level of the new direct measurements at FNAL and J-PARC. An additional data set of $20
~\textrm{fb}^{-1}$ at 
$\sqrt{s}$=3.773 GeV is most beneficial for the TFF measurements at BESIII for two reasons, On the one 
hand it increases the statistics at large invariant masses of the produced hadronic systems, and on the other hand it 
will allow to perform the first measurement of the doubly-virtual TFF, in the energy region most relevant for the 
$a_\mu$ calculations, with sufficient statistical precision to scrutinize hadronic models and provide valuable input to 
the new, data-driven theory approaches to $a_\mu^{\rm hLBL}$.

The scan measurements performed at BESIII have successfully allowed to determine baryon EM FFs at the respective 
thresholds and revealed puzzling features. In order to shed more light on these aspects, further detailed studies on 
enhanced data sets around the nucleon threshold at cms energies between $\sqrt{s} = (1.8, 2.0)$ GeV, the  
$\Lambda \bar{\Lambda}$ threshold at 2.2 GeV, and the $\Lambda_c \bar{\Lambda}_c$ threshold at 4.58 GeV are necessary.

So far, most emphasis has been put on the investigation of EM FFs of nucleons and the $\Lambda$ hyperon. An additional 
integrated luminosity of 100 pb$^{-1}$ at $\approx$ 2.5 GeV allows not only precision studies of the 
$\Lambda$, but also of $\Sigma^0$ with 6\% statistical
uncertainty, and $\Sigma^\pm$ with 4\% statistical
uncertainty, including the measurement of their effective form factors, their 
$|G_E/G_M|$ ratios, and the phase angle between $G_E$ and $G_M$ for $\Lambda$ and $\Sigma^+$ hyperons.

A high-statistics data set taken at 2.2 GeV in order to improve the understanding of the $\Lambda \bar{\Lambda}$ 
threshold behavior is also beneficial for studies of the $\phi(2170)$. It would allow us to perform PWAs of the $\phi\pi\pi$ 
final state. Around 2.4 GeV, a structure has been observed in the $\phi\pi\pi$ system. The existing data taken at 
BESIII at this energy suffer from small statistics. In order to shed light on the nature of this structure, it is 
necessary to collect more data with 20 ${\rm pb^{-1}}$ for each energy point around 2.4 GeV with 8\% uncertainty. 

In summary, Table \ref{tqcddata} summarizes proposed data in this chapter.

\begin{table}[tp]
{\centering \small
\caption{\label{tqcddata} Proposed data for $\tau$-QCD study. }
\begin{tabular}{l|l|l|l|l} \hline \hline
Energy        & Physics highlight     & Current data & Expected final data &
time (day) \\ \hline
1.8-2.0 GeV   & $R$, nucleon, resonances &  N/A
& 96 pb$^{-1}$ at 23 points & 66 \\  \hline

around 2.2324 GeV & $\Lambda \bar{\Lambda}$ threshold & one point 
& 40 pb$^{-1}$ at 4 points & 17 \\  \hline

2.35-2.83 GeV & $R$ \& resonances &  few points 
& 260 pb$^{-1}$ at 13 points & 60 \\  \hline

2.5 GeV       & hyperon  & 1 pb$^{-1}$ data
& 100 pb$^{-1}$ & 26 \\  \hline 

$J/\psi$ scan & phase   & 100 pb$^{-1}$ data & 1000 pb$^{-1}$ & 150 \\
\hline

\end{tabular}
}
\end{table}

%% file: Charm/charm.tex
\chapter[Charm physics]{Charm Physics}
\label{chapter:charm}

\input{Charm/charm_intro.tex}

\input{Charm/charm_meson_leptonic_decay.tex}
\input{Charm/charm_meson_hadronic_decay.tex}

\input{Charm/charmed_baryon.tex}
%\clearpage
\input{Charm/charmed_baryon_em.tex}

%\section{Summary}
%\label{sec:char_sum}
\input{Charm/charm_summary.tex}
\input{Charm/charm_bib.tex}

%% file: Charm/charm_intro.tex
\section{Introduction}

\begin{figure}[tp]
  \centering
  \subfigure[]{\includegraphics[width=3.0in]{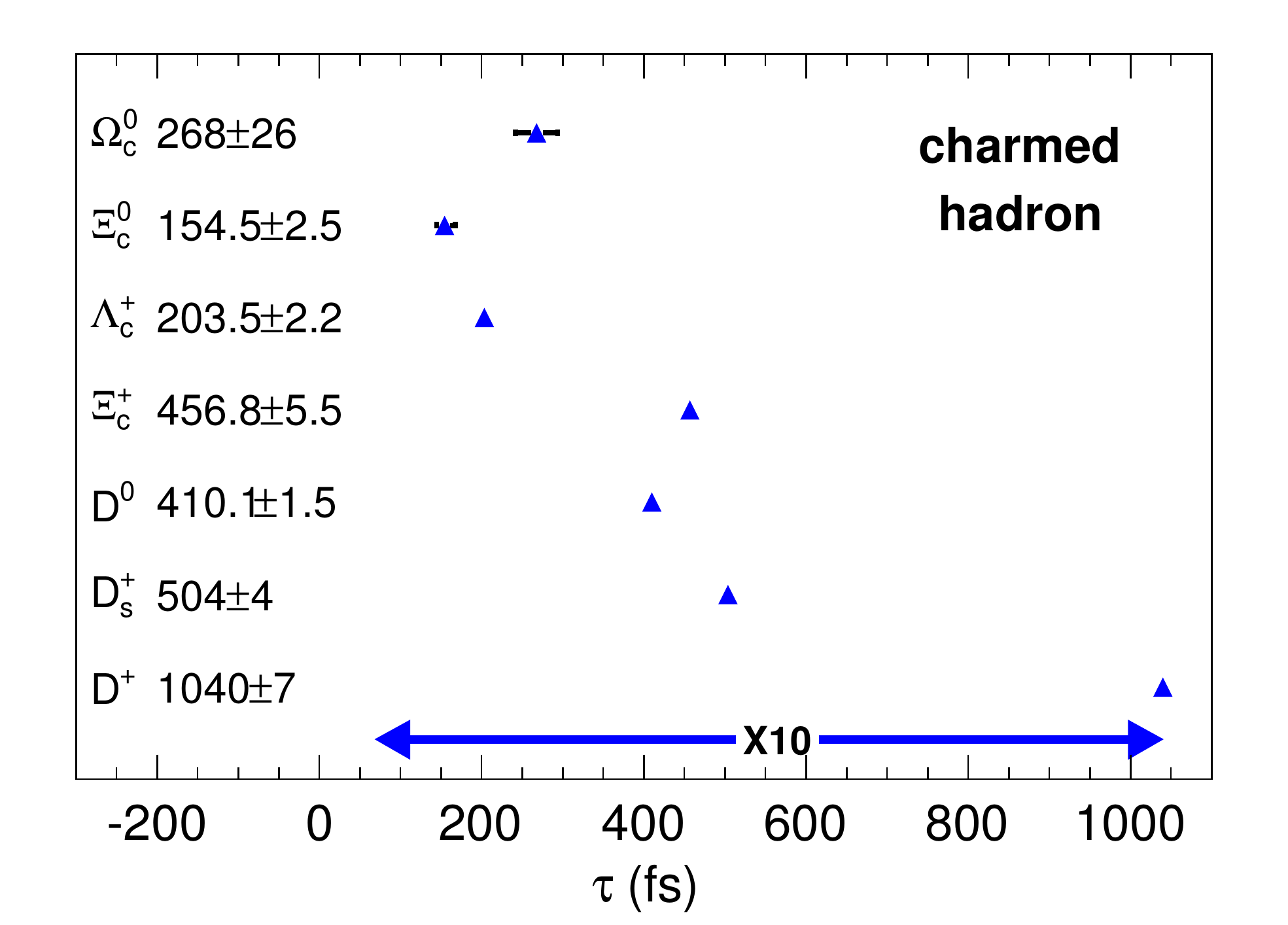}}
  \subfigure[]{\includegraphics[width=3.0in]{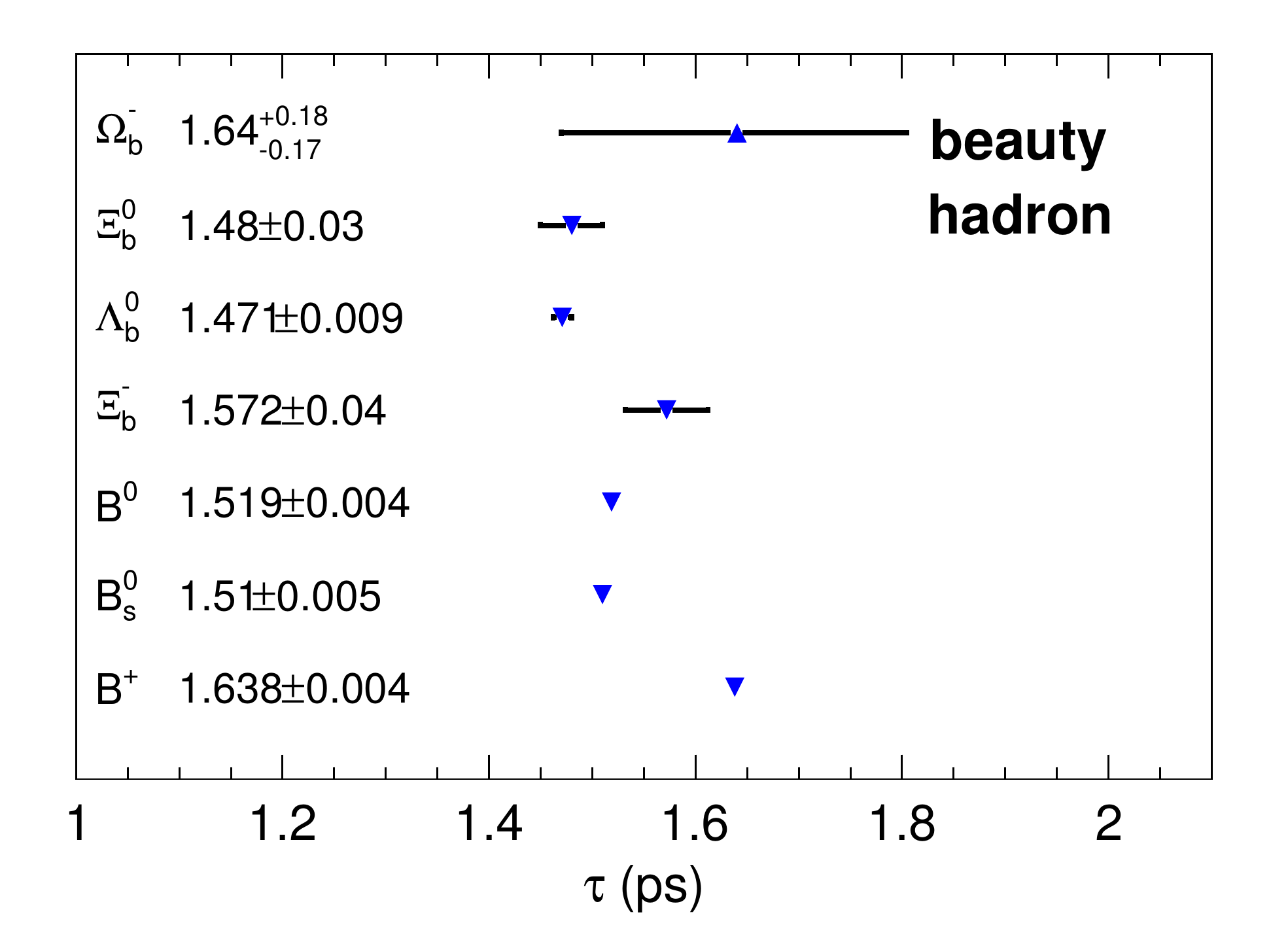}}
  \caption{(a) Comparison of the lifetimes for the charmed hadrons~\cite{pdg2016}. 
For the lifetimes of $\Omega_c^0$, $\Xi^{0}_{c}$ and $\Xi^{+}_{c}$, we take the recent measurement from LHCb~\cite{Aaij:2018dso}. (b) Comparison of the lifetimes for the beauty hadrons.
In the left figure, ``10$\times$" means that lifetimes differ by about one order of magnitude.}
\label{fig:charm-intro-life}
\end{figure}

The ground states of charmed hadrons, $\egeg$, $D^{0(+)}$, $D^+_s$, and $\Lambda_c^+$,
can only decay weakly and so precision studies of these charm decays provide important constraints on the weak interaction~\cite{Asner:2008nq}.
Furthermore, as the strong force is always involved in the decays of 
the charmed hadrons and the formation of the final-state hadrons,
precise measurements of the decay properties allow tests of non-perturbative quantum chromodynamics (QCD) calculations. The decay rates of the ground states of charmed hadrons are dominated by the weak decay of the charm quark. For comparison, the lifetimes of the charmed hadrons are shown in Fig.~\ref{fig:charm-intro-life}(a); it is 
surprising that their individual lifetimes differ by up to a factor of 10. As shown in Fig.~\ref{fig:charm-intro-life}(b), the ratio of $\tau(D^+)/\tau(D^0) = 2.54\pm 0.01$ is very different from that for the corresponding lifetimes in the beauty sector, $\tau(B^+)/\tau(B^0) = 1.076\pm 0.004$.  From these observations one can infer that deviations of the lifetime ratios from unity decrease with increasing heavy-flavor quark mass $m_Q$. Heavy flavor decays thus constitute an intriguing lab to study QCD. According to the data, the non-perturbative effects in the decays of the charmed hadrons are much more important than those in the beauty sector.
Comprehensive studies of the decays of charmed mesons and baryons will play an essential  role in furthering our understanding of strong interactions.  In particular, understanding the different decay mechanisms, such as weak-annihilation, $W$-exchange and final-state scattering, is essential for developing a complete theory of charmed hadron decays.  Therefore, the data at the BESIII experiment will have a leading role in understanding non-perturbative QCD.

BEPCII/BESIII produces charmed hadrons near their mass threshold; this allows
exclusive reconstruction of their decay products with well-determined kinematics.
Up to now, BESIII has collected data corresponding to the integrated luminosities of 2.9~fb$^{-1}$, 0.5~fb$^{-1}$, 3.2~fb$^{-1}$, and 0.6~fb$^{-1}$ at $\sqrt s=$3.773, 4.009, 4.178, and 4.600~GeV,
respectively, as well as data at $\sqrt s=$ 4.23, 4.26, and 4.36~GeV. Based on these data sets,
many world-leading results have been published. These include the first absolute branching fraction (BF) measurements of the $\Lambda_c^+$ baryon to hadronic and semi-leptonic (SL) final states, which are important milestones in the investigation of the charmed baryon sector;
the most accurate measurements of the Cabibbo-Kobayashi-Maskawa (CKM) matrix elements $|V_{cs}|$ and $|V_{cd}|$, which are essential inputs in tests of the CKM matrix unitarity; and the most precise measurements of the decay constants in leptonic decays, as well as the form factors in SL decays,  which are crucial measurements with which to calibrate LQCD for heavy quark studies.
Also, it is important to test lepton-flavor universality (LFU) by using the leptonic and SL charmed hadron decays.

However, improved knowledge of these charmed hadron decays is required to match the significantly improved LQCD calculations and to better understand the strong-force dynamics in the charm region. In addition, more precise measurements of the strong-phase difference between $D^{0}$ and $\bar{D}^{0}$ decays to final states used to measure the $\gamma$ angle (also known as $\phi_3$) of the CKM unitary triangle in $B^{+}\to D K^{+}$ decays are necessary ($D$ denotes either $D^{0}$ or $\bar{D}^{0}$); such improved measurements will prevent the $\gamma$ measurements at LHCb and Belle II from becoming systematically limited due to neutral $D$ strong-phase uncertainties.

This section describes the future pursuit of this rich charm programme. Specifically, we focus our discussions on the most important measurements that can be made with the proposed future data taking:
$D^{0(+)}$, $D^+_s$, and $\Lambda_c^+$ samples accumulated at $\sqrt s=$ 3.773, 4.178, and 4.64~GeV corresponding to integrated luminosities of 20~fb$^{-1}$, 6~fb$^{-1}$, and 5~fb$^{-1}$, respectively.
Furthermore, it will be interesting to study the baryons  $\Lambda_c$ and $\Sigma_c$ for the first time at threshold and at higher cms energies, for example, $e^+e^- \to \Lambda_c^+ \bar{\Sigma}_c^-$ and $\Lambda_c^+ \bar{\Sigma}_c \pi$. We also discuss possible energy upgrade of the BEPCII up to 5 GeV, so that the thresholds will be open for baryon pair productions of $e^+e^- \to \Sigma_c \bar{\Sigma}_c$ and $\Xi_c \bar{\Xi}_c$.

%% file: Charm/charm_meson_leptonic_decay.tex
\section{$D^{0(+)}$ and $D^+_s$ physics}
\label{sec:c_meson}

\subsection{Leptonic decays}
\label{subsec:pure-lep}

In the standard model (SM), the partial widths of the leptonic decays
$D_{(s)}^+ \to \ell^+\nu_\ell$ can be written as
\begin{equation}
\Gamma(D_{(s)}^+ \to \ell^+\nu_\ell)=
     \frac{G^2_F f^2_{D_{(s)}^+}} {8\pi}
      \mid V_{cd(s)} \mid^2
      m^2_\ell m_{D_{(s)}^+}
    \left (1- \frac{m^2_\ell}{m^2_{D_{(s)}^+}}\right )^2,
\label{eq01}
\end{equation}
\noindent
where $G_F$ is the Fermi coupling constant,
$f_{D_{(s)}^+}$ is the $D_{(s)}^+$ decay constant,
$|V_{cd(s)}|$ is the CKM matrix element~\cite{pdg2016}, and
$m_\ell$ $[m_{D_{(s)}^+}]$ is the lepton [$D_{(s)}^+$ meson] mass.
Using the lifetimes and the measured BFs of these decays,
one can determine the product $f_{D_{(s)}^+}|V_{cd(s)}|$.
By taking as input $f_{D_{(s)}^+}$ as calculated in LQCD, the value of $|V_{cd(s)}|$ can then be obtained. Alternatively, $|V_{cd(s)}|$ can be taken from
global fits to other CKM matrix elements and $f_{D_{(s)}^+}$ instead can be determined.

With the data sets in hand,
\bes3 has reported the improved measurements of the BFs of $D^+\to \mu^+\nu_\mu$,
$D^+\to \tau^+\nu_\tau$, and $D^+_s\to \mu^+\nu_\mu$,
as well as $f_{D^+_{(s)}}$ and $|V_{cs(d)}|$~\cite{bes3_muv,bes3_Ds_muv, Ablikim:2019rpl}.
The achieved precision is summarized in Table~\ref{tab:pure_LP}.
Experimental studies of
$D^+_s\to \tau^+\nu_\tau$ with $\tau^+\to\pi^+\nu_\tau$, $\tau^+\to e^+\nu_\tau\nu_e$,
$\tau^+\to\mu^+\nu_\tau\nu_\mu$, and $\tau^+\to \rho^+\nu_\tau$ are still ongoing.
Based on existing measurements at CLEO-c~\cite{cleo_dsmuv,cleo_dstauv2,cleo_dstauv3},
BaBar~\cite{babar_lv}, and Belle~\cite{belle_lv},
the expected signal yield of $D^+_s\to \tau^+\nu_\tau$ will be larger than
that of $D^+_s\to \mu^+\nu_\mu$, but will have more background and significant systematic uncertainties.
After weighting the measurements performed with different $\tau^+$ decays,
the sensitivity of the result for $D^+_s\to \tau^+\nu_\tau$
will be comparable to that made with $D^+_s\to \mu^+\nu_\mu$.
Thus, the measurement precision with $D^+_s\to \tau^+\nu_\tau$
can be estimated, as summarized in Table~\ref{tab:pure_LP}.

There are four reasons why the improved measurements of 
$D^+_{(s)}\to\ell^+\nu_\ell$ with 20 fb$^{-1}$ of data at 3.773 GeV
and 6 fb$^{-1}$ of data at 4.178 GeV are desirable.

\begin{sidewaystable}
\centering
\caption{\label{tab:pure_LP}\small
Expected precision of measurements of $D^+_{(s)}\to \ell^+\nu_\ell$
and those at Belle II.
For BESIII, some systematic uncertainties, which are statistically limited by control
samples, are expected to be improved. The total systematic uncertainty is
assumed to be comparable to the corresponding statistical uncertainty.
For $D^+\to \tau^+\nu_\tau$, only $\tau^+\to\pi\nu$ is used currently. 
However, more $\tau^+$ decay channels could be used.
For $D^+_{s}\to \tau^+\nu_\tau$, the expectation is based on 
$\tau^+\to \pi\nu$, $e\nu\nu$, $\mu\nu\nu$, and $\rho\nu$.
Considering that the LQCD uncertainty of $f_{D^+}$
has been reduced from 1.9\% to 0.2\%,
the $|V_{cd}|$ measured at \bes3 has been re-calculated, and is marked with $^*$.
Preliminary results are marked with $^{\dagger}$.
For Belle II, we assume that the systematic uncertainties can be reduced by
a factor of 2 compared to Belle's results, and the systematic uncertainty
for $D^+\to \mu^+\nu_\mu$ is the same as that of $D^+_s\to \mu^+\nu_\mu$.
} \vspace{0.1cm}
\small
\begin{tabular}{lcccc} \hline\hline
\multicolumn{1}{c}{} &\bes3 & \bes3 & Belle
& Belle II \\ \hline
\multicolumn{1}{c}{Luminosity} &2.9 fb$^{-1}$ at 3.773 GeV & 20 fb$^{-1}$ at 3.773 GeV & 1 ab$^{-1}$ at $\Upsilon(nS)$
& 50 ab$^{-1}$ at $\Upsilon(nS)$ \\ \hline
${\mathcal B}(D^+\to \mu^+\nu_\mu)$ & $5.1\%_{\rm stat.}\,1.6\%_{\rm syst.}$~\cite{bes3_muv} &$1.9\%_{\rm stat.}\,1.3\%_{\rm syst.}$ &-- &$3.0\%_{\rm stat.}\,1.8\%_{\rm syst.}$~\cite{Kou:2018nap}\\
$f_{D^+}$ (MeV) &$2.6\%_{\rm stat.}\,0.9\%_{\rm syst.}$ ~\cite{bes3_muv} &$1.0\%_{\rm stat.}\,0.8\%_{\rm syst.}$&-- &--  \\
$|V_{cd}|$ &$2.6\%_{\rm stat.}\,1.0\%_{\rm syst.}^*$ ~\cite{bes3_muv} &$1.0\%_{\rm stat.}\,0.8\%_{\rm syst.}^*$ &-- &--  \\
${\mathcal B}(D^+\to \tau^+\nu_\tau)$ &$20\%_{\rm stat.}\,10\%_{\rm syst.}$~\cite{Ablikim:2019rpl}&$8\%_{\rm stat.}\,5\%_{\rm syst.}$ &-- &--  \\
$\frac{{\mathcal B}(D^+\to \tau^+\nu_\tau)}{{\mathcal B}(D^+\to \mu^+\nu_\mu)}$&$20\%_{\rm stat.}\,13\%_{\rm syst.}$~\cite{Ablikim:2019rpl}&$8\%_{\rm stat.}\,5\%_{\rm syst.}$ &-- &--
\\ \hline \hline
\multicolumn{1}{c}{Luminosity} &3.2 fb$^{-1}$ at 4.178 GeV &6 fb$^{-1}$ at 4.178 GeV & 1 ab$^{-1}$ at $\Upsilon(nS)$
& 50 ab$^{-1}$ at $\Upsilon(nS)$ \\ \hline
${\mathcal B}(D^+_s\to \mu^+\nu_\mu)$ &$2.8\%_{\rm stat.}\,2.7\%_{\rm syst.}$~\cite{bes3_Ds_muv}&$2.1\%_{\rm stat.}\,2.2\%_{\rm syst.}$&$5.3\%_{\rm stat.}\,3.8\%_{\rm syst.}$&$0.8\%_{\rm stat.}\,1.8\%_{\rm syst.}$\\
$f_{D^+_s}$ (MeV) &$1.5\%_{\rm stat.}\,1.6\%_{\rm syst.}$~\cite{bes3_Ds_muv}&$1.0\%_{\rm stat.}\,1.2\%_{\rm syst.}$&-- &-- \\
$|V_{cs}|$ &$1.5\%_{\rm stat.}\,1.6\%_{\rm syst.}$~\cite{bes3_Ds_muv} &$1.0\%_{\rm stat.}\,1.2\%_{\rm syst.}$&-- &-- \\
$f_{D^+_s}/f_{D^+}$ &$3.0\%_{\rm stat.}\,1.5\%_{\rm syst.}$~\cite{bes3_Ds_muv} &$1.4\%_{\rm stat.}\,1.4\%_{\rm syst.}$ &-- &-- \\ \hline
${\mathcal B}(D^+_s\to \tau^+\nu_\tau)$ &$2.2\%_{\rm stat.}\,2.6\%_{\rm syst.}^{\dagger}$ &$1.6\%_{\rm stat.}\,2.4\%_{\rm syst.}$ &$3.7\%_{\rm stat.}\,5.4\%_{\rm syst.}$ &$0.6\%_{\rm stat.}\,2.7\%_{\rm syst.}$ \\

$f_{D^+_s}$ (MeV) &$1.1\%_{\rm stat.}\,1.5\%_{\rm syst.}^{\dagger}$&$0.9\%_{\rm stat.}\,1.4\%_{\rm syst.}$& & \\
$|V_{cs}|$ &$1.1\%_{\rm stat.}\,1.5\%_{\rm syst.}^{\dagger}$&$0.9\%_{\rm stat.}\,1.4\%_{\rm syst.}$&-- &-- \\ \hline

$\overline f^{\mu \&\tau}_{D^+_s}$ (MeV) &$0.9\%_{\rm stat.}\,1.0\%_{\rm syst.}^{\dagger}$&$0.6\%_{\rm stat.}\,0.9\%_{\rm syst.}$&$1.6\%_{\rm stat.}\,2.0\%_{\rm syst.}$&$0.3\%_{\rm stat.}\,1.0\%_{\rm syst.}$\\
$|\overline V_{cs}^{\mu \&\tau}|$ &$0.9\%_{\rm stat.}\,1.0\%_{\rm syst.}^{\dagger}$&$0.6\%_{\rm stat.}\,0.9\%_{\rm syst.}$&-- &-- \\ \hline \hline
$\frac{{\mathcal B}(D^+_s\to \tau^+\nu_\tau)}{{\mathcal B}(D^+_s\to \mu^+\nu_\mu)}$&$3.6\%_{\rm stat.}\,3.0\%_{\rm syst.}^{\dagger}$& $2.6\%_{\rm stat.}\,2.8\%_{\rm syst.}$&$6.4\%_{\rm stat.}\,5.2\%_{\rm syst.}$&$0.9\%_{\rm stat.}\,3.2\%_{\rm syst.}$
\\ \hline \hline
\end{tabular}
\end{sidewaystable}
%\end{table}

\begin{enumerate}
\item
{\bf \it Constraints on LQCD calculations:}

Figure~\ref{fig:fD} shows a comparison of $f_{D^+_{(s)}}$ measurements
by various experiments and the values evaluated in LQCD.
Focusing initially on $f_{D^{+}}$, it may be seen that \bes3 currently supplies the most precise individual measurement, which itself has a statistical uncertainty of around 2.5\%. 
In contrast to the LQCD uncertainty of 0.2\%, there is much room to be improved in experiment.
With 20~fb$^{-1}$ of data at 3.773~GeV, the relative statistical uncertainty on
$f_{D^+}$ can be reduced to approximately 1\%, which is still larger than the current systematic uncertainty.

\begin{figure}[tp]
  \centering
  \subfigure[]{\includegraphics[width=4in]{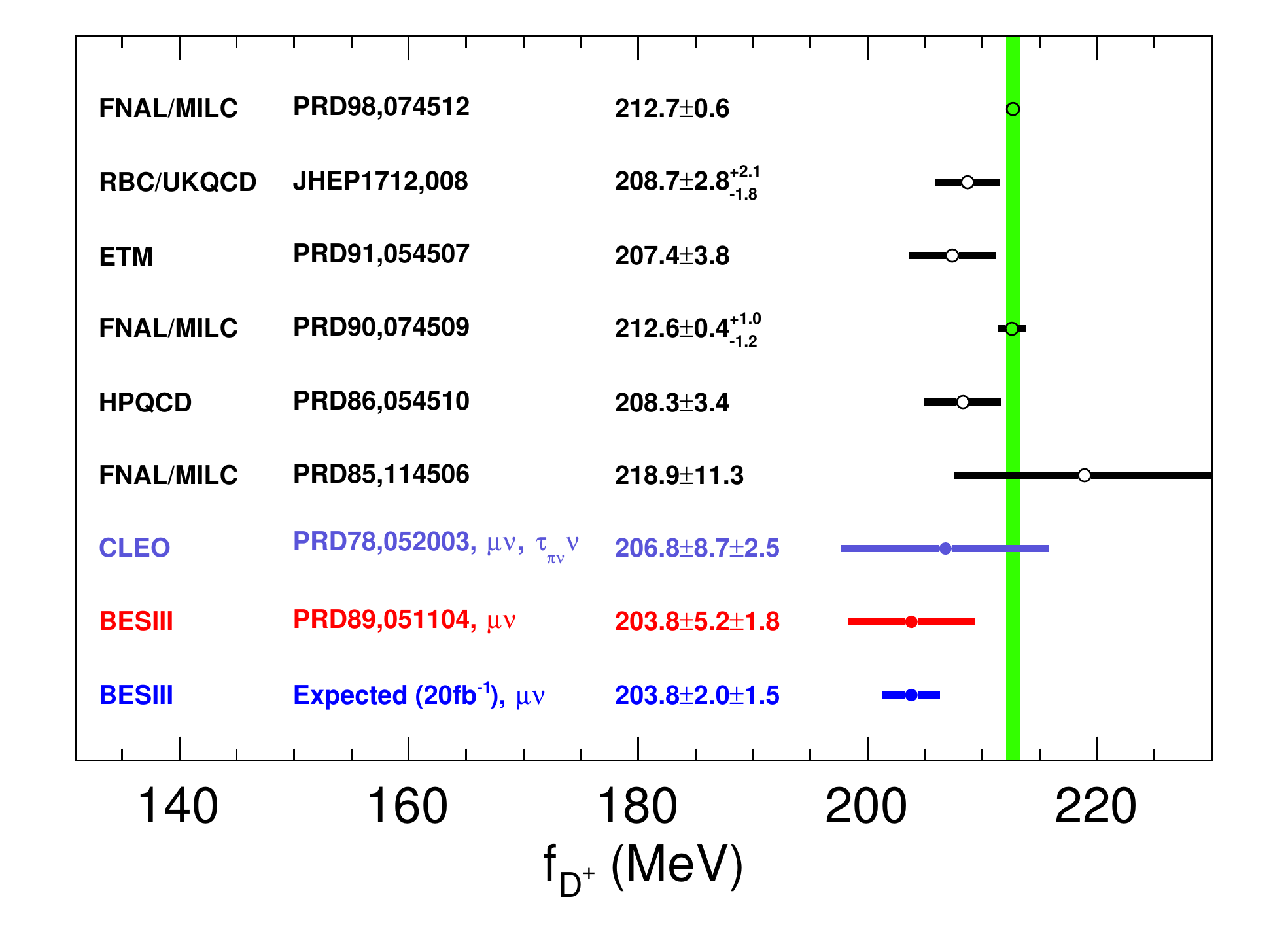}}
  \subfigure[]{\includegraphics[width=4in]{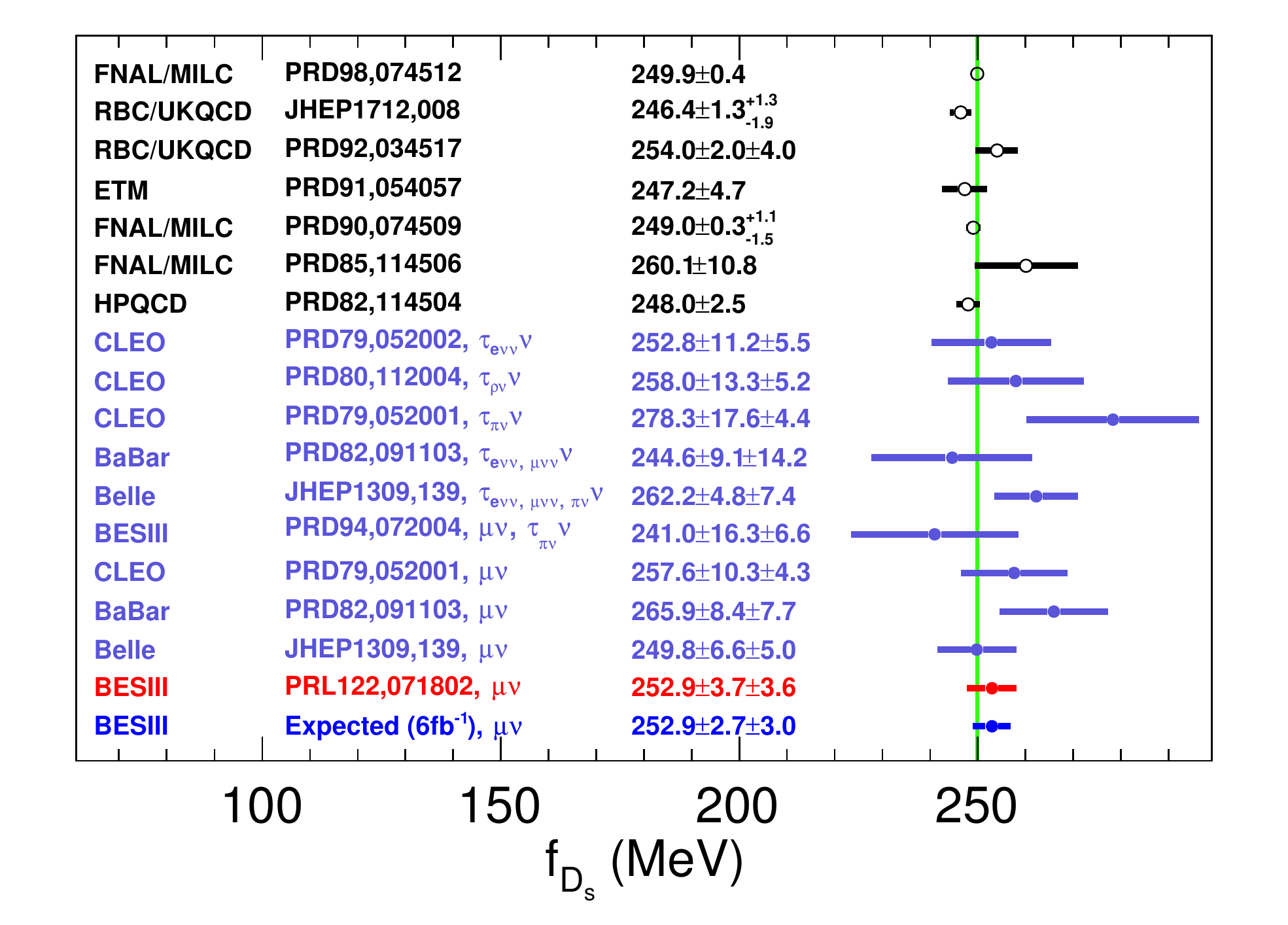}}
  \caption{\small Expected precision of the measurements of (a) $f_{D^+}$
using $D^+\to\mu^+\nu_\mu$ with 20 fb$^{-1}$ of data at 3.773 GeV and (b)
$f_{D^+_s}$ using $D^+_s\to\mu^+\nu_\mu$
with 6 fb$^{-1}$ of data at 4.178 GeV.
The green bands present the LQCD uncertainties~\cite{prd98_074512}.
The cirles and dots with error bars are the LQCD calculations and experimental
measurements, respectively. The value marked in red denotes the best
measurement, and the value marked in light blue denotes the expected
precision. 
}
\label{fig:fD}
\end{figure}

\item
{\bf \it Determinations of $|V_{cs(d)}|$:}

Alternatively, the LQCD calculations for $f_{D_{(s)}^{+}}$ may be taken as input, and the leptonic BF measurements used to confront weak physics. In the SM, quark-flavor mixing is described by the unitary $3\times 3$ CKM matrix
\begin{equation}
V_{\rm CKM}
= \left(
\begin{array}{ccc}
 V_{ud} &  V_{us} &  V_{ub} \\
 V_{cd} &  V_{cs} &  V_{cb} \\
 V_{td} &  V_{ts} &  V_{tb}  \\
\end{array} \right ).
\label{eq:ckm-m}
\end{equation}
\noindent
%\]
Any deviation from unitarity
would indicate new physics beyond the SM.
Improving the accuracy with which CKM matrix elements are determined
is one of the principal goals in flavor physics, as it will test the unitarity
of the CKM matrix with higher accuracy.

In the past decade, much progress has been achieved in LQCD calculations of
$f_{D^+_{(s)}}$.
The uncertainties of $f_{D^+_{(s)}}$ calculated in LQCD
have been reduced from (1-2)\% to the 0.2\% level~\cite{prd98_074512,prd91_054507}, thus providing precise information
to measure $|V_{cs}|$ and $|V_{cd}|$.
Comparison of the measured $|V_{cs(d)}|$ with different methods and experiments
are shown in Fig.~\ref{fig:Vcds}.
In the figure, the \bes3 result of $|V_{cd}|$ has been recalculated
with the latest LQCD calculation of $f_{D^+}=212.7\pm0.6$~MeV~\cite{prd98_074512}.
Currently, the average is dominated by the \bes3 measurements of
$D^+_{(s)}\to \ell^+\nu_\ell$ decays,
which have an uncertainty of 2.5\% (1.5\%) for $|V_{cd(s)}|$.
The statistical uncertainty of $|V_{cd}|$ is dominant,
whereas the statistical and systematic uncertainties of $|V_{cs}|$ are comparable.

The recalculated value of $|V_{cd}|$ is consistent with the
value of $|V_{cd}|=0.22522\pm0.00061$, obtained from a global fit to other CKM matrix element measurements that assumes unitarity in the SM, within $1.7\sigma$. With 20~fb$^{-1}$ of data at 3.773 GeV and 6~fb$^{-1}$ of data at 4.178 GeV,
the relative precision of the measurements of $|V_{cs}|$ and $|V_{cd}|$ with leptonic decays will both reach 1.0\%.

\begin{figure}[tp]
  \centering
  \subfigure[]{\includegraphics[width=4in]{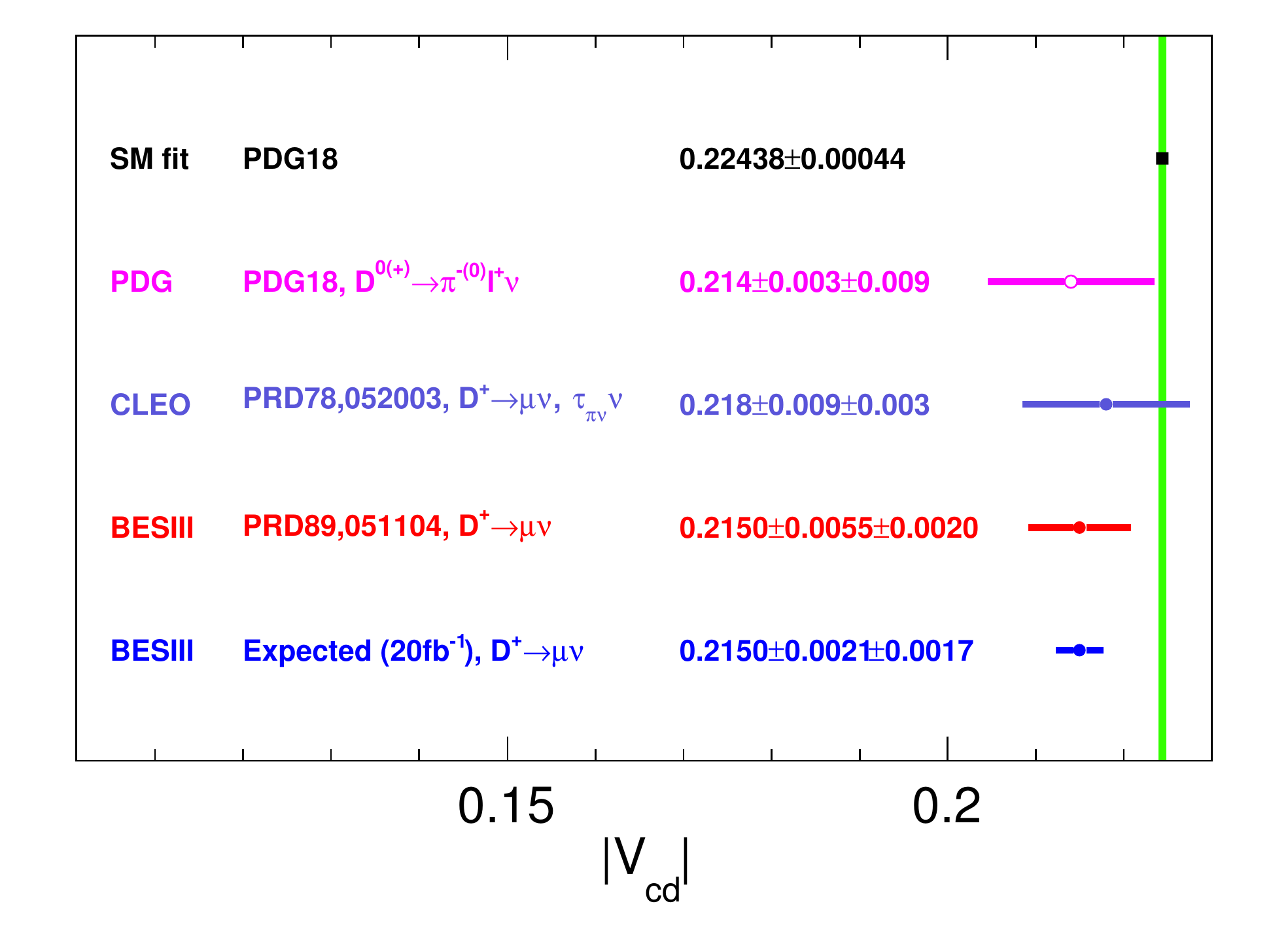}}
  \subfigure[]{\includegraphics[width=4in]{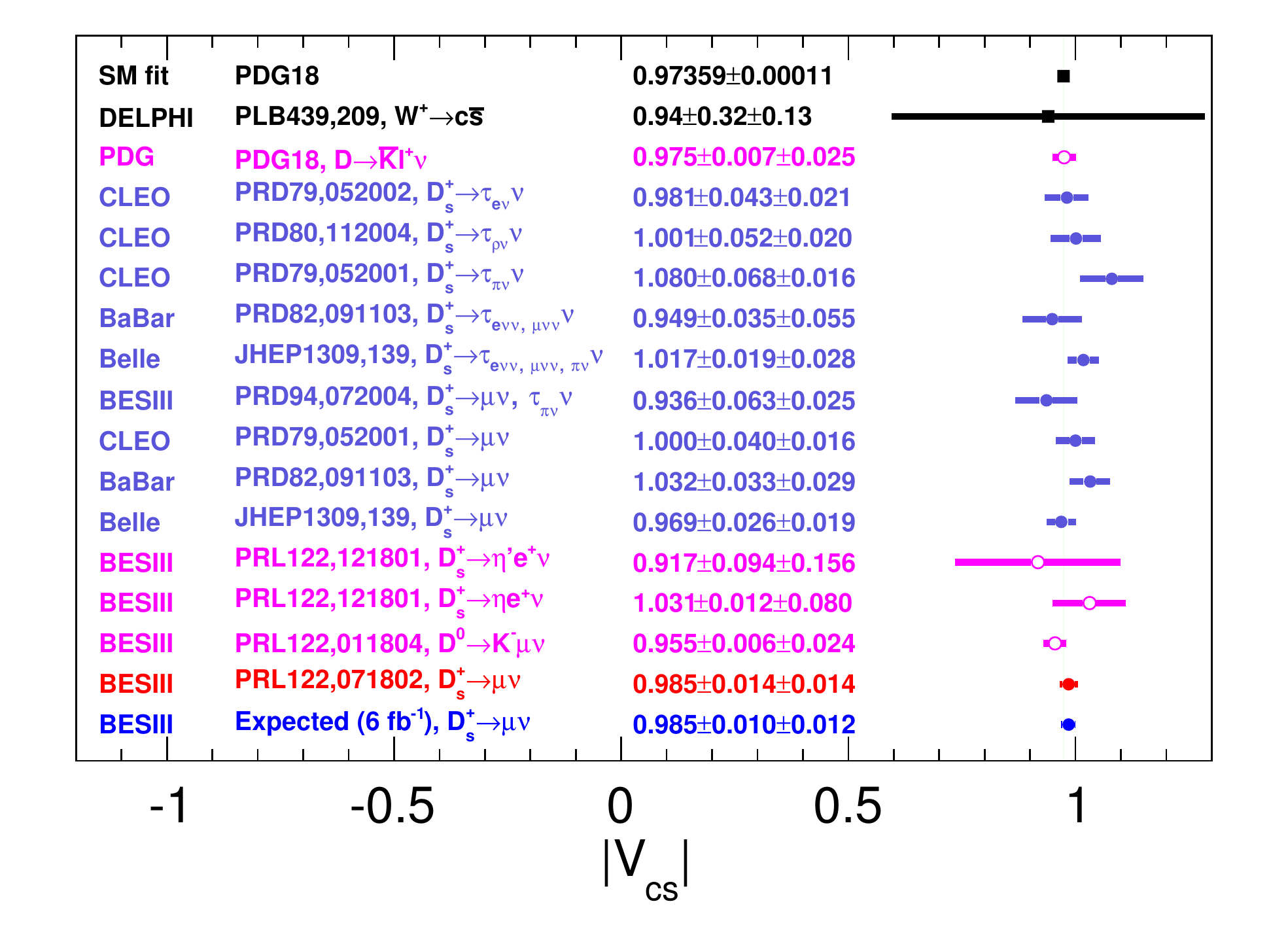}}
  \caption{\small Expected precision of the measurements of (a) $|V_{cd}|$
using $D^+\to\mu^+\nu_\mu$ with 20 fb$^{-1}$ of data at 3.773 GeV and (b)
$|V_{cs}|$ using $D^+_s\to\mu^+\nu_\mu$
with 6 fb$^{-1}$ of data at 4.178 GeV.
The green bands present the uncertainties of the average values from the
global fit in the SM~\cite{pdg2016}.
The circles, dots and rectangles with error bars are the values based on
SL $D$ decays, leptonic $D$ decays and other methods, respectively. 
The value marked in red denotes the best
measurement, and the value marked in light blue denotes the expected
precision. 
}
\label{fig:Vcds}
\end{figure}

\item
{\bf \it Tests on lepton flavor universality:}

In recent years, some hints of LFU violation have emerged in
some SL $B$-meson decays~\cite{babar_prl109_101802,babar_prd88_072012,belle_prl99_191807,belle_arxiv,belle_prd82_072005,lhcb_prl115_111803,lhcb_jhep02_104,lhcb_jhep09_179,lhcb_prl113_151601,belle_prl113_111801}.
Ref.~\cite{prd91_094009}
argues that  LFU violation may happen in $c\to s$ transitions due to an amplitude that includes a charged Higgs boson, that arises in a two-Higgs-doublet model, interfering with the SM amplitude involving a $W^{\pm}$ boson.  Therefore, it is important to test LFU with $D^+_{(s)}\to \ell^+\nu_\ell$ decays.

In the SM, the ratio of the partial widths of
$D^+_{(s)}\to \tau^+\nu_\tau$ and $D^+_{(s)}\to \mu^+\nu_\mu$
 is predicted to be
$$R_{D^+_{(s)}}=\frac{\Gamma(D^+_{(s)}\to \tau^+\nu_\tau)}{\Gamma(D^+_{(s)}\to \mu^+\nu_\mu)}=
\frac{m^2_{\tau^+}\left(1-\frac{m^2_{\tau^+}}{m^2_{D^+_{(s)}}} \right )^2}{m^2_{\mu^+} \left(1-\frac{m^2_{\mu^+}}{m^2_{D^+_{(s)}}} \right )^2}.$$
With the world average values of the masses of lepton and $D^+_{(s)}$~\cite{pdg2016},
one obtains $R_{D^+}=2.67$ and $R_{D^+_s}=9.74$ with negligible uncertainties.
The measured values of $R_{D^+_{(s)}}$ reported by \bes3 are $3.21\pm 0.64\pm0.43$ $(9.98\pm 0.52)$, which agree with the SM predicted values.
However, as previously noted, these measurements are statistically limited.
With 20 fb$^{-1}$ of data at 3.773 GeV the precision on $R_{D^{+}}$ will be 
statistically limited to about 8\%.  With 6~fb$^{-1}$ of data at 4.178~GeV the precision 
on $R_{D_{s}^{+}}$ will be systematically limited to about 3\%.

\item
{\bf \it Comparison with other experiments:}

The leptonic decays reconstructed at \bes3 and CLEO-c come from data sets accumulated  just above the open-charm threshold, where the
$D^0\bar D^0$ or $D^+D^-$ mesons are produced as a pair. The BFs can be determined by considering the yields for both the single-tagged and double-tagged events. With this method,
the background level and systematic uncertainties are lower than for the measurements at $B$ factories.

Currently, the best measurements of $D^+\to\ell^+\nu_\ell$ are from \bes3.
So far, no studies of $D^+\to\ell^+\nu_\ell$ have been reported by
BaBar, Belle, and LHCb, which may be due to the higher level of background any measurement at these experiments would encounter.
\bes3 is expected to provide unique data to improve the knowledge of $f_{D^+}$,
$|V_{cd}|$, and tests of LFU in $D^+\to\ell^+\nu_\ell$ decays in the next decade.

BaBar~\cite{babar_lv} and Belle~\cite{belle_lv} have reported the measurement of $D^+_s\to\ell^+\nu_\ell$
using $e^+e^-\to c\bar c\to DKX D^{*-}_s$ with $D^{*-}_s\to D^-_s\gamma$, performed
with about 0.5 and 1.0~ab$^{-1}$ of data taken around the $\Upsilon(4S)$, respectively.
The statistical and systematic uncertainties of these measurements of
$f_{D^+_{s}}$ and $|V_{cs}|$ at Belle are 1.6\% and 2.5\%, respectively.
The dominant systematic uncertainties are from
normalization, tag bias, particle identification, fit model and $D^+_s$ background.
With 50 ab$^{-1}$ of data at Belle II, the statistical uncertainties
of these measurements are expected to be reduced to 0.25\%, however it will be extremely challenging to reduce the systematics uncertainties to match this level. Hence it is expected that a measurement of $D_{s}^{+}\to \ell^+\nu_{\ell}$ performed at \bes3 with 6~fb$^{-1}$ of data at 4.178 GeV will have very significant weight in a future world average.
\end{enumerate}

\bes3 measurements of purely leptonic $D_{s}^{+}$ decays will be limited by systematic uncertainties to a large degree. This motivates the exploration of taking data at $\sqrt s=$ 4.009~GeV where only $D_s^{+}D_{s}^{-}$ production is possible; such a data set will have much reduced backgrounds in the double-tag method, thus improving the systematic uncertainties on the measurements. The drawback is the reduction in cross section by a factor of three with respect to operation at 4.178 GeV. Therefore, a data set of $20~\mathrm{fb}^{-1}$ would be required to match the statistical precision of 6~fb$^{-1}$ of data at 4.178~GeV across all measurements of $D^{+}_{s}$ decay. Such a large set corresponds to many years of data taking and as such cannot be considered as the highest priority among the various charm-physics data sets requested.

\subsection{Semileptonic decays}

In the SM, the weak and strong effects in SL $D^{0(+)}$ decays
can be well separated~\cite{Ivanov:2019nqd}.
Among them, the simplest case is $D^{0(+)}\to\bar{K}(\pi)\ell^+\nu_\ell$,
for which the differential decay rate can be simply written as
\begin{equation}
\frac{d\Gamma}{dq^2} =\frac{G_F^2}{24\pi^3}|V_{cs(d)}|^2
p_{K(\pi)}^3 |f_{+}^{K(\pi)}(q^2)|^2,
\end{equation}
where $G_F$ is the Fermi coupling constant, and 
$p_{K(\pi)}$ is the kaon (pion) momentum in the $D$ rest frame,
$f_{+}^{K(\pi)}(q^2)$ is the form factor of the hadronic weak
current depending on the square of the transferred four-momentum
$q=p_D-p_{K(\pi)}$.
From analyses of the dynamics in these decays,
one can obtain the product $f_{+}^{K(\pi)}(0)|V_{cs(d)}|$.
By taking the $f_{+}^{K(\pi)}(0)$ calculated in LQCD or $|V_{cd(s)}|$ from a
global fit assuming unitarity in the SM,
the value of either $|V_{cd(s)}|$ or $f_{+}^{K(\pi)}(0)$ can be obtained.

\begin{itemize}
\item
{\bf \it Form factors of SL decays:}

With the data sets in hand, \bes3 has reported improved measurements of
the absolute BFs and the form factors of
the SL decays
$D^{0(+)}\to \bar K e^+\nu_e$,
$D^{0(+)}\to \pi e^+\nu_e$~\cite{bes3_D0_kpiev,bes3_Dp_k0pi0ev},
$D^{0}\to K^-\mu^+\nu_\mu$~\cite{bes3_kmuv},
$D^+\to\eta e^+\nu_e$~\cite{bes3_etaev},
$D^+\to\bar K^{*0}e^+\nu_e$~\cite{bes3_Dp_kpiev},
$D^{0(+)}\to \rho^{-(0)}e^+\nu_e$~\cite{bes3_pipiev},
$D^+\to \omega e^+\nu_e$~\cite{bes3_Dp_omegaev},
$D^+_s\to K^{(*)0}e^+\nu_e$~\cite{panic2017_lilei} and
$D^+_s\to \eta^{(\prime)} e^+\nu_e$~\cite{ichep2018_cjc}.
Figure~\ref{fig:ff} shows comparison of the form factors
$f^K_+(0)$, and $f^\pi_+(0)$ measured by various experiments and those calculated in LQCD.
The \bes3 measurement of $f^K_+(0)$ is
dominated by systematic uncertainties, whereas other measurements
are dominated by statistical uncertainties.
The measurements of the form factors of
$D^0\to K^{*-}e^+\nu_e$,
$D^+\to \eta\mu^+\nu_\mu$,
$D^+_s\to \phi e^+\nu_e$,
$D^+_s\to f_0(980) e^+\nu_e$ and
$D^+_s\to \eta\mu^+\nu_\mu$
are still ongoing, but all will be statistically limited with the current data sets.

Measurements of SL $D^{0(+)}$ decays that contain a scalar or axial-vector meson in the final state,
$e.g.$, $D^{0(+)}\to \bar K_1(1270)e^+\nu_e$~\cite{cleo_k1ev} and $D^{0(+)}\to a_0(980)e^+\nu_e$~\cite{charm2016_dzl},
have been reported by CLEO-c and \bes3. However, these decay samples are
up to about 100 events, which are insufficient to perform form-factor measurements.

With 20 fb$^{-1}$ of data at 3.773 GeV,
all the form-factor measurements which are currently statistically limited
will be statistically improved by a factor of up to 2.6.
We also have opportunities to determine the form factors of
$D^0\to K_1^-(1270) e^+\nu_e$, $D^+\to \bar K_1^0(1270) e^+\nu_e$,
$D^+\to \eta^\prime e^+\nu_e$,
$D^0\to a_0^-(980) e^+\nu_e$, and $D^+\to a_0^0(980) e^+\nu_e$ for the first time.
In addition, studies of the semi-muonic decays of
$D^{0(+)}\to \bar{K}\mu^+\nu_\mu$ and $D^{0(+)}\to \pi\mu^+\nu_\mu$ will further improve the knowledge
of $f^K_+(0)$ and $f^\pi_+(0)$.

% Need to start from here

With 3.2 fb$^{-1}$ of data at 4.178 GeV,
the expected yield for each of the Cabibbo-favored (CF) SL $D^+_s$ decays is
about 1000, while that for each of the singly Cabibbo-suppressed (SCS) SL
$D^+_s$ decays is not more than 200.
In this case, all studies of the dynamics of SL $D^+_s$ decays
are restricted due to the limited data sets.
The measurements of the form factors in these decays will
be improved by a factor of up to 1.4 with 6~fb$^{-1}$ of data at 4.178~GeV.

\begin{figure}[tp]
  \centering
  \subfigure[]{\includegraphics[width=4in]{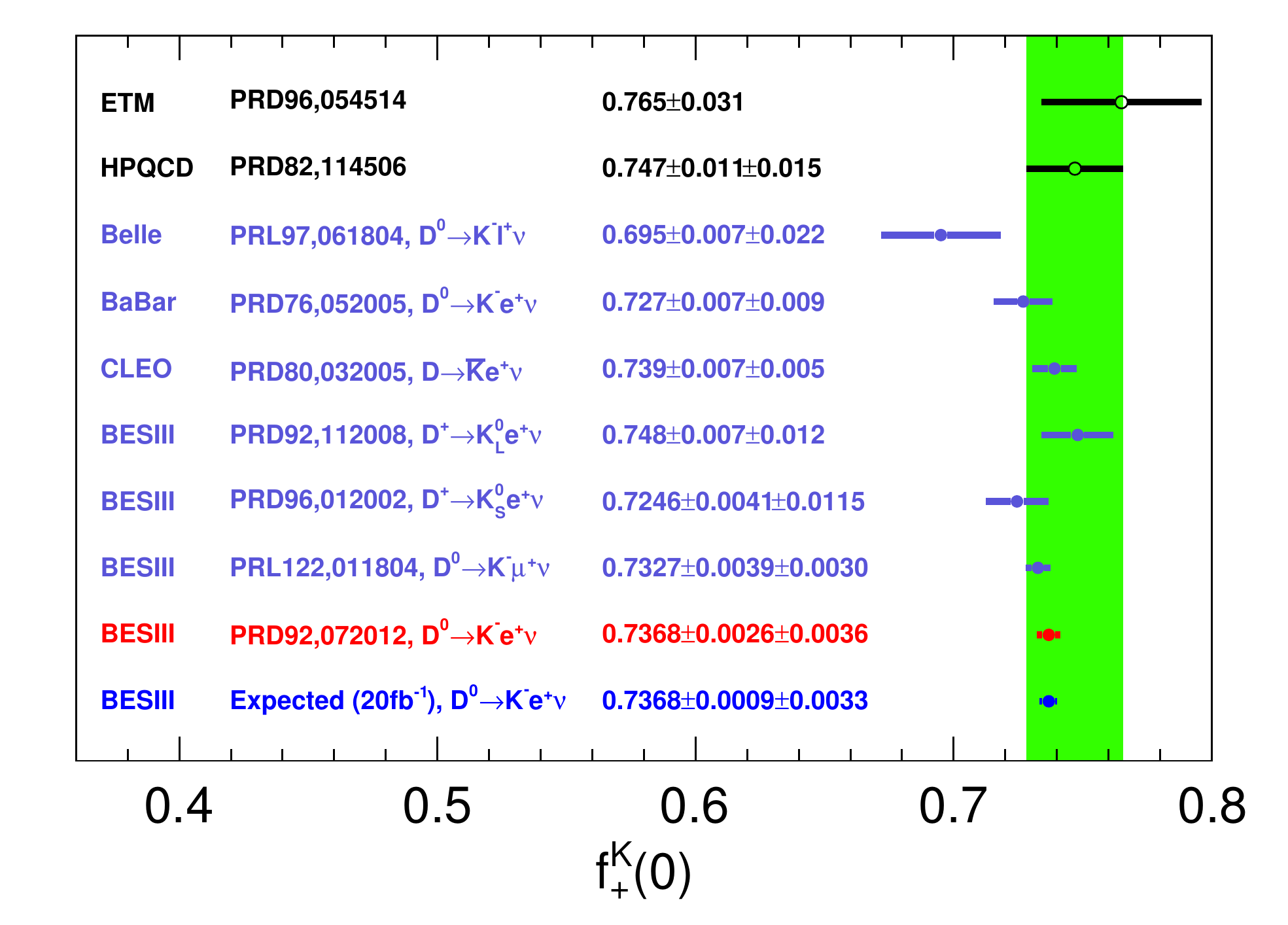}}
  \subfigure[]{\includegraphics[width=4in]{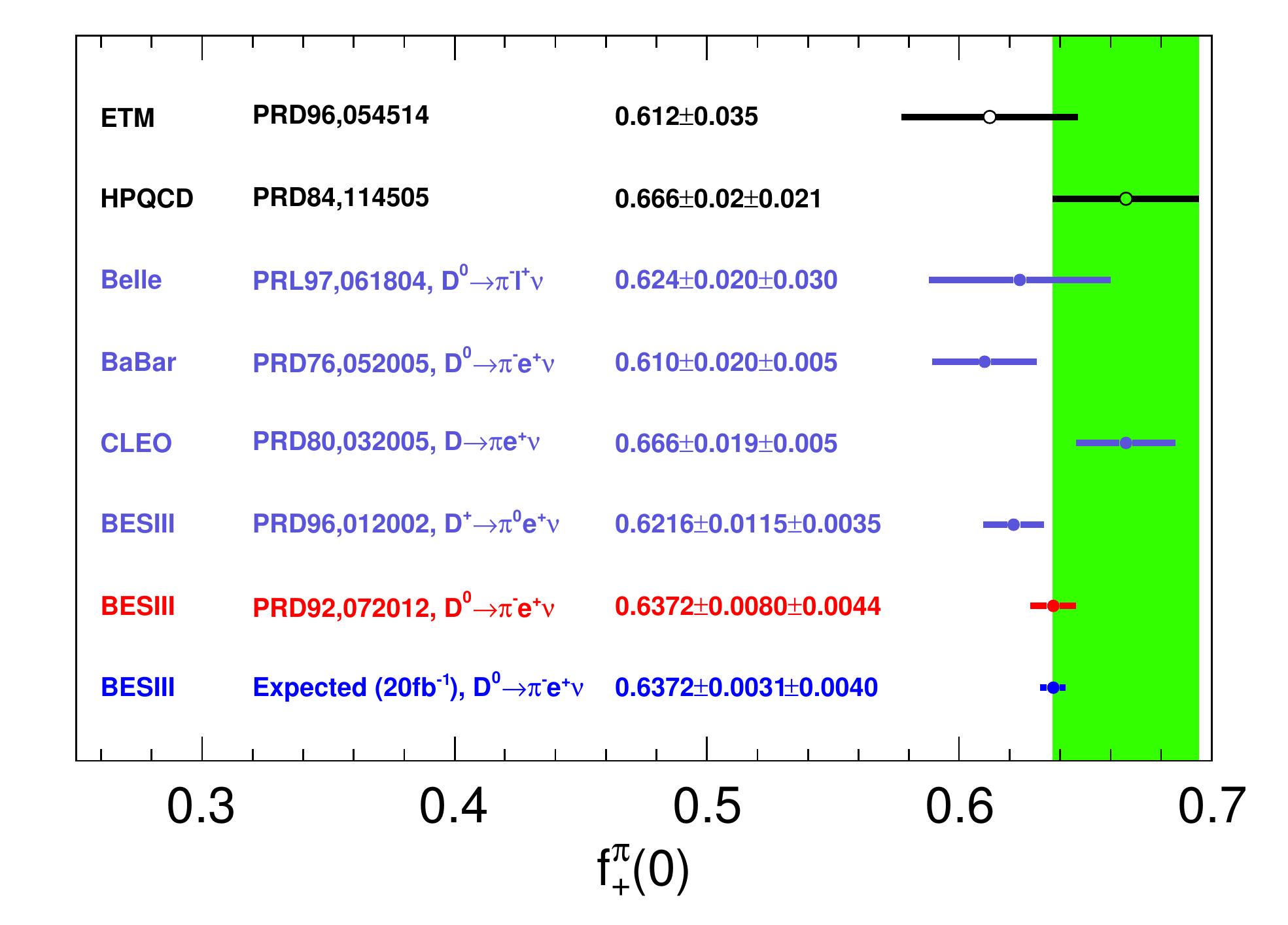}}
  \caption{\small Expected precision of the measurements of (a) $f^K_+(0)$
and (b) $f^\pi_+(0)$ with 20 fb$^{-1}$ of data at 3.773 GeV.
The green bands present the LQCD uncertainties~\cite{lqcd_fk,lqcd_fpi}.
The PDG value is combined from the results of from Belle, CLEO, BaBar and
BESIII.
The cirles and dots with error bars are the LQCD calculations and experimental
measurements, respectively. The value marked in red denotes the best
measurement, and the value marked in light blue denotes the expected
precision. 
}
\label{fig:ff}
\end{figure}

\item
{\bf \it Determinations of $|V_{cs(d)}|$:}

The CKM elements  $|V_{cs}|$ and $|V_{cd}|$ can also be determined
from analyses of the SL decays of $D^{0(+)}\to\bar K \ell^+\nu_\ell$ and $D^{0(+)}\to \pi\ell^+\nu_\ell$, when the values of the form factors are taken from LQCD. Results using this approach are included in Fig.~\ref{fig:Vcds}.
At present, the LQCD uncertainties
are 2.4\% for $f^K_+(0)$~\cite{lqcd_fk} and 4.4\% for $f^\pi_+(0)$~\cite{lqcd_fpi},
which are significantly larger than the associated experimental uncertainties, and therefore limit the determinations of $|V_{cs}|$ and $|V_{cd}|$ from this method.

In the coming decade, however,
the uncertainties of $f^K_+(0)$ and $f^\pi_+(0)$ calculated in LQCD
are expected to be reduced to the level of 1.0\% and 0.5\%~\cite{talk_beauty2014}, respectively.
Therefore, further improved measurements of $f^K_+(0)$ and $f^\pi_+(0)$
will play key roles in the determinations of $|V_{cs}|$ and $|V_{cd}|$.
With 20 fb$^{-1}$ of $\psi(3770)$ data,
the experimental uncertainties of the
measurements of $|V_{cs}|$ and $|V_{cd}|$ with SL $D^{0(+)}$ decays in electron channels
are expected to reach the 0.5\% level.
Studies of
$D^{0(+)}\to \bar{K}\mu^+\nu_\mu$ and $D^{0(+)}\to \pi\mu^+\nu_\mu$ will provide additional sensitivity.

In SL $D^+_s$ decays,
analyses of the dynamics of $D^+_s\to\eta^{(\prime)}e^+\nu_e$ decays with 6 fb$^{-1}$ of data at 4.178 GeV , where the $\eta-\eta'$ mixing is involved,
will provide complementary measurements of $|V_{cs}|$.
The statistical uncertainty is expected to reach the 2.0\% level; this will further improve the measurement precision of $|V_{cs}|$ at \bes3.

\item
{\bf \it Tests of lepton flavor universality:}
Previous measurements of the BFs of
$D^0\to\pi^-\mu^+\nu_\mu$ and $D^0\to\pi^-e^+\nu_e$~\cite{pdg2016},
resulted in the ratio of BFs ${\mathcal B}(D^0\to\pi^-\mu^+\nu_\mu)/{\mathcal B}(D^0\to\pi^-e^+\nu_e)=0.82\pm0.08$, which deviates from the SM prediction of $0.985\pm0.002$~\cite{epjc78_501} by
$2.1\sigma$. This possible hint of LFU violation has motivated
\bes3 to report more precise BF measurements to obtain the ratios ${\mathcal B}(D^0\to\pi^-\mu^+\nu_\mu)/{\mathcal B}(D^0\to\pi^-e^+\nu_e)=0.922\pm0.030\pm0.022$
and ${\mathcal B}(D^+\to\pi^0\mu^+\nu_\mu)/{\mathcal B}(D^+\to\pi^0e^+\nu_e)=0.964\pm0.037\pm0.026$~\cite{bes3_pimuv}.
These results are consistent with the SM predictions, within $1.7\sigma$ and $0.5\sigma$, respectively.
Much more accurate studies will be possible with 20 fb$^{-1}$ of data at 3.773 GeV, besides optimizations on the systematic uncertainties.

As Ref.~\cite{prd91_094009} pointed out,
due to the mediation of charged Higgs bosons in the two-Higgs-doublet model, LFU in $c\to s$ transitions may be violated.
In charm decays, the semi-tauonic decays involving a kaon in the final states are
kinematically forbidden.
Measurements of the ratios of the partial widths of $D^{0(+)}\to \bar{K} \mu^+\nu_\mu$
over those of $D^{0(+)}\to \bar{K} e^+\nu_e$ in different $q^2$ intervals
constitute a complementary test of LFU to those using semi-tauonic decays.
These measurements are currently statistically limited~\cite{bes3_kmuv,bes3_pimuv}, and will be significantly improved with 20 fb$^{-1}$ of data at 3.773 GeV.

\item
{\bf \it Studies on meson spectroscopy:}
Studies on intermediate resonances in hadronic final states in SL decays provide a clean environment 
to explore meson spectroscopy, as no other particles interfere. This corresponds to a much 
simpler treatment than those studies in $D^{0(+)}_{(s)}$ hadronic decays
or charmonium decays. For instance, in
$D^0\to K_1^-(1270) e^+\nu_e$ and $D^+\to \bar K_1^0(1270) e^+\nu_e$,
the single hadronic current allows the production of the axial-meson to be factorized,
which provides unique information on the structure of this lightest strange axial-vector mesons. 
At present, the world average values of the mass,  $1272\pm7$ MeV, and width,  $90\pm20$ MeV,
of the $\bar K_1(1270)$ have large uncertainties~\cite{pdg2016}. 
Furthermore, different analysis channels produce conflicting resonance parameters.
Based on the future BESIII data set, the precision of these parameters is expected to be improved by a factor of two to three.
Also, we have the opportunity to study the properties of
other particles, $e.g.$, $a_0^-(980)$ and $a_0^0(980)$ in SL decays.
In addition, it is possible to search for various SL $D^{0(+)}_{(s)}$
transitions into other scalar or axial-vector mesons, as listed in
Ref.~\cite{epjc77_587}.

\item
{\bf \it Comparison with other experiments:}
BaBar studied the SL decays
$D^0\to K^-e^+\nu_e$~\cite{babar_kev} and $D^0\to \pi^-e^+\nu_e$~\cite{babar_piev}
using the decay $D^0\to K^-\pi^+$ to normalize the measurements.
However, the BF of $D^0\to K^-\pi^+$ has an uncertainty of
1\% that is larger than the systematic uncertainties in the measurements
of $D^0\to K^-e^+\nu_e$ and $D^0\to \pi^-e^+\nu_e$ at \bes3.
Belle made absolute measurements of $D^0\to K^-\ell^+\nu_\ell$ and $D^0\to \pi^-\ell^+\nu_\ell$
by using the decay chain of $D^{(*)}_{\rm tag} D^{*-}_{\rm sig}X$ with 282 fb$^{-1}$ of
data taken around $\Upsilon(4S)$~\cite{belle_kpiev},
with precision summarized in Table~\ref{tab:singletagN_MC}.
So far, no other SL measurements have been reported by Belle.

\bes3 has the advantage of low background and small systematic uncertainty ($<1\%$)
for measurements of semi-electronic $D^{0(+)}$ decays, and with 20 fb$^{-1}$ of data at 3.773 GeV, will be competitive with Belle II.
On the other hand, the momenta of most muons in semi-muonic $D^{0(+)}$ decays produced at threshold
are lower than 0.5 GeV/$c$, which lies outside the detection ability of the muon counter, 
and hence makes the study of semi-muonic $D^{0(+)}$ decays challenging at \bes3.
However, improved and comprehensive measurements of dynamics of as many as semi-muonic
$D^{0(+)}$ decays into a pseudescalar meson are workable, as done in Refs.~\cite{bes3_kmuv,bes3_pimuv},
but will be challenging for those semi-muonic $D^{0(+)}$ transitions into a vector, scalar or axial-vector meson.

The SL decays can also be studied at LHCb and its upgrades, particularly in the muon channels.
However, the studies of the SL $D^{0(+)}_{(s)}$ decays involving electron and photon(s) in the final states
are more difficult in the environment of a hadron collider due to larger backgrounds.
\end{itemize}

\begin{sidewaystable}
%\begin{table}[tp]
%\begin{minipage}[t]{1.0\textwidth}
\centering
\caption{\label{tab:singletagN_MC}\small
Form-factor measurements of SL $D^{0(+)}$ decays.
For the decays involving a pseudoscalar or scalar meson, the
decay rate is parameterized by one form factor $f_+(0)$.
For the decays involving a vector or axial-vector meson,
the decay rate is parameterized by three form factors,
$V(0)$, $A_1(0)$, and $A_2(0)$, which has relations $r_V=V(0)/A_1(0)$ and $r_A=A_2(0)/A_1(0)$.
The $A_1(0)$ for $D^+\to \bar K^{*0}e^+\nu_e$ decay is reported due to low background and enough signals
at \bes3.
For Belle II, we assume that the systematic uncertainties can be reduced by
a factor of 2 compare to Belle's results.
The results marked in $^*$ for Belle (II) are based on both semi-electronic and semi-muonic decays.
}\vspace{0.1cm}
\footnotesize
\begin{tabular}{lcccc} \hline \hline
\multicolumn{1}{c}{} &\bes3 & \bes3 & Belle
& Belle II \\ \hline
\multicolumn{1}{c}{Luminosity} &2.9 fb$^{-1}$@3.773 GeV & 20 fb$^{-1}$@3.773 GeV & 0.28 ab$^{-1}$
& 50 ab$^{-1}$\\ \hline
$D^0\to K^-e^+\nu_e$&$0.4\%_{\rm stat.}\,0.5\%_{\rm syst.}$&$0.2\%_{\rm stat.}0.4\%_{\rm syst.}$ &$1.0\%_{\rm stat.}\,3.2\%_{\rm syst.}^*$ &$0.1\%_{\rm stat.}1.6\%_{\rm syst.}^*$ \\
$D^0\to K^-\mu^+\nu_\mu$&$0.5\%_{\rm stat.}\,0.4\%_{\rm syst.}$&$0.2\%_{\rm stat.}0.4\%_{\rm syst.}$ & & \\ \hline
$D^0\to \pi^-e^+\nu_e$&$1.3\%_{\rm stat.}\,0.7\%_{\rm syst.}$&$0.5\%_{\rm stat.}0.4\%_{\rm stat.}$ &$3.2\%_{\rm stat.}\,4.8\%_{\rm syst.}^*$ &$0.2\%_{\rm stat.}2.4\%_{\rm syst.}^*$ \\
$D^0\to \pi^-\mu^+\nu_\mu$& NA &$0.8\%_{\rm stat.}0.8\%_{\rm syst.}$ & & \\\hline
$D^0\to K^{*-}e^+\nu_e$& & & &\\
%$A_1(0)$& & &-- &--\\
$r_V$   &$5.0\%_{\rm stat.}2.0\%_{\rm syst.}$&$2.0\%_{\rm stat.}2.0\%_{\rm syst.}$ &-- &--\\
$r_A$   &$10.\%_{\rm stat.}2.0\%_{\rm syst.}$&$4.0\%_{\rm stat.}2.0\%_{\rm syst.}$ &-- &--\\ \hline \hline
$D^0\to a_0^-(980) e^+\nu_e$  & NA &$10.\%_{\rm stat.}5.0\%_{\rm syst.}$ &-- &--\\ \hline
$D^0\to K_1^-(1270) e^+\nu_e$ & NA &$10.\%_{\rm stat.}5.0\%_{\rm syst.}$ &-- &--\\ \hline \hline
$D^+\to \bar K^0e^+\nu_e$&$0.6\%_{\rm stat.}\,1.7\%_{\rm syst.}$&$0.2\%_{\rm stat.}1.0\%_{\rm syst.}$ &-- &--\\\hline
$D^+\to K^0_{\rm L} e^+\nu_e$&$0.9\%_{\rm stat.}\,1.6\%_{\rm syst.}$ &$0.4\%_{\rm stat.}1.0\%_{\rm syst.}$ &-- &--\\\hline
$D^+\to \bar K^0\mu^+\nu_\mu$ & NA &$0.3\%_{\rm stat.}1.0\%_{\rm syst.}$ &-- &--\\ \hline
$D^+\to \bar K^{*0}e^+\nu_e$& & & & \\
$A_1(0)$&$1.7\%_{\rm stat.}\,2.0\%_{\rm syst.}$&$0.7\%_{\rm stat.}1.0\%_{\rm syst.}$ &-- &--\\
$r_V$   &$4.0\%_{\rm stat.}\,0.5\%_{\rm syst.}$&$1.6\%_{\rm stat.}0.5\%_{\rm syst.}$ &-- &--\\
$r_A$   &$5.0\%_{\rm stat.}\,1.0\%_{\rm syst.}$&$2.0\%_{\rm stat.}1.0\%_{\rm syst.}$ &-- &--\\ \hline
$D^+\to \pi^0e^+\nu_e$&$1.9\%_{\rm stat.}\,0.5\%_{\rm syst.}$&$0.7\%_{\rm stat.}0.5\%_{\rm syst.}$ &-- &--\\ \hline
$D^+\to \pi^0\mu^+\nu_\mu$&NA &$1.0\%_{\rm stat.}1.0\%_{\rm syst.}$ &-- &--\\ \hline
$D^+\to \eta e^+\nu_e$&$4.5\%_{\rm stat.}2.0\%_{\rm syst.}$&$2.0\%_{\rm stat.}2.0\%_{\rm syst.}$ &-- &--\\\hline
$D^+\to \eta^\prime e^+\nu_e$     & NA &$10.\%_{\rm stat.}5.0\%_{\rm syst.}$ &-- &--\\ \hline
$D^+\to \omega e^+\nu_e$& & & &\\
%$A_1(0)$&NR & &-- &--\\
$r_V$   &$7.2\%_{\rm stat.}\,4.8\%_{\rm syst.}$&$3.0\%_{\rm stat.}2.0\%_{\rm syst.}$ &-- &--\\
$r_A$   &$14\%_{\rm stat.}\,5.0\%_{\rm syst.}$&$3.0\%_{\rm stat.}2.0\%_{\rm syst.}$ &-- &--\\ \hline \hline
$D^+\to a_0^0(980) e^+\nu_e$      & NA &$10.\%_{\rm stat.}5.0\%_{\rm syst.}$&-- &--\\  \hline
$D^+\to \bar K_1^0(1270) e^+\nu_e$& NA &$10.\%_{\rm stat.}5.0\%_{\rm syst.}$&-- &--\\  \hline
$D^{0(+)}\to \rho^{-(0)}e^+\nu_e$& & & & \\
%$A_1(0)$& & &-- &--\\
$r_V$   &$5.0\%_{\rm stat.}4.0\%_{\rm syst.}$ &$2.0\%_{\rm stat.}2.0\%_{\rm syst.}$ &-- &--\\
$r_A$   &$8.0\%_{\rm stat.}4.0\%_{\rm syst.}$ &$3.0\%_{\rm stat.}2.0\%_{\rm syst.}$ &-- &--\\ \hline \hline
\end{tabular}
\end{sidewaystable}
%\end{table}

%% file: Charm/charm_meson_hadronic_decay.tex
\subsection{Quantum-correlated measurements of $D^0$ hadronic decays}
\label{subsec:phi3}

The quantum correlation of the $D^0\bar D^0$ meson pair produced at $\psi(3770)$ provides a unique way to probe the amplitudes of the $D$ decays, the $D^0\bar D^0$  mixing parameters and potential
CP violation in $D^0$ decays~\cite{Xing:1996pn}. 
Furthermore, the determination of the strong-phase difference between the CF and doubly
Cabibbo-suppressed (DCS) amplitudes in the decay of quantum-correlated $D^0\bar D^0$ meson pairs has several motivations: understanding the non-perturbative QCD effects in the charm sector; serving as essential inputs to extract the angle $\gamma$ of the CKM unitarity triangle (UT); and relating the measured mixing parameters in hadronic decay $(x', y')$ to the mass and width difference parameters $(x, y)$~\cite{Amhis:2016xyh}.

The measurements of the CKM UT angles $\alpha$,
$\beta$, and $\gamma$ in $B$ decays are important to test CKM unitarity and search for
CP violation beyond the SM. Any discrepancy in the measurements of the unitarity triangle involving tree and loop dominated processes would indicate high-mass new physics within the loops.  Among the three CKM angles, $\gamma$,
where the current world-best measurement from LHCb is $(74.0^{+5.0}_{-5.8})^\circ$~\cite{lhcbgamma}, has particular importance, being the only
CP-violating observable that can be determined using tree-level decays. Degree-level precision on $\gamma$ will allow a rigorous comparison of the loop and tree-level determinations of the UT. The independence from loop diagrams means 
that the measurement of $\gamma$ has negligible theoretical uncertainty \cite{BROD}.
The precision measurement of $\gamma$ is one of the top priorities for the LHCb upgrade(s) and Belle II experiments.

The most precise method to measure $\gamma$ is
based upon the interference between $B^{+}\to\bar{D}^{0}K^{+}$
and $B^{+}\to D^{0}K^{+}$ decays~\cite{GLW, ADS, GGSZ}.
In the future, the statistical uncertainties of these measurements
will be greatly reduced by using the large $B$ meson samples recorded by LHCb and Belle II, and by extending these measurements to other similar $B$ modes such as
$B^{0}\to DK^{*0}$, where $D$ implies either $D^0$ or $\bar D^0$,
$B^{+}\to D^*K^{+}$, $B^+\to DK\pi\pi$, and $B^+\to DK^{*+}$~\cite{lhcb_note,Kou:2018nap}.
However, with increased statistical precision, limited knowledge of the strong phases of the $D$ decays will systematically restrict the overall sensitivity. Consequently, improved knowledge of the strong-phase related parameters in $D$ decays is essential to make measurements of $\gamma$ to degree-level precision.  
Strong-phase information in the following
$D$ decay modes has been obtained from the CLEO-c experiment and
has been used in the most recent $\gamma$ measurements:
\begin{itemize}
\item
measurement of the amplitude weighted average cosine and sine of the strong-phase difference, $c_{i}$ and $s_{i}$, where the index $i$  refers to the phase-space region of self-conjugate multi-body decays, such as $D\to K_{S}^{0}\pi^{+}\pi^{-}$
and $D\to K_{S}^{0}K^{+}K^{-}$~\cite{CLEO(2010)};
\item
measurement of the coherence factor and average strong-phase
difference in $D\to K^{\pm}\pi^{\mp}\pi^{+}\pi^{-}$ and
$D\to K^{\pm}\pi^{\mp}\pi^{0}$~\cite{T.Evans(2016)};
\item
measurement of the coherence factor and average strong-phase
difference in $D\to K_{S}^{0}K^{\pm}\pi^{\mp}$~\cite{CLEO(2012)};
\item
measurement of the CP-even content of $D\to\pi^{+}\pi^{-}\pi^{+}\pi^{-}$,
$D\to\pi^{+}\pi^{-}\pi^{0}$, and $D\to K^{+}K^{-}\pi^{0}$~\cite{S.Malde(2015)}; 
\item
measurement of the strong-phase difference in $D\to K^{\pm}\pi^{\mp}$~\cite{CLEO(2012)112001};
\item 
measurement of the CP-even content and strong-phase difference $c_{i}$ and $s_{i}$ in $D\to K_S \pi^{+}\pi^{-}\pi^0$~\cite{K:2017qxf}
.

\end{itemize}
Complementary constraints on the strong phase in $D\to K^{\pm}\pi^{\mp}\pi^{+}\pi^{-}$
decays have come from charm-mixing measurements at LHCb~\cite{LHCb(2016)}. It should be noted that the precision on the $D\to K^{\pm}\pi^{\mp}$ phase is dominated by the combination of charm-mixing measurements  at the LHCb, CDF experiments and the $B$ factories~\cite{Amhis:2016xyh} and so here the role of threshold data is less important. Recently, \bes3 reported the strong-phase measurement of
$D\to K^{\pm}\pi^{\mp}$ decays~\cite{BES(2014)}
and the preliminary results of the strong-phase measurements of
$D\to K_{S}^{0}\pi^{+}\pi^{-}$ decays~\cite{Lei(2019)} from current $\psi(3770)$ data with an integrated luminosity of 2.93 fb$^{-1}$.
The precision of these measurements demonstrated the powerful capabilities of \bes3 to determine strong-phase parameters accurately. However, all the current strong-phase measurements are still limited by the size of the $\psi(3770)$ data. Therefore, one of the most important goals of the \bes3 charm physics program is to improve the strong-phase measurements with a larger $\psi(3770)$ data.

For the existing determination of $\gamma$ that is made by combining the
CP-violation sensitive observables
from different $D$ modes in $B\to D^{(*)}K^{(*)}$ decays, the uncertainty arising from the CLEO-c inputs has been found to be about 2$^\circ$ for LHCb. The current \bes3 $\psi(3770)$ data is approximately four times larger than that of CLEO-c.  
The full analysis based on it gives in general a factor of 2.5(1.9) more precise results for $c_i$($s_i$), which reduces its contribution to the uncertainty on $\gamma$ to 0.7$^\circ$~\cite{Lei(2019)} at most.
As evident from \tablename~\ref{tab:exps}, this precision should be adequate for the LHCb Run-2 measurement,
but will not be sufficient for the future LHCb upgrade and Belle II era, in particular because these strong-phase uncertainties will be largely correlated between the two experiments for each mode.
To minimize the impact of the strong-phase measurement uncertainties
for these next generation experiments,  a much larger $\psi(3770)$ data at \bes3,
ideally corresponding to an integrated luminosity of 20~fb$^{-1}$, is essential, as BEPCII is the only machine working at the charm-threshold energy region. Furthermore, determining these parameters with radiative return events to the $\psi(3770)$ at Belle II will not be achievable with suitable precision even with a data set of 50~ab$^{-1}$. A 20~fb$^{-1}$
sample of $\psi(3770)$ data would lead to an uncertainty of approximately 
0.4$^\circ$ for the $\gamma$ measurement, which will be necessary for the goals of LHCb upgrade I and Belle II. 
Moreover, this improved precision will be essential to allow the even larger data of LHCb upgrade II \cite{Aaij:2244311} to be fully exploited in improving further the knowledge of $\gamma$ and allowing detailed comparison of the results obtained with different  decay modes.  A reasonable time frame for taking the $\psi(3770)$ data of 20 fb$^{-1}$ would be by 2025, when Belle II will have completed accumulation of its 50 ab$^{-1}$ of data and LHCb upgrade I will be mid-way through its period of operation.

\begin{table}[tp]
\centering
\caption{\small Expected $\gamma/\phi_3$ precision in the LHCb~\cite{lhcb_note} and Belle II~\cite{Kou:2018nap} experiments along with the timescales.}
\small\vspace{0.1cm}
\begin{tabular}{llll}
\hline\hline
Runs & Collected / Expected  & Year  & $\gamma / \phi_3$  \\ 
&integrated luminosity & attained  &  sensitivity  \\ \hline
LHCb Run-1 [7, 8 TeV] & 3 fb$^{-1}$  & 2012 & $8^\circ$  \\
LHCb Run-2 [13 TeV] & 6 fb$^{-1}$  & 2018 & $4^\circ$ \\
Belle II Run & 50 ab$^{-1}$ & 2025 & $1.5^\circ$ \\ 
LHCb upgrade I [14 TeV] & 50 fb$^{-1}$  & 2030 & $<1^\circ$ \\
LHCb upgrade II  [14 TeV] & 300 fb$^{-1}$  & ($>$)2035 & $< 0.4^\circ$ \\ \hline\hline
\end{tabular}
\label{tab:exps}
\end{table}

A synergy between the \bes3, LHCb, and Belle II experiments is the best way to accurately determine $\gamma$ in a manner that results in the  uncertainty on $\gamma$ being statistically rather than systematically limited. \tablename~\ref{tab:sf_mode} lists the decay modes of interest that can be measured at \bes3.
Furthermore, for the multi-body $D$ final states the phase-space binning schemes that could be employed are mentioned; increasing the number of bins improves the statistical sensitivity of the measurements as the amount of information loss relative to an unbinned method is reduced. The modes are listed in their approximate order of importance for LHCb measurements. For the Belle II experiment those modes with neutrals are more important due to the larger neutral reconstruction efficiency compared to LHCb; LHCb will have an advantage for two- and four-body $D$ decays containing only prompt charged pions and kaons in the final state.

In addition, it is worth noting  the ability of \bes3 to efficiently reconstruct $D$ modes containing a $K^{0}_{L}$ meson in the final state and determine the relevant strong-phases. These $K^{0}_{L}$ modes provide additional tags (for example Ref.~\cite{M.Nayak2015}) that increase the precision of other strong-phase parameter measurements at the $\psi(3770)$. Furthermore, it may be possible to use $K^{0}_{L}$ modes at Belle II to reconstruct $B^{+}\to DK^{+}$ given the anticipated improvements in $K^{0}_{L}$ reconstruction. Evidence of this is given by the fact that the determination of the UT angle $\beta$ at the $B$ factories benefited from included $B^{0}\to J/\psi K^{0}_{L}$ decays in the measurement.

\begin{table}[tp]
\centering
\caption{\small A priority-ordered list of strong-phase related measurements that
are important for precision measurements of $\gamma$ and indirect
CP violation in charm mixing~\cite{lhcb_note}.
For states that are not self-conjugate, the coherence factor, $R$, and average strong-phase
difference, $\delta$, can be measured~\cite{D.Atwood(2001)}. For self-conjugate states
there are two choices: either a measurement of the CP-even fraction, $F_{+}$~\cite{M.Nayak2015},
or a measurement of the amplitude weighted average cosine and sine of the strong-phase difference,
$c_{i}$ and $s_{i}$, where the index $i$ refers to the phase-space region of the given multi-body
decay~\cite{A.Giri(2003)}.
}\vspace{0.1cm}
\label{tab:sp}
%\begin{tabular}{L{3.cm} L{3.cm} L{9cm}}
\small
\begin{tabular}{lcl}
\hline\hline
Decay mode & Quantity of interest & Comments\\ \hline
& & \footnotesize{Binning schemes as those used in the CLEO-c} \\
$D\to K_{S}^{0}\pi^{+}\pi^{-}$ & $c_{i}$ and $s_{i}$  &  \footnotesize{analysis. With 20 fb$^{-1}$ of data at 3.773 GeV, it might be}\\
& &  \footnotesize{worthwhile to explore alternative binning.}\\ \hline
 & &  \footnotesize{Binning schemes as those used in the CLEO-c }\\
$D\to K_{S}^{0}K^{+}K^{-}$ & $c_{i}$ and $s_{i}$ &  \footnotesize{analysis. With 20 fb$^{-1}$ of data at 3.773 GeV, it might}\\
& &  \footnotesize{be worthwhile to explore alternative binning.} \\ \hline
$D\to K^{\pm}\pi^{\mp}\pi^{+}\pi^{-}$ & $R, \delta$ & \footnotesize{In bins guided by amplitude models, currently} \\
& & \footnotesize{under development by LHCb.}\\ \hline 
$D\to K^{+}K^{-}\pi^{+}\pi^{-}$ & $c_{i}$ and $s_{i}$ & \footnotesize{Binning scheme guided by the CLEO-c model~\cite{CLEO(2012)122002}} or\\
& & \footnotesize{potentially an improved model in the future.}\\ \hline
$D\to\pi^{+}\pi^{-}\pi^{+}\pi^{-}$ & $F_{+}$ or $c_{i}$ and $s_{i}$ & \footnotesize{Unbinned measurement of $F_{+}$. Measurements of } \\
 & & \footnotesize{$F_{+}$ in bins or $c_{i}$ and $s_{i}$ in bins could be explored.}\\ \hline
$D\to K^{\pm}\pi^{\mp}\pi^{0}$ & $R, \delta$ & \footnotesize{Simple 2-3 bin scheme could be considered.}\\ \hline 
$D\to K_{S}^{0}K^{\pm}\pi^{\mp}$ & $R, \delta$ & \footnotesize{Simple 2 bin scheme where one bin encloses} \\  
& & \footnotesize{the $K^{*}$ resonance.}\\ \hline
$D\to\pi^{+}\pi^{-}\pi^{0}$ & $F_{+}$ & \footnotesize{No binning required as $F_{+} \sim 1$.}\\ \hline 
 & & \footnotesize{Unbinned measurement of $F_{+}$ required.} \\
$D\to K_{S}^{0}\pi^{+}\pi^{-}\pi^{0}$ & $F_{+}$ or $c_{i}$ and $s_{i}$ &  \footnotesize{Additional measurements of $F_{+}$ or $c_{i}$ } \\
& & \footnotesize{and $s_{i}$ in bins could be explored.}\\ \hline 
$D\to K^{+}K^{-}\pi^{0}$ & $F_{+}$ & \footnotesize{Unbinned measurement required. Extensions to } \\
&  &  \footnotesize{binned measurements of either $F_{+}$ or $c_{i}$ and $s_{i}$.}\\ \hline 
$D\to K^{\pm}\pi^{\mp}$ & $\delta$ & \footnotesize{Of low priority due to good precision available }\\
& &  \footnotesize{through charm-mixing analyses.}\\
\hline\hline
\end{tabular}
\label{tab:sf_mode}
\end{table}

The measurements of strong phases using quantum-coherent analyses \cite{Gronau:2001nr} at BESIII provide important inputs to the LHCb and Belle II experiments to measure the $\DzDzb$ mixing parameters and
indirect CP violation in charm mixing.
The measurement methods, as described in Refs. \cite{C.Thomas(2012), S.Malde(2015)094032, S.Malde(2011)},
all use the charm strong-phase parameters.
For example, using the $c_i$ and $s_i$ results as inputs to time-dependent measurements of $D^0\to K_S \pip\pim$, the charm mixing parameters $x$ and $y$ can be determined in a model-independent way~\cite{Aaij:2015xoa}.
More precise strong-phase measurements listed in \tablename~\ref{tab:sp}
performed with the current and future $\psi(3770)$ data at \bes3,
will prevent these studies from being limited by strong-phase related uncertainties. Furthermore, the strong-phase measurements have recently been used in time-dependent
CP-violation measurements of $B^{0}\to D^{(*)0}\pi^{0}$ decays by Belle \cite{vorobiev(2016)} to make measurements of $\beta/\phi_1$ that are free of uncertainties related to penguin processes. Such measurements are very attractive at Belle II, so that they can be compared to the ``golden modes" $B^{0}\to J/\psi K^0$ to empirically determine 
the pollution with penguin diagrams in the determination of $\beta/\phi_1$.

%%%%%%%%%%%%%%%% Hai-Bo Li:  add a subsection on the CKMfitter global analysis 
%%%%%%%%%%%%%%%% 2018. 9. 28 

\subsection{Impact on CKM measurements}
\label{subsec:ckmfitter}

In the SM, quark-flavor mixing is described by the $3\times3$ CKM matrix $V_{\rm CKM}$
as shown in Eq.~(\ref{eq:ckm-m}) in Sec.~\ref{subsec:pure-lep}. Unitarity is the only, albeit powerful, constraint on $V_{\rm CKM}$. Without loss of generality, the $V_{\rm CKM}$ can be parameterized in terms of three mixing angles and one phase~\cite{Chau:1984fp}
\begin{equation}
V_{\rm CKM} = \left( \begin{matrix} c^{}_{12}c^{}_{13} & s^{}_{12}c ^{}_{13} &
s^{}_{13} e^{-i\delta} \cr -s^{}_{12}c^{}_{23}
-c^{}_{12}s^{}_{23}s^{}_{13} e^{i\delta} & c^{}_{12}c^{}_{23}
-s^{}_{12}s^{}_{23}s^{}_{13} e^{i\delta} & s^{}_{23}c^{}_{13} \cr
s^{}_{12}s^{}_{23} -c^{}_{12}c^{}_{23}s^{}_{13} e^{i\delta} &
-c^{}_{12}s^{}_{23} -s^{}_{12}c^{}_{23}s^{}_{13} e^{i\delta} &
c^{}_{23}c^{}_{13} \end{matrix} \right) \; ,
%       (2)
\end{equation}
where $c^{}_{ij} \equiv \cos\theta_{ij}$ and $s^{}_{ij} \equiv
\sin\theta_{ij}$ (for $ij=12,\,23$ and $13$). The irremovable phase
$\delta$ is the unique source of CP-violation in quark
flavor-changing processes within the SM. Following the observation of a hierarchy between the different matrix elements, Wolfenstein~\cite{Wolfenstein:1983yz} proposed an expansion of the CKM matrix in terms of the four parameters $\lambda$, $A$, $\rho$, and $\eta$, 
which are widely used in contemporary literature, and which is the parameterization employed in CKMfitter~\cite{Charles:2004jd}.

 \begin{figure}[htb]
  \centering

%\hspace{-0.7cm}

\includegraphics[scale=0.5]{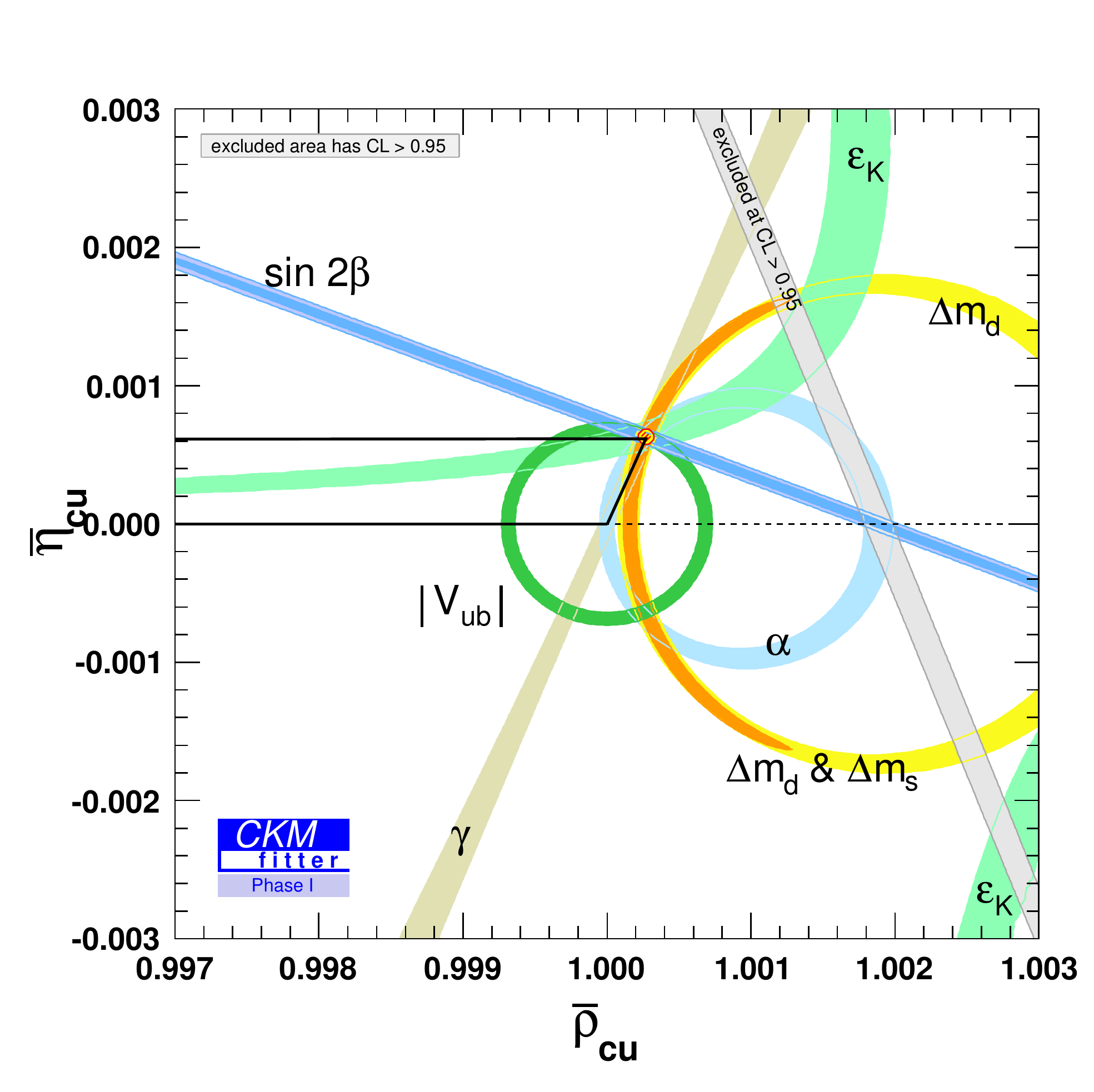}

%\vspace{-0.5cm}

\caption{Individual constraints and the global CKM fit on the
($\overline{\rho}_{cu},\overline{\eta}_{cu}$) plane according to our projections of the experimental status in
2025 (Phase I). The shaded areas have 95\% CL. Only a part of the $D$ meson UT is visible 
(in black solid lines): the two apices associated with large angles are shown, whereas
the missing corner is situated at the origin, far away on the left.}\label{DUT2025}
\end{figure}

\begin{figure}[htb]
  \centering

%\hspace{-0.7cm}

\includegraphics[scale=0.5]{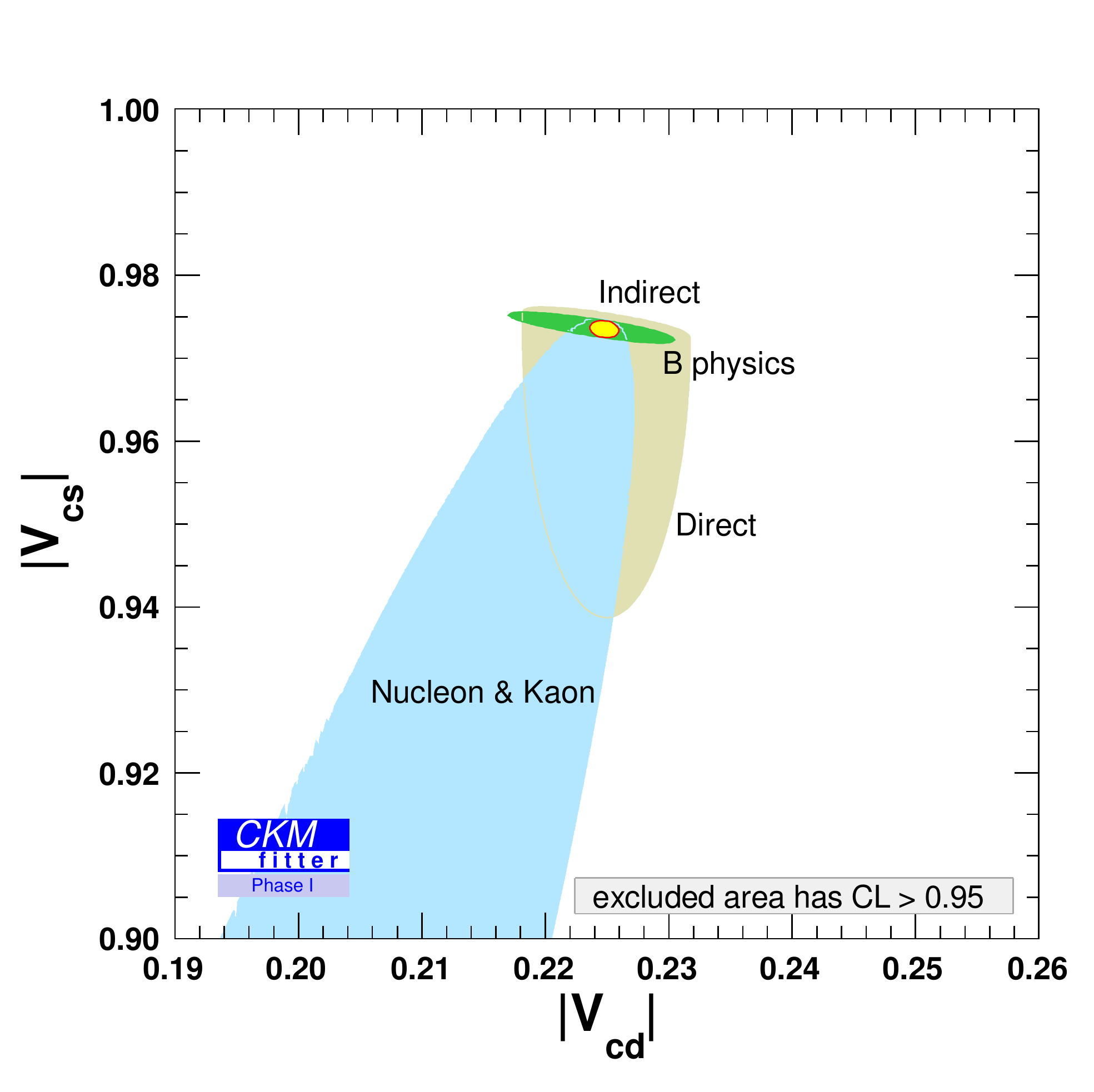}

%\vspace{-0.5cm}

\caption{Constraints in the ($|V_{cd}|$,$|V_{cs}|$) plane for the Phase I data. The indirect constraints (coming from $b$ transitions) are related to $|V_{cd}|$ and $|V_{cs}|$ through unitarity. The direct constraints combine leptonic and SL $D$ and $D_s$ decays from \bes3 experiment. The red hashed region of the global combination corresponds to 68\% CL.}\label{BESIII2025}
\end{figure}

The allowed region in the $\rho$ and $\eta$ space can be elegantly displayed by means of the UT described by the rescaled unitarity relation between the first and the third column of the CKM matrix ($i.e.$, corresponding to
the $b$-meson system).  The UT can be described in the complex $(\bar{\rho}, \bar{\eta})$ plane, where the apex is given by the following phase-convention independent definition~\cite{Buras:1994ec}
\begin{equation}
\bar{\rho} + i\bar{\eta} \equiv-\frac{V^{}_{ud}V^{*}_{ub}}{V^{}_{cd}V^{*}_{cb}}.
\end{equation}
We also propose to represent the combination of the CKM constraints in the plane that is relevant to the $D$ meson UT. In analogy with the exact and rephasing-invariant expression of $(\bar{\rho}, \bar{\eta})$ we define the coordinates of the apex of the $D$ meson UT~\cite{Asner:2008nq}
\begin{equation}
\overline\rho_{cu} +i\,\overline\eta_{cu} \equiv
-\frac{V_{ud}V_{cd}^*}{V_{us}V_{cs}^*},
\end{equation}
where $\overline\rho_{cu}=1+\mathcal{O}(\lambda^4)$ and
$\overline\eta_{cu}=\mathcal{O}(\lambda^4)$. 
One can see that this triangle has two sides with length very close to 1
and a small side of order ${\cal O}(\lambda^4)$, with angles 
$\alpha_{cu}=-\gamma$, $\beta_{cu}=\gamma +\pi +{\cal O}(\lambda^4)$ and
$\gamma_{cu}={\cal O}(\lambda^4)$.

To constrain these parameters, we consider a prospective exercise when BESIII will complete the taking of 20 fb$^{-1}$ data accumulated at the $D\bar{D}$ threshold~\cite{Asner:2008nq} and 6 fb$^{-1}$ data at 4.178 GeV around 2025 (Phase I). At that time, the collected data amount for LHCb will be 23 fb$^{-1}$~\cite{Bediaga:2018lhg}  and for CMS/ATLAS 300 fb$^{-1}$, which is after Run 3 and prior to the start of the HL-LHC. We also take into account the prospective accuracies for Belle II at 50 ab$^{-1}$~\cite{Kou:2018nap} already at Phase I. The individual constraints as well as the combination from the usual
observables are shown for 2025 in the $(\overline\rho_{cu},\overline\eta_{cu})$
plane in Fig.~\ref{DUT2025}. One can also transfer the constraints to the $(|V_{cs}|, |V_{cd}|)$ plane as shown in 
Fig.~\ref{BESIII2025}, which clearly shows the direct contribution from the \bes3 measurements
of $|V_{cs}|$ and $|V_{cd}|$, and allows precise tests of the consistency of CKM determination from 
different quark sectors. 

\begin{figure}[tp]
  \centering

%\hspace{-0.7cm}

\includegraphics[scale=0.5]{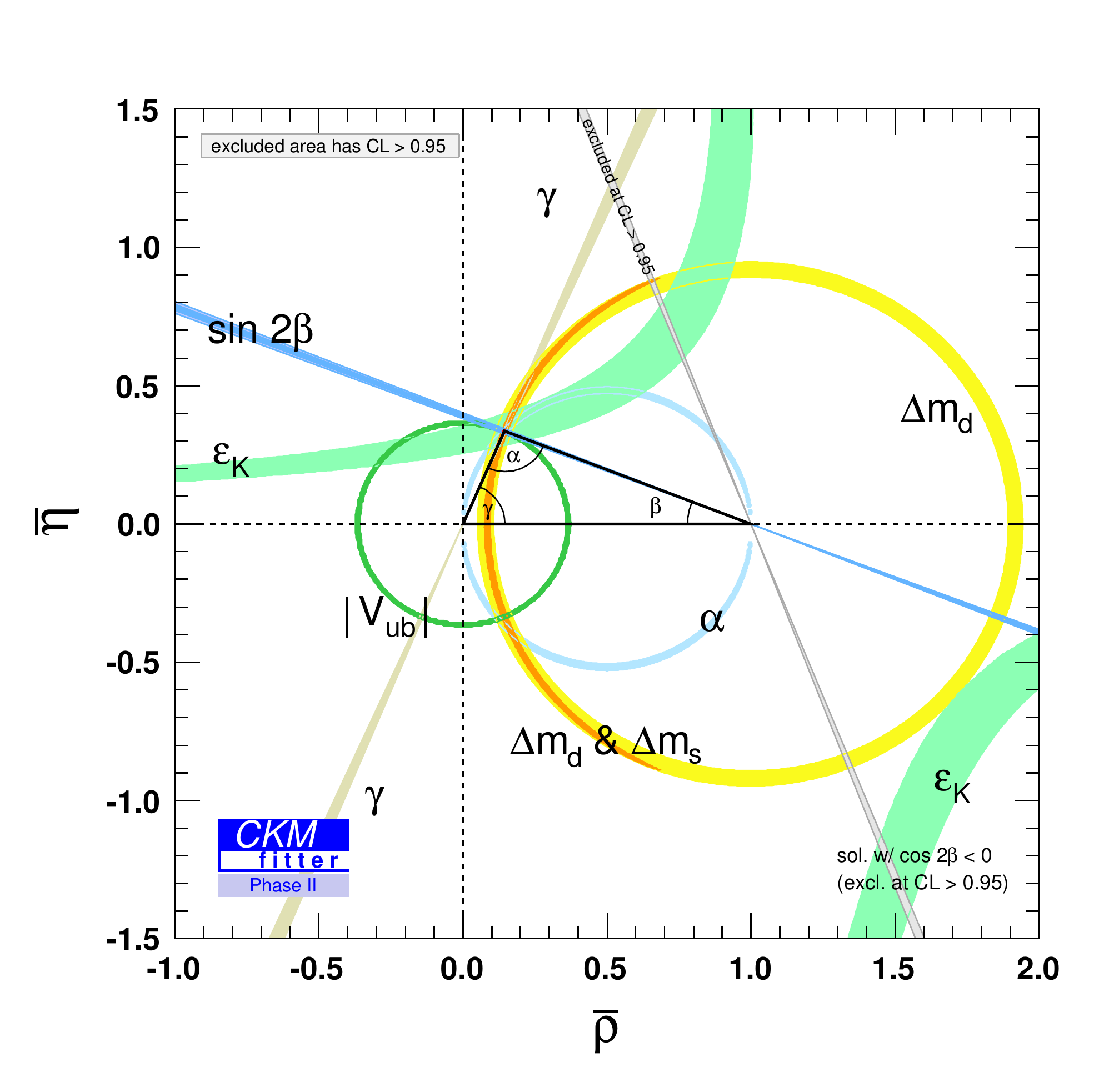}

%\vspace{-0.5cm}

\caption{Evolving constraints and the global fit in the ($\overline{\rho},\overline{\eta}$) plane with the anticipated improvements for the Phase II data ($300$ fb$^{-1}$ from LHCb~\cite{Bediaga:2018lhg}, 3000 fb$^{-1}$ from CMS/ATLAS, and 50 ab$^{-1}$ from Belle II~\cite{Kou:2018nap}). 
The $\gamma$ is expected to have an accuracy around 0.4 degrees (shown in Table~\ref{tab:exps}), thanks to \bes3 charm inputs with 20 fb$^{-1}$ data accumulated at the $D\bar{D}$ threshold~\cite{Asner:2008nq}.   The shaded
areas have 95\% CL.}\label{ckm06:fig_global}
\end{figure}

In the further Phase II beyond 2035, we assume larger data sets of 300 fb$^{-1}$
for LHCb and 3000 fb$^{-1}$ for CMS/ATLAS. The corresponding perspectives of individual constraints and the global
CKM fit on the $(\bar{\rho}, \bar{\eta})$ plane are shown in Fig.~\ref{ckm06:fig_global}, which involve both LHC and Belle II experiments for  the  CKM  measurements related to $B$ and $B_s$ mesons. 
We use available resources for the uncertainties~\cite{Bediaga:2018lhg,Kou:2018nap} assuming that all the measurements agree perfectly well within the SM. 
 As mentioned in Sec.~\ref{subsec:phi3}, thanks to the critical input with 20 fb$^{-1}$ data accumulated at the $D\bar{D}$ threshold~\cite{Asner:2008nq} from BESIII on charm strong phases, the knowledge of the angle $\gamma$ will be improved to 
$0.4^\circ$ or better, allowing for extremely precise tests of the CKM paradigm, and providing a sensitive probe for possible 
new physics contributions. 
 
%%%%%%%%%%%%%%%%%%%%%%%%%%
%%%%%%%%%%%%%%%%%%%%%%%%%%

\subsection{CP violation and $D$ mixing}

It is not competitive to search for CP violation and charm mixing parameters at BESIII,
since the achievable $D^0$ and $\bar D^0$ samples are less than those at LHCb and Belle II 
by more than 2-4 orders of magnitudes. 
However, \bes3 can uniquely explore the quantum coherence of the initial $\DzDzb$
state produced at $\psi(3770)$ to provide constraints
\cite{Ablikim:2015hih} on mixing and CP-violating parameters
\cite{Atwood:2002ak,Asner:2005wf} that have completely different
systematic uncertainties than those of LHCb or Belle II. 
For example,
decays of each $D$ in the $\DzDzb$ state into final states of the same
CP would immediately indicate CP-violation in charm, as the
$D$-states produced at $\psi(3770)$ have opposite CP,
\beq
\Gamma^{++}_{\DzDzb} = \left[
\left(x^2+y^2\right)\left(\cosh^2 a_m - \cos^2 \phi\right) \right] \Gamma^2(D \to f_+),
\eeq
where we employ two $CP=+$ states $f$, such as $f=\pi^+\pi^-$. Also,
$\phi = \arg(p/q)$, $R_m=|p/q|$, and $a_m=\log R_m$
\cite{Atwood:2002ak}.  In addition, superb kinematical
constraints might make \bes3 competitive in studies of CP-asymmetries
in some multi-body $D$ decays~\cite{Bigi:2011em}.

\label{sec:cpv_charm}

% %%%%%%%%%%%%%%%%%%%%%%%%%%%%
\subsection{CPT violation in charm mixing }

CPT is conserved in all local Lorentz-invariant theories, including SM and all its
commonly-discussed extensions. Yet, CPT violation might arise in
string theory or some extra-dimensional models with Lorentz-symmetry
violations in four dimensions. Time evolution studies of
CP-correlated $\DzDzb$ states are complementary to CPT-violation
studies at $B$-factories and the LHC.  While studies of those models
are not particularly well motivated at charm threshold,
CPT-violating effects can still be probed with the quantum-correlated 
$\DzDzb$ states at $\psi(3770)$~\cite{Kostelecky:2001ff}.

\subsection{Absolute measurement of hadronic decays}
\label{sec:precision_d}

The hadronic decay of charmed hadrons provides important information on not only the charmed hadron itself, but also on the light daughters of the hadron. Therefore, studying charm multi-body hadronic decays is a powerful tool for light hadron spectroscopy. Analogously, studying high-statistics hadronic decays of the charmed baryon $\Lambda_c^+$ with a data set of 5~fb$^{-1}$ produced at 4.64 GeV, as discussed in Sec.~\ref{sec:lambda_c}, is useful for improved understanding of light meson and baryon spectroscopy.

%\subsection{Two-body hadronic decays}

Experimental studies of two-body hadronic decays of $D^0$, $D^+$,
and $D^+_s$ mesons are very important for calibrating theoretical models. 
There are different final-state possibilities, namely $PP$, $VP$, $VV$, $SP$, $AP$, and $TP$, where $P$, $V$, $S$, $A$, and $T$ represent
pseudoscalar, vector, scalar, axial vector, and tensor mesons, respectively.
The simplest cases of 
the $PP$ decays of
$D^0$, $D^+$, and $D^+_s$ have been well investigated experimentally in recent years.
However, some of these measurements, particularly of DCS
decays, are still statistically limited. Studies of non-$PP$ final states require sophisticated amplitude analyses of multi-body decays of charmed mesons that 
are not well developed because of either the limited statistics or the high background level. 
Larger $D^{0(+)}$ and $D_s$ data samples are necessary to guarantee clean signal samples with sufficient yields to perform reliable amplitude analyses, so that all types of two-body process can be well understood.

At present, the sums of the BFs for the known
exclusive decays of $D^0$, $D^+$, and $D^+_s$ are all more than 80\%~\cite{pdg2016}.
However, there is still significant room to explore many unknown hadronic decays.
A 20 fb$^{-1}$ sample of data at the $\psi(3770)$ will allow the determination of many BFs of these missing decays to $K\pi\pi\pi$, $KK\pi\pi$, and $KK\pi\pi\pi$, as well as further exploration of the sub-structures in these decays.
Some of them, for example $D^0\to K^-\pi^+\pi^0\pi^0$ and
$D^+\to\bar K^0\pi^+\pi^0\pi^0$,
are expected to have a BF 
in the range 5 to 10\%~\cite{Barlag:1992ww}, which can be improved a lot with the advantage of BESIII in precise $\pi^0$ and photon
detections. 

LHCb has the ability to measure a large number of relative BFs of charm and beauty hadrons, on account of the high yields that result from the large cross section for heavy-flavor production in the forward region. Converting from the BF ratio to the absolute BF
incurs an additional uncertainty from the BF of
the reference mode, such as
$D^0\to K^-\pi^+$, $D^0\to K^-\pi^+\pi^+\pi^-$, $D^+\to K^-\pi^+\pi^+$, 
$D^+_s\to K^-K^+\pi^+$, and $\Lambda_c^+\to pK^-\pi^+$.
Absolute BF measurements
of the other charmed baryons, $\Xi_c^0$ and $\Xi_c^+$ are also strongly desired. Although it is
acknowledged that these modes would be very hard to measure with the current energy of the
collisions at \bes3, the importance of these measurements is stressed here in case future
developments could make them possible.
Relevant discussions on the prospects for measurements of charmed baryons are given in Sec.~\ref{sec:lambda_c}.

With 20 fb$^{-1}$ of data at 3.773 GeV and 6~fb$^{-1}$ around 4.178 GeV at \bes3,
many $D_{(s)}$ decays are expected to be measured with an uncertainty of about 1\%, which will then be limited by the systematic uncertainties related to particle reconstruction in the different final states.
Improved measurements of these absolute BFs at \bes3
will be highly beneficial to some key measurements at LHCb,
since it is expected that the uncertainty of the reference mode
will become the dominant uncertainty in several measurements.
One prime example is the LHCb measurement of
$B\to D^{*}\tau^+\nu_\tau$, which is used to test lepton universality~\cite{lhcb_note}.
\bes3 can make precise measurements of the BFs for
$D^0$, $D^+_s$, and $D^+$ inclusive decays to three charged pions and
to the neutral particles, and exclusive decays to final states with neutral kaons and pions
($\egeg$, $D_s^+\to \eta^\prime\pi^+\pi^0$, $D^+\to \bar K^0\pi^+\pi^+\pi^-\pi^0$,
and $D^{0(+)}\to\eta X$, $X$ denotes any possible particle combinations), which will have a significant impact on optimizing the background models. Another example is the determination of the CKM matrix elements $|V_{c(u)b}|$ via inclusive or exclusive SL $B$  decays,
such as $B\to D^{*}\ell^-\nu_\ell$ or $\Lambda^0_b \to \Lambda^+_c\bar{\ell}^-\nu_\ell$.
Precise measurements of absolute BFs of both
$D^{0(+)}_{(s)}$ and $\Lambda^+_c$ decays (Sec.~\ref{sec:lambda_c})
are crucial for improvement of the $|V_{c(u)b}|$ measurements.

Due to interference effects between the CF and DCS contributions, the BFs of
$D_{(s)}\to K^0_{S} X$ and $D_{(s)}\to K^0_{L} X$ are not expected to be equal.
The recoil-mass method, or that developed in Ref.~\cite{cleo_kspi}, allows the efficient reconstruction of $D_{(s)}\to K^0_{L} X$ decays at \bes3.
A difference of about 10\% has been observed in $D^0\to K^0_{S,L}\pi^0$ decays
at CLEO-c~\cite{cleo_kspi} and has been confirmed at \bes3~\cite{bes3_kspi}.
The two-body decays
$D^0\to K^0_{S,L} X$ ($X=\pi^0,\,\eta,\,\eta^\prime,\,\omega,\,\phi$)
are expected to have an asymmetry under the same mechanism~\cite{yufs},
which can be tested with the current $\psi(3770)$ data,
but the uncertainties will be dominated by statistical uncertainties. Furthermore, measurements of these modes are of interest in understanding the width difference in the neutral $D$ system~\cite{glwprime}.
With the future 20~fb$^{-1}$ and 6~fb$^{-1}$ of data taken
at $\sqrt s=$ 3.773 and 4.178 GeV,
these measurements will be performed with much better precision.

\newpage

%% file: Charm/charmed_baryon.tex
\section{Charmed baryons}
\label{sec:c_baryon}

\subsection{$\Lambda_c^+$ physics}
\label{sec:lambda_c}

Studies of charm baryons have been ongoing since 1975~\cite{Cazzoli:1975et}, with all ground states of singly charmed baryons, as well as some excited states, having been observed.
The constituents of the lightest charmed baryon ($\Lambda_c^+$)  are one diquark ($ud$) and one heavy charm quark ($c$), where, relative to the heavy quark, the light diquark is in a net quantum state of spin zero and isospin zero. 
In a naive spectator model the $\Lambda_c^+$ decays dominatedly through the weak amplitudes $c \to W^+ s$ and  $c \to W^+ d$
at leading order, which leads to a simpler theoretical description in non-perturbative models than in the case of charmed mesons.
Hence, studying $\Lambda_c^+$ decays allows a
deeper understanding of strong and weak interactions in the charm sector, which is complementary to that provided by charmed mesons.
In addition, the $\Lambda_c^+$ is the cornerstone of the charmed baryon spectra.
Improved knowledge of $\Lambda_c^+$ decays
is essential to the studies of the whole charmed baryon family.
Furthermore, this knowledge will provide important information to the studies of
beauty baryons that decay into final states involving $\Lambda_c^+$.

Compared to the significant progress in the studies of charmed
mesons ($D^0$, $D^+$, and $D^+_s$)  in both theory and experiment, the advancement of our understanding of charmed baryons has been relatively slow during the past 40 years.
Until 2014, no absolute measurements of the decay rates of the $\Lambda_c^+$ had been performed, with almost all these rates having been measured relative to the normalization mode $\Lambda_c^+\to pK^-\pi^+$, whose BF suffered from a large uncertainty of 25\%.
This overall situation has changed since 2014, when BESIII collected a data set of $e^+e^-$ annihilation at $\sqrt{s}=4.6$ GeV, corresponding to an integrated luminosity of 567 pb$^{-1}$.
At this energy, the $\Lambda_c^+\bar \Lambda_c^-$ pairs are produced in pairs with no accompanying hadrons. The total number of $\Lambda_c^+\bar \Lambda_c^-$ pairs produced is approximately 100,000.
This threshold data set provides a clean environment to systematically investigate the production and
decays of the $\Lambda_c^+$.  Precise BF measurements of $\Lambda_c^+\to pK^-\pi^+$ were reported by Belle~\cite{belle_pkpi} and BESIII~\cite{plb_bfs}; the combined precision of the $\Lambda_c^+\to pK^-\pi^+$ BF is 5.2\%, a five-fold reduction of the uncertainty with respect to the previous result. In addition, more analyses were implemented at BESIII, Belle, and LHCb. At BESIII a series of BF measurements have been reported, including
\begin{itemize}
\item
absolute BF measurement of
$\Lambda_c^+\to\Lambda \ell^+\nu_\ell$~\cite{plb_lev, Ablikim:2016vqd}, which motivated the first LQCD calculation of this channel~\cite{prl115_221805};
\item
absolute BF measurements of 12 Cabibbo-favored (CF) decays of the $\Lambda_c^+$~\cite{plb_bfs}, including
$\Lambda_c^+\to pK^0_{S}$, $pK^-\pi^+$, $pK^0_{S}\pi^0$,
$pK^0_{S}\pi^+\pi^-$, $\Lambda \pi^+$, $\Lambda \pi^+\pi^0$, $\Lambda
\pi^+\pi^+\pi^-$, $pK^-\pi^+\pi^0$, $\Sigma^0\pi^+$,
$\Sigma^+\pi^0$, $\Sigma^+\pi^+\pi^-$, and $\Sigma^+\omega$;
\item
studies of singly-Cabibbo-suppressed (SCS) decays
$\Lambda_c^+\to p\pi^+\pi^-$, $p K^+K^-$, $p\eta$, and $p\pi^0$~\cite{plb_ppipi,Ablikim:2017ors};
\item
observation of modes with a neutron in the final state, $\Lambda_c^+\to nK^0_{S}\pi^+$~\cite{Ablikim:2016mcr} and $\Sigma^-\pi^+\pi^+\pi^0$~\cite{Ablikim:2017iqd}.
\item
measurements of the absolute BFs of $\Lambda_c^+\to\Xi^0 K^+$, $\Xi^{*0} K^+$~\cite{Lc_XiK}, $\Sigma^+\eta$, $\Sigma^+\eta'$~\cite{Ablikim:2018czr}, $\Sigma(1385)^+\eta$ and $\Lambda\eta\pi^+$~\cite{Ablikim:2018byv}.
\item
studies of inclusive $\Lambda X$ decay~\cite{Lc_lambdax} and inclusive electronic decay~\cite{Lc_ex}.
\end{itemize}

Although much progress has been made since 2014, the knowledge of $\Lambda_c^+$ decays is
still very limited in comparison to that of charmed mesons. Firstly, only a single SL decay mode $\Lambda
\ell^+\nu_\ell$ has been observed. Secondly, the DCS modes have not been systematically studied. Thirdly, the known exclusive decays of the $\Lambda_c^+$ account for only about 60\% of the total BF. Many channels are still unobserved. In particular, information about decays that involve a neutron is minimal considering such decays should account for nearly half of the $\Lambda_c^+$ decay rate. A thorough investigation of these channels requires a much larger data set taken in the low-multiplicity environment obtained at threshold in $e^+e^-$ collisions.

\subsection{Prospects in $\Lambda_c^+$ physics}

The approved energy upgrade project of BEPCII will increase the collision energy up to 4.9 GeV. Together with commissioning of the top-up mode, we will be able to take 5 fb$^{-1}$ of data at 4.64 GeV, which corresponds to the energy point with the expected peaking cross section of the $\Lambda_c^+\bar{\Lambda}_c^-$ pairs. This data set will have more than 16 times the statistics of the current BESIII data set at 4.6 GeV, and will allow to improve the precision of the $\Lambda_c^+$ decay rates to a level comparable to those of the charmed mesons, as listed in \tablename~\ref{tab:pre}.  In addition, it will provide an opportunity to study many unexplored physics observables related to $\Lambda_c^+$ decays.  In particular, this advancement will boost our understanding of the non-perturbative effects in the charmed baryon sector.

\begin{sidewaystable}
%\end{multicols}
%\begin{table}[tp]
\centering
%\rotatebox{90}{
%\begin{minipage}{\textheight}
\small
\caption{\small Measured or projected precisions of charmed hadrons, along with the relative precision in parenthesis. For the future $\Lambda_c^+$ precision, it is estimated 
for 5 fb$^{-1}$ of data at $\sqrt{s}=4.64$ GeV.} \vspace{0.1cm}
\begin{tabular}{llll} \hline\hline
 & Leading hadronic decay & Typical two-body decay  & Leading SL decay  \\ \hline
\multirow{3}{*}{$\Lambda_c^+$}  &  $\BR(K^-p\pi^+)=$ &  $\BR(K^{0}_{S} p)=$  &  $\BR(\Lambda e^+\nu_e)=$  \\
 &  2014: $(5.0\pm1.3)\%~(26\%)$ &  2014: $(1.2\pm0.3)\%~(26\%)$  & 2014:  $(2.1\pm0.6)\%~(29\%)$  \\
         &  2017(w/ BESIII): $(6.35\pm0.33)\%~(5.2\%)$ &  BESIII: $(1.52\pm0.08)\%(~5.6\%)$  & BESIII:  $(3.63\pm0.43)\%~(12\%)$  \\
         &  5 fb$^{-1}$:  $\frac{\delta \BR }{\BR} < 2\%$ &  5 fb$^{-1}$:  $\frac{\delta \BR }{\BR} < 2\%$ &  5 fb$^{-1}$:  $\frac{\delta \BR }{\BR} \sim 3.3\%$ \\  \hline
$D^0$   &  $\BR(K^-\pi^+)~~~~\,\,=(3.89\pm0.04)\%~(1.0\%)$ & $\BR(K_{S}^{0}\pi^0)~\,=(1.19\pm0.04)\%~(3.4\%)$  &  $\BR(K^-e^+\nu_e)=(3.53\pm0.03)\%~(0.8\%)$  \\
$D^+$   &   $\BR(K^-\pi^+\pi^+)~=(8.98\pm0.28)\%~(3.1\%)$ &  $\BR(K_{S}^{0}\pi^+)~=(1.47\pm0.08)\%~(5.4\%)$  &  $\BR(K^0_{S} e^+\nu_e)\,=(4.41\pm0.07)\%~(1.5\%)$  \\
$D_s^+$   &  $\BR(K^-K^+\pi^+)=(5.45\pm0.17)\%~(3.8\%)$ &  $\BR(K_{S}^{0} K^+)=(1.40\pm0.05)\%~(3.6\%)$  &  $\BR(\phi e^+\nu_e)~~~=(2.39\pm0.23)\%~(9.6\%)$  \\ \hline\hline
\end{tabular}
\label{tab:pre}
%\end{minipage}
%}
%\end{table}
\end{sidewaystable}

%\begin{multicols}{2}

\begin{table*}[tp]
\centering
\caption{\small Expected rates of the SL modes and estimated precision for 5 fb$^{-1}$ of data at $\sqrt{s}=4.64$ GeV.} \vspace{0.1cm}
\small
\begin{tabular}{llc} \hline\hline
Mode  &  Expected rate (\%) & Relative uncertainty (\%)  \\ \hline
$\Lambda_c^+\to\Lambda\ell^+\nu_\ell$  &  3.6~\cite{epjc76_628,prl115_221805}  & 3.3 \\
$\Lambda_c^+\to\Lambda^* \ell^+\nu_\ell$  &  0.7~\cite{prc72_032005,prd93_014021}  & 10 \\
$\Lambda_c^+\to N K e^+\nu_e$   & 0.7~\cite{prc72_032005}  & 10 \\
$\Lambda_c^+\to \Sigma\pi \ell^+\nu_\ell$  & 0.7~\cite{prc72_032005}  & 10 \\
$\Lambda_c^+\to n e^+\nu_e$  & 0.2~\cite{epjc76_628,prd90_114033,prd93_056008}; 0.4~\cite{Meinel:2017ggx} & 17\\ \hline
\hline %
\end{tabular}
\label{tab:sl}
\end{table*}

A larger data set will guarantee the first absolute measurement of the form factors in the SL decay
$\Lambda_c^+\to \Lambda \ell^+\nu_\ell$, which is crucial for calibrating the various theoretical calculations.
Besides $\Lambda_c^+\to\Lambda\ell^+\nu_\ell$, more SL modes can be identified, such as those listed in \tablename~\ref{tab:sl}. According to the predicted rates in model calculations, the new CF modes $\Lambda_c^+\to p K^-e^+\nu_e$ and $\Sigma\pi e^+ \nu_e$ will be established for the first time with the double-tag technique. For the SCS mode,  studying $\Lambda_c^+\to n e^+\nu_e$ will be challenging due to the presence of two missing particles in the final state and the dominant $\Lambda_c^+\to \Lambda e^+\nu_e$ backgrounds. However, we still have the opportunity to identify the decay by taking advantage of the well constrained kinematics, the clean reaction environment and neutron shower information inside the electromagnetic calorimeter. Meanwhile, another SCS mode $\Lambda_c^+\to p \pi^-e^+\nu_e$ can be searched for in the enlarged data set.

The hadronic weak decay of a singly charmed baryon is expected to violate parity conservation. For instance, the two-body decay, $\Lambda_c^+\to\Lambda\pi^+$, proceeds via a $W$-interaction, $c\to W^+ + s$,  in which parity is not conserved. The $\Lambda$ and $\pi^+$  particles are allowed to be in an $S$- or $P$-wave state. The effects of parity violation are determined from the polarization of the charmed baryons, which is characterized by the angular distribution of the $\Lambda$ in the $\Lambda_c^+$ rest frame, taking the form of ${dN\over d\cos\theta_\Lambda}\propto 1+\alpha_{\Lambda\pi}\cos\theta_\Lambda$, where $\alpha_{\Lambda\pi}$ is the decay asymmetry parameter~\cite{Wang:2016elx,Cheng:2015iom}. 
In addition the decay asymmetry allows discrimination between different theoretical models, as listed in Ref.~\cite{Cheng:2015iom}. Some decay asymmetry parameters, $\egeg$, $\alpha_{\Lambda\pi}$ for $\Lambda_c^+\to\Lambda\pi^+$, $\alpha_{\Sigma^+\pi^0}$ for $\Lambda_c^+\to\Sigma^+\pi^0$, and $\alpha_{\Xi^-\pi^+}$ for $\Xi_c^0\to\Xi^-\pi^+$, have been studied previously, but with limited precision~\cite{pdg2016}.
Therefore, improved measurements are desirable, as they will shed light on the decay mechanism and allow searches for
CP asymmetries in the charmed baryon sector.
In addition, more decay asymmetry parameters in $\Lambda_c^+\to\Sigma^0\pi^+$, $p\bar{K}^0$, and $\Xi_c^0\to\Xi^-\pi^+$
can be accessed. Based on the sensitivity with the current \bes3 data set, a $16\times$ larger data set would result in an approximate precision of 4\% and 6\% for measuring $\alpha_{\Lambda\pi^+}$ and $\alpha_{\Sigma^0\pi^+}$, respectively.

The weak radiative decay $\Lambda_c^+ \to \gamma \Sigma^+$ is predicted to have a BF of 10$^{-5}$ to $10^{-4}$. 
The new data set will push the experimental sensitivity to 10$^{-5}$, which provides an opportunity to measure this process for the first time. Moreover, the SCS radiative decay $\Lambda_c^+ \to \gamma p$ can be searched for.

We will have better sensitivity to explore the SCS modes, which at present have limited precision or have not been studied before. Meanwhile, more modes with the neutron or $\Sigma^-$ in the final state can be accessed. This will significantly enhance our knowledge of the less well known decays, and will allow improved studies of $\Lambda_c^+\to p\pi^0$ and first searches for $n \pi^+$ and $n K^+$.

Thorough analysis of the involved intermediate states can be carried out in the decays to multi-body final states.  This can be achieved by implementing amplitude analyses of the copious hadronic decays, such as
$\Lambda_c^+\to pK^-\pi^+$, $pK^0_{S}\pi^0$,
$pK^0_{S}\pi^+\pi^-$, $\Lambda \pi^+\pi^0$, $\Lambda \pi^+\pi^+\pi^-$,
$pK^-\pi^+\pi^0$, and $\Sigma^+\pi^+\pi^-$. From these analyses, more two-body decay patterns of $\Lambda_c^+\to B^{\frac{3}{2}}P$ and $ B^{\frac{1}{2}}V$ can be extracted.
Here, $B^{\frac{3}{2}}$ and $B^{\frac{3}{2}}$ denote baryon states with
isospin $\frac{3}{2}$ and $\frac{1}{2}$, respectively.
Also, the $\Lambda_c^+$ decays, acting as an isospin filter, provide a good place to study light-hadron spectroscopy, such as the study  of $\Lambda^*$ and scalar meson states via the weak decays $\Lambda_c^+\to  \Sigma \pi \pi$, $N K \pi $, and $ \Lambda \pi^+ \eta$~\cite{Hyodo:2011js,Miyahara:2015cja,Xie:2016evi,Ablikim:2018byv}.

\subsection{$\Sigma_c$ and $\Xi_c$ physics}

Above 4.88 GeV, the $\Sigma_c$ baryon can be produced via $e^+e^- \to \Sigma_c \bar{\Lambda}_c^- \pi$. The width of the
$\Sigma_c^{0(++)}$ has been well determined by Belle, while only an upper limit on the width of the $\Sigma_c^{+}$
was determined. \bes3 will provide an improved width measurement of the $\Sigma_c^{+}$ via the dominant decay  $\Sigma_c^+ \to \Lambda_c^+\pi^0$, which is useful to test the overall understanding of the decay dynamics.
Moreover, \bes3 can also search for the decay of $\Sigma_c\to\Lambda_c^+\gamma$, which is theoretically expected to occur with a BF of approximately 1\%.
Above 4.74 GeV, the sum of the $\Lambda_c^+$ and $\Sigma_c^+$ masses, the isospin-violating EM reaction $e^+e^- \to \Sigma_c^+ \bar{\Lambda}_c^- $ happens. It is interesting to measure its cross section near threshold, whose ratio to the cross section of  $e^+e^- \to  \Lambda_c^+ \bar{\Lambda}_c^- $ will provide insight of the vaccum productions of $c\bar{c}$ and $s\bar{s}$ pairs. In addition, it is possible to study the $\Sigma_c$ baryon pair production at 4.91 GeV through  $e^+e^- \to \Sigma_c \bar{\Sigma}_c $. Relative to the production rate of $e^+e^- \to \Lambda_c^+ \bar{\Lambda}_c^-$, one can explore the mechanism of generating `good' spin-0 and `bad' spin-1 $u$-$d$ diquarks inside the $\Lambda_c$ and $\Sigma_c$, respectively, near threshold.

For the charmed baryons $\Xi^{0}_{c}$ and $\Xi^{+}_{c}$,
the relative uncertainties of the measured BFs are large, and most of the decays have not yet been studied experimentally~\cite{pdg2016}.
Belle performed first measurements of the absolute BFs of $\Xi^{0}_{c}$ and $\Xi^{+}_{c}$ decays in $\bar{B}{}^{-(0)}\to \bar{\Lambda}{}_c^- \Xi_c^{0(+)}$~\cite{Li:2018qak,Li:2019atu}. Their results are given as 
\begin{displaymath}
\begin{aligned}
\BR(\Xi_c^0 \to \Xi^-\pi^+) & =(1.80 \pm 0.50\pm 0.14)\%,\\
\BR(\Xi_c^0 \to \Lambda K^- \pi^+) &=(1.17 \pm 0.37\pm 0.09)\%,\\
\BR(\Xi_c^0 \to p K^- K^- \pi^+) &=(0.58 \pm 0.23\pm 0.05)\%,\\
\BR(\Xi_c^+ \to \Xi^- \pi^{+} \pi^+)&=(2.86 \pm 1.21\pm 0.38)\%, \\
\BR(\Xi_c^+ \to p K^- \pi^+)&=(0.45 \pm 0.21\pm 0.07)\%.
\end{aligned}
\end{displaymath}
So far, their statistical uncertainties are about 30$\sim$40\% and their systematic uncertainties are about 10\%. The statistical uncertainties can be suppressed to about 5\% at Belle II with 50 times more data set by 2025. However, the systematic uncertainties have less potential to be improved further. So the overall uncertainties will be below 10\%. 

If BEPCII has the possibility of increasing the cms energy above
4.95 GeV, which is just above the mass of $\Xi_c$ pairs, we will be able to perform absolute BF measurements of $\Xi_c$ decays in the same fashion as is done for the $\Lambda_c^+$. If we assume their production cross sections of $e^+e^-\to \Lambda_c^+ \bar{\Lambda}_c^-$ and $\Xi_c \bar{\Xi}_c$ are at the same level near mass threshold, the precisions at BESIII are expected to be competitive or superior to the Belle II results, depending on different modes. These precise absolute BF measurements will be extremely useful inputs for studying the $b$-baryon decays~\cite{lhcb_note}. Furthermore, many of the missing hadronic and SL decays of $\Xi_{c}$ can be studied for the first time. The data set taken at this energy can also be used to study the triplet states $\Sigma_c^{++}$,  $\Sigma_c^{+}$, and  $\Sigma_c^{0}$. 

%% file: Charm/charmed_baryon_em.tex
\subsection{The EM structure of charmed baryons}
\label{subsec:c_baryon_EM}

As discussed in Chapter \ref{chapter:qcd} of this White Paper, the \bes3 experiment is perfectly suited to perform precision studies of hyperon structure. The coming upgrades of  BEPCII opens up new avenues to study single-charm hyperons. In this section, we will discuss these possibilities and the prospects of performing tests of
CP violation in hyperon decays.

Let's consider a hyperon-antihyperon pair $Y\bar{Y}$, where $Y$ and $\bar{Y}$ have spin $1/2$, produced in $e^+e^-$ annihilation.  The differential cross section can be parameterized in terms of electric and magnetic form factors, $G_E(q^2)$ and $G_M(q^2)$, which are related to helicity flip and non-flip amplitudes, respectively. These are linear
combinations
constructed from Dirac and Fermi unconstrained form factors $F_1$ and $F_2$:
 $G_E=F_1+F_2$ and  $G_E=F_1+\tau F_2$, where
 $\tau = \frac{q^2}{4m^2_{Y}}$, and $m_{Y}$ is the $Y$ mass.
The differential cross section can be expressed as

\begin{equation}
\label{eta_R}
\frac{d\sigma}{d\cos\theta}\propto 1+\eta\cos^2\!\theta \; ,
\end{equation}
where 
$\theta$ is the polar angle of $\Lambda_c^+$ in the $e^+e^-$ center-of-msss system,
$-1\le\eta\le 1$ and is related to the form-factor ratio $R=|G_E/G_M|$ in the following way
\begin{equation}
R =\sqrt{\tau}\sqrt{\frac{1-\eta}{1+\eta}}.
\label{eta_def}
\end{equation}

The form factors are complex in the time-like region, $i.e.$,
they can be written as $G_M(q^2)=|G_M(q^2)|e^{i\Phi_M}$ and $G_E(q^2)=|G_E(q^2)|e^{i\Phi_E}$, from which the relative phase can be defined: $\Delta\Phi = \Phi_E-\Phi_M$.
This non-trivial relative phase has an impact on
spin projections and spin correlations
of the produced hyperons and could be detected experimentally via the
weak decay of hyperons.

The challenge in charmed baryon studies is the lack of prominent decay modes
with large BFs. All two-body decay channels have a BF of (1-2)\% \cite{pdg2016}. Consequently, even a large data
set yields only a few hundred reconstructed events per channel.  In particular,  only the single-tag method can be applied to produce an adequate sample.

The most straightforward mode to study the $\Lambda^+_c$ polarization is the
$\Lambda^+_c \to K_{S}^0 p$ channel. Then, the formalism outlined in Ref. \cite{Faldt:2017kgy} can be used, adapted for single-tag $\Lambda_c^+ \to K_{S}^0 p$ measurements by integrating over the antiproton angles:
\begin{equation}
%\begin{split}
{\cal{W}}({\boldsymbol{\xi}})=4\pi(1+\eta\cos^2\theta+\alpha_{{K_{S}^0}p}\sqrt{1-{\eta}^2}\sin({{\Delta\Phi}})(\sin\theta\cos\theta\sin\theta_1\sin\phi_1)),
\label{eq:diffincl1}
%\end{split}
\end{equation}
where $\boldsymbol{\xi}$ is a vector of the involved parameters and variables,
$\theta$ is the polar angle of $\Lambda_c^+$ in the $e^+e^-$ center-of-msss system,
$\theta_1$\,($\phi_1$) is the solid angle in the $\Lambda_c^+$\,($\bar \Lambda_c^-$) helicity system,
and $\alpha_{K_S^0 p}$ is the decay-asymmetry parameter.
The  disadvantage is that the decay asymmetry $\alpha_{K_S^0 p}$ is not known, so
only the product $\alpha_{K_S^0 p}\sin({{\Delta\Phi}})$ can be determined. However,
if one studies sequential decays such as $\Lambda^+_c \to \Lambda \pi^+, \Lambda \to
p\pi^-$, using the formalism outlined in Ref. \cite{Faldt:2017lam}, the asymmetry parameters $\beta$ and $\gamma$ are also accessible.

A measurement ~\cite{Ablikim:2019zwe} of $\Delta\Phi$ has been made using a data set collected by the \bes3 detector at $\sqrt{s}=4.6$ GeV (corresponding to $\tau=1.012$), using an integrated luminosity of $\mathcal{L}_{\textrm{int}}=0.6$ $\textmd{fb}^{-1}$.
The value reported is $\sin(\Delta\Phi)=-0.28\pm0.13\pm0.03$.
It is expected that the phase initially increases when $\sqrt{s}$ goes higher but the most
important improvement should come from larger statistics. Therefore, this motivates a high statistics study of the process $\eelclc$ at the energy providing the highest yield of $\Lambda_c$.
In particular, the reaction has maximum cross section at about 4.64 GeV, 
as expected from Belle data. This energy corresponds to a kinematic factor of $\tau=1.034$.

Another interesting study of the $\Lambda^+_c$ form factor would be with a kinematic factor of approximately $\tau=1.058$, which is the value at which the \bes3 measurement of the $\Lambda$ was made and yielded the result $\Delta \Phi = (37\pm12\pm7)^{\circ}$. Such a measurement will enable the first complete extraction of the EM form factor of a charmed hyperon, thus shedding light on the role of strangeness and charm in hadron structure, and provide a systematic comparison with the strange partner.

Several three-body decay channels have BFs five to seven times larger relative
to the two-body decays \cite{pdg2016}. In Ref.~\cite{Ablikim:2017lct}, it was found that for an integrated luminosity of 567~pb$^{-1}$ the number of tagged $\Lambda_c^+ \to pK^-\pi^+$ was around 3000, and similarly for the 
charge-conjugate $\bar{\Lambda}_c^- \to \bar{p}K^+\pi^-$. It may therefore be a better strategy to construct an angular observable for a three-body decay that is sensitive to the EM phase $\Delta\Phi$
\cite{Faldt:2017kgy,stefan,bigi}. With the data set proposed, such strategies can be pursued to give complementary measurements of $\Delta\Phi$.

Tests of CP violation can be made for two-body decays of charm hyperons as for strange hyperons; suitable observables for
CP tests have been derived, $\egeg$,  in Ref. \cite{hamann}. These asymmetries are common in terms of the asymmetry parameters measured separately for $\Lambda_c^{+}$ and $\bar{\Lambda}_c^{-}$, which can be done in the large data proposed above the threshold for production.

%% file: Charm/charm_summary.tex
%
\section{Summary}
In this chapter we have presented the physics studies that are possible with data corresponding to integrated luminosities of 20~fb$^{-1}$ at $\sqrt s=$ 3.773~GeV, 6~fb$^{-1}$ at $\sqrt s=$4.178~GeV, and 5~fb$^{-1}$ at $\sqrt s=$ 4.64~GeV; these cms energies correspond to the optimal values to accumulate samples of $D\bar{D}$, $D^{*+}_{s}D_{s}^{-}$, and $\Lambda^{+}_{c}\bar{\Lambda}^{-}_c$ events at threshold, respectively. Such samples at threshold allow a double-tag technique to be employed where the full event  can be reconstructed, even if it contains one undetected particle such as a $\nu$ or $K_{L}^{0}$ meson. These samples provide a unique environment to measure the absolute BFs of charmed hadrons to leptonic, SL and hadronic final states, with very low levels of background. Such measurements provide rigorous tests of QCD, CKM unitarity and LFU that complement similar studies of beauty hadrons. Furthermore, the 20~fb$^{-1}$ of data of coherent $\psi(3770)\to D^{0}\bar{D}^{0}$ events allow measurements of the strong-phase difference between the $D^{0}$ and $\bar{D}^{0}$ that are essential inputs to determining the UT angle $\gamma$ in a model-independent fashion from $B$ decays at LHCb and its upgrades, as well as at Belle~II. These strong-phase differences can only be determined using this data set to the required level of precision. In addition, these strong-phase measurements are important ingredients of model-independent measurements of $D^{0}\bar{D}^{0}$ mixing and searches for indirect
$\mathit{CP}$ violation in $D^{0}$ decay. Finally, there are additional studies of charmed baryons that can be performed related to their electromagnetic form factors and, if the BEPCII cms energy is upgraded, studies of the absolute BFs of $\Sigma_c$ and $\Xi_c$. Table~\ref{tab:prospect} presents the precision prospects on some key measurements in $D^{0(+)}$, $D^+_s$, and $\Lambda_c^+$ based on the proposed data set, and a comparison with Belle II.

\begin{table}[htp]
\centering
\caption{\label{tab:prospect}
Prospects on some key measurements at the future \bes3, and comparison with Belle II. `NA' means `not available'  and `--' means `no estimation'. }
\begin{small}
\begin{tabular}{lcccc}
\hline\hline
Observable & Measurement  & \bes3 & Belle II \\ \hline
$\mathcal B(D^+\to \ell^+\nu)$&$f_{D^+}|V_{cd}|$&1.1\%&1.4\% \\
$\mathcal B(D^+_s\to \ell^+\nu)$&$f_{D^+_s}|V_{cs}|$&1.0\%&1.0\% \\
$d\Gamma(D^{0(+)}\to\bar K\ell^+\nu)/d q^2$&$f^K_{+}(0)|V_{cs}|$&0.5\%&0.9\% \\
$d\Gamma(D^{0(+)}\to\pi\ell^+\nu)/d q^2$&$f^\pi_{+}(0)|V_{cd}|$&0.6\%&1.0\% \\
$d\Gamma(D^+_s\to\eta\ell^+\nu)/d q^2$&$f^\eta_{+}(0)|V_{cs}|$&0.8\%& -- \\
Strong phases in $D^0\bar D^0$&Constraint on $\gamma$ & $<0.4^\circ$ & N/A  \\
$\Lambda_c^+\to pK^-\pi^+$& $\mathcal B$ & 2\% & 3\%  \\
$\Lambda_c^+\to \Lambda \ell^+\nu$& $\mathcal B$ & 3.3\% & --  \\
\hline\hline
\end{tabular}
\end{small}
\end{table}

%% file: New_physics/new_physics.tex
\chapter[Exotic Decays and New Physics]{Exotic Decays and New
  Physics}

\label{chapter:newphy}
\input{New_physics/new_physics_main.tex}

\input{New_physics/bib_NP.tex}

%% file: New_physics/new_physics_main.tex
\section{Introduction}
\label{sec:intro}

With the discovery of the Higgs boson in 2012, the Standard
Model (SM) has been firmly established. However, there are many
compelling reasons to believe that the SM is not the ultimate theory,
and the search for physics beyond the SM is well motivated.

There is strong synergy between direct and indirect searches for New
Physics (NP).  To identify possible NP paradigms, results from both
low energy electron-positron colliders and high energy hadron
colliders are needed \cite{Zhu2014}. Studies performed at
electron-positron collider experiments such as \bes3 may indicate
hints of NP that could be directly probed at an energy frontier
experiment, or even make some discoveries directly.

With high luminosity, clean collision environment, and excellent
detector performance, the \bes3 experiment has great potential to
perform searches for NP. BESIII has already published some NP search
results based on the existing data sets. There are still some unique
opportunities worth further exploring. In general, NP searches at
\bes3 could be classified into three broad categories, which are
demonstrated in this chapter:

\begin{enumerate}\addtolength{\itemsep}{-0.5\baselineskip}
\item
Processes that are {\it allowed} in the SM.

NP searches of this type include testing relations among SM-allowed
processes that are known to hold only in the SM, but not necessarily
in models beyond the SM. Some examples have been discussed in depth by
previous chapters, such as testing CKM triangle relations, precision
measurement of $D_q \to \ell \bar \nu$ ($q=d$, $s$; $\ell=e$, $\mu$, $\tau$), precision QCD tests etc. We
focus on some additional topics,
such as weak decays of charmonium states in
Sec.~\ref{sec:charmoniumweak}, and rare radiative and rare leptonic
decays of $D$ mesons in Sec.~\ref{rare:charm}.
\item
Processes that are {\it forbidden} in the SM {\it at tree level}.

Processes that involve flavor-changing neutral current (FCNC)
interactions that change charm quantum number by one or two units do
not occur in the SM at tree level.
However, these transitions can happen in the SM at loop levels, which
makes them rare. Such processes can receive NP contributions from both
tree-level interactions mediated by new interactions, and loop
corrections with NP particles. Inclusive and exclusive transitions
mediated by $c \to u \gamma$ or $c \to u \ell \bar \ell$ will be
discussed in Sec.~\ref{rare:charm}.  Searches for violation of \cp
and other symmetries in baryon decays and in \DzDzb mixing will be
discussed in Sec.~\ref{sec:symbreaking:cpv}. Charged lepton flavor
violation decays will be discussed in Sec.~\ref{sec:clfv}.
\item
Processes that are {\it forbidden} in the SM.

Some processes, while allowed by space-time symmetries, are forbidden
in the SM. Even if allowed by NP, searching for these signatures
require high statistics. Their observation, however, would constitute
a high-impact discovery, as it would unambiguously point towards
physics beyond the SM. Examples include searches for the baryon
number-violating transitions as discussed in Sec.~\ref{cpv:lbdosc}
and Sec.~\ref{cpv:morelbd}, lepton number violating decays as covered
in Sec.~\ref{clfv:lnv}. Many well-motivated NP models predict the
existence of light, weakly-interacting particles. Since such light
particles are not part of the SM particle spectrum, the corresponding
processes do not occur in the SM. Such processes involving invisible
signatures are discused in Sec.~\ref{sec:inv}. Some further searches at the off-resonance energies,
where the electron and positron are not tuned to the $s$-channel resonance production of the charmonium states,
are discussed in Sec.~\ref{sec:offres}.

\end{enumerate}

With the accumulation of large data sets and possible increase of
luminosity and cms energy, as well as an ever-improving understanding
of the detector performance, \bes3 will have great potential in NP
searches in the coming years.

% Rare Decays
\input{New_physics/raredecay}
% Symmetry Breaking Processes
\input{New_physics/cpv}
\input{New_physics/bsmheavy}
% Light BSM Particle Searches
\input{New_physics/bsmlight}
\input{New_physics/offres}

%% file: New_physics/raredecay.tex
\newcommand{\obar}[1]{\mkern 1.5mu\overline{\mkern-1.5mu#1\mkern-1.5mu}\mkern 1.5mu}
\newcommand{\overbar}[1]{\mkern 2.5mu\overline{\mkern-2.5mu#1\mkern-0.0mu}\mkern 0.0mu}

%%%%%%%%%%%%%%%%%%%%%%%%%%%%%%%
\section{Rare decays of charmonia and charmed hadrons}

Experiments at the energy frontier may be able to probe NP through
direct production of new particles. These new degrees of freedom
could affect low energy observables.
Experiments at the intensity frontier can probe those new virtual
contributions via decays of charmonia and charmed hadrons, making them
complementary to direct searches at the energy frontier.

\input{New_physics/oniaweak}

% %%%%%%%%%%%%%%%%%%%%%%%%%%%%%%%
\subsection{Rare radiative and rare leptonic $D_{(s)}$ decays}\label{NPinRareDecays}
\label{rare:charm}

The decays of $D$ mesons that are mediated by quark-level FCNC
transitions $c \to u \gamma$ (rare radiative) and $c \to u \ell \ell$
(rare leptonic and semi-leptonic) only proceed at one loop in the SM.
The absence of a super-heavy down-type quark in the SM implies that
Glashow-Iliopoulos-Maiani cancellation mechanism is very effective,
making the charm sector of special interest in probing for NP. The
predicted short-distance (SD) contributions of the SM for FCNC in the charm
sector are well beyond the sensitivity of current experiments. Yet,
theoretical estimates suggest that the rates for FCNC processes could
be enhanced by long-distance (LD) effects by several orders of magnitude.

There are a number of interesting FCNC processes in the charm sector.
Such examples include $D\to h(h')\ell \ell^\prime$ and
$D\to h\nu\overbar\nu$, where
$h$ represents light hadron states and $\ell$ is a charged lepton. Such
decays could be interesting probes of NP, especially in light of
renewed interest to lepton flavor universality (LFU) studies in
$B$ decays. While SM interactions in general respect LFU, recent
experimental observations (see, $e.g.$, \cite{Aaij:2017vbb}) show hints
of LFU violation in rare semi-leptonic decays of $B$
mesons. Theoretically, such violation could come from new
lepton-flavor non-universal interactions \cite{Altmannshofer:2017yso},
which might also be detectable in $D$-decays
\cite{Dorsner:2017ufx}. Such interactions might also induce
lepton-flavor violating effects \cite{Glashow:2014iga} (see section
\ref{sec:clfv}), although this depends on a particular model of NP.

Even though the charm production rate in $e^+e^-$ collisions near the charm
threshold is lower than that at hadron colliders and $B$-factories,
\bes3 has the benefit of lower multiplicity
and the ability to impose powerful kinematic constraints, which can
deliver high purity for final states with invisible energy or photons.

Rare radiative decays (such as $D \to \rho \gamma$) are most likely
dominated by the LD SM contributions, which are quite difficult
to compute \cite{Burdman:1995te,Greub:1996wn,Burdman:2001tf}.
Yet, there are opportunities to study NP effects in rare radiative transitions.
These include a possibility that NP dominates the SM signal at least in
portions of the available phase space \cite{deBoer:2018buv},
using particular combinations of radiative transitions \cite{Fajfer:2000zx},
including CP-violating asymmetries \cite{Isidori:2012yx,deBoer:2017que},
or studying the photon's polarization patterns \cite{deBoer:2018zhz} that
could be more sensitive to NP contributions.

%%%%%%%%%%%%%%%%%%%%%%%
\subsubsection{Two-body rare decays with charged leptons}

The simplest rare leptonic decays, such as $D^0 \to \ell^+ \ell^-$,
have a very small SM contribution (both SD and LD ones), so they are
potentially very clean probes of NP amplitudes. In this section we
shall concentrate on the lepton-flavor conserving decays.

Experimentally, at present, there are only  upper
limits on $D^0 \to \ell^+ \ell^-$ decays~\cite{Amhis:2016xyh}.
Theoretically, all possible NP contributions to $c \to u \ell^+ \ell^-$
can be summarized in an effective Hamiltonian,
\beq\label{SeriesOfOperators2}
{\cal H}_{NP}^{rare}  =
\sum_{i=1}^{10}  {\rm \widetilde C}_i (\mu) ~ \widetilde Q_i,
\eeq
where ${\rm \widetilde C}_i$ are Wilson coefficients, and the $ \widetilde Q_i$ are the effective operators.
In this case, there are ten of them,
\bea
\begin{array}{l}
\widetilde Q_1 = (\overline{\ell}_L \gamma_\mu \ell_L) \
(\overline{u}_L \gamma^\mu
c_L)\ , \\
\widetilde Q_2 = (\overline{\ell}_L \gamma_\mu \ell_L) \
(\overline{u}_R \gamma^\mu
c_R)\ , \\
\widetilde Q_3 = (\overline{\ell}_L \ell_R) \ (\overline{u}_R c_L) \ ,
\end{array}
\qquad
\begin{array}{l}
\widetilde Q_4 = (\overline{\ell}_R \ell_L) \
(\overline{u}_R c_L) \ , \\
\widetilde Q_5 = (\overline{\ell}_R \sigma_{\mu\nu} \ell_L) \
( \overline{u}_R \sigma^{\mu\nu} c_L)\ ,\\
\phantom{xxxxx}
\end{array}
\label{SetOfOperatorsLL}
\eea
with five additional operators $\widetilde Q_6, \cdots, \widetilde Q_{10}$
that can be obtained from operators in Eq.~(\ref{SetOfOperatorsLL}) by
the substitutions $L \to R$ and $R \to L$.
It is worth noting that only eight operators contribute to
$D^0\to \ell^+\ell^-$, as
$\langle \ell^+ \ell^- | \widetilde Q_5 | D^0 \rangle =
\langle \ell^+ \ell^- | \widetilde Q_{10} | D^0 \rangle = 0$.
The most general $D^0 \to \ell^+ \ell^-$ decay amplitude can be written as
\beq\label{decayampl}
{\cal M} = {\bar u}({\bf p}_-, s_-) \left[ A + B \gamma_5
\right] v({\bf p}_+, s_+) \ \ ,
\eeq
which results in the branching fractions
\begin{eqnarray}\label{Dllgen}
& & {\cal B}_{D^0 \to \ell^+\ell^-} =
\frac{M_D}{8 \pi \Gamma_{\rm D}} \sqrt{1-\frac{4 m_\ell^2}{M_D^2}}
\left[ \left(1-\frac{4 m_\ell^2}{M_D^2}\right)\left|A\right|^2  +
\left|B\right|^2 \right] \ \  .
\end{eqnarray}
Any NP contribution described by
the operators of Eq.~(\ref{SetOfOperatorsLL}) gives for
the amplitudes $A$ and $B$,
\begin{eqnarray}
\left| A\right|  &=& G \frac{f_D M_D^2}{4 m_c} \left[\widetilde C_{3-8} +
\widetilde C_{4-9}\right]\ , \nonumber
 \\
\left| B\right|  &=& G \frac{f_D}{4} \left[
2 m_\ell \left(\widetilde C_{1-2} + \widetilde C_{6-7}\right)
+  \frac{M_D^2}{m_c}
\left(\widetilde C_{4-3} + \widetilde C_{9-8}\right)
\right]\ ,  \label{DlCoeff}
\end{eqnarray}
with $\widetilde C_{i-k} \equiv \widetilde C_i-\widetilde C_k$. Any NP
model that contributes to $D^0 \to \ell^+ \ell^-$ can be constrained
from the constraints on the Wilson coefficients in Eq.~(\ref{DlCoeff}).
It will be advantageous to study {\it correlations} of various
processes to isolate and constrain the NP contributions
\cite{Golowich:2009ii,Paul:2012ab}. Such correlations exist, for
instance, in \DzDzb mixing and rare decays~\cite{Golowich:2009ii}. In
general, one cannot predict the rare decay rate by knowing just the
mixing rate, even if both $x_D$ and ${\cal B}_{D^0 \to \ell^+\ell^-}$
are dominated by a given NP contribution. It is, however, possible for
a restricted subset of NP models~\cite{Golowich:2009ii}.  Predictions
for $D^0 \to \mu^+\mu^-$ branching fraction for $x_D \sim 1\%$ can
be found in Ref.\cite{Golowich:2009ii} with the definitions of NP
model parameters.

%%%%%%%%%%%%%%%%%%%%%%%

\begin{table}[!hbtp]
  \caption{\small The latest experimental upper limits on branching
    fractions (in units of $10^{-6}$) for the rare $D$ and $D_s$
    decays into $h(h')e^+e^-$. The expected BESIII sensitivities with the expected final charm data set listed in Sec.~\ref{chapter:sum} are also shown in the last column.}
\label{tab:np_raredecay_Dtohee}
\begin{center}
\begin{tabular}{l|rcccc}
  \hline\hline
  Decay & Upper limit  & Experiment & Year & Ref.  & BESIII Expected \\
  \hline
  $D^0\to \pi^0e^+e^-$	        &0.4		&\bes3	    &2018	  	&\cite{TheBESIIICollaboration2018a}	&0.1		\\
  $D^0\to \eta e^+e^-$	  	&0.3		&\bes3	    &2018	  	&\cite{TheBESIIICollaboration2018a}	&0.1		\\
  $D^0\to \omega e^+e^-$	&0.6		&\bes3	    &2018	  	& \cite{TheBESIIICollaboration2018a}	&0.2		\\
  $D^0\to K_S^0 e^+e^-$	  	&1.2	  	&\bes3	    &2018	  	&\cite{TheBESIIICollaboration2018a}	&0.5		\\
  $D^0\to \rho e^+e^-$	  	&124.0	&E791	    &2001	  	&\cite{E791_DTohee_2001}			&0.5		\\
  $D^0\to \phi e^+e^-$	  	&59.0	&E791	    &2001	  	&\cite{E791_DTohee_2001}			&0.5		\\
  $D^0\to \overbar{K}^{*0} e^+e^-$	
        &47.0	&E791		&2001	 &\cite{E791_DTohee_2001}			&0.5		\\
  \hline
  $ D^0\to \pi^+\pi^-e^+e^-$	&0.7		 &\bes3		&2018	 &\cite{TheBESIIICollaboration2018a}	&0.3		\\
  $D^0\to K^+K^- e^+e^-$	&1.1  	 &\bes3		&2018	 & \cite{TheBESIIICollaboration2018a}	&0.4		\\
  $D^0\to K^-\pi^+ e^+e^-$	&4.1	   	 &\bes3		&2018	 & \cite{TheBESIIICollaboration2018a}	&1.6		\\
  \hline
  $ D^+\to \pi^+e^+e^-$	  	&1.1		&BaBar		&2011	 &\cite{babar_D2hee_2011}			&0.12		\\
  $D^+\to K^+e^+e^-$	    	&1.0		&BaBar		&2011	 &\cite{babar_D2hee_2011}			&0.46		\\
  $D^+\to \pi^+\pi^0e^+e^-$	& 1.4	& \bes3		& 2018	 & \cite{TheBESIIICollaboration2018a}	&0.5		\\
  $D^+\to \pi^+K_S^0e^+e^-$	& 2.6	& \bes3		& 2018	 & \cite{TheBESIIICollaboration2018a}   &1.0		\\
  $D^+\to K_S^0K^+e^+e^-$	& 1.1		& \bes3		& 2018	 & \cite{TheBESIIICollaboration2018a}	&0.4		\\
  $D^+\to K^+\pi^0e^+e^-$	& 1.5		& \bes3		& 2018	 & \cite{TheBESIIICollaboration2018a}	&0.6		\\
  \hline
  $ D_s^+\to \pi^+e^+e^-$	& 13.0		&BaBar		&2011   &\cite{babar_D2hee_2011}      & 70.0		\\
  $D_s^+\to K^+e^+e^-$	    	& 3.7		&BaBar		&2011	 &\cite{babar_D2hee_2011}		&1.7		\\
  
  \hline
  \hline
\end{tabular}
\end{center}
\end{table}

\subsubsection{Three-body rare decays with charged leptons}

Theoretical predictions of the decay rates of di-lepton modes such as
$D\to h(h')e^+e^-$ are complicated due to the LD
contributions.  The rates with a lepton-pair mass in the non-resonant
regions could provide access to NP~\cite{Fajfer:2001sa,Fajfer:2012nr,deBoer:2015boa,Paul:2012ab}, at
least for some particular BSM models.
Table~\ref{tab:np_raredecay_Dtohee} shows some interesting
$D\to h(h')e^+e^-$ modes that can be studied at the \bes3 experiment,
and the corresponding experimental upper limits on the branching
fractions are also summarized.  Recent \bes3
paper~\cite{TheBESIIICollaboration2018a} has already improved some of
the limits by several orders of magnitude. With more data in future,
we expect to improve these limits and some new modes could also be
probed, depending on how background-free a given mode is. The expected BESIII sensitivities with
the expected final charm data set listed in Sec.~\ref{chapter:sum} are also shown in the last
column of Table~\ref{tab:np_raredecay_Dtohee}.

%%%%%%%%%%%%%%%%%%%%%%%%%%%%%%%%%%
\subsubsection{Three-body rare decays with neutrinos}

Neutral modes, such as $D^0\to \pi^0\nu\overbar{\nu}$, have never been
studied at charm threshold before. It is possible that the
LD SM effects are under better theoretical control in such
transitions \cite{Burdman:2001tf}. Belle reported a similar search for
the rare decays $B\to h^{(*)}\nu\overbar\nu$~\cite{belle_BTohvv_2007},
and BaBar searched for
$B^0\to \gamma \nu\overbar\nu$~\cite{babar_B0Togammavv_2004}.  With
$20\,{\rm fb}^{-1}$ data sample at $\sqrt{s}=3.773\,{\rm GeV}$, \bes3's
sensitivity of ${\cal B}(D^0\to\pi^0\nu\overbar\nu)$ measurement could reach $10^{-4}$ or better.

Radiative decay modes $D^0\to \gamma \nu\overbar\nu$ serve as physics
background to searches for light Dark Matter (DM), as described in Sec.~\ref{subsec:inv_heavy_meson}. This transition has an
unobservable small branching fraction of
${\cal B} (D^0\to \gamma \nu\overbar\nu) = 3.96\times 10^{-14}$ in the
SM~\cite{Badin:2010uh}, making the final state of a photon with
missing energy a suitable topology for searches for light DM at \bes3.

%% file: New_physics/oniaweak.tex
\subsection{Weak decays of charmonia states}
\label{sec:charmoniumweak}

The decays of $\psi(nS)$ ($n=1$, $2$) below the open-charm threshold are
dominated by the strong or electromagnetic interactions where the
intermediate gluons or virtual photons are produced by $c \bar{c}$
annihilation.  However, flavor-changing weak decays of these states
through virtual $W$ bosons are also possible in the SM framework.
%although the branching fractions are expected to be rather small,
%typically on the order of
For instance, the branching fractions of $J/\psi$ inclusive weak decays are
estimated to be of the order of $10^{-10}$~\cite{sanchis_1994}.  As
mentioned in Refs.~\cite{datta_1999, xmzhang_2001,
  hbli_2012,hill_1995}, the branching fractions of
$J/\psi\to D(\bar{D})X$ (with $X$ denoting any hadrons) can be
enhanced by new interactions.  Several NP
models, such as the top-color model, the minimal supersymmetric
standard model (MSSM) with $R$-parity violation and a general two-Higgs
doublet model (2HDM), allow $\psi(nS)$ flavor-changing processes
to occur with branching fractions around
$10^{-6}$~\cite{datta_1999,xmzhang_2001}.  The observation of an
anomalous production rate in $\psi(n\rm S)$ weak decays would be a strong hint of NP.

With the newly accumulated $10^{10}$ $J/\psi$ events,
we expect to improve the branching fraction measurements of
$\psi(nS)$ weak decays, including both hadronic and semi-leptonic weak
decays, by almost one order of magnitude, which will provide a more
stringent experimental test of the SM than previous searches, and
hence further constrain the parameter spaces of NP models. 
These weak decays can also be searched for in the expected $3\times10^{9}$ $\psi(3686)$ events

\subsubsection{\boldmath{$\psi(nS)\to D_{(s)}P/D_{(s)}V/D_{(s)}^*V$}}

Several theoretical calculations of the branching fractions of
two-body hadronic weak decays of
$J/\psi\to D_{(s)}P/D_{(s)}V/D_{(s)}^*V$, where $D$ represents a
charmed meson, $P$ and $V$ a pseudoscalar and vector meson,
respectively, are summarized in the last column of
Table~\ref{tab:np_raredecay_hadroniccharmonium}.  The charge conjugate
states are implicitly included.

\begin{table}[tbp]
  \caption{Predicted branching fractions and expected sensitivities with a
    sample of $10^{10}$ $J/\psi$ events of two-body hadronic weak
    decays of $J/\psi\to D_{(s)}P, D_{(s)}V$ and $D_{(s)}^*V$.}
    \label{tab:np_raredecay_hadroniccharmonium}
    \begin{center}
        \begin{tabular}{c|ll|cc}
            \hline\hline
            &Decay type & Example	& exp. sensitivity & predicted $\mathcal{B}$~\cite{dhir_2009, ylshen_2008, ymwang_2008, sharma_1999} \\ 
            & & 	&  $(\times10^{-6})$ & $(\times 10^{-10})$ \\ 
                       \hline
            $c\to s$ &$ D_{(s)}P$ 	&$J/\psi\to D_s^-\pi^+$			&9.9 	  &2.00~$\thicksim$~8.74 	\\
            &						 	   	&$J/\psi\to D^0K^0$				        &13.0	  &0.36~$\thicksim$~2.80	\\
            &$ D_{(s)}V$ 	&$J/\psi\to D_s^-\rho^+$		        &2.0 	  &12.60~$\thicksim$~50.50	\\
            &						 	   	&$J/\psi\to D^0K^{*0}$			        &0.38 	  &1.54~$\thicksim$~10.27	\\
            &$ D_{(s)}^*V$&$J/\psi\to D_s^{*-}\rho^+$	            &1.7	  &52.60		\\ \hline
            $c\to d$ &$ D_{(s)}P$	&$J/\psi\to D_s^-K^+$			&9.8	  &0.16~$\thicksim$~0.55	\\
            &						 		&$J/\psi\to D^-\pi^+$			        &0.21 	  &0.08~$\thicksim$~0.55	\\
            &						 		&$J/\psi\to D^0\eta$			        &0.72	  &0.016~$\thicksim$~0.070	\\
            &						 		&$J/\psi\to D^0\eta'$			        &0.25	  &0.003~$\thicksim$~0.004	\\
            &						 		&$J/\psi\to D^0\pi^0$			        &0.48	  &0.024~$\thicksim$~0.055	\\
            &$ D_{(s)}V$ 	&$J/\psi\to D_s^-K^{*+}$		        &5.4	  &0.82~$\thicksim$~2.79	\\
            &						 		&$J/\psi\to D^-\rho^+$			        &0.35 	  &0.42~$\thicksim$~2.20	\\
            &						 		&$J/\psi\to D^0\rho^0$			        &0.77	  &0.18~$\thicksim$~0.22	\\
            &						 		&$J/\psi\to D^0\omega$			        &0.35	  &0.16~$\thicksim$~0.18	\\
            &						 		&$J/\psi\to D^0\phi$			        &0.22	  &0.41~$\thicksim$~0.65	\\
            &$ D_{(s)}^*V$&$J/\psi\to D_s^{*-}K^{*+}$	            &4.5 	  &2.6		\\
            &						 		&$J/\psi\to D^{*-}\rho^{+}$			    &0.083	  &2.8		\\
            &						 		&$J/\psi\to D^{*-}K^{*+}$			    &0.027	  &9.6		\\
            \hline\hline
        \end{tabular}
    \end{center}
\end{table}

The BESII experiment searched for the hadronic decays
$J/\psi\to D_s^-\pi^+$, $J/\psi\to D^-\pi^+$, and $J/\psi\to D^0K^0$
and set upper limits at the order of $10^{-4} \sim 10^{-5}$ using a
sample of $5.8 \times 10^7$ $J/\psi$
events~\cite{bes_jpsihadronic_2008}.  \bes3 has searched for the rare
decays $J/\psi\to D_s^-\rho^+$ and $J/\psi\to D^0K^{*0}$ with a
 sample of $2.25\times 10^8$ $J/\psi$
events~\cite{bes3_jpsihadronic_2014}.  No signal was observed,
and upper limits at the 90\% C.L. were set on the branching fractions,
$\mathcal{B}(J/\psi\to D_s^-\rho^+) < 1.3\times 10^{-5}$ and
$\mathcal{B}(J/\psi\to D^0K^{*0}) < 2.5 \times 10^{-6}$.
%The \bes3 results are consistent with the predictions of the SM with the current sensitivities.
These results are several orders of magnitude above the SM predictions
and can be improved with larger data sets. The expected sensitivity
with a sample of $10^{10}$ $J/\psi$ events are estimated as listed in the third column of Table~\ref{tab:np_raredecay_hadroniccharmonium}.

%%%%%%%%%%%%%%%%%%%%%%%%%%%%%%%%
\subsubsection{\boldmath{$\psi(nS)\to D_{(s)}l^+\nu/D_{(s)}^{*}l^+\nu$}}

Semi-leptonic decays of $\psi(nS)$ mesons are induced by  weak
$c\to s$ or $c\to d$ transitions through a virtual intermediate $W$
boson. Theoretical calculations predict the branching fractions of
$J/\psi\to D_s^{(*)}l^+\nu$ and $J/\psi\to D^{(*)}l^+\nu$ to be
at the level of $10^{-9}$ and $10^{-10}$, respectively, by using QCD sum rules~\cite{ymwang_2008}, the covariant light-front quark model~\cite{ylshen_2008}, and the covariant constituent quark model ~\cite{Ivanov:2015woa}.
It is therefore interesting to search for semi-leptonic weak decays of $\psi(nS)$ states
in high intensity and low background experiments.

The BESII experiment searched for several semi-leptonic weak decays of the $J/\psi$.
Using $5.8 \times 10^7$ $J/\psi$ decay events, the upper limits at the
90\% C.L.
for $\mathcal{B}(J/\psi\to D_s^- e^+\nu)$ and
$\mathcal{B}(J/\psi\to D^- e^+\nu)$ were found to be
$3.6\times 10^{-5}$ and $1.2\times 10^{-5}$,
respectively~\cite{bes_jpsileptonic_2006}.  \bes3 has searched
for the decay $J/\psi\to D_s^-e^+\nu/D_s^{*-}e^+\nu$ with a much higher
sensitivity than previous analyses, based on a sample of
$2.25 \times 10^8$ $J/\psi$ events~\cite{bes3_jpsileptonic_2014}.  At
the 90\% C.L., the upper limits were determined to be
$\mathcal{B}(J/\psi\to D_s^-e^+\nu) < 1.3\times 10^{-6}$ and
$\mathcal{B}(J/\psi\to D_s^{*-}e^+\nu) < 1.8\times 10^{-6}$. Both are consistent with SM predictions, but can be improved
with more data.

%%%%%%%%%%%%%%%%%%%%%%%%%%%%%%%%%
\subsubsection{\boldmath{$\psi(nS)\to D^{(*)0}l^+l^- /D^{(*)0}\gamma$}}

The rates of $c\to u$ transitions of $\psi(nS)$ are predicted to
be tiny in the SM~\cite{datta_1999,ymwang_2009}. However, some NP
scenarios allow for larger FCNC transition rates. For example,
Ref. \cite{datta_1999} argues that the branching fraction of
$J/\psi\to DX_u$ (with $X_u$ denoting mesons containing the $u$ quark),
which is mediated by the $c\to u$ quark transition, could be enhanced
to be of order $10^{-6} -10^{-5}$.  Thus, an observation of FCNC in
the low-lying charmonium decays would indicate NP.

In practice, it is difficult to isolate pure $c\to u$ mediated transitions from $c\to s$
and $c\to d$ in hadronic weak decays of the type $\psi(nS)\to D^{(*)}X_u$.
Instead, theoretically (relatively) clean semi-leptonic or radiative rare decays
$\psi(nS)\to D^{(*)0}l^+l^-$ and $\psi(nS)\to D^{(*)0}\gamma$ should be employed.
The energy distributions of the final state photons or lepton pairs could be used as
kinematic constraints to identify those decays.

%%%Local Variables:
%%% mode: latex
%%% TeX-master: t
%%% End:

%% file: New_physics/cpv.tex
\def\st{\scriptstyle}
\def\sst{\scriptscriptstyle}
\def\mco{\multicolumn}
\def\epp{\epsilon^{\prime}}
\def\vep{\varepsilon}
\def\ra{\rightarrow}
\def\ppg{\pi^+\pi^-\gamma}
\def\vp{{\bf p}}
\def\ko{K^0}
\def\kb{\overline{K^0}}
\def\kl{K^{0}_{L}}
\def\al{\alpha}
\def\ab{\overline{\alpha}}
\def\Lamz{\Lambda^0}
\def\Lamb{\overline{\Lambda}^0}
\def\Lam{\Lambda}
\def\lam{\Lambda}
\def\ajpsi{\alpha_\psi}
\def\llb{\Lambda\bar\Lambda}
\def\lbar{\bar\Lambda}
\def\Sig{\Sigma}
\def\Cas{\Xi}
\def\Xim{\Xi^-}
\def\Xib{\overline{\Xi}^+}
\def\ra{\rightarrow}
\def\be{\begin{equation}}
\def\ee{\end{equation}}
\def\bea{\begin{eqnarray}}
\def\eea{\end{eqnarray}}

%%%%%%%%%%%%%%%%%%%%%%%%%%
\section{Symmetry test in hyperon decays}
\label{sec:symbreaking:cpv}

The Sakharov conditions for baryogengesis underline the role of CP
violation as one of the central pieces of the matter-antimatter
asymmetry puzzle~\cite{Sakharov:1967dj}.
However, the CKM mechanism for CP violation in the SM fails to fully explain the puzzle of the  matter-antimatter asymmetry by more than 10 orders-of-magnitude~\cite{Morrissey:2012db}.  This suggests that additional, heretofore undiscovered, CP violating processes exist, and has motivated
aggressive searches for new sources of CP violation in $b$-quark decays and neutrino oscillations~\cite{ref::pdg2016}.
CP violation in charm decay is very small and had not yet been found until the discovery in 2019 at LHCb.
LHCb finds non-zero CP violation in $D^0\to \pip\pim$ and $D^0\to K^+K^-$ decays with a significance of 5.3$\sigma$.  The time-integrated CP asymmetry is given as
\begin{eqnarray}
\Delta a_{\rm CP}&=&\frac{\Gamma(D\to \kk)-\Gamma(\bar{D}\to\kk)}{\Gamma(D\to \kk)+\Gamma(\bar{D}\to\kk)}-\frac{\Gamma(D\to\pip\pim)-\Gamma(\bar D\to\pip\pim)}{\Gamma(D\to\pip\pim)+\Gamma(\bar D\to\pip\pim)}  \nonumber \\
&=& (-0.154\pm0.029)\%,
\end{eqnarray}
where $D$($\bar{D}$) is a $D^0$($\bar{D}{}^0$) at time $t$=0~\cite{Aaij:2019kcg}.
This result is at the high end of theoretical estimates for its SM value, which ranges from $10^{-4}$ to $10^{-3}$. The LHCb result is intriguing, because it may be a sign of the long-sought-for-non-SM mechanism for CP violation. However, uncertainties in SM theoretical calculations for $\Delta a_{\rm CP}$ make it impossible to rule out this possibility. \bes3's current $10^{-3}\sim 10^{-2}$ level of sensitivity on charm meson CP violation is still more than one order of magnitude above the highest conceivable SM effects.
In addition to charmed mesons, \bes3 has accumulated huge data samples of strange and charmed baryons, thus
providing a unique opportunity to examine the strong dynamics of
strange/charm decays, and another route to probe the
phenomenon of CP violation.  In this section we will briefly discuss
strange/charmed baryon decays and outline various paths to the
observation of CP violation and baryon number violation (BNV).

%%%%%%%%%%%%%%%%%%%%%%
\subsection{Probing CP asymmetry in hyperon decays}
\label{sec:hyperon_np}

\bes3 has capability for testing CP symmetry in hyperon decays, produced
via $\jpsi\to B \bar{B}$ with $B \bar{B}$ denoting polarized,
quantum-entangled hyperon pairs, which adds an exciting new dimension
to the study of CP violations.

\begin{figure}[tbp]
  \centering
 \includegraphics[width=0.4\textwidth]{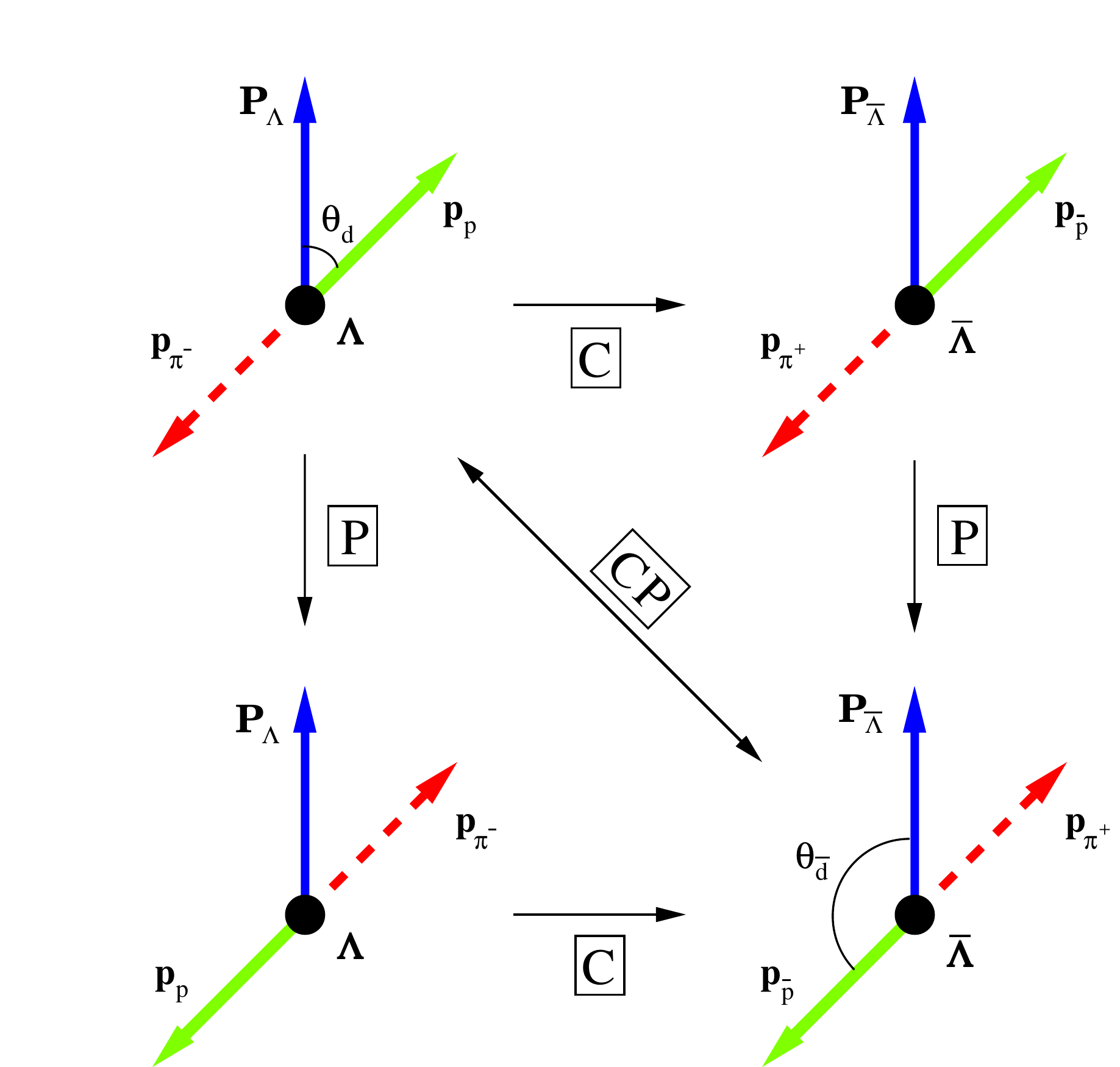}
 \caption{\small Illustration of the charge (C), parity (P) and CP transformation
   on the $\Lam \ra p\pi^-$ decays}
  \label{fig:lbdcp}
\end{figure}

A weak two-body decay of a spin one-half baryon under charge and parity
transformations is illustrated in Fig.~\ref{fig:lbdcp} for the
most prominent decay mode of $\Lam \ra p\pi^-$. The picture is drawn
in the $\Lambda$ rest frame and the spin polarization vector
${\bf P}_\Lambda$ is pointing upward. The daughter
proton emitted at angles $\theta_d$ and $\phi_d$ with respect to the
polarization vector ${\bf P}_\Lambda$ is transformed under CP into
an anti-proton emitted at angles $\pi-\theta_d$ and $-\phi_d$ with
respect to the polarization vector of the parent $\bar\Lam$.  In
addition, in the weak decay the polarization of the final proton could
have a transverse component (along the
${\bf p}_p\times{\bf p}_\Lambda$ vector). This component for the
antiproton will have opposite direction.

\subsubsection{Direct CP asymmetry}

The most straightforward  CP-odd observable is  the
difference between the partial decay rates for the decay
and the CP-transformed process:
 \vspace{8pt} \be \Delta =
\frac{\Gamma-\overline{\Gamma}}{\Gamma+\overline{\Gamma}} \ .
\label{eq:delta} \vspace{5pt} \ee Furthermore, the parity violation in
the hyperon decays allows much more sensitive tests to be constructed.  A
$1/2\to 1/2+0$ two-body nonleptonic weak decay can be described by
the partial width $\Gamma$ and decay parameters
$\alpha,\,\beta,\,\gamma$:
\begin{eqnarray}
\alpha=\frac{2\textrm{Re}(S^*P)}{|S|^2+|P|^2},\quad
\beta=\frac{2\textrm{Im}(S^*P)}{|S|^2+|P|^2}=\sqrt{1-\alpha^2}\sin\phi,\quad
\gamma=\sqrt{1-\alpha^2}\cos\phi.
\end{eqnarray}
Only two parameters are independent and it is convenient to use
$\alpha$ and $\phi$. The parameter $\alpha$ has a simple
interpretation due to the asymmetry of the angular distribution for the
daughter proton, as given by \be
\frac{d\Gamma}{d\Omega_d}\propto\frac{1}{4\pi}\left(1+\alpha
  {P}_\Lambda\cos\theta_d\right).  \ee

A comparison of the decay parameters of a hyperon and the
anti-hyperon leads to sensitive tests of CP symmetry. The CP-odd
observables are \vspace{8pt} \be \quad A = \frac {\al+\ab}{\al-\ab}
\ , \quad B = \frac {\beta+\overline{\beta}}{\beta-\overline{\beta}}
\ . \label{eq:ABB} \vspace{5pt} \ee Here $\al$ corresponds to
$\al_-$ in the $\Lambda\to p\pi^-$ decay, while $\ab$ corresponds to
$\al_+$ in the $\bar\Lambda\to \bar p\pi^+$ decay; and the quantity
$\beta$ can not be measured in the joint $J/\psi\to
\Lambda\bar\Lambda\to p \pi^-\bar p\pi^+$ decay. In general,
sensitivity of these asymmetries to CP violation scales as
$\Delta:A:B\sim 1:10:100$. The asymmetry $A$ has been studied in
$\bar p p$ experiments and at $e^+e^-$
colliders~\cite{Chauvat:1985fb,Barnes:1996si,Tixier:1988fv,Ablikim:2009ab}.
The present world average is $A=0.006\pm0.021$, while the CKM mechanism predicts a value of $A
\sim10^{-5}-10^{-4}$~\cite{Donoghue:1985ww,Donoghue:1986hh,Tandean:2002vy}.
Predictions in scenarios beyond the SM are given in
Refs.~\cite{He:1995na,luk98,Tandean:2003fr}. There  also exist dynamical
calculations in Refs.~\cite{Cheng:1991sn,Cheng:2018hwl}.
Experimental prospects on CP-symmetry tests for the charmed baryon
by determination of $\alpha$ are discussed in Ref.~\cite{Liu:2015qra}.

\input{New_physics/LLbar.tex}

\subsubsection{The triple-product asymmetry}
Apart from the above mentioned direct measurement, we can also exploit
the triple-product  asymmetry as a CP-violating observable.  Studies
of CP violation in $\Lambda$ and other hyperon decays using this
approach are proposed in Ref.~\cite{Bigi:2017eni}.  The direct
CP-violating asymmetry $\mathcal {A}_\textrm{dir}$ and the
triple-product asymmetry $\mathcal {A}_\textrm{T}$ depend on the weak phase
$\phi$ and the strong-phase $\delta$ as follows:
\begin{eqnarray}
\mathcal {A}_\textrm{dir} &\propto& \sin\phi\sin\delta, \nonumber\\
\mathcal {A}_\textrm{T} &\propto& \sin\phi\cos\delta.
\end{eqnarray}
The direct asymmetry only survives if there is a non-zero strong
phase, whereas the triple-product asymmetry with the strong phase
vanishes. Therefore the two methods are complementary, particularly if
the strong phase is unknown.

The triple product is defined as
$\vec v_1\cdot (\vec v_2 \times \vec v_3)$, where $\vec v$ can be a
three-momentum or a spin vector. Under time reversal the triple
product changes sign, and thus it is a $T$-violating signal. But
final-state interactions (FSI) can give a false CP-violation
signal. As a result, one must compare the channel to its
conjugate. For an illustration of this point, one may refer to
Refs.~\cite{Valencia:1988it, Bigi:2000yz}. Such considerations have led
to a corresponding proposal for \bes3~\cite{Kang:2009iy, Kang:2010td},
as well as in beauty
decays~\cite{Duraisamy:2013kcw,Bensalem:2002pz,Datta:2003mj,Gronau:2011cf,Gronau:2015gha},
and in the search for NP effects
\cite{DiSalvo:2012vd,Ajaltouni:2012zg}.

We illustrate the method with the process
$e^+e^-\to J/\psi\to\Lambda \bar\Lambda\to[p\pi^-] [\bar p
\pi^+]$ below. This process is transformed to itself under charge
conjugation. Since the polarization of the proton and the
antiproton is not measured, we have four independent vectors to
construct triple products. Two of them are ${\bf k}$ and
${\bf p}_\Lambda$ -- the three-momenta of incoming electron and
outgoing $\Lambda$ in the reaction cms frame. The remaining ones
are ${\bf q}_1$ and ${\bf q}_2$ -- the momenta of the proton and
antiproton in the $\Lambda$ and $\bar \Lambda$ rest frames,
respectively.  For example, defining
$C_T=({\bf p}_\Lambda\times {\bf q}_1)\cdot{\bf p}_{\Lambda}$, and
$\bar C_T=-({\bf p}_\Lambda\times {\bf q}_2)\cdot{\bf p}_{\Lambda}$,
we can define following triple-product asymmetries
\begin{eqnarray}\label{eq:TP}
\langle A_T  \rangle &=& \frac{N(C_T > 0) - N(C_T <0)}{N(C_T > 0)+ N(C_T <0)} \\
\langle \bar A_T  \rangle &=&  \frac{N(-\bar C_T > 0) - N(-\bar
C_T<0)}{N(-\bar C_T > 0) + N(-\bar C_T <0)} \; .
\end{eqnarray}
Therefore CPT invariance implies \begin{equation} {\mathcal A}_T =  \langle A_T  \rangle -
\langle \bar A_T \rangle \neq 0 \end{equation} is
a CP-odd observable, which can be measured merely through event counting.
% The considered $C_T$  variable is closely related to
% the term corresponding to the $I_z$ contribution in the general
% expression Eq.~\eqref{eqn:cxx}, which should be zero
% for strong and electromagnetic processes where
% parity is conserved.
Another  triple product is
$({\bf q}_1\times {\bf q}_2)\cdot{\bf p}_{\Lambda}$, related to the
$C_{x\bar z}-C_{z\bar x}$ spin correlation term.  This last triple
product allows limits on the $\Lambda$ electric-dipole moment
$d_\Lambda$ to be improved, as discussed in
Refs.~\cite{He:1992ng,He:1993ar}.

The CP asymmetries can also be studied locally as a function of the
kinematic variable $\cos\theta_\Lambda$, limiting the range of the
 triple-product values:
\begin{eqnarray} \mathcal{A}_T (d) &=&
\frac{N(C_T > |d|) - N(C_T < - |d|)}{N(C_T > |d|) +  N(C_T <-|d|)}.
\end{eqnarray}

\begin{center}
\begin{table}[tbp]
\caption{ The number of reconstructed events after considering the
  decay branching fractions, tracking and particle identification with
  $10^{10}$ $J/\psi$ events. The sensitivity is estimated without
  considering the possible background dilution, which is small at
  \bes3.  Systematic uncertainties are expected to be of the same
  order as the statistical ones.}
\renewcommand{\arraystretch}{1.2}
\begin{center}
\begin{tabular}{ccc}
\hline \hline Channel & Number of events  & Sensitivity on ${\cal
A}_T$\\ \hline $J/\psi\to\Lambda\bar\Lambda\to [p\pi^-] [\bar
p\pi^+]$ & $2.6\times
10^6$ & 0.06\%\\
$J/\psi\to\Sigma^+ \bar \Sigma^- \to [p\pi^0] [\bar p\pi^0]$ &
$2.5\times
10^6$ & 0.06\%\\
$J/\psi\to\Xi^0\bar\Xi^0\to [\Lambda\pi^-] [\bar \Lambda\pi^+]$
&$1.1\times 10^6$ &0.1\%\\
$J/\psi\to\Xi^- \bar\Xi^+\to [\Lambda\pi^0] [\bar \Lambda\pi^0]$
&$1.6\times 10^6$ &0.08\%\\
 \hline\hline
\end{tabular}
\label{EventNumber}
\end{center}
\end{table}
\end{center}

By considering the efficiency of the \bes3 detector, the number of observed
events  and the corresponding statistical error for various
channels are estimated. The expected sensitivity with the $10^{10}$
$J/\psi$ events is shown in Table~\ref{EventNumber}.

\begin{table}[tbp]
\caption{The BESIII sensitivities, which are presented as the estimated uncertainties multiplied by the square root of the observed signal yields, in the the spin entangled
   $\Xi$-$\bar\Xi$ system for the extracted parameters. Errors
    for the parameters of the charge conjugated decay modes are the
    same. The input values of the $\Xi$ parameters have only minor
    effect on the sensitivities.}
\renewcommand{\arraystretch}{1.2}
\scalebox{0.75}{
  \begin{tabular}{lrrrrrrrrrrrrr}  \hline\hline
&$\alpha_\Xi$& $\alpha_\Lambda$&$\phi_\Xi$&$\alpha_\psi$ &$\Delta\Phi$&$\left<\alpha_\Xi\right>$&$A_\Xi$&
$\left<\alpha_\Lambda\right>$&$A_\Lambda$&$\left<\alpha_\Xi\alpha_\Lambda\right>$&$A_{\Xi\Lambda}$&$\left<\phi_\Xi\right>$&$B_\Xi$\\
\hline
    $J/\psi\to\Xi^-\bar\Xi^+$ ($\Delta\Phi=0$)&$\phantom{-}2.0$&$\phantom{-}3.1$&$\phantom{-}5.8$&
    $\phantom{-}3.5$&$\phantom{-}6.0$&1.4&3.7&1.7&3.5&0.78&4.0&4.1&110\\
    $J/\psi\to\Xi^-\bar\Xi^+$ ($\Delta\Phi=\pi/2$)&1.9&2.8&5.4&3.0&13
    &1.4&3.5&1.6&3.1&0.76&3.9&3.8&100\\
    $e^+e^-\to\Xi^-\bar\Xi^+$    ($\alpha_\psi=1$)&$1.9$&$2.7$&$5.0$&$-$&$-$
    &1.3&3.4&1.4&3.1&0.76&4.0&3.5&96\\
    $\eta_c,\chi_{c0}\to\Xi^-\bar\Xi^+$ &$1.6$&$2.2$&$3.7$&$-$&$-$
    &1.1&2.9&1.0&2.6&0.72&3.9&2.6&71\\
\hline\hline
  \end{tabular}
  }
\label{tab:sigpara}
\end{table}

A wide range of CP tests in hyperon sector can also be
performed in a single measurement of the spin entangled $\Xi$-$\bar\Xi$
system~\cite{Adlarson:2019jtw}.  From the joint distributions for
$e^+e^-\to\Xi\bar\Xi$, it shows that the role of the
transverse polarization is fully replaced by the diagonal spin
correlations between the cascades. All decay parameters can be
determined simultaneously and the statistical uncertainties do not
depend on the size of the transverse polarization in the production
process. The BESIII sensitivities are shown in Table~\ref{tab:sigpara}
with correlations between parameters considered.  The results
practically do not change between the two extreme cases:
$\Delta\Phi=0$ and $\pi/2$.  The results for other decays:
$\psi(3686)\to\Xi^-\bar\Xi^+$ and $J/\psi,\psi(3686)\to\Xi^0\bar\Xi^0$
are similar. Also shown are the results for the
$e^+e^-\to \Xi^-\bar\Xi^+$ asymptotic case with $\alpha_\psi=0$ and
for a scalar charmonium decay to $\Xi\bar\Xi$. Contrary to
$e^+e^-\to \Lambda\bar\Lambda$, the polarization in the production
process plays practically no role.  The weak decay phases $\phi_\Xi$
and $\phi_{\bar\Xi}$ are not correlated with each other and with any
other parameters.

\subsection{Constraint on BNV from $\Lambda-\bar\Lambda$ oscillation}
\label{cpv:lbdosc}
The stability of ordinary matter implies baryon number ($B$) conservation. However, the observed fact of baryon asymmetry in the Universe shows that baryon number should be broken. There are many theoretical models in which
$B$ is not exactly conserved, with $B$ and lepton number ($L$)
violated simultaneously while conserving $B-L$.

It was pointed out long ago~\cite{ref::marshak} that a crucial test of
baryon number violation are neutron-antineutron ($n$-$\bar n$)
oscillations, and many corresponding experiments have been conducted~\cite{ref::pdg2016}. If $n$-$\bar n$ oscillation exists,
$\Lambda$-$\bar\Lambda$ oscillations may also take
place~\cite{ref::Luk}.  There is a proposal to search for
$\Lambda$-$\bar\Lambda$ oscillations in the decay of
$J/\psi\to\Lambda\bar\Lambda$~\cite{ref::prd81-051901r}. Until now,
however, there has not been any direct experimental searches for this
process.

At \bes3, the decay of $J/\psi\to pK^-\bar\Lambda$ has a very simple
final state and is almost background free, so it is well suited for
searching for $\Lambda-\bar\Lambda$ oscillations. Initially, $J/\psi$ decays into $pK^-\bar\Lambda$ final state
(defined as a right-sign event, as $\bar\Lambda\to \bar p \pi^+$ is detected along with $pK^-$), and then, with some $\bar\Lambda$
oscillating into $\Lambda$, the final state becomes $pK^-\Lambda$
(defined as a wrong-sign event, as $\Lambda\to p \pi^-$ is detected  with $pK^-$), so the probability of generating a
final $\Lambda$ from an initial $\bar\Lambda$ can be determined by the
ratio of wrong sign events over right sign events.

The time evolution of
the $\Lambda$-$\bar\Lambda$ oscillation can be described~\cite{ref::prd81-051901r} by a
Schroedinger-like equation,
\[
i\frac{\partial}{\partial t}\left(
\begin{array}{c}
 \Lambda(t)   \\
 \bar\Lambda(t)
\end{array}
\right)
=M
\left(
\begin{array}{c}
 \Lambda(t)   \\
 \bar\Lambda(t)
\end{array}
\right),
\]
where the matrix $M$ is Hermitian, 
\[
M=
\left(
\begin{array}{cc}
 m_{\Lambda}-\Delta E_{\Lambda}  & \delta m_{\Lambda\bar\Lambda}   \\
 \delta m_{\Lambda\bar\Lambda}  & m_{\bar\Lambda}-\Delta E_{\bar\Lambda}
\end{array}
\right),
\]
and $\delta m_{\Lambda\bar\Lambda}$ is the $\Delta B=2$ oscillation
parameter due to some NP effect,  $m_{\Lambda}$
($m_{\bar\Lambda}$) is the mass of the $\Lambda$($\bar\Lambda$)
baryon, and $\Delta E$ is the energy split due to an external magnetic field.
For an unbound $\Lambda$ propagating in a vacuum without an
external field, both $\Delta E_{\Lambda}$ and $\Delta E_{\bar\Lambda}$
are zero.

Starting with a beam of free $\Lambda$ particles, the probability $P(\bar\Lambda, t)$ of generating
a $\bar\Lambda$ after traveling some time $t$, is given by
$$P(\bar\Lambda, t)=\sin^2(\delta m_{\Lambda\bar\Lambda}\cdot t).$$
\bes3 can measure the time
integral of the probability, $i.e.$, the oscillation rate, by
$$P(\bar\Lambda)=\frac{\int^{T_{\rm max}}_{0}\sin^2(\delta m_{\Lambda\bar\Lambda}\cdot t)\cdot e^{-t/\tau_{\Lambda}}\cdot dt}{\int^{\infty}_{0}e^{-t/\tau_{\Lambda}}\cdot dt},$$
where $P(\bar\Lambda)$ is the time-integrated probability of
$\Lambda\to\bar\Lambda$, $\tau_{\Lambda}$ is the lifetime of
the $\Lambda$ baryon, and $T_{\rm max}$ is the maximum flight time.  If we assume $T_{\rm max}$ is large enough in the BESIII tracking system,  the oscillation parameter can be deduced as
$$\delta m_{\Lambda\bar\Lambda}=\sqrt{\frac{P(\bar\Lambda)}{2\cdot(\tau_{\Lambda}/\hbar)^2}}.$$
With the recently-accumulated data sample of $10^{10}$ $J/\psi$
events, the upper limit of the $\Lambda$-$\bar\Lambda$ oscillation
rate will be at $10^{-6}$ level (90\% C.L.) based on the analysis of
$J/\psi\to pK^-\bar\Lambda$ events, and the constraint on
$\delta m_{\Lambda\bar\Lambda}$ will be reduced to $10^{-16}$ MeV
(90\% CL). On the other hand, a time-dependent analysis of the produced $\Lambda$-$\bar\Lambda$ pairs from the $J/\psi$ decays can be investigated, taking advantage of their long mean flight distance of 7.6 cm in the detector.  So \bes3 will provide a very stringent
constraint on NP in this channel.

%%%%%%%%%%%%%%%%%%%%%%%%%%%%
\subsection{More symmetry violation in hyperon decays }
\label{cpv:morelbd}

Since \bes3 has a rather large hyperon data set,
$\Lambda$'s from hyperons can also be used to study symmetry
violations and search for NP, as was recently summarized
in a nice review~\cite{Li2016}.  Here we reproduce the summary table
(Table~\ref{tab:rare-exotics}) showing lepton or baryon-number
violating hyperon decays and their expected sensitivities with future
\bes3 data sets. More information can be found in the original
paper~\cite{Li2016}.

 \begin{table}[htbp]
  \begin{center}
    \caption{\small Lepton or baryon number violating hyperon decays and
      expected sensitivities with $10^{10}$ events on the $J/\psi$
      peak and $3\times 10^{9}$ events on the $\psi(3686)$ peak.  The current $J/\psi$ data are from CLAS as listed in
      PDG. ``-'' indicates ``not available'', $l = e$
      or $\mu$, and $M^{\pm}$ refers to the charged stable mesons
      ($M^\pm= \pi^\pm$ or $K^\pm$).  Each reaction shows evidence of
      $\Delta L = \pm 1$ or/and $\Delta B \neq 0$, and each reaction
      conserves electric charge and angular momentum~\cite{Li2016}. }
\vspace*{0.1cm}
 \label{tab:rare-exotics}
\begin{tabular}{@{}lllll}
\hline\hline
   Decay mode    & Current data &   Expected \bes3       &$\Delta L$  & $\Delta B$       \\
        & ${\cal B}$ ($\times 10^{-6}$) & ${\cal B}$   ($\times 10^{-6}$)        \\
              & & (at the 90\% C.L.)         \\
\hline
$\Lambda \to M^+ l^-$  &   $<0.4\sim 3.0$   & $<0.1$   & +1 & $-1$ \\
$\Lambda \to M^- l^+$  &    $<0.4 \sim 3.0$    &  $<0.1$ & $-1$ & $-1$ \\
$\Lambda \to K^0_{\rm S}  \nu $  &    $<20$   & $<0.6$   & $+1$ & $-1$ \\
$\Sigma^+ \to K^0_{\rm S}  l^+ $       & -   & $<0.2$  & $-1$ & $-1$ \\
$\Sigma^- \to K^0_{\rm S}  l^- $       & -   &  $<1.0$ & $+1$ & $-1$ \\
$\Xi^- \to K^0_{\rm S}  l^- $       & -   &$<0.2$   & $+1$ & $-1$ \\
$\Xi^0 \to M^+  l^-$ & -  &  $<0.1$& +1 & $-1$ \\
$\Xi^0 \to M^- l^+$  &  -   & $<0.1$  & $-1$ & $-1$ \\
$\Xi^0 \to K^0_{\rm S}  \nu $  &    -   & $<2.0$  & $+1$ & $-1$ \\
\hline\hline
\end{tabular}
\end{center}
\end{table}

%% file: New_physics/LLbar.tex
\begin{figure}[hbtp]
\begin{center}
  \includegraphics[width=7.5cm]{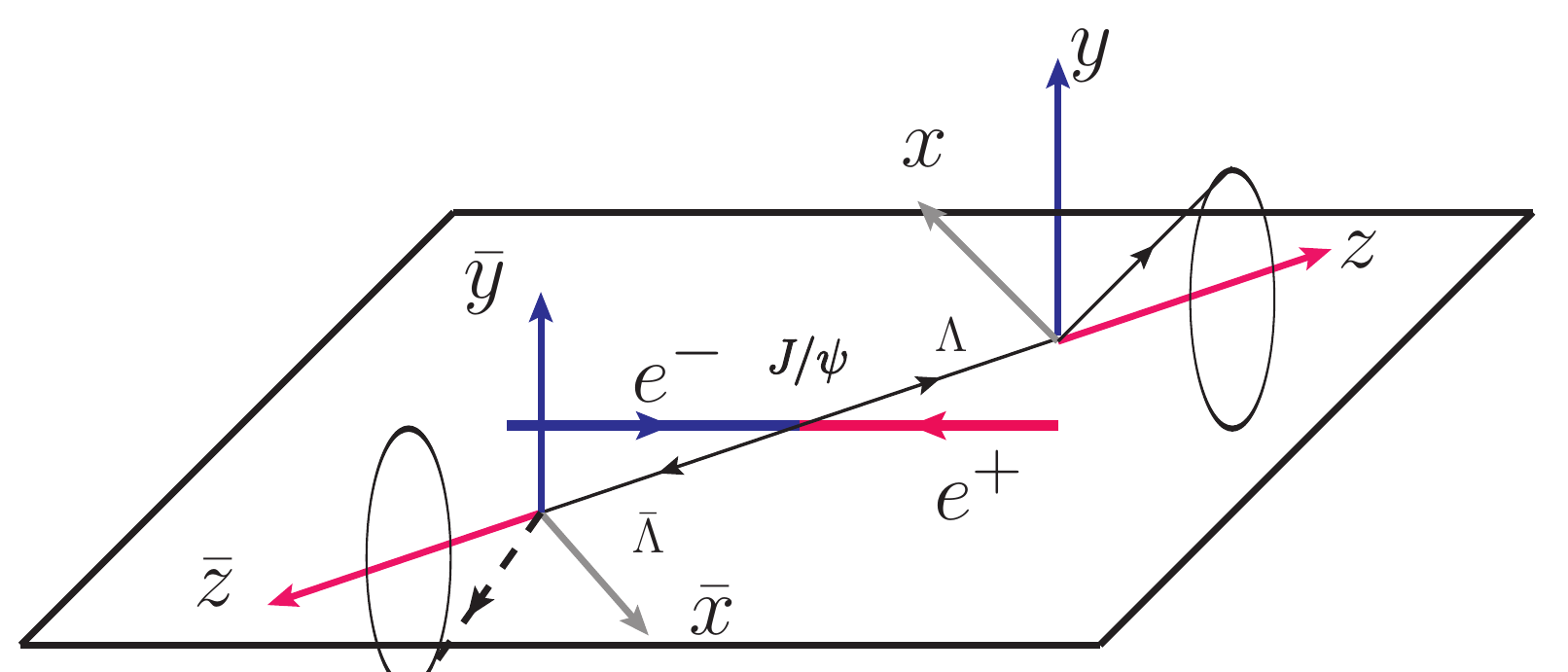}
  \caption{\small Definition of the helicity frames for $J/\psi \to \Lambda \bar{\Lambda} \to p \pi^{-} \bar{p} \pi^{+}$.}
  \label{HelAmp}
\end{center}
\end{figure}

At \bes3, $\jpsi$ mesons are produced in the annihilation of unpolarized
electron-positron beams and therefore the spin-density matrix of the
$\jpsi$ depends only on the scattering angle $\theta_\Lambda$ between
the electron beam direction and the $\Lambda$ momentum in the reaction
cms system.  The helicity frames to describe the
subsequent decay chains are shown in Fig.~\ref{HelAmp}.

The coherent production of $\Lambda/ \bar{\Lambda}$ pairs from the
decay $J/\psi \to \Lambda \bar{\Lambda}$ with subsequent weak decays
of the $\Lambda$ and $\bar{\Lambda}$ is a very simple spin-entangled
quantum system where the final state is specified by four real
parameters summarized in Table~\ref{definition}.

\begin{table}[tb]
\caption{\small List of kinematic variables (helicity angles) and parameters
used in the analysis of the decay chain.\label{definition}}
\begin{center}
\begin{tabular}{lccl}
\hline\hline
Decay & Coordinate system &Helicity angles  & Parameters \\\hline
$\jpsi\to\Lambda\bar\Lambda$ & cms & ($\theta_\Lambda,\phi_\Lambda$) & $\alpha_\psi$, $\Delta\Phi$\\
$\Lambda\to p\pi^-$ &$(x,~y,~z)$& ($\theta_1,\phi_1$) & $\alpha_-$\\
$\bar\Lambda\to \bar p\pi^+$ &$(\bar x,~\bar y,~\bar z)$& ($\theta_2,\phi_2$) & $\alpha_+$\\ \hline\hline
\end{tabular}
\end{center}
\end{table}

A recent \bes3 study with $1.3\times 10^9$ $J/\psi$ events has
observed a transverse spin polarization of $\Lam/\bar\Lam$ and the
phase between the hadronic form factors has been determined~\cite{Ablikim:2018zay}. The
polarization effect is illustrated by dividing the data sample into 50
$\cos\theta_\Lambda$ bins and plotting the $\cos\theta_\Lambda$
dependence of the moment
$\mu(\cos\theta_\Lambda)=1/N\sum_i^{N(\theta_\Lambda)}(\sin\theta_1^{i}\sin\phi_
1^{i}-\sin\theta_2^{i}\sin\phi_2^{i})$, where $N$ is the total number
of events in the data sample and $N(\theta_\Lambda)$ is the number of
events in a $\cos\theta_\Lambda$ bin. This dependence enables us to
extract simultaneously $\alpha_{-}=0.750\pm0.009\pm0.004$ and
$\alpha_+=-0.758\pm0.010\pm0.007$ and calculate the most precise value
for $A$ of $-0.006\pm0.012\pm0.007$, where in the propagation of the
statistical uncertainty a large value of the correlation coefficient
$\rho(\alpha_+,\alpha_-)=0.82$ is included.

New \bes3 data of $10^{10}$ $J/\psi$ events  will enable an improved measurement of the parameter
$A$.  In the future measurement the systematic uncertainty will be
significantly improved since this simple system permits several
internal consistency checks to be performed. In the general case of
two spin one-half particles we have 16 polarization and spin
correlation distributions to be fully described by only four global
parameters. In addition, the use of other monitoring and calibration
channels allows independent corrections for any bias to be determined.
In addition, if a similar polarization is observed also for other hyperons then
similar studies will be possible, {\it e.g.,} for the $\Xi$ hyperon with
its cascade decays $B$ asymmetry tests would be possible.

%% file: New_physics/bsmheavy.tex
\section{Charged Lepton Flavor (Number) Violation decays}
\label{sec:clfv}

%%%%%%%%%%%%%%%%%%%%%%
%\subsection{Introduction}
Charged Lepton Flavor Violation (CLFV) processes are highly suppressed
in the SM by finite but tiny neutrino masses. Their branching
fractions are predicted to be negligibly small -- and no such
reactions have been observed. Yet, there are various theoretical
models that predict the rates for CLFV transitions to be large enough
to be experimentally observable.  Examples include SUSY grand unified
theories~\cite{lfv1}, SUSY with a right-handed neutrino~\cite{lfv2},
gauge-mediated SUSY breaking~\cite{lfv3}, SUSY with vector-like
leptons~\cite{lfv4}, SUSY with $R$-parity violation~\cite{lfv5},
models with $Z^\prime$~\cite{lfv6}, or models with Lorentz
non-invariance~\cite{lfv7}. While the discovery of neutrino
oscillations has confirmed the existence of neutrino masses and LFV in
the neutral lepton sector, detection of LFV in the charged lepton
sector would provide direct evidence for NP.

Experimentally, searches for CLFV effects have been carried out in a
variety of ways, including decays of leptons, pseudoscalar, and
vector mesons, as well as in other processes.  In the charmonium
system, BESII obtained a limit
${\cal B}(J/\psi \to e \mu ) < 1.1 \times 10^{-6}$~\cite{lfv14} by
analyzing a data sample of 58 million $J/\psi$ events. \bes3 has so
far analyzed 255 million $J/\psi$ decays. Four events have been
observed, which is consistent with background estimate. As a result,
\bes3 set the upper limit of
${\cal B}(J/\psi \to e \mu ) < 1.6 \times 10^{-7}$ at the 90\%
C.L.~\cite{TheBESIIICollaboration2013}, which is currently the best
upper limit on CLFV in charmonium decays.  BESII also placed bounds on
${\cal B}(J/\psi \to \mu \tau) < 2. \times 10^{-6}$ and
${\cal B}(J/\psi \to e \tau) < 8.3 \times 10^{-6}$~\cite{lfv15}.  The
limits on these decay channels will be updated at \bes3 with the data
set that includes $10^{10}$ $J/\psi$ decays, so orders of magnitude
improvement is expected.

CLFV and lepton-number-violating (LNV) processes can also be probed in
$D$-meson decays at \bes3. No evidence has been found for the
$D$-meson decays with either CLFV or LNV. The present experimental
bounds on the branching fractions are generally set at the level of
$10^{-6}$ to $10^{-5}$ (with a notable exception of $D^0\to \mu e$,
where
${\cal B} (D^0\to \mu^\pm e^\mp) < 1.3 \times
10^{-8}$)~\cite{Aaij:2015qmj}. $D$ decays with LNV, such as $D^+ \to l^+ l^+ X^-$ and $D_s^+ \to l^+ l^+ X^-$, are
also forbidden in the minimal SM, but are possible if massive
neutrinos are Majorana particles.

However, for CLFV processes such as $D^0 \to l^+ l^{\prime-}$ and
$D^0 \to l^+ l^{\prime-} X$, LHCb will dominate the searches even if
\bes3 increases the available data set of $D$-meson decays by an order
of magnitude in the future.  On the other hand, \bes3 has some  potential to
search for LNV transitions such as $D^+ \to l^+ l^+ X^-$ and $D_s^+
\to l^+ l^+ X^-$, due to the clean environment and low charge
confusion rates.

%%%%%%%%%%%%
\subsubsection{Effective Lagrangian and NP models}\label{Intro_cCLFV}

NP models probed at \bes3 include those predicting
lepton-flavor violating interactions.  They can be probed in decays to
flavor-non-diagonal combination of final state leptons
$\ell_i = e,\mu,\tau$.  Due to a multitude of NP models contributing to
CLFV processes, it is advantageous to introduce an effective
Lagrangian that economically encodes all of these NP models. The details
of the models are encoded in Wilson coefficients, while quark-lepton
dynamical effects are described in a set of CLFV operators.

The effective Lagrangian ${\cal L}_{\rm eff}$ that involves CLFV can
in general be written as
\begin{equation}\label{eqn:Leff}
{\cal L}_{\rm eff}= {\cal L}_{\ell q} + {\cal L}_D + {\cal L}_{G} + \cdot \cdot \cdot,
\end{equation}
where ${\cal L}_D$ is a dipole part, ${\cal L}_{\ell q}$ is the part that contains four-fermion interactions, and
${\cal L}_{G}$ is a gluonic part~\cite{Hazard:2016fnc,Celis:2014asa}.

The dipole part ${\cal L}_{D}$ has been extremely well constrained in
purely leptonic CLFV decays of the type $\ell_1 \to \ell_2\gamma$
\cite{Hazard:2017udp,Raidal:2008jk}. It appears that any possible
contributions from ${\cal L}_{D}$ to charmed particle decays would be
four-to-six orders of magnitude smaller than the ones from other
sectors of ${\cal L}_{\rm eff}$, so it will be neglected in the following
discussion~\cite{Hazard:2016fnc}.

The four-fermion dimension-six lepton-quark part of the effective Lagrangian, Eq.~(\ref{eqn:Leff}) is
\cite{Hazard:2017udp}
\begin{eqnarray}\label{L4Fermion}
{\cal L}_{\ell q} = -\frac{1}{\Lambda^2} \sum_{q=u,c} \Big[
\left( C_{VR}^{qc \ell_1\ell_2} \ \overline\ell_1 \gamma^\mu P_R \ell_2 +
C_{VL}^{qc \ell_1\ell_2} \ \overline\ell_1 \gamma^\mu P_L \ell_2 \right) \ \overline q \gamma_\mu c &&
\nonumber \\
+ \
\left( C_{AR}^{qc \ell_1\ell_2} \ \overline\ell_1 \gamma^\mu P_R \ell_2 +
C_{AL}^{qc \ell_1\ell_2} \ \overline\ell_1 \gamma^\mu P_L \ell_2 \right) \ \overline q \gamma_\mu \gamma_5 c &&
\nonumber \\
+ \
m_2 m_c G_F \left( C_{SR}^{qc \ell_1\ell_2} \ \overline\ell_1 P_L \ell_2 +
C_{SL}^{qc \ell_1\ell_2} \ \overline\ell_1 P_R \ell_2 \right) \ \overline qc &&
\\
+ \
m_2 m_c G_F \left( C_{PR}^{qc \ell_1\ell_2} \ \overline\ell_1 P_L \ell_2 +
C_{PL}^{qc \ell_1\ell_2} \ \overline\ell_1 P_R \ell_2 \right) \ \overline q \gamma_5 c
\nonumber \\
+ \
m_2 m_c G_F \left( C_{TR}^{qc \ell_1\ell_2} \ \overline\ell_1 \sigma^{\mu\nu} P_L \ell_2 +
C_{TL}^{qc \ell_1\ell_2} \ \overline\ell_1 \sigma^{\mu\nu} P_R \ell_2 \right) \ \overline q \sigma_{\mu\nu} c
 &+& h.c. ~ \Big] .
\nonumber
\end{eqnarray}
Here  $P_{\rm R,L}=(1\pm \gamma_5)/2$ is the right (left) chiral projection operator. In general the Wilson coefficients would
be different for different lepton flavors $\ell_i$ and quark flavors $q=u,c$. This implies that decays of charmonium states and
$D$-mesons probe different terms (and different models of NP) in the effective CLFV Lagrangian.

The dimension seven gluonic operators in ${\cal L}_{G}$ appear for flavor-diagonal quark transitions,
{\it i.e.,} for $q=c$~\cite{Celis:2014asa,Petrov:2013vka},
\begin{eqnarray}\label{LGluon}
{\cal L}_{G} = -\frac{m_2 G_F}{\Lambda^2} \frac{\beta_L}{4\alpha_s} \Big[
\Big( C_{GR}^{\ell_1\ell_2} \ \overline\ell_1 P_L \ell_2 +
C_{GL}^{\ell_1\ell_2} \ \overline\ell_1 P_R \ell_2 \Big)  G_{\mu\nu}^a G^{a \mu\nu} &&
\nonumber \\
+ ~ \Big( C_{\bar G R}^{\ell_1\ell_2} \ \overline\ell_1 P_L \ell_2 +
C_{\bar G L}^{\ell_1\ell_2} \ \overline\ell_1 P_R \ell_2 \Big)  G_{\mu\nu}^a \widetilde G^{a \mu\nu}
 &+& h.c. \Big].
\end{eqnarray}
Here $\beta_L=-9 \alpha_s^2/(2\pi)$ is defined for the number of light active flavors, $L$, relevant to the scale
of the process, for which we take $\mu \approx 2$~GeV. All Wilson coefficients
should also be calculated at the same scale. $G_F$ is the Fermi constant and
$\widetilde G^{a \mu\nu} = (1/2) \epsilon^{\mu\nu\alpha\beta} G^a_{\alpha\beta}$ is a dual to the
gluon field strength tensor.

Each term in Eqs.~(\ref{L4Fermion}-\ref{LGluon}) can be separately probed in two-body decays of
charmonium states with different $J^{PC}$ and $D$-mesons, as their constrained kinematics only
selects operators with particular quantum numbers. As will be shown below, $J/\psi \to \bar\ell_1\ell_2$
decays select vector and tensor operators, while $\chi_c \to \bar\ell_1\ell_2$ and $\eta_c \to \bar\ell_1\ell_2$
only select scalar and pseudoscalar/axial operators in Eq.~(\ref{L4Fermion}).

%%%%%%%%%%%%%%%%%%%%%
\subsection{Decays of $J/\psi, \psi(3686) \to l_1 l_2, l_1 l_2
  \gamma$} \label{Spin1CLFVsection}

Experimental constraints on $J/\psi \to \ell_1 \ell_2$ branching fractions can be effectively converted
to bounds on Wilson coefficients of vector and tensor operators in Eq.~(\ref{L4Fermion}).
Those Wilson coefficients can then be related to model parameters of explicit realizations of possible
UV completions of the effective Lagrangian in Eq.~(\ref{eqn:Leff}). Examples of particular new
physics models include $Z^\prime$ scenarios~\cite{Nussinov:2000nm}, R-parity violating
supersymmetric models~\cite{Dreiner:2001kc,Dreiner:2006gu,Sun:2012yq}, and other
approaches~\cite{Abada:2015zea,Black:2002wh}.

The most general expression for the $J/\psi~(\mbox{or any~}\psi(nS)) \to \ell_1 \overline \ell_2$ decay
amplitude can be written as
\begin{eqnarray}\label{Spin1Amp}
{\cal A}(V \to \ell_1 \overline \ell_2) = \overline{u}(p_1, s_1) \left[
A_V^{\ell_1\ell_2} \gamma_\mu + B_V^{\ell_1\ell_2} \gamma_\mu \gamma_5
+ \frac{C_V^{\ell_1\ell_2}}{m_{V}} (p_2-p_1)_\mu
\right. ~~~~~~~~~~~~~~
\nonumber \\
\qquad + \left.
\frac{iD_V^{\ell_1\ell_2}}{m_{V} }(p_2-p_1)_\mu \gamma_5 \
\right] v(p_2,s_2) \ \epsilon^\mu(p),
\end{eqnarray}
where $V=J/\psi$ or $\psi(3686)$, and $A_V^{\ell_1\ell_2}$,
$B_V^{\ell_1\ell_2}$, $C_V^{\ell_1\ell_2}$, and $D_V^{\ell_1\ell_2}$
are dimensionless constants which depend on the underlying Wilson
coefficients of the effective Lagrangian of Eq.~(\ref{L4Fermion}) as
well as on hadronic effects associated with meson-to-vacuum matrix
elements or decay constants. We shall neglect dipole and tensor operator
contributions, which implies that $C_V^{\ell_1\ell_2}=D_V^{\ell_1\ell_2}=0$
\cite{Hazard:2016fnc}. The branching fractions of the vector $\psi$ states are calculated from
Eq.~(\ref{Spin1Amp}), which yield the ratio
\begin{eqnarray}\label{BRSpin1}
\frac{{\cal B}(\psi \to \ell_1 \overline \ell_2)}{{\cal B}(\psi \to e^+e^-)} &=&
\left(\frac{m_V \left(1-y^2\right)}{4\pi\alpha  f_\psi Q_q}\right)^2
\left[
\left|A_V^{\ell_1\ell_2}\right|^2 +  \left|B_V^{\ell_1\ell_2}\right|^2
\right].
\end{eqnarray}
Here $\alpha$ is the fine structure constant, $Q_c=2/3$ is the charge of the $c$-quark,
the mass of the lighter of the two leptons has been neglected, and $y=m_2/m_V$. The coefficients
$A_V^{\ell_1\ell_2}$ and $B_V^{\ell_1\ell_2}$ depend on the
initial state meson,
\begin{eqnarray}\label{VCoef1}
A_V^{\ell_1\ell_2} &=& ~~\frac{f_V m_V}{2 \Lambda^2}
\left(C_{VL}^{cc\ell_1\ell_2} + C_{VR}^{cc\ell_1\ell_2}\right),
\nonumber \\
B_V^{\ell_1\ell_2} &=& -\frac{f_V m_V}{2\Lambda^2}
\left(C_{VL}^{cc\ell_1\ell_2} - C_{VR}^{cc\ell_1\ell_2}\right).
\end{eqnarray}
The constraints on the Wilson coefficients also depend on the meson decay constants,
\begin{eqnarray}\label{DeConV}
\langle 0| \overline q \gamma^\mu q | V(p) \rangle &=& f_V m_V \epsilon^\mu (p)\,,
\end{eqnarray}
where $\epsilon^\mu(p)$ is the $V$-meson polarization vector, and $p$ is its momentum~\cite{Becirevic:2013bsa}.
The decay constants are $f_{J/\psi}=418\pm 9$ MeV and $f_{\psi(3686)}=294\pm 5$ MeV.
They are both known  experimentally from leptonic decays and
theoretically from lattice or QCD sum rule calculations.

Experimentally, there is an ongoing analysis based on an existing data
set of $1.3\times 10^{9}$ $J/\psi$ decays.  In this study the selection efficiencies
for both $J/\psi \to \mu \tau$ and $J/\psi \to e \tau$ are around
14\%.  Based on the same Cut and Count (CC) analysis technique, with
the assumption of similar efficiencies, we can make a projection of
future \bes3 sensitivity in this channel. By constructing cocktail
samples from the $1.3\times 10^9$ data set, and then performing toy MC
(pseudo-experiment) studies, the sensitivity with $10^{10}$  $J/\psi$ events
is evaluated.
The 90\% C.L. upper limits on the numbers of signal events are estimated
according to the recoiled $\tau$ mass distributions. With $10^{10}$
$J/\psi$ decays, the projected sensitivities of the two channels are
estimated to be both at $10^{-8}$ level. Such results would represent
an improvement of almost two orders of magnitudes compared to those
obtained at BESII. The systematic uncertainties have also been
estimated, and are expected to be sub-dominant even for this very
large sample.  With more advanced analysis techniques such as
multivariate analyses (MVA), the efficiencies could be increased with
decreased background levels which results in even better sensitivities.  The
current and future \bes3 constraints on
${\cal B}(J/\psi \to \ell_1 \ell_2)$ are summarized in
Table~\ref{tab:Vdecaylimits}.
Based on these projections, the
resulting constraints on the combination of Wilson coefficients and
NP scale $\Lambda$, both current and projected, can be found
in Table~\ref{tab:Constraints4fermion}.

\begin{table*}[tp]
  \caption{\label{tab:Vdecaylimits} \small
    Current~\cite{Hazard:2016fnc} and future \bes3 constraints on
    ${\cal B}(J/\psi \to \ell_1 \ell_2)$. The projections with CC and
    MVA methods provide the conservative and agressive estimations,
    respectively.}
\begin{center}
\begin{tabular}{cccc} \hline\hline
$\ell_1 \ell_2$ &$\mu \tau$ & $e \tau$ & $e \mu$  \\
\hline
Current upper limit &  $2.0 \times 10^{-6}$ & $8.3 \times 10^{-6}$ & $1.6 \times 10^{-7}$ \\
\bes3 projected (CC)  & $3.0 \times 10^{-8}$  & $4.5 \times 10^{-8}$ & $1.0 \times 10^{-8}$ \\
\bes3 projected (MVA) & $1.5 \times 10^{-8}$ & $2.5 \times10^{-8}$ & $6.0 \times 10^{-9}$ \\
  \hline\hline
\end{tabular}
\end{center}
\end{table*}
\begin{table*}[tp]
  \caption{\label{tab:Constraints4fermion} \small Constraints on the
    Wilson coefficients of four-fermion operators. Note that the
    constraints on the right-handed couplings $(L\to R)$ are the
    same. ``$-$'' means that no constraints are currently available,
    ``FPS'' means that the decay is forbidden by phase space, and
    ``N/A'' means that \bes3 sensitivity studies are yet to be
    performed. Current constraints are from~\cite{Hazard:2016fnc}.}
\begin{center}
  \begin{tabular}{ccccc} \hline \hline
    & Leptons &\multicolumn{2}{c}{Constraints}\\
    Wilson coeff (${\rm GeV}^{-2}$) & $\ell_1 \ell_2$ & Current &
                                                                  Projected
    \\ \hline
    $~$ & $\mu \tau$ & $5.5 \times 10^{-5}$ & $[5.0, 7.1] \times 10^{-6}$  \\
    $\left| C_{VL}^{cc\ell_1\ell_2}/\Lambda^2 \right|$ & $e \tau$
              & $1.1 \times 10^{-4}$ & $[6.5, 8.7] \times 10^{-6}$ \\
    $~$ & $e \mu$ & $1.0 \times 10^{-5}$ &  $[2.8, 3.7] \times 10^{-6}$  \\
    \hline
    $~$ & $\mu \tau$ & $-$ & $7.4 \times 10^{-4}$ \\
    $\left| C_{AL}^{cc\ell_1\ell_2}/\Lambda^2 \right|$ & $e \tau$
              & $-$ & $7.4 \times 10^{-4}$ \\
    $~$ & $e \mu$ & $-$ & N/A \\
    \hline
    $~$ & $\mu \tau$ & $-$ & $2.0$ \\
    $\left| C_{SL}^{cc\ell_1\ell_2}/\Lambda^2 \right|$ & $e \tau$
              & $-$ & $2.0$ \\
    $~$ & $e \mu$ & $-$ & N/A  \\
    \hline
    $~$ & $\mu \tau$ & FPS & FPS \\
    $\left| C_{AL}^{uc\ell_1\ell_2}/\Lambda^2 \right|$ & $e \tau$
              & $-$ & N/A \\
    $~$ & $e \mu$ & $1.3 \times 10^{-8}$ &  $2.2 \times 10^{-8}$\\
    \hline
    \hline
  \end{tabular}
\end{center}
\end{table*}

A promising approach for increasing the sensitivity of $J/\psi$ decays
to CLFV operators is to consider radiative charged lepton-flavor
violating (RCLFV) transitions. The addition of a photon to the final
state certainly reduces the number of the events available for studies
of CLFV decays. However, the data set of $J/\psi$'s accumulated by
\bes3 is huge, and this requirement also makes it possible for other
operators in ${\cal L}_{\rm eff}$ to contribute. Since the final state
kinematics is less constrained than in two-body decays, the
constraints on the Wilson coefficients of the effective Lagrangian
would depend on a set of $V \to \gamma$ form factors that are not very
well known~\cite{Hazard:2016fnc}.  To place meaningful constraints on
the Wilson coefficients from non-resonance $J/\psi$ RCLFV decays one
would need to employ the single-operator dominance hypothesis, {\it
  i.e.,} assume that only one operator contributes at a time. For the
axial, scalar, and pseudoscalar operators one
has~\cite{Hazard:2016fnc}
\begin{eqnarray} \label{3bodydecayrate}
\Gamma_A (J/\psi \to \gamma \ell_1 \overline \ell_2) &=& \frac{1}{18} \frac{\alpha Q_q^2}{\left(4 \pi\right)^2}
\frac{f_V^2 m_V^3}{\Lambda^4} \left[\left(C^{cc\ell_1\ell_2}_{AL}\right)^2+\left(C^{cc\ell_1\ell_2}_{AR}\right)^2 \right],
\nonumber \\
\Gamma_S(J/\psi \to \gamma \ell_1 \overline \ell_2) &=& \frac{1}{144} \frac{\alpha Q_q^2}{\left(4 \pi\right)^2}
\frac{f_V^2 G_F^2 m_V^7}{\Lambda^4} \left[\left(C^{cc\ell_1\ell_2}_{SL}\right)^2+\left(C^{cc\ell_1\ell_2}_{SR}\right)^2 \right] y^2,
\\
\Gamma_P (J/\psi \to \gamma \ell_1 \overline \ell_2) &=& \frac{1}{144} \frac{\alpha Q_q^2}{\left(4 \pi\right)^2}
\frac{f_V^2 G_F^2 m_V^7}{\Lambda^4} \left[\left(C^{cc\ell_1\ell_2}_{PL}\right)^2+\left(C^{cc\ell_1\ell_2}_{PR}\right)^2 \right] y^2.
\nonumber
\end{eqnarray}
The $J/\psi \to \gamma e \mu$ channel is experimentally challenging at \bes3,
so we focus on $J/\psi \to \gamma \mu \tau$ and
$J/\psi \to \gamma e \tau$, where there is an ongoing analysis involving
the current data set. If MVA were to be used, the efficiency would be
about 35\% for both channels. There is no detailed projection yet, but the
sensitivity to branching fractions could then reach $(1 - 3)\times 10^{-8}$.

%%%%%%%%%%%%%%%%%%%
\subsection{$\chi_c(\eta_c) \to l_1 l_2$ via photon tagging in $\psi(3686) \to
  \gamma \chi_c(\eta_c)$} \label{Spin0CLFVsection}

Similarly to probing operators with vector quantum numbers, as described in Sec.~\ref{Spin1CLFVsection}, the scalar and
pseudoscalar operators in Eq.~(\ref{L4Fermion}) can be probed in decays of scalar and pseudoscalar charmonia.
Although these states are not produced directly in $e^+e^-$ collisions, they can be studied in the radiative decays of vector charmonia~\cite{Hazard:2016fnc}.
Since at resonance
\begin{equation}
{\cal B}(V \to \gamma  \ell_1 \overline \ell_2) =
{\cal B}(V \to \gamma M) {\cal B}(M \to \ell_1 \overline \ell_2),
\end{equation}
states with large ${\cal B}(V \to \gamma M)$ can be used to probe ${\cal B}(M \to \ell_1 \overline \ell_2)$~\cite{Hazard:2016fnc}.
An example includes
\begin{eqnarray}\label{BranchRadc}
&&  {\cal B}(\psi(3686) \to \gamma \chi_{c0} ) = (9.99 \pm 0.27)\% \ ,
\nonumber \\
&& {\cal B}(\psi(3770) \to \gamma \chi_{c0}) = (0.73 \pm 0.09)\% \ ,
\nonumber
\end{eqnarray}
for probing scalar operators in the decays of the scalar states
$M=\chi_c$. \bes3 will have the highest sensitivity in the
$\psi(3686) \to \gamma \chi_{c0}$, and $\chi_{c0}\to \mu \tau$ and $\chi_{c0} \to e \tau$ decay channels.
It is estimated that the efficiency in ${\cal B}(\psi(3686) \to \gamma
\chi_{c0})$ could reach about 10\%.

If \bes3 could collect a data set with about $3\times 10^9$ $\psi(3686)$
events, the sensitivity for $\chi_{c0}\to \mu \tau$ and
$\chi_{c0} \to e \tau$ could reach $(1 - 3)\times 10^{-7}$. Similarly,
\begin{eqnarray}
&& {\cal B}(J/\psi \to \gamma \eta_c) = (1.7 \pm 0.4)\% \ ,
\nonumber \\
&& {\cal B}(\psi(3686) \to \gamma\eta_c) = (0.34 \pm 0.05)\% \ ,
\nonumber
\end{eqnarray}
for the pseudoscalar state $M=\eta_c$. \bes3 will be sensitive
to $J/\psi \to \gamma \eta_{c}$, $\eta_{c}\to \mu \tau$ and
$\eta_{c} \to e \tau$ decay channels.  With $10^{10}$ $J/\psi$
events and a detection efficiency of 10\%, the sensitivity to
$\eta_{c}\to \mu \tau$ and $\eta_{c} \to e \tau$ could reach
$(2 - 5)\times 10^{-7}$.

It must be pointed out that in both $\chi_{c0}$ and $\eta_{c}$ cases
the decay to $e \mu$ is more challenging, as both the level and
complexity of backgrounds are expected to be higher. It nevertheless
is still possible to probe these decay channels with the final \bes3
data set, although a separate dedicated analysis is needed.  Here it
is expected that \bes3 will be able to probe decay branching fractions
at the level of $10^{-7}$ as well.

The most general expressions for scalar and pseudoscalar decays
$M \to \ell_1 \overline \ell_2$ are both
\begin{eqnarray}\label{Spin0Amp}
{\cal A}(M\to \ell_1 \overline \ell_2) = \overline{u}(p_1, s_1) \left[
E_M^{\ell_1\ell_2}  + i F_M^{\ell_1\ell_2} \gamma_5
\right] v(p_2,s_2) \, ,
\end{eqnarray}
with $E_M^{\ell_1\ell_2}$ and $F_M^{\ell_1\ell_2}$ being dimensionless
constants for scalar $M=\chi_c$ or pseudoscalar $M=\eta_c$ decay
amplitudes, which depend on the Wilson coefficients of operators in
Eq.~(\ref{L4Fermion}) and decay constants. The corresponding branching
fraction is
\begin{eqnarray}\label{BRSpin0}
{\cal B}(M \to \ell_1 \overline \ell_2) = \frac{m_M}{8\pi \Gamma_M} \left(1-y^2\right)^2
\left[\left|E_M^{\ell_1\ell_2}\right|^2 + \left|F_M^{\ell_1\ell_2}\right|^2\right].
\end{eqnarray}
Here $\Gamma_M$ is the total width of the decaying state and $y = m_2/m_M$.
The generic expressions for the coefficients $E_M^{\ell_1\ell_2}$ and $F_M^{\ell_1\ell_2}$
for the pseudoscalar $M=P$ and scalar $M=S$ states are given in \cite{Hazard:2016fnc}. We can
simplify them by neglecting the contributions from gluonic operators of Eq.~(\ref{LGluon}), as
$\eta_c$ and $\chi_c$ are not expected to contain large gluonic components in their wave functions.
The Wilson coefficients of the gluonic operators are better probed in CLFV tau decays, where the low energy
theorems~\cite{Petrov:2013vka} or experimental data~\cite{Celis:2014asa} constrain gluonic matrix elements
in a model-independent manner.

With this, $P=\eta_c$ CLFV decays will be mainly sensitive to axial operator
contributions in ${\cal L}_{\ell q}$ of Eq.~(\ref{L4Fermion})~\cite{Hazard:2016fnc},
\begin{eqnarray}\label{PCoef}
E_P^{\ell_1\ell_2} &=& y \frac{m_P}{4 \Lambda^2} \left[
- i f_{P} \left[ 2 \left(C_{AL}^{cc\ell_1\ell_2}+C_{AR}^{cc\ell_1\ell_2}\right) - m_{P}^{2} G_{F}
\left( C_{PL}^{cc\ell_1\ell_2} + C_{PR}^{cc\ell_1\ell_2} \right) \right]
 \right],
\nonumber \\
F_P^{\ell_1\ell_2} &=& -y \frac{m_P}{4 \Lambda^2} \left[f_{P}
\left[ 2 \left(C_{AL}^{cc\ell_1\ell_2}-C_{AR}^{cc\ell_1\ell_2}\right) - m_P^2 G_F
 \left(C_{PL}^{cc\ell_1\ell_2}-C_{PR}^{cc\ell_1\ell_2}\right) \right]
 \right],
\end{eqnarray}
while scalar $S=\chi_c$ CLFV decays will uniquely probe scalar CLFV operators of Eq.~(\ref{L4Fermion}),
\begin{eqnarray}\label{SCoef1}
E_S^{\ell_1\ell_2} &=& iy f_{S} m_c \frac{m_S^2 G_F}{2 \Lambda^2}
\left(C_{SL}^{cc l_1 l_2} + C_{SR}^{cc l_1 l_2}\right),
\nonumber \\
F_S^{\ell_1\ell_2} &=& y f_{S} m_c \frac{m_S^2 G_F}{2 \Lambda^2}
\left(C_{SL}^{cc l_1 l_2} - C_{SR}^{cc l_1 l_2}\right),
\end{eqnarray}
where the decay constants are $f_{\eta_c} = (387\pm7)$
MeV~\cite{Becirevic:2013bsa}, and $f_{\chi_c} \approx 887$
MeV~\cite{Godfrey:2015vda}, for the pseudoscalar and scalar states,
respectively.
The resulting constraints on the combination of Wilson coefficients and
NP scale $\Lambda$, both current and projected, can be found in Table~\ref{tab:Constraints4fermion}.

%%%%%%%%%%%%%%%%%%%
\subsection{(radiative) Leptonic decays of $D^0 \to l_1 l_2, \gamma l_1 l_2$}

Studies of lepton flavor violation can also be performed with $D^0$
decays into $\ell_1 \ell_2$ and $\gamma \ell_1 \ell_2$ final
states. These decays involve FCNC transitions in both quark and lepton
currents, so the set of effective operators tested is given in
Eq.~(\ref{L4Fermion}) for $q=u$. We should emphasize that they are
different from the operators discussed in Sects.~\ref{Spin1CLFVsection} and \ref{Spin0CLFVsection}, although particular
NP models could give contributions to both sets of operators, in which
case the Wilson coefficients of Eq.~(\ref{L4Fermion}) for $q=c$ and
$q=u$ would depend on different parameters of the same NP model.

As $D$-mesons are pseudoscalar states, the branching fraction for
flavor off-diagonal leptonic decays of $D$-mesons is given by
Eq.~(\ref{BRSpin0}) with $M \to D$, so $\Gamma_D$ is the total width
of the $D^0$.  Calculating the form factors $E_D^{uc \ell_1\ell_2}$
and $F_D^{uc \ell_1\ell_2}$ for $\bar{D}^0$ ($q_1 q_2=cu$) states
yields~\cite{Hazard:2017udp}
\begin{eqnarray}\label{DMesonCoef}
E_D^{q_1 q_2\ell_1\ell_2} &=&  \frac{m_D f_{D} y}{2 \Lambda^2} \left[ \left(C_{AL}^{uc \ell_1\ell_2}+C_{AR}^{uc \ell_1\ell_2}\right) +
m_{D}^{2} G_{F} \left( C_{PL}^{q_1 q_2 \ell_1\ell_2} + C_{PR}^{q_1 q_2 \ell_1\ell_2} \right) \right],
\nonumber \\
F_D^{q_1 q_2\ell_1\ell_2} &=& i \frac{m_D  f_{D} y}{2 \Lambda^2}\left[\left(C_{AL}^{uc \ell_1\ell_2}-C_{AR}^{q_1 q_2 \ell_1\ell_2}\right) +
m_D^2 G_F \left(C_{PL}^{uc \ell_1\ell_2}-C_{PR}^{uc \ell_1\ell_2}\right) \right],
\nonumber \\
\end{eqnarray}
where $f_D=207.4\pm 3.8$ MeV is the  $D$-meson decay constant.

The resulting constraints on the combination of Wilson coefficients and
NP scale $\Lambda$, both current and projected, can be found in
Table~\ref{tab:Constraints4fermion}.
One should note that other operators can be probed in $D$-decays. To
access those, one can consider three-body decays of a $D$-meson into
$\gamma \ell_1 \ell_2$ ~\cite{Hazard:2017udp} or semileptonic CLFV
decays. While increasing the reach of experiments, it also complicates
theoretical interpretation of bounds, forcing one to introduce
additional hypotheses, such as single-operator dominance, as well as
model dependence. Using the results in Ref.~\cite{Hazard:2017udp}, a
set of constraints on CLFV operators can be obtained.
If \bes3 could take about $20\,{\rm fb}^{-1}$ data at the
$\psi(3770)$, the sensitivity for $D^0 \to \gamma e \mu$ could reach
$ (5-10)\times 10^{-7}$.

%%%%%%%%%%%%%%%%%%%
\subsection{CLFV and LNV $D_{(s)}$ decays with light mesons}
\label{clfv:lnv}

CLFV searches can include decays with light mesons in the final
states, such as $D \to \pi \ell_1 \ell_2$ or
$D_s \to K \ell_1 \ell_2$. A clear advantage in using those
transitions can be seen in that both charged and neutral $D$-mesons
can be used for analyses. Disadvantages include, just as in
$D \to \gamma \ell_1 \ell_2$, difficulties with the theoretical
interpretation of the experimental bounds (unless directly projected
onto particular NP models) and increased model dependence of the
bounds, as long-distance contributions become increasingly more
pronounced.

Finally, lepton-number violating processes such as
$D_{(s)}^+ \to \pi^- (K^-) \ell^+ \ell^+$ can also be probed at
BESIII. Especially, with 20 fb$^{-1}$ of $\psi(3770)$ data, BESIII could
improve the best upper limit to $4.6 \times 10^{-7}$ and
$2.3 \times 10^{-7}$ for $D^+ \to \pi^- e^+ e^+$ and
$D^+ \to K^- e^+ e^+$ respectively, extrapolated from the current
analysis.

%%% Local Variables:
%%% mode: latex
%%% TeX-master: t
%%% End:

%% file: New_physics/bsmlight.tex
%%%%%%%%%%%%%%%%%%%%%%%%%%%%%%%%
\section{Searches for light (invisible) NP particles}
\label{sec:inv}

Several well-motivated proposals of BSM physics include new degrees of
freedom (DOF) that do not interact with the SM particles
directly. Such DOFs constitute the so-called ``Dark Sector". Particles
that populate the Dark Sector could form a part or whole of the DM in our Universe. There are only a few interactions allowed
by SM symmetries that provide a portal from the SM sector into the
Dark Sector~\cite{Essig2013,Alexander2016}.  Depending on the masses,
such DM and portal DOFs could be probed at the low-energy
high-intensity frontier experiments, such as \bes3.  

%%%%%%%%%%%%%%%%%%%%%%%%%%%%%%%%
\subsection{Physics of the Dark Sector}
\label{sec:bsm:dsphy}

The presence of cold DM in our Universe provides the
most natural explanation for several observational puzzles. If DM has
a particle origin, it should be eventually detected in particle physics
experiments. In particular, if DM particles are light, with masses in
the keV-MeV range, as suggested by our understanding of small-scale
gravitational clustering in numerical simulations, they should be detectable
in the decays of heavy meson states at \bes3. These DM
particles could be fermions, scalars, or even vector bosons. In many
models they are produced in pairs, as such models feature $Z_2$
symmetry requirements for writing Lagrangians with such particles.

In other models light new particles can be produced not only in pairs,
but also individually, in which case they could serve the role of
mediators (portals) between the Dark Sector and the SM. A particular
motivation for such a scenario comes from the observations of
anomalous fluxes of cosmic-ray positrons. In 2008, the PAMELA
collaboration reported an excess of positrons above 10 GeV~\cite{ds2},
which has now been confirmed by many other experiments, such as ATIC
\cite{ds3}, Fermi-LAT~\cite{ds4} and AMS02
\cite{TheAMSCollaboration2013}. In a class of models, DM particles
with masses of ${\cal O}$(TeV) annihilate into a pair of light bosons
with masses of $\cal{O}$(GeV), which subsequently decay into charged
leptons~\cite{Arkani-Hamed2009,Pospelov2009}. An exchange of light
bosons would increase the dark-matter annihilation cross section,
allowing the observations of anomalous cosmic-ray positrons to be
explained. Moreover, if the mediator is light enough, no extra
anti-proton will be produced due to the kinematics. This feature is
consistent with the PAMELA antiprotons data. The light boson may be a
massive dark photon in the models with an extra U(1) gauge symmetry.

The dark photon field $V_\mu$ couples to the SM photon $A_\mu$ through kinetic
mixing~\cite{ds8},
\beq\label{KinMix}
{\cal L}_k =  -\frac{\kappa}{2} V_{\mu\nu} F^{\mu\nu},
\eeq
where $V_{\mu\nu}$ is the dark photon's field tensor
$V_{\mu\nu} = \partial_\mu V_\nu - \partial_\nu V_\mu$.
Note that this construction is gauge-invariant. The dark photon can acquire a
mass through the spontaneous symmetry breaking mechanism.
Some models predict that the mass of the dark photon is
at the scale from $\cal{O}$(MeV) to $\cal{O}$(GeV)~\cite{ds8,ds9}.
The kinetic mixing coupling $\kappa$ is taken to be very small.
Similarly to the case of neutral-meson mixing, the introduction of
the term mixing $A_\mu$ and $V_\mu$ implies that neither of the fields is
a mass eigenstate. Diagonalizing the Lagrangian~\cite{ds8,Aditya:2012ay}
introduces a small coupling $g$ between the new, weakly-interacting
vector field $V_\mu^\prime$, $g \approx \kappa e$, where $e$ is
the electric charge. Thus, dark photons can  be searched for in
the decays of charmed particles~\cite{Aditya:2012ay} and/or charmonia.

The structure of the Dark Sector can be complicated, possibly with a class
of light particles including scalars, pseudo-scalars, gauge bosons and fermions
at the GeV scale.  Since the interaction between the Dark Sector and
the SM sector is very weak, the constraints on the mixing parameter
and dark photon mass mainly come from the intensity frontier, such as
the measurements of electron and muon anomalous magnetic moments, low-energy electron-positron colliders, beam-dump experiments and fixed-target experiments~\cite{Essig2013,Alexander2016}.

%%%%%%%%%%%%%%%%%
\subsubsection{Light Dark Matter (LDM) with $Z_2$ symmetry}
\label{subsec:inv_heavy_meson}

Let us consider, as an example, the generic case of a complex neutral
scalar field $\chi_0$ describing LDM. The discussion of other spin
assignments of LDM and their effects in charm decays can be found in
\cite{Badin:2010uh}.  A generic effective Hamiltonian of
lowest dimension describing $c \to u \chi_0^* \chi_0$ interactions is
\beq
\label{ScalLagr} {\cal H}^{(s)}_{eff} =  2 \sum_i
\frac{C_i^{(s)}}{\Lambda^2} O_i,
\eeq
where $\Lambda$ is the scale associated with the particle(s) mediating interactions between the
SM and LDM fields, and $C_i^{(s)} $ are the relevant Wilson coefficients of the operators
\bea\label{ScalOper}
O_1 &=& m_c (\overline{u}_R c_L)(\chi_0^* \chi_0), \nonumber\\
O_2 &=& m_c (\overline{u}_L c_R)(\chi_0^* \chi_0), \\
O_3 &=&(\overline{u}_L  \gamma^{\mu} c_L)(\chi_0^* \stackrel{\leftrightarrow}{\partial}_{\mu}\chi_0), \nonumber \\
O_4 &=&(\overline{u}_R \gamma^{\mu} c_R)(\chi_0^* \stackrel{\leftrightarrow}{\partial}_{\mu}\chi_0), \nonumber
\eea
where
$\stackrel{\leftrightarrow}{\partial} =
(\stackrel{\rightarrow}{\partial}-\stackrel{\leftarrow}{\partial})/2$,
and $\chi_0^*$ is a conjugated state of $\chi_0$.
Operators $O_{3,4}$ disappear for real scalar LDM, in which case
$\chi_0^*=\chi_0$. It is implied in Eq.~(\ref{ScalOper}) that the
mediator of interactions between LDM and the SM fields is heavy,
$M_\Lambda > m_{D}$. The discussion can be modified for the case of a
light mediator (see~\cite{Badin:2010uh}).

The simplest decay that is mediated by Eq.~(\ref{ScalOper}) is
the transition $D^0 \to \chi_0 \chi_0$, which could contribute to
the invisible $D$-decay width,
\beq\label{Dphi}
{\cal B} (D^0 \to \chi_0 \chi_0) =
\frac{\left(C_1^{(s)}-C_2^{(s)}\right)^2}{4\pi M_{D} \Gamma_{D}}
\left(\frac{f_{D}
M_{D}^2 m_c}{\Lambda^2 (m_c +m_u)}\right)^2
\sqrt{1-4 x_\chi^2} ,
 \eeq
 where $x_\chi = {m_\chi}/{M_{D}}$ is a rescaled LDM mass. Clearly,
 this rate is not helicity-suppressed, so it could be a quite
 sensitive tool to determine LDM properties at \bes3. Current
 experimental constraints imply that
\beq\label{InvWilson}
\frac{\left(C_1^{(s)}-C_2^{(s)}\right)}{\Lambda^2} \leq 8 \times 10^{-8},
\eeq
for $m_\chi \ll m_D$, which corresponds to probing scales of NP over 3.5 TeV.
Those constraints could be improved with updated bounds on $D^0 \to$ invisible decays.
The constraints obtained in Eq.~(\ref{InvWilson}) can be used in constraining
parameters of particular models of light DM~\cite{Bird:2006jd,Badin:2010uh}.

Similarly, the decay width of the radiative decay $D^0 \rightarrow\chi_0^*\chi_0\gamma$ can be computed,
\begin{eqnarray} \label{GammaDphiphigamma}
{\cal B} (D^0 \to \chi_0^*\chi_0\gamma)&=& \frac{f_{D}^2 \alpha
C_3^{(s)}C^{(s)}_4 M_{D}^5}{6 \Lambda^4 \Gamma_{D}}
\left(\frac{F_{D}}{4\pi}\right)^2 \\
&\times&  \left(\frac{1}{6}\sqrt{1-4x_\chi^2}(1 - 16 x_\chi^2 - 12x_\chi^4)
-12x_\chi^4\log\frac{2 x_\chi}{1+\sqrt{1-4x_\chi^2}} \right).
\nonumber
\end{eqnarray}
We observe that Eq.~(\ref{GammaDphiphigamma}) does
not depend on $C^{(s)}_{1,2}$. This can be most easily seen from the
fact that the $D \to \gamma$ form factors of scalar and pseudoscalar
currents are zero.

%%%%%%%%%%%%%%%%%
\subsubsection{LDM without $Z_2$ symmetry}

If DM particles are very light, in the kev-MeV range, they do not need
to obey $Z_2$ symmetry, as their decays to pairs of SM states could be
either suppressed kinematically or by small coupling. This implies
that DM particles can be emitted and absorbed by SM particles.  Due to
their extremely small couplings to the SM particles, experimental
searches for such light states require large sample sizes and the
ability to resolve signals with missing energy.  This means that
\bes3 is ideally suited for such searches.  For example, such
particles could be searched for in leptonic $D/D_s \to \ell \bar\nu$
decays, which are helicity-suppressed in the SM. Additional emission
of a dark photon or an axion-like particle (ALP) $a$ could lift the
helicity suppression and change the energy spectrum of the lepton
\cite{Aditya:2012ay}. Similarly, flavor-changing transitions of the
type $D \to a \pi$ resulting in a two-body-like spectrum of pions in
$D \to \pi + invisible$ could become possible~\cite{Calibbi:2016hwq}.

Searches for dark photons can be performed in the decays of the charmonium states.
\bes3 has the largest $J/\psi$ data sample in the world. Associated production of dark photons with
other Dark Sector particles, such as the dark Higgs $h'$, is also possible and
has been studied~\cite{Li2010,Yin2009}, provided that  $m_{h'}$ is such that
production of this state is not kinematically suppressed.

%%%%%%%%%%%%%
\subsection{(radiative) Invisible decays of charmonia}
\label{subsec:inv_charmonium}

Invisible decays of quarkonium states might provide a window into what
may lie beyond the SM.  In the SM the invisible decays of the $J/\psi$
and other charmonium states are given by their decays into neutrino
final states,
\begin{equation}\label{PsiNeutrinos}
\frac{\Gamma(J/\psi \to \nu \bar{\nu})}{\Gamma(J/\psi \to e^+ e^-)} =
\frac{9 N_{\nu} G_F^2}{256 \pi^2 \alpha^2} M_{J/\psi}^4 \left(1 - \frac{8}{3} \sin^2{\theta_W}
\right)^2 \simeq 4.54 \times 10^{-7},
\end{equation}
where $G_F$ is the Fermi coupling, $\theta_W$ is the weak mixing
angle, and $N_\nu = 3$ is the number of light non-sterile
neutrinos. It is interesting to note that this result is about three
orders of magnitude smaller than the corresponding decay of an Upsilon
state~\cite{Chang:1997tq}. Similarly, the SM branching fraction of
radiative decays of $J/\psi$'s with missing energy is also very
tiny~\cite{Yeghiyan:2009xc}. Thus, we may neglect neutrino-background
effects when confronting theoretical predictions for $J/\psi$ decays
into invisible states with experimental data.  This implies that
invisible decays of charmonium states provide a great opportunity to
search for the glimpses of BSM physics.  In BSM scenarios the decay
rate might be enhanced either by new heavy particles modifying the
interactions of neutrinos and heavy quarks, or by opening new decay
channels into light DM states.

It is therefore possible to use invisible decays of charmonium states to put
constraints on various models of light DM, provided that DM states couple to
charmed quarks~\cite{McElrath:2005bp,Fayet:2009tv,Yeghiyan:2009xc,Fernandez:2014eja}.
Predictions for radiative decays with missing energy in light DM models
are available in Ref.~\cite{Fernandez:2015klv}.

Due to its superb kinematic-reconstruction capabilities, \bes3 can
search for the invisible and radiative invisible decays of $J/\psi$
and via decay chains
$\psi(3686) \to \pi^+\pi^- J/\psi (\to \mbox{invisible})$ using
$\pi^+\pi^-$ as a trigger, similar to the the invisible decay search
of the $\Upsilon(1S)$ at BaBar \cite{Aubert:2009ae}. With a data set
of $3\times 10^9$ $\psi(3686)$ events, the branching fraction of $J/\psi$
invisible decay could be probed to $3 \times 10^{-5}$.

%%%%%%%%%%%%%%%%%
\subsection{Invisible decays of $D$ mesons}
\label{subsec:inv_dmeson}

Decays of $D$ mesons into invisible final states can also provide
an excellent probe of light Dark Matter models. Similarly to
Sect.~\ref{subsec:inv_charmonium}, in the SM the invisible decays of
$D$ mesons are constituted by their decays into neutrino final states,
which can be computed. A major difference between the heavy quarkonium
and $D$ decays into neutrinos includes long-distance effects, which are
currently poorly known. However, they are not expected to dominate the
short-distance (SD) estimates by many orders of magnitude. The SD
contributions to the branching fractions for $D^0 \to \nu\overline{\nu}$
decays can be readily computed. One can immediately notice that the
left-handed structure of the Hamiltonian should result in helicity
suppression of those transitions. Assuming for neutrino masses that
$m_\nu \sim \sum_i m_{\nu_i} < 0.62$~eV~\cite{Goobar:2006xz}, where
$m_{\nu_i}$ is the mass of one of the neutrinos,
 \begin{equation}
{\cal B}(D^0\rightarrow \nu\overline{\nu})= \frac{G_F^2\alpha^2
f_D^2 m_D^3} {16 \pi^3 \sin^4 \theta_W\Gamma_{D^0}}
|V_{bc}V_{ub}^*|^2X(x_b)^2 x_\nu^2
\ \simeq  \ 1.1 \times10^{-30} \, .
 \end{equation}
 Here $x_\nu = m_\nu/m_{D}$ and $\Gamma_{D}=1/\tau_D$ is the total
 width of the $D^0$ meson. Such tiny rates imply that decays of heavy
 mesons into neutrino-antineutrino final states in the SM can be
 safely neglected as sources of background in the searches for DM in
 $D$ decays. Such helicity suppression in the final state can be
 overcome by adding a third particle, such as a photon, to the final
 state. In fact, the SM contribution to invisible width of a $D$-meson
 is dominated by $D$ decays into a four-neutrino final state with a
 branching fraction of
 ${\cal B}(D^0 \to 4\nu) =
 (2.96\pm0.39)\times10^{-27}$~\cite{Bhattacharya:2018}.  The resulting
 branching fractions for $D^0 \to \gamma\nu\bar\nu$ are also larger
 than $D^0\rightarrow \nu\overline{\nu}$ by orders of
 magnitude~\cite{Badin:2010uh, Aliev:1996sk}.

 The Belle collaboration has published a limit on the branching
 fraction of $D$ decays to invisible final states, setting it at
 $9.4 \times10^{-5}$ at the 90\% C.L.~\cite{Lai:2016uvj}. BESIII could
 improve this limit to the order of $10^{-6} \to 10^{-5}$ with the
 final charm data set.  Many NP models predict much larger branching
 fractions of $D$-mesons into the light DM states than that of
 $D \to \nu\bar\nu$.  This implies that the measurement of the
 invisible $D$-width provides a practically SM-background-free search
 for such states~\cite{Badin:2010uh}.  Searches for $D$ decays into a
 meson state and missing energy can also probe light DM
 states~\cite{McKeen:2009rm,Bird:2006jd}.

%%%%%%%%%%%%%%%%%
\subsection{Invisible decays of light mesons}
\label{subsec:inv_lmeson}

Decays of light mesons produced at \bes3 can also probe new light
particles~\cite{Dreiner:2009er}. In the case of dark photons, the
general rule is that if light mesons can decay into regular photons,
they could also decay into dark photons \cite{Reece2009,Li2010}. Since
low-energy electron-positron colliders produce numerous mesons, it is
also possible to investigate dark photons in the rare decays of
mesons. For instance, one can search for a resonance in the processes
$\phi \to \eta + V'$ and
$\pi/\eta \to \gamma + V' \to \gamma l^+ l^-$.

The invisible decay of $\eta$ and $\eta^\prime$ mesons has been
studied~\cite{Ablikim2013} with $2.25 \times 10^8$ $J/\psi$ events at \bes3, which
improved previous limits from BESII. Invisible decays of the vector
states, such as $\omega$ or $\phi$ are also
probed~\cite{Ablikim:2018liz}. The sensitivities of these results are
dominated by the available statistics. With $10^{10} J/\psi$ events
already recorded, the statistical uncertainty will be correspondingly
reduced. These expected upper limits could be further improved to
$2.8 \times 10^{-5}$ and $4.5 \times 10^{-5}$ for $\omega$ and $\phi$
invisible decays, respectively.

%%% Local Variables:
%%% mode: latex
%%% TeX-master: t
%%% End:

%% file: New_physics/offres.tex
%%%%%%%%%%%%%%%%%
\section{Off-resonance searches}
\label{sec:offres}

The clean collision environment and excellent detector performance
of \bes3 offer opportunities for NP searches in kinematical regions
where the energies of the electron and positron are {\it not} tuned to the
$s$-channel resonance production of the $\psi(nS)$ states.

%%%%%%%%%%%%%%%%%%%%%%%%%%%%%%%%
\subsection{Rare charm production: $e^+ e^- \to D^*(2007)$ }

As was mentioned in Sect. \ref{NPinRareDecays}, the rate of the simplest
FCNC decay, $D^0 \to \ell^+ \ell^-$, is helicity suppressed. This is
manifested by a decay amplitude that is proportional to the
mass of the final state lepton. This fact makes observation of the
branching fraction of $D^0 \to e^+ e^-$ a near-impossible task and
complicates the study of lepton flavor universality in charm transitions.

%%%%%%%%%%%%%%%%%%%%%%%%%%%%%%%%%%%
\begin{figure}[htbp]
\begin{center}
\includegraphics[width=6cm]{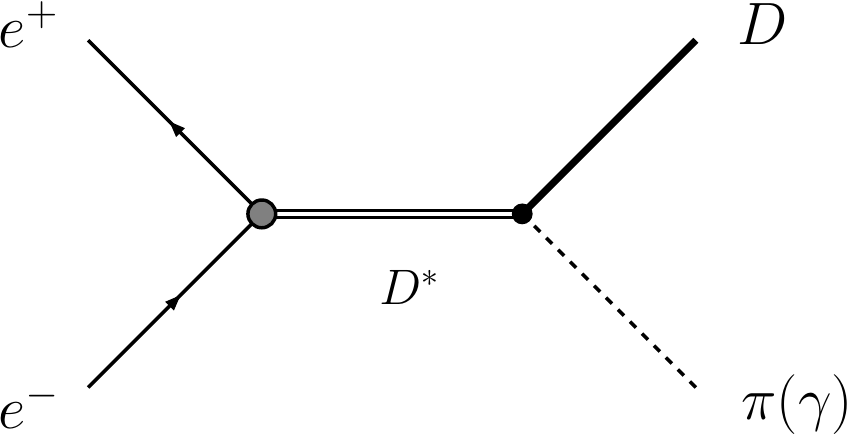}
\end{center}
\caption{\label{figure_Dstar} \small Probing the $ c\bar{u}\to e^+ e^-$ vertex
with $D^*(2007)^0$ resonant production in $e^+e^-$ collisions.}
\end{figure}
%%%%%%%%%%%%%%%%%%%%%%%%%%%%%%%%%%%

An interesting alternative to studies of $D$ decays is to measure
the corresponding {\it production} process $e^+e^- \to D^*$, as shown
in Fig.~\ref{figure_Dstar}~\cite{Khodjamirian:2015dda}. This is
possible if BEPCII takes data at a collision
energy corresponding to the mass of the $D^*$ meson,
$\sqrt{s} \approx 2007$ MeV. The production process, $e^+ e^- \to D^{*0}$,
which is inverse to the $D^* \to e^+ e^-$ decay, is rather rare, as it is
driven by FCNC. Yet, the produced $D^{*0}$ resonance, tagged
by a single charmed particle in the final state, will decay strongly
($D^{*0}\to D^0\pi^0$) or electro-magnetically ($D^{*0}\to D^0\gamma$)
with huge branching fractions of $(64.7\pm 0.9)\%$ and $(35.3\pm 0.9)\%$
respectively. This production mechanism, albeit very rare, has clear advantages for
NP studies compared to the $D^0 \to e^+e^-$ decay: the helicity
suppression is absent, and a richer set of effective operators can be
probed. It is also interesting to note that, contrary to other rare
decays of charmed mesons, long-distance SM contributions are under
theoretical control and contribute at the same order of magnitude as
the short-distance ones~\cite{Khodjamirian:2015dda}. Some
preliminary studies of this process have been done by CMD-3
at VEPP-2000 collider in Novosibirsk~\cite{Solodov:2017pyu}.

It is interesting to estimate the sensitivity of \bes3 to detect the process $e^+e^-\to D^*\to D\pi$.
It can be shown that~\cite{Khodjamirian:2015dda}, crudely, 
\begin{equation} \label{Off_res:constr}
{\cal B}_{D^* \to e^+e^-}\geq \left(\frac{1}{\epsilon \int Ldt } \right)
\times \frac{m_{D^*}^2}{12 \pi \ {\cal B}_{D^*\to D\pi}},
\end{equation}
assuming the beam energy resolution is smaller than the spread of the resonance cross section.
An average luminosity at the level of $L \approx 1.0\times 10^{32} ~{\rm cm}^{-2}s^{-1}$,
with a year ($\sim10^7\,s$) of running at the $D^*$ resonance yields $\int L dt = 1.0$ fb$^{-1}$.
Thus, the single-event sensitivity estimated from Eq.~(\ref{Off_res:constr}) implies that
\beq\label{eq:boundBR}
{\cal B }_{D^* \to e^+e^-}> 4\times 10^{-13}
\eeq
could be probed in an ideal way. 
However, in practice beam energy resolution around $\sqrt{s}=2$ GeV at BEPCII would be more than several hundreds keV, which is at least one order of magnitude wider than the $D^*$ resonance width. Hence, the realistic upper limit obtained with one-year data taking can be a few orders of magnitude worse than the number given in Eq.~(\ref{eq:boundBR}).

To estimate the NP scale sensitivity implied by Eq.~(\ref{eq:boundBR}), one can assume
single operator dominance with a Wilson coefficient $C$ to obtain
~\cite{Khodjamirian:2015dda}
\begin{equation}
\Lambda \sim \left(
\frac{1}{3\pi} \frac{m_{D^*}^3 f_{D^*}^2}{32 \Gamma_0} \frac{C^2}{{\cal B}(D^* \to e^+e^-)}
\right)^{1/4}.
\end{equation}
With the upper bound of Eq.~(\ref{eq:boundBR}), one notes that
observation of a single event in a year of running would probe NP
scales of the order of $\Lambda \sim 2.7$ TeV provided that
$C \sim 1$. Taking into account the current experimental bound,
${\cal B}_{D\to e^+e^-} = 7.9\times 10^{-8}$, one finds that only
scales $\Lambda \sim 200$ GeV are currently probed by $D\to e^+e^-$
decay.  It is the presence of the lepton mass factor that severely
limits the NP scale sensitivity in $D^0 \to e^+e^-$.

It should be noted that single-charm final states required for this
analysis can also be produced in non-leptonic weak decays of heavy
quarkonium states, such as $J/\psi \to D\pi$, discussed earlier in
Sect.~\ref{sec:charmoniumweak}, albeit in different kinematical
regions. Since heavy charmonium states lie far away from the energy
region required for this analysis, these transitions will {\it not}
produce any backgrounds for $e^+e^- \to D^* \to D\pi$, but can be used
to study experimental systematics associated with such final states
(see Sect.~\ref{sec:charmoniumweak}).

%%%%%%%%%%%%%%%%%%%%%%%%%%%
\subsection{Dark photon and dark Higgs searches}
\label{sec:bsm:dpdh}

Electron-positron colliders are suitable for probing dark photons
through either the direct production or rare decays of mesons (see
Sect. \ref{sec:inv}).  Dark photons, directly produced in $e^+ e^-$
annihilation could subsequently decay into charged
leptons,~\cite{Zhu2007,Reece2009,Fayet2007,Essig2009,Yin2009}, which
could be detected at \bes3. In comparison with the irreducible QED
background, dark-photon production is highly suppressed. To reduce the
background, precise reconstruction of the dark-photon mass and high
luminosity are important. Such analyses have been performed by
interpreting results from the BaBar
experiment~\cite{Reece2009,Bjorken2009a,ds15}.  The mixing parameter
$\kappa$ (see Sect. \ref{sec:bsm:dsphy} for definition) is constrained
for the dark photon with a mass of about 1 GeV. The limits can be
further improved at Belle II~\cite{ds17}. The potential reach of \bes3
has been discussed in Ref.~\cite{Li2010}, where 20 ${\rm fb}^{-1}$ of
data collected at $\psi(3770)$ is assumed.

At \bes3, the most promising channel to search for dark photons is
through the radiative decay
$e^{+}e^{-}\rightarrow \gamma V^{\prime\star} \rightarrow \gamma
e^{+}e^{-}$. The published \bes3 result\cite{Ablikim:2017aab} based on
\mbox{2.9\,fb$^{-1}$}  $\psi(3770)$ data is competitive with the upper limit
from BaBar \cite{BABAR2014} based on nine years running.

\bes3 searches would be also competitive if the dark photon decays
invisibly (or is detector-stable, {\it i.e.,} it mostly decays outside
of the detector volume).  An estimate of the possible constraints from
the existing data set shows \bes3 could reach or exceed the existing
BaBar limit. In the future, both the $\psi(3770)$ and $XYZ$ data sets
will be increased, so the exclusion limit will be further pushed down
by \bes3. However, it will be difficult to reach the relic density
limit, since it lies a further 2-4 orders of magnitude beyond.
Other data sets will provide an opportunity for studying other channels
that are feasible for dark-photon searches, such as
$J/\psi \to e^+ e^-+ V'$~\cite{Zhu2007} or $\psi(3686) \to \chi_c + V'$.

Another search strategy is to look for dark photons indirectly.
This applies to BSM models with light dark-Higgs particles $h'$. We
can search for a dark photon associated with a dark Higgs $h'$ through
the Higgs-strahlung $e^{+}e^{-}\rightarrow V^\prime h'$. Masses of
both the dark Higgs and the dark photon are unknown. The dark Higgs
$h'$ decay modes strongly depend on the relation between these two
masses.
\begin{enumerate}
\item{$m_{h'}> 2 \times m_{V'}$:} $h' \rightarrow V^\prime V^\prime $;
\item{$m_{V'} < m_{h'}< 2 \times m_{V'}$:} $h'\rightarrow  V^\prime
  V^{\prime\ast}$, where $V^{\prime\ast}$ decays into leptons;
\item{$m_{h'}<m_{V'}$:} $h'$ decays to lepton pairs or hadrons, or
  $h' \rightarrow \mbox{invisible}$.
\end{enumerate}

There may also be other light bosons in the dark sector, for instance
gauge bosons associated with extra non-Abelian
symmetries~\cite{Baumgart2009}.  The final states of the direct
production could contain more lepton pairs. In this case, it is easier
to extract the signals from large QED backgrounds via the
reconstruction of resonances. \bes3 has published two results from a
light Higgs search, exploiting $\psi(3686)$~\cite{Ablikim2012} and
$J/\psi$~\cite{TheBESIIICollaboration2016} decays respectively. These
results can also be interpreted to constrain the dark photon.  They
are still limited by statistics, so the sensitivity will be improved with more data.

%%%%%%%%%%%%%%%%%%%%%%%%
\subsection{Axion-Like particles}
\label{sec:bsm:other_alp}

Axion-like particles (ALP), commonly denoted as $a$, are present in
many possible extensions of the SM. ALPs arise as the Goldstone bosons
of theories with additional Peccei-Quinn symmetry (PQ)~\cite{Peccei,
  Weinberg}, which is a spontaneously broken global symmetry that is
anomalous with respect to the SM gauge interactions. The $a$ mainly
couples to a photon-pair via
$\mathcal{L} = \frac{g_{a\gamma \gamma}}{8}
\epsilon_{\mu\nu\alpha \beta} F^{\mu \nu}F^{\alpha \beta}a$. The
search for an Axion-like particle at BESIII can be performed via
$D \to K^{*}a$~\cite{Izaguirre2017}, ALP-strahlung ($e^+e^- \to \gamma a$) and
the photon fusion ($e^+e^- \to e^+e^-a$) processes~\cite{Dolan}. The
searches for Axion-like particles are neither yet explored by the
BESIII experiment nor any other $e^+e^-$ collider experiments. The
\mbox{Belle II} has a planned program on the searches for this kind of new
particle~\cite{Pietro}.

At BESIII, the search for an ALP can also be explored
via radiative decays of quarkonium states, such as
$J/\psi \to \gamma a$. The branching fraction of $J/\psi \to \gamma a$
can be computed as,
$\mathcal{B}(J/\psi \to \gamma a)=\frac{1}{8\pi \alpha} g_{a \gamma
  \gamma}^2 m_c^2 \mathcal{B}(J/\psi \to e^+e^-)$~\cite{Masso}.  The
large data-set collected by the BESIII at $J/\psi$ and $\psi(3686)$
provide a unique opportunity to explore the possibility of the
ALPs. One of the possible difficulties could arise how
to select two best photons for $a \to \gamma \gamma$
reconstruction. One of the easiest ways would be to combine all the
three possible combinations of di-photon invariant mass spectrum, and
then search for a narrow resonance  `$a$' signal.  While
assuming
$\mathcal{B}(J/\psi \to \gamma A^0) \times \mathcal{B}(A^0 \to \mu^+
\mu^-) = \mathcal{B}(J/\psi \to \gamma a) \times \mathcal{B}(a \to
\gamma \gamma)$, the $90\%$ C.L. upper limit on $g_{a \gamma \gamma}$
is expected to be within the range of $(8.7 - 115.9) \times 10^{-5}$
using the $J/\psi$ sample of $10^{10}$ events in the absence of any
significant signal.

%%%%%%%%%%%%%%%%%%%%%%%%%%
\subsection{Searches for fractionally charged particles}

The electric charges of all known elementary particles are either
zero, $\pm e$, $\pm \frac{1}{3} e$, or $\pm \frac{2}{3} e$, where $e$
is the magnitude of the electron's charge.  While it is known that
such charge assignments are required for the cancellation of
electroweak anomalies~\cite{Golowich:1990ki}, we can not explain why
this particular pattern of charges was chosen by Nature.  Furthermore,
quarks always combine into composite particles with charges $ne$,
where $n = 0, \pm1, \pm2$..., so all observable particles have zero
or integer charges in terms of $e$.  This seems quite natural but we
do not understand the physical law behind this simplicity.

Some BSM models predict the existence of fractionally charged
particles, such as leptoquarks~\cite{Hewett:1997ce}, modified QCD
\cite{DeRujula:1977wse,Slansky:1981tg,Caldi:1982dj}, etc. In the past
80 years, physicists have searched for fractionally charged particles
by employing various methods and technologies, such as particle
accelerators in both fixed target and collider modes, space science,
analyses of bulk matter and so on, without any confirmed
observation~\cite{Perl:2009zz}.

\begin{figure}[tbp]
\begin{center}
	\includegraphics[width=0.36\textwidth]{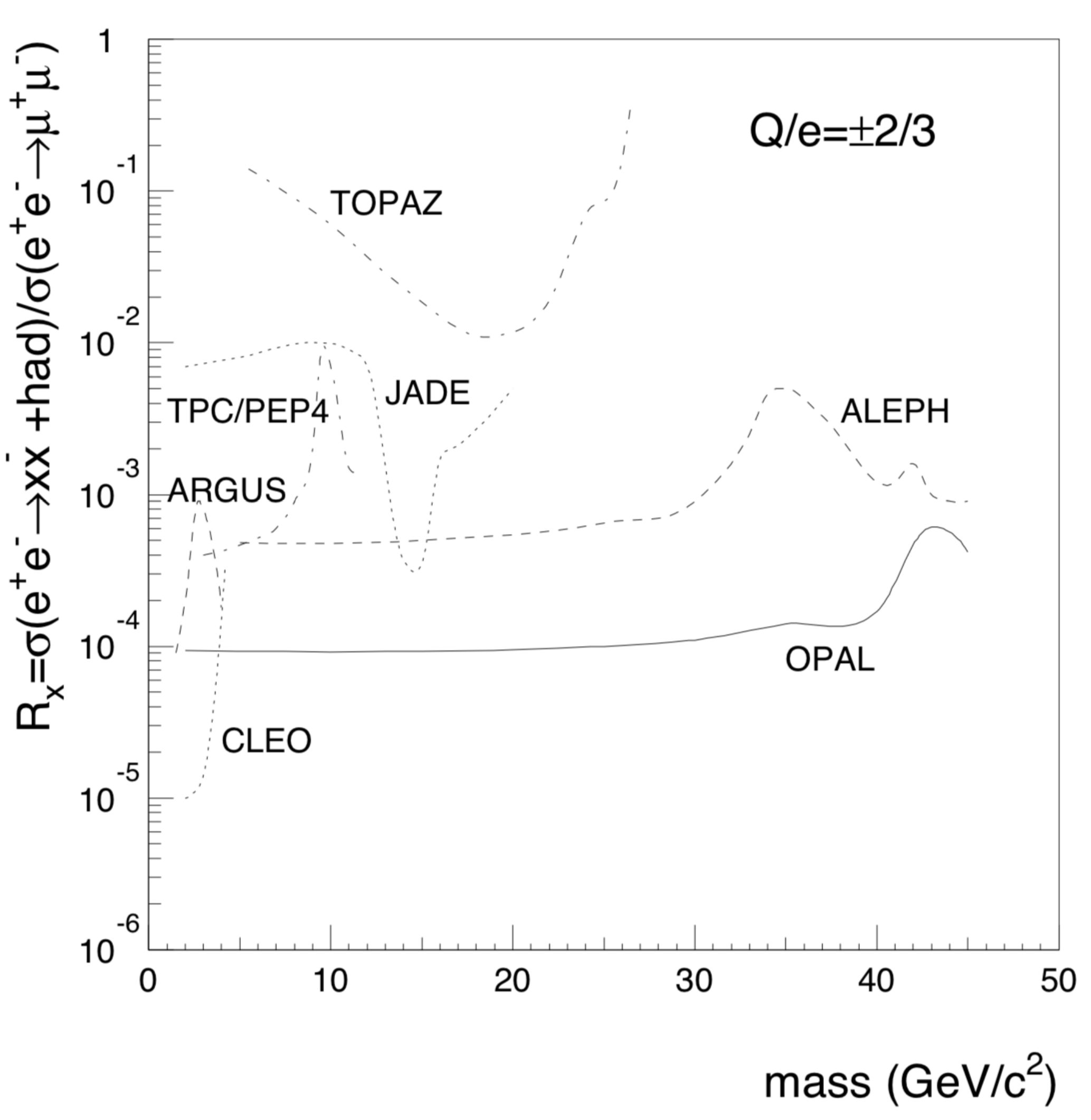}
	\includegraphics[width=0.55\textwidth]{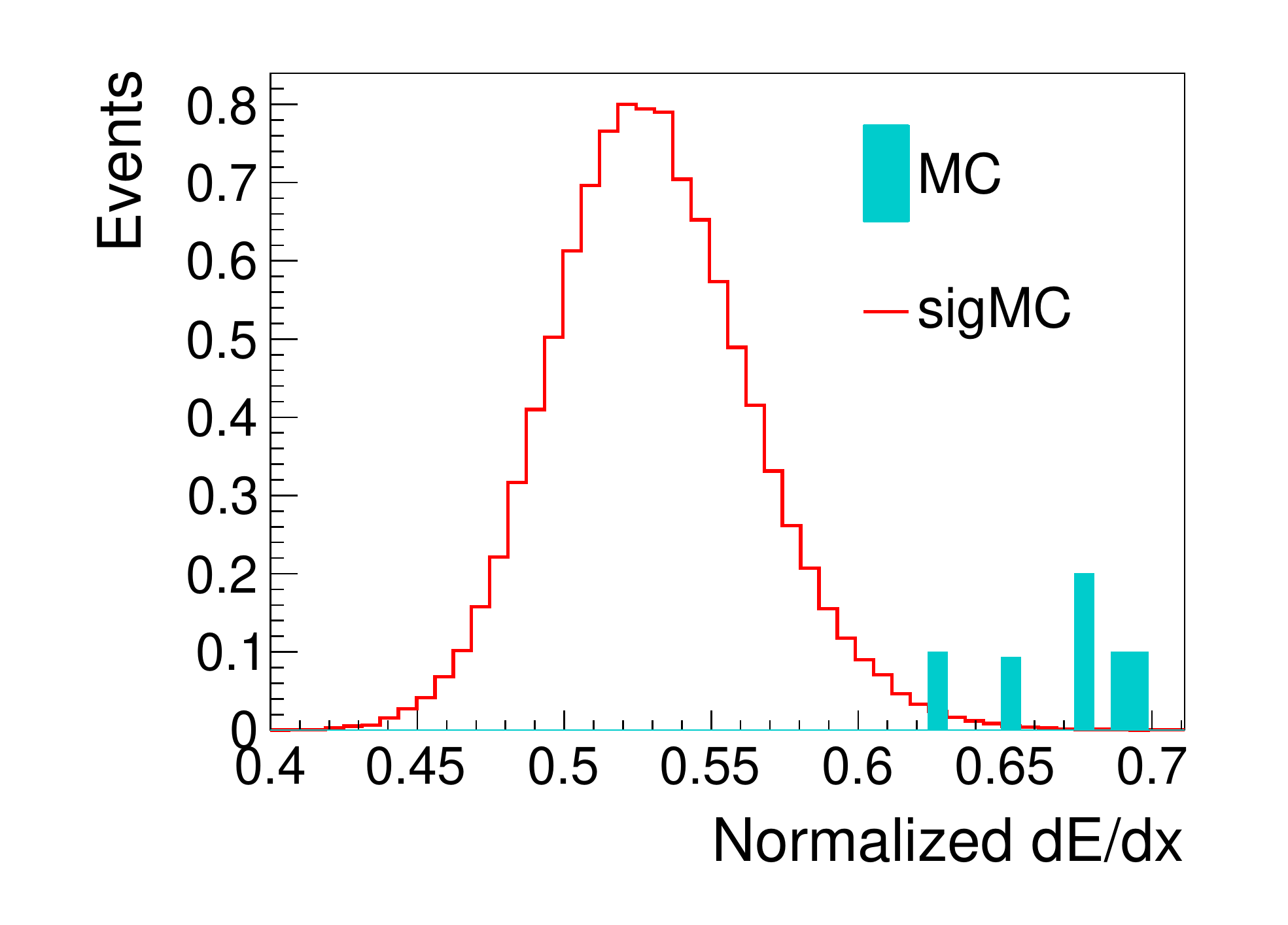}
	%\vskip 15pt
	\caption[]{ \small (Left) Existing limits of
          fractionally-charged particles in the low-mass region.
          (Right) A typical $dE/dx$ distribution in the signal region
          for $m(F) = 1.3~\rm{GeV}/c^2$ and $Q(F) = \frac{2}{3} e$, where the red
          histogram indicates signal and the cyan histogram shows the
          background estimation.  }
	\label{fig:SigdEdx}
\end{center}
\end{figure}

In collider experiments, the ionization energy loss $dE/dx$ can be
used as a signature to search for a fractionally charged particle $F$,
with low background.  The energy loss scales roughly as $(Q/e)^2$
where $Q$ is the charge of $F$.  Such searches have been performed at
by experiments at LEP, Tevatron, as well as earlier facilities, with
the assumed mass of $F$ larger than 45 GeV$/c^2$.  The existing search
results for low-mass fractionally charged particles are shown in
Fig.~\ref{fig:SigdEdx}(a), presented as the ratio of
$\sigma(e^+e^- \rightarrow F \bar{F} +had)/\sigma(e^+e^- \rightarrow
\mu^+\mu^-)$. The most sensitive low-mass search was performed by
CLEO~\cite{Bowcock:1989qj} and OPAL~\cite{Akers:1995az}. With
significant integrated luminosity in the tau-charm energy region,
\bes3 can search for fractionally charged particles in the low-mass
regime to improve the CLEO bounds.

\bes3's drift chamber can measure $dE/dx$ precisely, and the process
$e^+e^- \rightarrow F \bar{F}$ can be searched for by counting the
number of events in the corresponding $dE/dx$
distribution. Figure~\ref{fig:SigdEdx}(b) shows a typical signal
$dE/dx$ distribution from simulation of a given
$m(F) = 1.3 ~\rm{GeV}/c^2$ and $Q(F) = \frac{2}{3} e$, where only 0.57
background events are expected in the 2.9 fb$^{-1}$ $\psi(3770)$ data.
Using the Rolke method~\cite{Rolke2005}, the upper limit on
$\sigma(e^+e^- \rightarrow F^{+2/3}F^{-2/3})$ at the $90\%$ C.L. can be
extracted. Table~\ref{tab:Example} shows the expected \bes3
sensitivity for the cross section with varying mass assumptions. The
expected yields of signal and background are from MC simulation, and
only the statistical uncertainties are taken into account in the upper
limits. It is clear that \bes3 can provide improved constraints on the
mass of $F$ in the low mass region.

\begin{table}[tb]
  \begin{center}
    \caption{\small Projections of \bes3 limit on
      $\sigma(e^+e^- \to F^{+\frac{2}{3}}F^{-\frac{2}{3}})$ at $\sqrt{s}=$ 3.773
      $\rm{GeV}$ (with the statistical uncertainties only) at the 90\% C.L. }
    \label{tab:Example}
    \begin{tabular}{c  c  c  c  c  c  c}
      \hline \hline
      $m(F)$(GeV$/c^{2})$ & $N_{\rm bkg}$ & $N_{\rm obs}$(MC) &
                                                        $\epsilon$(signal) & $\sigma$ (fb) for 2.9 fb$^{-1}$&   $\sigma$ (fb) for 20 fb$^{-1}$ \\
      \hline
      0.3 &0    & 0    & 0.73     & 0.94 & 0.14\\
      1.3 &0.57 & 0.57 & 0.76     & 1.39 & 0.28\\
      1.7 &5.38 & 5.38 & 0.77     & 1.90 & 0.78\\

      \hline\hline
    \end{tabular}
  \end{center}
\end{table}

To facilitate the comparison with existing results, the expected \bes3
results can also be presented as the ratio
$R_F = \sigma(e^+e^- \rightarrow F \bar{F})/\sigma(e^+e^- \rightarrow
\mu^+\mu^-)$.  Using the information presented in
Table~\ref{tab:Example}, the \bes3 projection corresponds to
$R_F \approx 10^{-7}$. This result would be several orders of
magnitude lower than the existing data presented in
Fig.~\ref{fig:SigdEdx}(a).  Using all data samples with large
integrated luminosity ($\ge$500\,pb$^{-1}$) for each energy point at
\bes3, and by varying the charges of $F$, a set of better constraints
on the parameters of fractionally charged particles are expected.

%%% Local Variables:
%%% mode: latex
%%% TeX-master: t
%%% End:

%% file: Sum/sum.tex
\chapter[Summary]{Summary}
\label{chapter:sum}

\input{Sum/sum_main.tex}

\input{Sum/bib.tex}

%% file: Sum/sum_main.tex
The \bes3 experiment has been running very successfully since 2009, and has now collected around $30\,{\rm fb}^{-1}$ of
integrated luminosity at a variety of critical energy points. The physics output, to date, comprises more than 270 
papers in highly-ranked peer-reviewed journals, including one publication with more than 500 
citations~\cite{Ablikim:2013mio} and three others with more than 250 citations~\cite{Ablikim:2013wzq,Ablikim:2013emm,zc3885}. 
\bes3 is now recognized as one of the world's leading experiments in hadronic physics, and has made many significant 
measurements that are important entries in the Review of Particle Physics~\cite{PDG}.

Noteworthy areas of study have included the spectroscopy of hadronic states containing the charm quark, the spectroscopy of light-hadron states produced  in the decay of charmonia, and open-charm physics. Other significant results have been achieved in precision tests of the SM, checks of various predictions of QCD, and probes for new physics beyond the SM.

This outstanding physics output has been enabled through the interplay between a well-running accelerator and the superb capabilities of the \bes3 detector, which have allowed for new programs of operation and studies never previously attempted, for example high precision energy scans. Open questions have been answered, but new ones have arisen, which require higher precision measurements and even larger data sets. This requirement, therefore, provides compelling motivation for a future extended running program, accompanied by upgrades to both the machine and detector to deliver optimal performance. 

\bes3 has a crucial and unique role to play in the world-wide effort to explore and characterize the behavior of QCD in the non-perturbative regime, which is one of the least understood areas of the SM. \bes3 is the only hadron-physics experiment in the world that exploits electron-positron annihilation in the $\tau$-charm energy region. 

Globally new experiments are starting or planned. Belle II in Japan runs at a higher energy; the JLab experiments GlueX and CLAS12 in USA use photon and electron beams, respectively; LHCb at CERN studies proton-proton collisions, and PANDA in Germany will be a fixed-target experiment situated on an antiproton beam.  Given the complexity of the underlying physics, all these complementary approaches are necessary to tackle the unsolved problems in our understanding of the strong interaction. The active and planed experiments highlight the importance of the physics that BESIII studies and excels at.

With more data BESIII can address many critical questions regarding charm physics. 
The study of leptonic and semileptonic charmed-hadron decays can greatly improve our knowledge of decay constants, CKM matrix elements, form factors, and lepton flavor universality. LQCD, which plays a central role in these analyses, can be validated through stringent tests. Quantum-correlated $D^0\bar{D}{}^0$ production, accessible to \bes3 alone, permits strong-phase measurements which are necessary and invaluable input to $CP$-violation studies at other facilities. Recent \bes3 progress in mapping out decay modes of the $\Lambda_c$ will be expanded to include those channels currently missing, and new data involving $e^+e^-$ production of charmed baryon pairs will provide information on form factors in the time-like domain.

In Table~\ref{tab:data-set-sum} we present the current data samples available at \bes3 and those future samples needed to execute the physics program presented in this White Paper. To meet all requirements would mean another 12 years of operation assuming current BEPCII luminosity performance, and the collection by \bes3 of another $56\,{\rm fb}^{-1}$ of integrated luminosity. 

\begin{table}[tp]
\centering
\caption{\label{tab:data-set-sum}  List of data samples collected by
BESIII/BEPCII up to 2019, and the proposed samples for the remainder of the physics program. The most right column shows the number of required data taking days in current ($T_{\rm C}$) or upgraded ($T_{\rm U}$) machine. The machine upgrades include top-up implementation and beam current increase. }
\footnotesize
\begin{tabular}{c|c|c|c|c}
  \hline\hline
  Energy  & Physics motivations & Current data & Expected final data  & $T_{\rm C}$ / $T_{\rm U}$\\ 
   \hline \hline 
 1.8 - 2.0 GeV & $R$ values &  N/A   & 0.1 fb$^{-1}$ & 60/50 days \\ 
& Nucleon  cross-sections  &     & (fine scan)  &    \\   \hline
2.0 - 3.1 GeV  & $R$ values  & Fine scan  & Complete scan  & 250/180 days \\   
 &  Cross-sections  &   (20 energy points)   &   (additional points)  \\   \hline
 $J/\psi$ peak &  Light hadron \& Glueball  & 3.2 fb$^{-1}$   & 3.2 fb$^{-1}$  &  N/A \\  
&  $J/\psi$ decays  &  (10 billion)   &  (10 billion)  &  \\   \hline 
$\psi(3686)$ peak & Light hadron \& Glueball  &  0.67 fb$^{-1}$ & 4.5 fb$^{-1}$ & 150/90 days \\ 
&  Charmonium decays  &   (0.45 billion)    &   (3.0 billion)    \\   \hline
$\psi(3770)$ peak & $D^0/D^{\pm}$ decays  &  2.9 fb$^{-1}$  & 20.0 fb$^{-1}$  & 610/360 days  \\   \hline

  3.8 - 4.6 GeV  & $R$ values & Fine scan  & No requirement  & N/A \\  
 & $XYZ$/Open charm &   (105 energy points)  &     \\   \hline

 4.180 GeV  & $D_s$ decay  &  3.2 fb$^{-1}$  & 6 fb$^{-1}$ & 140/50 days   \\ 
& $XYZ$/Open charm &     &     \\   \hline
  &  $XYZ$/Open charm  &     &  \\
  4.0 - 4.6 GeV &  Higher charmonia  &  16.0 fb$^{-1}$     &  30 fb$^{-1}$  & 770/310 days\\
 &  cross-sections  &  at different  $\sqrt{s}$   &  at different  $\sqrt{s}$    \\   \hline
 4.6 - 4.9 GeV  & Charmed baryon/$XYZ$ & 0.56 fb$^{-1}$   &  15 fb$^{-1}$  & 1490/600 days \\ 
  &  cross-sections  &   at 4.6 GeV  &   at different  $\sqrt{s}$  \\   \hline
 4.74 GeV & $\Sigma^+_c \bar{\Lambda}^-_c$ cross-section &  N/A   &  1.0 fb$^{-1}$ & 100/40 days \\  \hline
4.91 GeV  & $\Sigma_c \bar{\Sigma}_c$ cross-section &  N/A   &  1.0 fb$^{-1}$ & 120/50 days \\  \hline
4.95 GeV  & $\Xi_c$ decays &  N/A   &  1.0 fb$^{-1}$ & 130/50 days\\ 
  \hline  \hline
\end{tabular}
\label{tab:finaldata}
\end{table}

The BESIII physics goals laid out herein will benefit greatly from improved BEPCII $e^+e^-$ collision luminosity delivery. The accelerator group is strongly encouraged to implement top-up injection and to operate BEPCII close to the peak instantaneous luminosity in order to achieve overall integrated luminosity improvement, and develop effective upgrade plan to further enhance the luminosity performance and energy range. With these efforts BESIII can achieve its science goals in less time than 12 years as estimated in this White Paper. It would be reasonable to conclude that BESIII can continue for 10 years with a high quality physics program. It is important that the collaboration allocates resources to evaluate the status of the detector system, to develop upgrade plans, to optimize the data taking strategy and to maximize the physics output of the experiment.

%% file: define/Acknowledgements.tex
\section*{Acknowledgements}
\addcontentsline{toc}{section}{Acknowledgements}

The authors thank the international review committee: Alexander E. Bondar
(Chair), Cesare Bini, Kuang-Ta Chao, Shaomin Chen, Marco Gersabeck, Marek Karliner,
Giovanni Passaleva and Yoshihide Sakai, for their great efforts in reviewing
the program and providing valuable comments and suggestions to improve this
White Paper. The BESIII collaboration thanks the staff of BEPCII and the IHEP computing center for their strong support. This work is supported in part by National Key Basic Research Program of China under Contract No. 2015CB856700; National Natural Science Foundation of China (NSFC) under Contracts Nos. 11335008, 11425524, 11625523, 11635010, 11735014, 11822506, 11935018; the Chinese Academy of Sciences (CAS) Large-Scale Scientific Facility Program; the CAS Center for Excellence in Particle Physics (CCEPP); Joint Large-Scale Scientific Facility Funds of the NSFC and CAS under Contracts Nos. U1532257, U1532258, U1732263; CAS Key Research Program of Frontier Sciences under Contracts Nos. QYZDJ-SSW-SLH003, QYZDJ-SSW-SLH040; 100 Talents Program of CAS; CAS PIFI; the Thousand Talents Program of China; INPAC and Shanghai Key Laboratory for Particle Physics and Cosmology; German Research Foundation DFG under Contracts Nos. Collaborative Research Center CRC 1044, FOR 2359; Istituto Nazionale di Fisica Nucleare, Italy; Koninklijke Nederlandse Akademie van Wetenschappen (KNAW) under Contract No. 530-4CDP03; Ministry of Development of Turkey under Contract No. DPT2006K-120470; National Science and Technology fund; The Knut and Alice Wallenberg Foundation (Sweden) under Contract No. 2016.0157; The Swedish Research Council; U. S. Department of Energy under Contracts Nos. DE-FG02-05ER41374, DE-SC-0010118, DE-SC-0012069; University of Groningen (RuG) and the Helmholtzzentrum fuer Schwerionenforschung GmbH (GSI), Darmstadt; the Ministry of Science and Higher Education of the Russian Federation, Agreement 14.W03.31.0026.

%% file: main.bbl
\begin{thebibliography}{9}
\addcontentsline{toc}{chapter}{Bibliography}

%%
%% Referece for %%
%%

\bibitem{Asner:2008nq} 
  D.~M.~Asner {\it et al.},
  %``Physics at BES-III,''
  Int.\ J.\ Mod.\ Phys.\ A {\bf 24}, S1 (2009).
%  [arXiv:0809.1869 [hep-ex]].
  %%CITATION = ARXIV:0809.1869;%%
  %272 citations counted in INSPIRE as of 27 May 2019

%\cite{Ablikim:2009aa}
\bibitem{Ablikim:2009aa}
  M.~Ablikim {\it et al.} (BESIII Collaboration),
  %``Design and Construction of the BESIII Detector,''
Nucl. Instrum. Methods Phys. Res., Sect. A {\bf 614}, 345  (2010).

\bibitem{Ablikim:2013mio1} 
M.~Ablikim {\it et al.} (BESIII Collaboration),
  %``Observation of a Charged Charmoniumlike Structure in $e^+e^-$ ? $?^+?^-$ J/? at $\sqrt{s}$ =4.26??GeV,''
  Phys.\ Rev.\ Lett.\  {\bf 110}, 252001 (2013).
  
  \bibitem{Ablikim:2013wzq1} 
  M.~Ablikim {\it et al.} (BESIII Collaboration),
  %``Observation of a Charged Charmoniumlike Structure $Z_c$(4020) and Search for the $Z_c$(3900) in $e^+e^- \to ?^+?^-h_c$,''
  Phys.\ Rev.\ Lett.\  {\bf 111}, 242001 (2013).
  
  \bibitem{Ablikim:2013emm1} 
  M.~Ablikim {\it et al.} (BESIII Collaboration),
  %``Observation of a charged charmoniumlike structure in $e^+e^- \to (D^{*} \bar{D}^{*})^{\pm} \pi^\mp$ at $\sqrt{s}=4.26$GeV,''
  Phys.\ Rev.\ Lett.\  {\bf 112}, 132001 (2014).

\bibitem{Ablikim:2013xfr1} 
  M.~Ablikim {\it et al.} (BESIII Collaboration),
  %``Observation of a charged $(D\bar{D}^{*})^\pm$ mass peak in $e^{+}e^{-} \to \pi D\bar{D}^{*}$ at $\sqrt{s} =$ 4.26 GeV,''
  Phys.\ Rev.\ Lett.\  {\bf 112}, 022001 (2014).

  \bibitem{Ablikim:2015flg1} 
  M.~Ablikim {\it et al.} (BESIII Collaboration),
  %``Measurements of absolute hadronic branching fractions of $\Lambda_{c}^{+}$ baryon,''
  Phys.\ Rev.\ Lett.\  {\bf 116}, 052001 (2016).
  
  \bibitem{Ablikim:2015prg1} 
  M.~Ablikim {\it et al.} (BESIII Collaboration),
  %``Measurement of the absolute branching fraction for $\Lambda^+_{c}\to \Lambda e^+\nu_e$,''
  Phys.\ Rev.\ Lett.\  {\bf 115}, 221805 (2015).
  
  \bibitem{ch1_reviews} For recent reviews, see
  N.~Brambilla, S.~Eidelman, C.~Hanhart, A.~Nefediev, C.~P.~Shen, C.~E.~Thomas, A.~Vairo and C.~Z.~Yuan,
  %``The $XYZ$ states: experimental and theoretical status and perspectives,''
  arXiv:1907.07583 [hep-ex];
  F.~K.~Guo, C.~Hanhart, U.~G.~Mei{\ss}ner, Q.~Wang, Q.~Zhao and B.~S.~Zou,
  %``Hadronic molecules,''
  Rev.\ Mod.\ Phys.\  {\bf 90}, 015004 (2018);
  H.~X.~Chen, W.~Chen, X.~Liu and S.~L.~Zhu,
  %``The hidden-charm pentaquark and tetraquark states,''
  Phys.\ Rept.\  {\bf 639}, 1 (2016);
  N.~Brambilla {\em et al.},
  %``Heavy quarkonium: progress, puzzles, and opportunities,''
  Eur.\ Phys.\ J.\ C {\bf 71}, 1534 (2011).
 
 \bibitem{klempt}
  For a review, see E.~Klempt and A.~Zaitsev,
  %``Glueballs, Hybrids, Multiquarks. Experimental facts versus QCD inspired concepts,''
  Phys.\ Rept.\  {\bf 454}, 1 (2007).
 

%\cite{Bai:1994zm}
\bibitem{Bai:1994zm}
  J.~Z.~Bai {\it et al.} (BES Collaboration),
  %``The BES detector,''
  Nucl. Instrum. Methods Phys. Res., Sect. A {\bf 344}, 319 (1994).
  %doi:10.1016/0168-9002(94)90081-7
  %%CITATION = doi:10.1016/0168-9002(94)90081-7;%%
  %228 citations counted in INSPIRE as of 21 Dec 2017

%\cite{Bai:2001dw}
\bibitem{Bai:2001dw}
  J.~Z.~Bai {\it et al.} (BES Collaboration),
  %``The BES upgrade,''
  Nucl. Instrum. Methods Phys. Res., Sect. A {\bf 458}, 627 (2001).
  %doi:10.1016/S0168-9002(00)00934-7
  %%CITATION = doi:10.1016/S0168-9002(00)00934-7;%%
  %236 citations counted in INSPIRE as of 21 Dec 2017

%\cite{Ye:1987nh}
\bibitem{Ye:1987nh}
  M.~H.~Ye and Z.~P.~Zheng,
  %``Bepc, The Beijing Electron Positron Collider,''
  Int.\ J.\ Mod.\ Phys.\ A {\bf 2}, 1707 (1987);
  %doi:10.1142/S0217751X87000880
  %%CITATION = doi:10.1142/S0217751X87000880;%%
  %3 citations counted in INSPIRE as of 21 Dec 2017
%\cite{Fang:1990wj}
%\bibitem{Fang:1990wj}
  S.~X.~Fang and S.~Y.~Chen,
  %``The Beijing Electron Positron Collider,''
  Part.\ Accel.\  {\bf 26}, 51 (1990).
  %%CITATION = PLACB,26,51;%%
  %3 citations counted in INSPIRE as of 21 Dec 2017

%\bibitem{part1:gaudi} G.~Barrand {\it et. al.}, Comput. Phys. Commun., 140,
%45-55 (2001).


\bibitem{BEPCII-news} https://phys.org/news/2016-04-bepcii-luminosity-world-11033cm2s.html

\bibitem{ETOF} P. Cao {\it et al.}, Nucl. Instrum. Methods Phys. Res., Sect. A (in press),  doi:10.1016/j.nima.2019.163053.

\bibitem{part1:dongmy}
M. Y. Dong {\it et al.}, Chin. Phys. C  {\bf 40}, 016001 (2016).

\bibitem{part1:cgem1}
 F.~Sauli,
  %``GEM: A new concept for electron amplification in gas detectors,''
  Nucl. Instrum. Methods Phys. Res., Sect. A {\bf 386}, 531 (1997).
  
\bibitem{part1:cgem11}
A. Amoroso {\it et al.},  Nucl. Instrum. Methods Phys. Res., Sect. A  {\bf 824}, 515 (2016).

\bibitem{part1:cgem2}
  G.~Mezzadri {\it et al.},
  %``Test beam results of a Cylindrical GEM detector for BESIII experiment,''
  PoS MPGD {\bf 2017}, 048 (2019).
%  [arXiv:1803.07489 [physics.ins-det]].
  
\bibitem{part1:cdc1}
  Y. J. Xie {\it et al.}, Chin. Phys. C {\bf 40}, 096003 (2016). 

\bibitem{ref:top-up} M. Aiba \textit{et al.}, Nucl. Instrum. Methods Phys. Res., Sect. A {\bf 880}, 98 (2018).
\bibitem{ref:crab-waist} P. Raimondi, talk presented at the 2nd Workshop on Super $B$ factory, LNF-INFN, Frascati, 2006.
\bibitem{ref:anton-tau} A. Bogomyagkov, talk presented at the 14th International Workshop on Tau Lepton Physics, IHEP, Beijing, 2016.
  
\end{thebibliography}

\begin{thebibliography}{9}
\addcontentsline{toc}{chapter}{Bibliography}

%%
%% Referece for %%
%%

\bibitem{Brambilla:2014jmp}
  N.~Brambilla {\it et al.},
  %``QCD and Strongly Coupled Gauge Theories: Challenges and Perspectives,''
  Eur.\ Phys.\ J.\ C {\bf 74},  2981 (2014).
%  doi:10.1140/epjc/s10052-014-2981-5
%  [arXiv:1404.3723 [hep-ph]].
  %%CITATION = doi:10.1140/epjc/s10052-014-2981-5;%%
  %209 citations counted in INSPIRE as of 13 Feb 2017
\bibitem{Meyer:2010ku}
  C.~A.~Meyer and Y.~Van Haarlem,
  %``The Status of Exotic-quantum-number Mesons,''
  Phys.\ Rev.\ C {\bf 82}, 025208 (2010).
%  doi:10.1103/PhysRevC.82.025208
%  [arXiv:1004.5516 [nucl-ex]].
%  %%CITATION = doi:10.1103/PhysRevC.82.025208;%%
  %67 citations counted in INSPIRE as of 13 Feb 2017
\bibitem{Crede:2008vw}
  V.~Crede and C.~A.~Meyer,
  %``The Experimental Status of Glueballs,''
  Prog.\ Part.\ Nucl.\ Phys.\  {\bf 63}, 74 (2009).
%  doi:10.1016/j.ppnp.2009.03.001
%  [arXiv:0812.0600 [hep-ex]].
  %%CITATION = doi:10.1016/j.ppnp.2009.03.001;%%
  %140 citations counted in INSPIRE as of 13 Feb 2017

\bibitem{Klempt:2007cp}
  E.~Klempt and A.~Zaitsev,
  %``Glueballs, Hybrids, Multiquarks. Experimental facts versus QCD inspired concepts,''
  Phys.\ Rept.\  {\bf 454}, 1 (2007).
%  doi:10.1016/j.physrep.2007.07.006
%  [arXiv:0708.4016 [hep-ph]].
  %%CITATION = doi:10.1016/j.physrep.2007.07.006;%%
  %533 citations counted in INSPIRE as of 13 Feb 2017
  \bibitem{Amsler:2004ps}
  C.~Amsler and N.~A.~Tornqvist,
  %``Mesons beyond the naive quark model,''
  Phys.\ Rept.\  {\bf 389}, 61 (2004).
%  doi:10.1016/j.physrep.2003.09.003
  %%CITATION = doi:10.1016/j.physrep.2003.09.003;%%
  %281 citations counted in INSPIRE as of 13 Feb 2017
  \bibitem{Godfrey:1998pd}
  S.~Godfrey and J.~Napolitano,
  %``Light meson spectroscopy,''
  Rev.\ Mod.\ Phys.\  {\bf 71}, 1411 (1999).
%  doi:10.1103/RevModPhys.71.1411
%  [hep-ph/9811410].
  %%CITATION = doi:10.1103/RevModPhys.71.1411;%%
  %206 citations counted in INSPIRE as of 13 Feb 2017
  \bibitem{Kopke:1988cs}
  L.~Kopke and N.~Wermes,
  %``J/psi Decays,''
  Phys.\ Rept.\  {\bf 174}, 67 (1989).
 % doi:10.1016/0370-1573(89)90074-4
  %%CITATION = doi:10.1016/0370-1573(89)90074-4;%%
  %177 citations counted in INSPIRE as of 14 Feb 2017
    \bibitem{bes3}
K.~T.~Chao and Y.~F. Wang, Int. J. Mod. Phys. A {\bf 24}, suppl. 1 (2009).
\bibitem{Bali:1993fb}
  G.~S.~Bali {\it et al.} (UKQCD Collaboration),
  %``A Comprehensive lattice study of SU(3) glueballs,''
  Phys.\ Lett.\ B {\bf 309}, 378 (1993).
  %%CITATION = doi:10.1016/0370-2693(93)90948-H;%%
  %455 citations counted in INSPIRE as of 13 Feb 2017
  \bibitem{Morningstar:1999rf}
  C.~J.~Morningstar and M.~J.~Peardon,
  %``The Glueball spectrum from an anisotropic lattice study,''
  Phys.\ Rev.\ D {\bf 60}, 034509 (1999).
  %%CITATION = doi:10.1103/PhysRevD.60.034509;%%
  %750 citations counted in INSPIRE as of 13 Feb 2017
  \bibitem{Chen:2005mg}
  Y.~Chen {\it et al.},
  %``Glueball spectrum and matrix elements on anisotropic lattices,''
  Phys.\ Rev.\ D {\bf 73}, 014516 (2006).
  %%CITATION = doi:10.1103/PhysRevD.73.014516;%%
  %350 citations counted in INSPIRE as of 13 Feb 2017

\bibitem{Sun:2017ipk}
  W.~Sun {\it et al.},
  %``Glueball spectrum from $N_f=2$ lattice QCD study on anisotropic lattices,''
  Chin.\ Phys.\ C {\bf 42}, 093103 (2018) .
  \bibitem{Gregory:2012hu}
  E.~Gregory  {\it et al.},
  %``Towards the glueball spectrum from unquenched lattice QCD,''
  JHEP {\bf 1210}, 170 (2012).
  %%CITATION = doi:10.1007/JHEP10(2012)170;%%
  %86 citations counted in INSPIRE as of 13 Feb 2017


  \bibitem{Gui:2012gx}
  L.~C.~Gui {\it et al.} (CLQCD Collaboration),
  %``Scalar Glueball in Radiative $J/\psi$ Decay on the Lattice,''
  Phys.\ Rev.\ Lett.\  {\bf 110},  021601 (2013).
    %%CITATION = doi:10.1103/PhysRevLett.110.021601;%%
  %35 citations counted in INSPIRE as of 13 Feb 2017
  \bibitem{Yang:2013xba}
  Y.~B.~Yang {\it et al.} (CLQCD Collaboration),
  %``Lattice Study of Radiative J/Š× Decay to a Tensor Glueball,''
  Phys.\ Rev.\ Lett.\  {\bf 111}, 091601 (2013).
  %%CITATION = doi:10.1103/PhysRevLett.111.091601;%%
  %12 citations counted in INSPIRE as of 13 Feb 2017
\bibitem{Gui:2019dtm}
  L.~C.~Gui, J.~M.~Dong, Y.~Chen and Y.~B.~Yang,
  %``Study of the pseudoscalar glueball in $J/\psi$ radiative decays,''
  arXiv:1906.03666 [hep-lat].

\bibitem{PDG}
 M.~Tanabashi {\it et al.} (Particle Data Group),
  %``Review of Particle Physics,''
  Phys.\ Rev.\ D {\bf 98},  030001 (2018).



\bibitem{Ablikim:2013hq}
  M.~Ablikim {\it et al.} (BESIII Collaboration),
  %``Partial wave analysis of $J/\psi \to \gamma \eta \eta$,''
  Phys.\ Rev.\ D {\bf 87}, 092009 (2013);
  Erratum: [Phys.\ Rev.\ D {\bf 87}, 119901 (2013)].
  %%CITATION = doi:10.1103/PhysRevD.87.092009, 10.1103/PhysRevD.87.119901;%%
  %38 citations counted in INSPIRE as of 13 Feb 2017
\bibitem{Ablikim:2018izx}
  M.~Ablikim {\it et al.} (BESIII Collaboration),
  %``Amplitude analysis of the $K_{S}K_{S}$ system produced in radiative $J/\psi$ decays,''
  Phys.\ Rev.\ D {\bf 98}, 072003 (2018).
   %%CITATION = doi:10.1103/PhysRevD.98.072003;%%
 \bibitem{Ablikim:2012ft}
  M.~Ablikim {\it et al.} (BESIII Collaboration),
  %``Study of the near-threshold ŠØ? mass enhancement in doubly OZI-suppressed J/Š×¡úŠÃŠØ? decays,''
  Phys.\ Rev.\ D {\bf 87}, 032008 (2013).
  %%CITATION = doi:10.1103/PhysRevD.87.032008;%%
  %25 citations counted in INSPIRE as of 13 Feb 2017
\bibitem{Ablikim:2015umt}
  M.~Ablikim {\it et al.} (BESIII Collaboration),
  %``Amplitude analysis of the ŠÐ$^0$ŠÐ$^0$ system produced in radiative J/Š× decays,''
  Phys.\ Rev.\ D {\bf 92} 052003 (2015);
   Erratum: [Phys.\ Rev.\ D {\bf 93},  039906 (2016)].
  %%CITATION = doi:10.1103/PhysRevD.92.052003, 10.1103/PhysRevD.93.039906;%%
  %21 citations counted in INSPIRE as of 03 Sep 2018


\bibitem{bibpiN} A. Etkin {\it et al}., Phys. Rev. Lett. {\bf 41}, 784 (1978); Phys. Lett. B {\bf 165}, 217 (1985); Phys. Lett. B {\bf 201}, 568 (1988).
\bibitem{Ablikim:2016hlu}
  M.~Ablikim {\it et al.} (BESIII Collaboration),
  %``Observation of pseudoscalar and tensor resonances in $J/\psi\to \gamma \phi \phi$,''
  Phys.\ Rev.\ D {\bf 93},  112011 (2016).
  %%CITATION = doi:10.1103/PhysRevD.93.112011;%%
  %5 citations counted in INSPIRE as of 13 Feb 2017

\bibitem{Masoni:2006rz}
  A.~Masoni, C.~Cicalo and G.~L.~Usai,
  %``The case of the pseudoscalar glueball,''  J.\ Phys.\ G {\bf 32}, R293 (2006).

\bibitem{BESIII:2012aa}
  M.~Ablikim {\it et al.} (BESIII Collaboration),
  %``First observation of $\eta(1405)$ decays into $f_{0}(980)\pi^0$,''
  Phys.\ Rev.\ Lett.\  {\bf 108}, 182001 (2012).
  %%CITATION = doi:10.1103/PhysRevLett.108.182001;%%
  %42 citations counted in INSPIRE as of 13 Feb 2017
  
\bibitem{Ablikim:2010aa}
  M.~Ablikim {\it et al.} (BESIII Collaboration),
  %``Study of $a_0^0(980) - f_0(980)$ mixing,''
  Phys.\ Rev.\ D {\bf 83}, 032003 (2011).

\bibitem{Ablikim:2018pik}
  M.~Ablikim {\it et al.} (BESIII Collaboration),
  %``Observation of $a^{0}_{0}(980)$-$f_{0}(980)$ Mixing,''
  Phys.\ Rev.\ Lett.\  {\bf 121}, 022001 (2018).  

 
  \bibitem{Wu:2011yx}
  J.~J.~Wu, X.~H.~Liu, Q.~Zhao and B.~S.~Zou,
  %``The Puzzle of anomalously large isospin violations in $\eta(1405/1475)\to 3\pi$,''
  Phys.\ Rev.\ Lett.\  {\bf 108}, 081803 (2012).
  %%CITATION = doi:10.1103/PhysRevLett.108.081803;%%
  %61 citations counted in INSPIRE as of 13 Feb 2017
 \bibitem{Ablikim:2018hxj}
  M.~Ablikim {\it et al.} (BESIII Collaboration),
  %``Study of $\eta(1475)$ and $X(1835)$ in radiative $J/\psi$ decays to $\gamma \phi$,''
  Phys.\ Rev.\ D {\bf 97}, 051101 (2018).
  %%CITATION = doi:10.1103/PhysRevD.97.051101;%%
  %2 citations counted in INSPIRE as of 03 Sep 2018
  \bibitem{Dudek:2011tt}
  J.~J.~Dudek {\it et al.},
  %``Isoscalar meson spectroscopy from lattice QCD,''
  Phys.\ Rev.\ D {\bf 83}, 111502 (2011).
  %%CITATION = doi:10.1103/PhysRevD.83.111502;%%
  %127 citations counted in INSPIRE as of 13 Feb 2017
  
%\bibitem{Chung:2002fz}
%  S.~U.~Chung, E.~Klempt and J.~G.~Korner,
%  %``SU(3) classification of p wave eta pi and eta-prime pi systems,''
%  Eur.\ Phys.\ J.\ A {\bf 15}, 539 (2002).

  \bibitem{Adams:2011sq}
  G.~S.~Adams {\it et al.} (CLEO Collaboration),
  %``Amplitude analyses of the decays $\chi_c1 -> \eta \pi^+ \pi^-$ and $\chi_c1 -> \eta' \pi^+ \pi^-$,''
  Phys.\ Rev.\ D {\bf 84}, 112009 (2011).
  %%CITATION = doi:10.1103/PhysRevD.84.112009;%%
  %26 citations counted in INSPIRE as of 13 Feb 2017
    \bibitem{Kornicer:2016axs}
  M.~Ablikim {\it et al.} (BESIII Collaboration),
  %``Amplitude analysis of the $\chi_{c1} \to \eta\pi^+\pi^-$ decays,''
Phys.\ Rev.\ D {\bf 95}, 032002 (2017).
  %%CITATION = ARXIV:1610.02479;%%
  %1 citations counted in INSPIRE as of 13 Feb 2017
\bibitem{Isgur:1985vy}
  N.~Isgur, R.~Kokoski and J.~Paton,
  %``Gluonic Excitations of Mesons: Why They Are Missing and Where to Find Them,''
  Phys.\ Rev.\ Lett.\  {\bf 54}, 869 (1985)
\bibitem{Page:1998gz}
  P.~R.~Page, E.~S.~Swanson and A.~P.~Szczepaniak,
  %``Hybrid meson decay phenomenology,''
  Phys.\ Rev.\ D {\bf 59}, 034016 (1999)
\bibitem{Huang:2010dc}
  P.~Z.~Huang, H.~X.~Chen and S.~L.~Zhu,
  %``The Strong Decay Patterns of the $1^{-+}$ Exotic Hybrid Mesons,''
  Phys.\ Rev.\ D {\bf 83}, 014021 (2011)
\bibitem{Chen:2010ic}
  H.~X.~Chen, Z.~X.~Cai, P.~Z.~Huang and S.~L.~Zhu,
  %``The Decay Properties of the 1^{-+} Hybrid State,''
  Phys.\ Rev.\ D {\bf 83}, 014006 (2011)
  
\bibitem{Jaffe:1976ig}
  R.~L.~Jaffe,
  %``Multi-Quark Hadrons. 1. The Phenomenology of (2 Quark 2 anti-Quark) Mesons,''
  Phys.\ Rev.\ D {\bf 15}, 267 (1977).
\bibitem{Alford:2000mm}
  M.~G.~Alford and R.~L.~Jaffe,
  %``Insight into the scalar mesons from a lattice calculation,''
  Nucl.\ Phys.\ B {\bf 578}, 367 (2000).
\bibitem{Maiani:2004uc}
  L.~Maiani, F.~Piccinini, A.~D.~Polosa and V.~Riquer,
  %``A New look at scalar mesons,''
  Phys.\ Rev.\ Lett.\  {\bf 93}, 212002 (2004).
\bibitem{Maiani:2007iw}
  L.~Maiani, A.~D.~Polosa and V.~Riquer,
  %``Structure of light scalar mesons from D(s) and D0 non-leptonic decays,''
  Phys.\ Lett.\ B {\bf 651}, 129 (2007).
\bibitem{Hooft:2008we}
  G.~'t Hooft, G.~Isidori, L.~Maiani, A.~D.~Polosa and V.~Riquer,
  %``A Theory of Scalar Mesons,''  P
  hys.\ Lett.\ B {\bf 662}, 424 (2008).
\bibitem{Weinstein:1990gu}
  J.~D.~Weinstein and N.~Isgur,
  %``K anti-K Molecules,''
  Phys.\ Rev.\ D {\bf 41}, 2236 (1990).
\bibitem{gluon} S.~Ishida {\it et al.}, in Proceedings of the 6th International Conference on Hadron Spectroscopy, Manchester, United Kingdom, 1995 (World Scientific, Singapore, 1995), p.454.
\bibitem{Achasov:1979xc}
  N.~N.~Achasov, S.~A.~Devyanin and G.~N.~Shestakov,
  %``The S* Delta0 Mixing as the Threshold Phenomenon,''
  Phys.\ Lett.\  {\bf 88B}, 367 (1979).
\bibitem{Wu:2007jh}
  J.~J.~Wu, Q.~Zhao and B.~S.~Zou,
  %``Possibility of measuring a0(980)-f0(980) mixing from J/psi ---> phi a0(980),''
  Phys.\ Rev.\ D {\bf 75}, 114012 (2007).
\bibitem{Hanhart:2007bd}
  C.~Hanhart, B.~Kubis and J.~R.~Pelaez,
  %``Investigation of a0-f0 mixing,''
  Phys.\ Rev.\ D {\bf 76}, 074028 (2007).
\bibitem{Wu:2008hx}
  J.~J.~Wu and B.~S.~Zou,
  %``Study a0(980)-f0(980) mixing from a0(980) -> f0(980) transition,''
  Phys.\ Rev.\  D {\bf 78}, 074017 (2008).

  
 \bibitem{x1835_bes2} M.~Ablikim {\it et al.} (BES Collaboration), Phys. Rev. Lett. {\bf95}, 262001 (2005).
\bibitem{x1835_bes3} M.~Ablikim {\it et al.} (BESIII Collaboration), Phys. Rev. Lett. {\bf106}, 072002 (2011).
\bibitem{x1835_qiny} M.~Ablikim {\it et al.} (BESIII Collaboration), Phys. Rev. Lett. {\bf115}, 091803 (2015).
\bibitem{xpp_bes2} J.~Z.~Bai {\it et al.} (BES Collaboration), Phys. Rev. Lett. {\bf91}, 022001 (2003).
\bibitem{xpp_bes3} M.~Ablikim {\it et al.} (BESIII Collaboration), Chin. Phys. C {\bf34}, 421 (2010).
\bibitem{xpp_cleo} J.~P.~Alexander {\it et al.} (CLEO Collaboration), Phys. Rev. D {\bf82}, 092002 (2010).
\bibitem{xpp_bes3pwa} M.~Ablikim {\it et al.} (BESIII Collaboration), Phys. Rev. Lett. {\bf108}, 112003 (2012).
\bibitem{Ablikim:2016itz}
  M.~Ablikim {\it et al.} (BESIII Collaboration),
  %``Observation of an anomalous line shape of the $\eta^{\prime}\pi^{+}\pi^{-}$ mass spectrum near the $p\bar{p}$ mass threshold in $J/\psi\rightarrow\gamma\eta^{\prime}\pi^{+}\pi^{-}$,''
  Phys.\ Rev.\ Lett.\  {\bf 117}, 042002 (2016).

\bibitem{pkl}  M.~Ablikim {\it  et al.} (BES Collaboration),
    Phys. Rev. Lett. {\bf 93}, 112002 (2004).

  \bibitem{Capstick:2000dk}
 S.~Capstick {\it et al.}, arXiv:hep-ph/0012238.
  %%CITATION = HEP-PH/0012238;%%

  \bibitem{Klempt:2009pi}
  E.~Klempt and J.~M.~Richard,
  %``Baryon spectroscopy,''
  Rev.\ Mod.\ Phys.\  {\bf 82}, 1095 (2010).
  %%CITATION = doi:10.1103/RevModPhys.82.1095;%%
  %209 citations counted in INSPIRE as of 13 Feb 2017
\bibitem{Capstick:2000qj}
  S.~Capstick and W.~Roberts,
  %``Quark models of baryon masses and decays,''
  Prog.\ Part.\ Nucl.\ Phys.\  {\bf 45}, S241 (2000).
  %%CITATION = doi:10.1016/S0146-6410(00)00109-5;%%
  %284 citations counted in INSPIRE as of 13 Feb 2017
\bibitem{Edwards:2011jj}
  R.~G.~Edwards, J.~J.~Dudek, D.~G.~Richards and S.~J.~Wallace,
  %``Excited state baryon spectroscopy from lattice QCD,''
  Phys.\ Rev.\ D {\bf 84}, 074508 (2011).
  %%CITATION = doi:10.1103/PhysRevD.84.074508;%%
  %253 citations counted in INSPIRE as of 13 Feb 2017
\bibitem{Zou:2000wg}
  B.~-S.~Zou,
 % \emph{$N^*$, $\Lambda^*$, $\Sigma^*$ and $\Xi^*$ resonances from $J/\psi$ and $\psi\prime$ decays},
  Nucl.\ Phys.\ A {\bf 684}, 330 (2001).
  %%CITATION = HEP-PH/0006039;%%

\bibitem{Zou:2001uc}
  B.~S.~Zou {\it et al.}  (BES Collaboration),
 % \emph{Baryon spectroscopy at Beijing Electron Positron Collider},
  PiN Newslett.~{\bf 16}, 174  (2002).
%  [{\tt hep-ph/0110264}].
  %%CITATION = HEP-PH/0110264;%%

\bibitem{Ablikim:2012zk}
  M.~Ablikim {\it et al.}  (BESIII Collaboration),
 % \emph{Observation of two new $N^*$ resonances in $\psi(3686) \rightarrow p\bar{p}\pi^0$},
  Phys.\ Rev.\ Lett.\  {\bf 110}, 022001 (2013).

  \bibitem{Battaglieri:2014gca}
  M.~Battaglieri {\it et al.},
  %``Analysis Tools for Next-Generation Hadron Spectroscopy Experiments,''
  Acta Phys.\ Polon.\ B {\bf 46}, 257 (2015).
  %%CITATION = doi:10.5506/APhysPolB.46.257;%%
  %18 citations counted in INSPIRE as of 13 Feb 2017
\bibitem{Berger:2010zza}
  N.~Berger, L.~Beijiang and W.~Jike,
  %``Partial wave analysis using graphics processing units,''
  J.\ Phys.\ Conf.\ Ser.\  {\bf 219}, 042031 (2010).
  %%CITATION = doi:10.1088/1742-6596/219/4/042031;%%
  %5 citations counted in INSPIRE as of 13 Feb 2017


\bibitem{Wang:2004du}
  J.~X.~Wang,
  %``Progress in FDC project,''
  Nucl. Instrum. Methods Phys. Res., Sect. A {\bf 534}, 241 (2004).
%
%\bibitem{Pevsner:1961pa}
%A.~Pevsner {\it et al.}, Phys. Rev. Lett. {\bf 7}, 421 (1961).
%
%\bibitem{Kalbfleisch:1964zz}
%G.~R. Kalbfleisch {\it et al.},Phys. Rev. Lett. {\bf 12}, 527 (1964).
%
%\bibitem{Goldberg:1964zza}
%M.~Goldberg {\it et~al.}, Phys. Rev. Lett. {\bf 12}, 546 (1964).

%\bibitem{Meissner:2010bna}
%  U.~G.~Meissner,
%  %``Isospin violation, light quark masses, and all that,''
%  Chin.\ Phys.\ C {\bf 34}, 1163 (2010).

\bibitem{Gasser:1983yg}
J.~Gasser and H.~Leutwyler, Annals Phys. {\bf 158}, 142 (1984).

\bibitem{Wess:1971yu}
J.~Wess and B.~Zumino, Phys. Lett. B {\bf 37}, 95 (1971).

\bibitem{Witten:1983tw}
E.~Witten, Nucl. Phys. B {\bf 223}, 422 (1983).

\bibitem{Bijnens:1989jb}
J.~Bijnens, A.~Bramon and F.~Cornet, Z. Phys. C {\bf 46}, 599 (1990).

\bibitem{Sakurai:1960ju}
J.~J. Sakurai, Annals Phys. {\bf 11}, 1 (1960).

\bibitem{Landsberg:1986fd}
L.~G. Landsberg, Phys. Rept. {\bf 128}, 301 (1985).

\bibitem{Kaiser:2000gs}
R.~Kaiser and H.~Leutwyler, Eur. Phys. J. C {\bf 17}, 623 (2000).

\bibitem{Ablikim:2012cn}
M.~Ablikim {\em et~al.} (BESIII Collaboration),  Chin. Phys. C {\bf 36}, 915 (2012).

\bibitem{Ablikim:2016fal}
M.~Ablikim {\em et~al.} (BESIII Collaboration), Chin. Phys. C {\bf 41}, 013001 (2017).

\bibitem{Ablikim:2016frj}
M.~Ablikim {\em et~al.} (BESIII Collaboration), Phys. Rev. Lett. {\bf 118}, 012001 (2017).

\bibitem{Ablikim:2015wnx}
M.~Ablikim {\em et~al.} (BESIII Collaboration), Phys. Rev.  D {\bf 92}, 012001 (2015).




%\bibitem{Ablikim:2015eos}
%M.~Ablikim {\em et~al.} (BESIII Collaboration),
%\newblock Phys. Rev. {\bf D92}, 051101 (2015).



%\bibitem{Ablikim:2016tuo}
%M.~Ablikim {\em et~al.} (BESIII Collaboration),
%\newblock Phys. Rev. {\bf D96}, 012005 (2017).

\bibitem{Ablikim:2014eoc}
M.~Ablikim {\em et~al.} (BESIII Collaboration),
Phys. Rev. Lett. {\bf 112}, 251801 (2014);
Addendum: [Phys.\ Rev.\ Lett.\  {\bf 113}, 039903 (2014)].


\bibitem{Ablikim:2018rho}
M.~Ablikim {\em et~al.} (BESIII Collaboration),
 Phys. Rev. Lett. {\bf 120}, 242003 (2018).




\bibitem{Kubis:2015sga} B. Kubis and J.  Plenter, Eur. Phys. J. C{\bf 75}, 283 (2015).

\bibitem{Stollenwerk:2011zz} F. Stollenwerk,  C. Hanhart, A. Kupsc, U. G.
                        Meissner and A. Wirzba,  Phys. Lett. B {\bf 707}, 184 (2012).



\bibitem{Hanhart:2016pcd} C. Hanhart , S. Holz, B.  Kubis and A. Kupść,
                        A. Wirzba and C. W. Xiao,
                  Eur. Phys. J.  C{\bf 77}, 98 (2017);   Erratum: Eur. Phys. J. C {\bf 78},450(2018).
\bibitem{Kubis:2009sb}
B. Kubis  and S. P. Schneider,
 Eur. Phys. J.  C{\bf 62}, 511 (2009).

%\bibitem{Isken:2017dkw}
%T.~Isken, B.~Kubis, S.~P. Schneider and P.~Stoffer,
%\newblock Eur. Phys. J. {\bf C77}, 489 (2017).

%\bibitem{Geng:2002ua}
%C.~Q. Geng, J.~N. Ng and T.~H. Wu,
%\newblock Mod. Phys. Lett. {\bf A17}, 1489 (2002).

%\bibitem{Gao:2002gq}
%D.~N. Gao,
%\newblock Mod. Phys. Lett. {\bf A17}, 1583 (2002).

%\bibitem{Ablikim:2012gf}
%M.~Ablikim {\em et~al.} (BESIII),
%\newblock Phys. Rev. {\bf D87}, 012009 (2013).

%\bibitem{Ablikim:2012vn}
%M.~Ablikim {\em et~al.} (BESIII),
%\newblock Phys. Rev. {\bf D87}, 032006 (2013).

%\bibitem{Ablikim:2016bjc}
%M.~Ablikim {\em et~al.} (BESIII),
%\newblock Phys. Rev. {\bf D93}, 072008 (2016).


%\bibitem{Adlarson:2016wkw}
%P. Adlarson {\em et~al.} (WASA-at-COSY),
%\newblock Phys. Lett. B {\bf 770}, 418 (2017).

\bibitem{theory1}
F. Niecknig, B. Kubis, and S. P. Schneider,
 Eur. Phys. J. C {\bf 72}, 2014 (2012).

\bibitem{theory2}
I. V. Danilkin  {\it et al.},
Phys. Rev. D {\bf 91}, 094029 (2015).

\bibitem{theory3}
C. Terschlusen, B. Strandberg, S. Leupold, and
F. Eichstadt,
Eur. Phys. J. A {\bf 49}, 116 (2013).

%\bibitem{Ablikim:2018yen}
%  M.~Ablikim {\it et al.} [BESIII Collaboration],
%  %``Dalitz Plot Analysis of the Decay $\omega \rightarrow \pi^{+}\pi^{-}\pi^{0}$,''
%  Phys.\ Rev.\ D {\bf 98}, 112007 (2018).

\bibitem{COMPASS}
F.~Gautheron {\it et al.} (COMPASS Collaboration),
  %``COMPASS-II Proposal,''
  SPSC-P-340, CERN-SPSC-2010-014.

\bibitem{GlueX}
P. Eugenio, PoS ConfinementX, 349  (2012).

\bibitem{PANDA}
M.~F.~M.~Lutz {\it et al.} (PANDA Collaboration),
  %``Physics Performance Report for PANDA: Strong Interaction Studies with Antiprotons,''
  arXiv:0903.3905 [hep-ex].

\bibitem{belle2}
 T.~Abe {\it et al.}  (Belle II Collaboration),
%%  ``Belle II Technical Design Report,''
 arXiv:1011.0352 [physics.ins-det];
 E.~Kou {\it et al.},
 %``The Belle II Physics book,''
 arXiv:1808.10567 [hep-ex].
 %%CITATION = ARXIV:1808.10567;%%

\end{thebibliography}

\begin{thebibliography}{9}
\addcontentsline{toc}{chapter}{Bibliography}

%%
%% Referece for %%
%%

%\cite{Lebed:2016hpi}

\bibitem{AppelPol} 
  T.~Appelquist and H.D.~Politzer,
  %``Physical Review Letters,''
Phys.\ Rev.\ Lett. {\bf 34}, 43 (1975).  
% doi:10.1103/PhysRevLett.34.43

\bibitem{Asner:2008nqch3} 
  D.~M.~Asner {\it et al.},
  %``Physics at BES-III,''
  Int.\ J.\ Mod.\ Phys.\ A {\bf 24}, S1 (2009).

\bibitem{pdg} 
  M.~Tanabashi {\it et al.} (Particle Data Group),
  %``Review of Particle Physics,''
  Phys.\ Rev.\ D {\bf 98},  030001 (2018).
%   doi:10.1103/PhysRevD.98.030001
  %%CITATION = doi:10.1103/PhysRevD.98.030001;%%
  %167 citations counted in INSPIRE as of 14 Sep 2018


  \bibitem{Pakhlova:2008vn} 
  G.~Pakhlova {\it et al.} (Belle Collaboration),
  %``Observation of a near-threshold enhancement in the e+e- ---> Lambda+(c) Lambda-(c) cross section using initial-state radiation,''
  Phys.\ Rev.\ Lett.\  {\bf 101}, 172001 (2008).

\bibitem{Wang:2007ea} 
  X.~L.~Wang {\it et al.} (Belle Collaboration),
  %``Observation of Two Resonant Structures in e+e- to pi+ pi- psi(2S) via Initial State Radiation at Belle,''
  Phys.\ Rev.\ Lett.\  {\bf 99}, 142002 (2007).
  

%\cite{Barnes:2005pb}
\bibitem{Barnes:2005pb}
  T.~Barnes, S.~Godfrey and E.~S.~Swanson,
  %``Higher charmonia,''
  Phys.\ Rev.\ D {\bf 72}, 054026 (2005).
  %doi:10.1103/PhysRevD.72.054026
 % [hep-ph/0505002].
  %%CITATION = doi:10.1103/PhysRevD.72.054026;%%
  %415 citations counted in INSPIRE as of 26 Oct 2016

%\cite{Cheung:2016bym}
\bibitem{Cheung:2016bym} 
  G.~K.~C.~Cheung {\it et al.} (Hadron Spectrum Collaboration),
  %``Excited and exotic charmonium, $D_s$ and $D$ meson spectra for two light quark masses from lattice QCD,''
  JHEP {\bf 1612}, 089 (2016)
  %doi:10.1007/JHEP12(2016)089
  %[arXiv:1610.01073 [hep-lat]].
  %%CITATION = doi:10.1007/JHEP12(2016)089;%%
  %35 citations counted in INSPIRE as of 07 Oct 2019


%\cite{Agashe:2014kda}
\bibitem{Agashe:2014kda} 
  K.~A.~Olive {\it et al.} (Particle Data Group),
  %``Review of Particle Physics,''
  Chin.\ Phys.\ C {\bf 38}, 090001 (2014) and 2015 update.
 % doi:10.1088/1674-1137/38/9/090001
  %%CITATION = doi:10.1088/1674-1137/38/9/090001;%%
  %7879 citations counted in INSPIRE as of 07 Oct 2019

%%%%%%%%%%%%%%%
%%
%%  from Kai Zhu, 2018/08/29   Hai-Bo Li 
%%%%%%%%
%\cite{PhysLett.B167.437}
\bibitem{PhysLett.B167.437}
  W.~E.~Caswell and G.~P.~Lepage,
  %``Effective Lagrangians for Bound State Problems in QED, QCD, and Other Field Theories,''
  Phys.\ Lett.\  B {\bf 167}, 437 (1986).
  %%doi:10.1016/0370-2693(86)91297-9
  %%CITATION = %%doi:10.1016/0370-2693(86)91297-9;%%
  %997 citations counted in INSPIRE as of 21 Aug 2018

%\cite{PhysRev.D46.4052}
\bibitem{PhysRev.D46.4052}
  G.~P.~Lepage, L.~Magnea, C.~Nakhleh, U.~Magnea and K.~Hornbostel,
  %``Improved nonrelativistic QCD for heavy quark physics,''
  Phys.\ Rev.\ D {\bf 46}, 4052 (1992).
  %%doi:10.1103/PhysRevD.46.4052
  %%[hep-lat/9205007].
  %%CITATION = %%doi:10.1103/PhysRevD.46.4052;%%
  %606 citations counted in INSPIRE as of 21 Aug 2018



%\cite{hep-ph/9407339}
\bibitem{hep-ph/9407339}
  G.~T.~Bodwin, E.~Braaten and G.~P.~Lepage,
  %``Rigorous QCD analysis of inclusive annihilation and production of heavy quarkonium,''
  Phys.\ Rev.\ D {\bf 51}, 1125 (1995);
  Erratum: [Phys.\ Rev.\ D {\bf 55}, 5853 (1997)].
  %%doi:10.1103/PhysRevD.55.5853, 10.1103/PhysRevD.51.1125
  %%[hep-ph/9407339].
  %%CITATION = %%doi:10.1103/PhysRevD.55.5853, 10.1103/PhysRevD.51.1125;%%
  %2161 citations counted in INSPIRE as of 21 Aug 2018

%\cite{EFT}
\bibitem{EFT}
  Alexey~A.~Petrov, Andrew~E.~Blechman,
  ``Effective Field Theories'',
  World Scientific Publishing Company Pte Limited, 2015,
  ISBN	9789814434935
  %Length	320 pages


%\cite{hep-ph/9707481}
\bibitem{hep-ph/9707481}
  A.~Pineda and J.~Soto,
  %``Effective field theory for ultrasoft momenta in NRQCD and NRQED,''
  Nucl.\ Phys.\ Proc.\ Suppl.\  {\bf 64}, 428 (1998).
  %%doi:10.1016/S0920-5632(97)01102-X
  %%[hep-ph/9707481].
  %%CITATION = %%doi:10.1016/S0920-5632(97)01102-X;%%
  %455 citations counted in INSPIRE as of 21 Aug 2018



%\cite{hep-ph/9907240}
\bibitem{hep-ph/9907240}
  N.~Brambilla, A.~Pineda, J.~Soto and A.~Vairo,
  %``Potential NRQCD: An Effective theory for heavy quarkonium,''
  Nucl.\ Phys.\ B {\bf 566}, 275 (2000).
  %%doi:10.1016/S0550-3213(99)00693-8
  %%[hep-ph/9907240].
  %%CITATION = %%doi:10.1016/S0550-3213(99)00693-8;%%
  %528 citations counted in INSPIRE as of 21 Aug 2018



%\cite{hep-ph/0410047}
\bibitem{hep-ph/0410047}
  N.~Brambilla, A.~Pineda, J.~Soto and A.~Vairo,
  %``Effective field theories for heavy quarkonium,''
  Rev.\ Mod.\ Phys.\  {\bf 77}, 1423 (2005).
  %%doi:10.1103/RevModPhys.77.1423
  %%[hep-ph/0410047].
  %%CITATION = %%doi:10.1103/RevModPhys.77.1423;%%
  %456 citations counted in INSPIRE as of 21 Aug 2018



%\cite{hep-ph/0601044}
\bibitem{hep-ph/0601044}
  Y.~P.~Kuang,
  %``QCD multipole expansion and hadronic transitions in heavy quarkonium systems,''
  Front.\ Phys.\ China {\bf 1}, 19 (2006).
  %%doi:10.1007/s11467-005-0012-6
  %% [hep-ph/0601044].
  %%CITATION = %%doi:10.1007/s11467-005-0012-6;%%
  %93 citations counted in INSPIRE as of 21 Aug 2018



%\cite{hep-ph/0412158}
\bibitem{hep-ph/0412158}
  N.~Brambilla {\it et al.} (Quarkonium Working Group),
  %``Heavy quarkonium physics,''
 arXiv:hep-ph/0412158.
  %%CITATION = HEP-PH/0412158;%%
  %853 citations counted in INSPIRE as of 21 Aug 2018

%\cite{PhysRevLett.51.963}
\bibitem{PhysRevLett.51.963} 
  M.~E.~B.~Franklin {\it et al.},
  %``Measurement of $\psi(3097)$ and $\psi^\prime$ (3686) Decays Into Selected Hadronic Modes,''
  Phys.\ Rev.\ Lett.\  {\bf 51}, 963 (1983).
  %doi:10.1103/PhysRevLett.51.963
  %%CITATION = doi:10.1103/PhysRevLett.51.963;%%
  %133 citations counted in INSPIRE as of 29 Sep 2019


%\cite{PhysRevLett.104.132002}
\bibitem{PhysRevLett.104.132002}
  M.~Ablikim {\it et al.} (BESIII Collaboration),
  %``Measurements of h_c(^1P_1) in psi' Decays,''
  Phys.\ Rev.\ Lett.\  {\bf 104}, 132002 (2010).
  %%doi:10.1103/PhysRevLett.104.132002
  %% [arXiv:1002.0501 [hep-ex]].
  %%CITATION = %%doi:10.1103/PhysRevLett.104.132002;%%
  %111 citations counted in INSPIRE as of 21 Aug 2018



%\cite{PhysRevLett.107.092001}
\bibitem{PhysRevLett.107.092001}
  M.~Ablikim {\it et al.} (BESIII Collaboration),
  %``Observation of $\chi_{c1}$ decays into vector meson pairs $\phi\phi$, $\omega\omega$, and $\omega\phi$,''
  Phys.\ Rev.\ Lett.\  {\bf 107}, 092001 (2011).
  %%doi:10.1103/PhysRevLett.107.092001
  %% [arXiv:1104.5068 [hep-ex]].
  %%CITATION = %%doi:10.1103/PhysRevLett.107.092001;%%
  %25 citations counted in INSPIRE as of 21 Aug 2018



%\cite{PhysRevLett.108.222002}
\bibitem{PhysRevLett.108.222002}
  M.~Ablikim {\it et al.} (BESIII Collaboration),
  %``Measurements of the mass and width of the $\eta_c$ using $\psi' -> \gamma \eta_c$,''
  Phys.\ Rev.\ Lett.\  {\bf 108}, 222002 (2012).
  %%doi:10.1103/PhysRevLett.108.222002
  %% [arXiv:1111.0398 [hep-ex]].
  %%CITATION = %%doi:10.1103/PhysRevLett.108.222002;%%
  %42 citations counted in INSPIRE as of 21 Aug 2018



%\cite{PhysRevLett.109.042003}
\bibitem{PhysRevLett.109.042003}
  M.~Ablikim {\it et al.} (BESIII Collaboration),
  %``First observation of the M1 transition $\psi(3686)\to \gamma\eta_c(2S)$,''
  Phys.\ Rev.\ Lett.\  {\bf 109}, 042003 (2012).
  %%doi:10.1103/PhysRevLett.109.042003
  %% [arXiv:1205.5103 [hep-ex]].
  %%CITATION = %%doi:10.1103/PhysRevLett.109.042003;%%
  %32 citations counted in INSPIRE as of 21 Aug 2018



%\cite{PhysRevLett.116.251802}
\bibitem{PhysRevLett.116.251802}
  M.~Ablikim {\it et al.} (BESIII Collaboration),
  %``Observation of $h_{c}$ radiative decay $h_{c} \rightarrow \gamma \eta'$ and evidence for $h_{c} \rightarrow \gamma \eta$,''
  Phys.\ Rev.\ Lett.\  {\bf 116},  251802 (2016).
  %%doi:10.1103/PhysRevLett.116.251802
  %% [arXiv:1603.04936 [hep-ex]].
  %%CITATION = %%doi:10.1103/PhysRevLett.116.251802;%%
  %9 citations counted in INSPIRE as of 21 Aug 2018

%%% added by Hai-Bo 
\bibitem{bes3C:pub-page} 
 http://english.ihep.cas.cn/chnl/245/index.html 

%\cite{Ablikim:2018ewr}
\bibitem{Ablikim:2018ewr} 
  M.~Ablikim {\it et al.} (BESIII Collaboration),
  %``First observations of $h_c \to$ hadrons,''
  Phys.\ Rev.\ D {\bf 99}, 072008 (2019).


%\cite{PhysRev.D87.012002}
\bibitem{PhysRev.D87.012002}
  M.~Ablikim {\it et al.} (BESIII Collaboration),
  %``Search for hadronic transition $��_{cJ} �� ��_c��^+��^-$ and observation of $��_{cJ} �� K\overline{K}�ЦЦ�$,''
  Phys.\ Rev.\ D {\bf 87},  012002 (2013).
  %%doi:10.1103/PhysRevD.87.012002
  %% [arXiv:1208.4805 [hep-ex]].
  %%CITATION = %%doi:10.1103/PhysRevD.87.012002;%%
  %57 citations counted in INSPIRE as of 21 Aug 2018



%\cite{hep-ph/0607278}
\bibitem{hep-ph/0607278}
  Y.~J.~Gao, Y.~J.~Zhang and K.~T.~Chao,
  %``Radiative decays of charmonium into light mesons,''
  Chin.\ Phys.\ Lett.\  {\bf 23}, 2376 (2006).
  %%doi:10.1088/0256-307X/23/9/008
  %% [hep-ph/0607278].
  %%CITATION = %%doi:10.1088/0256-307X/23/9/008;%%
  %18 citations counted in INSPIRE as of 21 Aug 2018



%\cite{1505.02665}
\bibitem{1505.02665}
  F.~Feng, Y.~Jia and W.~L.~Sang,
  %``Can Nonrelativistic QCD Explain the $\gamma\gamma^* \to \eta_c$  Transition Form Factor Data?,''
  Phys.\ Rev.\ Lett.\  {\bf 115},  222001 (2015).
  %%doi:10.1103/PhysRevLett.115.222001
  %% [arXiv:1505.02665 [hep-ph]].
  %%CITATION = %%doi:10.1103/PhysRevLett.115.222001;%%
  %13 citations counted in INSPIRE as of 21 Aug 2018



%\cite{1707.05758}
\bibitem{1707.05758}
  F.~Feng, Y.~Jia and W.~L.~Sang,
  %``Next-to-Next-to-Leading-Order QCD Corrections to the Hadronic width of Pseudoscalar Quarkonium,''
  Phys.\ Rev.\ Lett.\  {\bf 119},  252001 (2017).
  %%doi:10.1103/PhysRevLett.119.252001
  %% [arXiv:1707.05758 [hep-ph]].
  %%CITATION = %%doi:10.1103/PhysRevLett.119.252001;%%
  %5 citations counted in INSPIRE as of 21 Aug 2018



%\cite{PhysRevLett.105.261801}
\bibitem{PhysRevLett.105.261801}
  M.~Ablikim {\it et al.} (BESIII Collaboration),
  %``Evidence for psi' decays into gamma pi^0 and gamma eta,''
  Phys.\ Rev.\ Lett.\  {\bf 105}, 261801 (2010).
  %%doi:10.1103/PhysRevLett.105.261801
  %% [arXiv:1011.0885 [hep-ex]].
  %%CITATION = %%doi:10.1103/PhysRevLett.105.261801;%%
  %27 citations counted in INSPIRE as of 21 Aug 2018



%\cite{1505.03930}
\bibitem{1505.03930}
  K.~Zhu, X.~H.~Mo and C.~Z.~Yuan,
  %``Determination of the relative phase in $\psi'$ and $J/\psi$ decays into baryon and antibaryon,''
  Int.\ J.\ Mod.\ Phys.\ A {\bf 30},  1550148 (2015).
  %%doi:10.1142/S0217751X15501481
  %% [arXiv:1505.03930 [hep-ph]].
  %%CITATION = %%doi:10.1142/S0217751X15501481;%%
  %3 citations counted in INSPIRE as of 21 Aug 2018



%\cite{PhysRev.D88.112001}
\bibitem{PhysRev.D88.112001}
  M.~Ablikim {\it et al.} (BESIII Collaboration),
  %``Search for $��_c(2S)h_c �� p\overline{p}$ decays and measurements of the $��_{cJ} �� p\overline{p}$ branching fractions,''
  Phys.\ Rev.\ D {\bf 88},  112001 (2013).
  %%doi:10.1103/PhysRevD.88.112001
  %% [arXiv:1310.6099 [hep-ex]].
  %%CITATION = %%doi:10.1103/PhysRevD.88.112001;%%
  %4 citations counted in INSPIRE as of 21 Aug 2018


%\cite{PhysLett.A117.1}
\bibitem{PhysLett.A117.1} 
  N.~A.~Tornqvist,
  %``The Decay $J/\psi \to \Lambda \bar{\Lambda} \to \pi^- p \pi^+ \bar{p}$ as an {Einstein-Podolsky-Rosen} Experiment,''
  Phys.\ Lett.\ A {\bf 117}, 1 (1986).
  %doi:10.1016/0375-9601(86)90225-2
  %%CITATION = doi:10.1016/0375-9601(86)90225-2;%%
  %11 citations counted in INSPIRE as of 29 Sep 2019

  
%\cite{LKopke}
\bibitem{LKopke}
  L.~Kopke, talk presented at XXIIrd Intern Conf. on High Energy Physics, Berkeley, USA, 1986.

%\cite{PhysRev.D47.1744}
\bibitem{PhysRev.D47.1744} 
  X.~G.~He, J.~P.~Ma and B.~McKellar,
  %``CP violation in J / psi ---> Lambda anti-Lambda,''
  Phys.\ Rev.\ D {\bf 47}, 1744(R) (1993).
 % doi:10.1103/PhysRevD.47.R1744
  %[hep-ph/9211276].
  %%CITATION = doi:10.1103/PhysRevD.47.R1744;%%
  %12 citations counted in INSPIRE as of 29 Sep 2019

%\cite{PhysRev.D49.4548}
\bibitem{PhysRev.D49.4548} 
  X.~G.~He, J.~P.~Ma and B.~McKellar,
  %``CP violation in fermion pair decays of neutral boson particles,''
  Phys.\ Rev.\ D {\bf 49}, 4548 (1994)
  %doi:10.1103/PhysRevD.49.4548
  %[hep-ph/9310243].
  %%CITATION = doi:10.1103/PhysRevD.49.4548;%%
  %18 citations counted in INSPIRE as of 29 Sep 2019

  
%\cite{PhysRevLett.115.011803}
\bibitem{PhysRevLett.115.011803}
  M.~Ablikim {\it et al.} (BESIII Collaboration),
  %``Observation of the $\psi(1^3D_2)$ state in $e^+e^-\to\pi^+\pi^-\gamma\chi_{c1}$ at BESIII,''
  Phys.\ Rev.\ Lett.\  {\bf 115},  011803 (2015).
  %%doi:10.1103/PhysRevLett.115.011803
  %% [arXiv:1503.08203 [hep-ex]].
  %%CITATION = %%doi:10.1103/PhysRevLett.115.011803;%%
  %42 citations counted in INSPIRE as of 21 Aug 2018

%%%%%%%%%% End for from Zhu Kai  %%% Hai-Bo 


%\cite{Lebed:2016hpi}
\bibitem{reviews} 
  R.~F.~Lebed, R.~E.~Mitchell and E.~S.~Swanson,
  %``Heavy-Quark QCD Exotica,''
  Prog.\ Part.\ Nucl.\ Phys.\  {\bf 93}, 143 (2017).
  %doi:10.1016/j.ppnp.2016.11.003
  %[arXiv:1610.04528 [hep-ph]].
  %%CITATION = doi:10.1016/j.ppnp.2016.11.003;%%
  %47 citations counted in INSPIRE as of 08 Dec 2017
%\cite{Chen:2016qju}
\bibitem{Chen:2016qju} 
  H.~X.~Chen, W.~Chen, X.~Liu and S.~L.~Zhu,
  %``The hidden-charm pentaquark and tetraquark states,''
  Phys.\ Rept.\  {\bf 639}, 1 (2016).
  %doi:10.1016/j.physrep.2016.05.004
  %[arXiv:1601.02092 [hep-ph]].
  %%CITATION = doi:10.1016/j.physrep.2016.05.004;%%
  %278 citations counted in INSPIRE as of 06 Jul 2018
%\cite{Esposito:2016noz}
\bibitem{Esposito:2016noz} 
  A.~Esposito, A.~Pilloni and A.~D.~Polosa,
  %``Multiquark Resonances,''
  Phys.\ Rept.\  {\bf 668}, 1 (2016).
  %doi:10.1016/j.physrep.2016.11.002
  %[arXiv:1611.07920 [hep-ph]].
  %%CITATION = doi:10.1016/j.physrep.2016.11.002;%%
  %98 citations counted in INSPIRE as of 06 Jul 2018
%\cite{Guo:2017jvc}
\bibitem{Guo:2017jvc} 
  F.-K.~Guo, C.~Hanhart, U.-G.~Mei{\ss}ner, Q.~Wang, Q.~Zhao and B.-S.~Zou,
  %``Hadronic molecules,''
  Rev.\ Mod.\ Phys.\  {\bf 90}, 015004 (2018).
  %doi:10.1103/RevModPhys.90.015004
  %[arXiv:1705.00141 [hep-ph]].
  %%CITATION = doi:10.1103/RevModPhys.90.015004;%%
  %103 citations counted in INSPIRE as of 06 Jul 2018
%\cite{Ali:2017jda}
\bibitem{Ali:2017jda} 
  A.~Ali, J.~S.~Lange and S.~Stone,
  %``Exotics: Heavy Pentaquarks and Tetraquarks,''
  Prog.\ Part.\ Nucl.\ Phys.\  {\bf 97}, 123 (2017).
  %doi:10.1016/j.ppnp.2017.08.003
  %[arXiv:1706.00610 [hep-ph]].
  %%CITATION = doi:10.1016/j.ppnp.2017.08.003;%%
  %53 citations counted in INSPIRE as of 06 Jul 2018
%\cite{Olsen:2017bmm}
\bibitem{Olsen:2017bmm} 
  S.~L.~Olsen, T.~Skwarnicki and D.~Zieminska,
  %``Nonstandard heavy mesons and baryons: Experimental evidence,''
  Rev.\ Mod.\ Phys.\  {\bf 90}, 015003 (2018).
  %doi:10.1103/RevModPhys.90.015003
  %[arXiv:1708.04012 [hep-ph]].
  %%CITATION = doi:10.1103/RevModPhys.90.015003;%%
  %47 citations counted in INSPIRE as of 06 Jul 2018


%\cite{Choi:2003ue}
\bibitem{xdiscovery}
  S.~K.~Choi {\it et al.} (Belle Collaboration),
  %``Observation of a narrow charmonium - like state in exclusive B+- ---> K+- pi+ pi- J / psi decays,''
  Phys.\ Rev.\ Lett.\  {\bf 91}, 262001 (2003).
  % doi:10.1103/PhysRevLett.91.262001
  %% [hep-ex/0309032].
  %%CITATION = doi:10.1103/PhysRevLett.91.262001;%%
  %1514 citations counted in INSPIRE as of 14 Sep 2018

  

%\cite{Ablikim:2013dyn}
\bibitem{Ablikim:2013dyn}
  M.~Ablikim {\it et al.} (BESIII Collaboration),
  %``Observation of $e^+e^− \to \gamma X$(3872) at BESIII,''
  Phys.\ Rev.\ Lett.\  {\bf 112},  092001 (2014).
  %doi:10.1103/PhysRevLett.112.092001
  %% [arXiv:1310.4101 [hep-ex]].
  %%CITATION = doi:10.1103/PhysRevLett.112.092001;%%
  %67 citations counted in INSPIRE as of 26 Oct 2016


\bibitem{Guo:2013nza} 
  F.-K.~Guo, C.~Hanhart, U.-G.~Mei{\ss}ner, Q.~Wang and Q.~Zhao,
  %``Production of the X(3872) in charmonia radiative decays,''
  Phys.\ Lett.\ B {\bf 725}, 127 (2013).
  % doi:10.1016/j.physletb.2013.06.053
  %% [arXiv:1306.3096 [hep-ph]].


%\cite{Abe:2004zs}
\bibitem{xtoomegajpsi}
  K.~Abe {\it et al.} (Belle Collaboration),
  %``Observation of a near-threshold omega J/psi mass enhancement in exclusive B ---> K omega J/psi decays,''
  Phys.\ Rev.\ Lett.\  {\bf 94}, 182002 (2005).
  % doi:10.1103/PhysRevLett.94.182002
  %% [hep-ex/0408126].
  %%CITATION = doi:10.1103/PhysRevLett.94.182002;%%
  %454 citations counted in INSPIRE as of 14 Sep 2018

%\cite{Chilikin:2017evr}
\bibitem{x3860}
  K.~Chilikin {\it et al.} (Belle Collaboration),
  %``Observation of an alternative $\chi_{c0}(2P)$ candidate in $e^+ e^- \rightarrow J/\psi D \bar{D}$,''
  Phys.\ Rev.\ D {\bf 95}, 112003 (2017).
  % doi:10.1103/PhysRevD.95.112003
  %% [arXiv:1704.01872 [hep-ex]].
  %%CITATION = doi:10.1103/PhysRevD.95.112003;%%
  %19 citations counted in INSPIRE as of 14 Sep 2018


\bibitem{Guo:2012tv} 
  F.-K.~Guo and U.-G.~Mei{\ss}ner,
  %``Where is the \chi_{c0}(2P)?,''
  Phys.\ Rev.\ D {\bf 86}, 091501 (2012).
  % doi:10.1103/PhysRevD.86.091501
  %% [arXiv:1208.1134 [hep-ph]].

%\cite{Guo:2014ura}
\bibitem{Guo:2014ura} 
  F.~K.~Guo, U.~G.~Meißner and Z.~Yang,
  %``Production of the spin partner of the X(3872) in $e^+e^-$ collisions,''
  Phys.\ Lett.\ B {\bf 740}, 42 (2015)
  %doi:10.1016/j.physletb.2014.11.030
  %[arXiv:1410.4674 [hep-ph]].
  %%CITATION = doi:10.1016/j.physletb.2014.11.030;%%
  %8 citations counted in INSPIRE as of 30 Sep 2019

  
%\cite{Aubert:2005rm}
\bibitem{babary4260}
  B.~Aubert {\it et al.} (BaBar Collaboration),
  %``Observation of a broad structure in the $\pi^+ \pi^- J/\psi$ mass spectrum around 4.26-GeV/c$^2$,''
  Phys.\ Rev.\ Lett.\  {\bf 95}, 142001 (2005).
  % doi:10.1103/PhysRevLett.95.142001
  %% [hep-ex/0506081].
  %%CITATION = doi:10.1103/PhysRevLett.95.142001;%%
  %751 citations counted in INSPIRE as of 14 Sep 2018


%\cite{Liu:2012ze}
\bibitem{latticey4260}
  L.~Liu {\it et al.} (Hadron Spectrum Collaboration),
  %``Excited and exotic charmonium spectroscopy from lattice QCD,''
  JHEP {\bf 1207}, 126 (2012).
  % doi:10.1007/JHEP07(2012)126
  %% [arXiv:1204.5425 [hep-ph]].
  %%CITATION = doi:10.1007/JHEP07(2012)126;%%
  %191 citations counted in INSPIRE as of 14 Sep 2018


%\cite{Close:2005iz}
\bibitem{othery4260}
  F.~E.~Close and P.~R.~Page,
  %``Gluonic charmonium resonances at BaBar and BELLE?,''
  Phys.\ Lett.\ B {\bf 628}, 215 (2005).
  % doi:10.1016/j.physletb.2005.09.016
  %% [hep-ph/0507199].
  %%CITATION = doi:10.1016/j.physletb.2005.09.016;%%
  %235 citations counted in INSPIRE as of 14 Sep 2018
  
\bibitem{Wang:2013cya} 
  Q.~Wang, C.~Hanhart and Q.~Zhao,
  %``Decoding the riddle of $Y(4260)$ and $Z_c(3900)$,''
  Phys.\ Rev.\ Lett.\  {\bf 111}, 132003 (2013);
  % doi:10.1103/PhysRevLett.111.132003
  %% [arXiv:1303.6355 [hep-ph]];

\bibitem{Ablikim:2013mio:3}
  M.~Ablikim {\it et al.} (BESIII Collaboration),
  %``Observation of a Charged Charmoniumlike Structure in $e^+e^- \to \pi^+\pi^- J/\psi$ at $\sqrt{s}$ =4.26  GeV,''
  Phys.\ Rev.\ Lett.\  {\bf 110}, 252001 (2013).
  %doi:10.1103/PhysRevLett.110.252001
  %% [arXiv:1303.5949 [hep-ex]].
  %%CITATION = doi:10.1103/PhysRevLett.110.252001;%%
  %415 citations counted in INSPIRE as of 26 Oct 2016

\bibitem{Ablikim:2013wzq:3}
  M.~Ablikim {\it et al.} (BESIII Collaboration),
  %``Observation of a Charged Charmoniumlike Structure $Z_c$(4020) and Search for the $Z_c$(3900) in $e^+e^- \to \pi^+\pi^-h_c$,''
  Phys.\ Rev.\ Lett.\  {\bf 111},  242001 (2013).
  %doi:10.1103/PhysRevLett.111.242001
  %% [arXiv:1309.1896 [hep-ex]].
  %%CITATION = doi:10.1103/PhysRevLett.111.242001;%%
  %198 citations counted in INSPIRE as of 26 Oct 2016


%\cite{Choi:2007wga}
\bibitem{bellezc4430}
  S.~K.~Choi {\it et al.} (Belle Collaboration),
  %``Observation of a resonance-like structure in the pi+- psi-prime mass distribution in exclusive B ---> K pi+- psi-prime decays,''
  Phys.\ Rev.\ Lett.\  {\bf 100}, 142001 (2008).
  % doi:10.1103/PhysRevLett.100.142001
  %% [arXiv:0708.1790 [hep-ex]].
  %%CITATION = doi:10.1103/PhysRevLett.100.142001;%%
  %568 citations counted in INSPIRE as of 14 Sep 2018


%\cite{Ablikim:2013xfr}
\bibitem{Ablikim:2013xfr}
  M.~Ablikim {\it et al.} (BESIII Collaboration),
  %``Observation of a charged $(D\bar{D}^{*})^\pm$ mass peak in $e^{+}e^{-} \to \pi D\bar{D}^{*}$ at $\sqrt{s} =$ 4.26 GeV,''
  Phys.\ Rev.\ Lett.\  {\bf 112},  022001 (2014).
  %doi:10.1103/PhysRevLett.112.022001
  %% [arXiv:1310.1163 [hep-ex]].
  %%CITATION = doi:10.1103/PhysRevLett.112.022001;%%
  %152 citations counted in INSPIRE as of 26 Oct 2016

%\cite{Ablikim:2013emm}
\bibitem{Ablikim:2013emm:3}
  M.~Ablikim {\it et al.} (BESIII Collaboration),
  %``Observation of a charged charmoniumlike structure in $e^+e^- \to (D^{*} \bar{D}^{*})^{\pm} \pi^\mp$ at $\sqrt{s}=4.26$GeV,''
  Phys.\ Rev.\ Lett.\  {\bf 112},  132001 (2014).
  %doi:10.1103/PhysRevLett.112.132001
  %% [arXiv:1308.2760 [hep-ex]].
  %%CITATION = doi:10.1103/PhysRevLett.112.132001;%%
  %189 citations counted in INSPIRE as of 26 Oct 2016


%\cite{Ablikim:2016qzw}
\bibitem{pipijpsi} 
  M.~Ablikim {\it et al.} (BESIII Collaboration),
  %``Precise measurement of the $e^+e^-\to \pi^+\pi^-J/\psi$ cross section at center-of-mass energies from 3.77 to 4.60 GeV,''
  Phys.\ Rev.\ Lett.\  {\bf 118}, 092001 (2017).
  %doi:10.1103/PhysRevLett.118.092001
  %% [arXiv:1611.01317 [hep-ex]].
  %%CITATION = doi:10.1103/PhysRevLett.118.092001;%%
  %33 citations counted in INSPIRE as of 09 Dec 2017


%\cite{Ablikim:2015tbp}
\bibitem{Ablikim:2015tbp}
  M.~Ablikim {\it et al.} (BESIII Collaboration),
  %``Observation of $Z_c(3900)^{0}$ in $e^+e^-\to\pi^0\pi^0 J/\psi$,''
  Phys.\ Rev.\ Lett.\  {\bf 115},  112003 (2015).
  %doi:10.1103/PhysRevLett.115.112003
  %% [arXiv:1506.06018 [hep-ex]].
  %%CITATION = doi:10.1103/PhysRevLett.115.112003;%%
  %34 citations counted in INSPIRE as of 26 Oct 2016
%\cite{BESIII:2016adj}

\bibitem{BESIII:2016adj} 
  M.~Ablikim {\it et al.} (BESIII Collaboration),
  %``Evidence of Two Resonant Structures in $e^+ e^- \to \pi^+ \pi^- h_c$,''
  Phys.\ Rev.\ Lett.\  {\bf 118},  092002 (2017).
  %doi:10.1103/PhysRevLett.118.092002
  %% [arXiv:1610.07044 [hep-ex]].
  %%CITATION = doi:10.1103/PhysRevLett.118.092002;%%
  %57 citations counted in INSPIRE as of 31 Aug 2018



%\cite{Ablikim:2014qwy}
\bibitem{Ablikim:2014qwy}
  M.~Ablikim {\it et al.} (BESIII Collaboration),
  %``Study of $e^+e^-\to\omega\chi_{cJ}$ at center-of-mass energies from 4.21 to 4.42 GeV,''
  Phys.\ Rev.\ Lett.\  {\bf 114},  092003 (2015).
  %doi:10.1103/PhysRevLett.114.092003
  %% [arXiv:1410.6538 [hep-ex]].
  %%CITATION = doi:10.1103/PhysRevLett.114.092003;%%
  %31 citations counted in INSPIRE as of 26 Oct 2016

%\cite{Ablikim:2007gd}
\bibitem{Ablikim:2007gd} 
  M.~Ablikim {\it et al.} (BES Collaboration),
  %``Determination of the psi(3770), psi(4040), psi(4160) and psi(4415) resonance parameters,''
  eConf C {\bf 070805}, 02 (2007) [Phys.\ Lett.\ B {\bf 660}, 315 (2008)].
  % doi:10.1016/j.physletb.2007.11.100
  %% [arXiv:0705.4500 [hep-ex]].
  %%CITATION = doi:10.1016/j.physletb.2007.11.100;%%
  %104 citations counted in INSPIRE as of 16 Sep 2018
  
%\cite{Ablikim:2017oaf}
\bibitem{Ablikim:2017oaf} 
  M.~Ablikim {\it et al.} (BESIII Collaboration),
  %``Measurement of $e^{+}e^{-}\rightarrow \pi^{+}\pi^{-}\psi(3686)$ from 4.008 to 4.600~GeV and observation of a charged structure in the $\pi^{\pm}\psi(3686)$ mass spectrum,''
  Phys.\ Rev.\ D {\bf 96},  032004 (2017).
  % doi:10.1103/PhysRevD.96.032004
  %% [arXiv:1703.08787 [hep-ex]].
  %%CITATION = doi:10.1103/PhysRevD.96.032004;%%
  %20 citations counted in INSPIRE as of 16 Sep 2018


%\cite{Ablikim:2018vxx}
\bibitem{Ablikim:2018vxx} 
  M.~Ablikim {\it et al.} (BESIII Collaboration),
  %``Evidence of a resonant structure in the $e^+e^-\to \pi^+D^0D^{*-}$ cross section between 4.05 and 4.60 GeV,''
  Phys.\ Rev.\ Lett.\  {\bf 122}, 102002 (2019)
  %doi:10.1103/PhysRevLett.122.102002
  %[arXiv:1808.02847 [hep-ex]].
  %%CITATION = doi:10.1103/PhysRevLett.122.102002;%%
  %21 citations counted in INSPIRE as of 09 Oct 2019

\bibitem{Ablikim:2019zio} 
  M.~Ablikim {\it et al.} (BESIII Collaboration),
  %``Study of $e^+e^- \to \gamma \omega J/\psi$ and Observation of $X(3872) \to \omega J/\psi$,''
  Phys.\ Rev.\ Lett.\  {\bf 122}, 232002 (2019).


\bibitem{Ablikim:2019apl} 
  M.~Ablikim {\it et al.} [BESIII Collaboration],
  %``Cross section measurements of $e^+ e^-\to\omega\chi_{c0}$ form $\sqrt{s}=$ 4.178 to 4.278 GeV,''
  Phys.\ Rev.\ D {\bf 99}, 091103 (2019).



\bibitem{Cleven:2013mka} 
  M.~Cleven, Q.~Wang, F.~K.~Guo, C.~Hanhart, U.~G.~Mei{\ss}ner and Q.~Zhao,
  %``$Y(4260)$ as the first $S$-wave open charm vector molecular state?,''
  Phys.\ Rev.\ D {\bf 90}, 074039 (2014).

%\cite{Collaboration:2017njt}
\bibitem{Collaboration:2017njt} 
  M.~Ablikim {\it et al.} (BESIII Collaboration),
  %``Determination of the Spin and Parity of the $Z_c(3900)$,''
  Phys.\ Rev.\ Lett.\  {\bf 119},  072001 (2017).
  % doi:10.1103/PhysRevLett.119.072001
  %% [arXiv:1706.04100 [hep-ex]].
  %%CITATION = doi:10.1103/PhysRevLett.119.072001;%%
  %17 citations counted in INSPIRE as of 16 Sep 2018


%\cite{Swanson:2015bsa}
\bibitem{Swanson:2015bsa} 
  E.~S.~Swanson,
  %``Cusps and Exotic Charmonia,''
  Int.\ J.\ Mod.\ Phys.\ E {\bf 25}, 1642010 (2016).
  % doi:10.1142/S0218301316420106
  %% [arXiv:1504.07952 [hep-ph]].
  %%CITATION = doi:10.1142/S0218301316420106;%%
  %28 citations counted in INSPIRE as of 16 Sep 2018


\bibitem{belle2}
 T.~Abe {\it et al.} (Belle II Collaboration),
%%  ``Belle II Technical Design Report,''
 arXiv:1011.0352 [physics.ins-det];
 E.~Kou {\it et al.} (Belle II Collaboration),
 %``The Belle II Physics book,''
  PTEP {\bf 2019}, 123C01 (2019).


\bibitem{PBFB}   A.~J.~Bevan {\it et al.}
 (BaBar and Belle Collaborations),
 %``The Physics of the $B$ Factories,''
 Eur.\ Phys.\ J.\ C {\bf 74}, 3026 (2014).

%\cite{Aaij:2015eva}
\bibitem{lhcbx}
  R.~Aaij {\it et al.} (LHCb Collaboration),
  %``Quantum numbers of the $X(3872)$ state and orbital angular momentum in its $\rho^0 J\psi$ decay,''
  Phys.\ Rev.\ D {\bf 92},  011102 (2015).
  % doi:10.1103/PhysRevD.92.011102
  %% [arXiv:1504.06339 [hep-ex]].
  %%CITATION = doi:10.1103/PhysRevD.92.011102;%%
  %39 citations counted in INSPIRE as of 14 Sep 2018


%\cite{Abazov:2018cyu}
\bibitem{Abazov:2018cyu} 
  V.~M.~Abazov {\it et al.} (D0 Collaboration),
  %``Evidence for $Z_c^{\pm}(3900)$ in semi-inclusive decays of $b$-flavored hadrons,''
  Phys.\ Rev.\ D {\bf 98}, 052010 (2018).
  %doi:10.1103/PhysRevD.98.052010
  %[arXiv:1807.00183 [hep-ex]].
  %%CITATION = doi:10.1103/PhysRevD.98.052010;%%
  %14 citations counted in INSPIRE as of 09 Oct 2019

 %\cite{D0:2019zpb}
\bibitem{D0:2019zpb} 
  V.~M.~Abazov {\it et al.} (D0 Collaboration),
  %``Properties of $Z_c^{\pm}(3900)$ produced in $p \bar p$ collision,''
  Phys.\ Rev.\ D {\bf 100}, 012005 (2019).
  %doi:10.1103/PhysRevD.100.012005
  %[arXiv:1905.13704 [hep-ex]].
  %%CITATION = doi:10.1103/PhysRevD.100.012005;%%
  %1 citations counted in INSPIRE as of 09 Oct 2019


\end{thebibliography}

\begin{thebibliography}{9}
\addcontentsline{toc}{chapter}{Bibliography}


\bibitem{Asner:2008nq} 
  D.~M.~Asner {\it et al.},
  %``Physics at BES-III,''
  Int.\ J.\ Mod.\ Phys.\ A {\bf 24}, S1 (2009).
  
\bibitem{CEPCStudyGroup:2018ghi} 
  J.~B.~Guimar\~{a}es da Costa {\it et al.} (CEPC Study Group),
  %``CEPC Conceptual Design Report: Volume 2 - Physics & Detector,''
  arXiv:1811.10545 [hep-ex].

\bibitem{Abada:2019zxq} 
  A.~Abada {\it et al.} (FCC Collaboration),
  %``FCC-ee: The Lepton Collider : Future Circular Collider Conceptual Design Report Volume 2,''
  Eur.\ Phys.\ J.\ ST {\bf 228},  261 (2019).

\bibitem{Jegerlehner:2017gek}
  F.~Jegerlehner,
  %``The Anomalous Magnetic Moment of the Muon,''
  Springer Tracts Mod.\ Phys.\  {\bf 274}, 1 (2017).
  %Springer Tracts Mod.\ Phys.\  {\bf 274} (2017) pp.1.
  %doi:10.1007/978-3-319-63577-4
  %%CITATION = doi:10.1007/978-3-319-63577-4;%%
  %12 citations counted in INSPIRE as of 05 Oct 2018

\bibitem{ref_teubner18}
  A.~Keshavarzi, D.~Nomura and T.~Teubner,
  %``Muon $g-2$ and $\alpha(M_Z^2)$: a new data-based analysis,''
  Phys.\ Rev.\ D {\bf 97}, 114025 (2018).

\bibitem{ref_e821}
G.~W.~Bennet {\it et al.} (Muon $(g-2)$ Collaboration),
Phys. Rev. D {\bf 73}, 072003 (2006).

%\cite{Davier:2017zfy}
\bibitem{Davier:2017zfy}
  M.~Davier, A.~Hoecker, B.~Malaescu and Z.~Zhang,
  %``Reevaluation of the hadronic vacuum polarisation contributions to the Standard Model predictions of the muon $g-2$ and ${\alpha (m_Z^2)}$ using newest hadronic cross-section data,''
  Eur.\ Phys.\ J.\ C {\bf 77}, 827 (2017).


\bibitem{ref_e989a}
  B.~L.~Roberts, Chin. Phys. C {\bf 34}, 741 (2010).


\bibitem{fnal}
J. Grange {\it et al.}, arXiv:1501.06858.

\bibitem{jparc}
Tsutomu Mibe, Chin. Phys. C {\bf 34}, 745 (2010).

\bibitem{ref_stockinger}
D.~W.~Hertzog {\it et al.}, 
%''The physics case for the New Muon $(g-2)$ Experiment'',
arXiv:0705.4617.

\bibitem{M1_Prades:2009tw}
  J.~Prades, E.~de Rafael and A.~Vainshtein, Adv.\ Ser.\ Direct.\ High Energy Phys.\  {\bf 20}, 303 (2009).
  %``Hadronic Light-by-Light Scattering Contribution to the Muon Anomalous Magnetic Moment,''

\bibitem{M1_Colangelo:2014dfa}
  G.~Colangelo, M.~Hoferichter, M.~Procura and P.~Stoffer, JHEP {\bf 1409}, 091 (2014).
  %``Dispersive approach to hadronic light-by-light scattering,''

\bibitem{M1_Colangelo:2014pva}
  G.~Colangelo, M.~Hoferichter, B.~Kubis, M.~Procura and P.~Stoffer, Phys.\ Lett.\ B {\bf 738}, 6 (2014).
  %``Towards a data-driven analysis of hadronic light-by-light scattering,''

\bibitem{Colangelo:2015ama} 
  G.~Colangelo, M.~Hoferichter, M.~Procura and P.~Stoffer, JHEP {\bf 1509}, 074 (2015).
  %``Dispersion relation for hadronic light-by-light scattering: theoretical foundations,''

\bibitem{Colangelo:2017fiz} 
  G.~Colangelo, M.~Hoferichter, M.~Procura and P.~Stoffer, JHEP {\bf 1704}, 161 (2017).
  %``Dispersion relation for hadronic light-by-light scattering: two-pion contributions,''

\bibitem{Colangelo:2017qdm} 
  G.~Colangelo, M.~Hoferichter, M.~Procura and P.~Stoffer, Phys.\ Rev.\ Lett.\  {\bf 118}, 232001 (2017).
  %``Rescattering effects in the hadronic-light-by-light contribution to the anomalous magnetic moment of the muon,''

\bibitem{Pauk:2014rfa} 
  V.~Pauk and M.~Vanderhaeghen, Phys.\ Rev.\ D {\bf 90}, 113012 (2014).
  %``Anomalous magnetic moment of the muon in a dispersive approach,''

\bibitem{Pauk:2014jza}
  V.~Pauk and M.~Vanderhaeghen, arXiv:1403.7503.
  %``Two-loop massive scalar three-point function in a dispersive approach,''

\bibitem{Nyffeler:2016gnb} 
  A.~Nyffeler,
  %``Precision of a data-driven estimate of hadronic light-by-light scattering in the muon $g-2$: Pseudoscalar-pole contribution,''
  Phys.\ Rev.\ D {\bf 94}, 053006 (2016)
  
\bibitem{ref_whitepaper}
  T.~Blum, A.~Denig, I.~Logashenko, E.~de Rafael, B.~Lee Roberts, T.~Teubner and G.~Venanzoni,
  %``The Muon (g-2) Theory Value: Present and Future,''
  arXiv:1311.2198.

\bibitem{Colangelo:2018mtw} 
  G.~Colangelo, M.~Hoferichter and P.~Stoffer, JHEP {\bf 1902}, 6 (2019).
  %``Two-pion contribution to hadronic vacuum polarization,''
  
\bibitem{g-2theoryinitiative}
  Second Plenary Workshop of the $g-2$ Theory Initiative, 
  https://wwwth.kph.uni-mainz.de/g-2/ .

\bibitem{pdg2018}
M. Tanabashi {\it et al.} (Particle Data Group), Phys. Rev. D {\bf 98}, 030001 (2018).

%
%\bibitem{ref_teubner}
%  K.~Hagiwara, R.~Liao, A.~D.~Martin, D.~Nomura and T.~Teubner, J. Phys. G {\bf 38}, 085003 (2011).

\bibitem{ref_binner}
 S.~Binner, J.~H.~K\"uhn, K.~Melnikov, Phys. Lett. B {\bf 459}, 279 (1999).
 
 %\cite{Benayoun:1999hm}
\bibitem{Benayoun:1999hm}
  M.~Benayoun, S.~I.~Eidelman, V.~N.~Ivanchenko and Z.~K.~Silagadze,
  %``Spectroscopy at B factories using hard photon emission,''
  Mod.\ Phys.\ Lett.\ A {\bf 14}, 2605 (1999).
  
\bibitem{Druzhinin:2011qd}
  V.~P.~Druzhinin, S.~I.~Eidelman, S.~I.~Serednyakov and E.~P.~Solodov,
  %``Hadron Production via e+e- Collisions with Initial State Radiation,''
  Rev.\ Mod.\ Phys.\  {\bf 83}, 1545 (2011). 
%
%\bibitem{Rodrigo:2001jr}
%G.~Rodrigo {\it et al.}, Eur.\ Phys.\ J. C {\bf 22}, 81 (2001).
%
%\bibitem{Kuhn:2002xg}
%J.~H.~K\"uhn and G.~Rodrigo, Eur. Phys. J. C {\bf 25}, 215 (2002).
%``The radiative return at small angles: Virtual corrections,''
%
\bibitem{Rodrigo:2001kf}
G.~Rodrigo {\it et al.}, Eur.\ Phys.\ J. C {\bf 24}, 71 (2002).

\bibitem{Czyz:2008kw}
H.~Czy\.z, J.~H.~K\"uhn and A.~Wapienik,
Phys.\ Rev. D {\bf 77}, 114005 (2008).

\bibitem{Czyz:2009vj}
  H.~Czy\.z and J.~H.~K\"uhn,
  Phys.\ Rev. D {\bf 80}, 034035 (2009).


\bibitem{Ablikim:2015orh}
  M.~Ablikim {\it et al.} (BESIII Collaboration),
  %``Measurement of the $e^+ e^? \to \pi^+ \pi^?$ cross section between 600 and 900 MeV using initial state radiation,''
  Phys.\ Lett.\ B {\bf 753}, 629 (2016).
%


\bibitem{TMVA2007}
        A.~Hoecker, P.~Speckmayer, J.~Stelzer,
        J.~Therhaag, E.~von Toerne, and H.~Voss,
%        ``TMVA: Toolkit for Multivariate Data Analysis,''
        PoS A CAT 040 (2007).
 [physics/0703039].

\bibitem{ref_czyz_muons}
    F.~Campanario, H.~Czy\.z, J.~Gluza, M.~Gunia, T.~Riemann, G.~Rodrigo and V.~Yundin,
  %``Complete QED NLO contributions to the reaction $e^+e^- \to \mu^+\mu^-\gamma$ and their implementation in the event generator PHOKHARA,''
  JHEP {\bf 1402}, 114 (2014).
%  [arXiv:1312.3610 [hep-ph]].

\bibitem{ref_kloe8a}
  F.~Ambrosino {\it et al.}  (KLOE Collaboration),
  %``Measurement of sigma(e + e- ---> pi+ pi- gamma(gamma) and the dipion contribution to the muon anomaly with the KLOE detector,''
  Phys.\ Lett.\ B {\bf 670}, 285 (2009).
%  [arXiv:0809.3950 [hep-ex]].

\bibitem{ref_kloe11a}
F.~Ambrosino {\it et al.} (KLOE collaboration),
Phys. Lett. B {\bf 700}, 102 (2011).
%
\bibitem{ref_kloe12a}
F.~Ambrosino {\it et al.} (KLOE collaboration),
Phys. Lett. B {\bf 720}, 336 (2013).

\bibitem{ref_babar2pia}
B.~Aubert {\it et al.} (BaBar collaboration),
Phys. Rev. Lett. {\bf 103}, 231801 (2009).

\bibitem{Ablikim:2016xbg}
  M.~Ablikim {\it et al.} (BESIII Collaboration),
  %``Measurement of the leptonic decay width of $J/\psi$ using initial state radiation,''
  Phys.\ Lett.\ B {\bf 761}, 98 (2016).

\bibitem{ref_bes3dark}
M.~Ablikim {\it et al.} (BESIII Collaboration), Phys.\ Lett.\ B {\bf 774}, 252 (2017).

\bibitem{ref_cmd23pi}
  R.~R.~Akhmetshin {\it et al.}  (CMD-2 Collaboration),
  %``Measurement of omega meson parameters in pi+ pi- pi0 decay mode with CMD-2,''
  Phys.\ Lett.\ B {\bf 476}, 33 (2000).
%  [hep-ex/0002017].

\bibitem{ref_snd3pi}
  M.~N.~Achasov, K.~I.~Beloborodov, A.~V.~Berdyugin, A.~G.~Bogdanchikov, A.~V.~Bozhenok, A.~D.~Bukin, D.~A.~Bukin and T.~V.~Dimova {\it et al.},
  %``Study of the process e+ e- ---> pi+ pi- pi0 in the energy region s**(1/2) below 0.98-GeV,''
  Phys.\ Rev.\ D {\bf 68}, 052006 (2003).
%  [hep-ex/0305049].

\bibitem{ref_babar3pi}
B.~Aubert {\it et al.} (BaBar collaboration),
Phys. Rev. D {\bf 70}, 072004 (2004).

\bibitem{ref_dm23pi}
A.~Antonelli {\it et al.} (DM2 collaboration),
Z. Phys. C {\bf 56}, 15 (1992).

\bibitem{ref_cmd24pi}
  R.~R.~Akhmetshin {\it et al.}  (CMD-2 Collaboration),
  %``a(1)(1260) pi dominance in the process e+ e- ---> 4 pi at energies 1.05-GeV - 1.38-GeV,''
  Phys.\ Lett.\ B {\bf 466}, 392 (1999).
%  [hep-ex/9904024].

\bibitem{ref_snd4pi}
  M.~N.~Achasov {\it et al.}  (SND Collaboration),
  %``e+ e- --> 4pi processes investigation in the energy range 0.98-GeV to 1.38-GeV with SND detector,''
  BUDKER-INP-2001-34.

\bibitem{babar_4pi}
J.~P.~Lees {\it et al.} (BaBar Collaboration),
  %``Measurement of the ${e}^{+}{e}^{{-}}{\rightarrow}{{\pi}}^{+}{{\pi}}^{{-}}{{\pi}}^{0}{{\pi}}^{0}$ cross section using initial-state radiation at BaBar,''
Phys.\ Rev.\ D {\bf 96}, 092009 (2017).


\bibitem{rbes2}
J. Z. Bai {\it et al.} (BES Collaboration), Phys. Rev. Lett. {\bf 84}, 594 (2000); 
J. Z. Bai {\it et al.} (BES Collaboration), Phys. Rev. Lett. {\bf 88}, 101802 (2004); 
M. Ablikim {\it et al.} (BES Collaboration), Phys. Lett. B {\bf 641}, 145 (2006); 
M. Ablikim {\it et al.} (BES Collaboration), Phys. Lett. B {\bf 660}, 315 (2008); 
M. Ablikim {\it et al.} (BES Collaboration), Phys. Lett. B {\bf 677}, 239 (2009).

\bibitem{rkedr}
V. V. Anashin {\it et al.} (KEDR Collaboration), Phys. Lett. B {\bf 753}, 533 (2016);
V. V. Anashin {\it et al.} (KEDR Collaboration), Phys. Lett. B {\bf 770}, 174 (2017).  
  
\bibitem{Anashin:2018vdo}
  V.~V.~Anashin {\it et al.} (KEDR Collaboration),
  %``New precise measurement of $R_{\text{uds}}$ and $R$ between 3.08 and 3.72 GeV at the KEDR detector,''
 % arXiv:1805.06235.
    Phys.\ Lett.\ B {\bf 788}, 42 (2019).


\bibitem{Andersson:1999ui}
  B.~Andersson and H.~M.~Hu, arXiv:hep-ph/9910285. 

\bibitem{conexc}
  Ronggang Ping {\it et al.} Chin. Phys. C {\bf40}, 113002 (2016).



% meson transition form factors
\bibitem{M1_Jegerlehner:2013sja}
  F.~Jegerlehner,
  %``Application of Chiral Resonance Lagrangian Theories to the Muon $g-2$,''
  Acta Phys.\ Polon.\ B {\bf 44}, 2257 (2013).
 % [arXiv:1312.3978 [hep-ph]].
  %%CITATION = ARXIV:1312.3978;%%

\bibitem{Aubert:2006cy}
  B.~Aubert {\it et al.}  (BaBar Collaboration),
  %``Measurement of the eta and eta-prime transition form-factors at q**2 = 112-GeV**2,''
  Phys.\ Rev.\ D {\bf 74}, 012002  (2006).


\bibitem{BaBar:2018zpn} 
  J.~P.~Lees {\it et al.} (BaBar Collaboration),
  %``Measurement of the $\gamma^{\star}\gamma^{\star} \to \eta'$ transition form factor,''
  Phys.\ Rev.\ D {\bf 98}, 112002 (2018).

\bibitem{M1_Feldmann}
  T.~Feldmann, P.~Kroll, B.~Stech,
  %``Mixing and decay constants of pseudoscalar mesons,''
  Phys.\ Rev.\   D {\bf58}, 114006 (1998);
    T.~Feldmann,
  %``Quark structure of pseudoscalar mesons,''
  Int.\ J.\ Mod.\ Phys.\  A {\bf 15}, 159 (2000).

\bibitem{M1_Donoghue:1986wv}
  J.~F.~Donoghue, B.~R.~Holstein, Y.~C.~R.~Lin,
  %``Chiral Loops in pi0, eta0 ---> gamma gamma and eta eta-prime Mixing,''
  Phys.\ Rev.\ Lett.\  {\bf 55}, 2766 (1985).

\bibitem{M1_Leutwyler}
  H.~Leutwyler,
  %``On the 1/N expansion in chiral perturbation theory,''
  Nucl.\ Phys.\ Proc.\ Suppl.\  {\bf 64}, 223 (1998);
  R.~Kaiser and H.~Leutwyler,
  %``Large N(c) in chiral perturbation theory,''
  Eur.\ Phys.\ J.\  C {\bf 17}, 623 (2000).

\bibitem{M1_Behrend:1990sr}
  H.~J.~Behrend {\it et al.}  (CELLO Collaboration),
  %``A Measurement of the pi0, eta and eta-prime electromagnetic form-factors,''
  Z.\ Phys.\ C {\bf 49}, 401 (1991).

\bibitem{M1_Gronberg:1997fj}
  J.~Gronberg {\it et al.}  (CLEO Collaboration),
  %``Measurements of the meson - photon transition form-factors of light pseudoscalar mesons 
  % at large momentum transfer,''
  Phys.\ Rev.\ D {\bf 57} , 33 (1998). 
  
\bibitem{Aubert:2009mc}
  B.~Aubert {\it et al.}  (BaBar Collaboration),
  %``Measurement of the gamma gamma* ---> pi0 transition form factor,''
  Phys.\ Rev.\ D {\bf 80}, 052002  (2009).
  
\bibitem{Uehara:2012ag}
  S.~Uehara {\it et al.}  (Belle Collaboration),
  %``Measurement of $\gamma \gamma^* \to \pi^0$ transition form factor at Belle,''
  Phys.\ Rev.\ D {\bf 86}, 092007  (2012). 
  
\bibitem{Hoferichter:2018dmo}
  M.~Hoferichter, B.~L.~Hoid, B.~Kubis, S.~Leupold and S.~P.~Schneider, Phys.\ Rev.\ Lett.\  {\bf 121}, 112002 (2018);
  M.~Hoferichter, B.~L.~Hoid, B.~Kubis, S.~Leupold and S.~P.~Schneider, JHEP {\bf 1810}, 141 (2018).

\bibitem{Gerardin:2019vio}
  A.~G\'{e}rardin, H.~B.~Meyer and A.~Nyffeler, Phys.\ Rev.\ D {\bf 100},  034520 (2019).
  %``Lattice calculation of the pion transition form factor with $N_f=2+1$ Wilson quarks,''
  
\bibitem{Danilkin:2019mhd} 
  I.~Danilkin, C.~F.~Redmer and M.~Vanderhaeghen,
  %``The hadronic light-by-light contribution to the muon's anomalous magnetic moment,''
  Prog.\ Part.\ Nucl.\ Phys.\  {\bf 107}, 20 (2019).

\bibitem{Czyz:2018jpp}
  H.~Czyz and P.~Kisza, Comput.\ Phys.\ Commun.\ {\bf 234}, 245 (2019).
  %EKHARA 3.0: an update of the EKHARA Monte Carlo event generator

\bibitem{Hoferichter:2019nlq} 
  M.~Hoferichter and P.~Stoffer, JHEP {\bf 1907}, 073 (2019).
  %``Dispersion relations for $\gamma^*\gamma^*\to\pi\pi$: helicity amplitudes, subtractions, and anomalous thresholds,''

\bibitem{Danilkin:2019opj} 
  I.~Danilkin, O.~Deineka and M.~Vanderhaeghen, arXiv:1909.04158 [hep-ph].
  %``Dispersive analysis of the $\gamma^{*}\gamma^{*} \to \pi \pi$ process,''
  
\bibitem{Masuda:2015yoh}
 M.~Masuda {\it el al.} (Belle Collaboration), Phys.\ Rev.\ D {\bf 93}, 032003 (2016). 
 %Study of $\pi^0$ pair production in single-tag two-photon collisions


%\bibitem{M1_Benayoun:2014tra}
%  M.~Benayoun, J.~Bijnens, T.~Blum, I.~Caprini, G.~Colangelo, H.~Czy\.z, A.~Denig and C.~A.~Dominguez {\it et al.},
%  %``Hadronic contributions to the muon anomalous magnetic moment Workshop. $(g-2)_{\mu}$: Quo vadis? Workshop. Mini proceedings,''
%  arXiv:1407.4021.
%
%
%\bibitem{M1_Jegerlehner:2009ry}
%  F.~Jegerlehner and A.~Nyffeler,
%  %``The Muon g-2,''
%  Phys.\ Rept.\  {\bf 477}, 1 (2009);
%  %[arXiv:0902.3360 [hep-ph]].
%  %%CITATION = ARXIV:0902.3360;%%
%  F.~Jegerlehner,
%% {\it The anomalous magnetic moment of the muon,}
%  Springer Tracts Mod.\ Phys.\  {\bf 226}, 1 (2008).
%  %%CITATION = STPHB,226,1;%%
%
%
%
%\bibitem{M1_BPP}
%J.~Bijnens, E.~Pallante, J.~Prades,
%%``Hadronic light by light contributions to the muon g-2 in the large N(c)
%%limit,''
%Phys.\ Rev.\ Lett.\  {\bf {75}}, 1447  (1995)
%[Erratum-ibid.\  {\bf {75}}, 3781 (1995)];
%%J.~Bijnens, E.~Pallante and J.~Prades,
%%``Analysis of the Hadronic Light-by-Light Contributions to the Muon $g-2$,''
%Nucl.\ Phys.\ B {\bf {474}}, 379 (1996)
%%J.~Bijnens, E.~Pallante and J.~Prades,
%%``Comment on the pion pole part of the light-by-light contribution to the
%%muon g-2,''
%[Erratum-ibid.\ {\bf {626}}, 410 (2002)].
%
%\bibitem{M1_HKS}
%M.~Hayakawa, T.~Kinoshita, A.~I.~Sanda,
%%``Hadronic light by light scattering effect on muon g-2,''
%Phys.\ Rev.\ Lett.\  {\bf 75}, 790 (1995);
%%M.~Hayakawa, T.~Kinoshita and A.~I.~Sanda,
%%``Hadronic Light-by-light Scattering Contribution to Muon $ g - 2 $,''
%Phys.\ Rev.\ D {\bf {54}}, 3137  (1996).
%
%\bibitem{M1_HK}
%M.~Hayakawa, T.~Kinoshita,
%%``Pseudoscalar pole terms in the hadronic light-by-light scattering
%%contribution to muon g-2,''
%Phys.\ Rev.\ D {\bf 57}, 465 (1998)
%[Erratum-ibid.\ D {\bf {66}}, 019902 (2002)].
%%.~Hayakawa and T.~Kinoshita,
%%``Comment on the sign of the pseudoscalar pole contribution to the muon
%%g-2,''
%
%\bibitem{M1_MV}
%K.~Melnikov, A.~Vainshtein,
%%``Hadronic light-by-light scattering contribution to the muon anomalous
%%magnetic moment revisited,''
%Phys.\ Rev.\ D {\bf {70}}, 113006 (2004).
%
%
%

%%%% Baryon form factor %%%%
\bibitem{isgur}
  N.~Isgur, Proceedings of NSTAR 2000, JLAB-THY-00-20 (2000).
  
\bibitem{ashman}
  J. Ashman {\it et al.}, Phys. Lett. B. {\bf 206}, 206 (1988).

\bibitem{aidala}
  C.~A.~Aidala {\it et al.,} Rev. Mod. Phys. {\bf 85}, 655 (2013).

\bibitem{Miller}
  G.~A.~Miller, Phys. Rev. Lett. {\bf 99}, 112001 (2007).

\bibitem{ChPT1}
  S.~Weinberg, Nucl. Phys.  B {\bf 363}, 3 (1991).

\bibitem{ChPT2}
  R.~Machleidt and D.~R.~Entem, Phys. Rep. {\bf 503}, 1 (2011).

\bibitem{LQCD}
  K.~G.~Wilson, Phys. Rev. D {\bf 10}, 2445 (1974).

\bibitem{Skyrme}
  T. Skyrme, Proc. Roy. Soc. Lond. A {\bf 260}, 237 (1961).

\bibitem{Witten}
  G.~S.~Adkins, C.~R.~Nappi and E.~Witten, Nucl. Phys. B {\bf 228}, 552 (1982).

\bibitem{Punjabi}
  V.~Punjabi, C.~F. Perdrisat and M.~K. Jones, Eur. Phys. J. A {\bf 51}, 79 (2015).

\bibitem{granados}
  C.~Granados,~S.~Leupold and E.~Perotti, Eur. Phys. J. A {\bf 53}, 17 (2017).

\bibitem{dispersion}
  M.~A.~Belushkin {\it et al.,} Phys. Rev. {\bf C 75}, 035202 (2007).

\bibitem{analyticity}
  E.~C.~Titschmarsh, {\it The Theory of Functions}, Oxford University Press (1939).

\bibitem{Sakarov}
  A.~D.~Sakharov, Sov. Phys. Usp {\bf 34}, 375 (1991).

\bibitem{Sommerfeld}
  A.~Sommerfeld,  Ann. Phys. {\bf 403}, 257 (1931).

\bibitem{faldt}  
  G.~F\"{a}ldt, Eur.\ Phys.\ J.\ A {\bf 52}, 141 (2016).

\bibitem{faldtkupsc} 
  G.~F\"{a}ldt and A.~Kupsc, Phys.\ Lett.\ B {\bf 772}, 16 (2017).

\bibitem{Gakh}
  G.~I.~Gakh and E.~Tomasi-Gustafsson, Nucl.\ Phys.\ A {\bf 761}, 120 (2005).
  
\bibitem{egle_asy}
  E.~Tomasi-Gustafsson {\it et al.,} Phys. Lett. B {\bf 659}, 197 (2008).

\bibitem{polarimeter}
  A.~I.~Akhiezer and M.~P.~Rekalo, Sov. Phys. Dokl. {\bf 13}, 572 (1968);
  A.~Akhiezer, M.~P.~Rekalo, Sov.\ J.\ Part.\ Nucl. {\bf 4}, 277 (1974).

\bibitem{jlab}
  A.~J.~R.~Puckett {\it et al.} Phys. Rev. C {\bf 85}, 045203 (2012).

\bibitem{Rosenbluth}
  M.~Rosenbluth, Phys. Rev. {\bf 79}, 615 (1950).

\bibitem{guichon}
  P.~A.~M.~Guichon and M.~Vanderhaeghen, Phys. Rev. Lett. {\bf91} 142303 (2003).

\bibitem{melnitchouk}
  P.~G.~Blunden, W.~Melnitchouk and J.~A.~Tjon, Phys. Rev. Lett. {\bf 91} 142304 (2003).

\bibitem{nucleondis}
  S.~Pacetti, R.~Baldini Ferroli and E.~Tomasi-Gustafsson, Phys. Rept. {\bf 550}, 1 (2015), and references therein.

\bibitem{babarppbar}
  J.~P.~Lees {\it et al.} (BaBar Collaboration), Phys. Rev. D {\bf 87}, 092005 (2007).

\bibitem{ps170ppbar}
  G.~Bardin {\it et al.} (PS170 Collaboration) Nucl. Phys. B {\bf 411}, 3 (1994).

\bibitem{bes3ppbar2012}
  M.~Ablikim {\it et al.}(BESIII Collaboration), Phys. Rev. D {\bf 91}, 112004 (2015).
  

\bibitem{bes3ppbarisr}
  M.~Ablikim {\it et al.} (BESIII Collaboration), Phys. Rev. D {\bf 99}, 092002 (2019).


%protonfig
\bibitem{CMD3ppbar}
  R.~R.~Akhmetshin {\it et al.} (CMD-2 Collaboration), Phys. Lett. B {\bf 759}, 634 (2016).

\bibitem{BESppbar}
  M.~Ablikim {\it et al.} (BES  Collaboration), Phys. Lett. B {\bf 630}, 14 (2005).

\bibitem{feniceppbar}
  A.~Antonelli {\it et al.} (FENICE Collaboration), Nucl. Phys. B {\bf 517}, 3 (1998).

\bibitem{E760ppbar}
  T.~A.~Armstrong {\it et al.} (E760 Collaboration), Phys. Rev. Lett. {\bf 70}, 1212 (1993).

\bibitem{E835ppbar}
  M.~Ambrogiani {\it et al.} (E835 Collaboration), Phys. Rev. D {\bf 60}, 032002 (1999); 
  M.~Andreotti {\it et al.} (E835 Collaboration), Phys. Lett. B {\bf 559}, 20 (2003).

\bibitem{DM2ppbar}
  D.~Bisello {\it et al.} (DM2 Collaboration), Nucl. Phys. B {\bf 224}, 379 (1983);
  Z. Phys. C {\bf 48}, 23 (1990).

\bibitem{Egle}
  A.~Bianconi and E.~Tomasi-Gustafsson, Phys. Rev. Lett. {\bf 114}, 232301 (2015).

\bibitem{chiral}
  J.~Haidenbauer {\it et al.},  Nucl. Phys. A {\bf 929}, 102 (2014).

\bibitem{lattice}
  J.~R.~Green {\it et al.} Phys. Rev. D {\bf 90}, 074507 (2014); 
  B.~J\"ager {\it et al.}, PoS LATTICE {\bf 2013}, 272 (2014)

\bibitem{VMD}
  R.~Bijker and F.~Iachello, Phys. Rev. C {\bf 69}, 068201 (2004).

\bibitem{CQM}
  T.~Melde {\it et al.}, Phys. Rev. D {\bf 76}, 074020 (2007).

\bibitem{pQCD}
  S.~J.~Brodsky and G.~R.~Farrar, Phy. Rev. D {\bf 11}, 1309 (1975).
\bibitem{snd-nn}
  M. N. Achasov {\it et al.}, Phy. Rev. D {\bf 90}, 112007 (2014).

%\bibitem{fenice-nn}
%  A.~Antonelli {\it et al.} (FENICE Collaboration), Nucl. Phys. B {\bf 517}, 3 (1998).

\bibitem{elliskarliner}
  J.~R.~Ellis and M.~Karliner, New J. Phys. {\bf 4}, 18 (2002).

\bibitem{karlinernuss}
  M.~Karliner and S.~Nussinov, Phys. Lett. B {\bf 538}, 321-326 (2002).

\bibitem{hyperondm2}
  D.~Bisello {\it et al.} (DM2 Collaboration), Z. Phys. C {\bf 48}, 23 (1990).

\bibitem{babarllbar}
  B.~Aubert {\it et al.} (BaBar Collaboration), Phys. Rev. D {\bf 76}, 092006 (2007).

\bibitem{bes3hyp2012}
  M.~Ablikim {\it et al}.(BESIII Collaboration), Phys. Rev. D {\bf 97}, 032013 (2018).
  
\bibitem{cleocllbar}
  S.~Dobbs {\it et al}. Phys. Lett. B {\bf 739}, 90 (2014); {\it ibid.} Phys. Rev. D {\bf 96}, 092004 (2017).

\bibitem{anselmino}
  M.~Anselmino {\it et al.}, Rev. Mod. Phys. {\bf 65}, 1199 (1993).

\bibitem{wilczekjaffe}
  R.~Jaffe  and  F.~Wilczek,  Phys. Rev. Lett. {\bf 91}, 232003 (2003).

\bibitem{bes3hyp2015}
  M.~Ablikim {\it et al.} (BESIII Collaboration) Phys. Rev. Lett. {\bf 123}, 122003 (2019).


\bibitem{haidenbauerllbar} 
  J.~Haidenbauer and  U.-G.~Mei{\ss}ner, Phys. Lett. B {\bf 761}, 456 (2016). 


\bibitem{bellelc}
  G.~Pakhlova {\it et al.} (Belle Collaboration), Phys. Rev. Lett. {\bf 101} 172001 (2008).

\bibitem{bes3lc}
  M.~Ablikim {\it et al.} (BESIII Collaboration), Phys. Rev. Lett. {\bf 120}, 132001 (2018).

\bibitem{CMD2019}
  R.~R.~Akhmetshin {\it et al.} (CMD-3 Collaboration), Phys. Lett. B {\bf 794}, 64-68 (2019).
  
  
\bibitem{ref:cmd-3-ref}
  V.~F.~Dmitriev, A.~I.~Milstein and S.~G.~Salnikov, Phys. Rev. D {\bf 93}, 034033 (2016);
  A.~I.~Milstein, S.~G.~Salnikov, Nucl. Phys. A {\bf 966}, 54 (2017); 
  A.~I.~Milstein and S.~G.~Salnikov, Nucl. Phys. A {\bf 977}, 60 (2018).

%
%\bibitem{Meissner1}
%  J.~Haidenbauer, X.~W.~Kanga, U.-G.~Mei{\ss}ner, Nucl. Phys. A {\bf 929}, 102 (2014).

\bibitem{dai}
  L.-Y.~Dai {\it et al.,} Phys. Rev. {\bf D 96} (2017) 116001.

\bibitem{theoryfsi}
  O.~D.~Dalkarov, P.~A.~Khakhulin, and A.~Yu.~Voronin, Nucl. Phys. A {\bf 833}, 104 (2010).

\bibitem{theoryres}
  B.~El-Bennich {\it et al.}, Phys. Rev. C {\bf 79}, 054001 (2009);
  J.~Haidenbauer {\it et al.}, Phys. Lett. B {\bf 643}, 29 (2006).

\bibitem{theorycoulomb}
  R.~Baldini {\it et al.}, Eur. Phys. J. A {\bf 39}, 315 (2009);
  R.~Baldini {\it et al.}, Eur. Phys. J. A {\bf 48}, 33 (2012).

\bibitem{ref_diquark}
M. Anselmino {\it et al.}, Rev. Mod. Phys. {\bf 65}, 1199 (1993), and
references therein.


\bibitem{ref:Pacetti}
 S. Pacetti, talk presented at the Int. Workshop on Baryon Production in BESIII, Hefei, China, 2019.

\bibitem{bes3ppbarscan}
  M.~Ablikim {\it et al.} (BESIII Collaboration), 
  %``Measurement of proton electromagnetic form factors in $e^+e^- \to p\bar{p}$ in the energy region 2.00 - 3.08 GeV,''
  Phys.\ Rev.\ Lett.\  {\bf 124}, 042001 (2020).

%%%% fragmentation function %%%%

\bibitem{albino}
S. Albino, Rev. Mod. Phys. {\bf 82}, 2489 (2010), and references therein.

\bibitem{radici}
M. Radici, Nuovo Cim. C {\bf 035}, 69 (2012).

\bibitem{data}
F. Arleo, Eur. Phys. J. C {\bf 61}, 603 (2009).

\bibitem{DSS}
D. De Florian, R. Sassot and M. Stratmann, Phys. Rev. D {\bf 75}, 114010 (2007); D {\bf 76} 074033 (2007).

\bibitem{HKNS}
M. Hirai, S. Kumano, T. H. Nagai and K. Sudoh, Phys. Rev. D {\bf 75}, 094009 (2007).

\bibitem{AKK}
S. Albino, B. A. Kniehl and G. Kramer, Nucl. Phys. B {\bf 803}, 42 (2008).

\bibitem{kaonff}
E. Leader {\it et al.}, talk presented at 20th International Symposium on Spin Physics (SPIN2012) JINR, Dubna, Russia, 2012.

\bibitem{strangepuzzle}
E. Leader {\it et al.}, Phys. Rev. D {\bf 91}, 054017 (2015);
Phys. Rev. D {\bf 84}, 014002 (2011).

\bibitem{jlab12}
Jozef Dudek {\it et al.}, Eur. Phys. J. A {\bf 48}, 187 (2012).

\bibitem{eic}
A. Accardi {\it et al.}, Eur. Phys. J. A{\bf 62}, 218 (2016).

\bibitem{FFreview}
Isabella Garzia and Francesca Giordano, Eur. Phys. J. A {\bf 52}, 152 (2016).

\bibitem{collins}
J. C. Collins, Nucl. Phys. B {\bf 396}, 161 (1993).

\bibitem{peng}
P. Sun and F. Yuan, Phys. Rev. D {\bf 88}, 034016 (2013).

\bibitem{collinbes3}
M. Ablikim {\it et al.} (BESIII Collaboration), Phys. Rev. Lett. {\bf 116}, 042001 (2016).

%%%% Tau mass %%%%

\bibitem{Perl:1975bf}
M.~L.~Perl {\it et al.}, Phys.\ Rev.\ Lett.\  {\bf 35}, 1489 (1975).

\bibitem{PDG2012} 
J. Beringer {\it et al.} (Particle Data Group), Phys. Rev. D {\bf 86}, 010001 (2012).

\bibitem{LepUnivCor}
W.J. Marciano and A. Sirlin, Phys. Rev. Lett. {\bf 61}, 1815 (1988).

\bibitem{mntau}
R. Barate {\it et al.}, Eur. Phys. J. C {\bf 2}, 395 (1998).

\bibitem{Koide}
Y.~Koide, Phys. Rev. D {\bf 28}, 252 (1983); Mod. Phys. Lett. A {\bf 5}, 2319 (1990).

\bibitem{Perl:1977se}
M.~L.~Perl {\it et al.}, Phys.\ Lett.\ B {\bf 70}, 487 (1977).

\bibitem{Brandelik:1977xz}
R.~Brandelik {\it et al.} (DASP Collaboration), Phys.\ Lett.\ B {\bf 73}, 109 (1978).

\bibitem{Bartel:1978ii}
W.~Bartel {\it et al.}, Phys.\ Lett.\ B {\bf 77}, 331 (1978).

\bibitem{taudelco}
W.~Bacino {\it et al.}, Phys. Rev. Lett. {\bf41}, 13 (1978).

\bibitem{Blocker1982}
C. A. BLOCKER {\it et al.}, Phys.\ Lett.\ B {\bf 109}, 119 (1978).

\bibitem{tauargus}
H.~Albrecht {\it et al.} (ARGUS Collaboration), Phys. Lett. B {\bf 292}, 221 (1992).

\bibitem{taubes1}
J. Z.~Bai {\it et al.} (BES Collaboration), Phys. Rev. Lett. {\bf 69}, 3021 (1992).

\bibitem{taubes2}
J. Z.~Bai {\it et al.} (BES Collaboration), Phys. Rev. D {\bf 53}, 20 (1996).

\bibitem{taubes3}
J. Z.~Bai {\it et al.} (BES Collaboration), HEP\&NP {\bf 16}, 343 (1992).

\bibitem{taucleo1}
R.~Balest {\it et al.} (CLEO Collaboration), Phys. Rev. D {\bf 47}, 3671
(1993).

\bibitem{taucleo2}
A.~Anastassov {\it et al.} (CLEO Collaboration), Phys. Rev. D {\bf 55}, 2559 (1997);  Erratum: [Phys. Rev. D
{\bf 58}, 119904 (1998)].

\bibitem{tauopal}
G.~Abbiendi {\it et al.} (OPAL Collaboration), Phys. Lett. B {\bf 492}, 23 (2000).

\bibitem{taubelle}
K.~Belous {\it et al.} (Belle Collaboration), Phys. Rev. Lett. {\bf 99},
011801 (2007).

\bibitem{taukedr}
V. V.~Anashin {\it et al.} (KEDR Collaboration), J. Exp. The. Phys. Lett. {\bf 85}, 347 (2007).

\bibitem{taubabar}
B.~Aubert {\it et al.} (BaBar Collaboration), Phys. Rev. D {\bf 80}, 092005 (2009).

\bibitem{Ablikim:2014uzh}
M.~Ablikim {\it et al.} (BESIII Collaboration), Phys.\ Rev.\ D {\bf 90}, 012001
(2014).

\bibitem{PDG2000}
D. E. Groom {\it et al.} (Particle Data Group), Eur. Phys. J. C, {\bf 15}, 1 (2000).

\bibitem{PDG2014}
K. A. Olive {\it et al.} (Particle Data Group), Chin. Phys. C, {\bf 38}, 090001 (2014).

\bibitem{PDG1996}
R. M. Barnett {\it et al.} (Particle Data Group), Phys. Rev. D, {\bf 54}, 1 (1996).

\bibitem{PDG1994}
L. Montanet {\it et al.} (Particle Data Group), Phys. Rev. D, {\bf 50}, 1173 (1994).

\bibitem{Bogomyagkov}
A.~Bogomyagkov {\it et al.}, the 9th European Particle Accelerator Conference (EPAC 2004),
Lucerne, Switzerland, 2004.

\bibitem{principle}
Xiao-Hu Mo {\it et al.}, Chin. Phys. C {\bf 32}, 995 (2008).

\bibitem{bems2009}
M. N. Achasov {\it et al.}, Nucl. Phys. B (Proc. Suppl.) {\bf 189}, 366 (2009).

\bibitem{bems2010}
X. H. Mo, E.V. Abakumova, M.N. Achasov et al., Chin. Phys. C {\bf 34}, 912 (2010).

\bibitem{bems}
E.V. Abakumova {\it et al.}, Nucl. Instrum. Methods Phys. Res., Sect. A {\bf 659}, 21 (2011).

\bibitem{bems2}
J. Y. Zhang {\it et al.}, Nucl. Phys. B (Proc. Suppl.) {\bf 225}, 309
(2012). 

\bibitem{Mo:2015gla}
X.~H.~Mo, Int.\ J.\ Mod.\ Phys.\ A {\bf 30}, 1550149 (2015).

\bibitem{wangyk2007}
Y. K.~Wang, X.H.~Mo, C.Z.~Yuan, J.P.~Liu, Nucl. Instrum. Methods Phys. Res., Sect. A {\bf 583}, 
479 (2007).

\bibitem{wangyk2009}
Y. K.~Wang, J.Y.~Zhang, X.H.~Mo, C.Z.~Yuan, Chin. Phys. C {\bf 33}, 501 (2009).

\bibitem{wangbq2012}
B. Q.~Wang, X.H.~Mo, C.Z.~Yuan, Int. J. Mod. Phys. A {\bf 27}, 1250150 (2012).

\bibitem{wangbq2013}
B. Q.~Wang, X.H.~Mo, Chin. Phys. C {\bf 37}, 026202 (2013). 

\bibitem{moxh2016mpla}
X. H. Mo, Mod. Phys. Lett. A {\bf 31}, 1650196 (2016).

\bibitem{cleo2nd} 
J. Bartelt {\it et al.} (CLEO Collaboration), Phys.\ Rev.\ Lett.\  {\bf 76},
(1996) 4119.

\bibitem{babar2nd} 
P. del Amo Sanchez {\it et al.} (BaBar Collaboration), Phys. Rev. D {\bf 83},
032002 (2011).

\bibitem{babar3k} 
B. Aubert {\it et al.} (BaBar Collaboration), Phys.\ Rev.\ Lett.\  {\bf 100}, 011801 (2008).

\bibitem{bell3k} 
M. J. Lee {\it et al.} (Belle Collaboration), Phys. Rev. D {\bf 81},
113007 (2010).

\bibitem{zhang}
J. Y. Zhang {\it et al.}, Chin. Phys. C {\bf 40}, 076001 (2016).

%%%% relative phase %%%%

%\cite{Ablikim:2018ege}
\bibitem{Ablikim:2018ege} 
  M.~Ablikim {\it et al.} (BESIII Collaboration),
  %``Measurement of the phase between strong and electromagnetic amplitudes of $J/\psi$ decays,''
  Phys. Lett. B {\bf 791}, 375 (2019).
  %%CITATION = ARXIV:1808.02166;%%

%\cite{Seth:2013eaa}
\bibitem{Seth:2013eaa} 
  K.~K.~Seth, S.~Dobbs, A.~Tomaradze, T.~Xiao and G.~Bonvicini,
  %``First Measurement of the Electromagnetic Form Factor of the Neutral Kaon at a Large Momentum Transfer and the Effect of $SU(3)$ Breaking,''
  Phys.\ Lett.\ B {\bf 730}, 332 (2014).
%  doi:10.1016/j.physletb.2014.02.003
%  [arXiv:1307.6587 [hep-ex]].
  %%CITATION = doi:10.1016/j.physletb.2014.02.003;%%
  %9 citations counted in INSPIRE as of 29 Oct 2018
%\cite{Seth:2012nn}
\bibitem{Seth:2012nn} 
  K.~K.~Seth, S.~Dobbs, Z.~Metreveli, A.~Tomaradze, T.~Xiao and G.~Bonvicini,
  %``Electromagnetic Structure of the Proton, Pion, and Kaon by High-Precision Form Factor Measurements at Large Timelike Momentum Transfers,''
  Phys.\ Rev.\ Lett.\  {\bf 110}, 022002 (2013). 
%  doi:10.1103/PhysRevLett.110.022002
%  [arXiv:1210.1596 [hep-ex]].
  %%CITATION = doi:10.1103/PhysRevLett.110.022002;%%
  %40 citations counted in INSPIRE as of 29 Oct 2018


\bibitem{Ferroli:2016jri} 
  R.~B.~Ferroli, A.~Mangoni and S.~Pacetti,
  %``$G$-parity violating amplitudes in the $J/\psi \to \pi^+ \pi^-$ decay,''
  %arXiv:1611.04437.
Phys. Rev. C {\bf 98}, 045210 (2018).

\bibitem{2015zhu}
K. Zhu, X. H. Mo, C. Z. Yuan, Int. J. Mod. Phys. A {\bf 30}, 1550148 (2015).

%%%% phi(2170) %%%%

\bibitem{zhu2008}
Shi-Lin Zhu, Int. J. of Mod. Phys. E {\bf17}, 283 (2008), and references therein.

\bibitem{chen2016}
Hua-Xing Chen {\it et al.}, Phys. Rept. {\bf 639}, 1 (2016), and references therein.

\bibitem{olsen2018}
Stephen Lars Olsen {\it et al.}, Rev.\ Mod.\ Phys.\  {\bf 90}, 015003 (2018), and references therein.

\bibitem{strange}
T. Barnes, {\it et al.}, Phys. Rev. D {\bf 68}, 054014 (2003).


\bibitem{2170babar}
J. P. Lees {\it et al.} (BaBar Collaboration), Phys. Rev. D {\bf 86}, 012008 (2012); 
B. Aubert {\it et al.} (BaBar Collaboration), Phys. Rev. D {\bf 76}, 012008 (2007).

\bibitem{2170belle}
C. P. Shen {\it et al.} (Belle Collaboration), Phys. Rev. D {\bf 80}, 031101(R) (2009).

\bibitem{2017bes}
M. Ablikim {\it et al.} (BES Collaboration), Phys. Rev. Lett. {\bf 100}, 012003 (2008); 
M. Ablikim {\it et al.} (BESIII Collaboration), Phys. Rev. D {\bf 95}, 052017 (2015).

\bibitem{2017bes3}
M. Ablikim {\it et al.} (BESIII Collaboration), 
%arXiv:1709.04323[hep-ex].
Phys. Rev. D {\bf 99}, 012014 (2019).

\bibitem{2017ding}
G.~J.~Ding and M.~L.~Yan, Phys. Lett. B {\bf 657}, 49 (2007).

\bibitem{2017wang}
X.~Wang {\it et al.}, Phys. Rev. D {\bf 85}, 074024 (2012).

\bibitem{2017afonin}
S.~S.~Afonin and I.~V.~Pusenkov, Phys.\ Rev.\ D {\bf 90}, 094020 (2014).

\bibitem{2019pang}
Cheng-Qun Pang, Phys.\ Rev.\ D {\bf 99}, 074015 (2019).

\bibitem{2017ding2}
G.~J.~Ding and M.~L.~Yan, Phys. Lett. B {\bf 650}, 390 (2007).

\bibitem{2017wang2}
Z.~G.~Wang, Nucl. Phys. A {\bf 791}, 106 (2007).

\bibitem{2017chen}
H.~X.~Chen {\it et al.}, Phys. Rev. D {\bf 78}, 034012 (2008).

\bibitem{2017drenska}
N.~V.~Drenska, R.~Faccini and A.~D.~Polosa, Phys. Lett. B {\bf 669}, 160 (2008).

\bibitem{2019ke}
Hong-Wei Ke and Xue-Qian Li, Phys. Rev. D {\bf 99}, 036014 (2019).

\bibitem{2017zhao}
L.~Zhao {\it et al.}, Phys. Rev. D {\bf 87}, 054034 (2013).

\bibitem{2017deng}
C.~Deng {\it et al.}, Phys. Rev. D {\bf 88}, 074007 (2013).

\bibitem{2017dong}
Yubing Dong {\it et al.}, Phys. Rev. D {\bf 96}, 074027 (2017).

\bibitem{2017oset}
A.~Martinez Torres {\it et al.}, Phys. Rev. D {\bf 78}, 074031 (2008); 
S.~Gomez-Avila, M.~Napsuciale and E.~Oset, Phys. Rev. D {\bf 79}, 034018 (2009).

\bibitem{2010bes}
M. Ablikim {\it et al.} (BES Collaboration), Phys. Lett. B {\bf 685}, 27 (2010).

\bibitem{1999page}
Philip R. Page {\it et al.}, Phys. Rev. D {\bf 59}, 034016 (1999).

\bibitem{phieta2007}
B. Aubert {\it et al.} (BaBar Collaboration), Phys. Rev. D {\bf 77}, 092002 (2007).

\bibitem{phietaprime}
B. Aubert {\it et al.} (BaBar Collaboration), Phys. Rev. D {\bf 76}, 092005 (2007).

\bibitem{cpcshen}
C. P. Shen and C. Z. Yuan, Chin. Phys. C {\bf 34}, 1045 (2010).

\bibitem{2018chen}
H. X. Chen {\it et al.}, Phys. Rev. D {\bf 98}, 014011 (2018).

%\bibitem{fourkaon}
%A. Martinez Torres {\it et al.} Phys. Rev. D {\bf 78}, 074031 (2008).

\bibitem{milena2017}
M. Piotrowska {\it et al.} Phys. Rev. D {\bf 96}, 054033 (2017).

\bibitem{Lichard2018}
Peter Lichard, Phys. Rev. D {\bf 98}, 113011 (2018).

\end{thebibliography}

\begin{thebibliography}{9}
\addcontentsline{toc}{chapter}{Bibliography}

%%%%%%%%%%%%%%%%%%%%%%%%%%%%%%%%%%%%%%%%%%%%%%%%%%%%%%%%%%%%
%% Reference for charm section %%
%%%%%%%%%%%%%%%%%%%%%%%%%%%%%%%%%%%%%%%%%%%%%%%%%%%%%%%%%%%%

\bibitem{Asner:2008nq}
 D.~M.~Asner {\it et al.}, Int. J. Mod. Phys. A {\bf 24}, S1 (2009).

\bibitem{pdg2016}
M. Tanabashi {\it et al.} (Particle Data Group), Phys. Rev. D {\bf 98}, 030001 (2018).

\bibitem{Aaij:2018dso}
R.~Aaij {\it et al.} (LHCb Collaboration), Phys. Rev. Lett. {\bf 121}, 092003 (2018).


\bibitem{bes3_muv}
M. Ablikim {\it et al.} (BESIII Collaboration), Phys. Rev. D {\bf 89}, 051104 (2014).

\bibitem{Ablikim:2019rpl} 
  M.~Ablikim {\it et al.} (BESIII Collaboration),
  %``Observation of the leptonic decay $D^+ \to \tau^+ \nu_\tau$,''
  Phys.\ Rev.\ Lett.\  {\bf 123}, 211802 (2019).

\bibitem{bes3_Ds_muv}
M. Ablikim {\it et al.} (BESIII Collaboration), Phys. Rev. Lett. {\bf 122},
071802 (2019).

\bibitem{cleo_dsmuv}
J. P. Alexander {\it et al.} (CLEO Collaboration), Phys. Rev. D {\bf 79}, 052001 (2009).

\bibitem{cleo_dstauv2}
P. U. E. Onyisi {\it et al.} (CLEO Collaboration), Phys. Rev. D {\bf 79}, 052002 (2009).

\bibitem{cleo_dstauv3}
P. Naik {\it et al.} (CLEO Collaboration), Phys. Rev. D {\bf 80}, 112004 (2009).

\bibitem{babar_lv}
P. del Amo Sanchez {\it et al.} (BaBar Collaboration), Phys. Rev. D {\bf 82}, 091103 (2010).

\bibitem{belle_lv}
A. Zupanc {\it et al.} (Belle Collaboration), JHEP {\bf 1309}, 139 (2013).

\bibitem{prd98_074512}
A. Bazavov {\it et al.} (Fermilab Lattice and MILC Collaboration), Phys. Rev. D {\bf 98}, 074512 (2018).

\bibitem{prd91_054507}
N. Carrasco {\it et al.} (ETM Collaboration), Phys. Rev. D {\bf 91}, 054507, (2015).

\bibitem{babar_prl109_101802}
J. P. Lees {\it et al.} (BaBar Collaboration), Phys. Rev. Lett. {\bf 109}, 101802 (2012).

\bibitem{babar_prd88_072012}
J. P. Lees {\it et al.} (BaBar Collaboration), Phys. Rev. D {\bf 88}, 072012 (2013).

\bibitem{belle_prl99_191807}
A. Matyja {\it et al.} (Belle Collaboration), Phys. Rev. Lett. {\bf 99}, 191807 (2007).

\bibitem{belle_arxiv}
I. Adachi {\it et al.} (Belle Collaboration), arXiv:0910.4301 [hep-ex].

\bibitem{belle_prd82_072005}
A. Bozek {\it et al.} (Belle Collaboration), Phys. Rev. D {\bf 82}, 072005 (2010).

\bibitem{lhcb_prl115_111803}
R. Aaij {\it et al.} (LHCb Collaboration), Phys. Rev. Lett. {\bf 115}, 111803 (2015).

\bibitem{lhcb_jhep02_104}
R. Aaij {\it et al.} (LHCb Collaboration), JHEP {\bf 1602}, 104 (2016).

\bibitem{lhcb_jhep09_179}
R. Aaij {\it et al.} (LHCb Collaboration), JHEP {\bf 1509}, 179 (2015).

\bibitem{lhcb_prl113_151601}
R. Aaij {\it et al.} (LHCb Collaboration), Phys. Rev. Lett. {\bf 113}, 151601 (2014).

\bibitem{belle_prl113_111801}
S. Wehle {\it et al.} (Belle Collaboration), Phys. Rev. Lett. {\bf 113}, 111801 (2017).

\bibitem{prd91_094009}
S. Fajfer, I. Nisandzic, and U. Rojec, Phys. Rev. D {\bf 91}, 094009 (2015).

%\cite{Ivanov:2019nqd}
\bibitem{Ivanov:2019nqd} 
  M.~A.~Ivanov, J.~G.~Körner, J.~N.~Pandya, P.~Santorelli, N.~R.~Soni and C.~T.~Tran,
  %``Exclusive semileptonic decays of D and D$_{s}$ mesons in the covariant confining quark model,''
  Front.\ Phys.\ (Beijing) {\bf 14}, 64401 (2019).

\bibitem{bes3_D0_kpiev}
M. Ablikim {\it et al.} (BESIII Collaboration), Phys. Rev. D {\bf 92}, 072012 (2015).

\bibitem{bes3_Dp_k0pi0ev}
M. Ablikim {\it et al.} (BESIII Collaboration), Phys. Rev. D {\bf 96}, 012002 (2016).

\bibitem{bes3_kmuv}
M. Ablikim {\it et al.} (BESIII Collaboration), Phys. Rev. Lett. {\bf 122},
011804 (2019).

\bibitem{bes3_etaev}
M. Ablikim {\it et al.} (BESIII Collaboration), Phys. Rev. D {\bf 97}, 092009 (2018).

\bibitem{bes3_Dp_kpiev}
M. Ablikim {\it et al.} (BESIII Collaboration), Phys. Rev. D {\bf 94}, 032001 (2016).

\bibitem{bes3_pipiev}
M. Ablikim {\it et al.} (BESIII Collaboration), Phys. Rev. Lett. {\bf 122},
062001 (2019).

\bibitem{bes3_Dp_omegaev}
M. Ablikim {\it et al.} (BESIII Collaboration), Phys. Rev. D {\bf 92}, 071101 (2015).

\bibitem{panic2017_lilei}
M. Ablikim {\it et al.} (BESIII Collaboration), Phys. Rev. Lett. {\bf 122},
061801 (2019).

\bibitem{ichep2018_cjc}
M. Ablikim {\it et al.} (BESIII Collaboration), Phys. Rev. Lett. {\bf 122}, 
121801 (2019).

%Jiangchuan Chen (For BESIII Collaboration), Proceeding of ICHEP2018, arXiv:1812.00406;
%Youhua Yang (For BESIII Collaboration), Proceeding of CKM2018, arXiv:1812.00320.

\bibitem{cleo_k1ev}
M. Artuso {\it et al.} (Belle Collaboration), Phys. Rev. Lett. {\bf 99}, 191801 (2007).

\bibitem{charm2016_dzl}
M. Ablikim {\it et al.} (BESIII Collaboration), Phys. Rev. Lett. {\bf 121}, 081802 (2018).

\bibitem{lqcd_fk}
H. Na {\it et al.} (HPQCD Collaboration), Phys. Rev. D {\bf 82}, 114506 (2010).

\bibitem{lqcd_fpi}
H. Na {\it et al.} (HPQCD Collaboration), Phys. Rev. D {\bf 84}, 114505 (2011).

\bibitem{talk_beauty2014}
Aida X. EI-Khadra, talk presented at Beauty2014.

\bibitem{epjc78_501}
L. Riggio, G. Salerno, and S. Simula, Eur. Phys. J. C {\bf 78} 501 (2018).

\bibitem{bes3_pimuv}
M. Ablikim {\it et al.} (BESIII Collaboration), Phys. Rev. Lett. {\bf 121}, 171803 (2018).

\bibitem{epjc77_587}
H. Y. Cheng and X. W. Kang, Eur. Phys. J. C {\bf 77}, 587 (2017) 
[Erratum: Eur. Phys. J. C {\bf 77}, 863 (2017)].

\bibitem{babar_kev}
B. Aubert {\it et al.} (BaBar Collaboration), Phys. Rev. D {\bf 76}, 052005 (2007).

\bibitem{babar_piev}
J. P. Lees {\it et al.} (BaBar Collaboration), Phys. Rev. D {\bf 91}, 052022 (2015).

\bibitem{belle_kpiev}
L. Widhalm {\it et al.} (Belle Collaboration), Phys. Rev. Lett. {\bf 97}, 061804 (2006).

\bibitem{Xing:1996pn}
Z.~Z.~Xing, Phys. Rev. D {\bf 55}, 196 (1997).

\bibitem{Amhis:2016xyh} Y.~Amhis {\it et al.} (HFLAV Collaboration), Eur. Phys. J. C {\bf 77}, 895 (2017).

\bibitem{lhcbgamma}
LHCb Collaboration, Technical Report,  
%''Update of the LHCb combination of the CKM angle $\gamma$'', 
LHCb-CONF-2018-002 (2018).

\bibitem{BROD}
J. Brod and J. Zupan, JHEP {\bf 1401}, 051  (2014).

\bibitem{GLW}
M.~Gronau and D.~London,  Phys. Lett. B {\bf 253}, 483 (1991);
M.~Gronau and D.~Wyler, Phys. Lett. B {\bf 265}, 172 (1991).

\bibitem{ADS}
D. Atwood, I. Dunietz, and A. Soni, Phys. Rev. Lett {\bf 78}, 3257 (1997);
Phys. Rev {\bf D 63},  036005 (2001).

\bibitem{GGSZ}
A.~Giri, Y.~Grossman, A.~Soffer, and J.~Zupan,  Phys. Rev. D {\bf 68} (2003).

\bibitem{lhcb_note} 
S. S. Malde, Technical Report, LHCb-PUB-2016-025 (2016).

\bibitem{Kou:2018nap}
E.~Kou {\it et al.} (Belle II Collaboration), PTEP {\bf 2019}, 123C01 (2019).

\bibitem{CLEO(2010)}
J. Libby {\it et al.} (CLEO collaboration), Phys. Rev. D {\bf 82}, 112006 (2010).

\bibitem{T.Evans(2016)}
T. Evans {\it et al.}, Phys. Lett. B {\bf 757}, 520 (2016); Erratum: [Phys.\ Lett.\ B {\bf 765}, 402 (2017)].

\bibitem{CLEO(2012)}
J. Insler {\it et al.} (CLEO Collaboration), Phys. Rev. D {\bf 85}, 092016 (2012).

\bibitem{S.Malde(2015)}
S. Malde {\it et al.}, Phys. Lett. B {\bf 747}, 9 (2015).

\bibitem{CLEO(2012)112001}
D. M. Asner {\it et al.} (CLEO Collaboration), Phys. Rev. D {\bf 86}, 112001 (2012).

\bibitem{K:2017qxf} 
  P.~K.~Resmi, J.~Libby, S.~Malde and G.~Wilkinson,
  %``Quantum-correlated measurements of $D\to K^{0}_{\rm S}\pi^{+}\pi^{-}\pi^{0}$ decays and consequences for the determination of the CKM angle $\gamma$,''
  JHEP {\bf 1801}, 082 (2018).

\bibitem{LHCb(2016)}
R. Aaij {\it et al.} (LHCb Collaboration), Phys. Rev. Lett. {\bf 116}, 241801 (2016).

\bibitem{BES(2014)}
M. Ablikim {\it et al.} (BESIII Collaboration), Phys. Lett. B {\bf 734}, 227 (2014).

\bibitem{Lei(2019)}
  M.~Ablikim {\it et al.},
  %``Model-independent determination of the relative strong-phase difference between $D^0$ and $\bar{D}^0\rightarrow K^0_{S,L}\pi^+\pi^-$ and its impact on the measurement of the CKM angle $\gamma/\phi_3$,''
  arXiv:2003.00091 [hep-ex].

\bibitem{Aaij:2244311} R.~Aaij {\it et al.} (LHCb Collaboration), Technical Report,
%''Expression of Interest for a Phase-II LHCb Upgrade: Opportunities in flavour physics, and beyond, in the HL-LHC era'', 
CERN-LHCC-2017-003 (2017).

\bibitem{D.Atwood(2001)}
D. Atwood, I. Dunietz, and A. Soni, Phys. Rev. D {\bf 63}, 036005 (2001).


\bibitem{M.Nayak2015}
M. Nayak {\it et al.}, Phys. Lett. B {\bf 740}, 1 (2015).


\bibitem{A.Giri(2003)} A. Giri, Y. Grossman, A. Soffer, and J. Zupan,
Phys. Rev. D {\bf 68}, 054018 (2003).


\bibitem{CLEO(2012)122002}
M. Artuso {\it et al.} (CLEO Collaboration), Phys. Rev. D {\bf 85}, 122002 (2012).


\bibitem{Gronau:2001nr}
M.~Gronau, Y.~Grossman and J.~L. Rosner,
%{Measuring D0 - anti-D0 mixing and relative strong phases
%at a charm factory}.
Phys. Lett. B {\bf 508}, 37 (2001).


\bibitem{C.Thomas(2012)}
C. Thomas and G. Wilkinson, JHEP {\bf 1210}, 185 (2012).

\bibitem{S.Malde(2015)094032}
S. Malde, C. Thomas, and G. Wilkinson, Phys. Rev. D {\bf 91}, 094032 (2015).


\bibitem{S.Malde(2011)}
S. Malde and G. Wilkinson, Phys. Lett. B {\bf 701}, 353 (2011).

\bibitem{Aaij:2015xoa} 
  R.~Aaij {\it et al.} (LHCb Collaboration),
  %``Model-independent measurement of mixing parameters in D$^{0}$ ? K$_{S}^{0}$ ?$^{+}$?$^{?}$ decays,''
  JHEP {\bf 1604}, 033 (2016)



\bibitem{vorobiev(2016)}
V. Vorobiev {\it et al.} (Belle Collaboration), Phys. Rev. D {\bf 94}, 052004 (2016).


\bibitem{Chau:1984fp}
L.~L.~Chau and W.~Y.~Keung, Phys. Rev. Lett.  {\bf 53}, 1802 (1984).

\bibitem{Wolfenstein:1983yz}
L.~Wolfenstein, Phys. Rev. Lett.  {\bf 51}, 1945 (1983).

\bibitem{Charles:2004jd}
J.~Charles {\it et al.} (CKMfitter Group), Eur. Phys. J. C {\bf 41}, 1 (2005).

\bibitem{Buras:1994ec}
A.~J.~Buras, M.~E.~Lautenbacher, and G.~Ostermaier, Phys. Rev. D {\bf 50}, 3433 (1994).

\bibitem{Bediaga:2018lhg}
I.~Bediaga {\it et al.} (LHCb Collaboration),
%``Physics case for an LHCb Upgrade II - Opportunities in flavour physics, and beyond, in the HL-LHC era'',
arXiv:1808.08865 [hep-ex].

\bibitem{Ablikim:2015hih}
M.~Ablikim {\it et al.} (\bes3 Collaboration),
%{Measurement of $y_{CP}$ in $D^0-\overline{D}^0$
%oscillation using quantum correlations in $e^+e^-\to
%D^0\overline{D}^0$ at $\sqrt{s}$ = 3.773\,GeV}.
Phys. Lett. B {\bf 744}, 339 (2015).

\bibitem{Atwood:2002ak}
D.~Atwood and A.~A. Petrov,
%{Lifetime differences in heavy mesons with time independent
%measurements}.
Phys. Rev. D {\bf 71}, 054032 (2005).

\bibitem{Asner:2005wf}
D.~M. Asner and W.~M. Sun,
%{Time-independent measurements of D0 - anti-D0 mixing and
%relative strong phases using quantum correlations}.
Phys. Rev. D {\bf 73}, 034024 (2006);
[Erratum: Phys. Rev.D {\bf 77}, 019901 (2008)].


\bibitem{Bigi:2011em}
I.~I. Bigi and A.~Paul,
%{On CP Asymmetries in Two-, Three- and Four-Body D Decays}.
JHEP {\bf 03}, 021 (2012).

\bibitem{Kostelecky:2001ff}
V.~A.~Kostelecky,
%{Formalism for CPT, T and Lorentz violation in neutral meson oscillations}.
Phys. Rev. D {\bf 64}, 076001 (2001).


\bibitem{Barlag:1992ww}
S.~Barlag {\it et al.} (ACCMOR Collaboration), Z. Phys. C {\bf 55}, 383 (1992).

\bibitem{cleo_kspi}
Q. He {\it et al.} (CLEO Collaboration), Phys. Rev. Lett. {\bf 100}, 091801 (2008).

\bibitem{bes3_kspi}
Wenjing Zheng (for the BESIII Collaboration), PoS CHARM2016 (2016) 075.

\bibitem{yufs}
Fusheng Yu, presented at 2016 workshop of BESIII Charm Hadron physics,
Dec. 27-28, 2016, Beijing, China.

\bibitem{glwprime}
T. Gershon, J. Libby, and G. Wilkinson, Phys. Lett.  B {\bf 750}, 338 (2015).

\bibitem{Cazzoli:1975et}
E.~G.~Cazzoli {\it et al.}, Phys. Rev. Lett. {\bf 34},  1125 (1975).

\bibitem{belle_pkpi}
A. Zupanc {\it et al.} (Belle Collaboration), Phys. Rev. Lett. {\bf 113}, 042002  (2014).

\bibitem{plb_bfs}
M. Ablikm {\it et al.} (BESIII Collaboration), Phys. Rev. Lett. {\bf 116},  052001 (2016).

\bibitem{plb_lev}
M. Ablikm {\it et al.} (BESIII Collaboration), Phys. Rev. Lett. {\bf 115}, 221805 (2015).

\bibitem{Ablikim:2016vqd}
M.~Ablikim {\it et al.} (BESIII Collaboration), Phys. Lett. B {\bf 767}, 42  (2017).

\bibitem{prl115_221805}
S. Meinel, Phys. Rev. Lett. {\bf 118}, 082001 (2017).

\bibitem{plb_ppipi}
M. Ablikm {\it et al.} (BESIII Collaboration), Phys. Rev. Lett. {\bf 117},  232002 (2016).

\bibitem{Ablikim:2017ors}
M.~Ablikim {\it et al.} (BESIII Collaboration), Phys. Rev. D {\bf 95}, 111102 (2017).

\bibitem{Ablikim:2016mcr}
M.~Ablikim {\it et al.} (BESIII Collaboration), Phys. Rev. Lett. {\bf 118}, 112001 (2017).

\bibitem{Ablikim:2017iqd}
M.~Ablikim {\it et al.} (BESIII Collaboration), Phys. Lett. B {\bf 772}, 388 (2017).

\bibitem{Lc_XiK}
M.~Ablikim {\it et al.} (BESIII Collaboration), Phys. Lett. B {\bf 783}, 200 (2018).

\bibitem{Ablikim:2018czr} 
  M.~Ablikim {\it et al.} (BESIII Collaboration),
  %``Evidence for the decays of $\Lambda^+_{c}\to\Sigma^+\eta$ and $\Sigma^+\eta^\prime$,''
  Chin.\ Phys.\ C {\bf 43}, 083002 (2019).


\bibitem{Ablikim:2018byv} 
  M.~Ablikim {\it et al.} (BESIII Collaboration),
  %``Measurement of the absolute branching fractions of $\Lambda_{c}^{+}\to\Lambda\eta\pi^{+}$ and $\Sigma(1385)^{+}\eta$,''
  Phys.\ Rev.\ D {\bf 99}, 032010 (2019).

\bibitem{Lc_lambdax}
M.~Ablikim {\it et al.} (BESIII Collaboration), Phys. Rev. Lett. {\bf 121}, 062003 (2018).

\bibitem{Lc_ex}
M.~Ablikim {\it et al.} (BESIII Collaboration), Phys. Rev. Lett. {\bf 121}, 251801 (2018).

\bibitem{epjc76_628}
R. N. Faustov and V. O. Galkina, Eur. Phys. J. C {\bf 76}, 628 (2016).

\bibitem{prc72_032005}
M. Pervin {\it et al.}, Phys. Rev. C {\bf 72}, 035201 (2005).

\bibitem{prd93_014021}
N. Ikeno and E. Oset, Phys. Rev. D {\bf 93}, 014021 (2016).

\bibitem{prd90_114033}
T. Gutsche {\it et al.}, Phys. Rev. D {\bf 90}, 114033 (2014).

\bibitem{prd93_056008}
Cai-Dian L\"u, Wei Wang, and Fu-Sheng Yu, Phys. Rev. D {\bf 93}, 056008 (2016).

\bibitem{Meinel:2017ggx} 
  S.~Meinel,
  %``$\Lambda_c \to N$ form factors from lattice QCD and phenomenology of $\Lambda_c \to n \ell^+ \nu_\ell$ and $\Lambda_c \to p \mu^+ \mu^-$ decays,''
  Phys.\ Rev.\ D {\bf 97}, 034511 (2018).

\bibitem{Wang:2016elx}
D.~Wang, R.~G.~Ping, L.~Li, X.~R.~Lyu, and Y.~H.~Zheng, Chin. Phys. C {\bf 41}, 023106 (2017).

\bibitem{Cheng:2015iom}
H.~Y.~Cheng, Front. Phys. (Beijing) {\bf 10}, 101406 (2015).

\bibitem{Hyodo:2011js}
T.~Hyodo and M.~Oka, Phys. Rev. C {\bf 84}, 035201 (2011).

\bibitem{Miyahara:2015cja}
K.~Miyahara, T.~Hyodo, and E.~Oset, Phys. Rev. C {\bf 92}, 055204 (2015).

\bibitem{Xie:2016evi}
J.~J.~Xie and L.~S.~Geng, Eur. Phys. J. C {\bf 76}, 496 (2016).

\bibitem{Li:2018qak} 
  Y.~B.~Li {\it et al.} (Belle Collaboration),
  %``First Measurements of Absolute Branching Fractions of the $\Xi_c^0$ Baryon at Belle,''
  Phys.\ Rev.\ Lett.\  {\bf 122}, 082001 (2019).
  
  \bibitem{Li:2019atu} 
  Y.~B.~Li {\it et al.} (Belle Collaboration),
  %``First measurements of absolute branching fractions of the $\Xi_c^+$ baryon at Belle,''
    Phys.\ Rev.\ D {\bf 100}, 031101 (2019).
  

%\bibitem{mass} S.~D\"urr \textit{et al.}, Science \textbf{322} (5905),1224 (2008).

%\bibitem{spin} J.~Ashman \textit{et al.}, Phys. Lett. B \textbf{206}, 206 (1988).

%\bibitem{radius} R.~Pohl, Nature \textbf{466} (7303), 364 (2010).

%\bibitem{carlos} C.~Granados \textit{et al.}, Eur. Phys. J. A. \textbf{53}, 117 (2017).

%\bibitem{ffreview} V. Punjabi, C. F. Perdrisat, and M. K. Jones, Eur. Phys. J. A \textbf{51}, 79 (2015).

%\bibitem{theoryFF} S. Pacetti, R. Baldini-Ferroli, and E. Tomasi-Gustafsson, Phys. Rep. \textbf{550-551}, 1 (2015).

%\bibitem{Sakharov} A.~Sakharov, J. Exp. Th. Phys. Lett. \textbf{5}, 24 (1967).

%\bibitem{lhcb} R. Aaij \textit{et al.}, Nature Phys. \textbf{13}, 391 (2017).

%\bibitem{Ablikim:2009ab} M.~Ablikim {\it et al.} (BES Collaboration), Phys. Rev. D {\bf 81}, 012003 (2010).

%\bibitem{theorypolarization} A. Z. Dubnickova {\it et al.}, Nuovo Cim. A {\bf 109}, 241 (1996).

%\bibitem{Faldt:2016qee} G.~F\"aldt, Eur. Phys. J. A {\bf 52}, 141 (2016).

\bibitem{Faldt:2017kgy} G.~F\"aldt and A.~Kupsc, Phys. Lett. B {\bf 772}, 16 (2017).

\bibitem{Faldt:2017lam} G. F\"aldt, Phys. Rev. D {\bf 97}, 053002 (2018).

\bibitem{Ablikim:2019zwe} 
  M.~Ablikim {\it et al.} (BESIII Collaboration),
  %``Measurements of Weak Decay Asymmetries of $\Lambda_c^+\to pK_S^0$, $\Lambda\pi^+$, $\Sigma^+\pi^0$, and $\Sigma^0\pi^+$,''
 Phys.\ Rev.\ D {\bf 100}, 072004 (2019).

\bibitem{Ablikim:2017lct} 
  M.~Ablikim {\it et al.} (BESIII Collaboration),
  %``Precision measurement of the $e^{+}e^{-}~\rightarrow~\Lambda_{c}^{+} \bar{\Lambda}_{c}^{-}$ cross section near threshold,''
  Phys.\ Rev.\ Lett.\  {\bf 120}, 132001 (2018).

\bibitem{stefan}
S. Leupold, private communication (2017).

\bibitem{bigi}
S. Bianco \textit{et al.}, Nuovo Cim. {\bf 26N7}, 1 (2003).

\bibitem{hamann}
N. Hamann {\it et al.}, Technical Report, CERN-SPSLC-92-19 (1992).



\end{thebibliography}

\begin{thebibliography}{10}
\addcontentsline{toc}{chapter}{Bibliography}

\bibitem{Zhu2014}
S.-H.~Zhu,
%{A New Paradigm: Role of Electron-positron and Hadron Colliders}.
arxiv:1410.2042.

%%%psiweak
\bibitem{sanchis_1994} M. A. Sanchis, Z. Phys. C {\bf 62}, 271 (1994).
\bibitem{datta_1999} A. Datta, P. O'Donnell, S. Pakvasa, X.M. Zhang, Phys. Rev. D {\bf 60}, 014011 (1999).
\bibitem{xmzhang_2001} X.M. Zhang, High Energy Phys. Nucl. Phys. {\bf 25}, 461 (2001).
\bibitem{hbli_2012} H.-B. Li and S.-H. Zhu, Chin. Phys. C {\bf 36}, 932 (2012).
\bibitem{hill_1995} C. Hill, Phys. Lett. B {\bf 345}, 483 (1995).
\bibitem{dhir_2009} R. Dhir, R.C. Verma and A. Sharma, Adv. High Energy Phys. {\bf 2013}, 706543  (2013).

\bibitem{ymwang_2008} Y.-M. Wang, H. Zhou, Z.-T. Wei, X.-Q. Li and C.-D. L\"{u}, Eur. Phys. J. C  {\bf 55}, 607 (2008).
\bibitem{ylshen_2008} Y.-L. Shen, Y.-M. Wang, Phy. Rev. D  {\bf 78}, 074012 (2008).

  
\bibitem{sharma_1999} K.K. Sharma, R.C. Verma, Int. J. Mod. Phys. A  {\bf 14}, 937 (1999).


\bibitem{bes_jpsihadronic_2008} M.~Ablikim {\it et al.} (BES Collaboration), Phys. Lett. B {\bf 663}, 297 (2008).

\bibitem{bes3_jpsihadronic_2014} M.~Ablikim {\it et al.} (\bes3 Collaboration), Phys. Rev. D  {\bf 89}, 071101 (2014).


\bibitem{Ivanov:2015woa} 
  M.~A.~Ivanov and C.~T.~Tran,
  %``Exclusive decays J/??D$_{(s)}^{(*)?}?^+?_?$ in a covariant constituent quark model with infrared confinement,''
  Phys.\ Rev.\ D {\bf 92}, 074030 (2015).

\bibitem{bes_jpsileptonic_2006} M.~Ablikim {\it et al.} (BES Collaboration), Phys. Lett. B {\bf 639}, 418 (2006).

\bibitem{bes3_jpsileptonic_2014} M.~Ablikim {\it et al.} (\bes3 Collaboration), Phys. Rev. D
  {\bf 90}, 112014 (2014).

\bibitem{ymwang_2009} Y.-M. Wang {\it et al.}, J. Phys. G {\bf 36}, 105002 (2009).
%%% endpsiweak


% \bibitem{Golowich:2007ka}
% E.~Golowich, J.~Hewett, S.~Pakvasa and A.~A. Petrov,
% %{Implications of $D^0$ - $\bar{D}^0$ Mixing for New Physics}.
% Phys. Rev. D {\bf 76}, 095009 (2007).

% \bibitem{Gedalia:2009kh}
% O.~Gedalia, Y.~Grossman, Y.~Nir and G.~Perez,
% %{Lessons from Recent Measurements of D0 - anti-D0 Mixing}.
% Phys. Rev. D {\bf 80}, 055024 (2009).

% \bibitem{Nir:1993mx}
% Y.~Nir and N.~Seiberg,
% %{Should squarks be degenerate?}
% Phys. Lett. B {\bf 309}, 337 (1993).

% \bibitem{Blum:2009sk}
% K.~Blum, Y.~Grossman, Y.~Nir and G.~Perez,
% %{Combining K0 - anti-K0 mixing and D0 - anti-D0 mixing to
% %constrain the flavor structure of new physics}.
% Phys. Rev. Lett. {\bf 102}, 211802 (2009).

% \bibitem{Isidori:2010kg}
% G.~Isidori, Y.~Nir and G.~Perez,
% %{Flavor Physics Constraints for Physics Beyond the Standard Model}.
% Ann. Rev. Nucl. Part. Sci. {\bf 60}, 355(2010).


\bibitem{Aaij:2017vbb}
R.~Aaij {\it et al.} (LHCb Collaboration),
%{Test of lepton universality with $B^{0} \rightarrow K^{*0}\ell^{+}\ell^{-}$ decays}.
JHEP {\bf 08}, 055 (2017).

\bibitem{Altmannshofer:2017yso}
W.~Altmannshofer, P.~Stangl and D.~M. Straub,
%{Interpreting Hints for Lepton Flavor Universality Violation}.
Phys. Rev. D {\bf 96}, 055008 (2017).

\bibitem{Dorsner:2017ufx}
I.~Dorsner, S.~Fajfer, D.~A. Faroughy and N.~Kosnik,
%{The role of the $S_3$ GUT leptoquark in flavor
%universality and collider searches}.
JHEP {\bf 10}, 188 (2017).

\bibitem{Glashow:2014iga}
S.~L. Glashow, D.~ Guadagnoli and K.~Lane,
%{Lepton Flavor Violation in $B$ Decays?}
Phys. Rev. Lett. {\bf 114}, 091801(2015).

\bibitem{Burdman:1995te}
G.~Burdman, E.~Golowich, J.~L.~Hewett and S.~Pakvasa,
%{Radiative weak decays of charm mesons}.
Phys. Rev. D {\bf 52}, 6383 (1995).

\bibitem{Greub:1996wn}
C.~Greub, T.~Hurth, M.~Misiak and D.~Wyler,
%{The c $\to$ u gamma contribution to weak radiative charm decay}.
Phys. Lett. B {\bf 382}, 415 (1996).

\bibitem{Burdman:2001tf}
G.~Burdman, E.~Golowich, J.~L.~Hewett and S.~Pakvasa,
%{Rare charm decays in the standard model and beyond}.
Phys. Rev. D {\bf 66}, 014009 (2002).

\bibitem{deBoer:2018buv}
S.~de~Boer and G.~Hiller,
%{Null tests from angular distributions in $D \to P_1 P_2
%l^+l^-$, $l=e,\mu$ decays on and off peak}.
Phys. Rev. D {\bf 98}, 035041 (2018).


\bibitem{Fajfer:2000zx}
S.~Fajfer, S.~Prelovsek, P.~Singer and D.~Wyler,
%{A Possible arena for searching new physics: The Gamma (D0
%$\to$ rho0 gamma) / Gamma (D0 $\to$ omega gamma) ratio}.
Phys. Lett. B {\bf 487}, 81 (2000).

\bibitem{Isidori:2012yx}
G.~Isidori and J.~F.~Kamenik,
%{Shedding light on CP violation in the charm system via D
%to V gamma decays}.
Phys. Rev. Lett. {\bf 109}, 171801 (2012).

\bibitem{deBoer:2017que}
S.~de~Boer and G.~Hiller,
%{Rare radiative charm decays within the standard model and beyond}.
JHEP {\bf 08}, 091 (2017).

\bibitem{deBoer:2018zhz}
S.~de~Boer and G.~Hiller,
%{The photon polarization in radiative $D$ decays, phenomenologically}.
Eur. Phys. J. C {\bf 78}, 188 (2018).

\bibitem{Amhis:2016xyh}
Y.~Amhis {\it et al.} (Heavy Flavor Averaging Group),
% {Averages of $b$-hadron, $c$-hadron and $\tau$-lepton properties as  of summer 2016}.
Eur. Phys. J. C {\bf 77}, 895 (2017).

\bibitem{Golowich:2009ii}
E.~Golowich, J.~Hewett, S.~Pakvasa and A.~A.~Petrov,
%{Relating D0-anti-D0 Mixing and D0 $\to$ l+ l- with New Physics}.
Phys. Rev. D {\bf 79}, 114030 (2009).

\bibitem{Paul:2012ab}
A.~Paul, A.De~La~Puente and I.~I.~Bigi,
%{Manifestations of warped extra dimension in rare charm
%decays and asymmetries}.
Phys. Rev. D {\bf 90}, 014035 (2014).

\bibitem{Fajfer:2001sa}
S.~Fajfer, Sasa Prelovsek and P.~Singer,
%{Rare charm meson decays D $\to$ P lepton+ lepton- and c
%$\to$ u lepton+ lepton- in SM and MSSM}.
Phys. Rev. D {\bf 64}, 114009 (2001).

\bibitem{Fajfer:2012nr}
S.~Fajfer and N.~Kosnik,
%{Resonance catalyzed CP asymmetries in D $\to$ P$l^+l^-$}.
Phys. Rev. D {\bf 87}, 054026 (2013).

\bibitem{deBoer:2015boa}
S.~de~Boer and G.~Hiller,
%{Flavor and new physics opportunities with rare charm
%decays into leptons}.
Phys. Rev. D {\bf 93}, 074001 (2016).


%%% DRaredecay
%\bibitem{cleo_D2hee_1996} CLEO Collaboration, Phys. Rev. Lett. 76, 3065-3069 (1996).
\bibitem{TheBESIIICollaboration2018a} M.~Ablikim {\it et al.}  (\bes3 Collaboration), Phys. Rev. D {\bf 97}, 072015 (2018).
\bibitem{E791_DTohee_2001}   E.~M.~Aitala {\it et al.} (E791 Collaboration), Phys. Rev. Lett. {\bf 86}, 3969 (2001).
\bibitem{babar_D2hee_2011} J.~P.~Lees {\it et al.} (BaBar Collaboration), Phys. Rev. D {\bf 84}, 072006 (2011).
\bibitem{belle_BTohvv_2007} K. F. Chen {\it et al.} (Belle Collaboration),  Phys. Rev. Lett. {\bf 99}, 221802 (2007).
\bibitem{babar_B0Togammavv_2004}  J.~P.~Lees {\it et al.} (BaBar Collaboration), Phys. Rev. Lett. {\bf  93}, 091802 (2004).
  %%% endDRaredecay


\bibitem{Badin:2010uh}
A.~Badin and A.~A.~Petrov,
%{Searching for light Dark Matter in heavy meson decays}.
Phys. Rev. D {\bf 82}, 034005 (2010).

%%cpv
\bibitem{Sakharov:1967dj}
A.~D. Sakharov,
Pisma Zh. Eksp. Teor. Fiz.  {\bf 5}, 32 (1967);
Usp. Fiz. Nauk  {\bf 161}, 61 (1991).


\bibitem{Morrissey:2012db} 
  D.~E.~Morrissey and M.~J.~Ramsey-Musolf,
  %``Electroweak baryogenesis,''
  New J.\ Phys.\  {\bf 14}, 125003 (2012).


\bibitem{ref::pdg2016} M. Tanabashi {\it et al.} (Particle Data Group), Phys. Rev. D, {\bf 98}, 030001 (2018).

\bibitem{Aaij:2019kcg} 
  R.~Aaij {\it et al.} (LHCb Collaboration),
  %``Observation of $C\!P$ violation in charm decays,''
Phys.\ Rev.\ Lett.\  {\bf 122}, 211803 (2019).

\bibitem{Chauvat:1985fb}
  P.~Chauvat {\it et al.} (R608 Collaboration),
  %``Test of {CP} Invariance in $\Lambda^0$ Decay,''
  Phys.\ Lett.\  B {\bf 163}, 273 (1985).
%%  doi:10.1016/0370-2693(85)90236-9

%\cite{Barnes:1996si}
\bibitem{Barnes:1996si}
  P.~D.~Barnes {\it et al.},
  %``Observables in high statistics measurements of the reaction anti-p p --> anti-Lambda Lambda,''
  Phys.\ Rev.\ C {\bf 54}, 1877 (1996).

%\cite{Tixier:1988fv}
\bibitem{Tixier:1988fv}
  M.~H.~Tixier {\it et al.} (DM2 Collaboration),
  %``Looking at {CP} Invariance and Quantum Mechanics in $J/\psi \to \Lambda \bar{\Lambda}$ Decay,''
  Phys.\ Lett.\ B {\bf 212}, 523 (1988).


\bibitem{Ablikim:2009ab}
M.~Ablikim {\it et al.} (\bes3 Collaboration),
%{Measurement of the asymmetry parameter for the decay
%$\bar\Lambda \to \bar p\pi^+$}.
Phys. Rev. D  {\bf 81}, 012003 (2010).

%\cite{Donoghue:1985ww}
\bibitem{Donoghue:1985ww}
  J.~F.~Donoghue and S.~Pakvasa,
  %``Signals of {CP} Nonconservation in Hyperon Decay,''
  Phys.\ Rev.\ Lett.\  {\bf 55}, 162 (1985).


\bibitem{Donoghue:1986hh}
J.~F. Donoghue, X.-G.~He and S.~Pakvasa,
%{Hyperon Decays and CP Nonconservation}.
Phys. Rev. D {\bf 34}, 833 (1986).

\bibitem{Tandean:2002vy} 
  J.~Tandean and G.~Valencia,
  %``CP violation in hyperon nonleptonic decays within the standard model,''
  Phys.\ Rev.\ D {\bf 67}, 056001 (2003).

\bibitem{He:1995na}
X.-G.~He and G.~Valencia,
%{CP violation in Lambda $\to$ p pi- beyond the Standard Model}.
Phys. Rev. D {\bf 52}, 5257 (1995).

\bibitem{luk98} K.~B.~Luk, arXiv:hep-ex/9803002.

\bibitem{Tandean:2003fr} 
  J.~Tandean,
  %``New physics and CP violation in hyperon nonleptonic decays,''
  Phys.\ Rev.\ D {\bf 69}, 076008 (2004).

\bibitem{Cheng:1991sn}
H.-Y. Cheng and B.~Tseng,
%{Nonleptonic weak decays of charmed baryons}.
Phys. Rev D  {\bf 46}, 1042 (1992);
[Erratum: Phys. Rev. D  {\bf 55},1697 (1997)].

\bibitem{Cheng:2018hwl}
H.-Y.~Cheng, X.-W.~Kang and F.~Xu,
%{Singly Cabibbo-suppressed hadronic decays of $\Lambda_c^+$}.
Phys. Rev. D  {\bf 97}, 074028 (2018).

  % %\cite{Faldt:2017kgy}
% \bibitem{Faldt:2017kgy}
%   G.~F$\ddot{\textrm{a}}$ldt and A.~Kupsc,
%   %``Hadronic structure functions in the $e^+ e^- \rightarrow \bar{\Lambda} \Lambda$ reaction,''
%   Phys.\ Lett.\ B {\bf 772}, 16 (2017).

% \bibitem{DataSet8} G. F$\ddot{\textrm{a}}$ldt, Eur. Phys. J. A {\bf52}, 141 (2016);
% G. F$\ddot{\textrm{a}}$ldt, Eur. Phys. J. A {\bf51}, 74 (2015).



\bibitem{Liu:2015qra}
J. Liu, R.-G. Ping and H.-B. Li,
 J. Phys. G  {\bf 42}, 095002 (2015).

\bibitem{Ablikim:2018zay} 
  M.~Ablikim {\it et al.} (BESIII Collaboration),
  %``Polarization and Entanglement in Baryon-Antibaryon Pair Production in Electron-Positron Annihilation,''
  Nature Phys.\  {\bf 15}, 631 (2019).

\bibitem{Bigi:2017eni}
I.~I. Bigi, X.-W. Kang and H.-B. Li,
Chin. Phys. C {\bf 42}, 013101 (2018).

\bibitem{Valencia:1988it}
G.~Valencia,
Phys. Rev. D {\bf 39}, 3339 (1989).

\bibitem{Bigi:2000yz}
I.~I. Bigi and A.~I. Sanda,
%CP violation
%2000,
%Camb.
%Monogr.
Part. Phys. Nucl. Phys. Cosmol.  {\bf 9}, 1 (2009).

\bibitem{Kang:2009iy}
X.-W. Kang and H.-B. Li,
Phys. Lett. B {\bf 684}, 137 (2010).

\bibitem{Kang:2010td}
X.-W. Kang, H.-B. Li, G.-R. Lu and A.~Datta,
Int. J. Mod. Phys. A {\bf 26}, 2523 (2011).

\bibitem{Duraisamy:2013kcw}
M.~Duraisamy and A.~Datta,
JHEP {\bf 09}, 005(2013).

\bibitem{Bensalem:2002pz}
W.~Bensalem, A.~Datta and D.~London,
Phys. Lett. B  {\bf 538}, 309 (2002).

\bibitem{Datta:2003mj}
A.~Datta and D.~London,
Int. J. Mod. Phys. A {\bf 19}, 2505 (2004).

\bibitem{Gronau:2011cf}
M.~Gronau and J.~L. Rosner,
Phys. Rev. D {\bf 84}, 096013 (2011).

\bibitem{Gronau:2015gha}
M.~Gronau and J.~L. Rosner,
Phys. Lett. B {\bf 749}, 104 (2015).

\bibitem{DiSalvo:2012vd}
E.~Di~Salvo and Z.~J. Ajaltouni,
Mod. Phys. Lett. A  {\bf 28},135004 (2013).

\bibitem{Ajaltouni:2012zg}
Z.~J. Ajaltouni and E.~Di~Salvo,
Int. J. Mod. Phys. A  {\bf 27}, 1250086(2012).

%\cite{He:1992ng}
\bibitem{He:1992ng}
  X.~G.~He, J.~P.~Ma and B.~McKellar,
  %``CP violation in J / psi $\to$ Lambda anti-Lambda,''
  Phys.\ Rev.\ D {\bf 47}, R1744 (1993).

%\cite{He:1993ar}
\bibitem{He:1993ar}
  X.~G.~He, J.~P.~Ma and B.~McKellar,
  %``CP violation in fermion pair decays of neutral boson particles,''
  Phys.\ Rev.\ D {\bf 49}, 4548 (1994).


\bibitem{Adlarson:2019jtw} 
  P.~Adlarson and A.~Kupsc,
  %``CP symmetry tests in the cascade-anticascade decay of charmonium,''
  arXiv:1908.03102 [hep-ph].


%%% zmg
\bibitem{ref::marshak} R. N. Mohapatra and R. E. Marshak, Phys. Rev. Lett. {\bf 44}, 1316 (1980).
\bibitem{ref::Luk} K.-B. Luk, in International Workshop on the Search for Baryon and Lepton Number Violations, LBNL, 2007.
\bibitem{ref::prd81-051901r} X. W. Kang, H. B. Li and G. R. Lu, Phys. Rev. D {\bf 81}, 051901(R) (2010).
%%% endzmg

\bibitem{Li2016}
H.-B. Li,
%{Prospects for rare and forbidden hyperon decays at \bes3}.
Frontiers of Physics {\bf 12}, 121301 (2017).
%
%\bibitem{Gronau:2001nr}
%M.~Gronau, Y.~Grossman and J.~L. Rosner,
%%{Measuring D0 - anti-D0 mixing and relative strong phases
%%at a charm factory}.
%Phys. Lett. B {\bf 508}, 37 (2001).
%


%%% clfv
\bibitem{lfv1} S. Dimopoulos and H. Georgi, Nucl.\ Phys.\ B  {\bf 193}, 150 (1981); N. Sakai, Z. Phys. C {\bf 11}, 153 (1981).
\bibitem{lfv2} F. Borzumati and A. Masiero, Phys. Rev. Lett.  {\bf 57}, 961 (1986).
\bibitem{lfv3} M. Dine, Y. Nir and Y. Shirman, Phys. Rev. D  {\bf 55}, 1501 (1997); S. L. Dubovsky and D. S. Gorbunov, Phys. Lett. B {\bf 419}, 223 (1998).
\bibitem{lfv4} R. Kitano and K. Yamamoto, Phys. Rev. D {\bf 62}, 073007 (2000).
\bibitem{lfv5} J. E. Kim and D. G. Lee, Phys. Rev. D {\bf 56}, 100 (1997); K. Huitu {\it et al.}, Phys. Lett. B {\bf 430}, 355 (1998);  A. Faessler {\it et al.}, Nucl.\ Phys.\ B  {\bf  587}, 25 (2000); M. Chaichian and K. Huitu, Phys. Lett. B  {\bf 384}, 157 (1996).
\bibitem{lfv6} J. Bernabeu, E. Nardi and D. Tommasini, Nucl.\ Phys.\ B  {\bf  409}, 69 (1993).
\bibitem{lfv7} S. Coleman and S. L. Glashow, Phys. Rev. D {\bf 59}, 116008 (1999).
\bibitem{lfv14} J. Z. Bai {\it et al.} (BES Collaboration), Phys. Lett. B {\bf 561}, 49 (2003)


\bibitem{TheBESIIICollaboration2013}
M.~Ablikim {\it et al.}  (\bes3 Collaboration), %{Search for the lepton flavor violation process Jpsi to emu at \bes3}.
Phys. Rev. D {\bf 87}, 112007 (2013).

\bibitem{lfv15} M. Ablikim {\it et al.} (\bes3 Collaboration), Phys. Lett. B {\bf 598}, 172 (2004).
%%% endclfv

\bibitem{Aaij:2015qmj}
  R.~Aaij {\it et al.} (LHCb Collaboration),
  %``Search for the lepton-flavour violating decay $D^0 \to e^\pm\mu^\mp$,''
  Phys.\ Lett.\ B {\bf 754}, 167 (2016).

\bibitem{Hazard:2016fnc}
D.~E. Hazard and A.~A. Petrov,
%{Lepton flavor violating quarkonium decays}.
Phys. Rev. D {\bf 94}, 074023 (2016).

\bibitem{Celis:2014asa}
A.~Celis, V.~Cirigliano and E.~Passemar,
%{Model-discriminating power of lepton flavor violating
%$\tau$ decays}.
Phys. Rev. D {\bf 89}, 095014 (2014).

\bibitem{Hazard:2017udp}
D.~E.~Hazard and A.~A.~Petrov,
%{Radiative lepton flavor violating B, D and K decays}.
Phys.Rev. D {\bf 98}, 015027 (2018).

\bibitem{Raidal:2008jk}
M.~Raidal {\it et al.},
%{Flavour physics of leptons and dipole moments}.
Eur. Phys. J. C  {\bf 57}, 13 (2008).

\bibitem{Petrov:2013vka}
A.~A.~Petrov and D.~V.~Zhuridov,
%{Lepton flavor-violating transitions in effective field
%theory and gluonic operators}.
Phys. Rev. D {\bf 89}, 033005 (2014).

\bibitem{Nussinov:2000nm}
S.~Nussinov, R.~D. Peccei and X.~M. Zhang,
%{On unitarity based relations between various lepton family
%violating processes}.
Phys. Rev. D {\bf 63}, 016003 (2001).

\bibitem{Dreiner:2001kc}
H.~K. Dreiner, G.~Polesello and M.~Thormeier,
%{Bounds on broken R parity from leptonic meson decays}.
Phys. Rev. D {\bf 65}, 115006 (2002).

\bibitem{Dreiner:2006gu}
H.~K. Dreiner, M.~Kramer and B.~O'Leary,
Phys. Rev. D  {\bf 75}, 114016 (2007).

\bibitem{Sun:2012yq}
K.-S.~Sun, T.-F.~Feng, T.-J.~Gao and S.-M.~Zhao,
Nucl. Phys. B  {\bf 865}, 486 (2012).

\bibitem{Abada:2015zea}
A.~Abada, D.~Becirevic, M.~Lucente and O.~Sumensari,
Phys. Rev. D {\bf 91}, 113013 (2015).

\bibitem{Black:2002wh}
D.~Black, T.~Han, H.-J.~He and M.~Sher,
Phys. Rev. D {\bf 66}, 053002 (2002).

\bibitem{Becirevic:2013bsa}
D.~Becirevic, G.~Duplancic, B.~Klajn, B.~Melic and F.~Sanfilippo,
%{Lattice QCD and QCD sum rule determination of the decay constants of $\eta_c$, J/$\psi$ and $h_c$ states}.
Nucl. Phys. B {\bf 883}, 306 (2014).

\bibitem{Godfrey:2015vda}
S.~Godfrey and H.~E. Logan,
%{Probe of new light Higgs bosons from bottomonium $\chi_{b0}$ decay}.
Phys. Rev. D {\bf 93}, 055014 (2016).

%%%%dark sector
\bibitem{Essig2013}
R.~Essig {\it et al.},
%{Dark Sectors and New, Light, Weakly-Coupled Particles}.
arXiv:1311.0029.

\bibitem{Alexander2016}
J.~Alexander {\it et al.}, arXiv:1608.08632.
%{Dark Sectors 2016 Workshop: Community Report}.


%%%darksec1
\bibitem{ds2} O.~Adriani {\it et al.} (PAMELA Collaboration), Nature {\bf 458}, 607 (2009).

\bibitem{ds3} J. Chang  {\it et al.}, Nature {\bf 456}, 362 (2008).
\bibitem{ds4} A. A. Abdo {\it et al.} (Fermi-LAT Collaboration), Phys. Rev. Lett.  {\bf 102}, 181101 (2009).
%%% enddarksec1

\bibitem{TheAMSCollaboration2013}
M.~Aguilar {\it et al.} (AMS Collaboration),
Phys. Rev. Lett.  {\bf 110}, 141102 (2013).

\bibitem{Arkani-Hamed2009}
N.~Arkani-Hamed, D.~Finkbeiner, T.~Slatyer and N.~Weiner,
Phys. Rev. D {\bf 79}, 015014 (2009).

\bibitem{Pospelov2009}
M.~Pospelov and A.~Ritz,
Phys. Lett. B  {\bf 671}, 391 (2009).

%%%darksec2
\bibitem{ds8} N. Arkani-Hamed and N. Weiner, JHEP {\bf 0812}, 104 (2008).
\bibitem{ds9} C. Cheung, J. T. Ruderman, L.-T. Wang and I. Yavin, Phys. Rev. D {\bf 80}, 035008 (2009).
%%%enddarksec2


\bibitem{Aditya:2012ay}
Y.~G. Aditya, K.~J. Healey and A.~A. Petrov,
Phys. Lett. B {\bf 710}, 118 (2012).

\bibitem{Bird:2006jd}
C.~Bird, R.~V. Kowalewski and M.~Pospelov,
%{Dark matter pair-production in b $\to$ s transitions}.
Mod. Phys. Lett. A  {\bf 21}, 457 (2006).

\bibitem{Calibbi:2016hwq}
L.~Calibbi, F.~Goertz, D.~Redigolo, R.~Ziegler and J.~Zupan,
%{Minimal axion model from flavor}.
Phys. Rev. D  {\bf 95}, 095009 (2017).

\bibitem{Li2010}
H.-B. Li and T.~Luo,
%{Probing dark force at BES-III/BEPCII}.
Phys. Lett. B {\bf 686}, 249 (2010).


\bibitem{Yin2009}
P.-F.~Yin, J.~Liu and S.-H.~Zhu,
%{Detecting light leptophilic gauge boson at \bes3 detector}.
Phys. Lett. B   {\bf 679}, 362 (2009).

\bibitem{Chang:1997tq}
L.~N. Chang, O.~Lebedev and J.~N. Ng,
Phys. Lett. B {\bf 441}, 419 (1998).

\bibitem{Yeghiyan:2009xc}
G.~K. Yeghiyan.
Phys. Rev. D {\bf 80}, 115019 (2009).

\bibitem{McElrath:2005bp}
B.~McElrath,
Phys. Rev. D {\bf 72}, 103508 (2005).

\bibitem{Fayet:2009tv}
P.~Fayet,
Phys. Rev. D {\bf 81}, 054025 (2010).

\bibitem{Fernandez:2014eja}
N.~Fernandez, J.~Kumar, I.~Seong and P.~Stengel,
Phys. Rev. D {\bf 90}, 015029 (2014).

\bibitem{Fernandez:2015klv}
N.~Fernandez, I.~Seong and P.~Stengel,
%{Constraints on Light Dark Matter from Single-Photon Decays
%of Heavy Quarkonium}.
Phys. Rev. D {\bf 93}, 054023 (2016).

\bibitem{Aubert:2009ae}
B.~Aubert {\it et al.},
%{A Search for Invisible Decays of the Upsilon(1S)}.
Phys. Rev. Lett.  {\bf 103}, 251801 (2009).


\bibitem{Goobar:2006xz}
A.~Goobar, S.~Hannestad, E.~Mortsell and H.~Tu,
JCAP   {\bf 0606}, 019(2006).

\bibitem{Bhattacharya:2018} B. Bhattacharya, C. Grant and A. Petrov,
  arXiv:1809.04606.


\bibitem{Aliev:1996sk}
T.~M. Aliev, A.~Ozpineci and M.~Savci,
%{Rare B $\to$ neutrino anti-neutrino gamma decay in light cone QCD sum rule}.
Phys. Lett. B  {\bf 393}, 143 (1997).

\bibitem{Lai:2016uvj}
Y.~T. Lai {\it et al.},
%{Search for $D^{0}$ decays to invisible final states at Belle}.
Phys. Rev. D  {\bf 95}, 011102 (2017).

\bibitem{McKeen:2009rm}
D.~McKeen,
%{WIMPless Dark Matter and Meson Decays with Missing Energy}.
Phys. Rev. D  {\bf 79}, 114001 (2009).

\bibitem{Dreiner:2009er}
H.~K. Dreiner, S.~Grab, D.~Koschade, M.~Kramer, B.~O'Leary and U.~Langenfeld,
%{Rare meson decays into very light neutralinos}.
Phys. Rev. D  {\bf 80}, 035018 (2009).

\bibitem{Reece2009}
M.~Reece and L.-T.~Wang,
%{Searching for the light dark gauge boson in GeV-scale experiments}.
JHEP   {\bf 07}, 051 (2009).



\bibitem{Ablikim2013}
 M.~Ablikim {\it et al.} (\bes3 Collaboration),
%{Search for $\eta$ and $\eta^{'}$ invisible decays in J/$\psi\to \phi \eta$ and $\phi \eta^{'}$}.
Phys. Rev. D   {\bf 87}, 012009 (2013).


\bibitem{Ablikim:2018liz}
M.~Ablikim {\it et al.} (\bes3 Collaboration),
%{Search for invisible decays of $\omega$ and $\phi$ with $J/\psi$ data at \bes3}.
Phys. Rev. D  {\bf 98},  2001 (2018).

%%% offres
\bibitem{Khodjamirian:2015dda}
  A.~Khodjamirian, T.~Mannel and A.~A.~Petrov,
  JHEP {\bf 1511}, 142 (2015).


% \bibitem{deBoer:2016dcg}
%   S.~de Boer, B.~Muller and D.~Seidel,
%   JHEP {\bf 1608}, 091 (2016)


% \bibitem{Grinstein:2015aua}
%   B.~Grinstein and J.~M.~Camalich,
%   Phys.\ Rev.\ Lett.\  {\bf 116}, no. 14, 141801 (2016)
% %%% endoffres


\bibitem{Solodov:2017pyu}
E.~P. Solodov {\it et al.}
% %{Study of the e+e- $\to$ hadrons reactions with CMD-3
% %detector at VEPP-2000 collider}.
{\em PoS}, EPS-HEP2017:407, 2017.

% \bibitem{Patrignani:2016xqp}
% C.~Patrignani {\it et al.}
% %{Review of Particle Physics}.
% {\em Chin. Phys.}, C40(10):100001, 2016.

%\bibitem{Grinstein:2015aua}
%Benjamin Grinstein and J.~Martin~Camalich.
%{Weak Decays of Excited B Mesons}.
%Phys. Rev. Lett. 116(14):141801, 2016.

\bibitem{Zhu2007}
S.-H.~Zhu,
%{U boson at the BES III detector}.
Phys. Rev. D   {\bf  75}, 1 (2007).

\bibitem{Fayet2007}
P.~Fayet,
%{U-boson production in e+e- annihilations, $\psi$ and
%$\Upsilon$ decays and light dark matter}.
Phys. Rev. D   {\bf 75}, 115017 (2007).

\bibitem{Essig2009}
R.~Essig, P.~Schuster and N.~Toro,
%{Probing dark forces and light hidden sectors at low-energy
%e+e- colliders}.
Phys. Rev. D {\bf 80}, 42 (2009).

\bibitem{Bjorken2009a}
J.~Bjorken, R.~Essig, P.~Schuster and N.~Toro,
%{New fixed-target experiments to search for dark gauge forces}.
Phys. Rev. D   {\bf 80}, 1 (2009).

%%%darksec3
\bibitem{ds15} B. Aubert {\it et al.} (BaBar Collaboration),
  Phys. Rev. Lett.  {\bf 103}, 081803 (2009).
\bibitem{ds17} B. O'Leary {\it et al.} (SuperB Collaboration), arXiv:1008.1541.
\bibitem{Ablikim:2017aab} M. Ablikim {\it et al.} (\bes3 Collaboration), Phys. Lett. B {\bf 774}, 252 (2017).
%%% enddarksec3


\bibitem{BABAR2014}
J.~P.~Lees {\it et al.} (BaBar Collaboration),
%{Search for a dark photon in e+e- collisions at BABAR}.
Phys. Rev. Lett.   {\bf 113}, 201801 (2014).

\bibitem{Baumgart2009}
M.~Baumgart, C.~Cheung, J.~T.~Ruderman, L.-T.~Wang and I.~Yavin.
%{Non-abelian dark sectors and their collider signatures}.
JHEP   {\bf 04}, 014 (2009).

\bibitem{Ablikim2012}
 M.~Ablikim {\it et al.} (\bes3 Collaboration),
%{Search for a light exotic particle in J/$\psi$ radiative decays}.
Phys. Rev. D   {\bf 85}, 092012 (2012).

\bibitem{TheBESIIICollaboration2016}
 M.~Ablikim {\it et al.} (\bes3 Collaboration),
%{Search for a light CP-odd Higgs boson in radiative decays of JPsi}.
Phys. Rev. D {\bf 93}, 052005 (2016).

\bibitem{Peccei}
 R.~D.~Peccei and H.~Quinn, Phys. Rev. Lett. {\bf 38}, 1440 (1977); Phys. Rev. D {\bf 16}, 1791 (1977). 

\bibitem{Weinberg}
  S.~Weinberg, Phys. Rev. Lett. {\bf 40}, 223 (1978);
  F.~Wilczek, Phys. Rev. Lett. {\bf 40}, 279 (1978).

\bibitem{Izaguirre2017}
E.~Izaguirre, T.~Lin and B.~Shuve,
Phys. Rev. Lett. {\bf 118}, 111802 (2017).


\bibitem{Dolan}
  M.~J. Dolan, T.~Ferber, C.~Hearty, F.~Kahlhoefer and K.~Schmidt-Hoberg,
  JHEP {\bf 12}, 92 (2017).
  
\bibitem{Pietro}  G.~D.~Pietro,
  arXiv:1808:00776. 
  
\bibitem{Masso}
  E.~Masso and R.~Toldra,
  Phys. Rev. D {\bf 52}, 1755 (1995).


\bibitem{Golowich:1990ki}
E.~Golowich and P.~B. Pal,
Phys. Rev. D {\bf 41}, 3537 (1990).


\bibitem{Hewett:1997ce}
J.~L. Hewett and T.~G. Rizzo,
Phys. Rev. D {\bf 56}, 5709 (1997).

\bibitem{DeRujula:1977wse}
A.~De~Rujula, R.~C. Giles and R.~L. Jaffe,
Phys. Rev. D {\bf 17}, 285 (1978).

\bibitem{Slansky:1981tg}
R.~Slansky, J.~T. Goldman and G.~L. Shaw,
Phys. Rev. Lett. {\bf 47}, 887 (1981).

\bibitem{Caldi:1982dj}
D.~G. Caldi and S.~Nussinov,
Phys. Rev. D {\bf 28}, 3138 (1983).

\bibitem{Perl:2009zz}
  M.~L.~Perl, E.~R.~Lee and D.~Loomba,
  Ann.\ Rev.\ Nucl.\ Part.\ Sci.\  {\bf 59}, 47 (2009).


% \bibitem{Ackerstaff:1998si}
% K.~Ackerstaff {\it et al.} (OPAL Collaboration),
% Phys. Lett. B {\bf 433}, 195 (1998).

\bibitem{Bowcock:1989qj}
T.~J.~V. Bowcock {\it et al.} (CLEO Collaboration),
%{Search for the Production of Fractionally Charged
%Particles in $e^+ e^-$ Annihilations at $\sqrt{s}=10$.5 {GeV}}.
Phys. Rev. D {\bf 40}, 263 (1989).

\bibitem{Akers:1995az}
  R.~Akers {\it et al.} (OPAL Collaboration),
  Z.\ Phys.\ C {\bf 67}, 203 (1995).


% \bibitem{Abbiendi:2003yd}
% G.~Abbiendi {\it et al.} (OPAL Collaboration),
% Phys. Lett. B {\bf 572}, 8 (2003).


% \bibitem{Abreu:1996py}
% P.~Abreu {\it et al.} (DELPHI Collaboration),
% Phys. Lett. B {\bf 396}, 315 (1997).

% \bibitem{Buskulic:1992mr}
% D.~Buskulic {\it et al.} (ALEPH Collaboration),
% %{Search for particles with unexpected mass and charge in Z decays}.
% Phys. Lett. B {\bf 303}, 198 (1993).

% \bibitem{Abe:1992vr}
% F.~Abe {\it et al.} (CDF Collaboration),
% %{Limits on the production of massive stable charged particles}.
% Phys. Rev. D {\bf 46}, 1889 (1992).

% \bibitem{Acosta:2002ju}
% D.~Acosta {\it et al.}(CDF Collaboration),
% %{Search for long-lived charged massive particles in
% %$\bar{p}p$ collisions at $\sqrt{s} = 1.8$ TeV}.
% Phys. Rev. Lett. {\bf  90}, 131801 (2003).


\bibitem{Rolke2005}
W.~A. Rolke, A.~M. L{\'{o}}pez and J.~Conrad,
%{Limits and confidence intervals in the presence of
%nuisance parameters}.
Nucl. Instrum. Methods Phys. Res., Sect. A {\bf 551}, 493 (2005).

\end{thebibliography}

\begin{thebibliography}{9}
\addcontentsline{toc}{chapter}{Bibliography}
\bibitem{Ablikim:2013mio} 
M.~Ablikim {\it et al.} (BESIII Collaboration),
  %``Observation of a Charged Charmoniumlike Structure in $e^+e^-$ ? $?^+?^-$ J/? at $\sqrt{s}$ =4.26??GeV,''
  Phys.\ Rev.\ Lett.\  {\bf 110}, 252001 (2013).
  
  \bibitem{Ablikim:2013wzq} 
  M.~Ablikim {\it et al.} (BESIII Collaboration),
  %``Observation of a Charged Charmoniumlike Structure $Z_c$(4020) and Search for the $Z_c$(3900) in $e^+e^- \to ?^+?^-h_c$,''
  Phys.\ Rev.\ Lett.\  {\bf 111}, 242001 (2013).
  
  \bibitem{Ablikim:2013emm} 
  M.~Ablikim {\it et al.} (BESIII Collaboration),
  %``Observation of a charged charmoniumlike structure in $e^+e^- \to (D^{*} \bar{D}^{*})^{\pm} \pi^\mp$ at $\sqrt{s}=4.26$GeV,''
  Phys.\ Rev.\ Lett.\  {\bf 112}, 132001 (2014).
  
  \bibitem{zc3885}
  M.~Ablikim {\it et al.} (BESIII Collaboration),
  %``Observation of a charged (DD*bar)- mass peak in e+e- --> pi+ (DD*bar)- at Ecm=4.26 GeV,''
  Phys.\ Rev.\ Lett.\  {\bf 112}, 022001 (2014).
  
 \bibitem{PDG}
 M.~Tanabashi {\it et al.} (Particle Data Group),
  %``Review of Particle Physics,''
  Phys.\ Rev.\ D {\bf 98},  030001 (2018).
  
  \end{thebibliography}
